M. I. Mishchenko, V. K. Rosenbush, N. N. Kiselev,
D. F. Lupishko, V. P. Tishkovets, V. G. Kaydash,
I. N. Belskaya, Y. S. Efimov, N. M. Shakhovskoy

# POLARIMETRIC REMOTE SENSING OF SOLAR SYSTEM OBJECTS

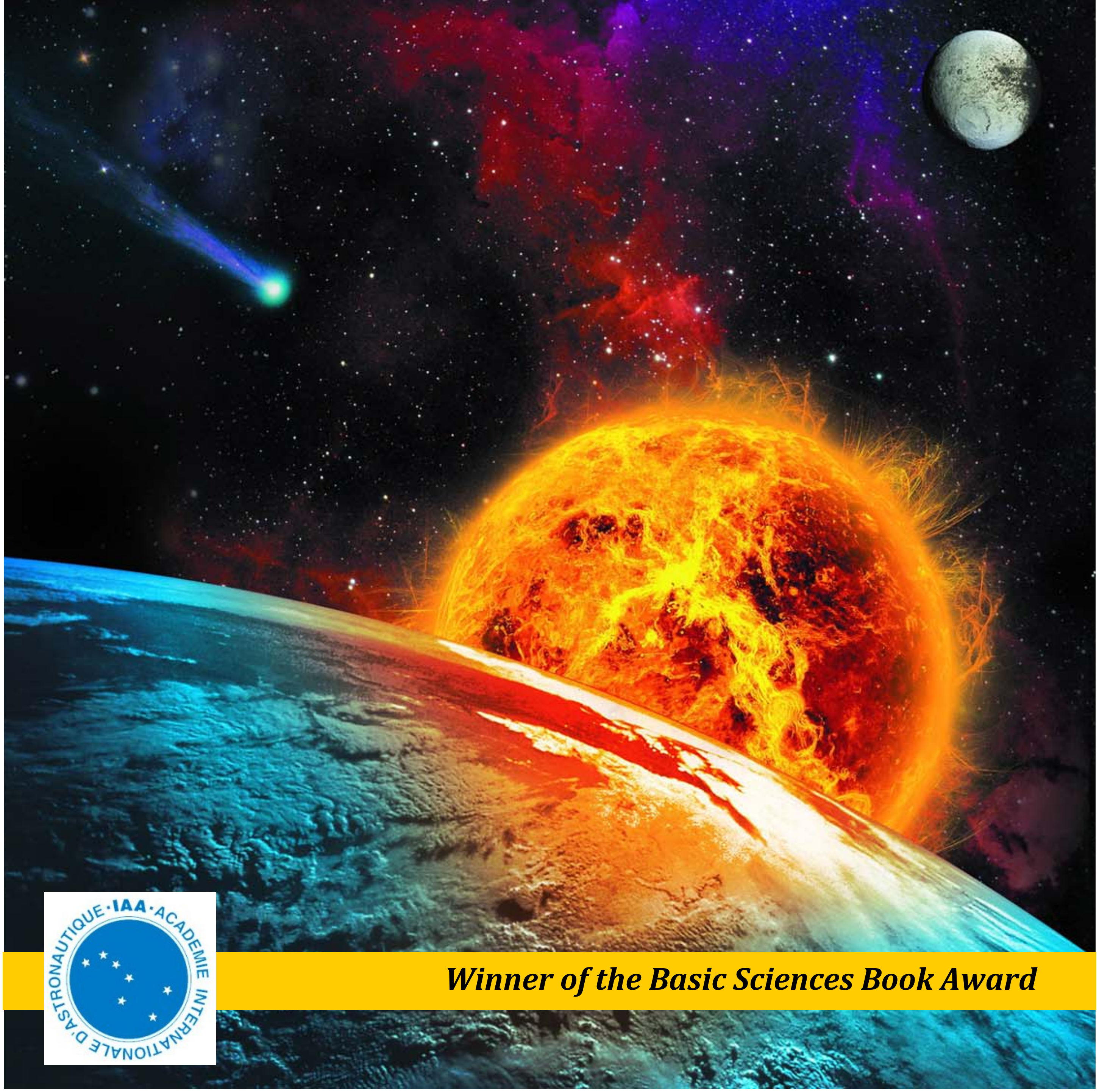

IAA · ACADEMIE INTERNATIONALE D'ASTRONAUTIQUE

*Winner of the Basic Sciences Book Award*

NATIONAL ACADEMY OF SCIENCES OF UKRAINE
MAIN ASTRONOMICAL OBSERVATORY

NATIONAL AERONAUTICS AND SPACE
ADMINISTRATION OF THE USA
GODDARD INSTITUTE FOR SPACE STUDIES

НАЦІОНАЛЬНА АКАДЕМІЯ НАУК УКРАЇНИ
ГОЛОВНА АСТРОНОМІЧНА ОБСЕРВАТОРІЯ

НАЦІОНАЛЬНЕ УПРАВЛІННЯ АЕРОНАВТИКИ
ТА ДОСЛІДЖЕННЯ КОСМІЧНОГО ПРОСТОРУ США
ҐОДДАРДІВСЬКИЙ ІНСТИТУТ КОСМІЧНИХ ДОСЛІДЖЕНЬ

М. І. Міщенко, В. К. Розенбуш, М. М. Кисельов,
Д. Ф. Лупішко, В. П. Тишковець, В. Г. Кайдаш,
І. М. Бельська, Ю. С. Єфімов, М. М. Шаховський

# ДИСТАНЦІЙНЕ ЗОНДУВАННЯ ОБ'ЄКТІВ СОНЯЧНОЇ СИСТЕМИ ПОЛЯРИМЕТРИЧНИМИ ЗАСОБАМИ

*ПРОЕКТ
«УКРАЇНСЬКА НАУКОВА КНИГА
ІНОЗЕМНОЮ МОВОЮ»*

M. I. Mishchenko, V. K. Rosenbush, N. N. Kiselev,
D. F. Lupishko, V. P. Tishkovets, V. G. Kaydash,
I. N. Belskaya, Y. S. Efimov, N. M. Shakhovskoy

# POLARIMETRIC REMOTE SENSING OF SOLAR SYSTEM OBJECTS

This book outlines the basic physical principles and practical methods of polarimetric remote sensing of Solar System objects and summarizes numerous advanced applications of polarimetry in geophysics and planetary astrophysics. In the first chapter we present a complete and rigorous theory of electromagnetic scattering by disperse media directly based on the Maxwell equations and describe advanced physically based modeling tools. This is followed, in Chapter 2, by a theoretical analysis of polarimetry as a remote-sensing tool and an outline of basic principles of polarimetric measurements and their practical implementations. In Chapters 3 and 4, we describe the results of extensive ground-based, aircraft, and spacecraft observations of numerous Solar System objects (the Earth and other planets, planetary satellites, Saturn's rings, asteroids, trans-Neptunian objects, and comets). Theoretical analyses of these data are used to retrieve optical and physical characteristics of planetary surfaces and atmospheres as well as to identify a number of new phenomena and effects.

This monograph is intended for science professionals, educators, and graduate students specializing in remote sensing, astrophysics, atmospheric physics, optics of disperse and disordered media, and optical particle characterization.

Книга містить основні фізичні принципи і практичні методи поляриметричного дистанційного зондування об'єктів Сонячної системи та підсумовує їхнє застосування в геофізиці й планетній астрофізиці. В першому розділі книги представлено завершену строгу теорію розсіяння електромагнітних хвиль дисперсними середовищами на основі рівнянь Максвела й описано сучасні фізично обґрунтовані методи теоретичного моделювання. В другому розділі представлено теоретичний аналіз поляриметрії як методу дистанційного зондування, а також описано теоретичні основи і принципи вимірювання поляризованого випромінювання, ґрунтуючись на яких створено унікальну прецизійну апаратуру для спостережень. В третьому та четвертому розділах проведено аналіз великого обсягу наземних та аерокосмічних спостережень, на основі якого визначено оптичні та фізичні характеристики поверхонь і атмосфер багатьох тіл Сонячної системи (Земля та інші планети, супутники планет, кільця Сатурна, астероїди, транснептунові об'єкти й комети) та відкрито цілий ряд нових явищ і ефектів.

Призначено для науковців, викладачів, аспірантів та студентів, які спеціалізуються в дистанційному зондуванні, астрофізиці, атмосферній фізиці, оптиці дисперсних і випадкових середовищ та оптичній діагностиці частинок.

# Foreword
# to the project «Ukrainian Scientific Book in a Foreign Language»

*It is a great pleasure to introduce to the Ukrainian and international scientific communities the first volume of a new series of scientific monographs initiated by the Publications Council of the National Academy of Sciences of Ukraine (NASU). Although the literal English translation of the title of this series ("Ukrainian Scientific Book in a Foreign Language") may sound a bit inelegant to a sophisticated English speaker, it makes perfect sense from the perspective of what this series is intended to accomplish. Indeed, the main idea is to provide as broad an international access to major accomplishments of leading Ukrainian scientists as possible. Of course, the "foreign language" will in most cases be the English language.*

*Being a professional astronomer, I am especially pleased that the subject of this first volume is astrophysics of the Solar System or, more specifically, remote sensing of the Solar System using polarimetric techniques. The book is authored by eight Ukrainian astrophysicists and an American scientist of Ukrainian origin, all of whom are internationally recognized experts in this area of research. They represent four astronomical organizations of Ukraine (the Main Astronomical Observatory of NASU, the Institute of Astronomy of the Kharkiv V. N. Karazin National University, the Crimean Astrophysical Observatory, and the Institute of Radioastronomy of NASU) as well as the Goddard Institute for Space Studies of the National Aeronautics and Space Administration of the USA.*

*This book is the result of multi-decadal theoretical and experimental investigations which have contributed profoundly to the solution of a wide range of fundamental and applied problems of science: from the origin and evolution of the Solar System to the detection and characterization of potentially hazardous asteroids, monitoring of global climate changes on Earth, and ecological control of the terrestrial atmosphere. I am confident that the publication of this monograph will serve to summarize the current status of polarimetric remote sensing of the Solar System as well as to facilitate further progress in this important scientific discipline.*

Yaroslav S. Yatskiv

Head of the Scientific Publishing Council of
the National Academy of Sciences of Ukraine

Kyiv

*To our families and friends*

# Contents

# Editorial

Aerosol and cloud particles exert a strong influence on the regional and global climates of the Earth and other planets. Microscopic particles forming the regolith surfaces of many Solar System bodies and cometary atmospheres have a strong and often controlling effect on many ambient physical and chemical processes. Moreover, they are "living witnesses" of the history of the formation and evolution of the Solar System and can tell us much about the events that have taken place over the past ~5 billion years in the circumsolar part of the Universe. Thus, detailed and accurate knowledge of the physical and chemical properties of such particles has the utmost scientific and practical importance.

More often than not it is impossible to collect samples of such particles and subject them to a laboratory test. Therefore, in most cases one has to rely on theoretical analyses of remote measurements of the electromagnetic radiation scattered by the particles. Fortunately, the scattering and absorption properties of small particles often exhibit a strong dependence on their size, shape, orientation, and refractive index. This factor makes remote sensing an extremely useful and often *the only practicable* means of physical and chemical particle characterization in geophysics and planetary astrophysics.

For a long time remote-sensing studies had relied on measurements of only the scattered intensity and its spectral dependence. Eventually, however, it has become widely recognized that polarimetric characteristics of the scattered radiation contain much more accurate and specific information about such important properties of particles as their size, morphology, and chemical composition.

The progress in polarimetric remote-sensing research has always been hampered by the fact that the human eye is "polarization blind" and responds only to the intensity of light impinging on the retina. As a consequence, to give a simple definition of polarization readily intelligible to a non-expert is almost as difficult as to describe color to a color-blind person. However, continuing progress in electromagnetic scattering theory coupled with great advances in the polarization measurement capability has resulted in overwhelming examples of the immense practical power of polarimetric remote sensing which are no longer possible to ignore. As a result of persistent research efforts, polarimetry has become one of the most informative, accurate, and efficient means of remote sensing. Many prominent scientists have contributed to the foundation of polarimetry as a major remote-sensing discipline in geophysics and planetary astrophysics. An important and often decisive role in this process has been played by the authors of this monograph.

The main objective of this book is to summarize the contributions to the field of polarimetric remote sensing of Solar System bodies which we and our colleagues have made over the past four decades. We believe that doing this is useful because a specialized monograph on this subject appears to be lacking and also because of the systematic and comprehensive nature of our collective research activities. Specifically, our work has included the development of a complete and rigorous theory of electromagnetic scattering by disperse media and accurate, physically based modeling tools required for the analysis of polarimetric measurements. We have advanced theoretical fundamentals and principles of measurements of polarized radiation and used them to design unique and precise instrumentation. Based on analyses of extensive ground-based, aircraft, and spacecraft observations, we have determined, for the first time, the optical and physical characteristics of surfaces and atmospheres of numerous Solar System objects (such as planets, satellites, Saturn's rings, asteroids, trans-Neptunian objects, and comets) as well as discovered a significant number of new phenomena and effects. We have also helped develop and implement unprecedented polarimetric techniques for the remote sensing of aerosol and cloud particles in the terrestrial atmosphere from aircraft and orbiting satellites.

This monograph is not intended as an exhaustive review of all the countless applications of polarimetry in geophysics and astrophysics. It is also not intended as a college-level textbook similar to those by Stephens (1994) and Мороженко (2004). We believe, however, that it can form the basis of a specialized graduate-level course in polarimetric remote sensing. With this purpose in mind, we have made the book as self-contained as the practical limitations imposed by the publisher have permitted. Important supplements to this book are the monographs by Mishchenko et al. (2002a, 2006b) and Hovenier et al. (2004), where detailed theoretical derivations and further information can be found. The first of them is in the public domain and is available in the PDF format at http://www.giss.nasa.gov/staff/mmishchenko/books.html.

Based on the specific character of this book, the reference list was not intended to be comprehensive. The reader will also notice that some publications included in the reference list exist only in the Russian or the Ukrainian language. As such, they may not be particularly useful to those who do not speak either language. In several cases, however, there is an official English translation of the Russian or Ukrainian original. The corresponding bibliographic information is included in the reference list to the extent possible.

Following Mishchenko et al. (2002a, 2006b), we denote vectors using the Times bold font and matrices using the Arial bold font. Unit vectors are denoted by a caret, whereas dyads and dyadics are denoted by the symbol ↔. The Times italic font is reserved for scalar quantities, important exceptions being the square root of minus one, the differential sign, and the base of natural logarithms, which are denoted by Times roman characters i, d, and e, respectively. Another exception is the relative refractive index, which is denoted by a sloping sans serif $m$. For the reader's convenience, the list of most important acronyms and abbreviations is included at the end of the book.

Finally, it is worth mentioning that the optical particle-characterization techniques developed by these authors have found extensive applications in many other areas of science and technology such as nanoscience, nanotechnology, biomedicine, nanobiotechnology, biophotonics, chemistry of colloids and suspensions, solid-state physics, laser physics, etc. In fact, our publications have been cited in more than 330 different journals, including *Biophysical Journal*, *Cancer Research*, *Cytometry*, *Journal of Colloid and Interface Science*, *Journal of Dairy Science*, *Journal of Nanoscience and Nanotechnology*, *Journal of Polymer Science*, *Metamaterials*, *Nanobiotechnology*, *Nanotechnology*, *Progress in Energy and Combustion Science*, *Solar Energy Materials and Solar Cells*, etc. We view this as a convincing example of how relatively modest investments in remote-sensing and astrophysics research can generate a great practical return highly beneficial to the entire society. We also hope that this factor will help increase the usefulness of our book and broaden its readership.

Consistent with the above rationale, this monograph is intended for science professionals, educators, and graduate students specializing in remote sensing, astrophysics, atmospheric physics, optics of disperse and disordered media, optical particle characterization, biomedicine, and nanoscience.

*M. I. Mishchenko, V. K. Rosenbush, N. N. Kiselev,*
Editors

*Kyiv – New York*

# Acknowledgments

We are grateful to Academician Yaroslav S. Yatskiv for suggesting the idea of the English edition of this monograph and for his generous support during all stages of this project. We also appreciate helpful and highly professional assistance from the staff of the academic publisher “Akademperiodyka”.

It is a great pleasure to thank the following colleagues for numerous discussions and productive long-term collaborations: Kirill Antoniuk, Yuri Barabanenkov, Antonella Barucci, Matthew Berg, Anatoli Borovoi, Oleg Bugaenko, Brian Cairns, Barbara Carlson, Alberto Cellino, Jacek Chowdhary, Petr Chýlek, Carl Codan, Anthony Davis, Viktor Degtyarev, Zhanna Dlugach, Helmut Domke, Oleg Dubovik, Bryan Fafaul, Sonia Fornasier, Igor Geogdzhayev, Roald E. Gershberg, James Hansen, Otto Hasekamp, Steven Hill, Alfons Hoeskstra, Ronald Hooker, Helmuth Horvath, James Hough, Joop Hovenier, Vsevolod Ivanov, Klaus Jockers, Michael Kahnert, Nikolai Khlebtsov, Alexander Kokhanovsky, Sergej Kolesnikov, Ludmilla Kolokolova, Miroslav Kocifaj, Theodore Kostiuk, Andrew Lacis, Kuo-Nan Liou, Pavel Litvinov, Li Liu, Kari Lumme, Andreas Macke, Daniel Mackowski, Hal Maring, Pinar Mengüç, Alexander Morozhenko, Karri Muinonen, Olga Muñoz, Elena Petrova, Vilppu Piirola, William Rossow, Yuri Shkuratov, Larry Travis, Cornelis van der Mee, Feodor Velichko, Sergei Velichko, Gorden Videen, Tõnu Viik, Nikolai Voshchinnikov, Warren Wiscombe, Thomas Wriedt, Ping Yang, Edgard Yanovitskij, Maksim Yurkin, Nadia Zakharova, and Eleonora Zege. We are grateful to Michael A’Hearn, David Shleicher, Tony Farnham, and Rita Shulz for providing a set of narrowband spectral filters which we have used extensively in observations of comets. Last but not least, we thank our families for their unconditional and devoted support and understanding.

Our research has been sponsored by many grants from the National Academy of Sciences of Ukraine, the National Aeronautics and Space Administration of the USA, the Academy of Sciences of the Former USSR, INTAS, the US Civilian Research & Development Foundation, and the National Polar-orbiting Operational Environmental Satellite System program of the USA. M.I.M. acknowledges continual support from the NASA Radiation Sciences Program managed by Hal Maring and from the NASA Glory Mission project.

*I. N. Belskaya, Y. S. Efimov, V. G. Kaydash,*
*N. N. Kiselev, D. F. Lupishko, M. I. Mishchenko,*
*V. K. Rosenbush, N. M. Shakhovskoy, V. P. Tishkovets*

*Kharkiv – Kyiv – Nauchny – New York*

# Introduction

Despite the immense diversity of applications of remote sensing in geophysics and planetary astrophysics, each application has the same basic stages summarized in the following block diagram:

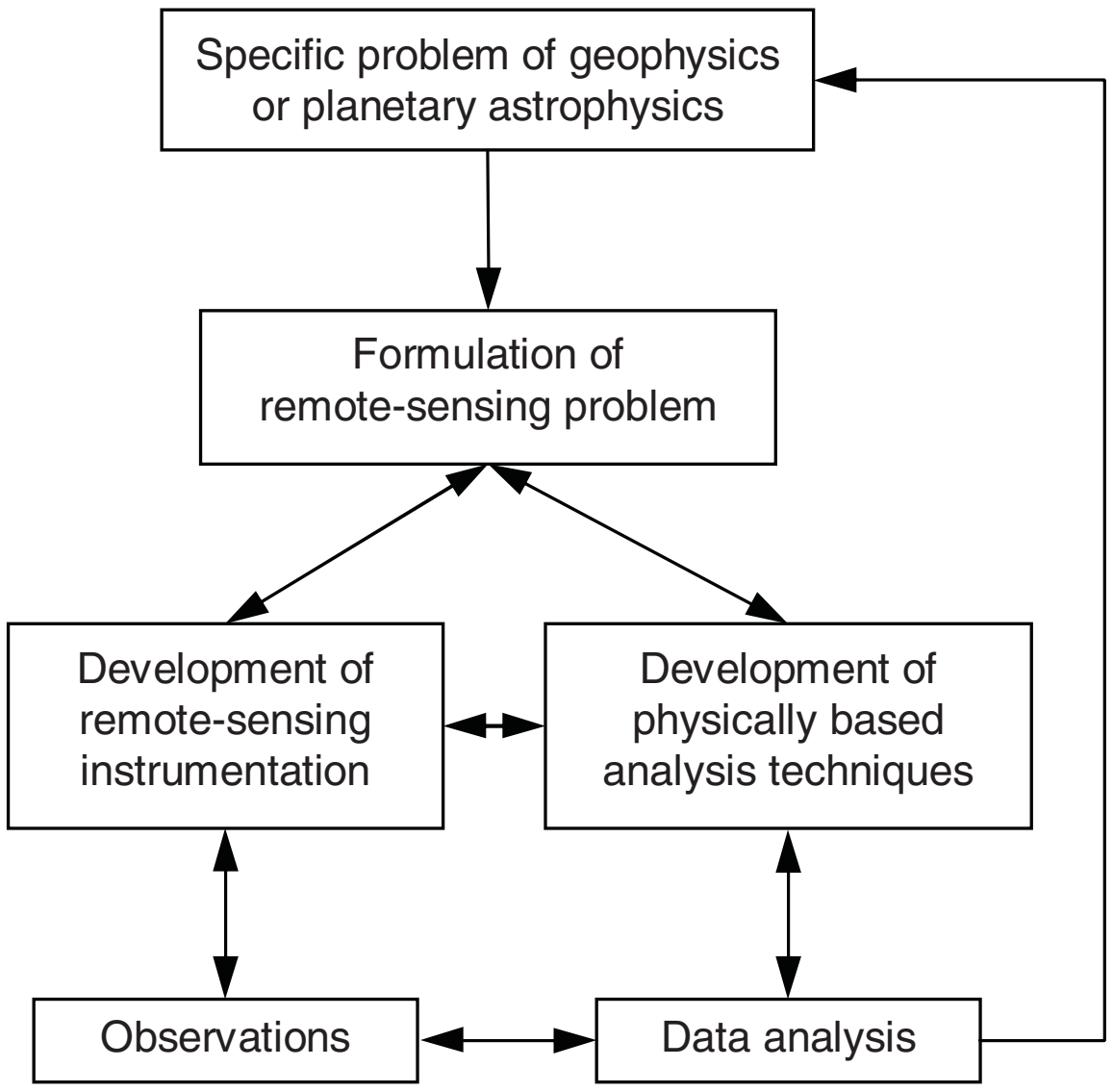

The specific character of a particular geophysical or astrophysical problem ultimately defines the complexity of the requisite remote-sensing approach. The double-headed arrows in the diagram signify the inherent interdependence of all the basic stages. For example, newly performed observations may demonstrate the need for improved instrumentation, while difficulties encountered at the data-analysis stage may ultimately necessitate a complete reformulation of the remote-sensing problem whenever the remote-sensing methodology adopted proves to be inadequate.

The main objectives of this monograph are (i) to discuss in detail all basic stages of polarimetric remote sensing and (ii) to demonstrate that polarimetry is a powerful (and often the only adequate) remote-sensing methodology which can greatly outperform remote sensing techniques based on measurements of only intensity. The main theme of the book is remote sensing of objects composed of (or containing) small particles, such as atmospheres of the Earth, other planets, and comets as well as planetary particulate surfaces.

Accordingly, in Chapter 1 we present a complete and rigorous theory of electromagnetic scattering by disperse media directly based on classical electromagnetics. We discuss numerically exact computer solutions as well as asymptotically accurate analytical solutions of the Maxwell equations (such as the *T*-matrix approach and the microphysical theories of radiative transfer and coherent backscattering) and describe state-of-the-art physically based modeling tools.

Chapter 2 begins with an illustrative theoretical analysis of polarimetry as a remote-sensing tool. It then outlines basic principles of polarimetric measurements and measurement-error budgeting and gives an overview of several practical hardware implementations of the polarimetric technique.

In two final chapters, we describe the results of our extensive photopolarimetric observations of numerous Solar System objects such as planets, planetary satellites, asteroids, and comets. We demonstrate how theoretical analyses of these data can be used to retrieve optical and physical characteristics of particles constituting planetary surfaces and atmospheres as well as to identify a number of new physical phenomena and effects.

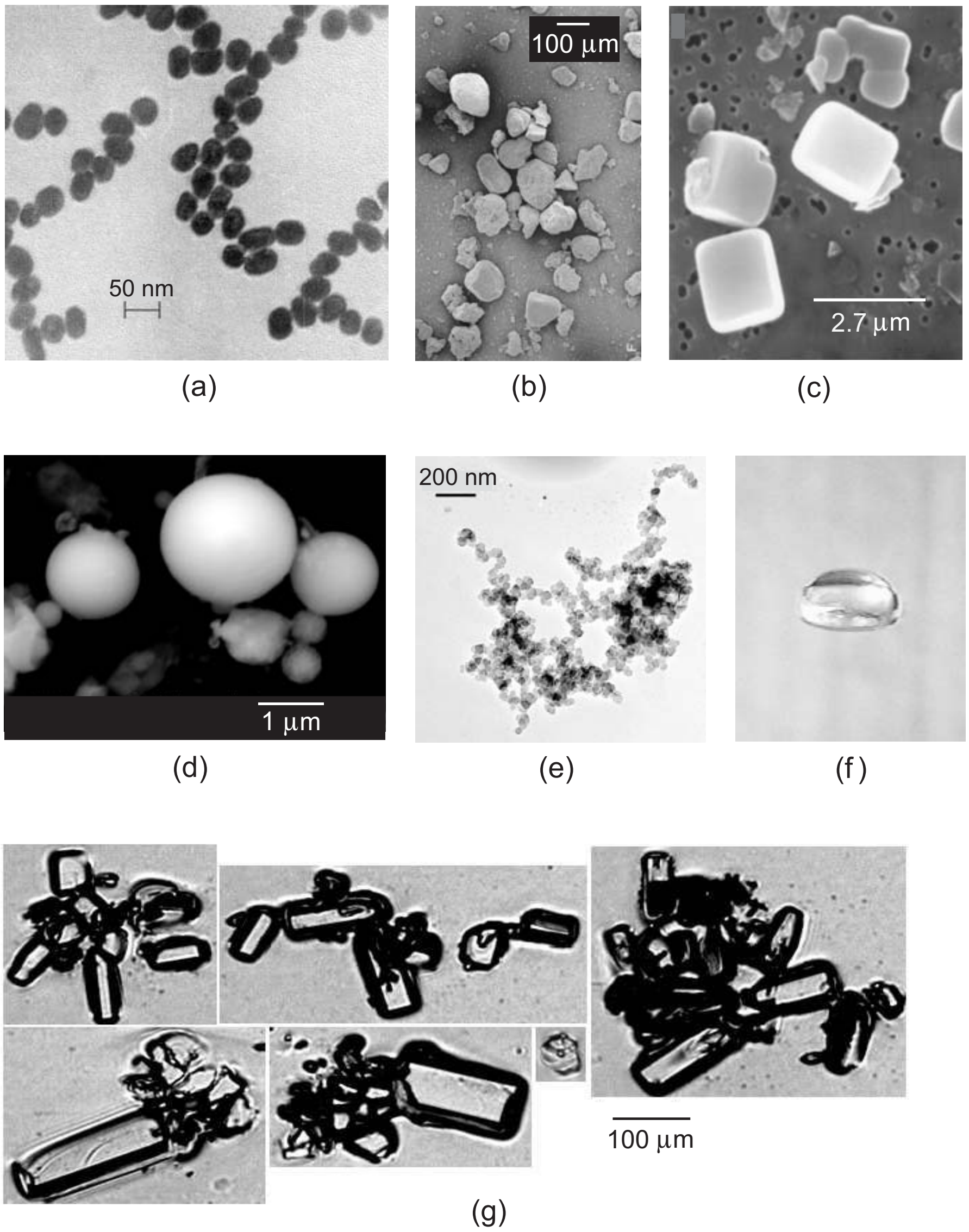

**Plate 1.1.** Examples of man-made and natural small particles. (a) 40-nm-diameter gold particles (after Khlebtsov et al. 1996). (b) Sahara desert sand (after Volten et al. 2001). (c) Dry sea-salt particles (after Chamaillard et al. 2003). (d) Fly ash particles (after Ebert et al. 2002). (e) A soot aggregate (after Li et al. 2003). (f) A 6-mm-diameter falling raindrop. (g) Cirrus cloud crystals (after Arnott et al. 1994).

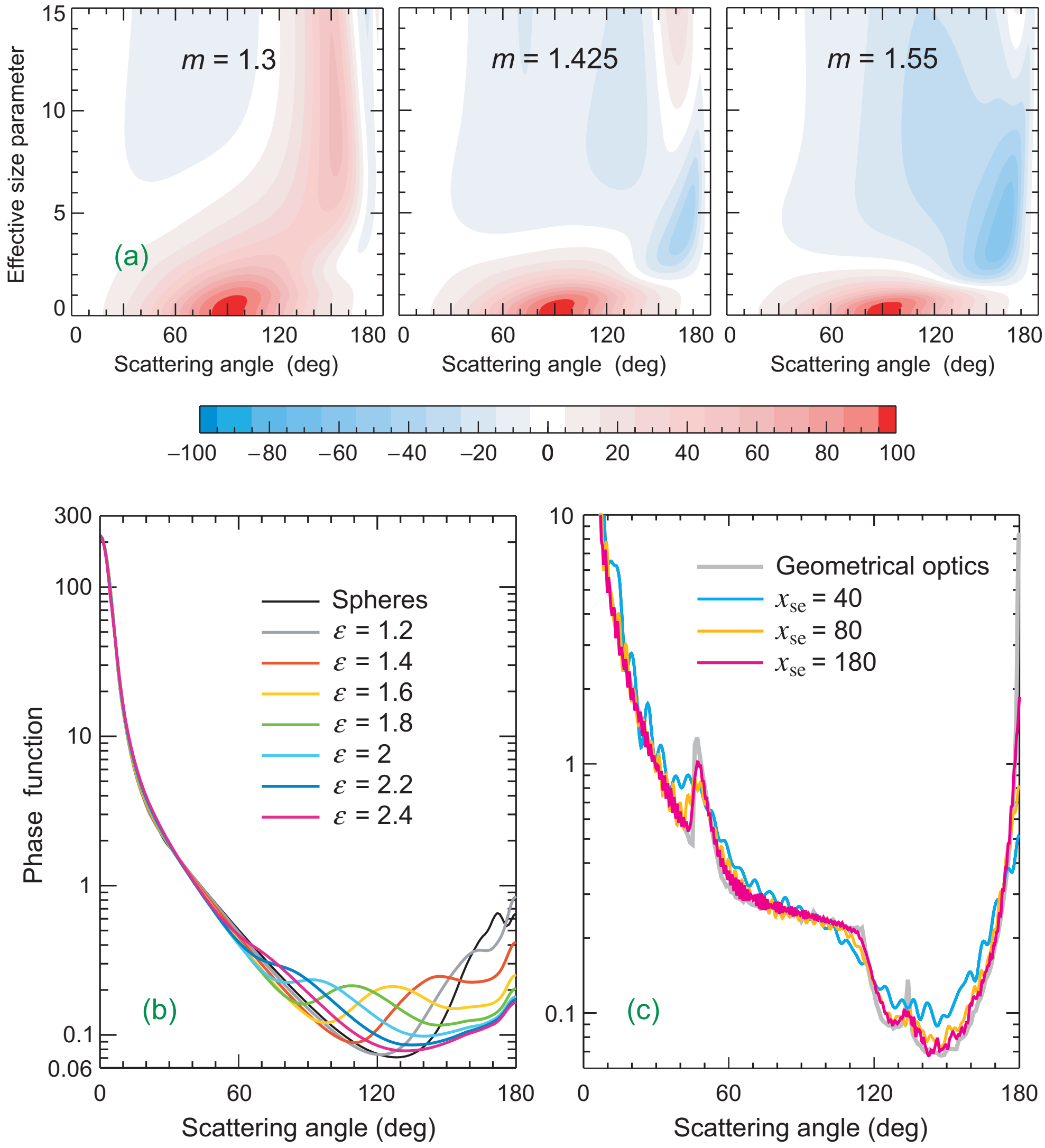

**Plate 1.2.** (a) Degree of linear polarization for unpolarized incident light (in %) versus scattering angle and effective size parameter for polydisperse spherical particles with relative refractive indices 1.3, 1.425, and 1.55. (b) *T*-matrix computations of the phase function for micrometer-sized polydisperse spheres and randomly oriented surface-equivalent prolate spheroids with aspect ratios ranging from 1.2 to 2.4 at a wavelength of 443 nm. The relative refractive index is fixed at 1.53 + i0.008. (c) Geometrical-optics and *T*-matrix phase functions for monodisperse, randomly oriented circular cylinders with surface-equivalent-sphere size parameters $x_{se}$ = 40, 80, and 180. The relative refractive index is $1.311 + i0.311\times10^{-8}$ and is typical of water ice at visible wavelengths.

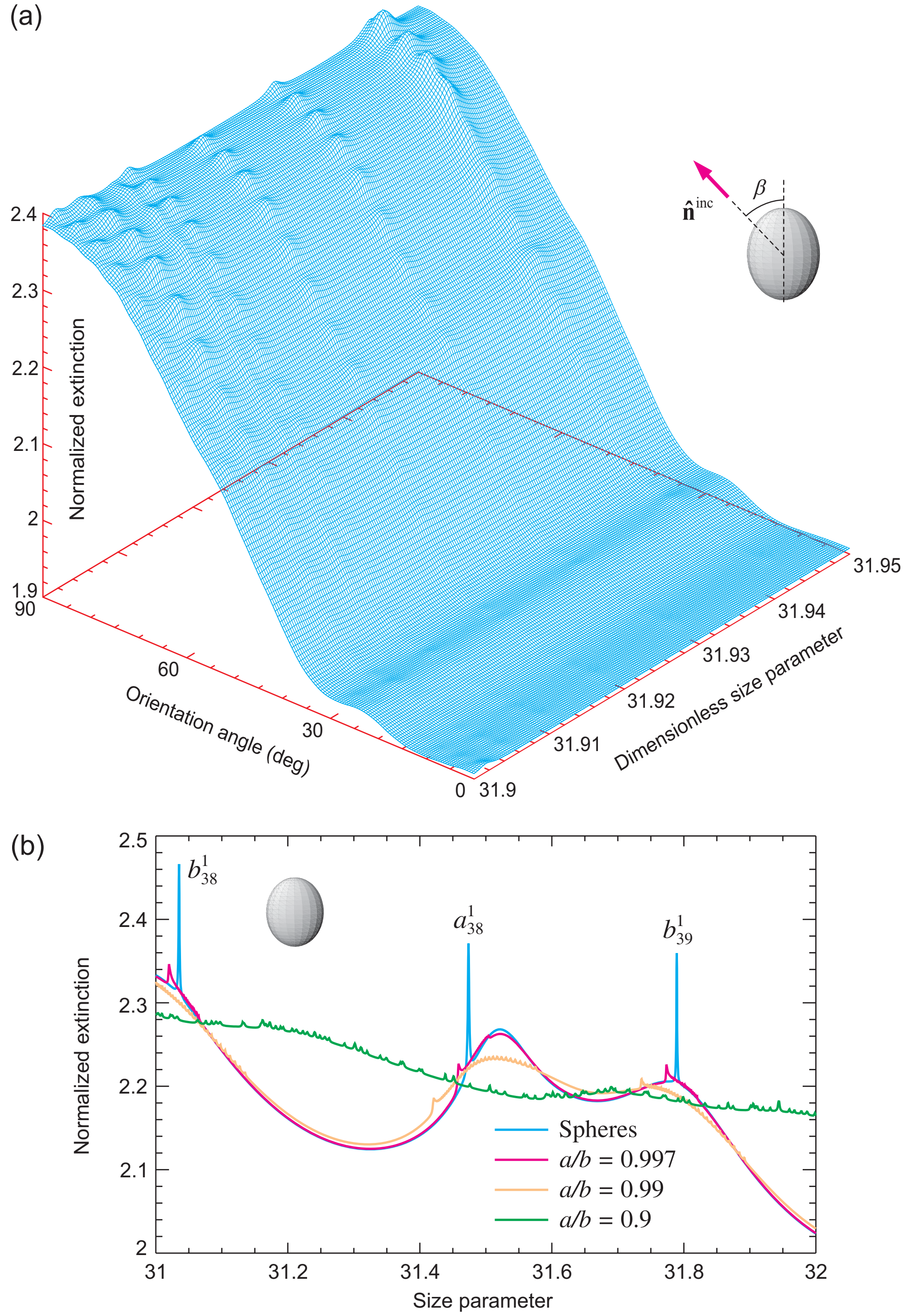

**Plate 1.3.** (a) Normalized extinction versus $x_{ev}$ and orientation angle for a prolate spheroid with $m = 1.4$ and $a/b = 0.9$. (b) Normalized extinction versus $x_{ev}$ for a sphere and a prolate spheroid in random orientation. The results are shown for several values of the spheroid's semi-axis ratio $a/b$. The relative refractive index is fixed at 1.4.

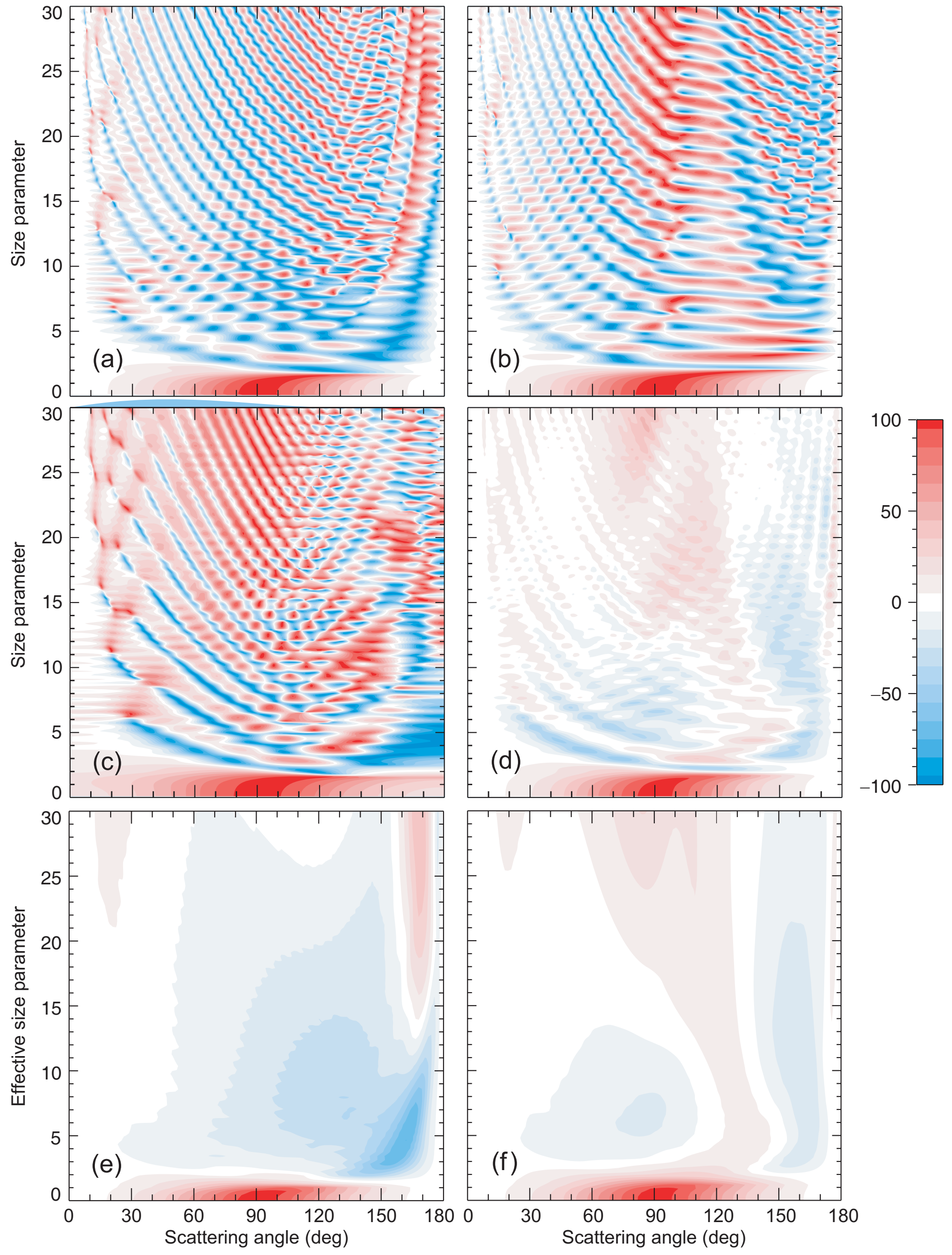

**Plate 1.4.** (a–c) $-F_{12}/F_{11}$ (in %) versus $\Theta$ and $x_{se}$ for a spherical particle (a) and a surface-equivalent oblate spheroid in a fixed orientation (b,c). The rotation axis of the spheroid is oriented along the incidence direction (b) or perpendicularly to the scattering plane (c). (d) $-\langle F_{12}\rangle_\xi/\langle F_{11}\rangle_\xi$ for monodisperse, randomly oriented oblate spheroids. (e,f) $-\langle F_{12}\rangle_\xi/\langle F_{11}\rangle_\xi$ versus $\Theta$ and effective surface-equivalent-sphere size parameter for polydisperse spheres and polydisperse, randomly oriented oblate spheroids. The refractive index of all particles is 1.53 + i0.008 and the spheroid semi-axis ratio is $a/b = 1.7$.

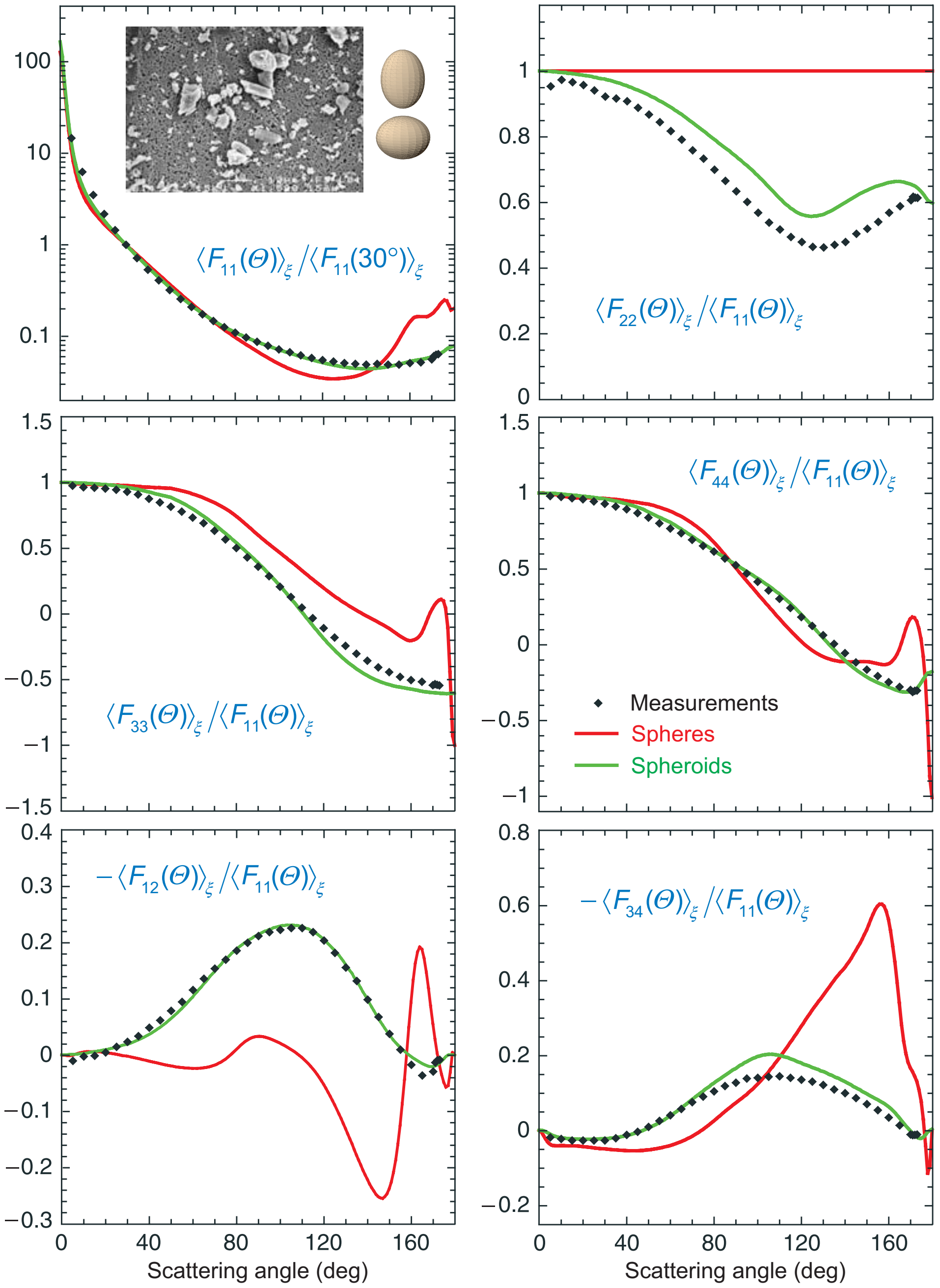

**Plate 1.5.** Diamonds depict the results of laboratory measurements of the ensemble-averaged scattering matrix for micrometer-sized feldspar particles at a wavelength of 633 nm. The green curves show the result of fitting the laboratory data with *T*-matrix results computed for a shape distribution of polydisperse, randomly oriented prolate and oblate spheroids. The real and model particle shapes are contrasted in the inset. The red curves show the results computed for volume-equivalent polydisperse spherical particles.

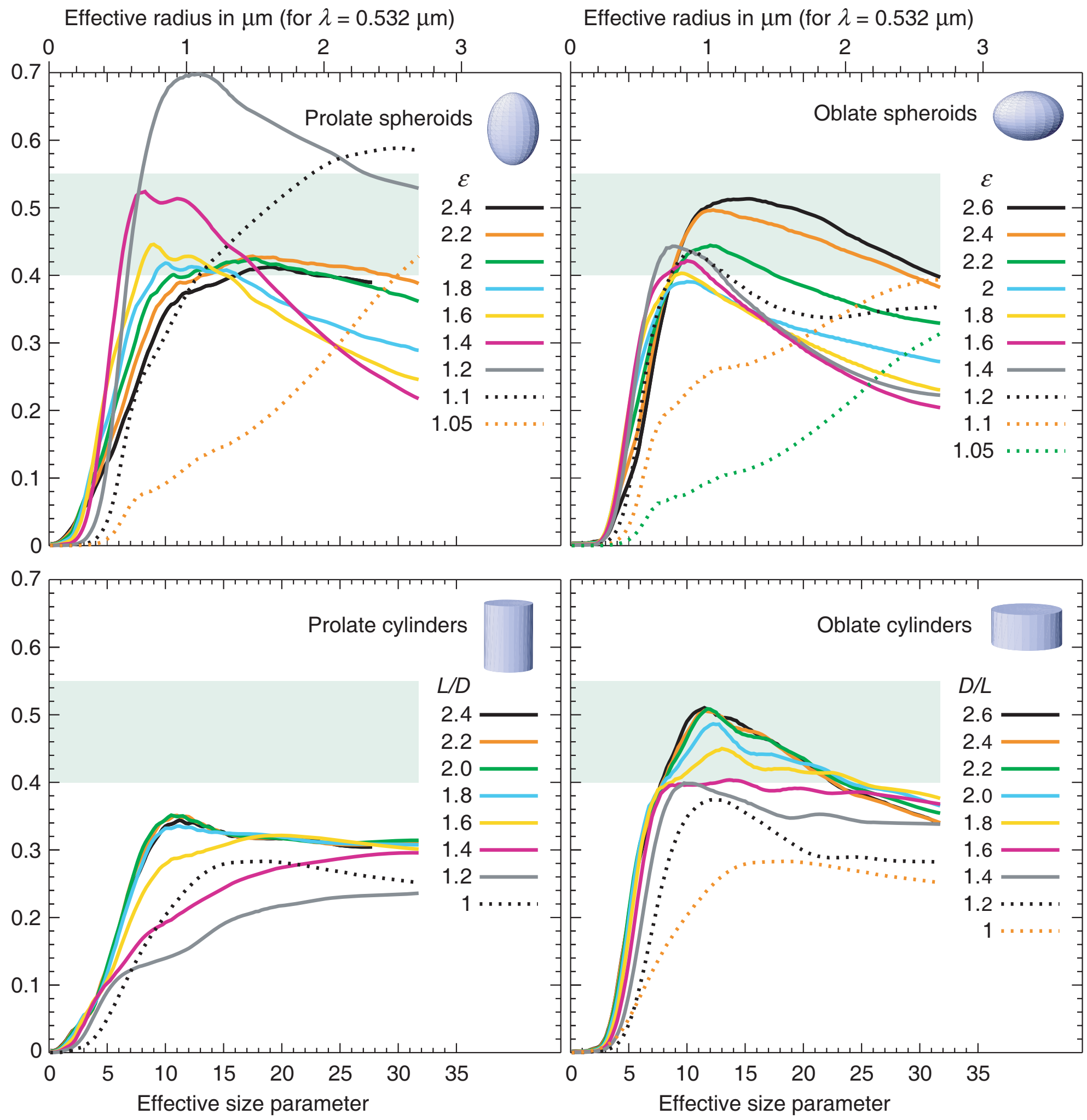

**Plate 1.6.** Linear depolarization ratio (vertical axes) versus effective surface-equivalent-sphere size parameter for polydisperse, randomly oriented ice spheroids with aspect ratios ranging from 1.05 to 2.6 and circular cylinders with various length-to-diameter or diameter-to-length ratios. The relative refractive index is 1.311. The light-green bands show the range of highest depolarization ratios typically observed for anthropogenic cirrus clouds in the form of aircraft condensation trails (Freudenthaler et al. 1996). The upper horizontal axes convert effective size parameters to effective radii assuming the wavelength $\lambda = 2\pi/k_1 = 0.532$ µm.

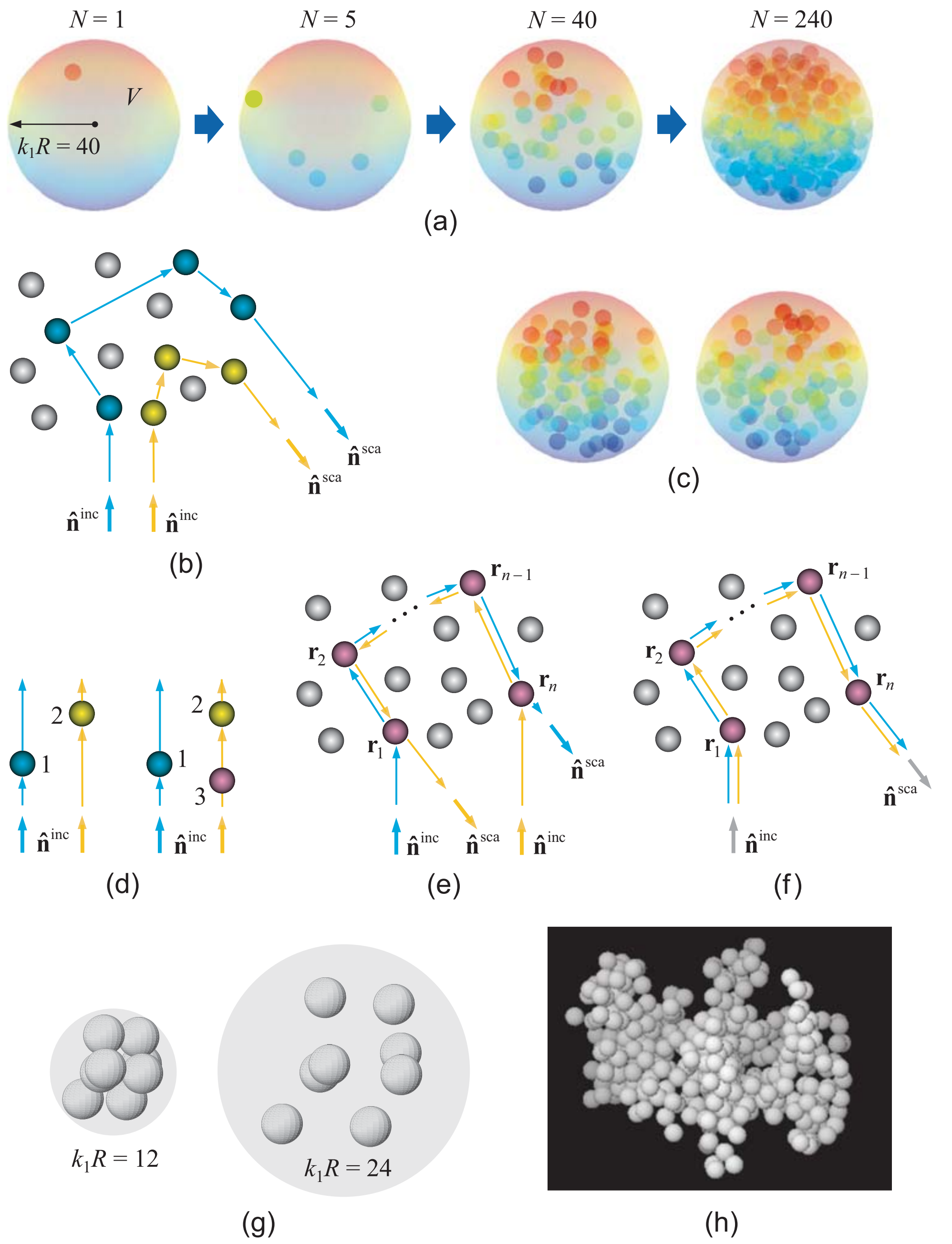

**Plate 1.7.** (a) Spherical scattering volume $V$ filled with $N$ randomly positioned particles. (b) Interference origin of speckle. (c) Two random realizations of the 80-particle group. (d) Forward-scattering interference. (e) Interference origin of coherent backscattering. (f) Interference origin of the diffuse background. (g) Spherical volumes filled with eight randomly positioned, identical particles. (h) Fractal aggregate composed of 334 identical spherical monomers.

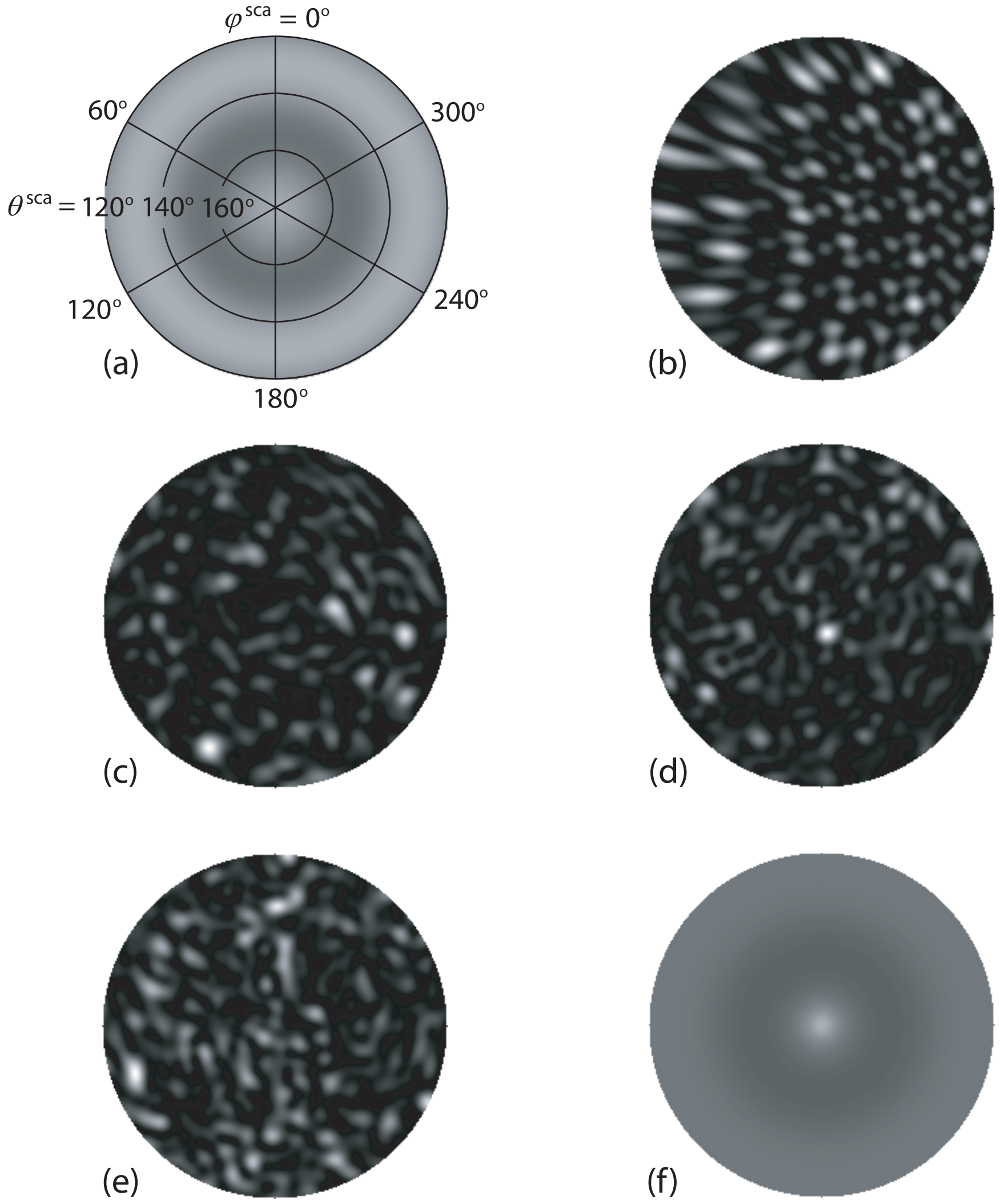

**Plate 1.8.** Angular distribution of scattered intensity in the far-field zone of the spherical volume $V$ filled with $N$ particles. (a) $N$ = 1, fixed orientation. (b) $N$ = 5, fixed orientation. (c) $N$ = 20, fixed orientation. (d) and (e) $N$ = 80, fixed orientation. (f) $N$ = 80, random orientation. The gray scale is individually adjusted in order to maximally reveal the details of each scattering pattern. Panel (a) also shows the angular coordinates used for all panels.

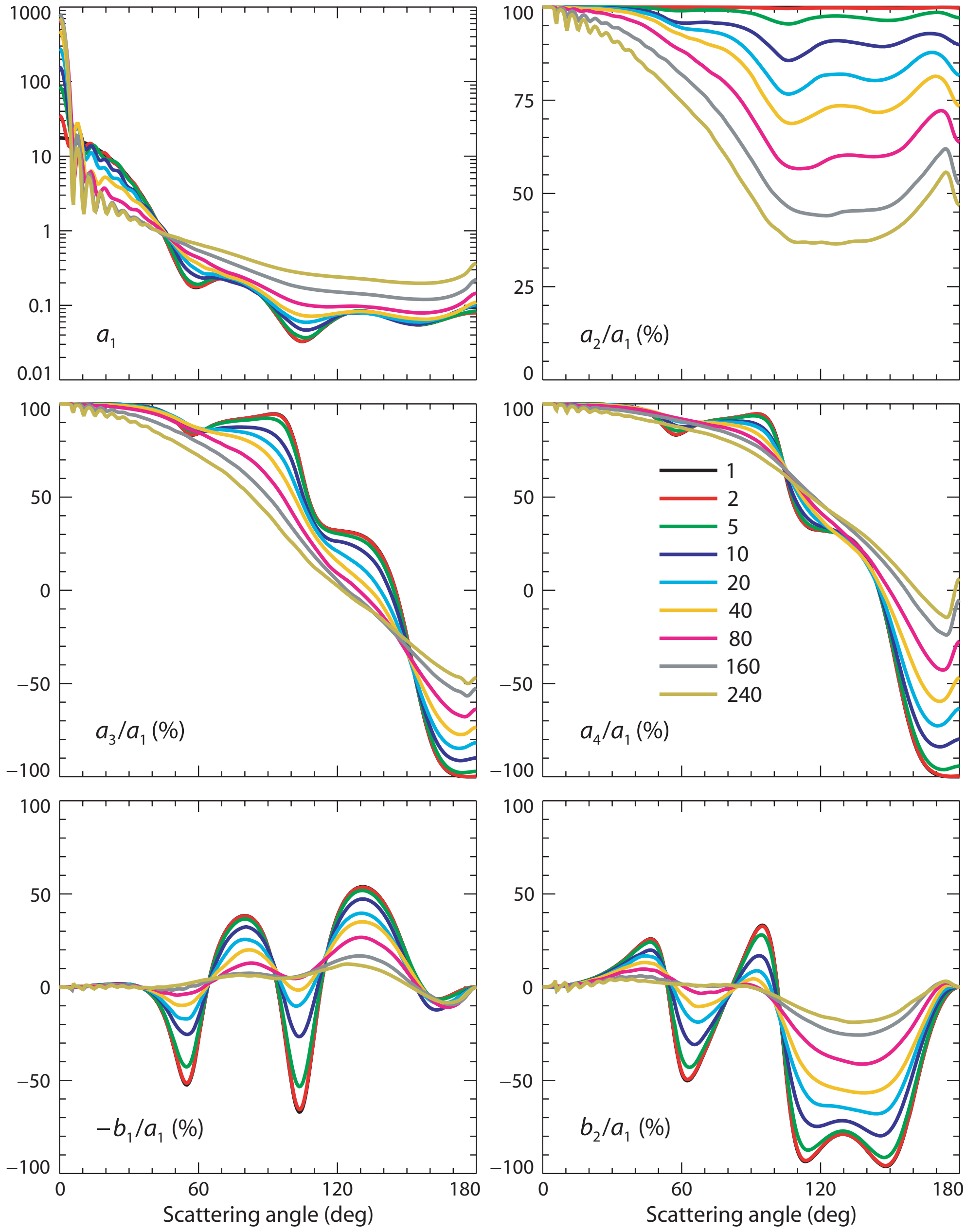

**Plate 1.9.** Elements of the normalized Stokes scattering matrix computed for a $k_1R = 40$ spherical volume of discrete random medium filled with $N = 1, \ldots, 240$ particles.

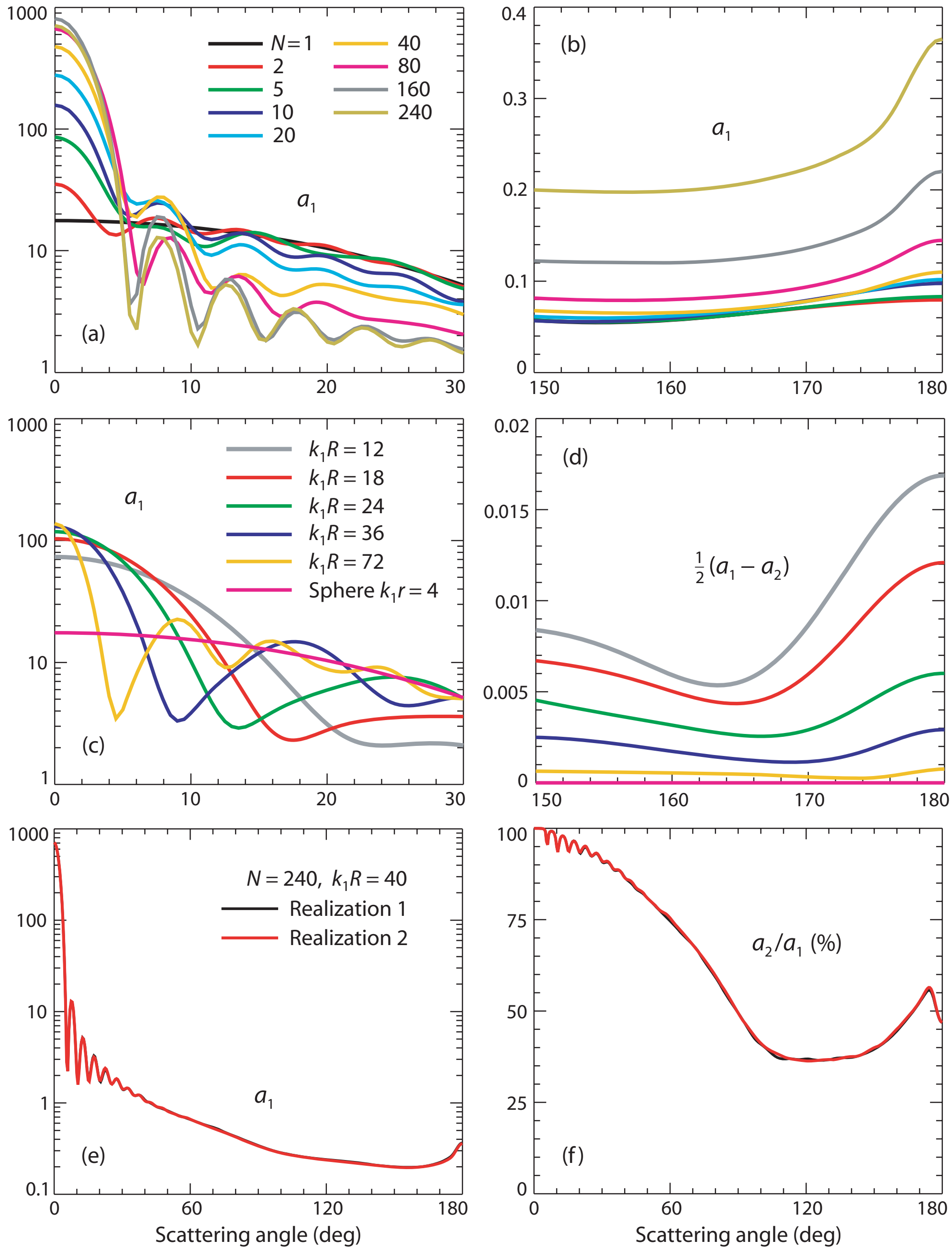

**Plate 1.10.** Scattering properties of a spherical volume of discrete random medium.

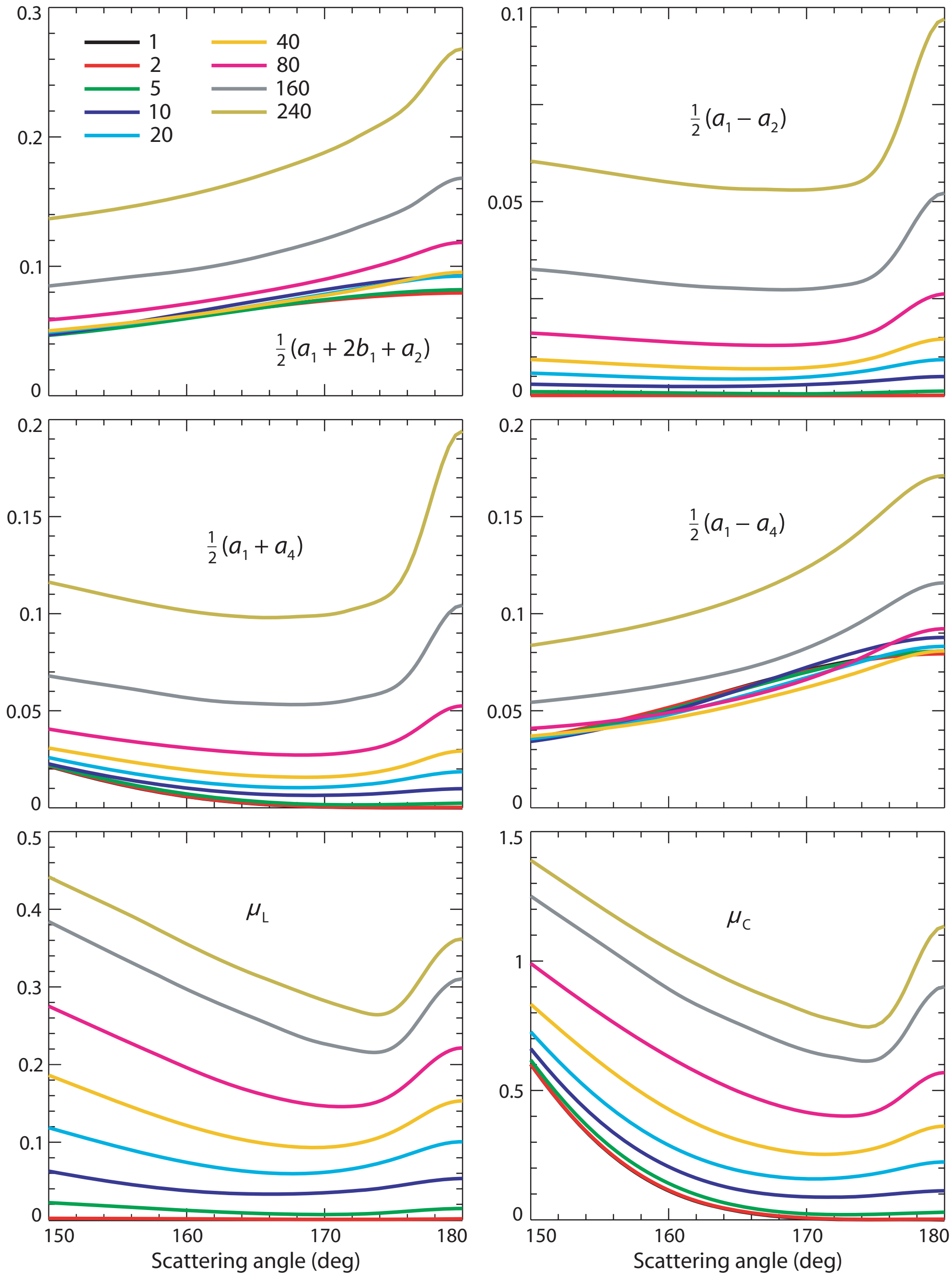

**Plate 1.11.** Polarization characteristics of backscattered light computed for a $k_1R = 40$ spherical volume of discrete random medium filled with $N = 1, \ldots, 240$ particles.

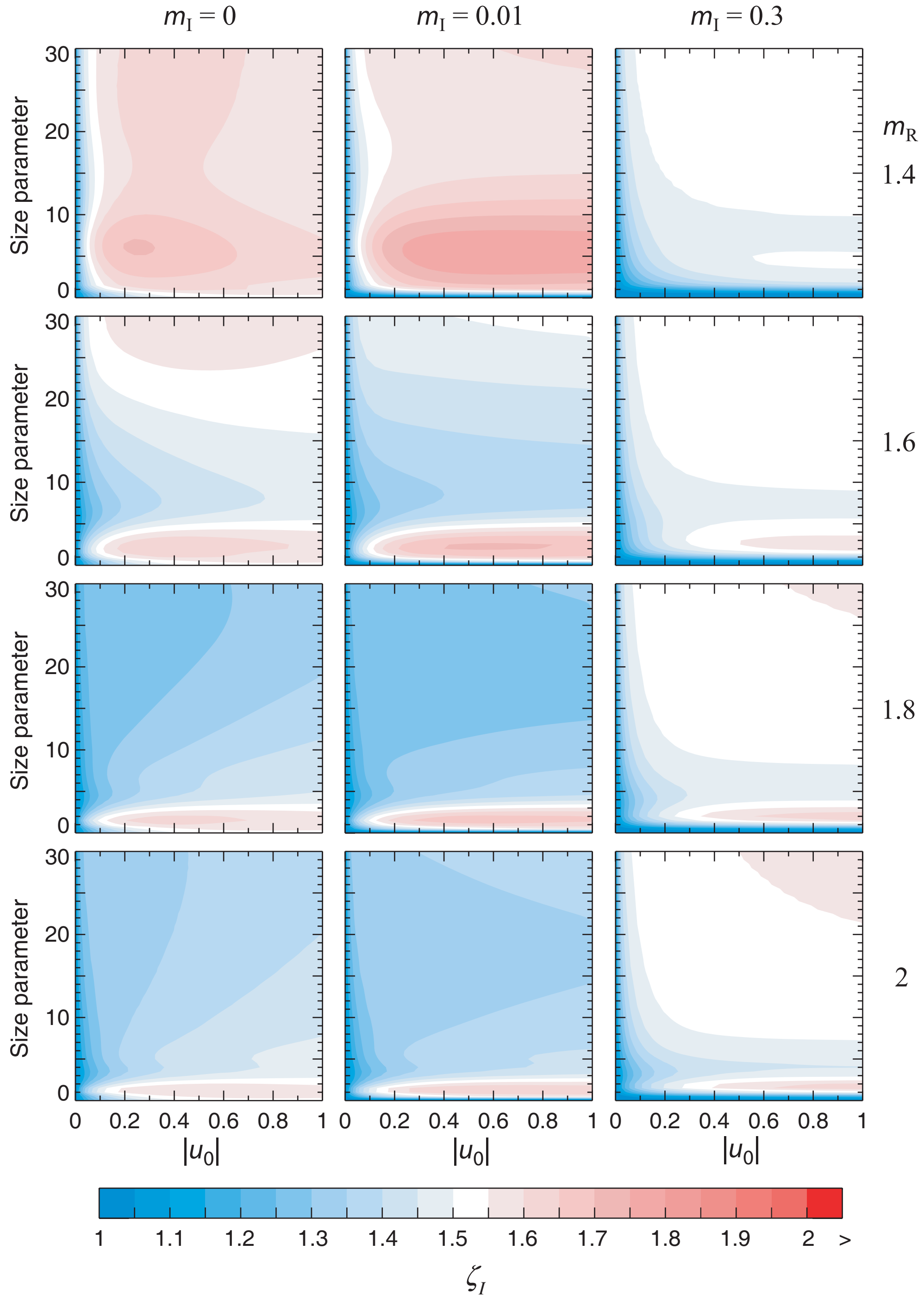

**Plate 1.12.** Enhancement factor $\zeta_I$.

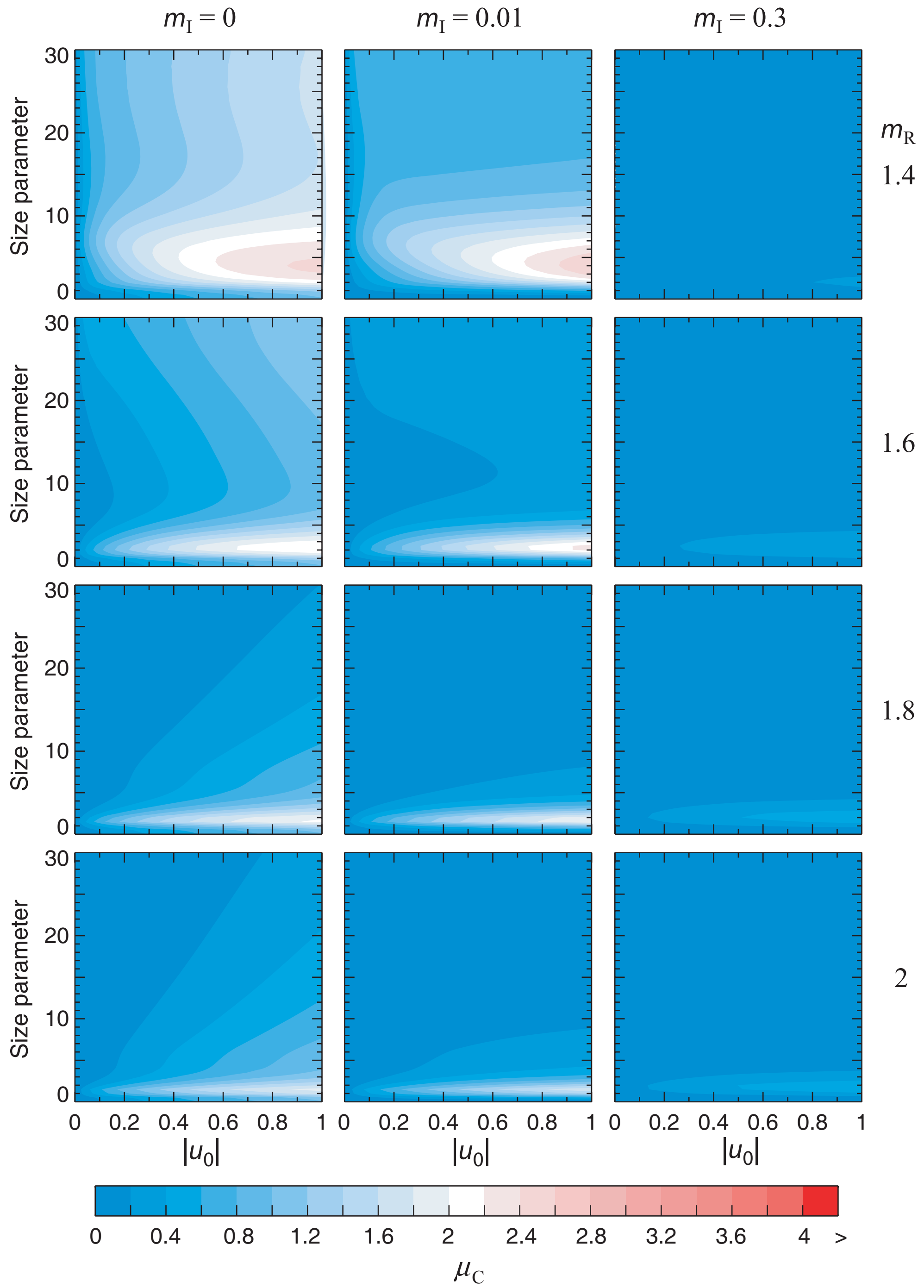

**Plate 1.13.** Polarization ratio $\mu_C$.

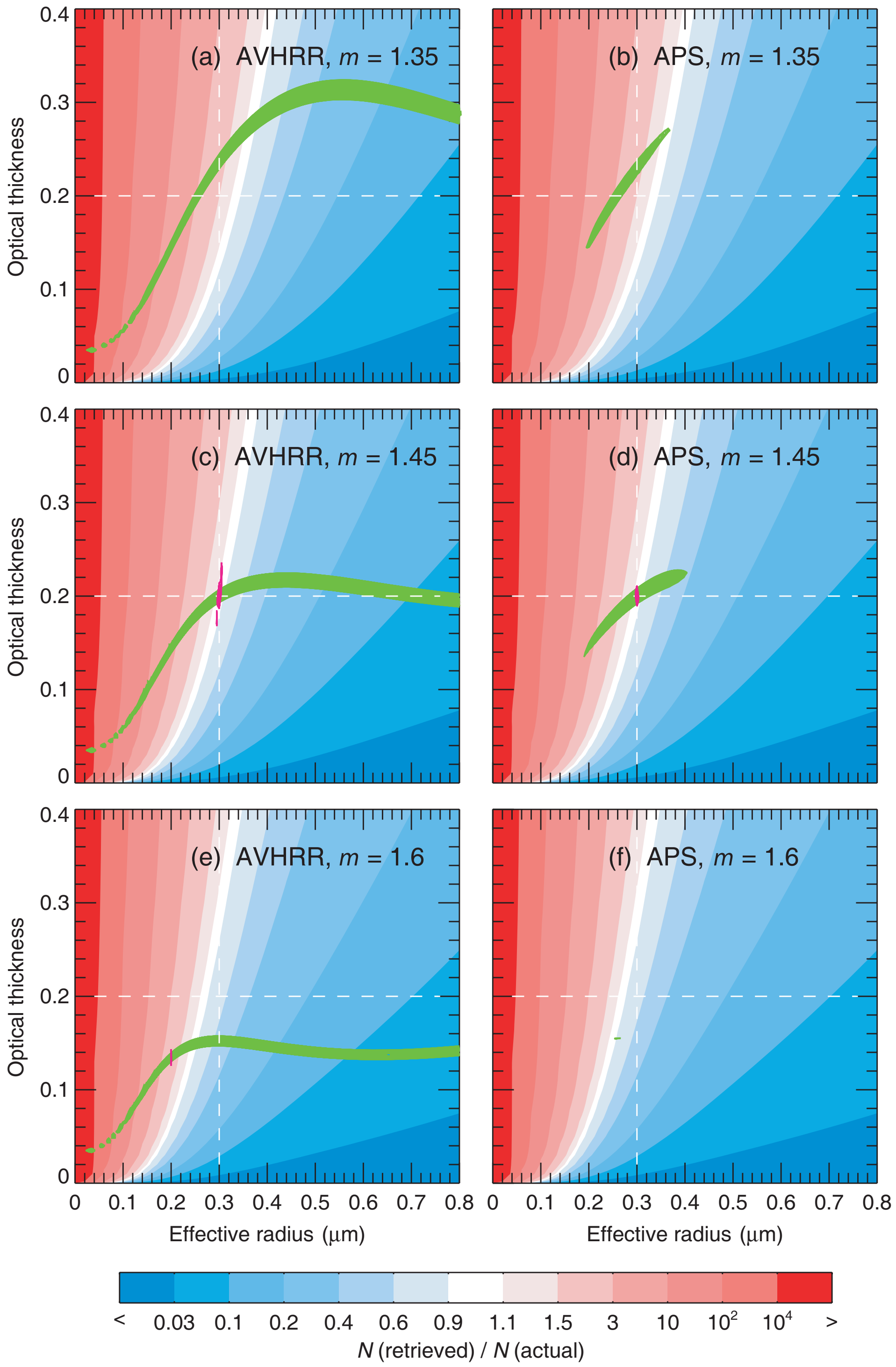

**Plate 2.1.** Modeling different types of aerosol retrievals (see text).

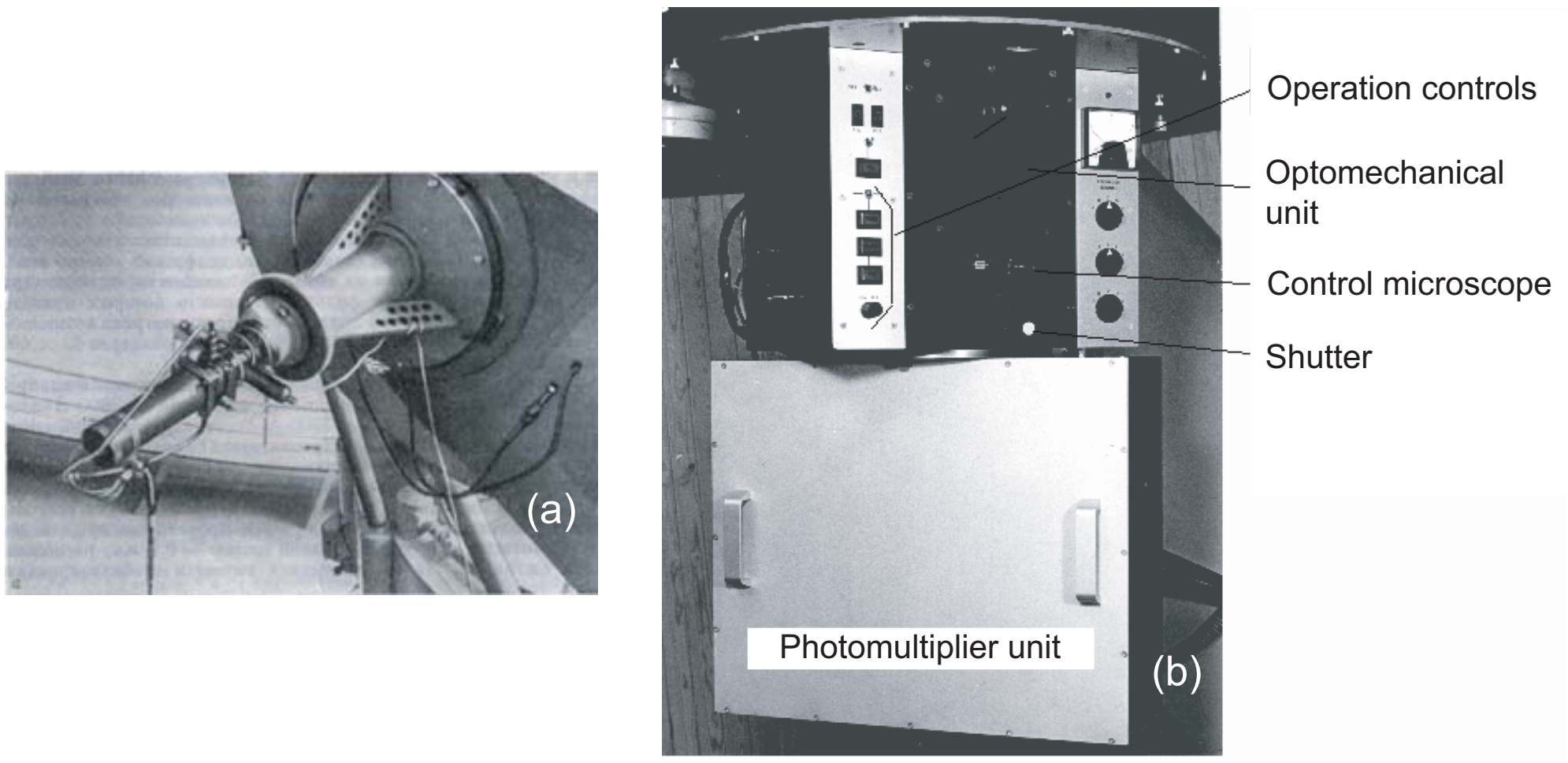

**Plate 2.2.** (a) The first CrAO polarimeter mounted on the telescope ZTSh. (b) Five-channel photopolarimeter.

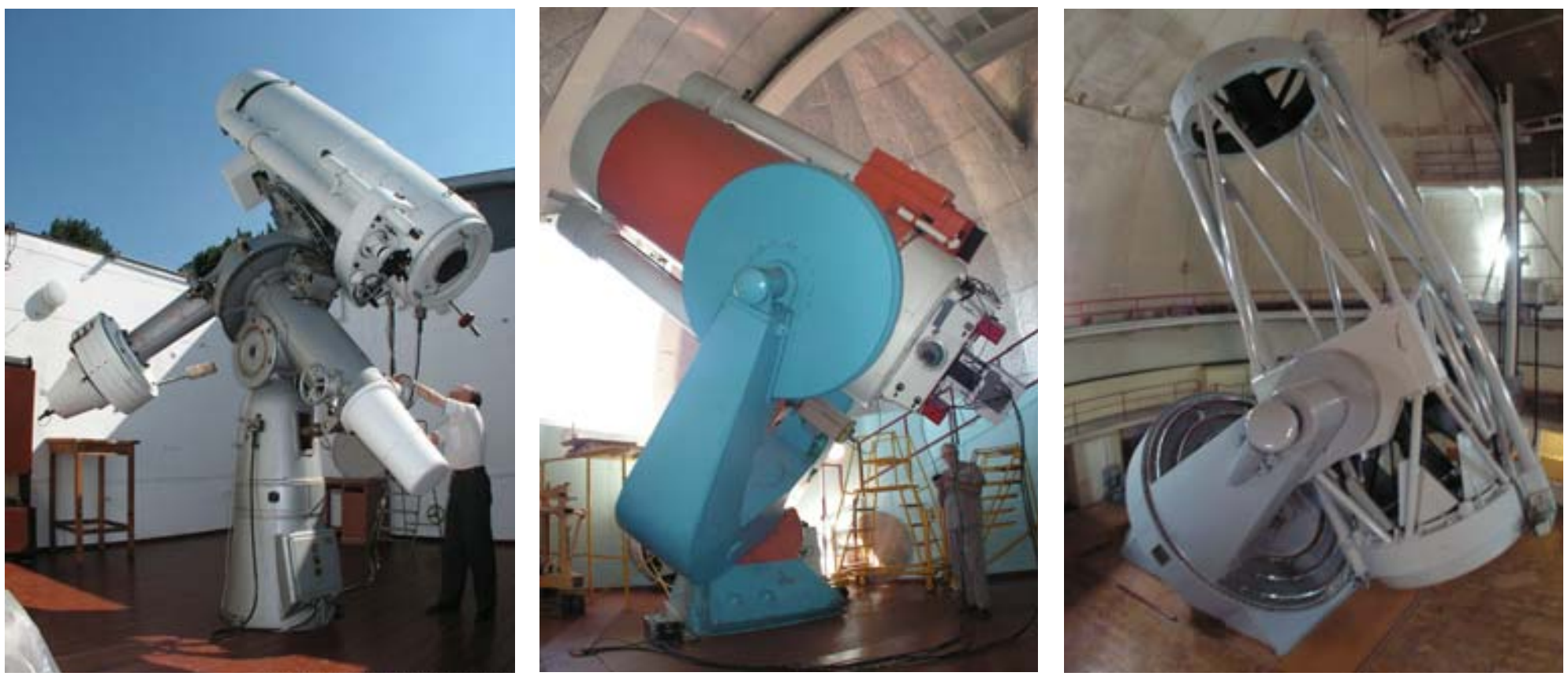

AZT-8 AZT-11 ZTSh

**Plate 2.3.** CrAO telescopes used for photopolarimetric observations.

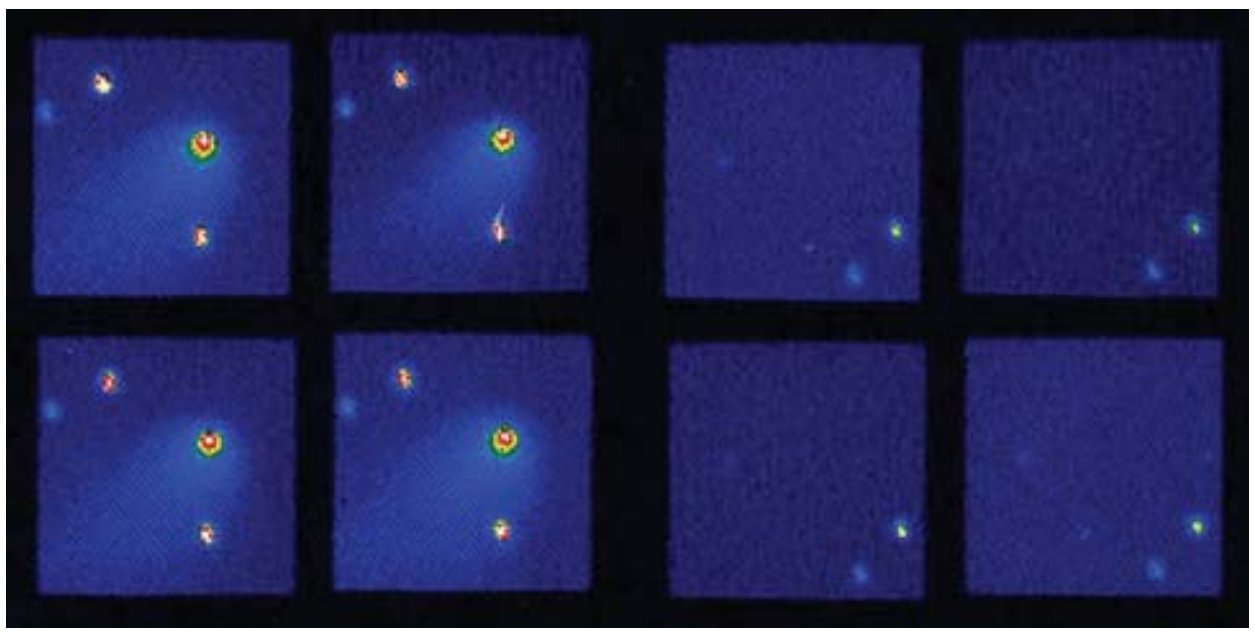

**Plate 2.4.** Polarization images of comet 21P/Ashbrook–Jackson and the background sky.

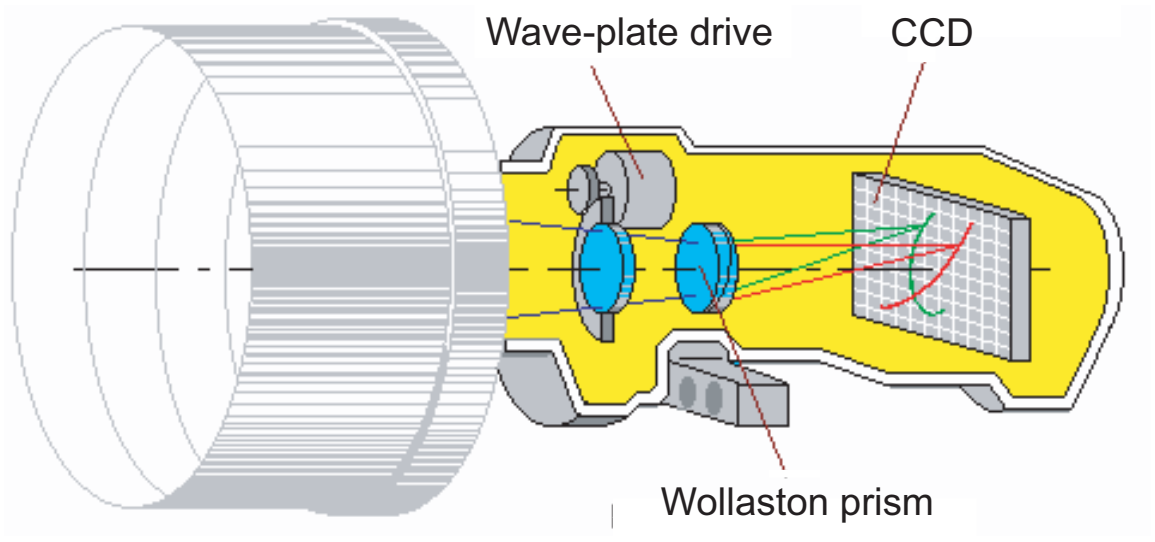

**Plate 2.5.** Optical scheme of the UV spectropolarimeter with an original Wollaston prism.

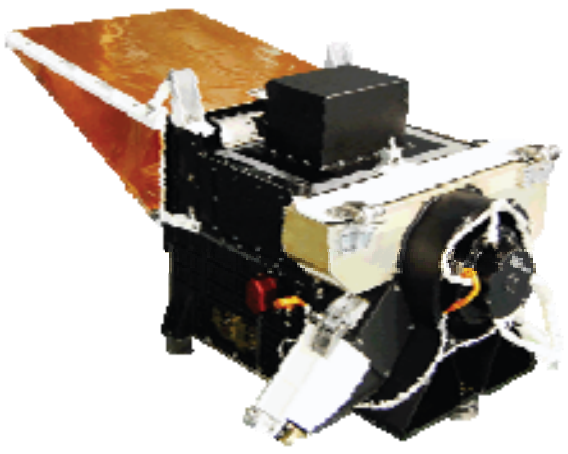

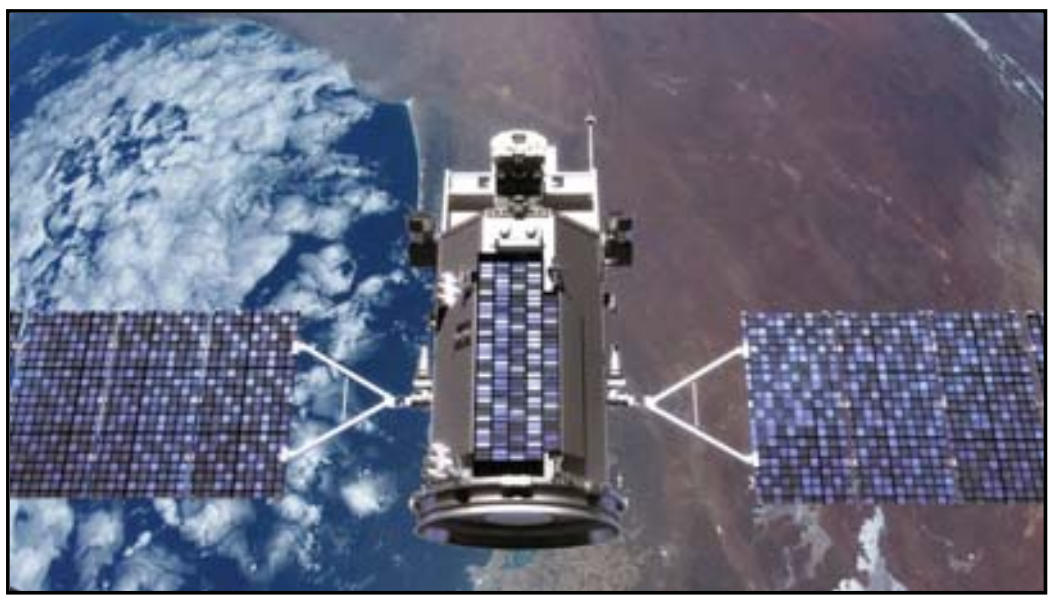

**Plate 2.6.** Orbital photopolarimeter APS (left) and artist's rendition of the NASA orbital satellite Glory (right).

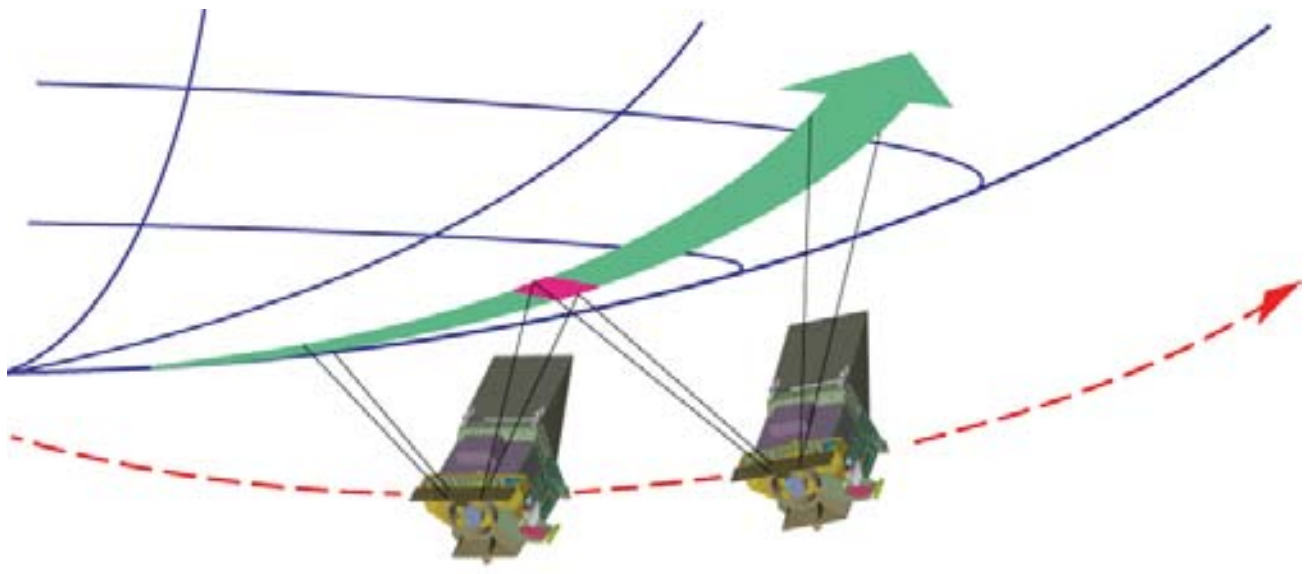

**Plate 2.7.** Along-track multiangle APS measurements via 360° scanning from the sun-synchronous polar-orbiting Glory spacecraft.

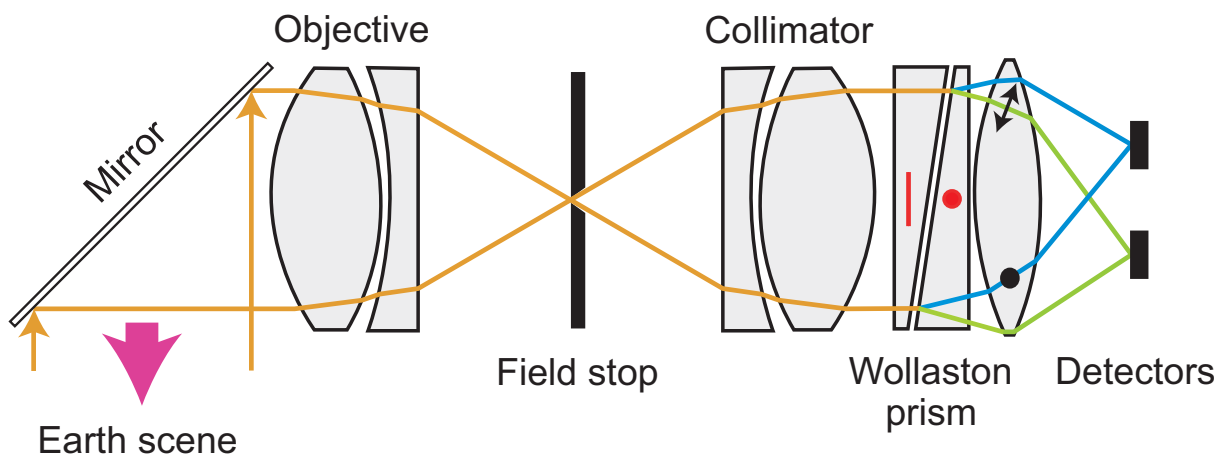

**Plate 2.8.** APS optical approach for polarization measurement. Red markings show the orientations of the optical axes of the birefringent crystals forming the Wollaston prism. Orange lines show ray paths undergoing the split into orthogonal polarizations as indicated by the green and blue lines.

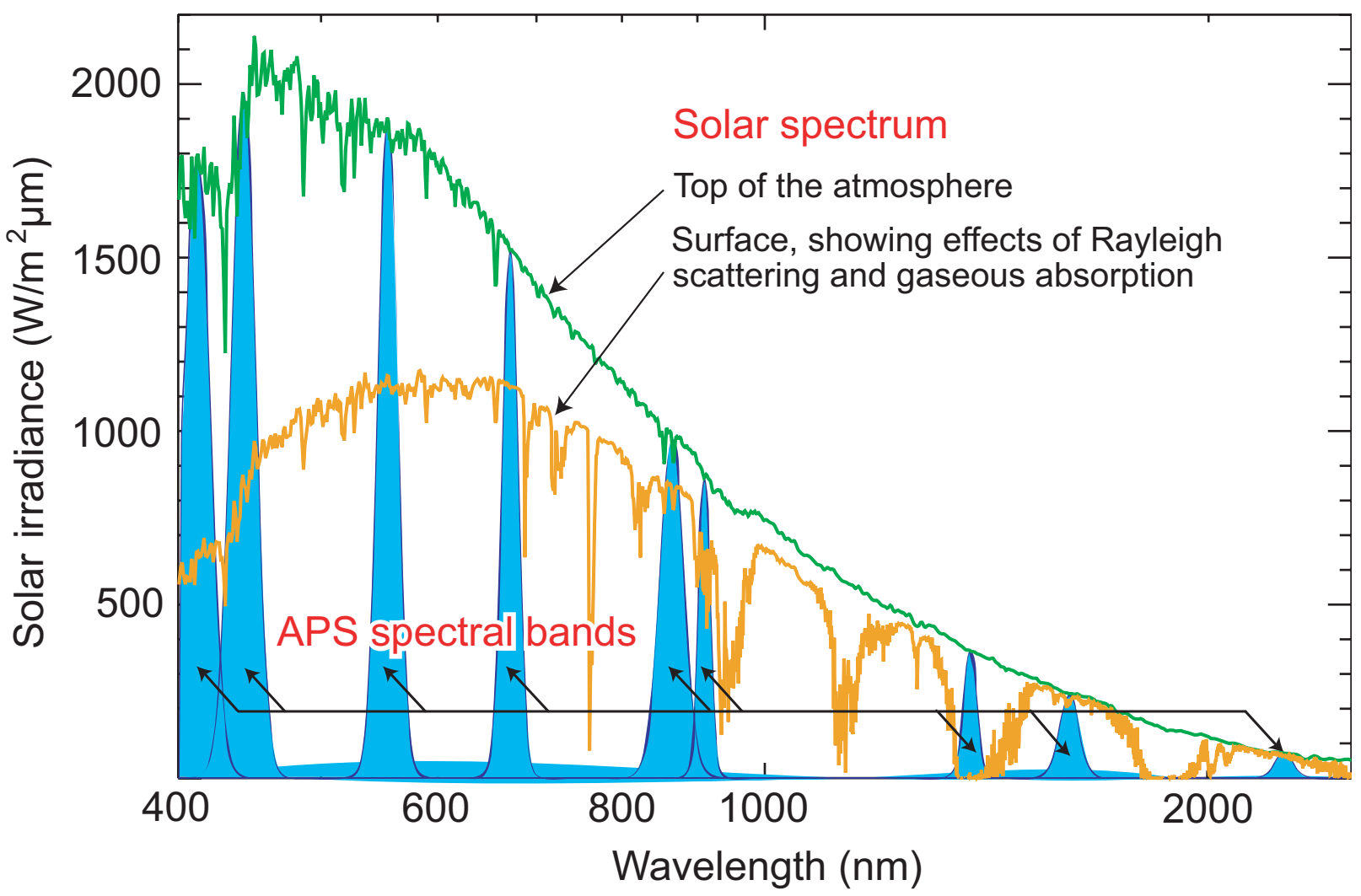

**Plate 2.9.** Locations of the APS spectral channels.

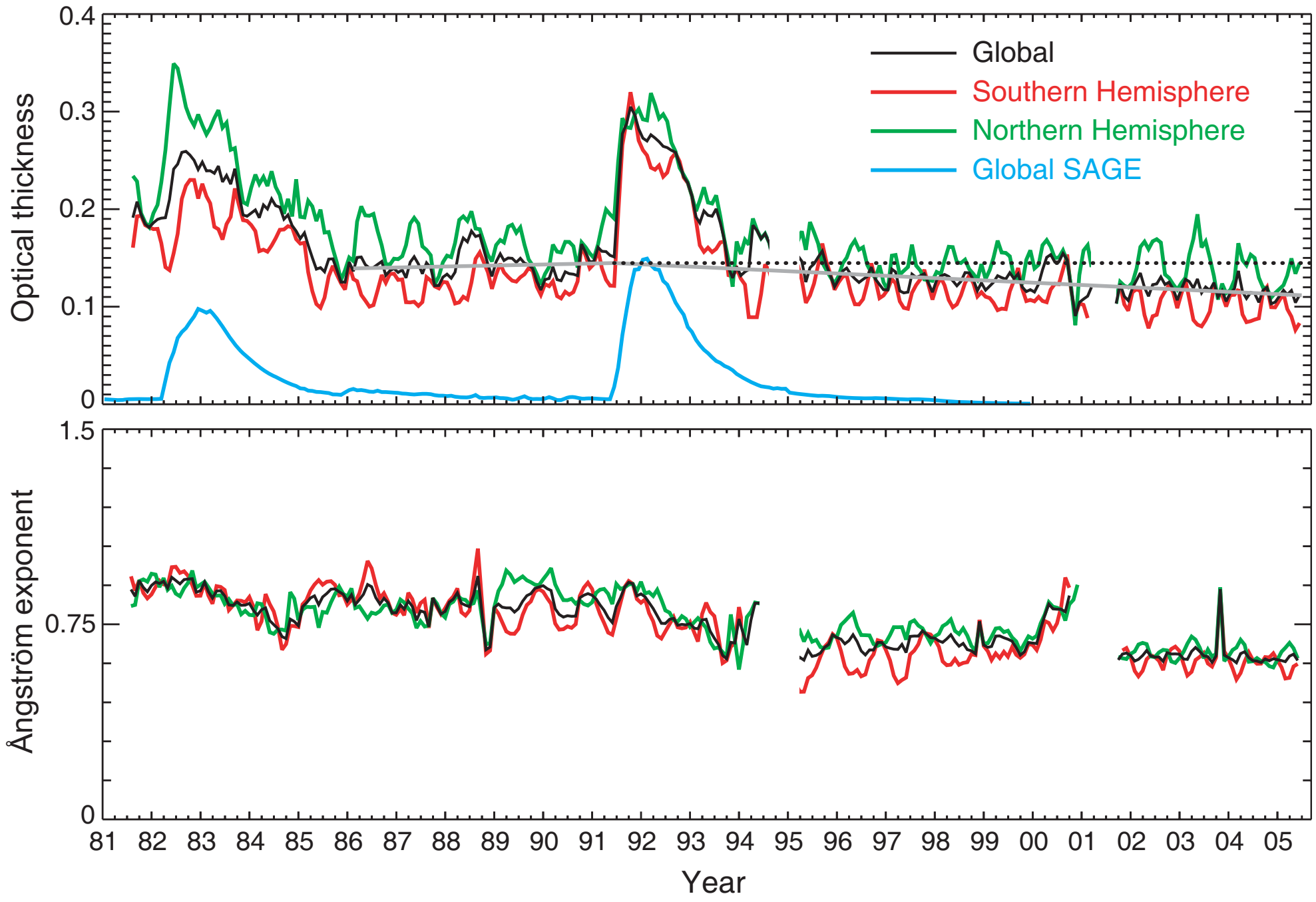

**Plate 3.1.** Global and hemispherical monthly averages of the AOT and Ångström exponent over the oceans for the period August 1981 – June 2005 derived with the latest version of the GACP retrieval algorithm. The blue curve depicts the SAGE record of the globally averaged stratospheric AOT. The solid grey lines show pre- and post-Pinatubo linear regressions. The black dotted line represents the June 1991 pre-Pinatubo regression level.

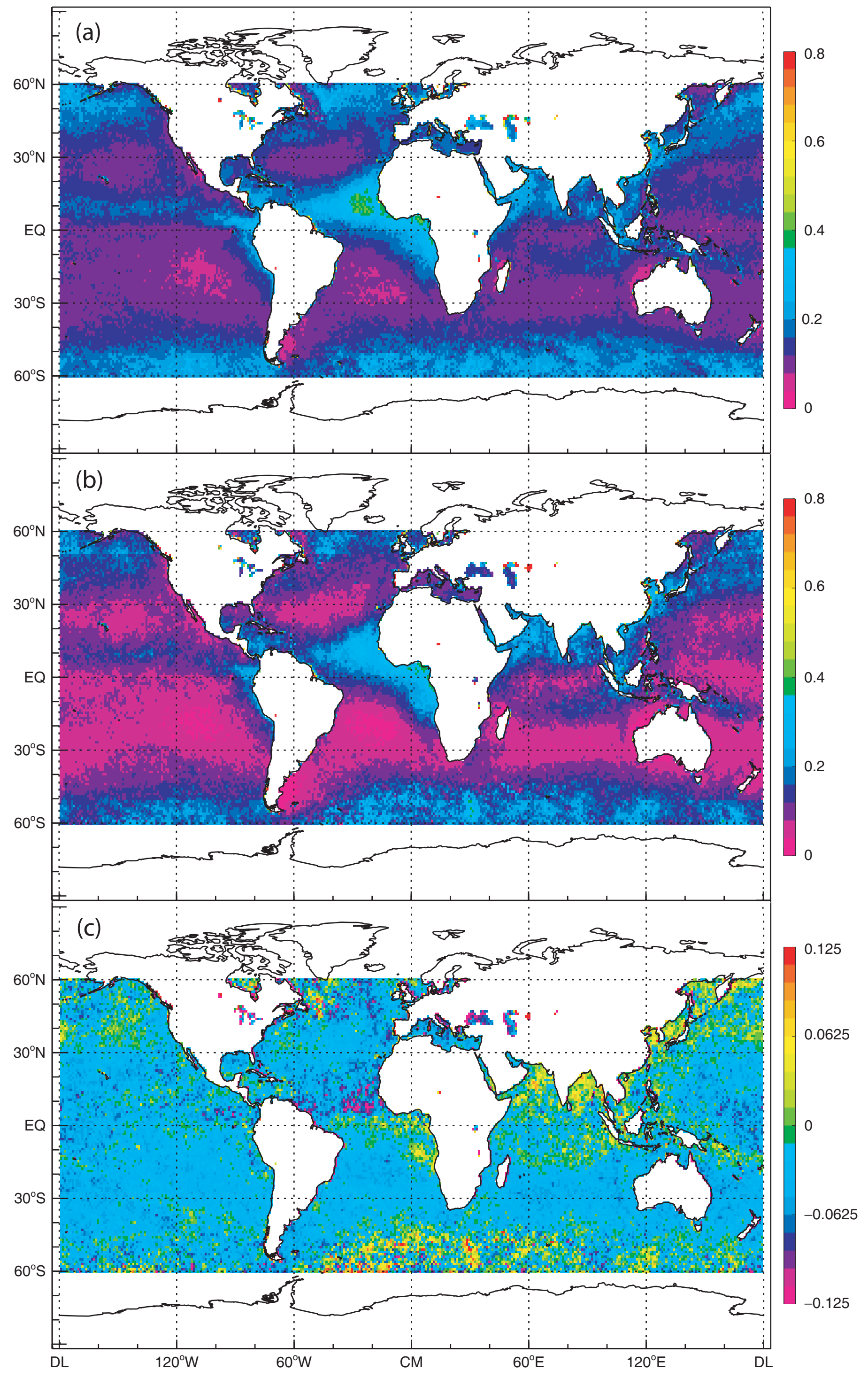

**Plate 3.2**. (a) AOT averaged over the period July 1988 – June 1991. (b) AOT averaged over the period July 2002 – June 2005. (c) Difference between the AOTs in panels (b) and (a).

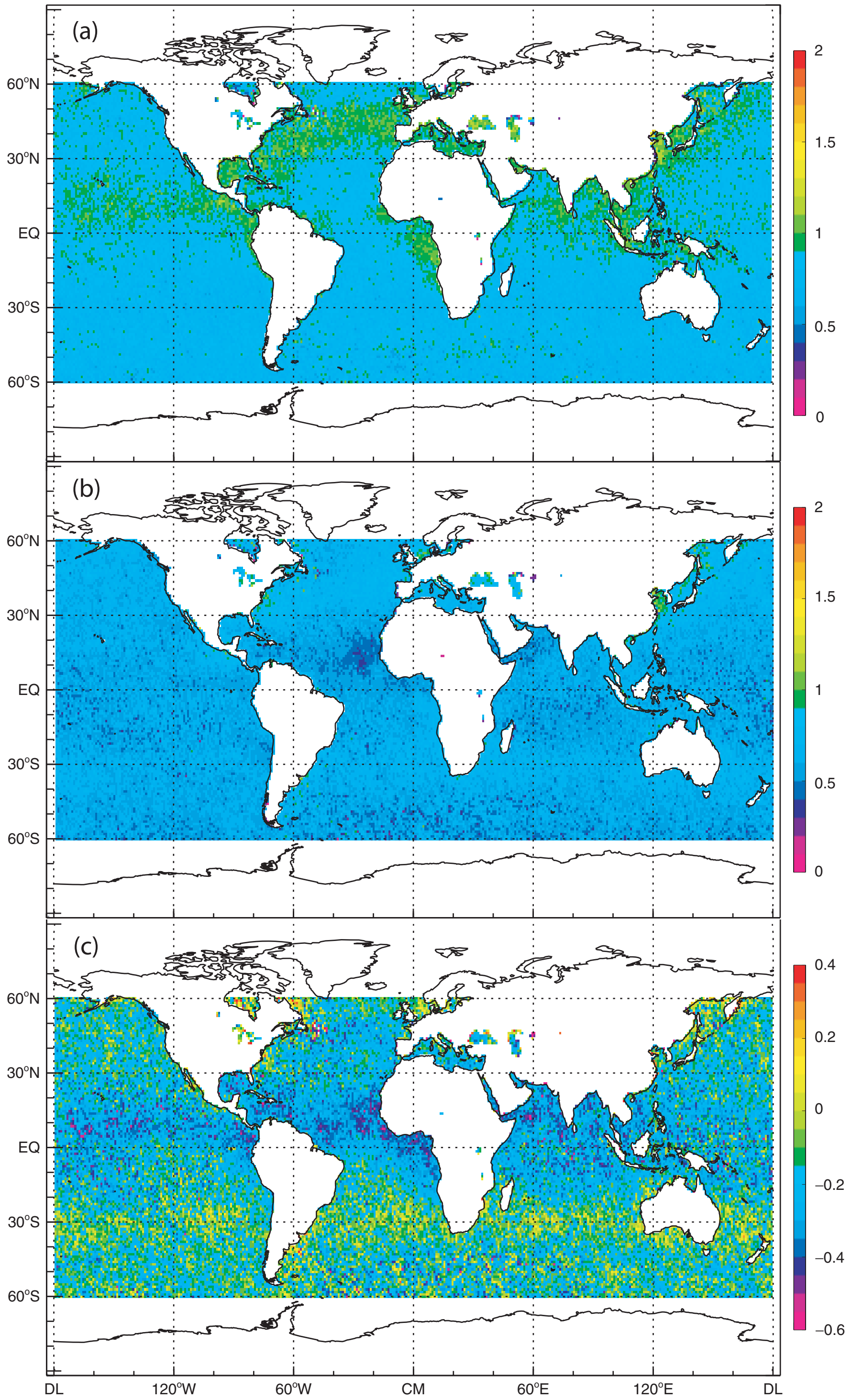

**Plate 3.3**. Ångström exponent averaged over the periods July 1988 – June 1991 (a) and July 2002 – June 2005 (b). (c) Difference between the Ångström exponents in panels (b) and (a).

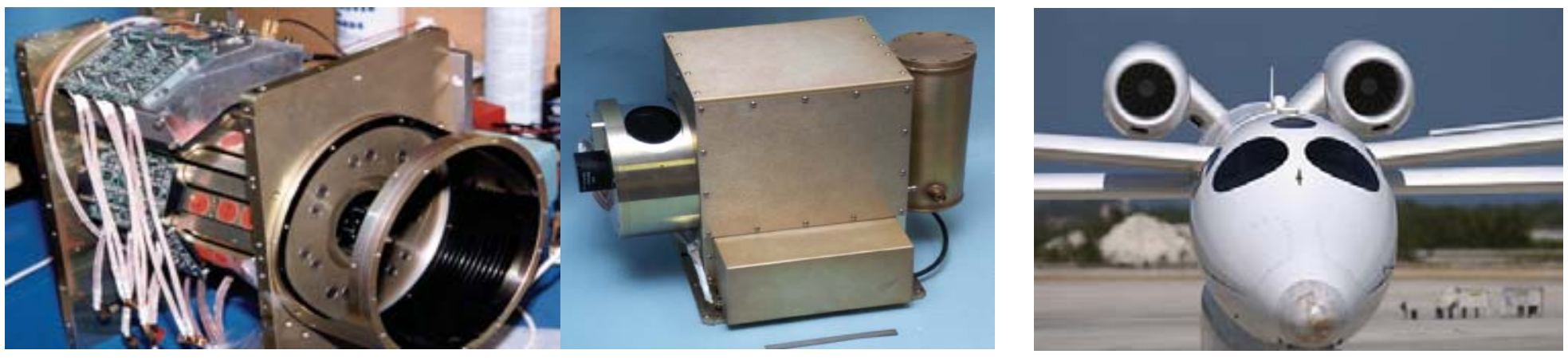

**Plate 3.4.** Polarimeter RSP (left) designed for remote-sensing studies of aerosols and clouds from high-flying NASA aircraft such as the Proteus (right).

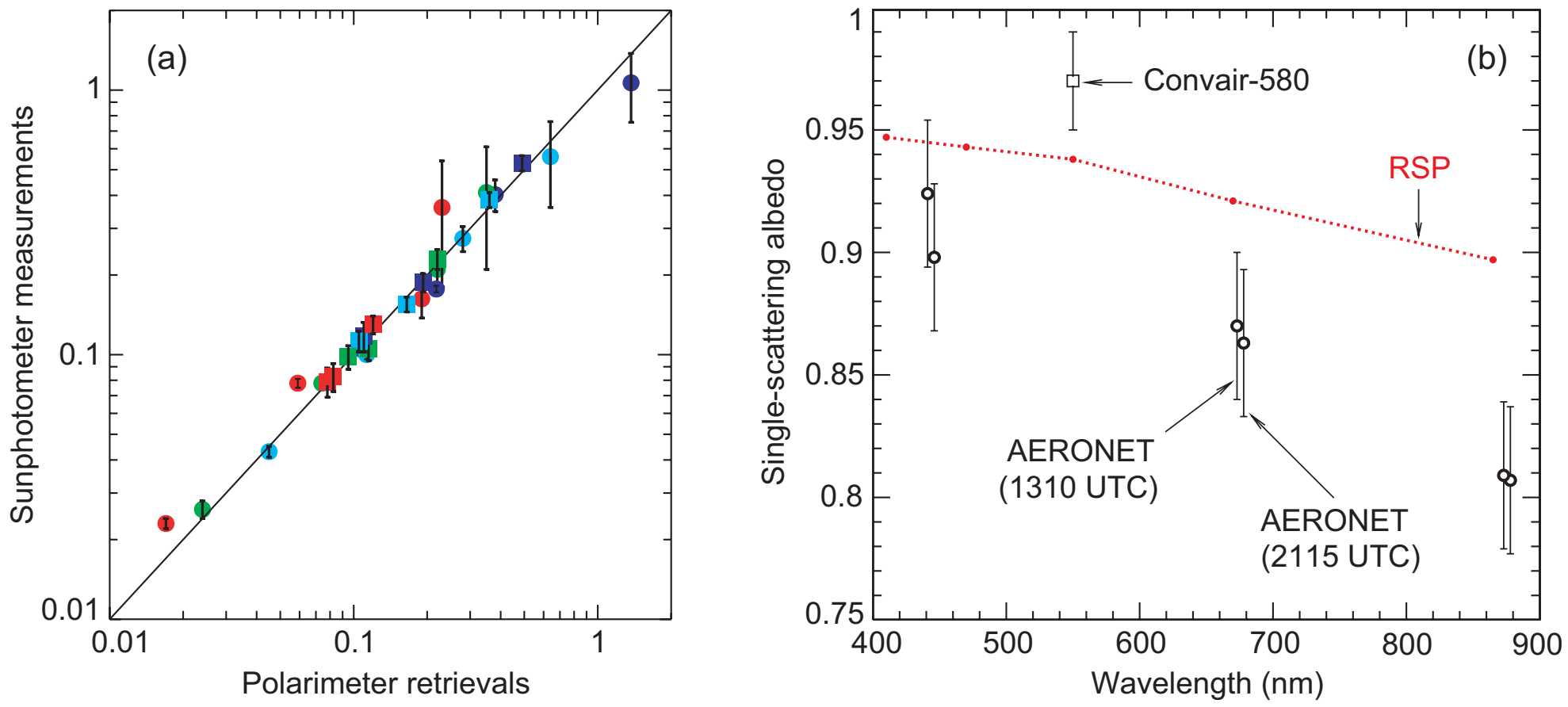

**Plate 3.5.** (a) Optical thickness comparison. Sunphotometer measurements at 410/443, 500, 673, and 865 nm shown as blue, turquoise, green, and red symbols, respectively, are compared with RSP retrievals for the same wavelength. The circular symbols are for retrievals over land while the square symbols are for retrievals over ocean. Error bars are only shown for the sunphotometer measurements. (b) Single-scattering albedos as a function of wavelength. The red dotted line shows the best-estimate values retrieved from RSP data. Also included are estimates from data collected during Convair-580 flight 1874 and from the AERONET data.

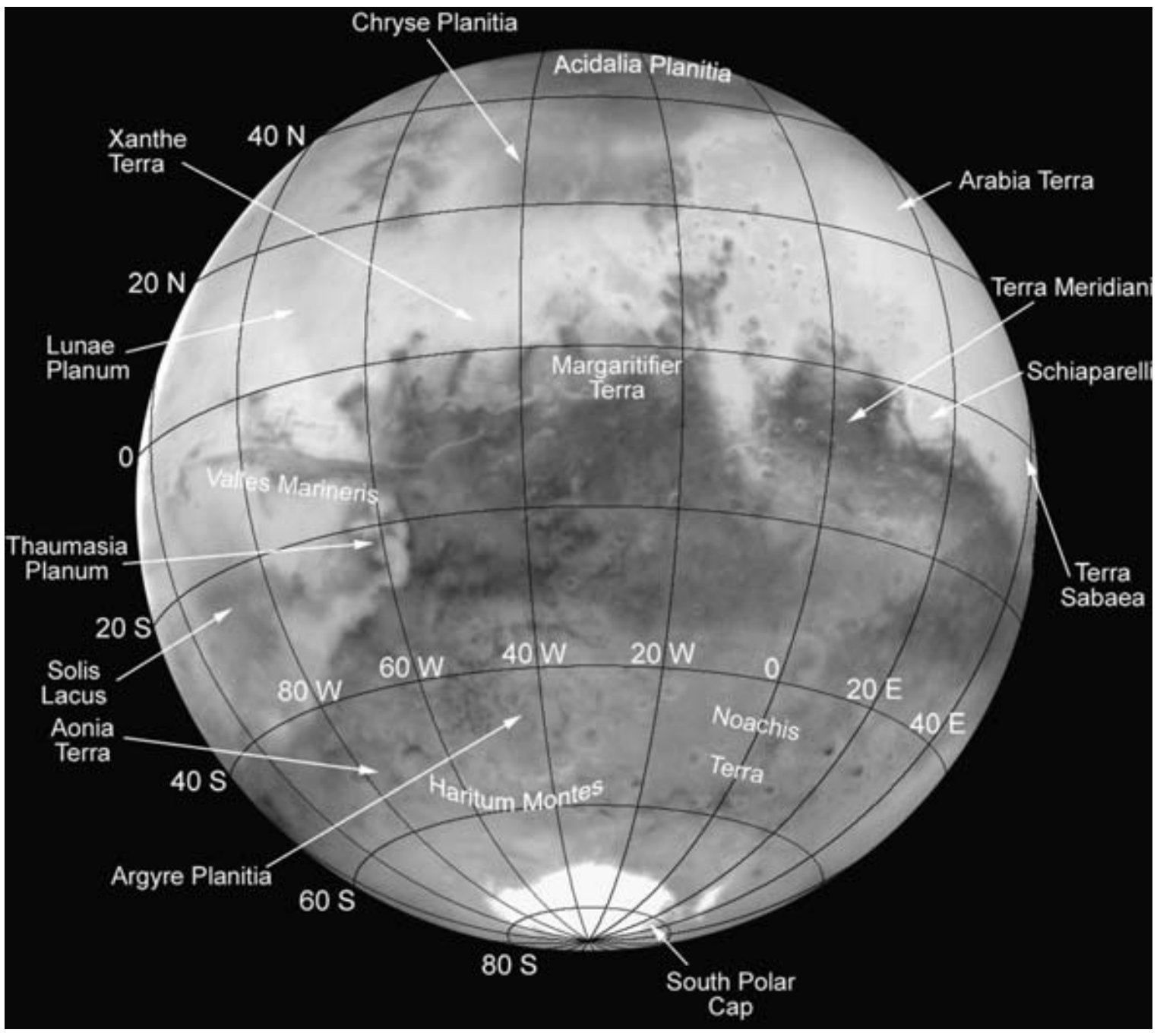

**Plate 3.6.** The map of the martian hemisphere centered at ~19°S and ~34°W as represented by an image acquired in the red spectral band (filter F658W).

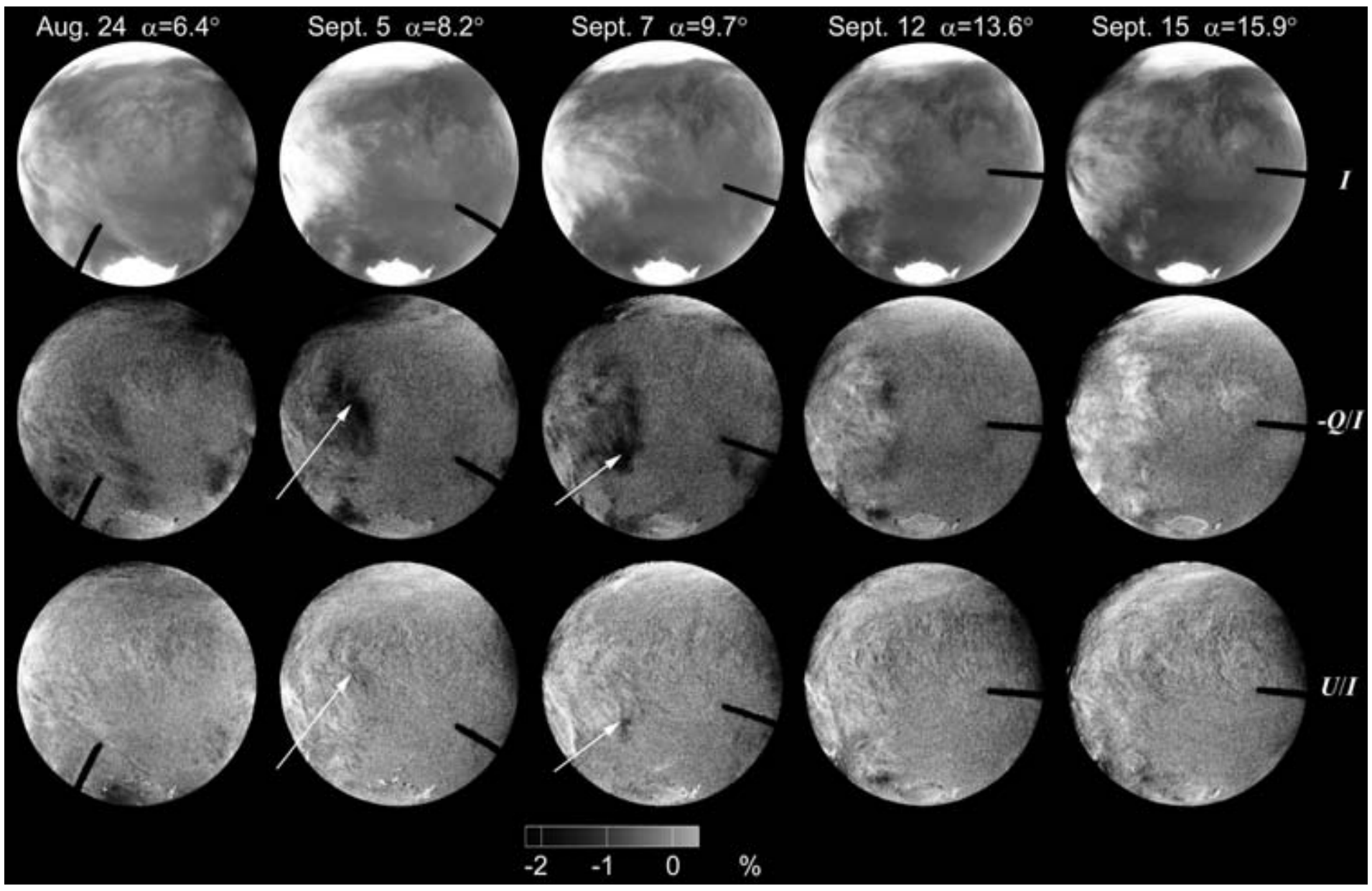

**Plate 3.7.** Intensity *I* and ratios −*Q*/*I* and *U*/*I* for Mars on all five observation dates measured with the filter F330W. The Stokes parameters are defined with respect to the photometric-equator-related reference frame. The black coronographic finger shadows parts of the detector. The arrows indicate the polarimetric transient effect.

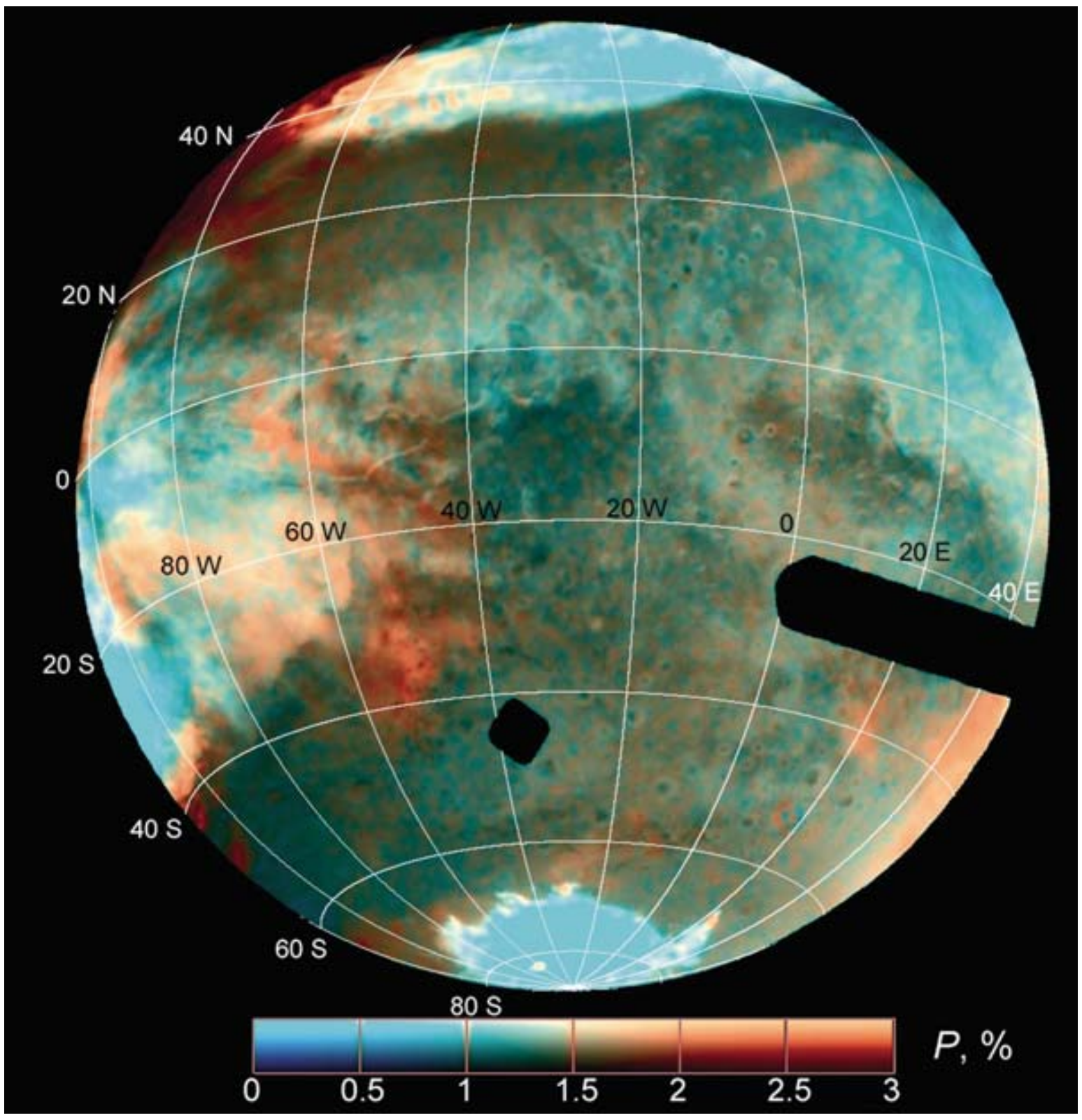

**Plate 3.8.** Color-coded distribution of the degree of linear polarization $(Q^2+U^2)^{1/2}/I$ over the martian disk is superposed on the brightness image (filter F435W; September 7, 2003).

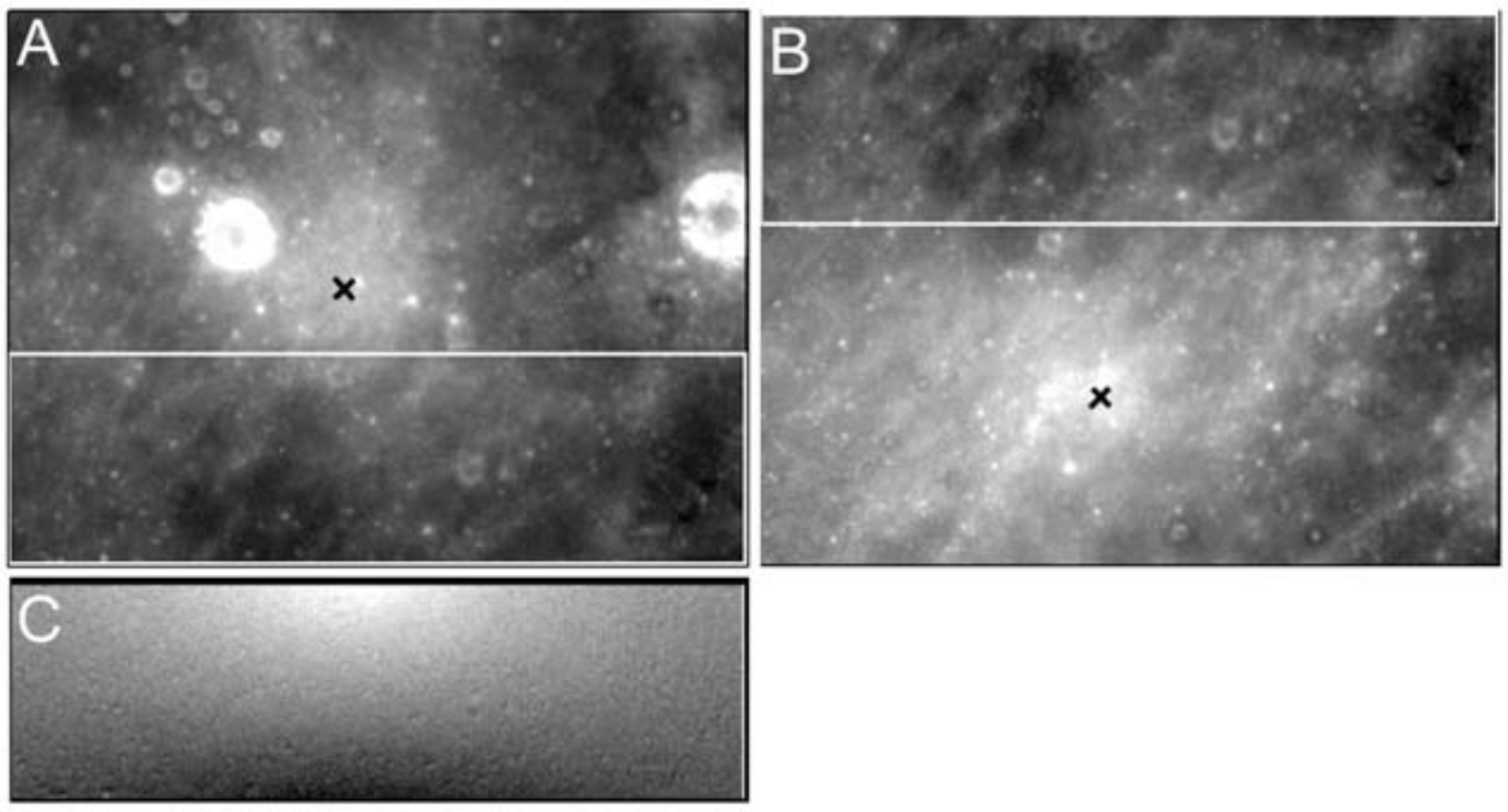

**Plate 3.9.** Images obtained with the UVVis camera onboard the Clementine spacecraft. The south is above. Craters Bruce and Blagg in the north-eastern part of Sinus Medii are seen in the source image (A). The adjacent part of Sinus Medii is shown in (B). The ratio of the source images (C) reveals a pattern caused by a mutual shift of the opposition spot by ~1°. The crosses denote the opposition point $\alpha = 0°$.

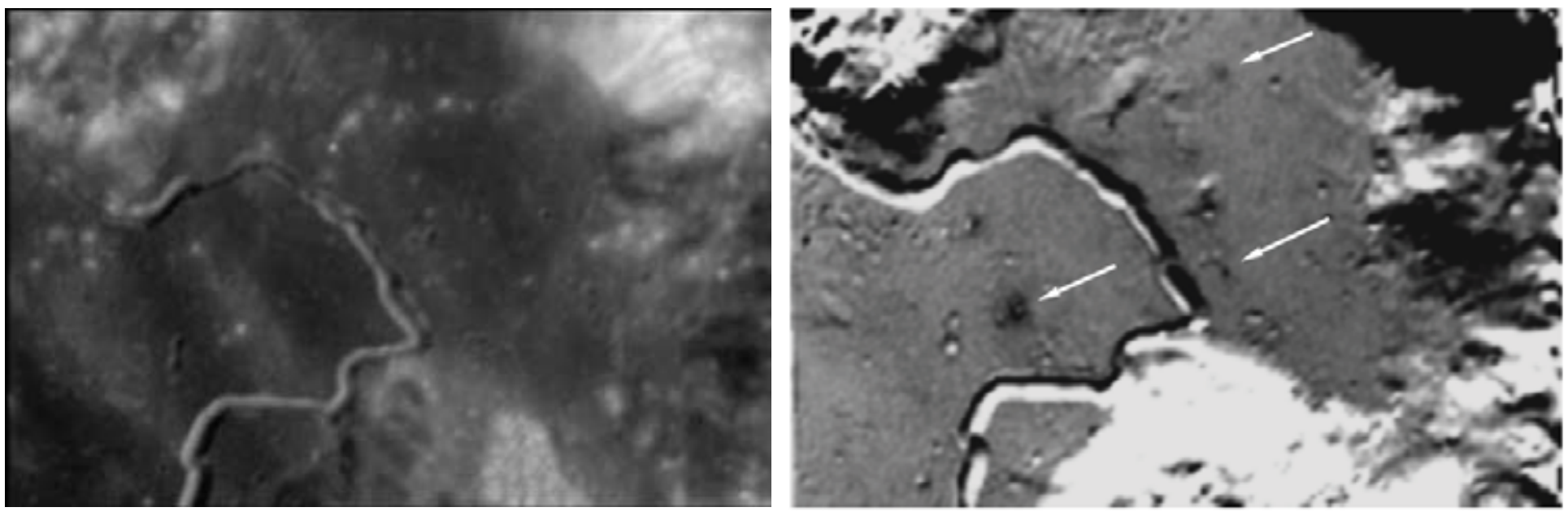

(a) (b)

**Plate 3.10.** (a) Region of the lunar surface surrounding the Apollo-15 landing site. (b) The corresponding map of the phase-function slope distribution. Darker shades represent flatter phase functions. The arrows indicate the most characteristic photometric anomalies.

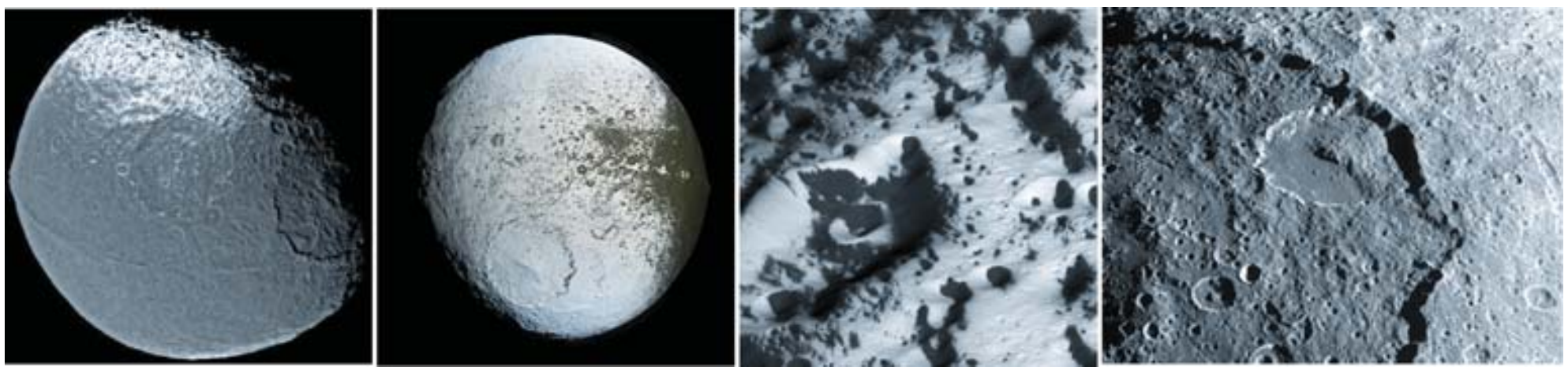

**Plate 3.11.** The left-hand image shows the trailing hemisphere of Iapetus covered with ancient impact craters. Essentially along the equator one can see a unique geological structure in the form of a mountain ridge which is more that 13 km tall, ~20 km wide, and at least 1,300 km long. The two central images show the distribution of bright and dark materials over the surface of Iapetus. The right-hand image shows a gigantic landslide indicative of the presence of fine-dispersed and porous surface material. The images were obtained from the Cassini spacecraft (courtesy NASA).

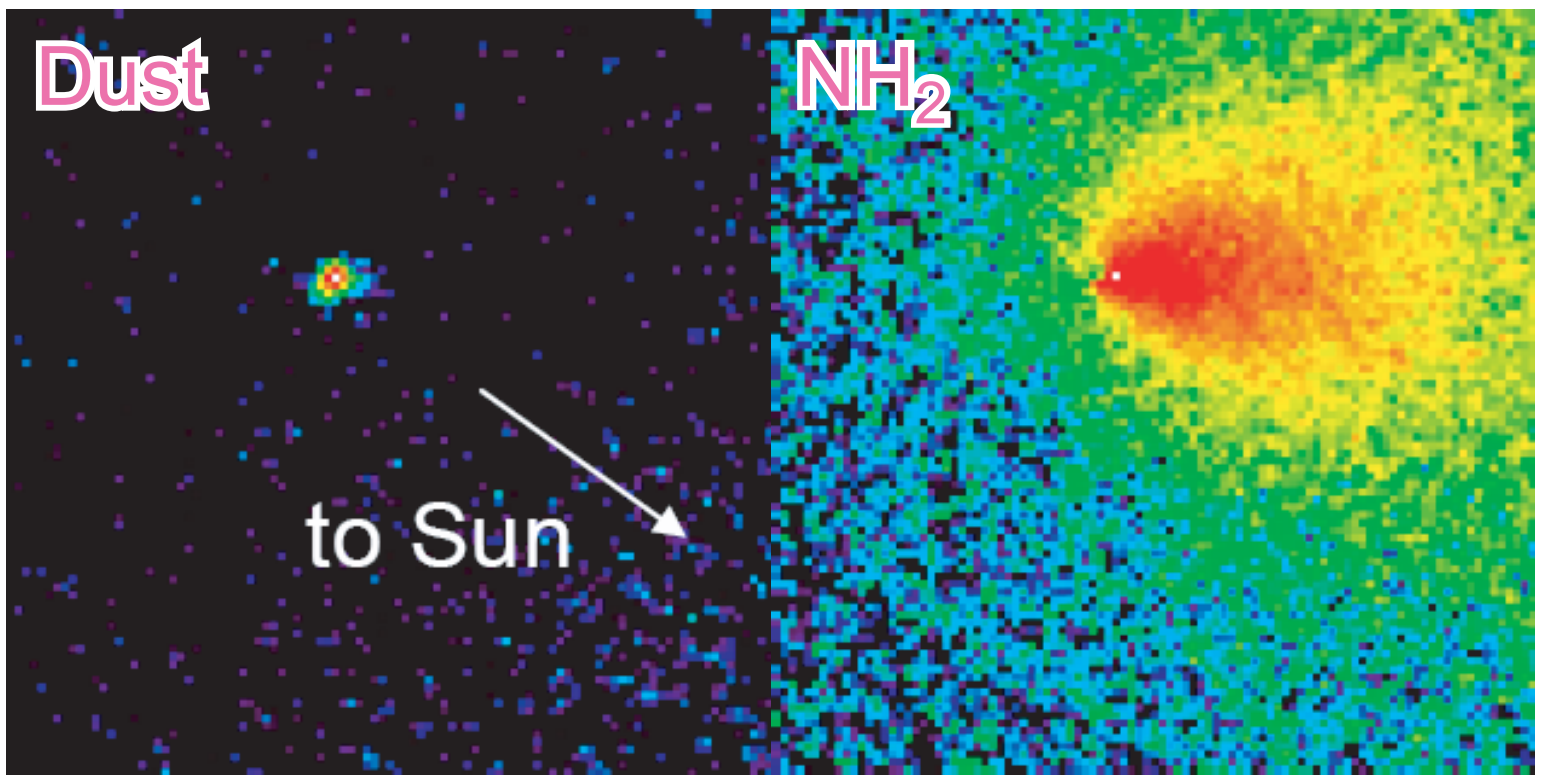

**Plate 4.1.** Optical images of dust and $NH_2$ in comet Encke obtained by K. Jockers, T. Bonev, and G. Borisov on the 2-m telescope of the Bulgarian National Astronomical Observatory (Rozhen) on 21 November 2003.

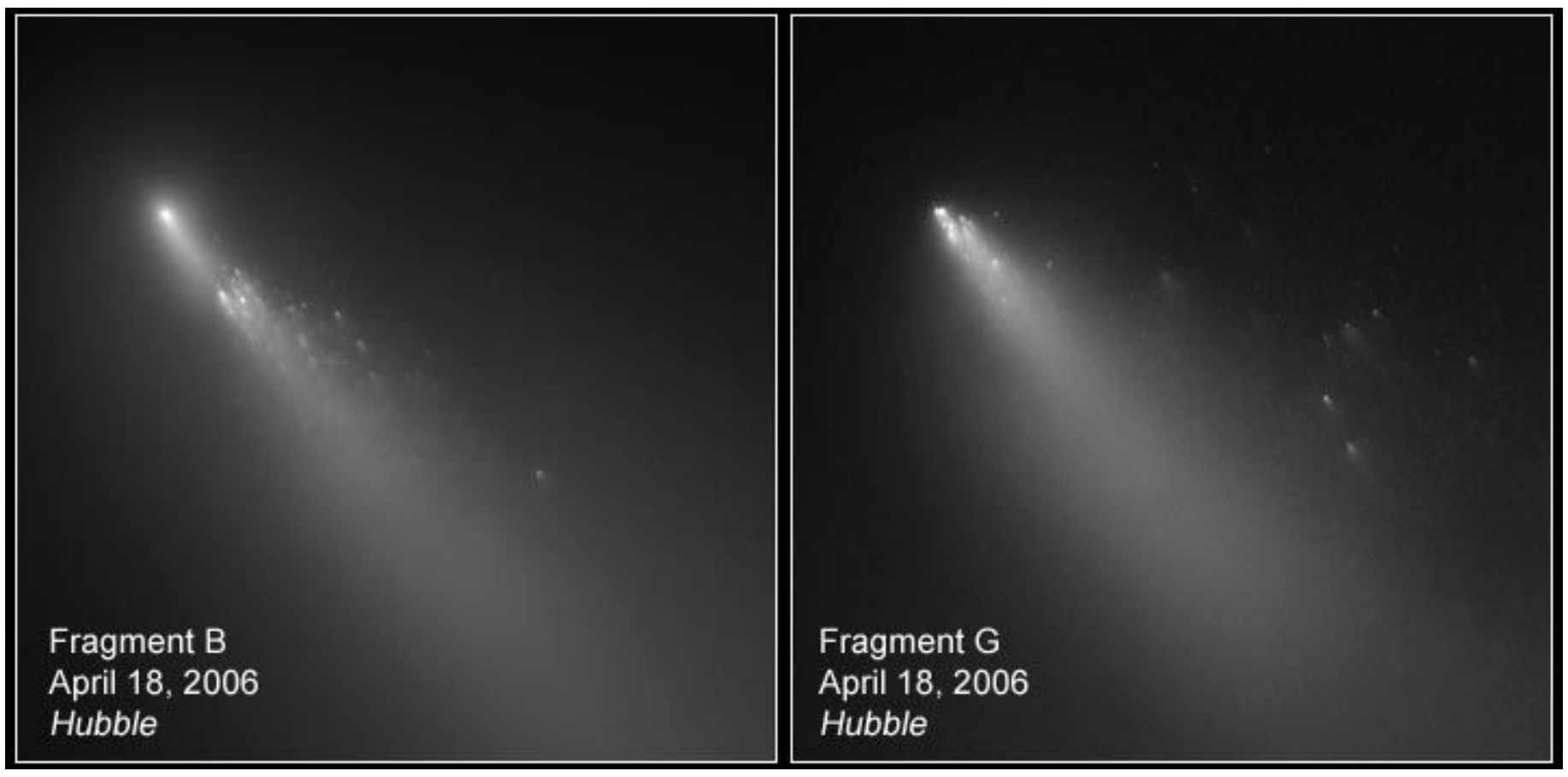

**Plate 4.2.** HST images documenting the fragmentation of the B and G subnuclei of comet Schwassmann–Wachmann 3 (courtesy NASA).

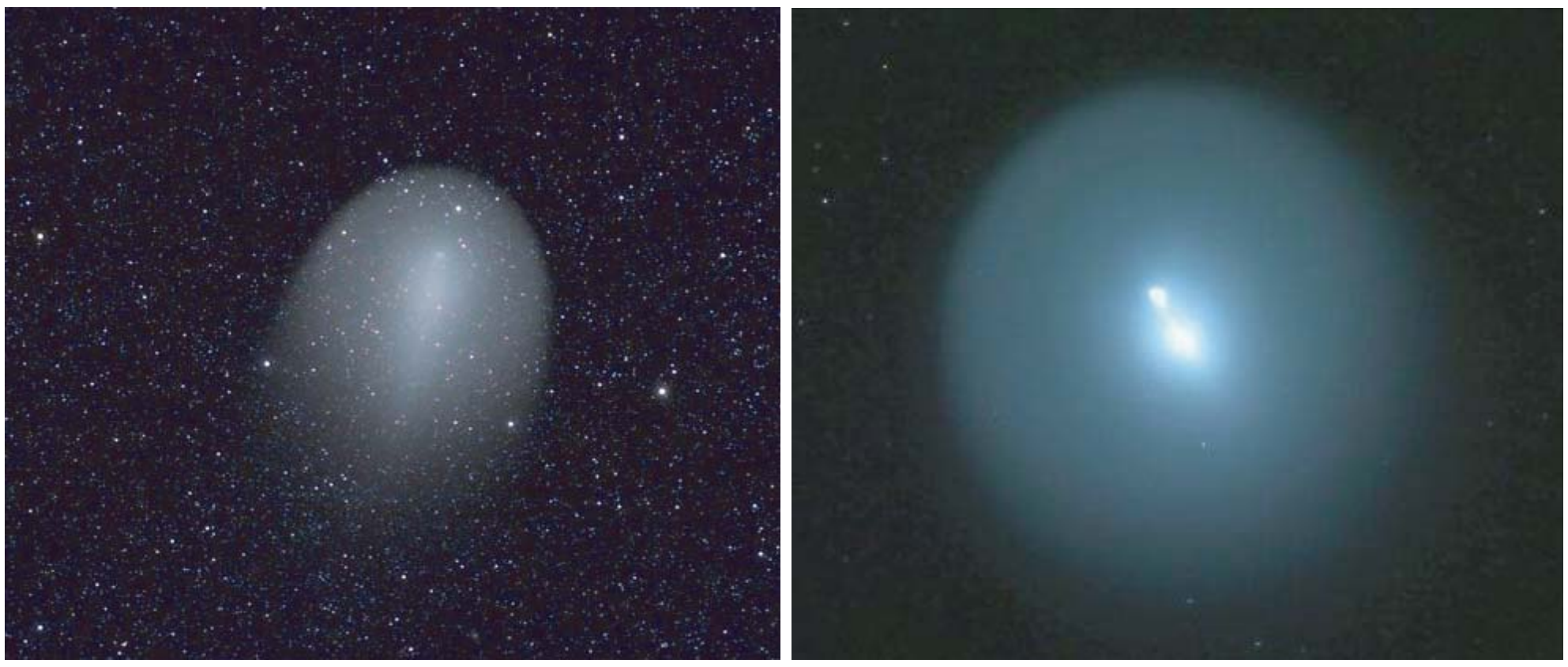

**Plate 4.3.** Images of comet Holmes taken on 3 and 27 November 2007 during a superoutburst (http://uk.geocities.com/martinmobberley/Comets.html; courtesy M. P. Mobberley).

# 1

# Electromagnetic scattering by discrete random media: unified microphysical theory

Correct and reliable interpretation of polarimetric remote-sensing observations of the Earth and other Solar System bodies is only possible in the framework of a rigorous, physically based theory of electromagnetic scattering by particles and particle groups. Therefore, one of the main objectives of our research has been the development of a unified microphysical approach to electromagnetic scattering based on direct solutions of the Maxwell equations with analytical or numerically exact computer techniques. Accordingly, the theory described below is based only on three fundamental premises: the macroscopic Maxwell equations, their integral form (viz., the rigorous vector form of the Foldy–Lax equations), and the numerically exact *T*-matrix technique for the computation of the scattered field. By using these well established principles as the starting point, we have addressed a wide range of electromagnetic scattering problems and made the theory directly applicable to the analysis of actual remote-sensing observations.

This chapter is intended to provide an accessible outline of the basic underlying theory as well as the more recent fundamental developments. It discusses elastic electromagnetic scattering by individual particles as well as random many-particle groups and summarizes the unified microphysical approach to the phenomena of radiative energy transport and weak localization (WL) of electromagnetic waves. In particular, we clarify the exact meaning of such fundamental concepts as single and multiple scattering, discuss the effect of physical parameters of particles on their single-scattering characteristics, demonstrate how the theories of radiative transfer (RT) and WL originate in the Maxwell equations, and expose and correct certain misconceptions of the traditional phenomenological approach to RT.

## 1.1. Basic assumptions

A terrestrial cloud consisting of randomly positioned and randomly moving water droplets or ice crystals is but one example of so-called discrete (or particulate) random media. The phenomenological theory of RT treats a cloud as a fictitious continuous medium in which the primary building unit is a vaguely defined elementary (or differential) volume element. In contrast, the microphysical theories of RT and WL account for the actual existence of particles as discrete inclusions with a refractive index different from that of the surrounding medium. Another fundamen-tal difference is that the microphysical approach explicitly starts with the Maxwell equations as basic physical laws governing the process of interaction of electromagnetic radiation with matter and invokes no *ad hoc* physical concepts and laws not already contained in classical electromagnetics. The word "microphysical" then serves to emphasize the direct traceability of the RT and WL theories from fundamental physics not afforded by the phenomenological approach.

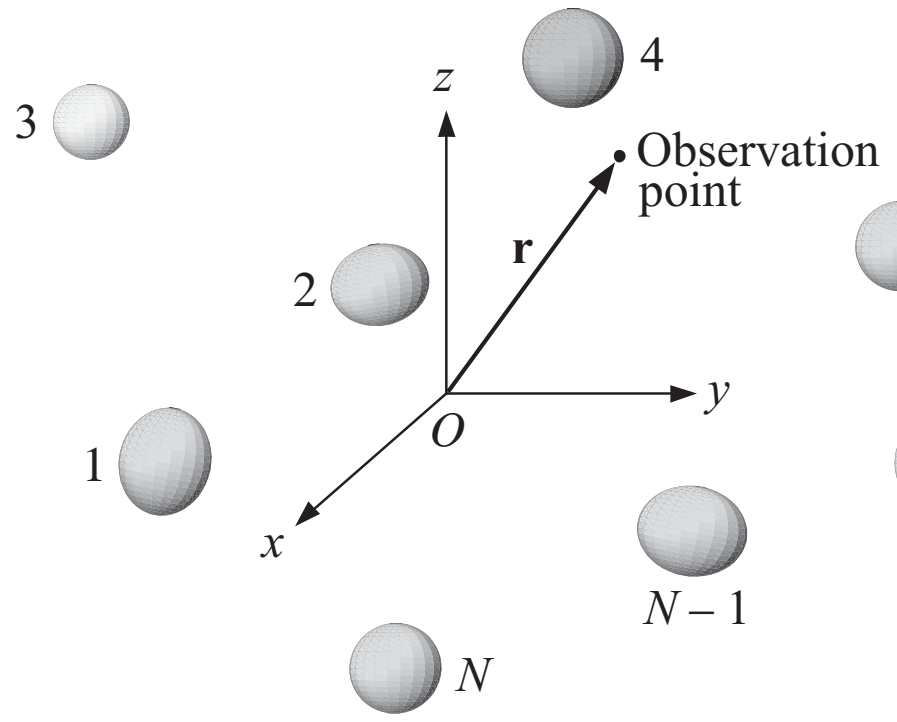

**Fig. 1.1.** Scattering object in the form of a group of $N$ discrete particles.

Specifically, the unified microphysical theory of electromagnetic scattering by particles and particle groups rests on the following well-defined assumptions intended to formulate the overall problem in strict physical terms:

1. At each moment in time, the entire scattering object (e.g., a cloud of water droplets or a regolith surface) can be represented by a specific spatial configuration of a number $N$ of discrete finite particles (Fig. 1.1). The unbounded host medium surrounding the scattering object is homogeneous, linear, isotropic, and nonabsorbing. Each particle is sufficiently large so that its atomic structure can be ignored and the particle can be characterized by optical constants appropriate to bulk matter. In electromagnetic terms, the presence of a particle means that the optical constants inside the particle volume are different from those of the surrounding host medium.

2. The entire scattering object is illuminated by either:

(i) a plane electromagnetic wave given by

$$\left.\begin{aligned} \mathbf{E}^{\mathrm{inc}}(\mathbf{r},t) &= \mathbf{E}_0^{\mathrm{inc}} \exp(\mathrm{i}\mathbf{k}^{\mathrm{inc}}\cdot\mathbf{r} - \mathrm{i}\omega t) \\ \mathbf{H}^{\mathrm{inc}}(\mathbf{r},t) &= \mathbf{H}_0^{\mathrm{inc}} \exp(\mathrm{i}\mathbf{k}^{\mathrm{inc}}\cdot\mathbf{r} - \mathrm{i}\omega t) \end{aligned}\right\} \quad \mathbf{r}\in\Re^3 \tag{1.1}$$

with constant amplitudes $\mathbf{E}_0^{\mathrm{inc}}$ and $\mathbf{H}_0^{\mathrm{inc}}$, where $\mathbf{E}$ is the electric and $\mathbf{H}$ the magnetic field, $t$ is time, $\mathbf{r}$ is the position vector, $\omega$ is the angular frequency, $\mathbf{k}^{\mathrm{inc}}$ is the real-valued wave vector, $\mathrm{i} = (-1)^{1/2}$, and $\Re^3$ denotes the entire three-dimensional space; or

(ii) a quasi-monochromatic parallel beam of light of infinite lateral extent given by the following formulas:

$$\left.\begin{aligned} \mathbf{E}^{\mathrm{inc}}(\mathbf{r},t) &= \mathbf{E}_0^{\mathrm{inc}}(t)\exp(\mathrm{i}\mathbf{k}^{\mathrm{inc}}\cdot\mathbf{r}-\mathrm{i}\omega t) \\ \mathbf{H}^{\mathrm{inc}}(\mathbf{r},t) &= \mathbf{H}_0^{\mathrm{inc}}(t)\exp(\mathrm{i}\mathbf{k}^{\mathrm{inc}}\cdot\mathbf{r}-\mathrm{i}\omega t) \end{aligned}\right\} \quad \mathbf{r}\in\Re^3, \tag{1.2}$$

where fluctuations in time of the complex amplitudes of the electric and magnetic fields, $\mathbf{E}_0^{\mathrm{inc}}(t)$ and $\mathbf{H}_0^{\mathrm{inc}}(t)$, around their respective mean values occur much more slowly than the harmonic oscillations of the time factor $\exp(-\mathrm{i}\omega t)$.

This restriction explicitly excludes other types of illumination such as a focused laser beam of finite lateral extent or a pulsed beam.

3. Nonlinear optics effects are excluded by assuming that the optical constants of both the scattering object and the surrounding medium are independent of the electric and magnetic fields.

4. It is assumed that electromagnetic scattering is elastic. This means that the scattered light has the same frequency as the incident light, which excludes inelastic scattering phenomena such as Raman and Brillouin scattering as well as the specific consideration of the small Doppler shift of frequency of the scattered light relative to that of the incident light due the movement of the particles with respect to the source of illumination.

5. It is assumed that any significant changes in the scattering object (e.g., changes in particle positions and/or orientations with respect to the laboratory reference frame) occur (i) over time intervals $T$ much longer than the period of time-harmonic oscillations of the electromagnetic field, $T \gg 2\pi/\omega$; and (ii) much more slowly than temporal changes of the amplitudes $\mathbf{E}_0^{\mathrm{inc}}(t)$ and $\mathbf{H}_0^{\mathrm{inc}}(t)$.

6. The phenomenon of thermal emission is excluded. The validity of this assumption depends on the combination {object's temperature; wavelength}. For example, it is usually valid for objects at room or lower temperature and for short-wave infrared and shorter wavelengths.

## 1.2. The macroscopic Maxwell equations

The assumptions listed in the preceding section imply that all fields and sources are time harmonic and allow one to fully describe the total electromagnetic field at any moment in time everywhere in space as the solution of the so-called frequency-domain macroscopic differential Maxwell equations (Stratton 1941). The specific dependence of the optical constants on spatial coordinates and the corresponding boundary conditions at any moment are fully defined by the instantaneous geometrical configuration of the $N$ particles (Fig. 1.1).

Specifically, it is convenient to factor out the time-harmonic dependence of the electric and magnetic fields: $\mathbf{E}(\mathbf{r},t)=\exp(-\mathrm{i}\omega t)\,\mathbf{E}(\mathbf{r})$ and $\mathbf{H}(\mathbf{r},t)=\exp(-\mathrm{i}\omega t)\,\mathbf{H}(\mathbf{r})$. The frequency-domain monochromatic Maxwell curl equations describing the scattering problem in terms of the electric and magnetic field amplitudes $\mathbf{E}(\mathbf{r})$ and $\mathbf{H}(\mathbf{r})$ can then be written as follows:

$$\left.\begin{aligned} \nabla\times\mathbf{E}(\mathbf{r}) &= \mathrm{i}\omega\mu_0\mathbf{H}(\mathbf{r}) \\ \nabla\times\mathbf{H}(\mathbf{r}) &= -\mathrm{i}\omega\varepsilon_1\mathbf{E}(\mathbf{r}) \end{aligned}\right\} \quad \mathbf{r}\in V_{\mathrm{EXT}}, \tag{1.3}$$

$$\left.\begin{aligned}\nabla\times\mathbf{E}(\mathbf{r}) &= \mathrm{i}\omega\mu_0\mathbf{H}(\mathbf{r})\\ \nabla\times\mathbf{H}(\mathbf{r}) &= -\mathrm{i}\omega\varepsilon_2(\mathbf{r},\omega)\mathbf{E}(\mathbf{r})\end{aligned}\right\}\quad \mathbf{r}\in V_{\mathrm{INT}}. \tag{1.4}$$

In these equations, $V_{\mathrm{INT}}$ is the cumulative "interior" volume occupied by the scattering object; $V_{\mathrm{EXT}}$ is the infinite exterior region such that $V_{\mathrm{INT}}\cup V_{\mathrm{EXT}}=\Re^3$; the host medium and the scattering object are assumed to be nonmagnetic; $\mu_0$ is the permeability of a vacuum; $\varepsilon_1$ is the real-valued electric permittivity of the host medium; and $\varepsilon_2(\mathbf{r},\omega)$ is the complex permittivity of the scattering object. Since the first relations in Eqs. (1.3) and (1.4) yield the magnetic field provided that the electric field is known everywhere, the solution of Eqs. (1.3) and (1.4) is usually sought in terms of only the electric field.

Although the amplitudes $\mathbf{E}(\mathbf{r})$ and $\mathbf{H}(\mathbf{r})$ do not depend on time explicitly, they can change in time implicitly if the incident light is quasi-monochromatic and/or as a consequence of temporal variability of the scattering object. However, such changes occur much more slowly than the time-harmonic oscillations described by the factor $\exp(-\mathrm{i}\omega t)$, which explains the common practical applicability of the frequency-domain Maxwell equations.

It should of course be recognized that macroscopic electromagnetics ignores the discreteness of matter forming the scattering object and operates with continuous sources of fields. Therefore, its predictions can fall short in cases where quantum effects are important. Even so, the quantum theory can often be used to determine the macroscopic electromagnetic properties of bodies consisting of very large numbers of atoms (Akhiezer and Peletminskii 1981). This approach works for particles larger than about 50 Å (Huffman 1988), thereby implying a very wide range of validity of macroscopic electromagnetics. Thus, our use of macroscopic electromagnetics as the point of departure is founded on the well-established fact that this theory follows directly from more fundamental physical theories as a consequence of well-characterized and verifiable approximations. In other words, the equations of classical macroscopic electromagnetics are accepted here essentially as basic axioms valid in a wide and well-defined range of relevant situations. The reader will see that this approach allows the development of a self-contained and self-consistent scattering theory in which the need to invoke alternative physical concepts and laws is completely obviated.

## 1.3. Electromagnetic scattering

The fundamental solution of the Maxwell equations in the form of a time-harmonic plane electromagnetic wave, Eq. (1.1), represents the transport of electromagnetic energy from one point to another and embodies the concept of a perfectly monochromatic parallel beam of light. In particular, a plane electromagnetic wave propagates in an infinite nonabsorbing medium without a change in its intensity or polarization state (see Fig. 1.2a). However, the presence of a finite object modifies the electromagnetic field that would otherwise exist in the unbounded homogeneous space. This modification is called *electromagnetic scattering*.

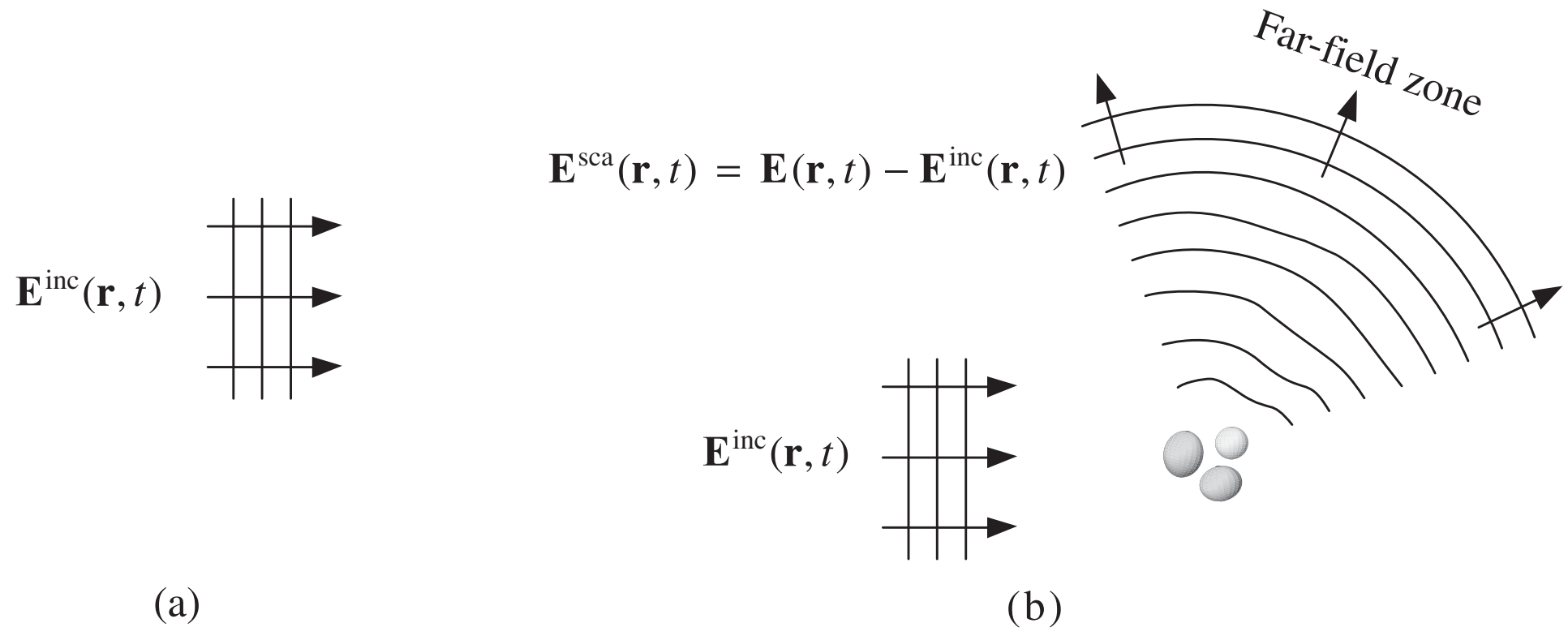

**Fig. 1.2.** Scattering by a fixed finite object. In this case the object consists of three disjoint, stationary particles.

The difference between the total field in the presence of the object, $\mathbf{E}(\mathbf{r},t)$, and the total field that would exist in the absence of the object, $\mathbf{E}^{inc}(\mathbf{r},t)$, can be thought of as the field scattered by the object, $\mathbf{E}^{sca}(\mathbf{r},t)$ (Fig. 1.2b). In other words, the total field in the presence of the object is represented as the vector sum of the respective "incident" and "scattered" fields:

$$\mathbf{E}(\mathbf{r},t) = \mathbf{E}^{inc}(\mathbf{r},t) + \mathbf{E}^{sca}(\mathbf{r},t). \tag{1.5}$$

The reader should recognize that the separation of the total field into the incident and scattered fields according to Eq. (1.5) is a purely mathematical procedure. This means that classical electromagnetic scattering is not a physical process per se but rather an abbreviated way to state that the total field computed in the presence of an object is different from that computed in the absence of the object (Mishchenko 2008a, 2009; Mishchenko and Travis 2008). To "describe electromagnetic scattering" then means to quantify the difference between the two fields as a function of the object's physical properties.

This explains why any measurement of electromagnetic scattering is always reduced to the measurement of certain optical observables first in the absence of the object and then in the presence of the object. The differences between the readings of detectors of electromagnetic energy quantify the scattering and absorption properties of the object and can often be interpreted in order to infer the object's microphysical properties. The first measurement stage can sometimes be implicit (e.g., it is often bypassed by assuming that the reading of a detector in the absence of the scattering object is zero) or is called "detector calibration", but this does not change the two-stage character of any scattering measurement.

We have already mentioned that the practical applicability of the frequency-domain formalism implies the stationarity of the electromagnetic field over a time interval long compared with the period of time-harmonic oscillations. Therefore, this formalism can be used to describe scattering of quasi-monochromatic as well as monochromatic light.

An especially transparent description of electromagnetic scattering is afforded by the so-called volume integral equation (VIE) which follows from the frequency-domain macroscopic Maxwell equations and is exact (Saxon 1955a):

$$\mathbf{E}(\mathbf{r}) = \mathbf{E}^{\rm inc}(\mathbf{r}) + k_1^2 \int_{V_{\rm INT}} d\mathbf{r}' \vec{G}(\mathbf{r},\mathbf{r}') \cdot \mathbf{E}(\mathbf{r}')[m^2(\mathbf{r}')-1]$$

$$= \mathbf{E}^{\rm inc}(\mathbf{r}) + k_1^2 \left( \vec{I} + \frac{1}{k_1^2}\nabla \otimes \nabla \right) \cdot \int_{V_{\rm INT}} d\mathbf{r}' \,\mathbf{E}(\mathbf{r}') \frac{\exp(ik_1|\mathbf{r}-\mathbf{r}'|)}{4\pi|\mathbf{r}-\mathbf{r}'|}[m^2(\mathbf{r}')-1], \quad (1.6)$$

where $\mathbf{r} \in \Re^3$, the common time-harmonic factor $\exp(-i\omega t)$ is omitted, $k_1 = |\mathbf{k}^{\rm inc}| = \omega(\varepsilon_1\mu_0)^{1/2}$ is the wave number in the host medium, $m(\mathbf{r}') = [\varepsilon_2(\mathbf{r}',\omega)/\varepsilon_1]^{1/2}$ is the refractive index of the interior relative to that of the host exterior medium, $\vec{G}(\mathbf{r},\mathbf{r}')$ is the free space dyadic Green's function, $\vec{I}$ is the identity dyadic, and $\otimes$ is the dyadic product sign. One can see that the VIE expresses the total field everywhere in space in terms of the total internal field. If the scattering object is absent (i.e., $m(\mathbf{r}') \equiv 1$), then the total field is identically equal to the incident field. Otherwise the total field contains a scattering component given by the second term on the right-hand side of Eq. (1.6). Since the internal field is not known in general, it must be found by solving the VIE either analytically or numerically.

The VIE makes explicit two fundamental facts. First, the phenomenon of electromagnetic scattering is not limited to the case of the incident field in the form of a plane wave. In fact, it encompasses any incident field as long as the latter satisfies the Maxwell equations, e.g., spherical and cylindrical electromagnetic waves.

Second, irrespective of the morphological structure of the scattering object the latter remains a single, unified scatterer. Although the human eye may classify the scattering object as a "collection of discrete particles", the incident field always perceives the object as one scatterer in the form of a specific spatial distribution of the relative refractive index. The latter point can be made even more explicit by expressing the scattered electric field in terms of the incident field:

$$\mathbf{E}^{\rm sca}(\mathbf{r}) = \int_{V_{\rm INT}} d\mathbf{r}' \vec{G}(\mathbf{r},\mathbf{r}') \cdot \int_{V_{\rm INT}} d\mathbf{r}'' \vec{T}(\mathbf{r}',\mathbf{r}'') \cdot \mathbf{E}^{\rm inc}(\mathbf{r}''), \quad \mathbf{r} \in \Re^3, \quad (1.7)$$

where $\vec{T}$ is the so-called dyadic transition operator of the scattering object (Tsang et al. 1985). Substituting Eq. (1.7) in Eq. (1.6) yields the following equation for $\vec{T}$:

$$\vec{T}(\mathbf{r},\mathbf{r}') = k_1^2[m^2(\mathbf{r})-1]\,\delta(\mathbf{r}-\mathbf{r}')\vec{I}$$

$$+ k_1^2[m^2(\mathbf{r})-1] \int_{V_{\rm INT}} d\mathbf{r}'' \vec{G}(\mathbf{r},\mathbf{r}'') \cdot \vec{T}(\mathbf{r}'',\mathbf{r}'), \quad \mathbf{r},\mathbf{r}' \in V_{\rm INT}, \quad (1.8)$$

where $\delta(\mathbf{r})$ is the three-dimensional delta function.

It has not been proven mathematically that Eq. (1.8) has a solution and that this solution is unique. Therefore, one has to believe in the existence and uniqueness of solution based on what is usually called "simple physical considerations". However, the indisputable advantage of Eqs. (1.7) and (1.8) is that $\vec{T}$ is the property of the scattering object only and is independent of the incident field. Furthermore, $\vec{T}$ provides a complete description of electromagnetic scattering by the object for any incident time-harmonic field. We will see later that the concept of dyadic transition operator plays a central role in the theory of multiple scattering.

The ubiquitous presence of electromagnetic scattering in varying environments explains its fundamental importance in the modeling of electromagnetic energy transport for various science and engineering applications. This is also true of the situations in which electromagnetic scattering is induced artificially and used for particle characterization (e.g., Hoekstra et al. 2007; Лопатин и др. 2004). The exact theoretical and numerical techniques for the computation of the electromagnetic field scattered by a finite fixed object composed of one or several particles are many and are reviewed thoroughly by Wriedt et al (1999), Mishchenko et al. (2000a, 2002a), Doicu et al. (2000, 2006), Kahnert (2003), Babenko et al. (2003), and Wriedt (2009). All of these techniques have certain practical limitations in terms of the object's morphology and its size relative to the incident wavelength and cannot be used yet to describe electromagnetic scattering by large multi-particle objects such as atmospheric clouds, particulate surfaces, and particle suspensions. This makes imperative the use of well-characterized approximate solutions that do not require unrealistic computer resources while being sufficiently accurate for specific applications. One of the main objectives of this chapter is to demonstrate that the microphysical theories of RT and WL are two such approximations.

## 1.4. Far-field and near-field scattering

A fundamental property of the dyadic Green's function is the following asymptotic behavior:

$$\overleftrightarrow{G}(\mathbf{r},\mathbf{r}') \underset{r\to\infty}{\rightarrow} (\overleftrightarrow{I} - \hat{\mathbf{r}} \otimes \hat{\mathbf{r}}) \frac{\exp(\mathrm{i}k_1 r)}{4\pi r} \exp(-\mathrm{i}k_1 \hat{\mathbf{r}} \cdot \mathbf{r}'), \tag{1.9}$$

where $r = |\mathbf{r}|$, and $\hat{\mathbf{r}} = \mathbf{r}/r$. Placing the origin of the laboratory coordinate system $O$ close to the geometrical center of the scattering object (see Fig. 1.3a) and substituting Eqs. (1.1) and (1.9) in Eq. (1.7) yields (Mishchenko et al. 2002a, 2006b)

$$\mathbf{E}^{\mathrm{sca}}(\mathbf{r}) \underset{r\to\infty}{\rightarrow} \frac{\exp(\mathrm{i}k_1 r)}{r} \mathbf{E}_1^{\mathrm{sca}}(\hat{\mathbf{n}}^{\mathrm{sca}}) = \frac{\exp(\mathrm{i}k_1 r)}{r} \overleftrightarrow{A}(\hat{\mathbf{n}}^{\mathrm{sca}}, \hat{\mathbf{n}}^{\mathrm{inc}}) \cdot \mathbf{E}_0^{\mathrm{inc}}, \quad \hat{\mathbf{n}}^{\mathrm{sca}} \cdot \mathbf{E}_1^{\mathrm{sca}}(\hat{\mathbf{n}}^{\mathrm{sca}}) = 0. \tag{1.10}$$

Here, $\hat{\mathbf{n}}^{\mathrm{inc}} = \mathbf{k}^{\mathrm{inc}}/k_1$ is a unit vector in the incidence direction, $\hat{\mathbf{n}}^{\mathrm{sca}} = \hat{\mathbf{r}}$ is a unit vector in the scattering direction, and $\overleftrightarrow{A}$ is the so-called scattering dyadic such that

$$\hat{\mathbf{n}}^{\mathrm{sca}} \cdot \overleftrightarrow{A}(\hat{\mathbf{n}}^{\mathrm{sca}}, \hat{\mathbf{n}}^{\mathrm{inc}}) = \overleftrightarrow{A}(\hat{\mathbf{n}}^{\mathrm{sca}}, \hat{\mathbf{n}}^{\mathrm{inc}}) \cdot \hat{\mathbf{n}}^{\mathrm{inc}} = \mathbf{0}, \tag{1.11}$$

where $\mathbf{0}$ is a zero vector. The scattering dyadic has the dimension of length and describes the scattering of a plane electromagnetic wave in the so-called *far-field zone* of the object. It follows from Eq. (1.10) that the propagation of the scattered electromagnetic wave is radial and away from the object. Also, the electric and magnetic field vectors vibrate in the plane perpendicular to the propagation direction and their amplitudes decay inversely with distance from the object.

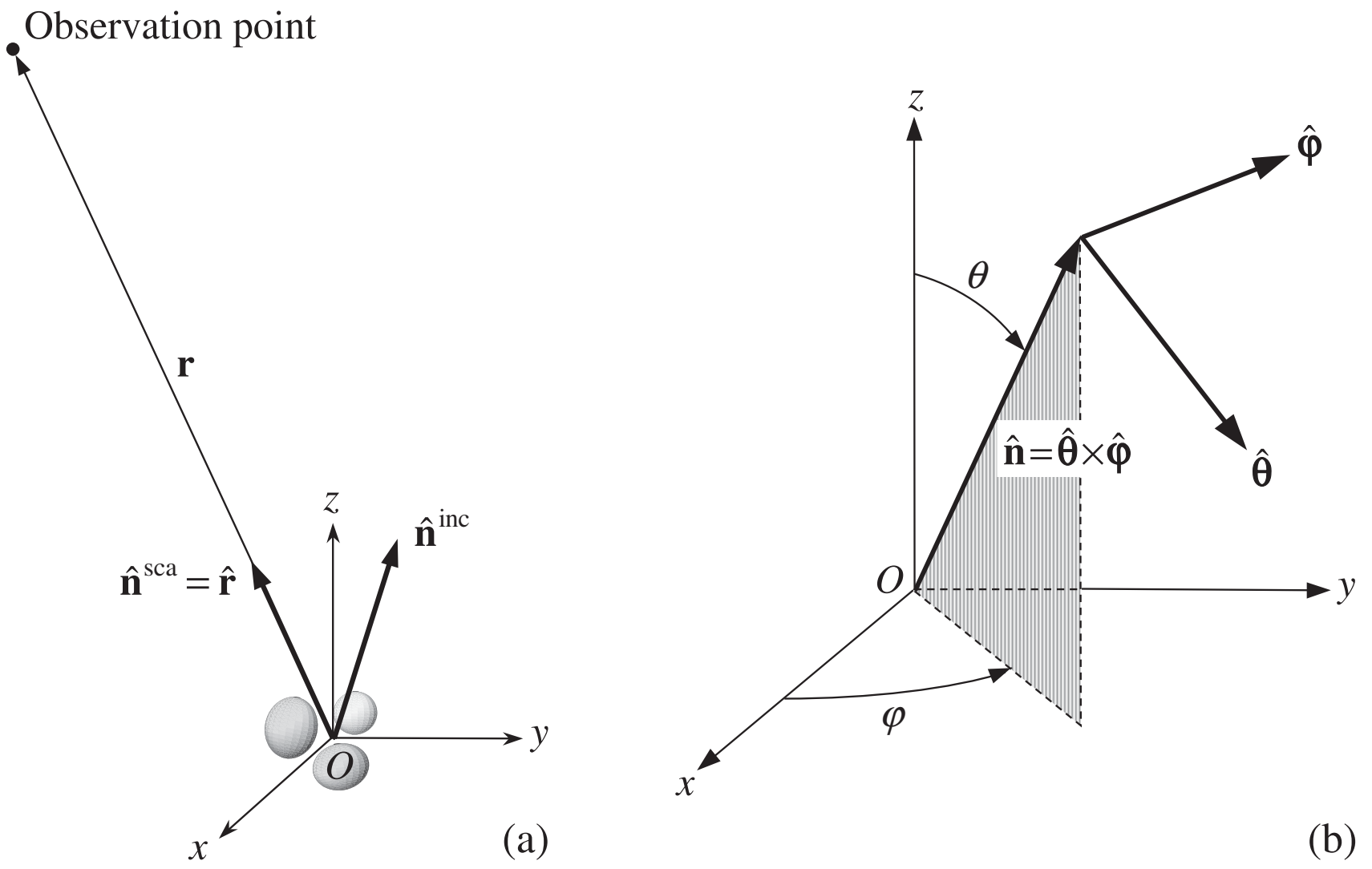

**Fig. 1.3.** (a) Scattering in the far-field zone of the entire object. (b) Right-handed spherical coordinate system.

The main convenience of the far-field approximation is that it reduces the scattered field to a simple outgoing spherical wave (Fig. 1.2b). Furthermore, Eq. (1.11) shows that only four out of the nine components of the scattering dyadic are independent in the spherical polar coordinate system centered at the origin (Fig. 1.3a). It is therefore convenient to introduce the $2 \times 2$ so-called amplitude scattering matrix $\mathbf{S}$ which describes the transformation of the $\theta$- and $\varphi$-components of the incident plane wave into the $\theta$- and $\varphi$-components of the scattered spherical wave:

$$\mathbf{E}^{\text{sca}}(r\hat{\mathbf{n}}^{\text{sca}}) = \frac{\exp(\mathrm{i}k_1 r)}{r} \mathbf{S}(\hat{\mathbf{n}}^{\text{sca}}, \hat{\mathbf{n}}^{\text{inc}}) \mathbf{E}_0^{\text{inc}}. \tag{1.12}$$

Here $\mathbf{E}$ denotes a two-element column formed by the $\theta$- and $\varphi$-components of the electric field vector:

$$\mathbf{E} = \begin{bmatrix} E_\theta \\ E_\varphi \end{bmatrix}; \tag{1.13}$$

$\theta \in [0, \pi]$ is the polar angle measured from the positive $z$-axis; and $\varphi \in [0, 2\pi)$ is the azimuth angle measured from the positive $x$-axis in the clockwise direction when looking in the direction of the positive $z$-axis (Fig. 1.3b). The amplitude scattering matrix also has the dimension of length and is a function on the incidence and scattering directions as well as of the size, morphology, composition, and orientation of the scattering object with respect to the laboratory coordinate system. It also depends on the choice of the origin of the coordinate system relative to the object. If known, the amplitude scattering matrix yields the scattered and thus the total field, thereby providing a complete description of the scattering pattern in the far-field zone.

The detailed conditions defining the far-field zone are as follows (Mishchenko 2006a):

$$k_1(r-a) \gg 1, \tag{1.14}$$

$$r \gg a, \tag{1.15}$$

$$r \gg \frac{k_1 a^2}{2}, \tag{1.16}$$

where $a$ is the radius of the smallest circumscribing sphere of the entire scattering object centered at $O$. The inequality (1.14) requires that the distance from any point inside the object to the observation point be much greater than the wavelength $\lambda = 2\pi/k_1$. This ensures that at the observation point, the partial field scattered by any differential volume element of the scattering object develops into an outgoing spherical wavelet. The inequality (1.15) requires the observation point to be located at a distance from the object much greater than the object's size. This ensures that when the partial wavelets generated by different elementary volume elements the object arrive at the observation point, they propagate in essentially the same scattering direction and are equally attenuated by the factor 1/distance. Finally, as a consequence of the inequality (1.16) the surfaces of constant phase of the partial wavelets generated by the elementary volume elements forming the object coincide locally when they reach the far-field observation point, and the wavelets form a single outgoing spherical wave. This implies that the entire object is effectively treated as a point-like scatterer located at the origin of the laboratory coordinate system.

The conditions (1.14)–(1.16) are often satisfied for sufficiently small (size parameters $k_1 a \lesssim 10^4$) isolated single-particle scatterers. The exact or approximate computation of the amplitude scattering matrix for such particles from the Maxwell equations is also often possible, which explains the widespread use of the amplitude scattering matrix as a single-particle electromagnetic characteristic.

However, there are many practical situations in which the conditions (1.14)–(1.16) are profoundly violated. A typical example is remote sensing of water clouds in the terrestrial atmosphere using detectors of electromagnetic radiation mounted on aircraft or satellite platforms. Such detectors usually measure radiation coming from a small part of a cloud and do not "perceive" the entire cloud as a single point-like scatterer (detector 1 in Fig. 1.4). Furthermore, the notion of the far-field zone of the cloud becomes completely meaningless if a detector is placed inside the cloud (detector 2). It is thus clear that to model theoretically the response of such "near-field" detectors one has to use scattering characteristics other than the scattering dyadic or the amplitude scattering matrix.

## 1.5. Actual observables

Because of high frequency of the time-harmonic oscillations, traditional optical instruments cannot measure the electric and magnetic fields associated with the incident and scattered waves. Indeed, it follows from

$$\frac{1}{T}\int_{t}^{t+T} \mathrm{d}t' \exp(-\mathrm{i}\omega t') \underset{T \gg 2\pi/\omega}{=} 0 \tag{1.17}$$

that accumulating and averaging a signal proportional to the electric or the magnetic field over a time interval $T$ long compared with the period of oscillations would yield a zero net result. Therefore, optical instruments usually measure quantities that have the dimension of energy flux and are defined in such a way that the time-harmonic factor $\exp(-\mathrm{i}\omega t)$ vanishes upon multiplication by its complex-conjugate counterpart: $\exp(-\mathrm{i}\omega t)[\exp(-\mathrm{i}\omega t)]^* \equiv 1$. This means that in order to make the theory applicable to analyses of actual optical observations, the scattering phenomenon must be characterized in terms of carefully chosen derivative quantities that can be measured directly. This explains the key importance of the concept of an actual *optical observable* to the discipline of light scattering by particles (Cloude 2009).

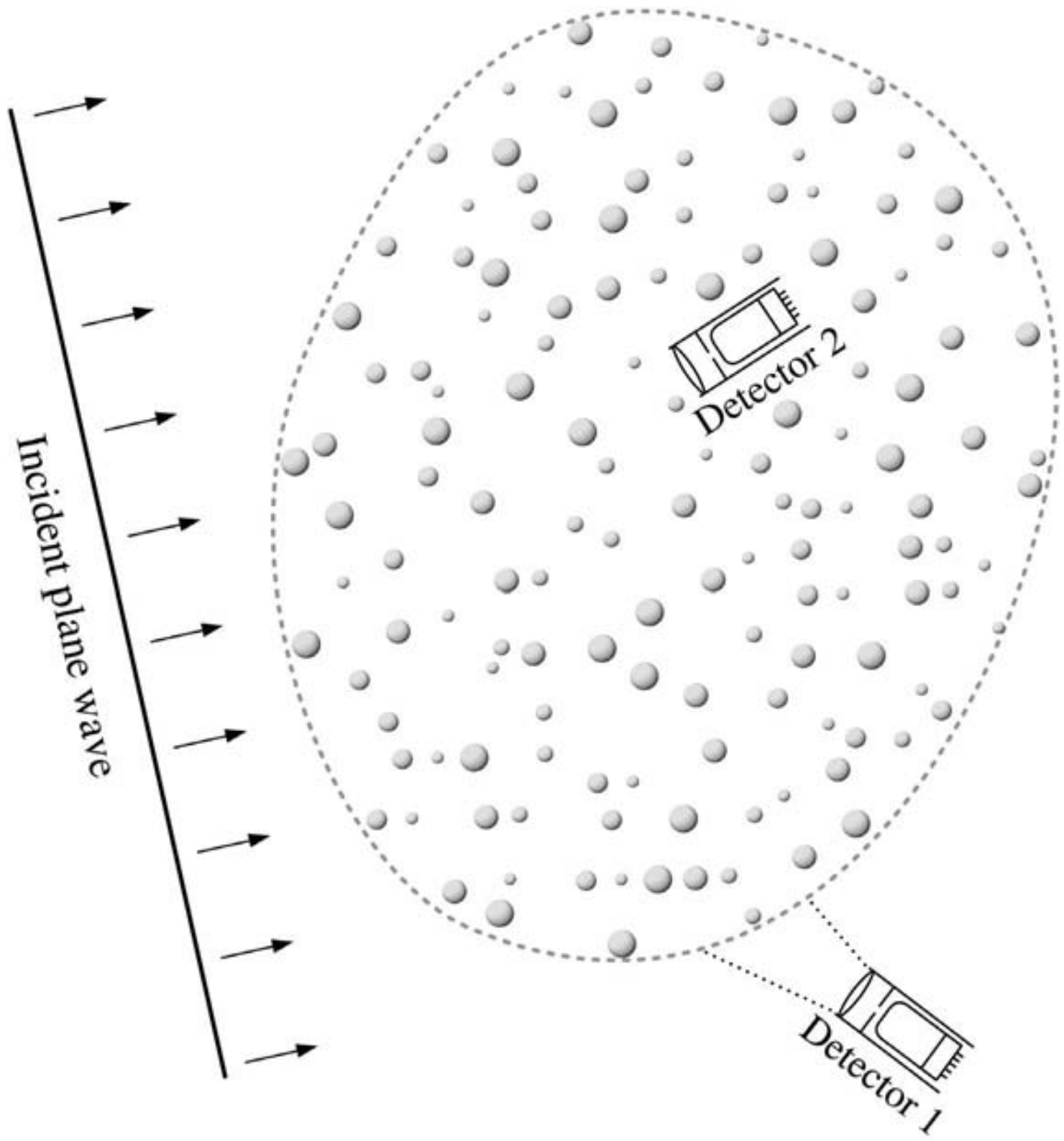

**Fig. 1.4.** Near-field scattering by a large multi-particle group.

Although one can always define the magnitude and the direction of the electromagnetic energy flux at any point in space in terms of the Poynting vector (Rothwell and Cloud 2009), the latter carries no information about the polarization state of the incident and scattered fields. The conventional approach to ameliorate this problem dates back to Stokes (1852). He proposed using four real-valued quan-tities, $I$, $Q$, $U$, and $V$, which have the dimension of monochromatic energy flux ($\mathrm{W m^{-2}}$) and fully characterize a *transverse* electromagnetic wave inasmuch as it is subject to practical optical analysis. (By definition, the electric and magnetic field vectors of a transverse electromagnetic wave vibrate in the plane perpendicular to the propagation direction.) These quantities, traditionally called the Stokes parameters, form the four-element so-called Stokes column vector $\mathbf{I}$ and carry information about both the total intensity, $I$, and the polarization state of the wave. The ellipsometric content of the Stokes parameters is discussed in detail by Mishchenko et al. (2002a, 2006b).

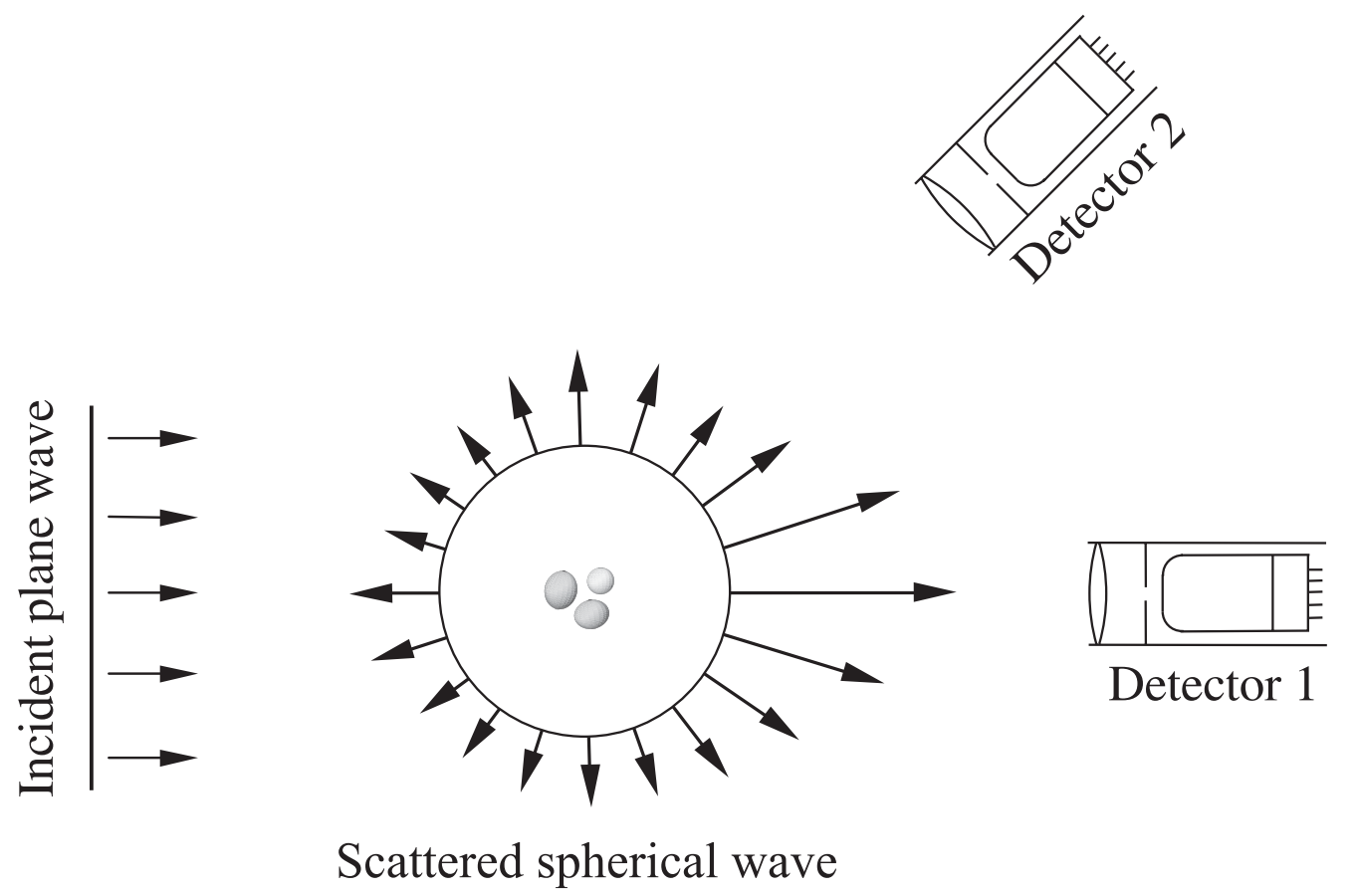

**Fig. 1.5.** Definition of the extinction and phase matrices.

In the case of far-field scattering, the transversality of both the incident plane wave and the scattered spherical wave allows one to define the corresponding sets of Stokes parameters:

$$\mathbf{I}^{\mathrm{inc}} = \begin{bmatrix} I^{\mathrm{inc}} \\ Q^{\mathrm{inc}} \\ U^{\mathrm{inc}} \\ V^{\mathrm{inc}} \end{bmatrix} = \frac{1}{2}\sqrt{\frac{\varepsilon_1}{\mu_0}} \begin{bmatrix} E_{0\theta}^{\mathrm{inc}}(E_{0\theta}^{\mathrm{inc}})^* + E_{0\varphi}^{\mathrm{inc}}(E_{0\varphi}^{\mathrm{inc}})^* \\ E_{0\theta}^{\mathrm{inc}}(E_{0\theta}^{\mathrm{inc}})^* - E_{0\varphi}^{\mathrm{inc}}(E_{0\varphi}^{\mathrm{inc}})^* \\ -E_{0\theta}^{\mathrm{inc}}(E_{0\varphi}^{\mathrm{inc}})^* - E_{0\varphi}^{\mathrm{inc}}(E_{0\theta}^{\mathrm{inc}})^* \\ \mathrm{i}[E_{0\varphi}^{\mathrm{inc}}(E_{0\theta}^{\mathrm{inc}})^* - E_{0\theta}^{\mathrm{inc}}(E_{0\varphi}^{\mathrm{inc}})^*] \end{bmatrix}, \tag{1.18}$$

$$\mathbf{I}^{\mathrm{sca}}(r\hat{\mathbf{n}}^{\mathrm{sca}}) = \begin{bmatrix} I^{\mathrm{sca}} \\ Q^{\mathrm{sca}} \\ U^{\mathrm{sca}} \\ V^{\mathrm{sca}} \end{bmatrix} = \frac{1}{r^2}\frac{1}{2}\sqrt{\frac{\varepsilon_1}{\mu_0}} \begin{bmatrix} E_{1\theta}^{\mathrm{sca}}(E_{1\theta}^{\mathrm{sca}})^* + E_{1\varphi}^{\mathrm{sca}}(E_{1\varphi}^{\mathrm{sca}})^* \\ E_{1\theta}^{\mathrm{sca}}(E_{1\theta}^{\mathrm{sca}})^* - E_{1\varphi}^{\mathrm{sca}}(E_{1\varphi}^{\mathrm{sca}})^* \\ -E_{1\theta}^{\mathrm{sca}}(E_{1\varphi}^{\mathrm{sca}})^* - E_{1\varphi}^{\mathrm{sca}}(E_{1\theta}^{\mathrm{sca}})^* \\ \mathrm{i}[E_{1\varphi}^{\mathrm{sca}}(E_{1\theta}^{\mathrm{sca}})^* - E_{1\theta}^{\mathrm{sca}}(E_{1\varphi}^{\mathrm{sca}})^*] \end{bmatrix}. \tag{1.19}$$

Then the response of a well-collimated polarization-sensitive detector of light can be described in terms of the $4\times4$ so-called phase and extinction matrices.

Specifically, detector 2 in Fig. 1.5 collects only the scattered light, and its polarized reading is fully characterized by the product of the phase matrix $\mathbf{Z}$ and the Stokes column vector of the incident wave:

$$\textbf{Signal 2} = \Delta S\,\mathbf{I}^{\text{sca}}(r\hat{\mathbf{n}}^{\text{sca}}) = \frac{\Delta S}{r^2}\mathbf{Z}(\hat{\mathbf{n}}^{\text{sca}},\hat{\mathbf{n}}^{\text{inc}})\mathbf{I}^{\text{inc}}, \quad \hat{\mathbf{n}}^{\text{sca}} \neq \hat{\mathbf{n}}^{\text{inc}}, \tag{1.20}$$

where $\Delta S$ is the acceptance area of the detector. In other words, the phase matrix realizes the transformation of the Stokes parameters of the incident wave into the Stokes parameters of the scattered wave. The elements of the phase matrix have the dimension of area and are quadratic combinations of the elements of the amplitude scattering matrix $\mathbf{S}(\hat{\mathbf{n}}^{\text{sca}},\hat{\mathbf{n}}^{\text{inc}})$ (Mishchenko et al. 2002a).

Unlike detector 2, detector 1 in Fig. 1.5 is facing the incident light, and its polarized reading consists of three parts:

1. the one due to the incident light;
2. the one due to the forward-scattered light; and
3. the one due to the interference of the incident wave and the wave scattered by the object in the exact forward direction:

$$\textbf{Signal 1} = \int_{\Delta S} \mathrm{d}S\,\mathbf{I}(r\hat{\mathbf{r}})$$

$$= \Delta S\,\mathbf{I}^{\text{inc}} + \frac{\Delta S}{r^2}\mathbf{Z}(\hat{\mathbf{n}}^{\text{inc}},\hat{\mathbf{n}}^{\text{inc}})\mathbf{I}^{\text{inc}} - \mathbf{K}(\hat{\mathbf{n}}^{\text{inc}})\mathbf{I}^{\text{inc}} \tag{1.21}$$

$$= \Delta S\,\mathbf{I}^{\text{inc}} + \boldsymbol{O}(r^{-2}) - \mathbf{K}(\hat{\mathbf{n}}^{\text{inc}})\mathbf{I}^{\text{inc}}, \tag{1.22}$$

where $\boldsymbol{O}(r^{-2})$ is a $4\times4$ matrix with elements vanishing at infinity as $r^{-2}$ (Mishchenko et al. 2002a). The third part is described by minus the product of the extinction matrix $\mathbf{K}$ and the Stokes column vector of the incident wave. The elements of the extinction matrix have the dimension of area and are linear combinations of the elements of the forward-scattering amplitude matrix $\mathbf{S}(\hat{\mathbf{n}}^{\text{inc}},\hat{\mathbf{n}}^{\text{inc}})$ (Mishchenko et al. 2002a). Equations (1.21) and (1.22) represent the most general form of the so-called optical theorem.

The situation depicted in Fig. 1.5 is, in many respects, the embodiment of the concept of light scattering. Indeed, it demonstrates that in the absence of the object, detector 2 would measure no signal, while the signal measured by detector 1 would be proportional to the Stokes column vector of the incident light. In the presence of the object, the readings of both detectors change. The reading of detector 2 is now proportional to the Stokes column vector of the scattered wave, while the polarized signal measured by detector 1 is modified in two ways. First, the total measured intensity is attenuated as a combined result of the scattering of electromagnetic energy by the object in all directions and, possibly, the transformation of electromagnetic energy into other forms of energy (such as heat) inside the object. Second, the modification rates for the four Stokes parameters of the measured signal can be different. This effect is typical of objects lacking perfect spherical symmetry and is called dichroism (Dolginov et al. 1995). Thus, to describe far-field scattering means, in effect, to quantify the differences between the readings of detectors 1 and 2 in the presence of the object and in the absence of the object. This quantification can be fully achieved in terms of the phase and extinction matrices which depend on object's characteristics such as size, shape, refractive index, and orientation and can be readily computed provided that the amplitude scattering matrix is known.

The near field is not, in general, a transverse electromagnetic wave. Therefore, to characterize the response of the "near-field" detectors shown in Fig. 1.4, one must define quantities other than the Stokes parameters and the extinction and phase matrices. Still the actual observables must be defined in such a way that they can be measured by an optical device ultimately recording the flux of electromagnetic energy. We will see in later sections how this is done in the framework of the theories of RT and WL.

Although we have been assuming so far that the infinite medium surrounding the scattering object is nonabsorbing, the above formalism affords a straightforward generalization to the case of an absorbing host provided that one consistently operates with quantities representing actual optical observables (Mishchenko 2007).

## 1.6. Derivative far-field characteristics

There are several derivative quantities that are often used to describe various manifestations of electromagnetic scattering in the far-field zone of the object. The product of the extinction cross section and the intensity of the incident plane wave yields the total attenuation of the electromagnetic power measured by detector 1 in Fig. 1.5 owing to the presence of the particle. This means that, in general, the extinction cross section depends on the polarization state and propagation direction of the incident wave and is given by (Mishchenko et al. 2002a, 2006b)

$$C_{\mathrm{ext}}(\hat{\mathbf{n}}^{\mathrm{inc}}) = \frac{1}{I^{\mathrm{inc}}}[K_{11}(\hat{\mathbf{n}}^{\mathrm{inc}})I^{\mathrm{inc}} + K_{12}(\hat{\mathbf{n}}^{\mathrm{inc}})Q^{\mathrm{inc}} + K_{13}(\hat{\mathbf{n}}^{\mathrm{inc}})U^{\mathrm{inc}} + K_{14}(\hat{\mathbf{n}}^{\mathrm{inc}})V^{\mathrm{inc}}]. \tag{1.23}$$

The product of the scattering cross section and the intensity of the incident plane wave yields the total far-field power scattered by the particle in all directions. We thus have (Mishchenko et al. 2002a, 2006b)

$$\begin{aligned} C_{\mathrm{sca}}(\hat{\mathbf{n}}^{\mathrm{inc}}) &\underset{r\to\infty}{=} \frac{r^2}{I^{\mathrm{inc}}}\int_{4\pi} \mathrm{d}\hat{\mathbf{r}}\, I^{\mathrm{sca}}(r\hat{\mathbf{r}}) \\ &= \frac{1}{I^{\mathrm{inc}}}\int_{4\pi} \mathrm{d}\hat{\mathbf{r}}[Z_{11}(\hat{\mathbf{r}},\hat{\mathbf{n}}^{\mathrm{inc}})I^{\mathrm{inc}} + Z_{12}(\hat{\mathbf{r}},\hat{\mathbf{n}}^{\mathrm{inc}})Q^{\mathrm{inc}} \\ &\qquad + Z_{13}(\hat{\mathbf{r}},\hat{\mathbf{n}}^{\mathrm{inc}})U^{\mathrm{inc}} + Z_{14}(\hat{\mathbf{r}},\hat{\mathbf{n}}^{\mathrm{inc}})V^{\mathrm{inc}}], \end{aligned} \tag{1.24}$$

where $\mathrm{d}\hat{\mathbf{r}}$ is a differential solid-angle element centered around $\hat{\mathbf{r}}$. This means that $C_{\mathrm{sca}}$ also depends on the polarization state as well as on the propagation direction of the incident wave. The absorption cross section is defined as the difference between the extinction and scattering cross sections:

$$C_{\mathrm{abs}}(\hat{\mathbf{n}}^{\mathrm{inc}}) = C_{\mathrm{ext}}(\hat{\mathbf{n}}^{\mathrm{inc}}) - C_{\mathrm{sca}}(\hat{\mathbf{n}}^{\mathrm{inc}}) \geq 0. \tag{1.25}$$

All optical cross sections have the dimension of area. Finally, the dimensionless single-scattering albedo is defined as the ratio of the scattering and extinction cross sections:

$$\varpi(\hat{\mathbf{n}}^{\text{inc}}) = \frac{C_{\text{sca}}(\hat{\mathbf{n}}^{\text{inc}})}{C_{\text{ext}}(\hat{\mathbf{n}}^{\text{inc}})} \le 1. \tag{1.26}$$

A particular case of the phase matrix is the scattering matrix defined by

$$\mathsf{F}(\Theta) = \mathsf{Z}(\theta^{\text{sca}} = \Theta, \varphi^{\text{sca}} = 0; \theta^{\text{inc}} = 0, \varphi^{\text{inc}} = 0), \quad \Theta \in [0, \pi], \tag{1.27}$$

where $\Theta$, traditionally called the scattering angle, is the angle between the incidence and scattering directions. It is easy to see that the scattering matrix relates the Stokes parameters of the incident and scattered waves defined with respect to the same so-called scattering plane, i.e., the plane through the incidence and scattering directions (van de Hulst 1957; Mishchenko et al. 2002a; Hovenier et al. 2004).

An important phenomenon is the radiation force exerted on the scattering object. In the case of an incident plane wave, this force is given by

$$\begin{aligned}\mathbf{F} = \frac{1}{c}\hat{\mathbf{n}}^{\text{inc}} C_{\text{ext}} I^{\text{inc}} - \frac{1}{c}\int_{4\pi} \mathrm{d}\hat{\mathbf{r}}\,\hat{\mathbf{r}}[Z_{11}(\hat{\mathbf{r}}, \hat{\mathbf{n}}^{\text{inc}}) I^{\text{inc}} + Z_{12}(\hat{\mathbf{r}}, \hat{\mathbf{n}}^{\text{inc}}) Q^{\text{inc}} \\ + Z_{13}(\hat{\mathbf{r}}, \hat{\mathbf{n}}^{\text{inc}}) U^{\text{inc}} + Z_{14}(\hat{\mathbf{r}}, \hat{\mathbf{n}}^{\text{inc}}) V^{\text{inc}}],\end{aligned} \tag{1.28}$$

where $c$ is the speed of light (Mishchenko 2001). Obviously, the radiation force depends on the polarization state of the incident light. An additional component of the radiation force can be caused by emitted electromagnetic radiation provided that the absolute temperature of the object is sufficiently high (Mishchenko et al. 2002a).

## 1.7. Foldy–Lax equations

As we have already mentioned, many theoretical techniques based on a direct solution of the differential Maxwell equations or their integral counterparts are applicable to an arbitrary fixed finite object, be it a single physical body or a cluster consisting of several distinct components, either touching or spatially separated. These techniques are based on treating the object as a single scatterer and yield the total scattered field. However, if the object is a multi-particle group, such as a cloud of water droplets, then it is often convenient to represent the total scattered field as a vector superposition of partial fields scattered by the individual particles. This means that the total electric field at a point $\mathbf{r}$ is written as follows:

$$\mathbf{E}(\mathbf{r}) = \mathbf{E}^{\text{inc}}(\mathbf{r}) + \sum_{i=1}^{N} \mathbf{E}_i^{\text{sca}}(\mathbf{r}), \quad \mathbf{r} \in \Re^3, \tag{1.29}$$

where $N$ is the number of particles in the group and $\mathbf{E}_i^{\text{sca}}(\mathbf{r})$ is the $i$th partial scattered electric field.

The partial scattered fields can be found by solving vector so-called Foldy–Lax equations (FLEs) which follow directly from the VIE and are exact (Babenko et al. 2003; Mishchenko et al. 2006b). Specifically, the $i$th partial scattered field is given by the formula

$$\mathbf{E}_i^{\mathrm{sca}}(\mathbf{r}) = \int_{V_i} \mathrm{d}\mathbf{r}' \vec{G}(\mathbf{r}, \mathbf{r}') \cdot \int_{V_i} \mathrm{d}\mathbf{r}'' \vec{T}_i(\mathbf{r}', \mathbf{r}'') \cdot \mathbf{E}_i(\mathbf{r}''), \tag{1.30}$$

where $V_i$ is the volume occupied by the $i$th particle, $\mathbf{E}_i(\mathbf{r}'')$ is the electric field "exciting" particle $i$, and each of the $N$ dyadics $\vec{T}_i$ can be found by solving individually the following equation:

$$\vec{T}_i(\mathbf{r}, \mathbf{r}') = k_1^2[m_i^2(\mathbf{r}) - 1]\delta(\mathbf{r} - \mathbf{r}')\vec{I}$$
$$+ k_1^2[m_i^2(\mathbf{r}) - 1] \int_{V_i} \mathrm{d}\mathbf{r}'' \vec{G}(\mathbf{r}, \mathbf{r}'') \cdot \vec{T}_i(\mathbf{r}'', \mathbf{r}'), \quad \mathbf{r}, \mathbf{r}' \in V_i. \tag{1.31}$$

Comparison with Eq. (1.8) shows that $\vec{T}_i$ for each $i$ is, in fact, the dyadic transition operator of particle $i$ with respect to the fixed laboratory coordinate system computed in the absence of all the other particles. Thus, the $N$ dyadic transition operators are totally independent of each other. However, the "exciting" fields are interdependent and must be found by solving the following system of $N$ linear integral equations:

$$\mathbf{E}_i(\mathbf{r}) = \mathbf{E}^{\mathrm{inc}}(\mathbf{r}) + \sum_{j(\neq i)=1}^{N} \int_{V_j} \mathrm{d}\mathbf{r}' \vec{G}(\mathbf{r}, \mathbf{r}') \cdot \int_{V_j} \mathrm{d}\mathbf{r}'' \vec{T}_j(\mathbf{r}', \mathbf{r}'') \cdot \mathbf{E}_j(\mathbf{r}''), \quad \mathbf{r} \in V_i, \quad i = 1, \ldots, N. \tag{1.32}$$

In general, the FLEs (1.29)–(1.32) are equivalent to Eqs. (1.7)–(1.8). However, the fact that $\vec{T}_i$ for each $i$ is an individual property of the $i$th particle computed as if this particle were alone allows one to introduce the mathematical concept of multiple scattering. This will be the subject of the following section.

One specific, numerically exact approach to solve the FLEs is the so-called superposition $T$-matrix method which involves the expansion of the various electric fields in vector spherical wave functions (VSWFs) centered either at the common origin of the entire scattering object or at the individual particle origins (see Section 1.12). This technique is especially efficient in application to groups of spherically symmetric particles and will be used in Section 1.17 to illustrate the scattering effects that can or cannot be described by the theories of RT and WL.

## 1.8. Multiple scattering

Let us rewrite Eqs. (1.29), (1.30), and (1.32) in the following compact operator form:

$$E = E^{\mathrm{inc}} + \sum_{i=1}^{N} \hat{G}\hat{T}_i E_i, \tag{1.33}$$

$$E_i = E^{\mathrm{inc}} + \sum_{j(\neq i)=1}^{N} \hat{G}\hat{T}_j E_j, \tag{1.34}$$

where

$$\hat{G}\hat{T}_j E_j = \int_{V_j} d\mathbf{r}' \ddot{G}(\mathbf{r}, \mathbf{r}') \cdot \int_{V_j} d\mathbf{r}'' \ddot{T}_j(\mathbf{r}', \mathbf{r}'') \cdot \mathbf{E}_j(\mathbf{r}''). \tag{1.35}$$

Iterating Eq. (1.34) yields

$$E_i = E^{\text{inc}} + \sum_{j(\neq i)=1}^{N} \hat{G}\hat{T}_j E^{\text{inc}} + \sum_{\substack{j(\neq i)=1 \\ l(\neq j)=1}}^{N} \hat{G}\hat{T}_j \hat{G}\hat{T}_l E^{\text{inc}} + \sum_{\substack{j(\neq i)=1 \\ l(\neq j)=1 \\ m(\neq l)=1}}^{N} \hat{G}\hat{T}_j \hat{G}\hat{T}_l \hat{G}\hat{T}_m E^{\text{inc}} + \cdots, \tag{1.36}$$

whereas the substitution of Eq. (1.36) in Eq. (1.33) gives what can be interpreted as an order-of-scattering expansion of the total electric field:

$$E = E^{\text{inc}} + E^{\text{sca}}, \tag{1.37}$$

$$E^{\text{sca}} = \sum_{i=1}^{N} \hat{G}\hat{T}_i E^{\text{inc}} + \sum_{\substack{i=1 \\ j(\neq i)=1}}^{N} \hat{G}\hat{T}_i \hat{G}\hat{T}_j E^{\text{inc}} + \sum_{\substack{i=1 \\ j(\neq i)=1 \\ l(\neq j)=1}}^{N} \hat{G}\hat{T}_i \hat{G}\hat{T}_j \hat{G}\hat{T}_l E^{\text{inc}} + \cdots. \tag{1.38}$$

Indeed, the dyadic transition operators are independent of each other, and each of them can be interpreted as a unique and complete electromagnetic identifier of the corresponding particle. Therefore, $\hat{G}\hat{T}_i E^{\text{inc}}$ could be interpreted as the partial scattered filed at the observation point generated by particle $i$ in response to the excitation by the incident field only, $\hat{G}\hat{T}_i \hat{G}\hat{T}_j E^{\text{inc}}$ is the partial field generated by the same particle in response to the excitation caused by particle $j$ in response to the excitation by the incident field, etc. Thus, the first term on the right-hand side of Eq. (1.38) could be interpreted as the sum of all single-scattering contributions, the second term is the sum of all double-scattering contributions, etc. The first term on the right-hand side of Eq. (1.37) represents the unscattered (i.e., incident) field. This order-of-scattering interpretation of Eqs. (1.37) and (1.38) is visualized in Fig. 1.6.

We will see very soon that Eqs. (1.37) and (1.38) constitute a very fruitful way of re-writing the original FLEs and that the use of the "multiple scattering" terminology is a convenient and compact way of illustrating their numerous consequences. It is important to recognize, however, that besides being an interpretation and visualization tool, the mathematical concept of multiple scattering does not represent an actual physical phenomenon (Mishchenko 2008a, 2009). For example, the term $\hat{G}\hat{T}_i \hat{G}\hat{T}_j \hat{G}\hat{T}_l E^{\text{inc}}$ on the right-hand side of Eq. (1.38) cannot be interpreted by saying that a "light ray" (or a "localized blob of energy") approaches particle $l$, gets scattered by particle $l$ towards particle $j$, approaches particle $j$, gets scattered by particle $j$ towards particle $i$, approaches particle $i$, gets scattered by particle $i$ towards the observation point, and finally arrives at the observation point. Indeed, it follows from Eq. (1.32) that all mutual particle–particle "excitations" occur simultaneously and are not temporally discrete and ordered events. The purely mathematical character of the multiple-scattering interpretation of Eq. (1.38) becomes especially transparent upon realizing that this equation is quite general and can be applied not only to a multi-particle group but also to a single body wherein the latter is subdivided arbitrarily into $N$ non-overlapping adjacent geometrical regions $V_i$.

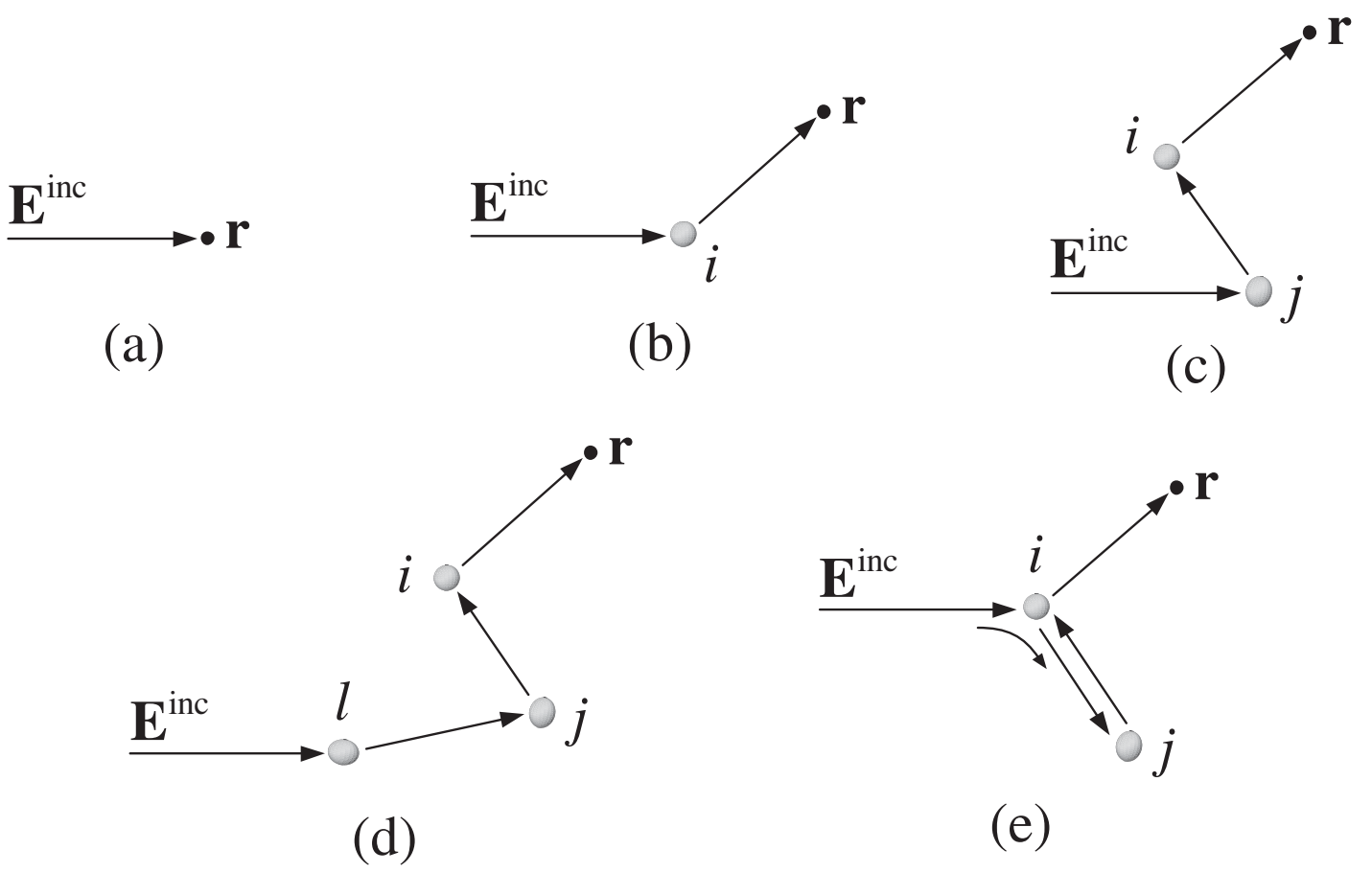

**Fig. 1.6.** (a) Unscattered (incident) field; (b) single scattering; (c) double scattering; (d,e) triple scattering.

It is convenient to represent the mathematical order-of-scattering expansion (1.37) – (1.38) of the total electric field using the diagram method. In Fig. 1.7, the arrows represent the incident field, the symbol —● denotes the "multiplication" of a field by a $\hat{G}\hat{T}$ dyadic according to Eq. (1.35), and a dashed curve indicates that two "scattering events" involve the same particle. The first five terms on the right-hand side of the diagrammatic expression in Fig. 1.7 describe, respectively, the (cumulative) contributions of the five scattering scenarios illustrated in Fig. 1.6.

## 1.9. Far-field Foldy–Lax equations

We have seen in Section 1.4 that as a direct consequence of Eqs. (1.7) and (1.9), the behavior of the scattered field becomes much simpler in the far-field zone of the scattering object. Since the structure of Eqs. (1.30) and (1.32) is analogous to that of Eq. (1.7), one can expect a similar simplification of the FLEs upon making the following two assumptions:

1. The $N$ particles forming the group are separated widely enough that each of them is located in the far-field zones of all the other particles.
2. The observation point is located in the far-field zone of any particle forming the group.

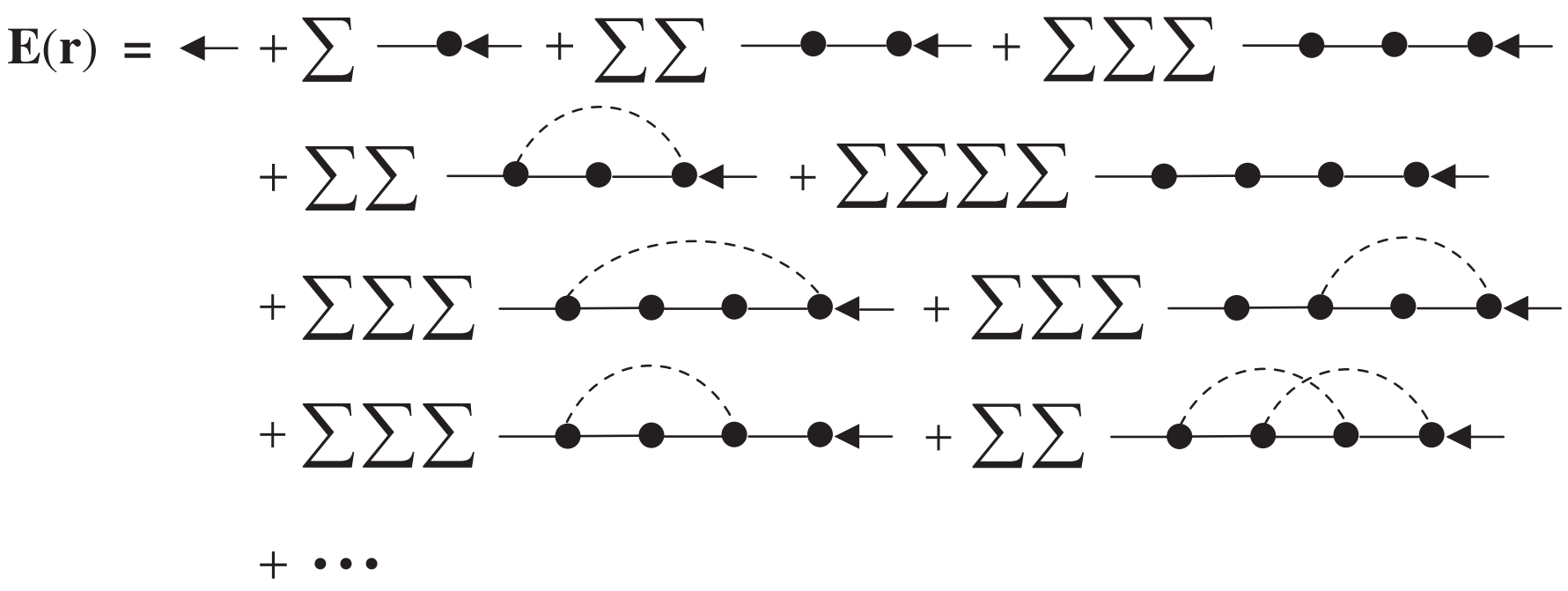

**Fig. 1.7.** Diagrammatic representation of Eqs. (1.37) and (1.38).

Indeed, the contribution of the $j$th particle to the field exciting the $i$th particle in Eq. (1.32) can now be represented as a simple outgoing spherical wavelet centered at the origin of particle $j$. The radius of curvature of this wavelet at the origin of particle $i$ is much larger than the size of particle $i$ so that the wavelet can be considered as *locally* plane. The scattering of this wavelet by particle $i$ can then be described in terms of the corresponding scattering dyadic $\vec{A}_i$ (see Eq. (1.10)). As a consequence, the system of integral FLEs is converted into a system of algebraic equations (Mishchenko et al. 2006b).

Specifically, assuming that the incident field is a plane electromagnetic wave propagating in the direction $\hat{\mathbf{n}}^{\text{inc}}$, we have for the total field at a point $\mathbf{r}$ located in the far-field zone of any particle:

$$\mathbf{E}(\mathbf{r}) = \mathbf{E}^{\text{inc}}(\mathbf{r}) + \sum_{i=1}^{N} G(r_i)\vec{A}_i(\hat{\mathbf{r}}_i, \hat{\mathbf{n}}^{\text{inc}}) \cdot \mathbf{E}^{\text{inc}}(\mathbf{R}_i) + \sum_{i=1}^{N} G(r_i) \sum_{j(\neq i)=1}^{N} \vec{A}_i(\hat{\mathbf{r}}_i, \hat{\mathbf{R}}_{ij}) \cdot \mathbf{E}_{ij}, \tag{1.39}$$

where $G(r) = \exp(\mathrm{i}k_1 r)/r$, $r_i$ is the distance between the origin of particle $i$ and the observation point, $\hat{\mathbf{r}}_i$ is the unit vector directed from the origin of particle $i$ towards the observation point, $\mathbf{R}_i$ is the position vector of the $i$th particle origin, and $\hat{\mathbf{R}}_{ij}$ is the unit vector directed from the origin of particle $j$ towards the origin of particle $i$ (see Fig. 1.8). Equation (1.39) shows that the total field at any point located sufficiently far from any particle in the group is the superposition of the incident plane wave and $N$ spherical waves generated by and centered at the $N$ particles. The amplitudes of the particle–particle "excitations" $\mathbf{E}_{ij}$ are found from the following system of $N(N-1)$ linear algebraic equations:

$$\mathbf{E}_{ij} = G(R_{ij})\vec{A}_j(\hat{\mathbf{R}}_{ij}, \hat{\mathbf{n}}^{\text{inc}}) \cdot \mathbf{E}^{\text{inc}}(\mathbf{R}_j) + G(R_{ij}) \sum_{l(\neq j)=1}^{N} \vec{A}_j(\hat{\mathbf{R}}_{ij}, \hat{\mathbf{R}}_{jl}) \cdot \mathbf{E}_{jl}, \tag{1.40}$$

$$i, j = 1, \ldots, N, \quad j \neq i,$$

where $R_{ij}$ is the distance between the origins of particles $j$ and $i$.

This system is much simpler than the original system of FLEs and can, in principle, be solved on a computer provided that $N$ is not too large. The expression for the order-of-scattering expansion of the total field also becomes much simpler:

$$\mathbf{E} = \mathbf{E}^{\text{inc}} + \sum_{i=1}^{N} \vec{B}_{ri0} \cdot \mathbf{E}_i^{\text{inc}} + \sum_{i=1}^{N} \sum_{j(\neq i)=1}^{N} \vec{B}_{rij} \cdot \vec{B}_{ij0} \cdot \mathbf{E}_j^{\text{inc}}$$

$$+\sum_{i=1}^{N}\sum_{j(\neq i)=1}^{N}\sum_{l(\neq j)=1}^{N}\vec{\vec{B}}_{rij}\cdot\vec{\vec{B}}_{ijl}\cdot\vec{\vec{B}}_{jl0}\cdot\mathbf{E}_l^{\rm inc}$$

$$+\sum_{i=1}^{N}\sum_{j(\neq i)=1}^{N}\sum_{l(\neq j)=1}^{N}\sum_{m(\neq l)=1}^{N}\vec{\vec{B}}_{rij}\cdot\vec{\vec{B}}_{ijl}\cdot\vec{\vec{B}}_{jlm}\cdot\vec{\vec{B}}_{lm0}\cdot\mathbf{E}_m^{\rm inc}+\cdots, \tag{1.41}$$

where

$$\mathbf{E} = \mathbf{E}(\mathbf{r}), \quad \mathbf{E}^{\rm inc} = \mathbf{E}^{\rm inc}(\mathbf{r}), \quad \mathbf{E}_i^{\rm inc} = \mathbf{E}^{\rm inc}(\mathbf{R}_i), \tag{1.42}$$

$$\vec{\vec{B}}_{ri0} = G(r_i)\,\vec{\vec{A}}_i(\hat{\mathbf{r}}_i, \hat{\mathbf{n}}^{\rm inc}), \tag{1.43}$$

$$\vec{\vec{B}}_{rij} = G(r_i)\,\vec{\vec{A}}_i(\hat{\mathbf{r}}_i, \hat{\mathbf{R}}_{ij}), \tag{1.44}$$

$$\vec{\vec{B}}_{ij0} = G(R_{ij})\,\vec{\vec{A}}_j(\hat{\mathbf{R}}_{ij}, \hat{\mathbf{n}}^{\rm inc}), \tag{1.45}$$

$$\vec{\vec{B}}_{ijl} = G(R_{ij})\,\vec{\vec{A}}_j(\hat{\mathbf{R}}_{ij}, \hat{\mathbf{R}}_{jl}). \tag{1.46}$$

A remarkable feature of the above formulas is that now the role of the unique electromagnetic identifier of each particle is assumed by the corresponding scattering dyadic, that is, the same quantity that would completely describe far-field scattering by this particle if it were alone rather than a member of the group. Although the dyadic transition operator is the most general scattering property of a particle, the scattering dyadic or, equivalently, the amplitude scattering matrix have been used so frequently to describe far-field scattering by a particle that they have become almost synonymic with the words "single scattering". This appears to add some no-toriety to the order-of-scattering interpretation of Eq. (1.41). One should not forget, however, that Eq. (1.41) is just an approximate version of Eqs. (1.37) and (1.38) and does not make multiple scattering a real physical phenomenon.

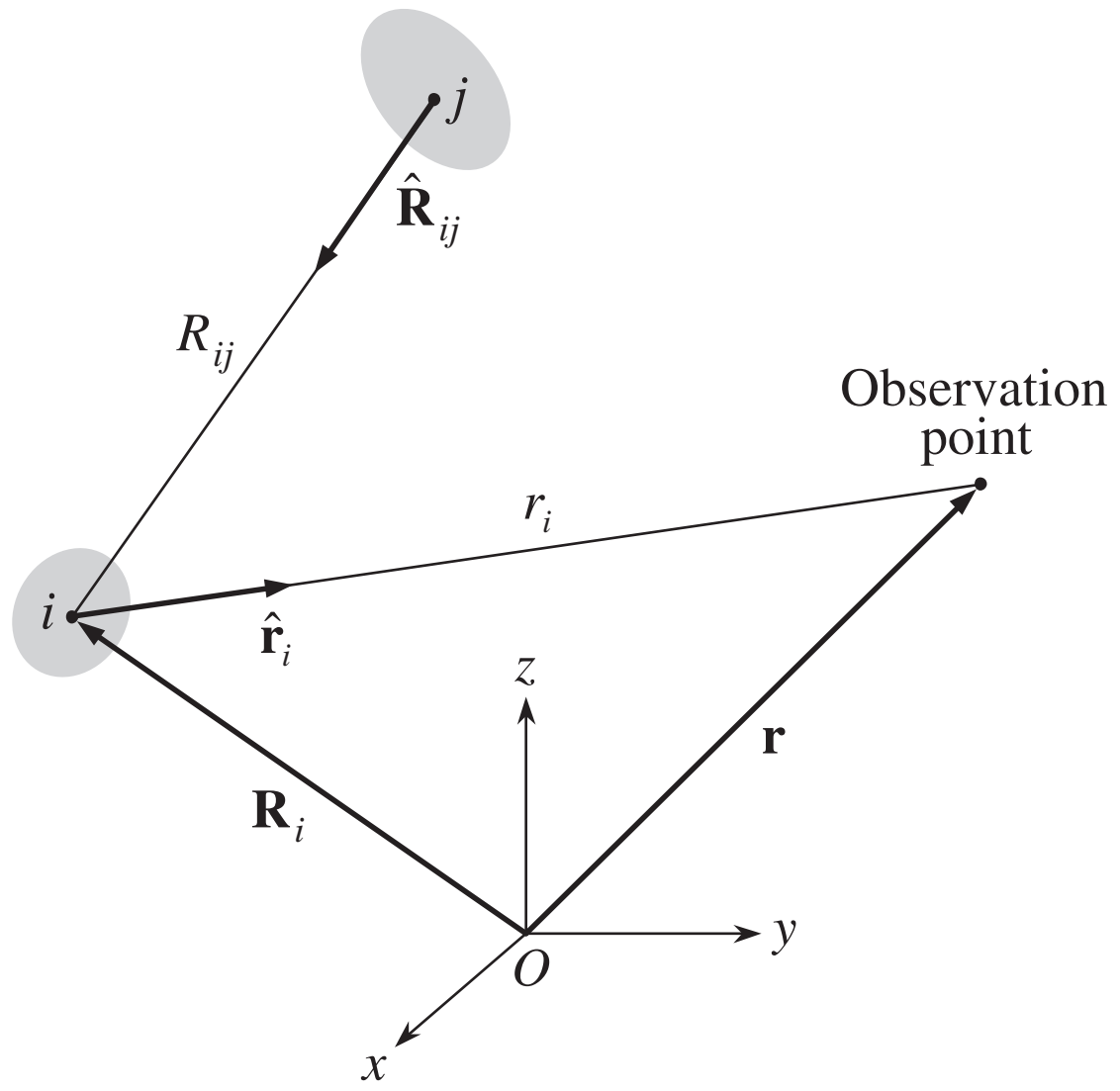

**Fig. 1.8.** Scattering contribution caused by particle *i* in response to "excitation" by particle *j*.

The diagrammatic formula shown in Fig. 1.7 can also represent Eq. (1.41) provided that now the symbol —● denotes the premultiplication of a field by a $\vec{B}$ dyadic.

## 1.10. Ergodicity

Most of our previous discussion has been based on the assumption that the configuration of the scattering object with respect to the laboratory coordinate system is fixed. However, quite often one has to deal with an object in the form of a multiparticle group in which the particles are randomly rotating and moving relative to each other. The particles may even change there sizes and shapes owing to evaporation, sublimation, condensation, or melting. Important examples of such "stochastic" scattering objects are atmospheric clouds consisting of water droplets and/or ice crystals, plumes of aerosol particles, and various particle suspensions. The physical and chemical processes controlling the temporal evolution of such objects can be extremely complex and convoluted.

Although a random group can be described at any given moment in terms of a specific fixed particle configuration, any measurement takes a finite amount of time during which the group goes through an infinite sequence of evolving discrete configurations. Sometimes the result of the measurement can be modeled numerically by solving the Maxwell equations for many time-sequential discrete configurations and then taking the average. A far more practical approach in most cases is based on the assumption of ergodicity. Specifically, all further discussion will be based on the following two fundamental premises:

1. The scattering object can be adequately characterized at any moment in time by a finite set of physical parameters.

2. The scattering object is sufficiently variable in time and the time interval necessary to take a measurement is sufficiently long that averaging the scattering signal over this interval is essentially equivalent to averaging the signal over an appropriate analytical probability distribution of the physical parameters characterizing the scattering object.

In other words, we will assume that averaging over time for one specific realization of a random scattering process is equivalent to ensemble averaging.

To better appreciate the meaning of ergodicity, let us consider the measurement of a scattering characteristic $A$ of a cloud of spherical water droplets. This characteristic depends on time implicitly by being a function of time-dependent physical parameters of the cloud such as the coordinates and radii of all the constituent particles. The full set of particle positions and radii will be denoted collectively by $\psi$ and determines the state of the entire cloud at a moment in time. In order to interpret the measurement of $A[\psi(t)]$ accumulated over a period of time extending from $t = t_0$ to $t = t_0 + T$, one needs a way of predicting theoretically the average value

$$\overline{A} = \frac{1}{T}\int_{t_0}^{t_0+T} \mathrm{d}t\, A[\psi(t)]. \tag{1.47}$$

As we have already mentioned, the temporal evolution of the cloud of water droplets is described by an intricate system of equations representing the various physical and chemical processes in action. To incorporate the solution of this system of equations for each moment of time into the theoretical averaging procedure (1.47) can be a formidable task and is almost never done. Instead, averaging over time is replaced by ensemble averaging based on the following rationale.

Although the coordinates and sizes of water droplets in the cloud change with time in a specific way, the range of instantaneous states of the cloud captured by the detector during the measurement becomes representative of that captured over an infinite period of time provided that $T$ is sufficiently long. We thus have

$$\overline{A} \approx \lim_{\tau\to\infty} \frac{1}{\tau}\int_{t_0}^{t_0+\tau} \mathrm{d}t\, A[\psi(t)] = \langle A\rangle_t. \tag{1.48}$$

Notice now that the infinite integral in Eq. (1.48) can be expected to "sample" every physically realizable state $\psi$ of the cloud. Furthermore, this sampling is statistically representative in that the number of times each state is sampled is large and tends to infinity in the limit $\tau\to\infty$. Most importantly, the cumulative contribution of a cloud state $\psi$ to $\langle A\rangle_t$ is independent of the specific moments in time when this state actually occurred in the process of the temporal evolution of the cloud. Rather, it depends on how many times this state was sampled. Therefore, this cumulative contribution can be thought of as being proportional to the probability of occurrence of the state $\psi$ at *any* moment of time. This means that instead of specifying the state of the cloud at each moment $t$ and integrating over all $t$, one can introduce an appropriate time-independent probability density function $p(\psi)$ and integrate over the entire physically realizable range of cloud states:

$$\langle A\rangle_t \approx \int \mathrm{d}\psi\, p(\psi) A(\psi) = \langle A\rangle_\psi, \tag{1.49}$$

where

$$\int \mathrm{d}\psi\, p(\psi) = 1. \tag{1.50}$$

Equation (1.49) is the mathematical expression of the principle of ergodicity introduced above. Processes such as Brownian motion and turbulence often help to establish a significant degree of randomness of particle positions and orientations, which seems to explain why theoretical predictions based on the ergodic hypothesis agree well with experimental data (Berne and Pecora 1976). The practical meaning of ergodicity in the framework of the RT theory will be discussed in Section 1.18.

### 1.11. Scattering by a small random particle group

Consider now a random group of $N \geq 1$ particles viewed from a distance much greater than the entire size of the group. Let us assume that $N$ is sufficiently small and the particles are separated widely enough that the second term on the right-hand side of Eq. (1.32) is always much smaller than the first term. This means that each particle is "excited" only by the external incident field, which is the gist of the *single-scattering approximation* (SSA). Let us also assume that (i) the scattering signal is accumulated over a time interval long enough to establish full ergodicity of the group, and (ii) particle positions are uncorrelated and statistically random. It is then straightforward to show that the time averages of the phase and extinction matrices of the entire group in Eqs. (1.20)−(1.22) can be calculated as follows:

$$\langle \mathbf{Z}(\hat{\mathbf{n}}^{\mathrm{sca}}, \hat{\mathbf{n}}^{\mathrm{inc}}) \rangle_t = N \langle \mathbf{Z}_1(\hat{\mathbf{n}}^{\mathrm{sca}}, \hat{\mathbf{n}}^{\mathrm{inc}}) \rangle_\xi, \tag{1.51}$$

$$\langle \mathbf{K}(\hat{\mathbf{n}}^{\mathrm{inc}}) \rangle_t = N \langle \mathbf{K}_1(\hat{\mathbf{n}}^{\mathrm{inc}}) \rangle_\xi \tag{1.52}$$

(Mishchenko et al. 2004b). Here, $\langle \mathbf{Z}_1(\hat{\mathbf{n}}^{\mathrm{sca}}, \hat{\mathbf{n}}^{\mathrm{inc}}) \rangle_\xi$ and $\langle \mathbf{K}_1(\hat{\mathbf{n}}^{\mathrm{inc}}) \rangle_\xi$ are the single-particle phase and extinction matrix, respectively, computed with respect to the particle-centered coordinate system and averaged over all physically realizable particle states $\xi$ in the group. The state of a particle indicates collectively its size, refractive index, shape, orientation, etc, i.e., all physical characteristics except the position. Obviously, the time averages of the extinction, scattering, and absorption cross sections and the scattering matrix of the entire random particle group can be expressed in terms of the respective ensemble-averaged single-particle cross sections:

$$\langle C_{\mathrm{ext}}(\hat{\mathbf{n}}^{\mathrm{inc}}) \rangle_t = N \langle C_{\mathrm{ext},1}(\hat{\mathbf{n}}^{\mathrm{inc}}) \rangle_\xi, \tag{1.53}$$

$$\langle C_{\mathrm{sca}}(\hat{\mathbf{n}}^{\mathrm{inc}}) \rangle_t = N \langle C_{\mathrm{sca},1}(\hat{\mathbf{n}}^{\mathrm{inc}}) \rangle_\xi, \tag{1.54}$$

$$\langle C_{\mathrm{abs}}(\hat{\mathbf{n}}^{\mathrm{inc}}) \rangle_t = N \langle C_{\mathrm{abs},1}(\hat{\mathbf{n}}^{\mathrm{inc}}) \rangle_\xi, \tag{1.55}$$

$$\langle \mathbf{F}(\Theta) \rangle_t = N \langle \mathbf{F}_1(\Theta) \rangle_\xi. \tag{1.56}$$

The single-scattering albedo of the group is given by the ratio of the ensemble-averaged single-particle scattering and extinction cross sections:

$$\varpi(\hat{\mathbf{n}}^{\mathrm{inc}}) = \frac{\langle C_{\mathrm{sca},1}(\hat{\mathbf{n}}^{\mathrm{inc}}) \rangle_\xi}{\langle C_{\mathrm{ext},1}(\hat{\mathbf{n}}^{\mathrm{inc}}) \rangle_\xi}. \tag{1.57}$$

Finally, the time-averaged radiation force exerted on the group is given by

$$\langle \mathbf{F} \rangle_t = N \langle \mathbf{F}_1 \rangle_\xi. \tag{1.58}$$

The simplicity of the SSA makes it very attractive. The qualitative and quantitative conditions of applicability of the SSA are discussed in detail in Mishchenko et al. (2004b, 2007e). It is shown that one often needs large inter-particle distances and low packing densities in order to make the SSA sufficiently accurate.

### 1.12. Lorenz–Mie theory and *T*-matrix method

In order to compute the scattered field or any far-field optical observable introduced in the preceding sections, one must solve the macroscopic Maxwell equations with a rigorous analytical or numerically exact computer method. Given the virtually unlimited variability of particles and spatial particle configurations encountered in natural and anthropogenic environments, as illustrated by Plate 1.1, such a computation can be a rather nontrivial problem. A fairly detailed summary of the existing theoretical techniques can be found in the monograph by Mishchenko et al. (2000a) as well as in the reviews by Kahnert (2003) and Wriedt (2009). In this section we will briefly consider only two methods which have found extensive practical applications.

In the framework of the classical Lorenz–Mie theory applicable to a single homogeneous spherical particle (Mie 1908; Stratton 1941; Mishchenko and Travis 2008), the incident, internal, and scattered fields are expanded in suitable sets of VSWFs. The expansion coefficients of an incident plane wave can be computed analytically, whereas those of the internal and scattered fields are determined by satisfying the boundary conditions on the sphere surface. Owing to the orthogonality of the VSWFs, each expansion coefficient of either the internal or the scattered field is determined separately. This makes the Lorenz–Mie theory extremely efficient and numerically exact. Several computer implementations of this solution are available on the World Wide Web. Section 5.10 of Mishchenko et al. (2002a) provides a detailed user guide to the public-domain Lorenz–Mie code posted at http://www.giss.nasa.gov/staff/mmishchenko. Using a recursive procedure, one can generalize the Lorenz–Mie theory and treat concentric multi-layer spheres (e.g., Babenko et al. 2003 and references therein).

The Lorenz–Mie solution can also be extended to a cluster of non-overlapping spheres by using the translation addition theorem for the VSWFs. The total field scattered by a multi-sphere cluster is expressed as a superposition of individual fields scattered from each sphere. The external electric field illuminating the cluster and the individual fields scattered by the component spheres are expanded in VSWFs with origins at the individual sphere centers. The orthogonality of the VSWFs in the sphere boundary conditions is exploited by applying the translation addition theorem in which a VSWF centered at one sphere origin is re-expanded about another sphere origin. This procedure ultimately results in a matrix equation for the scattered-field expansion coefficients of each sphere. Numerical solution of this equation for the specific incident wave gives the individual scattered fields and thereby the total scattered field (Fuller and Mackowski 2000).

Alternatively, inversion of the cluster matrix equation gives sphere-centered transition matrices (or *T* matrices) that transform the expansion coefficients of the incident wave into the expansion coefficients of the individual scattered fields. In the far-field zone of the entire aggregate, the individual scattered-field expansions can be transformed into a single expansion centered at a single origin inside the cluster. This procedure gives the *T* matrix that transforms the incident-wave expansion coefficients into the single-origin expansion coefficients of the total scattered field and can be used in the analytical averaging of scattering characteristics over cluster orientations (Mackowski and Mishchenko 1996).

The *T*-matrix method (TMM) was proposed by Waterman (1971) and is a direct generalization of the Lorenz–Mie theory. Consider scattering of a plane electromagnetic wave

$$\mathbf{E}^{\text{inc}}(\mathbf{r}) = \mathbf{E}_0^{\text{inc}} \exp(\mathrm{i}k_1 \hat{\mathbf{n}}^{\text{inc}} \cdot \mathbf{r}), \quad \mathbf{E}_0^{\text{inc}} \cdot \hat{\mathbf{n}}^{\text{inc}} = 0 \tag{1.59}$$

by an arbitrary finite scattering object in the form of a single particle or a fixed aggregate. The incident and scattered fields are expanded in VSWFs as follows:

$$\mathbf{E}^{\text{inc}}(\mathbf{r}) = \sum_{n=1}^{\infty} \sum_{m=-n}^{n} [a_{mn} \, \mathrm{Rg}\mathbf{M}_{mn}(k_1\mathbf{r}) + b_{mn} \, \mathrm{Rg}\mathbf{N}_{mn}(k_1\mathbf{r})], \tag{1.60}$$

$$\mathbf{E}^{\text{sca}}(\mathbf{r}) = \sum_{n=1}^{\infty} \sum_{m=-n}^{n} [p_{mn} \, \mathbf{M}_{mn}(k_1\mathbf{r}) + q_{mn} \, \mathbf{N}_{mn}(k_1\mathbf{r})], \quad r > r_>, \tag{1.61}$$

where $r_>$ is the radius of the smallest circumscribing sphere of the scatterer centered at the origin of the laboratory coordinate system. The mathematical properties of the VSWFs appearing in Eqs. (1.60) and (1.61) are well known. The functions $\mathrm{Rg}\mathbf{M}_{mn}$ and $\mathrm{Rg}\mathbf{N}_{mn}$ are regular (finite) at the origin, while the use of the outgoing functions $\mathbf{M}_{mn}$ and $\mathbf{N}_{mn}$ in Eq. (1.61) ensures that the transverse component of the scattered electric field decays as $1/r$, whereas the radial component decays faster than $1/r$ with $r \to \infty$.

The expansion coefficients of the incident plane wave are given by simple analytical expressions. Owing to the linearity of the Maxwell equations and constitutive relations, the relation between the scattered-field expansion coefficients $p_{mn}$ and $q_{mn}$ on the one hand and the incident field expansion coefficients $a_{mn}$ and $b_{mn}$ on the other hand must be linear and is given by the transition matrix $\mathbf{T}$ as follows:

$$p_{mn} = \sum_{n'=1}^{\infty} \sum_{m'=-n'}^{n'} (T^{11}_{mnm'n'} a_{m'n'} + T^{12}_{mnm'n'} b_{m'n'}), \tag{1.62}$$

$$q_{mn} = \sum_{n'=1}^{\infty} \sum_{m'=-n'}^{n'} (T^{21}_{mnm'n'} a_{m'n'} + T^{22}_{mnm'n'} b_{m'n'}). \tag{1.63}$$

In compact matrix notation,

$$\begin{bmatrix} \mathbf{p} \\ \mathbf{q} \end{bmatrix} = \mathbf{T} \begin{bmatrix} \mathbf{a} \\ \mathbf{b} \end{bmatrix} = \begin{bmatrix} \mathbf{T}^{11} & \mathbf{T}^{12} \\ \mathbf{T}^{21} & \mathbf{T}^{22} \end{bmatrix} \begin{bmatrix} \mathbf{a} \\ \mathbf{b} \end{bmatrix}, \tag{1.64}$$

which means that the column vector of the expansion coefficients of the scattered field is obtained by multiplying the *T* matrix and the column vector of the expansion coefficients of the incident field.

Equation (1.64) is the cornerstone of the *T*-matrix approach. Indeed, if the *T* matrix is known then one can find the scattered field everywhere outside the smallest circumscribing sphere of the object, including the far-field zone. A fundamental feature of the *T*-matrix approach is that the *T* matrix depends only on the physical and geometrical characteristics of the scattering particle (such as particle size relative to the wavelength, shape (morphology), relative refractive index, and orientation with respect to the laboratory reference frame) and is completely independent of the propagation directions and polarization states of the incident and scattered fields. This means that the $T$ matrix need be computed only once and then can be used in calculations for any directions of incidence and scattering and for any polarization state of the incident field.

The TMM was initially developed by Waterman (1971) for a single homogeneous object and was generalized to a multilayered scatterer and an arbitrary cluster of nonspherical particles by Peterson and Ström (1973, 1974). For a homogeneous or concentrically layered sphere, all TMM formulas reduce to those of the Lorenz–Mie theory. In the case of a cluster composed of spherical components, the TMM reduces to the multi-sphere Lorenz–Mie theory mentioned above.

The decisive advantage of the TMM is the use of special functions with well known and convenient analytical properties. As a consequence, the $T$ matrix possesses a number of general analytical properties, the most important of which are the rotation and translation transformation rules.

Let us first consider a fixed laboratory coordinate system $L$ and a coordinate system $P$ affixed to the object. Both coordinate systems have a common origin inside the object. Let the Euler angles $\alpha$, $\beta$, and $\gamma$ transform the coordinate system $L$ into the coordinate system $P$. Then it can be shown (Tsang et al. 1985) that

$$T^{kl}_{mnm'n'}(L;\alpha,\beta,\gamma) = \sum_{m_1 m_2} D^{n}_{mm_1}(\alpha,\beta,\gamma)\, T^{kl}_{m_1 n m_2 n'}(P)\, D^{n'}_{m_2 m'}(-\gamma,-\beta,-\alpha), \quad (1.65)$$

where $D^{n}_{mm'}$ are Wigner $D$ functions (Varshalovich et al. 1988).

Analogously, consider the same scattering problem in two coordinate systems which have identical spatial orientations but different origins. Let the vector $\mathbf{r}_{21}$ connect the origin of coordinate system 1 with that of coordinate system 2. Then the respective $T$ matrices are expressed in terms of one another by the following linear relationship:

$$\begin{bmatrix} \mathbf{T}^{11}(2) & \mathbf{T}^{12}(2) \\ \mathbf{T}^{21}(2) & \mathbf{T}^{22}(2) \end{bmatrix} = \begin{bmatrix} \mathrm{Rg}\mathbf{A}(k_1\mathbf{r}_{21}) & \mathrm{Rg}\mathbf{B}(k_1\mathbf{r}_{21}) \\ \mathrm{Rg}\mathbf{B}(k_1\mathbf{r}_{21}) & \mathrm{Rg}\mathbf{A}(k_1\mathbf{r}_{21}) \end{bmatrix} \begin{bmatrix} \mathbf{T}^{11}(1) & \mathbf{T}^{12}(1) \\ \mathbf{T}^{21}(1) & \mathbf{T}^{22}(1) \end{bmatrix}$$
$$\times \begin{bmatrix} \mathrm{Rg}\mathbf{A}(-k_1\mathbf{r}_{21}) & \mathrm{Rg}\mathbf{B}(-k_1\mathbf{r}_{21}) \\ \mathrm{Rg}\mathbf{B}(-k_1\mathbf{r}_{21}) & \mathrm{Rg}\mathbf{A}(-k_1\mathbf{r}_{21}) \end{bmatrix} \quad (1.66)$$

(Mishchenko et al. 1996c). The matrices on both sides of the $T$ matrix on the right-hand side of Eq. (1.66) consist of coefficients which afford an efficient numerical computation.

The relation (1.65) allows one to drastically simplify orientational averaging in Eqs. (1.51)–(1.58) by doing most of the work analytically. For example, the extinction cross section averaged over the uniform orientation distribution of a nonspherical particle is given by the following remarkably simple formula:

$$\langle C_{\mathrm{ext},1}\rangle_{\xi} = -\frac{2\pi}{k_1^2}\mathrm{Re}\sum_{nm}[T^{11}_{mnmn}(P) + T^{22}_{mnmn}(P)] \quad (1.67)$$

(Mishchenko 1990b; Хлебцов 1991). The corresponding formula for the orientation average of the scattering cross section is hardly more complicated:

$$\langle C_{\text{sca},1}\rangle_\xi = \frac{2\pi}{k_1^2} \sum_{nmn'm'kl} |T^{kl}_{mnm'n'}(P)|^2 \qquad (1.68)$$

(Мищенко 1991; Хлебцов 1991). Analytical averaging of the scattering matrix over orientations in Eq. (1.56) is also possible (Mishchenko 1991e; Khlebtsov 1992; Mackowski and Mishchenko 1996). This approach has been found to be much more efficient than the direct numerical averaging over the Euler angles and helps reduce the requisite computer time by a factor of 50–200. Furthermore, a simple procedure developed by Mishchenko (1993a) allows one to control automatically the convergence of computations upon increasing the size of the $T$ matrix in unit steps, which makes the computer calculations even more efficient. Other examples of applying Eq. (1.65) are the analytical formula for the orientation-averaged radiation force for randomly oriented particles (Mishchenko 1991d) and that for the extinction matrix averaged over an axially symmetric orientation distribution of nonspherical particles (Mishchenko 1991c; Borghese et al. 2007).

The computation of the $T$ matrix for a scattering object in the form of a simple single-body particle is usually based on the so-called extended boundary condition method (EBCM). The $T$ matrix $\mathbf{T}(P)$ is expressed as $-(\text{Rg}\mathbf{Q})\,\mathbf{Q}^{-1}$, where the elements of the matrices $\mathbf{Q}$ and $\text{Rg}\mathbf{Q}$ are obtained numerically by integrating vector products of VSWFs over the object's surface using an appropriate quadrature formula. The numerical computation of the $T$ matrix is often problematic because the vector products of VSWFs can vary by orders of magnitude over the particle surface, thereby resulting in the loss of numerical accuracy. Furthermore, the numerical inversion of the matrix $\mathbf{Q}$ is an ill posed problem and can also be unstable.

It has been demonstrated in Mishchenko and Travis (1994a) and Mishchenko et al. (2002a) that extremely efficient ways of improving the convergence of numerical EBCM calculations are the following:

- the computation of the elements of the matrix $\mathbf{Q}$ and its inverse using extended-precision FORTRAN variables (real variables of the type REAL*16);
- the inversion of the matrix $\mathbf{Q}$ using a special version of the LU factorization procedure.

Despite their remarkable simplicity, these recipes have helped increase the range of applications of the TMM quite dramatically.

A representative collection of public-domain $T$-matrix computer programs has been available at http://www.giss.nasa.gov/staff/mmishchenko since 1996 and has been used in more than 600 peer-reviewed publications. These programs can be used in computations for axially symmetric particles and clusters of spherical monomers. Typical examples are spheroids (Mishchenko and Travis 1994b,c), finite circular cylinders (Mishchenko et al. 1996b), Chebyshev particles (Mishchenko and Lacis 2003), osculating spheres (Mishchenko and Videen 1999), spheres cut by a plane (Mishchenko and Lacis 2003), and clusters of spherical particles with touching or separated components (Mishchenko et al. 1995b). In all cases the scatterers can be randomly (Mishchenko and Travis 1998) or preferentially (Mishchenko 2000) oriented. Further references can be found in Mishchenko et al. (2004c, 2007f, 2008).

All $T$-matrix programs have been thoroughly tested (Hovenier et al. 1996) against the separation of variables method for spheroids (Farafonov et al. 1996; Voshchinnikov 2004). Very high numerical accuracy of the $T$-matrix codes has allowed us to generate benchmark results with five and more accurate decimals which can be used for testing other numerically exact and approximate approaches (Mishchenko 1991b,e; Mishchenko et al. 1996b). Extensive timing tests have shown that the numerical efficiency of these $T$-matrix codes is unparalleled, especially in computations for randomly oriented particles.

## 1.13. Spherically symmetric particles

The aim of this and the following three sections is to use the general theory outlined above as well as illustrative numerical results to discuss far-field scattering and absorption properties of individual particles (Sections 1.5 and 1.6) and small random particle groups (Section 1.11).

It follows from the Lorenz–Mie theory (or its generalizations for radially inhomogeneous particles) that the extinction, scattering, and absorption cross sections, the single-scattering albedo, and the radiation force for a spherically symmetric particle are independent of the direction of propagation and polarization state of the incident light. Furthermore, the extinction matrix is diagonal and given by

$$\mathbf{K}(\hat{\mathbf{n}}^{\mathrm{inc}}) \equiv \mathbf{K} = \mathrm{diag}[C_{\mathrm{ext}}, C_{\mathrm{ext}}, C_{\mathrm{ext}}, C_{\mathrm{ext}}]. \tag{1.69}$$

The phase matrix satisfies the symmetry relations (Hovenier 1969)

$$\mathbf{Z}(\theta^{\mathrm{sca}}, \varphi^{\mathrm{inc}}; \theta^{\mathrm{inc}}, \varphi^{\mathrm{sca}}) = \mathbf{Z}(\theta^{\mathrm{sca}}, -\varphi^{\mathrm{sca}}; \theta^{\mathrm{inc}}, -\varphi^{\mathrm{inc}}) = \mathbf{\Delta}_{34}\,\mathbf{Z}(\theta^{\mathrm{sca}}, \varphi^{\mathrm{sca}}; \theta^{\mathrm{inc}}, \varphi^{\mathrm{inc}})\mathbf{\Delta}_{34}, \tag{1.70}$$

$$\mathbf{Z}(\pi - \theta^{\mathrm{sca}}, \varphi^{\mathrm{sca}}; \pi - \theta^{\mathrm{inc}}, \varphi^{\mathrm{inc}}) = \mathbf{\Delta}_{34}\mathbf{Z}(\theta^{\mathrm{sca}}, \varphi^{\mathrm{sca}}; \theta^{\mathrm{inc}}, \varphi^{\mathrm{inc}})\mathbf{\Delta}_{34}, \tag{1.71}$$

where $\mathbf{\Delta}_{34} = \mathrm{diag}[1, 1, -1, -1]$. Also, it depends only on the difference between the azimuthal angles of the scattering and incidence directions rather than on their specific values:

$$\mathbf{Z}(\hat{\mathbf{n}}^{\mathrm{sca}}, \hat{\mathbf{n}}^{\mathrm{inc}}) = \mathbf{Z}(\theta^{\mathrm{sca}}, \theta^{\mathrm{inc}}, \varphi^{\mathrm{sca}} - \varphi^{\mathrm{inc}}). \tag{1.72}$$

Finally, the scattering matrix has a very simple block-diagonal structure with only four independent elements (van de Hulst 1957):

$$\mathbf{F}(\Theta) = \begin{bmatrix} F_{11}(\Theta) & F_{12}(\Theta) & 0 & 0 \\ F_{12}(\Theta) & F_{11}(\Theta) & 0 & 0 \\ 0 & 0 & F_{33}(\Theta) & F_{34}(\Theta) \\ 0 & 0 & -F_{34}(\Theta) & F_{33}(\Theta) \end{bmatrix}. \tag{1.73}$$

In addition,

$$F_{33}(0) = F_{11}(0), \qquad F_{33}(\pi) = -F_{11}(\pi), \tag{1.74}$$

$$F_{12}(0) = F_{12}(\pi) = 0, \qquad F_{34}(0) = F_{34}(\pi) = 0. \tag{1.75}$$

The above results apply also to the ensemble-averaged phase, extinction, and scattering matrices and the optical cross sections of a small group of spherically symmetric particles.

Given the high efficiency of computer codes based on the Lorenz–Mie theory (e.g., Wiscombe 1980), computer calculations for spherical particles, monodisperse as well as polydisperse, have become quite routine. Representative Lorenz–Mie results have been discussed in detail by Hansen and Travis (1974) and Mishchenko et al. (2002a), which allows us to largely skip this subject. However, since polarimetry is the main focus of this book, we will facilitate the discussion in Chapters 2–4 by giving an illustrative example demonstrating the sensitivity of polarization to particle size and refractive index. Plate 1.2a shows the ratio $-\langle F_{12}\rangle_\xi / \langle F_{11}\rangle_\xi$, called the degree of linear polarization for unpolarized incident light, computed for a moderately wide size distribution of spherical particles, where $\langle ... \rangle_\xi$ in this case denotes averaging over particle radii. The effective size parameter is defined as $x_{\text{eff}} = k_1 r_{\text{eff}}$, where the effective radius $r_{\text{eff}}$ is the ratio of the third to the second moment of the size distribution (Hansen and Travis 1974). It is obvious that polarization is indeed quite a sensitive function of particle microphysical characteristics and can change not only its absolute value but even the sign with minute variations in particle size and/or refractive index. Plate 1.2a explains why multispectral measurements of polarization in a wide range of scattering angles can contain very specific and accurate information on particle size and composition. In fact, diagrams similar to those in Plate 1.2a helped Hansen and Hovenier (1974) to analyze ground-based polarimetric observations of Venus and determine the size distribution, shape (sphericity), and chemical composition of Venus cloud particles with extreme accuracy.

### 1.14. General effects of nonsphericity and orientation

The discussion in this section applies equally to a fixed nonspherical particle or a small random particle group in which particles are preferentially oriented. Note that the class of "nonspherical" particles includes, as a particulate case, spherical as well as nonspherical particles made of an optically active material. Then, in general,

- the $4 \times 4$ extinction matrix $\mathbf{K}$ or $\langle \mathbf{K} \rangle_\xi$ does not degenerate to a direction- and polarization-independent scalar extinction cross section;
- the (ensemble-averaged) extinction, scattering, and absorption cross sections, the single-scattering albedo, and the radiation force depend on the propagation direction and polarization state of the incident beam;
- the scattering matrix $\mathbf{F}$ or $\langle \mathbf{F} \rangle_\xi$ does not have the simple block-diagonal structure of Eq. (1.73): all 16 elements of the scattering matrix can be nonzero and depend on the orientation of the particle(s) with respect to the scattering plane rather than on only the scattering angle;
- Eqs. (1.74)–(1.75) are not valid; and
- the phase matrix $\mathbf{Z}$ or $\langle \mathbf{Z} \rangle_\xi$ depends on the specific values of the azimuthal angles of the incidence and scattering directions rather than on their difference and does not obey the symmetry relations (1.70) and (1.71).

Thus, any of these effects can directly indicate the presence of perfectly or preferentially oriented particles lacking perfect spherical symmetry.

Plate 1.3a gives an example of the dependence of the extinction cross section on the incidence direction (Mishchenko and Lacis 2003). It depicts the results of *T*-matrix computations of the dimensionless normalized extinction $\widetilde{C}_{\rm ext} = C_{\rm ext}/(\pi r_{\rm ev}^2)$ for a prolate spheroid with a semi-axis ratio of $a/b = 0.9$, where $r_{\rm ev}$ is the radius of the equal-volume sphere, $b$ is the spheroid semi-axis along the axis of rotation, and $a$ is the semi-axis in the perpendicular direction. $\widetilde{C}_{\rm ext}$ is plotted as a function of the size parameter $x_{\rm ev} = k_1 r_{\rm ev}$ and the angle $\beta$ between the spheroid axis of rotation and the incidence direction. The relative refractive index of the spheroid is fixed at 1.4, and the incident light is assumed to be unpolarized. The significant overall increase of the extinction cross section $C_{\rm ext}$ with increasing $\beta$ can be explained by the growing area of the spheroid's geometrical projection on the plane perpendicular to the incidence direction. The numerous local maxima in the surface plot of extinction are manifestations of so-called morphology-dependent resonances (MDRs) (Chýlek 1976; Hill and Benner 1988; Mishchenko et al. 2002a) which will be discussed in more detail below.

Panels (a–c) of Plate 1.4 illustrate the shape and orientation dependence of the ratio $-F_{12}/F_{11}$. The degree of linear polarization is plotted as a function of the scattering angle $\Theta$ and surface-equivalent-sphere size parameter $x_{\rm se} = k_1 r_{\rm se}$, where $r_{\rm se}$ is the radius of the surface-equivalent sphere. These panels reveal intricate distributions of the areas of positive and negative polarization first displayed in this manner for monodisperse particles by Hansen and Travis (1974). Each complex so-called "butterfly structure" is a superposition of countless MDRs of varying shape, width, and amplitude and a component caused by the interference of the incident and scattered fields. With the exception of the region of Rayleigh scattering ($x_{\rm se} \lesssim 1$), the three panels are vastly different. In particular, the polarization patterns for the same spheroids but in two different orientations resemble each other no more than either pattern resembles that for the surface-equivalent spheres. The results shown in panel (c) obviously violate the equalities (1.75) and thus cannot be attributed mistakenly to spherically symmetric particles. However, the specific spheroid orientation used to create panel (b) does not cause a violation of Eq. (1.75), which shows that in some situations Eq. (1.75) alone cannot be used to distinguish unambiguously between a spherically symmetric particle and a preferentially oriented nonspherical particle.

## 1.15. Mirror-symmetric ensembles of randomly oriented particles: general traits

Consider now a small random group of particles such that the distribution of particle orientations during the measurement is statistically uniform. Furthermore, we assume that each particle in the group has a plane of symmetry and/or is accompanied by its mirror counterpart. Then most of the results of Section 1.13 apply (van de Hulst 1957; Mishchenko et al. 2002a). Specifically, the ensemble-averaged extinction, scattering, and absorption cross sections and the single-scattering albedo are independent of the propagation direction and polarization state of the incident light. The ensemble-averaged extinction matrix is diagonal and given by

$$\langle \mathbf{K}(\hat{\mathbf{n}}^{\text{inc}}) \rangle_\xi \equiv \langle \mathbf{K} \rangle_\xi = \text{diag}[\langle C_{\text{ext}} \rangle_\xi, \langle C_{\text{ext}} \rangle_\xi, \langle C_{\text{ext}} \rangle_\xi, \langle C_{\text{ext}} \rangle_\xi]. \tag{1.76}$$

The ensemble-averaged phase matrix satisfies the symmetry relations

$$\begin{aligned} \langle \mathbf{Z}(\theta^{\text{sca}}, \varphi^{\text{inc}}; \theta^{\text{inc}}, \varphi^{\text{sca}}) \rangle_\xi &= \langle \mathbf{Z}(\theta^{\text{sca}}, -\varphi^{\text{sca}}; \theta^{\text{inc}}, -\varphi^{\text{inc}}) \rangle_\xi \\ &= \mathbf{\Delta}_{34} \langle \mathbf{Z}(\theta^{\text{sca}}, \varphi^{\text{sca}}; \theta^{\text{inc}}, \varphi^{\text{inc}}) \rangle_\xi \mathbf{\Delta}_{34}, \end{aligned} \tag{1.77}$$

$$\langle \mathbf{Z}(\pi - \theta^{\text{sca}}, \varphi^{\text{sca}}; \pi - \theta^{\text{inc}}, \varphi^{\text{inc}}) \rangle_\xi = \mathbf{\Delta}_{34} \langle \mathbf{Z}(\theta^{\text{sca}}, \varphi^{\text{sca}}; \theta^{\text{inc}}, \varphi^{\text{inc}}) \rangle_\xi \mathbf{\Delta}_{34} \tag{1.78}$$

and also depends only on the difference between the azimuthal angles of the scattering and incidence directions rather than on their specific values:

$$\langle \mathbf{Z}(\hat{\mathbf{n}}^{\text{sca}}, \hat{\mathbf{n}}^{\text{inc}}) \rangle_\xi = \langle \mathbf{Z}(\theta^{\text{sca}}, \theta^{\text{inc}}, \varphi^{\text{sca}} - \varphi^{\text{inc}}) \rangle_\xi. \tag{1.79}$$

The ensemble-averaged scattering matrix has a similar block-diagonal structure, but now has six rather than four independent elements:

$$\langle \mathbf{F}(\Theta) \rangle_\xi = \begin{bmatrix} \langle F_{11}(\Theta) \rangle_\xi & \langle F_{12}(\Theta) \rangle_\xi & 0 & 0 \\ \langle F_{12}(\Theta) \rangle_\xi & \langle F_{22}(\Theta) \rangle_\xi & 0 & 0 \\ 0 & 0 & \langle F_{33}(\Theta) \rangle_\xi & \langle F_{34}(\Theta) \rangle_\xi \\ 0 & 0 & -\langle F_{34}(\Theta) \rangle_\xi & \langle F_{44}(\Theta) \rangle_\xi \end{bmatrix}. \tag{1.80}$$

The equalities (1.74) are no longer valid, but the properties (1.75) are preserved:

$$\langle F_{12}(0) \rangle_\xi = \langle F_{12}(\pi) \rangle_\xi = 0, \qquad \langle F_{34}(0) \rangle_\xi = \langle F_{34}(\pi) \rangle_\xi = 0. \tag{1.81}$$

The equalities (1.81) are illustrated by the *T*-matrix results for randomly oriented monodisperse spheroids shown in Plate 1.4d.

Despite the similarity of the matrices (1.73) and (1.80), the identities $\langle F_{22}(\Theta) \rangle_\xi \equiv \langle F_{11}(\Theta) \rangle_\xi$ and $\langle F_{44}(\Theta) \rangle_\xi \equiv \langle F_{33}(\Theta) \rangle_\xi$ do not hold in general, which is well illustrated by the results of laboratory measurements for natural feldspar particles shown in Plate 1.5 (Volten et al. 2001). As a consequence, measurements of the linear backscattering depolarization ratio

$$\delta_{\text{L}} = \frac{\langle F_{11}(\pi) \rangle_\xi - \langle F_{22}(\pi) \rangle_\xi}{\langle F_{11}(\pi) \rangle_\xi + \langle F_{22}(\pi) \rangle_\xi}, \quad 0 \le \delta_{\text{L}} \le 1 \tag{1.82}$$

and the closely related circular backscattering depolarization ratio (Mishchenko and Hovenier 1995)

$$\delta_{\text{C}} = \frac{\langle F_{11}(\pi) \rangle_\xi + \langle F_{44}(\pi) \rangle_\xi}{\langle F_{11}(\pi) \rangle_\xi - \langle F_{44}(\pi) \rangle_\xi} = \frac{2\delta_{\text{L}}}{1 - \delta_{\text{L}}} \ge 2\delta_{\text{L}} \tag{1.83}$$

are among the most reliable means of detecting particle nonsphericity (Sassen 2000). We will demonstrate later, however, that these ratios are not necessarily good quantitative indicators of the *degree* of particle nonsphericity.

## 1.16. Mirror-symmetric ensembles of randomly oriented particles: quantitative traits

Of course, besides the qualitative distinctions discussed in the preceding section, there can be significant quantitative differences between specific scattering properties of randomly oriented nonspherical particles and "equivalent" (e.g., volume-equivalent or surface-equivalent) spheres. We begin by discussing the effects of nonsphericity and random orientation on MDRs. Plate 1.3b summarizes the results of numerically exact *T*-matrix computations for monodisperse spheres and volume-equivalent, randomly oriented spheroids with a relative refractive index of 1.4 in the range of size parameters affected by three super-narrow Lorenz–Mie MDRs $b_{38}^1$, $a_{38}^1$, and $b_{39}^1$ as well as three broader resonance features (Mishchenko and Lacis 2003). We follow the notation introduced by Chýlek et al. (1978) which implies, for example, that $b_{38}^1$ is the first resonance generated by the $b_{38}$ partial Lorenz–Mie coefficient as the sphere size parameter increases from zero. Now the direction-independent normalized extinction is defined as the ratio $\widetilde{C}_{\text{ext}} = \langle C_{\text{ext}} \rangle_\xi / (\pi r_{\text{ev}}^2)$, where $\langle C_{\text{ext}} \rangle_\xi$ is the orientation-averaged extinction cross section. It is seen that increasing the aspect ratio of the spheroid $\varepsilon$ (i.e., the ratio of the largest to the smallest particle dimensions) rapidly reduces the height of the extinction peaks. It is in fact remarkable that the deformation of a sphere by as little as one hundredth of a wavelength essentially annihilates the super-narrow MDRs. A secondary effect of increasing asphericity is to shift the resonances to smaller size parameters. It takes significantly larger asphericities to suppress the broader MDRs. An interesting feature of the curve for $a/b = 0.9$ is the minute high-frequency ripple superposed on a slowly and weakly varying background. This ripple is absent in the curves for the nearly spherical spheroids and is the contribution of additional natural frequencies of oscillation of distinctly aspherical spheroids with specific orientations relative to the incident beam. This effect is well seen in Plate 1.3a.

The smoothing effect of averaging over orientations of a nonspherical particle on MDRs is also well seen from the comparison of Plates 1.4b, 1.4c, and 1.4d. Averaging over sizes reinforces this effect. This is demonstrated by Plates 1.4e and 1.4f which parallel Plates 1.4a and 1.4d, respectively, and show the ratio $-\langle F_{12}(\Theta) \rangle_\xi / \langle F_{11}(\Theta) \rangle_\xi$ computed for a modified power law size distribution of spheres and randomly oriented surface-equivalent spheroids (Mishchenko et al. 1996c). The resulting polarization patterns are now smooth enough to derive conclusions regarding the likely quantitative effects of nonsphericity of natural polydisperse particles. Among such effects are the bridge of positive polarization at side-scattering angles and a negative polarization branch at backscattering angles measured previously for narrow size distributions of nearly cubically shaped NaCl particles with mean size parameters ranging from 3.1 to 19.9 (Perry et al. 1978) as well as for many other types of mineral particles (Volten et al. 2001; Muñoz and Volten 2006; Shkuratov et al. 2007).

We have already mentioned that backscattering depolarization measurements are widely used for detecting and characterizing nonspherical particles (Sassen 2000). *T*-matrix results depicted in Plate 1.6 demonstrate indeed that wavelength-sized particles can cause large depolarization ratios. An interesting trait of essen-tially all the curves shown in this plate is a rapid increase in $\delta_L$ as the effective size parameter increases from 0 to about 10. Moreover, maximal $\delta_L$ values for most shapes are observed at size parameters close to and sometimes slightly smaller than 10. Unfortunately, the *T*-matrix results show no obvious relationship between $\delta_L$ and the particle aspect ratio. Even spheroids with aspect ratios as small as 1.05 (a 2.5% deviation from the perfect spherical shape) produce strong depolarization. In fact, the largest $\delta_L$ values are generated by prolate spheroids with aspect ratios as small as 1.2 (a 10% deviation from a sphere). Furthermore, $\delta_L$ for spheroids and, especially, cylinders tends to saturate with increasing aspect ratio. These results suggest that although a non-zero $\delta_L$ value is an unequivocal indication of particle nonsphericity, it is not necessarily a measure of the degree of deviation of the particle shape from that of a perfect sphere.

Despite the significant progress in our ability to model scattering by nonspherical particles, direct theoretical computations for many types of natural and artificial particles with sizes comparable to and greater than the wavelength (Plate 1.1) remain highly problematic. Therefore, there have been several attempts to simulate the scattering and absorption properties of actual particles using simple model shapes. These attempts have been based on the realization that in addition to size and orientation averaging as discussed above, averaging over shapes may also prove to be necessary in many cases. Indeed, quite often ensembles of natural and artificial particle exhibit a vast variety of shapes, which makes questionable the utility of a single model shape (however "irregular" it may look to the human eye) in the representation of scattering properties of an ensemble.

To illustrate this point, Plate 1.2b shows the phase functions, defined as

$$a_1(\Theta) = \frac{4\pi \langle F_{11}(\Theta)\rangle_\xi}{\langle C_{\mathrm{sca}}\rangle_\xi}, \tag{1.84}$$

computed for polydisperse, randomly oriented prolate spheroids with varying aspect ratios (Mishchenko et al. 1997b). The phase function describes the angular distribution of the scattered intensity provided that the incident light is unpolarized and satisfies the following normalization condition:

$$\frac{1}{2}\int_0^\pi \mathrm{d}\Theta \sin\Theta \, a_1(\Theta) = 1 \tag{1.85}$$

(cf. Eq. (1.24)). Plate 1.2b demonstrates indeed that even after size and orientation averaging, each spheroidal shape produces a unique, shape-specific scattering pattern, whereas laboratory and *in situ* measurements for real nonspherical particles usually show smooth, rather featureless phase functions. However, the green curves in Plate 1.5 (Dubovik et al. 2006) show that *shape mixtures* of polydisperse, randomly oriented prolate and oblate spheroids can provide a good quantitative fit to the results of accurate laboratory measurements of the scattering matrix for natural irregular particles.

These examples lead to two important conclusions. First of all, they provide evidence that the often observed smooth scattering-angle dependence of the elements of the scattering matrix for natural and artificial ensembles of nonspherical particles is largely caused by the diversity of particle shapes in the ensemble. Secondly, they suggest that at least some scattering properties of ensembles of irregular particles can be adequately modeled using polydisperse shape mixtures of simple particles such as spheroids. These two conclusions form the gist of the so-called statistical approach according to which particles chosen for the purpose of ensemble averaging need not be in one-to-one morphological correspondence with the actual particle ensemble and may have relatively simple shapes (Hill et al. 1984; Mishchenko et al. 1997b). Needless to say, forming representative mixtures of less regular particles then spheroids should be expected to eventually provide an even better model of electromagnetic scattering by many natural and artificial particle ensembles (e.g., Bi et al. 2009; Zubko et al. 2009).

Contrasting the green and the respective red curves in Plate 1.5 provides a good illustration of the typical nonspherical–spherical differences in the elements of the scattering matrix discussed in detail by Mishchenko et al. (2002a). For example, several theoretical and laboratory analyses of the phase-function patterns for volume- or surface-equivalent spherical and nonspherical particles have revealed the following five distinct scattering-angle ranges:

- nonsphere $\approx$ sphere from $\Theta = 0°$ to $\Theta \sim 15° - 20°$;
- nonsphere $>$ sphere from $\Theta \sim 15° - 20°$ to $\Theta \sim 35°$;
- nonsphere $<$ sphere from $\Theta \sim 35°$ to $\Theta \sim 85°$; (1.86)
- nonsphere $\gg$ sphere from $\Theta \sim 85°$ to $\Theta \sim 150°$;
- nonsphere $\ll$ sphere from $\Theta \sim 150°$ to $\Theta = 180°$.

Although the specific boundaries of these regions can shift with particle shape and relative refractive index (e.g., Mugnai and Wiscombe 1989; Mishchenko et al. 2002a), the enhanced side-scattering and suppressed backscattering appear to be rather universal characteristics of nonspherical particles.

The degree of linear polarization for unpolarized incident light, $-\langle F_{12}(\Theta)\rangle_\xi / \langle F_{11}(\Theta)\rangle_\xi$, tends to be positive at scattering angles around $100° - 120°$ for nonspherical particles. Whereas $\langle F_{22}(\Theta)\rangle_\xi / \langle F_{11}(\Theta)\rangle_\xi \equiv 1$ for spherically symmetric scatterers, the same ratio for nonspherical particles deviates significantly from the value 1 and exhibits strong backscattering depolarization. Similarly,

$$\langle F_{33}(\Theta)\rangle_\xi / \langle F_{11}(\Theta)\rangle_\xi \equiv \langle F_{44}(\Theta)\rangle_\xi / \langle F_{11}(\Theta)\rangle_\xi$$

for spherically symmetric particles, whereas the ratio $\langle F_{44}(\Theta)\rangle_\xi / \langle F_{11}(\Theta)\rangle_\xi$ for nonspherical particles tends to be greater than the ratio $\langle F_{33}(\Theta)\rangle_\xi / \langle F_{11}(\Theta)\rangle_\xi$ at most scattering angles, especially in the backscattering direction. The ratios $\langle F_{34}(\Theta)\rangle_\xi / \langle F_{11}(\Theta)\rangle_\xi$ for spherical and nonspherical particles also reveal significant quantitative differences, especially at large scattering angles.

Unlike ensembles of irregularly shaped particles, regular nonspherical shapes may cause pronounced angular features in the elements of the scattering matrix, especially as the particle size starts to exceed the wavelength of the incident light. This is well illustrated by Plate 1.2c which shows the results of exact *T*-matrix computations of the phase function for randomly oriented compact circular cylinders with surface-equivalent-sphere size parameters ranging from 40 to 180 (Mishchenko and Macke 1999). As the size parameter increases, the *T*-matrix phase functions develop such typical geometrical-optics features as the 46° halo and the strong and narrow backscattering peak seen in the grey curve. Such pronounced phase-function features caused by regular ice-crystal shapes are responsible for many spectacular atmospheric-optics displays and affect the results of cirrus-cloud remote sensing (Волковицкий и др. 1984; Mishchenko et al. 1996a; Liou 2002; Yang and Liou 1998, 2006). In many cases, however, various imperfections of the ice-crystal shape and/or multiple internal inclusions (e.g., in the form of air bubbles) destroy sharp geometrical-optics features such as halos and cause smooth and featureless phase functions similar to that depicted by diamonds in the upper left panel of Plate 1.5 (Macke et al. 1996a,b; Labonnote et al. 2001).

In most cases nonspherical–spherical differences in the optical cross sections and the single-scattering albedo are not nearly as significant as those in the scattering matrix elements. The same is true of the asymmetry parameter defined as

$$g = \langle \cos\Theta \rangle = \frac{1}{2}\int_0^{\pi} \mathrm{d}\Theta \sin\Theta \, a_1(\Theta) \cos\Theta. \tag{1.87}$$

This does not mean, however, that the effects of nonsphericity on the integral scattering and absorption characteristics are always negligible or unimportant. A good example of particles characterized by integral radiometric properties vastly different from those of volume-equivalent spheres are clusters composed of large numbers of small monomers such as soot aggregates (Sorensen 2001; Liu and Mishchenko 2005, 2007; Liu et al. 2008; Moosmüller et al. 2009).

## 1.17. Multiple scattering by random particulate media: exact results

After having discussed the single-scattering properties of individual particles and small particle groups, we switch to the subject of electromagnetic scattering by many-particle groups. The far-field order-of-scattering expansion (1.41) coupled with the principle of ergodicity provides the foundation necessary to develop the microphysical theories of RT and WL. However, before discussing these inherently *approximate* theories, in this section we will use *numerically exact* results in order to develop an understanding of what further assumptions and approximations will be necessary and what specific scattering effects these theories may or may not be expected to encompass. To this end, we will analyze *T*-matrix results computed for a macroscopic volume filled with randomly distributed wavelength-sized particles (Mishchenko et al. 2007d, 2009b,c; Mishchenko and Liu 2007, 2009). We have already mentioned that, for practical reasons, the superposition TMM discussed in Section 1.12 cannot be used yet to handle random media consisting of very large numbers of particles such as clouds, colloids, and powder surfaces. However, it does provide the potential to model rather complex particulate systems and thereby simulate the effect of randomness of particle positions as well as the onset and evolution of various "multiple-scattering" effects with increasing number of particles in a statistically homogeneous volume of discrete random medium.

### *1.17.1. Static and dynamic light scattering*

As we have explained above, in order to simulate measurements of light scattering by a rapidly changing object one needs to solve the Maxwell equations repeatedly for a representative set of distinct object configurations. After the set of solutions of the Maxwell equations has been obtained, one has a choice of

- analyzing the statistical information content of differences in the individual solutions; or
- applying an averaging procedure and thereby isolating the static component of the scattering pattern.

These two approaches are known as *dynamic* and *static* light scattering (Berne and Pecora 1976; Brown 1993; Mishchenko et al. 2006b).

### *1.17.2. Fixed configurations of randomly positioned particles: speckle*

Let us assume that a number $N$ of identical spherical particles are distributed randomly throughout a spherical volume $V$ with a radius $R$ much greater than the particle radius $a$, as shown in Plate 1.7a. The size parameter of the particles is fixed at $k_1a = 4$, while the size parameter of the spherical volume is fixed at $k_1R = 40$. The refractive index of the particles relative that of the surrounding medium is 1.32. The large spherical volume $V$ is illuminated by a plane electromagnetic wave. The incidence direction coincides with the positive direction of the $z$-axis of the laboratory reference frame and the meridional plane of the incidence direction coincides with the $xz$ half-plane with $x \geq 0$ (i.e., $\varphi^{\text{inc}} = 0$ according to Fig. 1.3). The angular distribution and polarization state of the scattered light in the far-field zone of the entire volume is described by the Stokes phase matrix $\mathbf{Z}$ according to Eq. (1.20).

Let us first assume that the incident light is circularly polarized in the counter-clockwise sense when viewed in the direction of propagation, which implies that $V^{\text{inc}} = I^{\text{inc}}$ and $Q^{\text{inc}} = U^{\text{inc}} = 0$. Panels (a−d) of Plate 1.8 show the far-field angular distributions of the intensity $I^{\text{sca}}$ scattered in the backward hemisphere by the large spherical volume filled with $N$ = 1, 5, 20, and 80 particles. The individual particle positions within $V$ were chosen using a random coordinate generator, but otherwise they are fixed. The scattering pattern for $N$ = 1 is rather smooth and perfectly azimuthally symmetric, as it should be for a single wavelength-sized spherical particle. However, panels (b−d) demonstrate typical speckle patterns.

The origin of the speckle can be explained as follows. Equations (1.38) and (1.9) suggest that at the distant observation point, the partial field due to any particle sequence contributing to the right-hand side of Eq. (1.38) becomes an outgoing spherical wavelet centered at the last particle of the sequence. For example, the term $\hat{G}\hat{T}_i\hat{G}\hat{T}_j\hat{G}\hat{T}_l E^{\text{inc}}$ becomes a spherical wavelet centered at particle $i$ since the left-most dyadic Green's function takes the following asymptotic form:

$$\begin{aligned} \overleftrightarrow{G}(\mathbf{r},\mathbf{r}') &= \overleftrightarrow{G}(\mathbf{r}-\mathbf{R}_i,\mathbf{r}'-\mathbf{R}_i) \\ &\underset{r_i\to\infty}{\longrightarrow} (\overleftrightarrow{I} - \hat{\mathbf{r}}_i \otimes \hat{\mathbf{r}}_i)\frac{\exp(\mathrm{i}k_1 r_i)}{4\pi r_i}\exp[-\mathrm{i}k_1\hat{\mathbf{r}}_i\cdot(\mathbf{r}'-\mathbf{R}_i)], \quad \mathbf{r}'\in V_i, \end{aligned} \tag{1.88}$$

where the notation is explained in Fig. 1.8, and we have made use of the translational invariance property of $\vec{G}$. This occurs irrespective of whether the particles are densely packed or sparsely distributed. The Stokes parameters of the scattered light (Eq. (1.19)), can be directly expressed in terms of the elements of the scattering coherency dyad $\ddot{\rho}^{\text{sca}} = \mathbf{E}^{\text{sca}} \otimes (\mathbf{E}^{\text{sca}})^*$:

$$\mathbf{I}^{\text{sca}} = \frac{1}{2}\sqrt{\frac{\varepsilon_1}{\mu_0}} \begin{bmatrix} \hat{\boldsymbol{\theta}}^{\text{sca}} \cdot \ddot{\rho}^{\text{sca}} \cdot \hat{\boldsymbol{\theta}}^{\text{sca}} + \hat{\boldsymbol{\varphi}}^{\text{sca}} \cdot \ddot{\rho}^{\text{sca}} \cdot \hat{\boldsymbol{\varphi}}^{\text{sca}} \\ \hat{\boldsymbol{\theta}}^{\text{sca}} \cdot \ddot{\rho}^{\text{sca}} \cdot \hat{\boldsymbol{\theta}}^{\text{sca}} - \hat{\boldsymbol{\varphi}}^{\text{sca}} \cdot \ddot{\rho}^{\text{sca}} \cdot \hat{\boldsymbol{\varphi}}^{\text{sca}} \\ -\hat{\boldsymbol{\theta}}^{\text{sca}} \cdot \ddot{\rho}^{\text{sca}} \cdot \hat{\boldsymbol{\varphi}}^{\text{sca}} - \hat{\boldsymbol{\varphi}}^{\text{sca}} \cdot \ddot{\rho}^{\text{sca}} \cdot \hat{\boldsymbol{\theta}}^{\text{sca}} \\ \mathrm{i}\,(\hat{\boldsymbol{\varphi}}^{\text{sca}} \cdot \ddot{\rho}^{\text{sca}} \cdot \hat{\boldsymbol{\theta}}^{\text{sca}} - \hat{\boldsymbol{\theta}}^{\text{sca}} \cdot \ddot{\rho}^{\text{sca}} \cdot \hat{\boldsymbol{\varphi}}^{\text{sca}}) \end{bmatrix}, \tag{1.89}$$

where $\hat{\boldsymbol{\theta}}^{\text{sca}}$ and $\hat{\boldsymbol{\varphi}}^{\text{sca}}$ are the polar-angle and azimuth-angle unit vectors such that $\hat{\mathbf{n}}^{\text{sca}} = \hat{\boldsymbol{\theta}}^{\text{sca}} \times \hat{\boldsymbol{\varphi}}^{\text{sca}}$. The dyadic product of the right-hand side of Eq. (1.38) and its complex-conjugate counterpart is the sum of an infinite number of terms, each describing the result of interference of two spherical wavelets centered at the end particles of a pair of particle sequences.

One such pair of particle sequences is shown in Plate 1.7b. If the interference of the corresponding spherical wavelets is constructive (destructive) then it serves to increase (decrease) the total intensity scattered in the direction $\hat{\mathbf{n}}^{\text{sca}}$. The total scattered intensity is the sum of the interference results contributed by all possible pairs of particle sequences. The typical angular width of each interference maximum or minimum is proportional to $1/k_1 R$, whereas the number of these maxima and minima grows swiftly with increasing $N$. These two factors explain the increasingly spotty appearance of the scattering patterns in Plates 1.8b–d.

Of course, the speckle pattern depends not only on the number of particles $N$ but also on the specific way they are arranged with respect to the laboratory coordinate system. This is illustrated by Plates 1.8d and 1.8e computed for two different random 80-particle configurations shown in Plate 1.7c.

### *1.17.3. Static scattering*

Plates 1.8d and 1.8e illustrate the range of variability of the speckle pattern that can be expected upon even minute changes in a random multi-particle configuration. Obviously, neither the speckle pattern nor its variability are reproduced by the classical theories of RT and WL, which indicates that *neither theory describes the instantaneous state of electromagnetic radiation* in a discrete random medium. Instead, both theories fall in the realm of static scattering and describe the result of averaging the relevant optical observables over a significant period of time or, equivalently, over a significant range of random particle positions.

To illustrate this fundamental point, one needs an efficient way of averaging the computed scattering signal over very many configurations of the $N$-particle group. A brute-force solution would be to use a random coordinate generator repeatedly to create a large number of different $N$-particle configurations and then average numerically the corresponding individual $T$-matrix results. The more effective approach used here is to create only one random $N$-particle configuration and then average over all possible orientations of this configuration with respect to the labo-ratory coordinate system. This procedure yields an infinite continuous set of random realizations of the $N$-particle group and takes full advantage of the highly efficient orientation averaging procedure afforded by the superposition $T$-matrix method.

Plate 1.8f shows the result of averaging the speckle pattern over the uniform orientation distribution of the 80-particle configuration used to compute Plate 1.8d. One can see that with the exception of a notable backscattering peak, the speckle structure is essentially gone. This is not surprising. Indeed, each speckle element is the result of constructive or destructive interference of two wavelets scattered along specific particle sequences such as those shown in Plate 1.7b. The phase difference between the wavelets changes randomly as the particles move, so that the average result of the interference is zero. However, there are certain pairs of wavelets that interfere constructively irrespective of particle positions and thereby are responsible for the residual scattering pattern. We will demonstrate below that the backscattering intensity peak seen in Plate 1.8f as well as the smooth intensity background are in fact caused by special classes of such wavelet pairs.

In what follows, we employ the mathematical concept of multiple scattering to interpret various effects of increasing the number of particles filling the scattering volume on the static scattering patterns obtained by averaging over all orientations of a random $N$-particle configuration with respect to the laboratory reference frame. We make a simplifying assumption that $\varphi^{\mathrm{sca}} = 0$ and define the scattering direction in terms of the scattering angle $\Theta = \theta^{\mathrm{sca}}$. Then the far-field scattering pattern can be conveniently described in terms of the so-called normalized Stokes scattering matrix (Mishchenko et al. 2002a, 2006b) given by

$$\tilde{\mathbf{F}}(\Theta) = \frac{4\pi\langle \mathbf{F}(\Theta)\rangle_\xi}{\langle C_{\mathrm{sca}}\rangle_\xi} = \begin{bmatrix} a_1(\Theta) & b_1(\Theta) & 0 & 0 \\ b_1(\Theta) & a_2(\Theta) & 0 & 0 \\ 0 & 0 & a_3(\Theta) & b_2(\Theta) \\ 0 & 0 & -b_2(\Theta) & a_4(\Theta) \end{bmatrix} \tag{1.90}$$

(cf. Eq. (1.84)). The block-diagonal structure of this matrix was confirmed by the numerically exact $T$-matrix results and is largely caused by averaging over the uniform orientation distribution of a multi-particle group coupled with sufficient randomness of particle positions throughout the scattering volume. Each scattering matrix element denoted in Eq. (1.90) by a zero has been found to be at least an order of magnitude smaller than the smallest non-zero element (in the absolute-value sense).

The upper left-hand panel of Plate 1.9 vividly demonstrates several fundamental consequences of increasing the number of particles in the scattering volume. First, the constructive interference of light singly scattered by the component particles in the exact forward direction causes a strong and narrow forward-scattering enhancement (Ivanov et al. 1970; Mishchenko et al. 2004b). This feature is further detailed in Plate 1.10a and explained in Plate 1.7d. It can be called forward-scattering localization of electromagnetic waves. Indeed, the exact forward-scattering direction is unique in that the phases of the wavelets singly forward-scattered by all the particles in the volume are exactly the same irrespective of the specific particle positions (see the left-hand panel of Plate 1.7d). In the absence of multiple scattering, the constructive interference of these wavelets would lead to an increase of the forward-scattering phase function $a_1(0°)$ by a factor of $N$. This increase does occur for $N = 2$ and 5 (Plate 1.10a), but then it slows down, and by the time $N$ reaches the value 160 the $a_1(0°)$ value saturates. This behavior can be interpreted in terms of a multiple-scattering effect whereby particle 3 (see the right-hand panel of Plate 1.7d) "shades" particle 2 by attenuating the incident field exciting particle 2. In fact, we will see in the following section that it is this multiple-scattering effect that leads to the exponential extinction law in the framework of the RT theory.

Second, the phase functions at scattering angles $\Theta > 170°$ start to develop a backscattering enhancement which becomes quite pronounced for $N \geq 160$ (see Plate 1.10b). This feature is a typical manifestation of WL of electromagnetic waves in the backscattering direction, otherwise known as the coherent backscattering (CB) effect. (The term "weak localization of electromagnetic waves" was introduced by solid-state physicists in order to draw an analogy with the effect of weak localization of electrons in dirty metals (Sheng 2006; Akkermans and Montambaux 2007)). The standard explanation of WL is illustrated in Plate 1.7e and is as follows. The conjugate wavelets scattered along the same string of $n$ particles but in opposite directions interfere in the far-field zone, the interference being constructive or destructive depending on the respective phase difference,

$$\Delta = k_1(\mathbf{r}_n - \mathbf{r}_1) \cdot (\hat{\mathbf{n}}^{\text{inc}} + \hat{\mathbf{n}}^{\text{sca}}). \tag{1.91}$$

If the observation direction $\hat{\mathbf{n}}^{\text{sca}}$ is far from the exact backscattering direction given by $-\hat{\mathbf{n}}^{\text{inc}}$ then the average effect of interference of the conjugate wavelets scattered along various strings of particles is zero, owing to randomness of particle positions. However, at exactly the backscattering direction, $\hat{\mathbf{n}}^{\text{sca}} = -\hat{\mathbf{n}}^{\text{inc}}$, the phase difference between the conjugate paths involving *any* string of particles is identically equal to zero, and the interference is always constructive and causes an intensity peak.

The third consequence of increasing $N$ is that the phase functions at scattering angles $30° \leq \Theta \leq 170°$ become progressively smooth and featureless, thereby causing the "diffuse" intensity background clearly identifiable in Plate 1.8f. The major contributor to the background intensity is another class of wavelet pairs as illustrated in Plate 1.7f. Now the wavelet scattered along a string of $n$ particles "interferes with itself". Since the corresponding phase difference is exactly equal to zero irrespective of particle positions, the self-interference is always constructive for any string of particles. Therefore, the contribution of this class of wavelet pairs survives the ensemble averaging for any $\hat{\mathbf{n}}^{\text{inc}}$ and $\hat{\mathbf{n}}^{\text{sca}}$. The smoothness of the background intensity can be interpreted as a typical result of increasing amount of multiple scattering whereby light undergoing many scattering events "forgets" the initial incidence direction $\hat{\mathbf{n}}^{\text{inc}}$ and is more likely to contribute equally to all "exit" directions $\hat{\mathbf{n}}^{\text{sca}}$.

As we have already mentioned, the degree of linear polarization of the scattered light for unpolarized incident light is given by the ratio $-b_1/a_1$. The bottom left-hand panel of Plate 1.9 shows that the most obvious effect of increasing $N$ is to smooth out the low-frequency oscillations in the polarization curve for the single wavelength-sized sphere and, on average, to make polarization more neutral. The standard multiple-scattering explanation of this trait is that the main contribution to the second Stokes parameter, $Q^{\text{sca}}$, comes from the first order of scattering, whereas light scattered many times (Plate 1.7f) becomes largely unpolarized (Hansen and Travis 1974; Mishchenko et al. 2006b).

The ratio $a_2/a_1$ is identically equal to unity for scattering by a single sphere (cf. Eqs. (1.73) and (1.90)). Therefore, the rapidly growing deviation of this ratio from 100% for $N \geq 5$, Plate 1.9, can again be interpreted qualitatively as a direct consequence of the strengthened depolarizing effect of multiple scattering. Similarly, $a_3(\Theta) \equiv a_4(\Theta)$ and $a_3(180°)/a_1(180°) = -1$ for single scattering by a spherically symmetric particle, but multiple scattering in particle groups with $N \geq 5$ causes an increasingly significant violation of these equalities.

If the incident light is polarized linearly in the *xz*-plane then $Q^{\text{inc}} = I^{\text{inc}}$ and $U^{\text{inc}} = V^{\text{inc}} = 0.$ The corresponding angular distributions of the co-polarized,

$$\frac{1}{2}(I^{\text{sca}} + Q^{\text{sca}}) \propto \frac{1}{2}[a_1(\Theta) + 2b_1(\Theta) + a_2(\Theta)], \tag{1.92}$$

and cross-polarized,

$$\frac{1}{2}(I^{\text{sca}} - Q^{\text{sca}}) \propto \frac{1}{2}[a_1(\Theta) - a_2(\Theta)], \tag{1.93}$$

scattered intensities are shown in Plate 1.11. Also depicted are the same-helicity,

$$\frac{1}{2}(I^{\text{sca}} + V^{\text{sca}}) \propto \frac{1}{2}[a_1(\Theta) + a_4(\Theta)], \tag{1.94}$$

and opposite-helicity,

$$\frac{1}{2}(I^{\text{sca}} - V^{\text{sca}}) \propto \frac{1}{2}[a_1(\Theta) - a_4(\Theta)], \tag{1.95}$$

scattered intensities for the case of incident light polarized circularly in the counter-clockwise direction when looking in the direction of propagation ($Q^{\text{inc}} = U^{\text{inc}} = 0$ and $V^{\text{inc}} = I^{\text{inc}}$). All of these quantities exhibit WL in the form of backscattering peaks growing in amplitude with *N*.

By far the most definitive demonstration of the onset of the CB effect is provided by the $(a_1 - a_2)/2$ and $(a_1 + a_4)/2$ curves in Plate 1.11. Indeed, the corresponding single-particle curves show no backscattering enhancement whatsoever, so the backscattering peaks that develop with increasing *N* (and thus with increasing amount of multiple scattering) can be attributed unequivocally to WL. Plate 1.11 also depicts the angular profiles of the so-called linear, $\mu_{\text{L}}$, and circular, $\mu_{\text{C}}$, polarization ratios defined as the ratio of the cross-polarized to co-polarized scattered intensities and the ratio of the same-helicity to the opposite-helicity scattered intensities:

$$\mu_{\text{L}} = \frac{I^{\text{sca}} - Q^{\text{sca}}}{I^{\text{sca}} + Q^{\text{sca}}} = \frac{a_1(\Theta) - a_2(\Theta)}{a_1(\Theta) + 2b_1(\Theta) + a_2(\Theta)}, \tag{1.96}$$

$$\mu_{\rm C} = \frac{I^{\rm sca} + V^{\rm sca}}{I^{\rm sca} - V^{\rm sca}} = \frac{a_1(\Theta) + a_4(\Theta)}{a_1(\Theta) - a_4(\Theta)}. \tag{1.97}$$

These quantities are used widely in radar and lidar remote sensing (Ulaby and Elachi 1990; Ostro 1993; Stephens 1994) because they vanish at the exact backscattering direction if multiple scattering is insignificant and the scattering particles are spherically symmetric. (The reader may recall that in the framework of the far-field approximation, $\mu_{\rm L}$ was denoted as $\delta_{\rm L}$ and called the linear depolarization ratio, while $\mu_{\rm C}$ was denoted as $\delta_{\rm C}$ and called the circular depolarization ratio in Section 1.15. We will keep this terminological distinction since, in general, the far-field approximation becomes inapplicable in the case of the RT and WL theories.) Our results demonstrate convincingly that multiple scattering causes an increasingly significant deviation of $\mu_{\rm L}(180°)$ and $\mu_{\rm C}(180°)$ from zero, while WL causes pronounced backscattering peaks in the $\mu_{\rm L}$ and $\mu_{\rm C}$ angular profiles.

The interference mechanism implies that the angular widths of the forward-scattering and CB peaks must be proportional to $1/k_1R$. To verify this, we have performed computations assuming that the number of $k_1r = 4$ particles is fixed at $N = 8$ while the size parameter of the volume is varied from $k_1R = 12$ to 72 in steps of 6. The arrangement of the eight particles inside the $k_1R = 12$ volume is random but such that each particle is in contact with at least one other particle. The other ten particulate volumes with $k_1R = 18, 24, \ldots, 72$ are obtained by uniformly scaling all particle coordinates of the $k_1R = 12$ volume while keeping the particle size fixed. This procedure is illustrated in Plate 1.7g showing the original $k_1R = 12$ volume and the derivative $k_1R = 24$ volume. The corresponding *T*-matrix results are depicted in Plates 1.10c,d and demonstrate indeed that the widths of both peaks decrease with increasing inter-particle separation, thus corroborating their interference nature. The nearly constant amplitude of the forward-scattering peak and the rapidly decreasing amplitude of the backscattering peak testify again that these features are caused by single and multiple scattering, respectively. Indeed, the single-scattering term does not and the multiple-scattering terms do contain $1/(\text{inter-particle distance})$ factors in the far-field order-of-scattering expansion (1.41).

Although the results shown in Plates 1.9–1.11 are based on averaging over orientations of only one random *N*-particle configuration, they can be expected to be statistically representative of all possible realizations of the *N*-particle group, at least for large *N* (Mishchenko et al. 2007d). This is well demonstrated by Plates 1.10e and 1.10f computed for two different realizations of a 240-particle group occupying a $k_1R = 40$ spherical volume.

Thus, the three fundamental classes of wavelet pairs illustrated in Plates 1.7d–1.7f are the main contributors to the scattering patterns shown in Plates 1.9–1.11. In the following sections we will see how and to what extent they are incorporated in the theories of RT and WL.

## 1.18. Radiative transfer theory

The early history of the phenomenological theory of RT describing electromagnetic energy transport in macroscopic media composed of sparsely and randomly distributed, elastically scattering particles is described by Иванов (1991). He traces the origin of the simplest form of the RT equation (RTE)—no account of polarization, idealized isotropically scattering particles—to papers by Lommel (1887) and Chwolson (1889). Unfortunately, these early publications have been hardly noticed, and the first introduction of the RTE has traditionally been attributed to the paper by Schuster (1905).

Gans (1924) was the first to account for the polarization nature of light in the context of the phenomenological RT theory. However, he analyzed only the special case of a plane-parallel Rayleigh-scattering medium illuminated by perpendicularly incident light and considered only the first two components of the Stokes column vector. The case of arbitrary illumination and arbitrary polarization was first studied by Chandrasekhar (1950), but his analysis was again limited to Rayleigh-scattering particulate media. Finally, Розенберг (1955) introduced the most general form of the RTE, the so-called vector RTE (VRTE), which fully accounts for the polarization nature of light and is applicable to scattering media composed of arbitrarily shaped and arbitrarily oriented particles.

Since its inception, the RT theory has had a remarkable history of practical applications in numerous areas of atmospheric radiation (Соболев 1972; Hansen and Travis 1974; van de Hust 1980; Lenoble 1985, 1993; Goody and Yung 1989; Liou 1992, 2002; Yanovitskij 1997; Kokhanovsky 2003; Marshak and Davis 2005; Bohren and Clothiaux 2006; Zdunkovski et al. 2007), remote sensing (Ishimaru 1978; Ulaby and Elachi 1990; Stephens 1994), oceanography (Mobley 1994; Thomas and Stamnes 1999), image transfer (Zege et al. 1991), astrophysics (Hansen and Hovenier 1974; Dolginov et al. 1995), biomedicine (Khlebtsov et al. 2002; Tuchin et al. 2006; Tuchin 2007), and engineering (Viskanta and Mengüç 1987; Siegel and Howell 2002; Modest 2003). At the same time, it has also had a long history of confusing and even misleading accounts of its fundamental principles. Indeed, the palette of phenomenological derivations of the RTE encountered in various monographs, textbooks, and reviews is quite rich, which by itself is a sign of a serious problem. On one hand, most of the derivations are rather short and either present the RTE as a trivial consequence of energy conservation or expect the reader to accept the RTE as a fundamental experimental law implicitly supplementing other basic physical principles such as the laws of classical and quantum electrodynamics. On the other hand, there are derivations which rise to the level of a philosophical essay in which the RTE emerges as an allegedly logical outcome of a multi-page discourse almost devoid of formulas but full of ill-defined "collective effects", "elementary volume elements", and "incoherent light rays". Some of the derivations even invoke the concept of photons as "localized particles of light", "discrete blobs of energy without phases", or "corpuscles that are moving according to the laws of classical mechanics". As such, they imply that the notorious wave–particle duality of light somehow manifests itself in the scattering process that is fully controlled by the macroscopic Maxwell equations.

No matter how realistic the various phenomenological accounts of RT may look at first sight (Preisendorfer 1965; Mobley 1994), they inevitably fall apart upon scrutiny of their physical foundation (Apresyan and Kravtsov 1996). It is, therefore, not surprising that quite recently, Mandel and Wolf (1995) stated that "in spite of the extensive use of the theory of radiative energy transfer, no satisfactory derivation of its basic equation from electromagnetic theory has been obtained up to now". Furthermore, the phenomenological accounts completely overlook the fundamental link between RT and WL. Most importantly, they conceal the irrefutable fact that as long as scattering occurs without frequency redistribution and the particles are macroscopic and can be characterized by a refractive index, the RTE describes multiple scattering of classical electromagnetic waves and, as such, must be derived directly from the macroscopic Maxwell equations via a series of well defined and reproducible analytical steps (Боровой 1966; Barabanenkov 1975; Tsang et al. 1985).

This ambiguous situation has finally changed, and a complete derivation of the general VRTE directly from the macroscopic Maxwell equations for the case of elastically scattering discrete random media has been published (Mishchenko 2002, 2003; Mishchenko et al. 2006b). This derivation can be used to clarify the role and physical meaning of the various quantities entering the VRTE, establish a direct link between the theories of RT and WL, cross-examine the terminologies used in the traditional phenomenological and the new microphysical approaches, and identify and correct certain misconceptions of the phenomenological approach. These are the four principal goals of the following sections. Since the emphasis is on the fundamentals of the multiple scattering, RT, and WL theories, specific analytical and numerical techniques for solving the VRTE (such as the adding, discrete ordinate, and Monte Carlo methods) are not discussed in detail.

The preceding sections make us well prepared to proceed with the outline of the RT theory by considering the near-field scattering of a plane electromagnetic wave by a large group of $N$ particles randomly distributed throughout a large 3D volume $V$ (Fig. 1.4). What follows is a brief sketch of the theory detailed in Chapter 8 of Mishchenko et al. (2006b).

In accordance with the above discussion, the derivation of the VRTE involves several fundamental premises and approximations. The first one is to assume that each particle is located in the far-field zones of all the other particles and that the observation point is also located in the far-field zones of all the particles forming the scattering medium. As we have seen in Section 1.9, this assumption leads to a dramatic simplification of the FLEs wherein the latter are converted from a system of volume integral equations into a system of linear algebraic equations. However, it also limits the applicability of the final result by requiring that the particles in the scattering medium are not closely spaced, a condition that is nonetheless met in many natural circumstances.

The order-of-scattering form of the far-field FLEs (1.41) allows one to represent the total electric field at a point in space as a sum of contributions arising from light-scattering paths going through all possible particle sequences. The second major assumption in the derivation of the VRTE, called the Twersky approximation (Twersky 1964; Mishchenko et al. 2006b), is that all paths going through a particle more than once can be neglected. It is straightforward to demonstrate that doing this is justified provided that the number of particles in the scattering volume, $N$, is very large. Thus, instead of the diagrammatic equation depicted in Fig. 1.7 we will work with a simplified version depicted in Fig. 1.9.

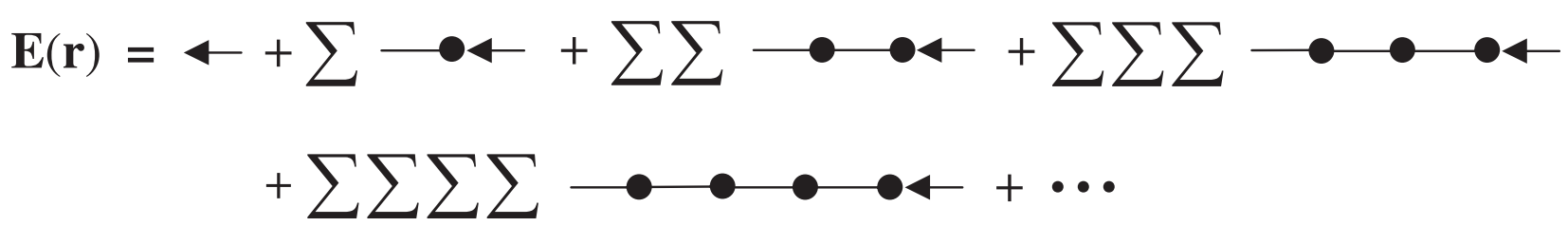

**Fig. 1.9.** The Twersky approximation.

The third major assumption is that of full ergodicity, which allows one to replace averaging over time by averaging over particle positions and states according to the discussion in Section 1.10.

The fourth major assumption is that

- the position and state of each particle are statistically independent of each other and of those of all the other particles; and
- the spatial distribution of the particles throughout the medium is random and statistically uniform.

As one might expect, this assumption leads to a major simplification of all analytical derivations. The practical meaning of ergodicity and uniformity will be discussed at the end of this section.

The next major step is the characterization of the multiply scattered radiation by the coherency dyadic

$$\overleftrightarrow{C}(\mathbf{r}) = \langle \mathbf{E}(\mathbf{r},t) \otimes \mathbf{E}^{*}(\mathbf{r},t)\rangle_{t} \approx \langle \mathbf{E}(\mathbf{r}) \otimes \mathbf{E}^{*}(\mathbf{r})\rangle_{\mathbf{R},\xi}, \qquad (1.98)$$

where the subscripts $\mathbf{R}$ and $\xi$ denote averaging over all particle coordinates and states, respectively. The coherency dyadic is appropriately defined as a nonvanishing quantity (Section 1.5). Because of the averaging over particle coordinates, $\overleftrightarrow{C}(\mathbf{r})$ is a continuous function of the position vector. Furthermore, as we will see later, the coherency dyadic allows the definition of derivative quantities which are observable directly.

The Twersky expansion of the coherency dyadic is depicted diagrammatically in Fig. 1.10. To classify the different terms entering the expanded expression inside the angular brackets on the right-hand side of this equation, we will use the notation illustrated in Fig. 1.11a. In this particular case, the upper and lower scattering paths go through different particles. However, the two paths can involve one or more common particles, as shown in panels (c) and (d) by using the dashed connectors. Furthermore, if the number of common particles is two or more, they can enter the upper and lower paths in the same order, as in panel (c), or in the reverse order, as in panel (d). Panel (e) shows a mixed diagram in which two common particles appear in the same order and two other common particles appear in the reverse order. The contribution of this diagram to the coherency dyadic is simply

$$[\vec{\vec{B}}_{rij} \cdot \vec{\vec{B}}_{ijk} \cdot \vec{\vec{B}}_{jkl} \cdot \vec{\vec{B}}_{kl0} \cdot \mathbf{E}_l^{\mathrm{inc}}] \otimes [\vec{\vec{B}}_{rik} \cdot \vec{\vec{B}}_{ikj} \cdot \vec{\vec{B}}_{kjl} \cdot \vec{\vec{B}}_{jl0} \cdot \mathbf{E}_l^{\mathrm{inc}}]^*. \tag{1.99}$$

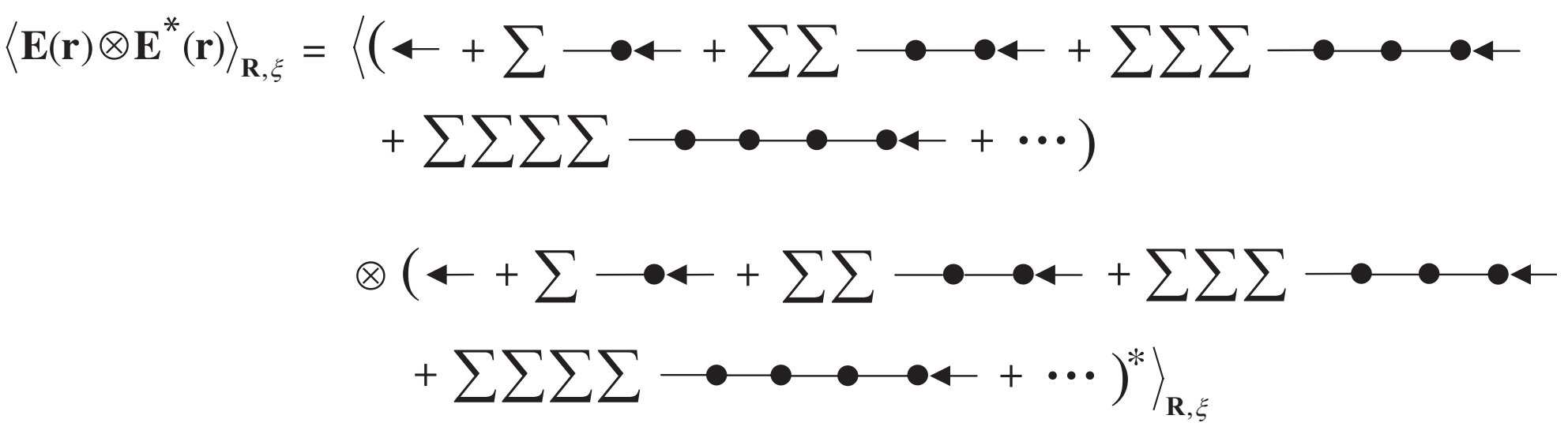

**Fig. 1.10.** The Twersky expansion of the coherency dyadic.

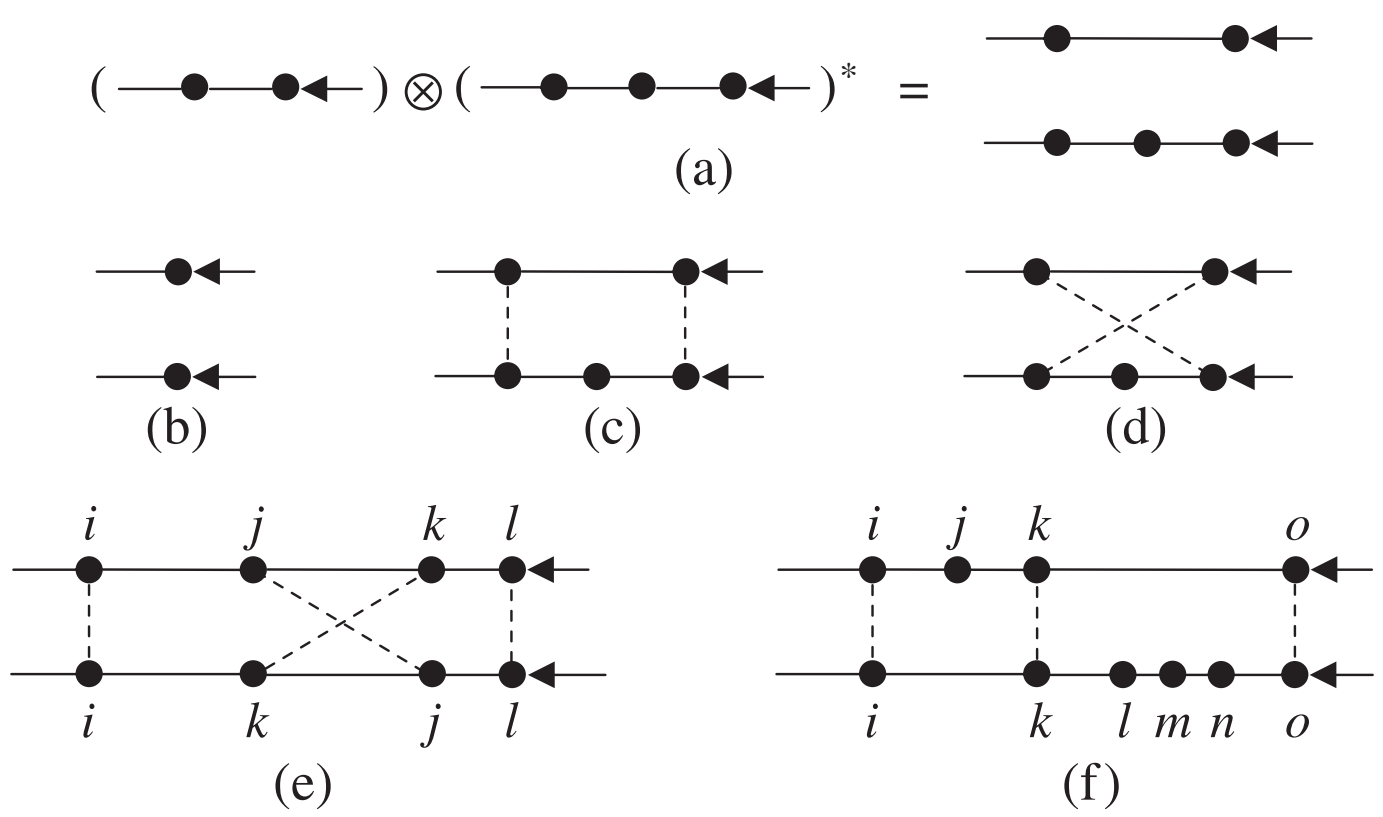

**Fig. 1.11.** Classification of terms in the Twersky expansion of the coherency dyadic.

By the nature of the Twersky approximation, neither the upper path nor the lower path can go through a particle more than once. Therefore, no particle can be the origin of more than one connector.

The next major assumption in the derivation of the VRTE is that all diagrams with crossing connectors can be neglected. The rationale for making this assumption can be illustrated by considering the contribution of the term depicted in Fig. 1.11e to the coherency dyadic. Indeed, by substituting Eqs. (1.43)–(1.46) in Eq. (1.99), we see that the resulting expression includes a rapidly oscillating exponential factor $\exp[\mathrm{i}k_1(R_{ij} + R_{kl} - R_{ik} - R_{jl})]$. This factor causes the contribution of this term to vanish upon averaging over the positions of particles $j$ and $k$ within the volume $V$ provided that all linear dimensions of the volume are much greater than the wavelength of the incident light. However, there is a class of diagrams with crossing connectors which can give a non-vanishing contribution to the coherency dyadic. This class will be discussed in the following section.

Let us now consider the contribution of the diagrams with no crossing connectors like the one shown in Fig. 1.11f. The presence of unconnected particle $j$ in the upper scattering path causes an exponential factor $\exp[\mathrm{i}k_1(R_{ij} + R_{jk})]$, which oscillates rapidly everywhere in $V$ except along the straight line connecting the origins of particles $i$ and $k$, where this factor is constant. The stationary-phase evaluation of the integral describing the average over all positions of particle $j$ yields a very important result: the only effect particle $j$ has in the context of this specific diagram is to attenuate the field generated by particle $k$ and exciting particle $i$ and to potentially cause dichroism. Similarly, particles *l*, *m*, and *n* have any effect only when they all are positioned along the straight line connecting the origins of particles *k* and *o*, and this effect is again to cause attenuation and, possibly, dichroism.

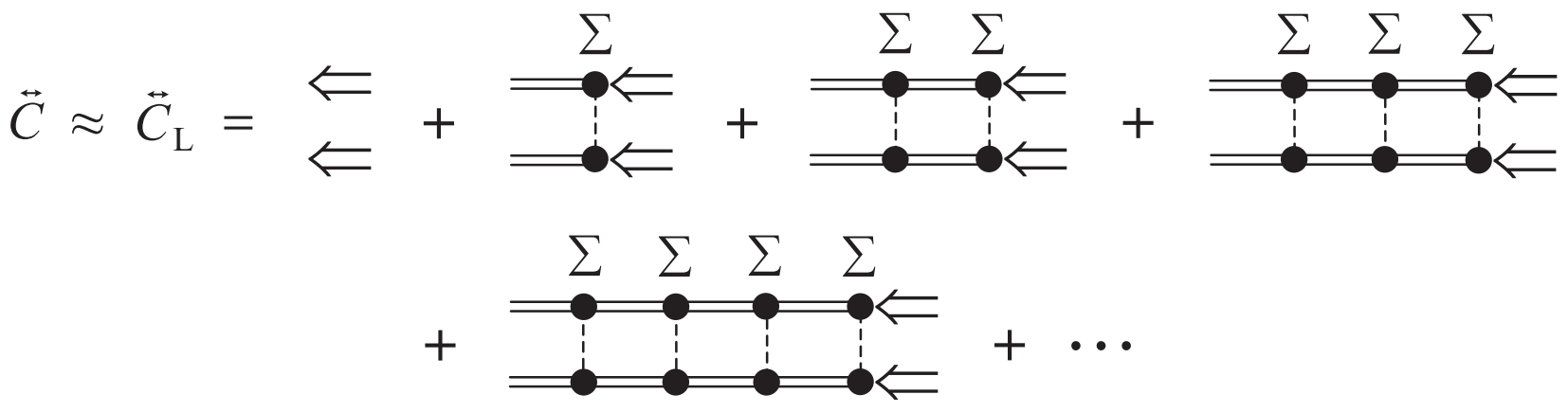

**Fig. 1.12.** Ladder approximation for the coherency dyadic.

Careful analytical evaluation of the cumulative position- and state-averaged contribution of all diagrams with vertical connectors coupled with the assumption that *N* is very large leads to the equation depicted diagrammatically in Fig. 1.12 (Mishchenko et al. 2006b). The symbol $\Leftarrow$ denotes the incident field attenuated by the unconnected particles on its way to the observation point or to the right-most connected particle, the double lines denote similar attenuation by unconnected particles of a wave propagating from one connected particle to another, and the symbols Σ denote both the summation over all appropriate particles and the averaging over the particle positions and states. Owing to their appearance, the diagrams on the right-hand side of this equation are called ladder diagrams. Therefore, this diagrammatic formula can be called the ladder approximation for the coherency dyadic. Obviously, the diagram visualized in Plate 1.7f and discussed in the preceding section belongs to the class of ladder diagrams.

The expanded expression for the ladder coherency dyadic has the form of an angular decomposition in terms of the so-called ladder specific coherency dyadic $\overleftrightarrow{\Sigma}_{\mathrm{L}}(\mathbf{r}, \hat{\mathbf{q}})$:

$$\overleftrightarrow{C}_{\mathrm{L}}(\mathbf{r}) = \int_{4\pi} \mathrm{d}\hat{\mathbf{q}}\, \overleftrightarrow{\Sigma}_{\mathrm{L}}(\mathbf{r}, \hat{\mathbf{q}}), \tag{1.100}$$

where the integration is performed over all propagation directions as specified by the unit vector $\hat{\mathbf{q}}$. Furthermore, it is straightforward to show that the specific coherency dyadic satisfies an integral RTE (Mishchenko et al. 2006b). The ladder specific coherency dyadic can, in turn, be used to define the so-called specific intensity column vector,

$$\tilde{\mathbf{I}}(\mathbf{r}, \hat{\mathbf{q}}) = \begin{bmatrix} \tilde{I}(\mathbf{r}, \hat{\mathbf{q}} \\ \tilde{Q}(\mathbf{r}, \hat{\mathbf{q}}) \\ \tilde{U}(\mathbf{r}, \hat{\mathbf{q}}) \\ \tilde{V}(\mathbf{r}, \hat{\mathbf{q}}) \end{bmatrix} = \frac{1}{2}\sqrt{\frac{\varepsilon_1}{\mu_0}} \begin{bmatrix} \hat{\boldsymbol{\theta}}(\hat{\mathbf{q}}) \cdot \overleftrightarrow{\Sigma}_{\mathrm{L}}(\mathbf{r}, \hat{\mathbf{q}}) \cdot \hat{\boldsymbol{\theta}}(\hat{\mathbf{q}}) + \hat{\boldsymbol{\varphi}}(\hat{\mathbf{q}}) \cdot \overleftrightarrow{\Sigma}_{\mathrm{L}}(\mathbf{r}, \hat{\mathbf{q}}) \cdot \hat{\boldsymbol{\varphi}}(\hat{\mathbf{q}}) \\ \hat{\boldsymbol{\theta}}(\hat{\mathbf{q}}) \cdot \overleftrightarrow{\Sigma}_{\mathrm{L}}(\mathbf{r}, \hat{\mathbf{q}}) \cdot \hat{\boldsymbol{\theta}}(\hat{\mathbf{q}}) - \hat{\boldsymbol{\varphi}}(\hat{\mathbf{q}}) \cdot \overleftrightarrow{\Sigma}_{\mathrm{L}}(\mathbf{r}, \hat{\mathbf{q}}) \cdot \hat{\boldsymbol{\varphi}}(\hat{\mathbf{q}}) \\ -\hat{\boldsymbol{\theta}}(\hat{\mathbf{q}}) \cdot \overleftrightarrow{\Sigma}_{\mathrm{L}}(\mathbf{r}, \hat{\mathbf{q}}) \cdot \hat{\boldsymbol{\varphi}}(\hat{\mathbf{q}}) - \hat{\boldsymbol{\varphi}}(\hat{\mathbf{q}}) \cdot \overleftrightarrow{\Sigma}_{\mathrm{L}}(\mathbf{r}, \hat{\mathbf{q}}) \cdot \hat{\boldsymbol{\theta}}(\hat{\mathbf{q}}) \\ \mathrm{i}[\hat{\boldsymbol{\varphi}}(\hat{\mathbf{q}}) \cdot \overleftrightarrow{\Sigma}_{\mathrm{L}}(\mathbf{r}, \hat{\mathbf{q}}) \cdot \hat{\boldsymbol{\theta}}(\hat{\mathbf{q}}) - \hat{\boldsymbol{\theta}}(\hat{\mathbf{q}}) \cdot \overleftrightarrow{\Sigma}_{\mathrm{L}}(\mathbf{r}, \hat{\mathbf{q}}) \cdot \hat{\boldsymbol{\varphi}}(\hat{\mathbf{q}})] \end{bmatrix}, \tag{1.101}$$

which also satisfies an integral RTE. Finally, the latter can be converted into the following classical integro-differential form:

$$\hat{\mathbf{q}}\cdot\nabla\tilde{\mathbf{I}}(\mathbf{r},\hat{\mathbf{q}}) = -n_0\langle\mathbf{K}_1(\hat{\mathbf{q}})\rangle_\xi\tilde{\mathbf{I}}(\mathbf{r},\hat{\mathbf{q}}) + n_0\int_{4\pi}\mathrm{d}\hat{\mathbf{q}}'\langle\mathbf{Z}_1(\hat{\mathbf{q}},\hat{\mathbf{q}}')\rangle_\xi\tilde{\mathbf{I}}(\mathbf{r},\hat{\mathbf{q}}'). \quad (1.102)$$

As before (Section 1.11), $\langle\mathbf{K}_1(\hat{\mathbf{q}})\rangle_\xi$ and $\langle\mathbf{Z}_1(\hat{\mathbf{q}},\hat{\mathbf{q}}')\rangle_\xi$ are the single-particle extinction and phase matrix, respectively, averaged over all particle states, while $n_0 = N/V$ is the particle number density. The specific intensity column vector can be decomposed into the coherent and diffuse parts,

$$\tilde{\mathbf{I}}(\mathbf{r},\hat{\mathbf{q}}) = \delta(\hat{\mathbf{q}}-\hat{\mathbf{n}}^{\mathrm{inc}})\mathbf{I}_{\mathrm{c}}(\mathbf{r}) + \tilde{\mathbf{I}}_{\mathrm{d}}(\mathbf{r},\hat{\mathbf{q}}), \quad (1.103)$$

each satisfying its own VRTE:

$$\mathbf{n}^{\mathrm{inc}}\cdot\nabla\mathbf{I}_{\mathrm{c}}(\mathbf{r}) = -n_0\langle\mathbf{K}_1(\hat{\mathbf{n}}^{\mathrm{inc}})\rangle_\xi\mathbf{I}_{\mathrm{c}}(\mathbf{r}), \quad (1.104)$$

$$\hat{\mathbf{q}}\cdot\nabla\tilde{\mathbf{I}}_{\mathrm{d}}(\mathbf{r},\hat{\mathbf{q}}) = -n_0\langle\mathbf{K}_1(\hat{\mathbf{q}})\rangle_\xi\tilde{\mathbf{I}}_{\mathrm{d}}(\mathbf{r},\hat{\mathbf{q}}) + n_0\int_{4\pi}\mathrm{d}\hat{\mathbf{q}}'\langle\mathbf{Z}_1(\hat{\mathbf{q}},\hat{\mathbf{q}}')\rangle_\xi\tilde{\mathbf{I}}_{\mathrm{d}}(\mathbf{r},\hat{\mathbf{q}}')$$

$$+\, n_0\langle\mathbf{Z}_1(\hat{\mathbf{q}},\mathbf{n}^{\mathrm{inc}})\rangle_\xi\mathbf{I}_{\mathrm{c}}(\mathbf{r}). \quad (1.105)$$

$\mathbf{I}_{\mathrm{c}}$ reduces to the Stokes column vector of the incident wave at the illuminated boundary of the medium, but is subject to exponential attenuation and, possibly, the effect of dichroism inside the medium.

The VRTE (1.102) becomes considerably simpler in the case of a plane-parallel, macroscopically isotropic and mirror-symmetric scattering medium (Hovenier et al. 2004; Mishchenko et al. 2006b):

$$u\frac{\mathrm{d}\tilde{\mathbf{I}}(\tau,\hat{\mathbf{q}})}{\mathrm{d}\tau} = -\tilde{\mathbf{I}}(\tau,\hat{\mathbf{q}}) + \frac{1}{\langle C_{\mathrm{ext},1}\rangle_\xi}\int_{4\pi}\mathrm{d}\hat{\mathbf{n}}'\langle\mathbf{Z}_1(\hat{\mathbf{q}},\hat{\mathbf{q}}')\rangle_\xi\tilde{\mathbf{I}}(\tau,\hat{\mathbf{q}}'), \quad (1.106)$$

where $\mathrm{d}\tau = n_0\langle C_{\mathrm{ext},1}\rangle_\xi\mathrm{d}z$ is the differential element of the optical depth, $\langle C_{\mathrm{ext},1}\rangle_\xi$ is the average extinction cross section per particle, and $u = -\cos\theta$ is the direction cosine. The $z$-axis of the laboratory right-handed coordinate system is assumed to be perpendicular to the plane boundaries of the medium and directed outwards.

The most important corollaries of the microphysical derivation of the VRTE are the following (Mishchenko et al. 2006b).

1. The derivation of the VRTE does not need fundamental physical laws other than those already contained in the classical frequency-domain macroscopic electromagnetics. In particular, the ill-defined concepts of collective effects, elementary volume elements, incoherent light rays, and photons as localized particles of light have no relevance whatsoever to the transfer of electromagnetic radiation in elastically scattering discrete random media. It is, in fact, remarkable that although the VRTE (1.102) has the formal mathematical structure of a kinetic equation describing particle transport, it follows directly from the electromagnetic *wave* theory.

2. The VRTE is derived by keeping only one class of wavelet pairs illustrated by Plate 1.7f and Fig. 1.11f. The effect of unconnected particles is reduced to exponential attenuation and dichroism.

3. In the context of the RT theory, the scattering properties of particles are specified in terms of the extinction and phase matrices rather than in terms of the scattering dyadic or the scattering amplitude matrix. Each particle with its individual extinction and phase matrices is effectively replaced with an average particle having the extinction and phase matrices obtained by averaging over the states of all the particles.

4. In the framework of the exact FLEs, the source of multiple scattering is the constant-amplitude incident field in Eq. (1.38). In the framework of the approximate RT theory, this role is effectively (or, better to say, "mathematically") assumed by the exponentially attenuated coherent (or "unscattered") part of the specific intensity column vector described by Eq. (1.104).

5. Averaging over all particle positions makes $\mathbf{I}_c$ and $\tilde{\mathbf{I}}_d$ continuous functions of the position vector of the observation point $\mathbf{r}$ and also makes $\tilde{\mathbf{I}}_d$ a continuous function of the propagation direction $\hat{\mathbf{q}}$.

6. For the same reason, $\tilde{\mathbf{I}}$ differs from the Stokes column vector of a transverse electromagnetic wave, $\mathbf{I}$, in that it has the dimension of monochromatic radiance, $\mathrm{Wm^{-2}sr^{-1}}$, rather than the dimension of monochromatic energy flux, $\mathrm{Wm^{-2}}$. The reader can see readily that this particular dimension of $\tilde{\mathbf{I}}$ is a direct consequence of the definition (1.98), the angular decomposition of $\vec{C}_\mathrm{L}(\mathbf{r})$ according to Eq. (1.100), and the definition (1.101).

7. The VRTE is an inherently vector equation. The frequently used scalar version of the VRTE is obtained by artificially replacing the specific intensity vector by its first element (i.e., the specific intensity) and the extinction and phase matrices by their respective $(1,1)$ elements. As such, the scalar approximation has no compelling physical justification besides being easier to solve and providing acceptable accuracy in some (but not all!) cases (Chandrasekhar 1950; Kattawar and Adams 1990; Mishchenko et al. 1994; Lacis et al. 1998; Mishchenko et al. 2006b).

8. The VRTE remains valid if the incident light is a parallel quasi-monochromatic beam.

The integral form of the VRTE can be used to clarify the physical meaning of the coherent Stokes column vector $\mathbf{I}_c$ and the diffuse specific intensity column vector $\tilde{\mathbf{I}}_d$. The fundamental difference between these two quantities is that the former describes a monodirectional whereas the latter describes an uncollimated flow of electromagnetic energy according to Eq. (1.103). In particular, the first element of the coherent Stokes column vector, i.e., the coherent intensity $I_c(\mathbf{r})$, is the electromagnetic power per unit area of a small surface element $\Delta S$ perpendicular to the incidence direction $\hat{\mathbf{n}}^{\mathrm{inc}}$, whereas the first element of the diffuse specific intensity column vector (i.e., the diffuse specific intensity $\tilde{I}_d(\mathbf{r}, \hat{\mathbf{q}})$) is the electromagnetic power per unit area of a small surface element $\Delta S$ perpendicular to $\hat{\mathbf{q}}$ per one steradian of a small solid angle $\Delta\Omega$ centered around $\hat{\mathbf{q}}$ (see Fig. 1.13).

This interpretation of $\mathbf{I}_c(\mathbf{r})$ and $\tilde{\mathbf{I}}_d(\mathbf{r}, \hat{\mathbf{q}})$ implies that both quantities can be measured by appropriately placed and oriented detectors of electromagnetic energy flux. Indeed, the instantaneous direction of the electromagnetic energy flow is given by the Poynting vector and changes rapidly inside a discrete random medium owing to changing particle positions. Therefore, at any given moment in time, a well-collimated detector of electromagnetic energy flux placed inside the medium may or may not register any signal depending on the specific instantaneous orientation of the Poynting vector. Averaging over time (or, equivalently, over particle positions) ensures that the reading of the detector is always non-zero and is a continuous function of its orientation. The fact that the specific intensity column vector can be both computed theoretically by solving the VRTE and measured with a suitable optical device explains the practical usefulness of the RT theory in countless applications in various branches of science and engineering.

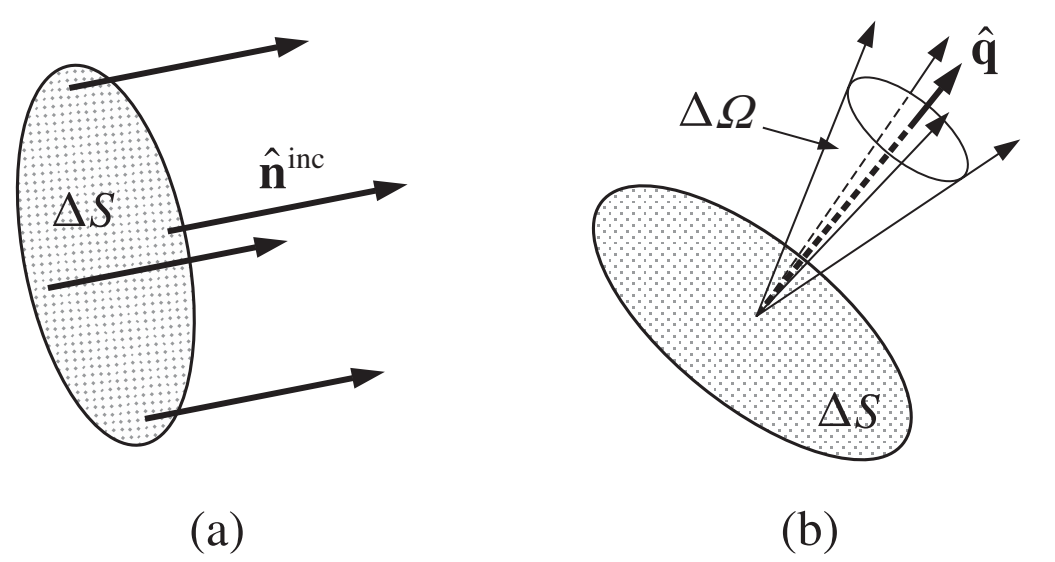

**Fig. 1.13.** Physical meaning of (a) coherent intensity and (b) specific intensity.

Since the microphysical derivation of the RTE involves statistical averaging over particle states and positions, neither the coherent Stokes column vector nor the diffuse specific intensity column vector characterize the instantaneous distribution of the radiation field inside the scattering medium. Instead, they characterize the directional flow of electromagnetic radiation averaged over a sufficiently long period of time. This conclusion is consistent with the discussion in the preceding section. The minimal averaging time necessary to ensure ergodicity may be different for different scattering systems, but the following is always true: the longer the averaging time the more accurate the theoretical prediction based on the RTE. Accumulating a signal over an extended period of time is often used to improve the accuracy of a measurement by reducing the effect of random noise. However, the situation with the RT theory is fundamentally different since averaging the signal over an extended period of time is necessary to ensure the very applicability of the RTE.

Although the microphysical derivation of the RTE rests on several fundamental premises discussed above, most of them appear to be quite realistic in a great variety of applications. However, the assumptions of ergodicity and spatial uniformity deserve further analysis since they may appear to be too restrictive for the RTE to be useful.

The meaning of the assumptions of ergodicity and uniformity is illustrated in Fig. 1.14. The detector of electromagnetic energy has a finite acceptance area $\Delta S$ and an angular aperture small enough to resolve the angular variability of the diffuse radiation field (e.g., ~1°). Both define the part of the scattering volume $V$ bounded schematically by the dotted lines in Fig. 1.14; this part will be called the acceptance volume. According to the integral form of the RTE, all energy recorded by the detector comes directly from the particles contained in the acceptance volume. The energy exciting each particle can be either the (attenuated) incident light or the light scattered by the other particles. The light scattered by a particle from the acceptance volume towards the detector can be attenuated by other particles located closer to the detector.

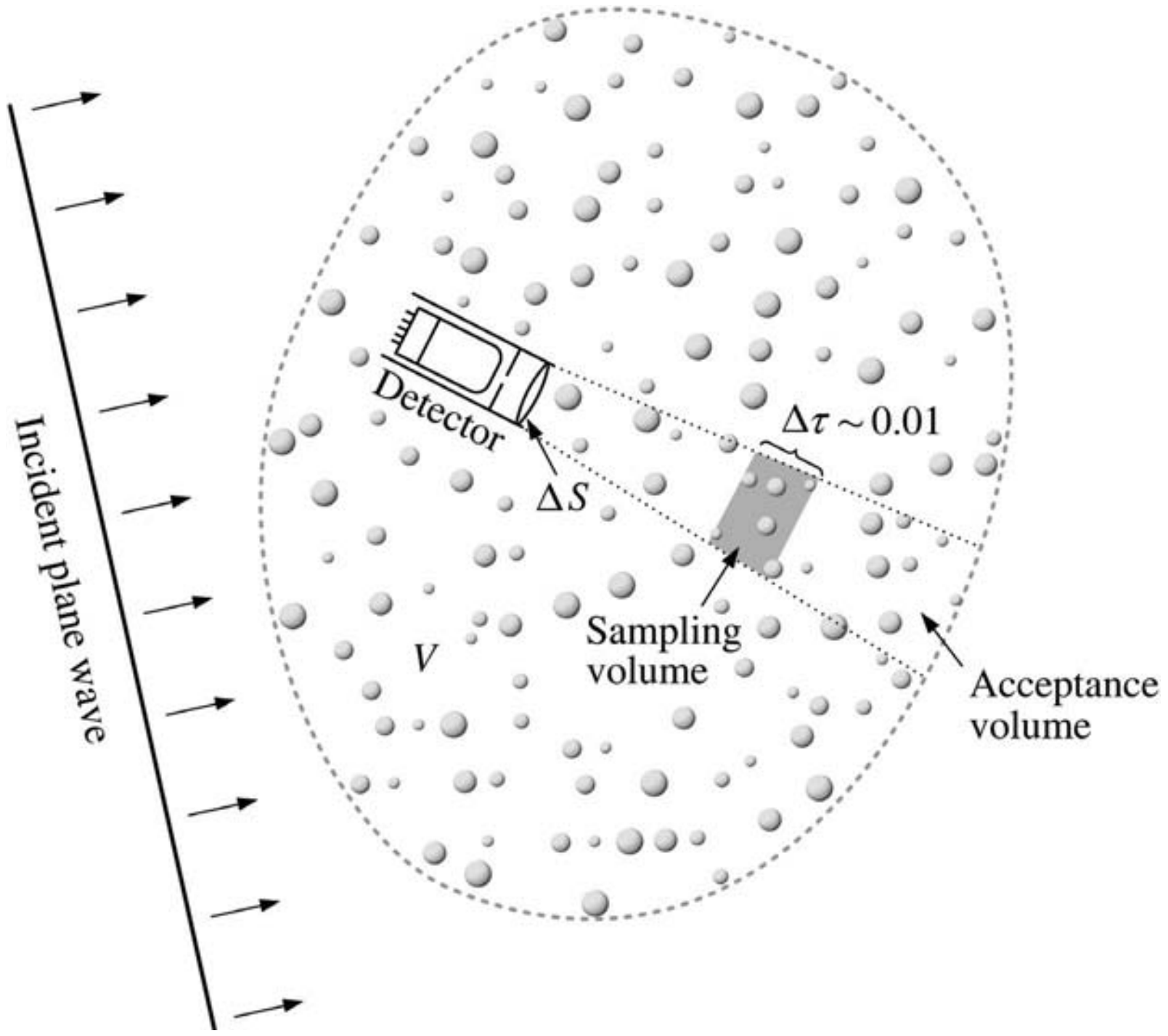

**Fig. 1.14.** Practical meaning of the assumptions of ergodicity and uniformity.

Let us assume that the detector accumulates the signal over a time interval $\Delta t$ and subdivide the acceptance volume into a number of sampling volumes such that their optical thickness $\Delta\tau$ along the line of sight of the detector is very small (~0.01). One of these sampling volumes is shown schematically in Fig. 1.14. Obviously, the contribution of a particle to the detector signal is essentially independent of the specific particle position in the sampling volume. Therefore, the *strict* meaning of the assumptions of ergodicity and statistical uniformity of particle positions within the scattering volume $V$ is that each particle visits each sampling volume during the measurement interval $\Delta t$.

In reality, however, the scattering volume $V$ contains many particles of the same type. Therefore, the *practical* meaning of ergodicity and uniformity is that particles of each type visit each sampling volume during the measurement interval $\Delta t$ a number of times statistically representative of the total number of such particles in the entire scattering volume. Obviously, this requirement is significantly softer and can be expected to be met in many actual circumstances.

There are several efficient techniques for the practical numerical solution of the VRTE and its approximate (and physically unjustified) scalar version such as the adding–doubling method (Hansen and Travis 1974; Hovenier et al. 2004), the invariant imbedding method (Mishchenko 1990a), the successive orders of scattering technique (van de Hulst 1980; Zhai et al. 2009), the discrete ordinates technique (Rozanov and Kokhanovsky 2006), the Monte Carlo method (Марчук и др. 1976), etc. Many of them rely on the assumption of a plane-parallel scattering medium and are summarized in Lenoble (1985).

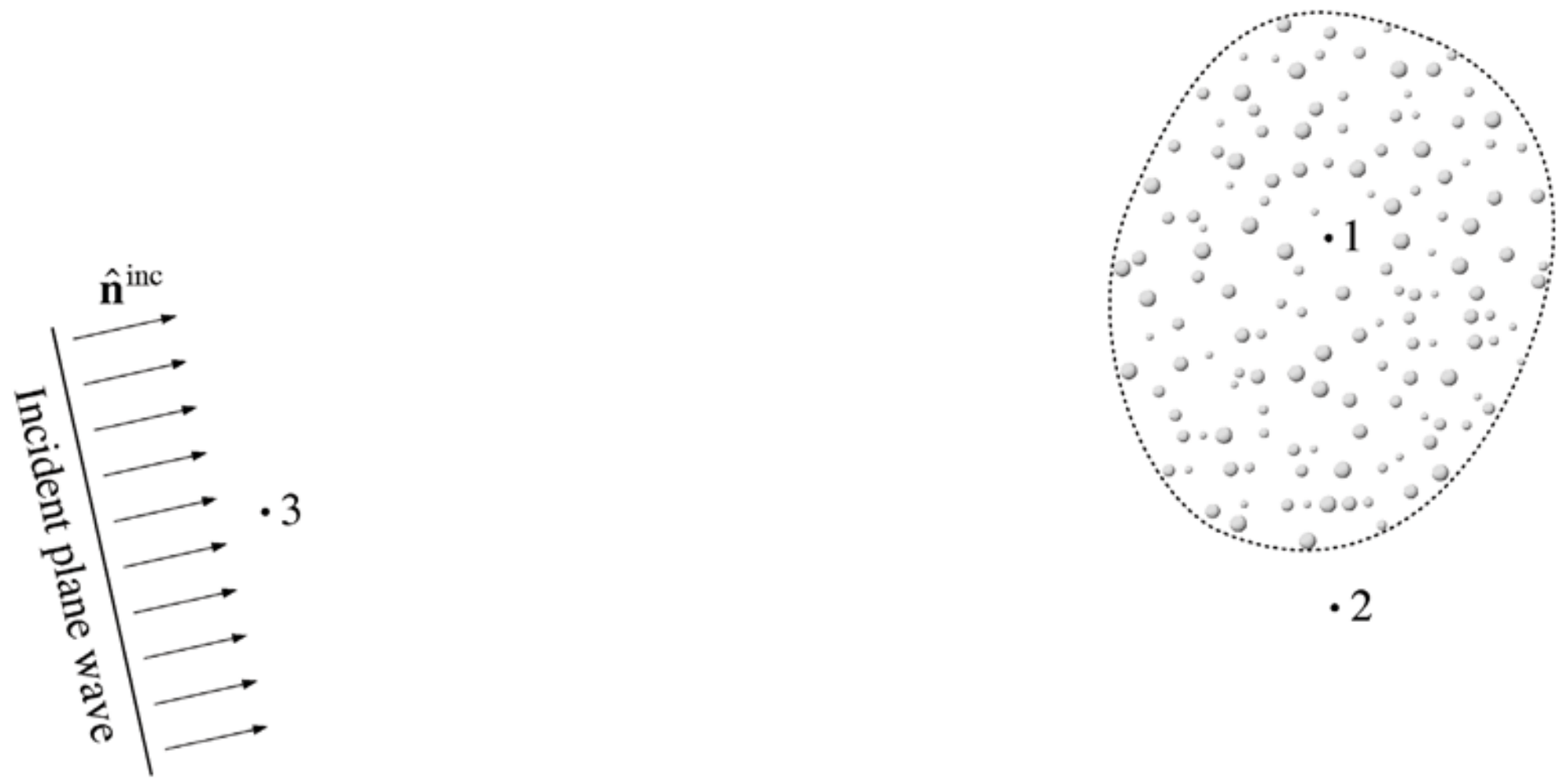

**Fig. 1.15.** Scattering of a plane electromagnetic wave by a volume of discrete random medium filled with sparsely distributed particles.

The above multiple-scattering formalism and the microphysical theory of RT can be generalized to the case of an absorbing host medium surrounding the particles (Mishchenko 2008b,c). Again, an essential ingredient of this generalization is the use of quantities ultimately representing actual optical observables.

## 1.19. Weak localization

Consider again a scattering object in the form of a large group of discrete, randomly and sparsely distributed particles as shown schematically in Fig. 1.15. The object is illuminated by a plane electromagnetic wave propagating in the direction $\hat{\mathbf{n}}^{\rm inc}$. The reader should recall that the RTE is derived by neglecting all diagrams with crossing connectors in the diagrammatic representation of the coherency dyadic. Following the line of reasoning outlined in the previous section, one may indeed conclude that upon statistical averaging the contribution of all the diagrams of the type illustrated in Fig. 1.16 must vanish at near-field observation points located either inside the object (observation point 1 in Fig. 1.15) or outside the object (observation point 2).

However, we have already discussed in Section 1.17.3 that there is an exception corresponding to the situation when the observation point is in the far-field zone of the scattering object and is located within its "back-shadow" (observation point 3 in Fig. 1.15). Then the class of diagrams illustrated by Plate 1.7e and Figs. 1.16c–e gives a nonzero contribution (Гнедин и Долгинов 1963) that causes the WL effect.

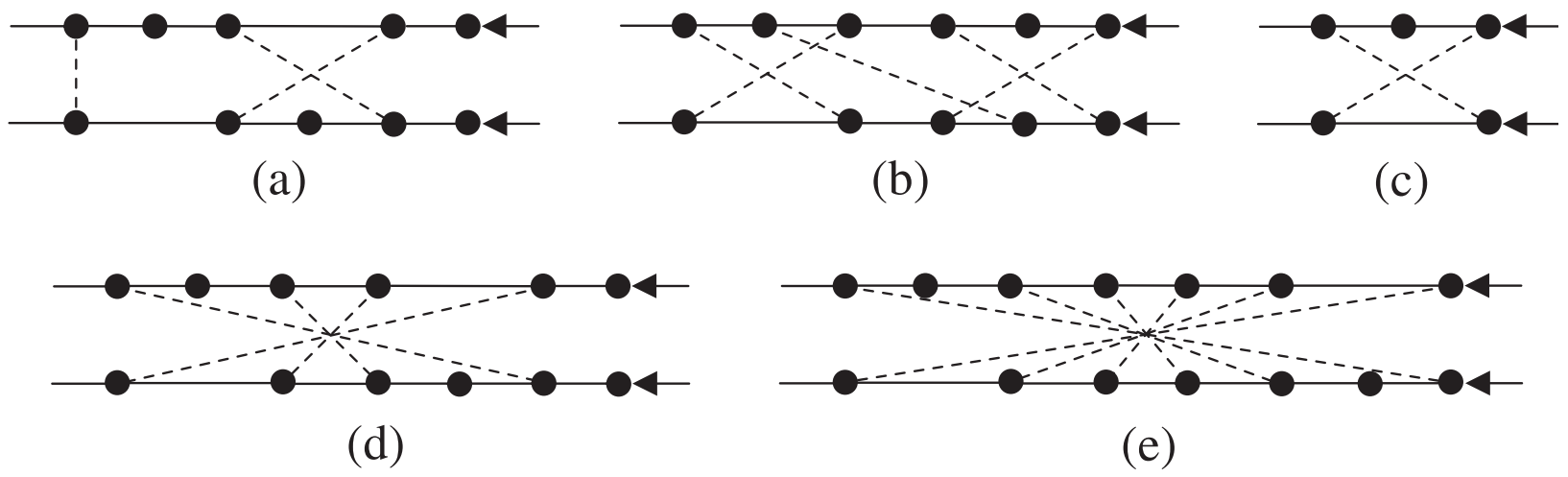

**Fig. 1.16.** Diagrams with crossing connectors.

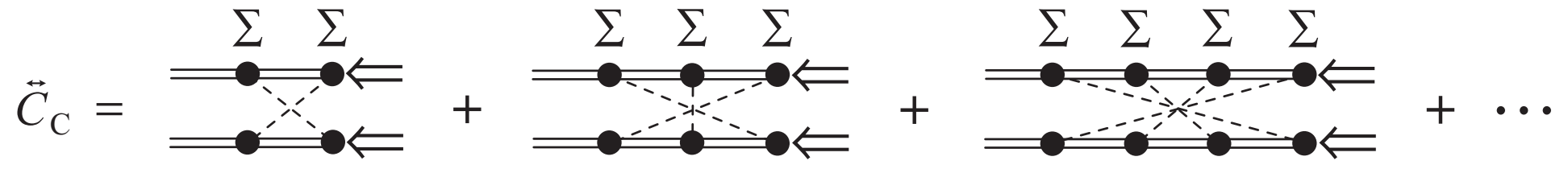

**Fig. 1.17.** The cyclical part of the coherency dyadic.

These diagrams are called maximally crossed since they can be drawn in such a way that all connectors cross at one point.

The expression for the cumulative contribution of all maximally crossed (or cyclical) diagrams to the coherency dyadic at an observation point can be derived using the diagrammatic technique introduced in the preceding section. The final result can be summarized by the diagrammatic expression shown in Fig. 1.17. As before, the symbol Σ denotes both the summation over all appropriate particles and the statistical averaging over the particle states and positions, whereas the double lines account for the effect of exponential attenuation and, possibly, dichroism. It is very instructive to compare Fig. 1.17 with Fig. 1.12 since this comparison reveals quite vividly the morphological difference between the participating diagrams. The total coherency dyadic is now approximated by the following expression:

$$\overleftrightarrow{C} \approx \overleftrightarrow{C}_\mathrm{L} + \overleftrightarrow{C}_\mathrm{C}. \tag{1.107}$$

The inclusion of the cyclical diagrams makes the computation of the coherency dyadic much more involved and limits the range of problems that can be solved accurately. In particular, no closed-form equation similar to the VRTE has been derived to describe the CB contribution to the specific coherency dyadic $\overleftrightarrow{\Sigma}_\mathrm{C}(\mathbf{r}, \hat{\mathbf{q}})$. However, the reciprocal nature of each single-scattering event leads to an interesting exact result: all characteristics of the CB effect at the exact backscattering direction can be rigorously expressed in terms of the solution of the VRTE (Section 1.25.1). This result as well as other theoretical and numerical approaches to the problem of WL are reviewed by Barabanenkov et al. (1991), Kuz'min and Romanov (1996), van Rossum and Niuwenhuizen (1999), Lenke and Maret (2000), Muinonen (2004), and Mishchenko et al. (2006b).

The angular width of the CB effect is inversely proportional to $k_1\langle l\rangle$, where $\langle l\rangle$ is the average distance between the end particles of the various particle sequences such as those shown in Fig. 1.7e. Factors limiting $\langle l\rangle$ and thereby increasing the angular width of the various manifestations of CB are absorption by particles and a finite size of the scattering medium. The finite-size effect is well illustrated by Plate 1.10d. For optically thick media, a good proxy for $\langle l \rangle$ is the so-called transport mean free path $l_{\mathrm{tr}}$ (Ishimaru 1978).

The angular width of the various CB features for objects such as water clouds is extremely small and can hardly be observed with passive instruments measuring the scattered sunlight since in this case $l_{\mathrm{tr}}$ is many orders of magnitude greater than the wavelength. However, CB can affect substantially the results of active observations of clouds with lidars and radars (e.g., Kobayashi et al. 2007) since these instruments observe electromagnetic radiation scattered in exactly the backscattering direction.

The situation is different for densely packed particulate media, in which case $l_{\mathrm{tr}}$ can be comparable to the wavelength of the incident light. As a consequence, CB can be detectable not only with active instruments or specifically designed laboratory equipment (e.g., Labeyrie et al. 2000; Gross et al. 2007; Psarev et al. 2007) but even in telescopic observations of sunlight scattered by surfaces of high-albedo solar system bodies (e.g., Rosenbush et al. 2002a; Mishchenko et al. 2006a, and references therein). It is important to recognize, however, that the very concept of wave phase applies only to transverse waves such as plane and spherical waves. Therefore, the interference explanation of WL is implicitly based on the assumption that each particle in any particle string (Plate 1.7e), is located in the far-field zones of the previous and the following particle. This assumption can often be violated in densely packed particulate media. However, the presence of strong CB peaks in the exact $T$-matrix results obtained for 240 densely packed particles (Plates 1.10b and 1.11) indicates that the wavelets scattered along strings of widely separated particles still provide a significant contribution to the total scattered signal.

Monostatic radars use the same antenna to transmit and receive electromagnetic waves. Therefore, radar measurements of particulate media are inevitably affected by WL. Several Solar System objects have been found to generate radar returns quite uncharacteristic of bare solid surfaces. For example, the icy Galilean satellites of Jupiter exhibit both high radar reflectivities and circular polarization ratios exceeding one (Ostro 1993). Similar radar echoes have been detected in radar observations of the poles of Mercury (Harmon et al. 1994). These measurements have been interpreted in terms of multiple scattering, including WL, of electromagnetic waves by voids or rocks imbedded in a transparent layer of ice (Mishchenko 1992c; Hapke 1993).

The interference explanation of WL assumes that the observer is located in the far-field zone of the entire scattering medium. In reality, the various manifestations of CB can be observed at distances shorter than those dictated by Eq. (1.16). Specifically, the distance $d$ from the scattering medium to the observation point must satisfy the following inequality (Mishchenko et al. 2006b):

$$d \gg \frac{1}{2} k_1 \langle l \rangle^2. \tag{1.108}$$

However, the requirement (1.108) can still be rather demanding if the scattering medium is composed of nonabsorbing, wavelength-sized or larger particles and its minimal dimension is greater than $l_{\mathrm{tr}}$.

## 1.20. Forward-scattering interference

Similarly to CB, the forward-scattering localization of electromagnetic waves discussed in Section 1.17.3 and illustrated in Plate 1.7d is an expressly far-field scattering effect and as such is not accounted for by the RTE. Indeed, it can be readily shown that the contribution of the diagrams of the type shown in Fig. 1.11b evaluated at a near-field observation point does not vanish only when both particles are positioned along the same straight line parallel to the incidence direction and going through the observation point. This non-vanishing contribution is ultimately included in the exponentially attenuated coherent Stokes column vector $\mathbf{I}_c$. In order to observe the forward-scattering interference effect directly, the observation point must be located in the far-field zone, i.e., at a distance $r$ from the scattering volume satisfying the inequalities (1.14)−(1.16). This factor makes the RTE a rather robust approximation.

## 1.21. "Independent" scattering

We have seen before that at any moment in time, the incident electromagnetic wave perceives the entire multi-particle group as a unified, albeit morphologically complex, scatterer. We have also witnessed how the RTE emerges from the Maxwell equations as a consequence of several assumptions (such as wide inter-particle separation) and, in the final analysis, contains single-particle extinction and phase matrices averaged over the particle states. However, a traditional (and incorrect!) way of dealing with the problem of multiple scattering and RT has been to proceed in exactly the opposite direction:

- by first considering the scattering properties of each particle in total isolation from all the other particle by solving individually the macroscopic Maxwell equations (e.g., by using the Lorenz–Mie theory);
- then considering widely separated, randomly positioned particles forming a particle group as "independent scatterers" characterized by the previously determined individual extinction and phase matrices;
- then considering "incoherent single scattering" by the "independently scattering particles" occupying an imaginary "elementary scattering volume";
- then considering "incoherent multiple scattering" by the "elementary volume elements"; and finally
- by speculating how the "single-scattering properties" of the individual particles and the "elementary volume elements" can change as a consequence of hypothetical "packing density" effects.

It is thus clear that the notion of "independent scattering" has been very important to the discipline of light scattering by disperse media. Several definitions of "independently scattering particles" have been given in the literature (e.g., van de Hulst 1957; Cartigny et al. 1986; Tien 1988; Kumar and Tien 1990; Ivezić and Mengüç 1996; Liou 2002; Mishchenko et al. 2002a, 2006b; Martin 2006, and references therein). Some of these definitions may be rather vague and some of them actually refer to the SSA (Agarwal and Mengüç 1991; Mishchenko et al. 2006b, 2007e). (It should be kept in mind that the specific conditions of applicability of the SSA and the RT approximation in terms of the minimal average inter-particle separation may be somewhat different (Mishchenko et al. 2006b).) The common intent of these definitions has been to ensure that observable consequences of scattering by a disperse medium are described in terms of the extinction and phase matrices of the individual particles, i.e., the *quantities describing single-particle transformations of Stokes parameters rather than electric fields*.

The microphysical approach outlined above makes it quite clear that a particle is an independent scatterer only when it is completely alone. Particles forming a group cannot be independent scatterers irrespective of how widely they are separated and how randomly they are distributed since the forward-scattering interference and CB effects are ubiquitous and cannot be described in terms of individual-particle extinction and phase matrices. In the case of large rarefied objects such as terrestrial water clouds, the forward-scattering and CB intensity peaks are extremely narrow, contain a negligible fraction of the total scattered energy, and are hardly observable, thereby making the RT theory a very good quantitative descriptor of many actual observables. We have seen, however, that even in the limited context of the RT theory, particles are not characterized by their individual extinction and phase matrices. Instead, each actual particle is replaced by an imaginary "average particle" characterized by the ensemble-averaged extinction and phase matrices (Мищенко 2008).

We must, therefore, conclude that the term "independent scattering" has little heuristic value and can, in fact, be quite misleading. Perhaps the least ambiguous, albeit still undesirable, way to use this term is in application to randomly positioned particles located in the far-field zones of each other, since the randomness and far-field conditions are necessary in the microphysical theories of RT and WL.

## 1.22. Radiative transfer in gaseous media

It is well known that multiple scattering can be caused not only by particles but also by density and anisotropy fluctuations in rarified molecular media such as gases (Smoluchowski 1908). This type of scattering is traditionally called Rayleigh scattering and is thoroughly reviewed by Fabelinskii (1968) and Kuz'min et al. (1994). Each density and/or anisotropy fluctuation can be considered a particle in the sense of causing the electric permittivity (or, in general, the electric permittivity tensor) in a small volume element to be different from that of the surrounding medium. As long as such volume elements are located in the far-field zones of each other, the microphysical approach outlined in Section 1.18 remains applicable, thereby leading to the classical RTE describing multiple Rayleigh scattering (Chandrasekhar 1950). The specific form of the corresponding extinction and phase matrices depends on the type of gas (or gas mixture) and on factors such as gaseous pressure and temperature.

It is sometimes claimed that the actual cause of Rayleigh scattering are randomly positioned and randomly moving individual molecules rather than electric permittivity fluctuations. However, molecules can often be separated by distances much smaller than the wavelength, thereby grossly violating the far-field zone assumption used to derive the RTE. Of course, one cannot exclude completely the possibility that the RTE can be derived without the far-field zone assumption, but until and unless this has been done, it is more prudent to attribute Rayleigh scattering to molecular fluctuations rather than to the individual molecules.

Quite often a gaseous medium contains randomly distributed macroscopic particles. Typical examples are aerosols and cloud particles suspended in a planetary atmosphere. Obviously, the RTE still remains applicable provided that the particles and the density/anisotropy fluctuations are located in the far-field zones of each other. The phase and extinction matrices entering the VRTE are obtained by straightforward averaging over the gas–particle mixture.

## 1.23. Energy conservation

An interesting and practically important property of the VRTE is that it satisfies precisely the energy conservation law. Indeed, using the vector identity $\mathbf{a}\cdot\nabla f = \nabla\cdot(\mathbf{a}f) - f\,\nabla\cdot\mathbf{a}$, where $f$ is any scalar function of spatial coordinates, and taking into account that $\hat{\mathbf{q}}$ is a constant vector, we can rewrite Eq. (1.102) in the form

$$\nabla\cdot[\hat{\mathbf{q}}\,\tilde{\mathbf{I}}(\mathbf{r},\hat{\mathbf{q}})] = -n_0\langle\mathbf{K}_1(\mathbf{r},\hat{\mathbf{q}})\rangle_\xi\tilde{\mathbf{I}}(\mathbf{r},\hat{\mathbf{q}}) + n_0\int_{4\pi}\mathrm{d}\hat{\mathbf{q}}'\langle\mathbf{Z}_1(\hat{\mathbf{q}},\hat{\mathbf{q}}')\rangle_\xi\,\tilde{\mathbf{I}}(\mathbf{r},\hat{\mathbf{q}}'). \quad (1.109)$$

Let us now introduce the flux density vector as

$$\mathbf{F}(\mathbf{r}) = \int_{4\pi}\mathrm{d}\hat{\mathbf{q}}\,\hat{\mathbf{q}}\,\tilde{I}(\mathbf{r},\hat{\mathbf{q}}). \quad (1.110)$$

Obviously, the product $\hat{\mathbf{p}}\cdot\mathbf{F}(\mathbf{r})\mathrm{d}S$ gives the amount and the direction of the net flow of power through a surface element d$S$ normal to $\hat{\mathbf{p}}$ (see Fig. 1.18). Integrating both sides of Eq. (1.109) over all propagation directions $\hat{\mathbf{q}}$ yields (Mishchenko et al. 2006b)

$$-\nabla\cdot\mathbf{F}(\mathbf{r}) = n_0\int_{4\pi}\mathrm{d}\hat{\mathbf{q}}\,\langle C_{\mathrm{abs},1}(\hat{\mathbf{q}})\rangle_\xi\,\tilde{I}(\mathbf{r},\mathbf{q}), \quad (1.111)$$

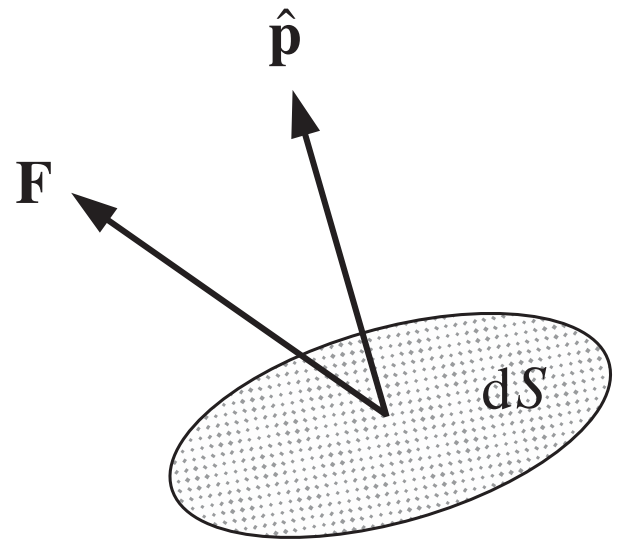

**Fig. 1.18.** Electromagnetic power flow through an elementary surface element.

where, as before, $\langle C_{\text{abs},1}(\hat{\mathbf{q}})\rangle_\xi$ is the ensemble-averaged absorption cross section per particle. The physical meaning of this formula is very transparent: the net inflow of electromagnetic power per unit volume is equal to the total power absorbed per unit volume. If the particles forming the scattering medium are nonabsorbing so that $\langle C_{\text{abs},1}(\hat{\mathbf{q}})\rangle_\xi = 0$, then the flux density vector is divergence-free:

$$\nabla \cdot \mathbf{F}(\mathbf{r}) = 0. \tag{1.112}$$

This is a manifestation of the conservation of the power flux, which means that the amount of electromagnetic energy entering a volume element per unit time is equal to the amount of electromagnetic energy leaving the volume element per unit time.

The previous discussion clearly shows that the VRTE follows from the Maxwell equations only after several simplifying assumptions have been made. Therefore, the fact that the VRTE fully complies with the energy conservation law is as much troubling as it is encouraging. Indeed, attempting to improve the accuracy of RTE predictions by including the maximally crossed diagrams appears to destroy energy conservation by adding the "surplus" energy contained in the CB intensity peak. It remains unclear whether this additional energy is "taken" from the far wings of the backscattering peak, which would imply that the contribution of the maximally crossed diagrams to the specific intensity at certain reflection directions may be negative (cf. van Tiggelen et al. 1995). It is also possible that the negative contribution restoring energy conservation is supplied by all the other diagram types not accounted for by the microphysical RT and WL theories.

## 1.24. Discussion

The above discussion shows that the theories of RT and WL follow from the macroscopic frequency-domain Maxwell equations as a consequence of several well-defined approximations. This does not mean, of course, that all of these approximations are mandatory and that the RT and WL theories cannot be derived under less restrictive assumptions. However, until and unless the latter has been done, it is prudent to consider the approximations introduced in Sections 1.18 and 1.19 as necessary.

An instructive way to look at the microphysical approach to RT and WL is in terms of the famous classification of the methods used to solve problems of wave propagation in random media into two categories, called "honest" and "dishonest" (Keller 1962). Specifically, let us assume that a wave is described by a vector-valued function $\mathbf{u}(\mathbf{r}, t)$ of the position vector and time. As in Section 1.10, we denote by $\psi$ the full set of parameters defining the state of the entire scattering medium at a moment in time. According to Keller,

> "in an honest method the solution $\mathbf{u}(\mathbf{r}, t, \psi)$ is first determined for each value of $\psi$. The solution may sometimes be found exactly and explicitly, but more often it is expressed in the form of a series in some parameter, or as a sequence of iterates, or by some other approximation procedure. In the process of solving for $\mathbf{u}(\mathbf{r}, t, \psi)$ randomness plays no role and therefore it provides no advantage. The second step is to compute the mean

> value of $\mathbf{u}(\mathbf{r}, t, \psi)$, as well as its variance and other statistics, from the explicit expression. In this step randomness may have the helpful effect of yielding simpler expressions for the statistics of $\mathbf{u}$ than those for $\mathbf{u}$ itself. In a dishonest method randomness is utilized before $\mathbf{u}(\mathbf{r}, t, \psi)$ is determined. In all cases probability is introduced before $\mathbf{u}$ is determined and an unproved assumption is made about some statistical property of the random wave motion. The assumption usually simplifies the problem so that it becomes solvable."

The reader can easily recognize that the microphysical approach to RT and WL described above belongs to the category of "honest" methods.

Figure 1.19 provides a schematic summary of the microphysical theories of RT and WL and classifies their place within the broader context of classical macroscopic electromagnetics. It also helps formulate problems that still await solution.

First of all, by using the macroscopic frequency-domain Maxwell equations as the starting point, we have completely excluded from consideration such phenomena as emission and frequency redistribution as well as situations involving finite-beam or pulsed illumination. These areas of electromagnetic energy transfer remain purely phenomenological (e.g., Иванов 1969; Oxenius 1986; Нагирнер 2001; Hanel et al. 2003; Mätzler 2006; Wehrse and Kalkofen 2006; Ito et al. 2007) and invoke the RTE without strict derivation from first physical principles.

Another challenging subject is RT in stochastic heterogeneous media composed of widely separated yet spatially correlated particles. For example, it has been suggested (Shaw et al. 2002; Knyazikhin et al. 2005; Marshak et al. 2005) that cloud droplets belonging to a particular size range may tend to form groups of spatially correlated particles (clusters) imbedded in an otherwise homogeneous cloud. It was shown by Mishchenko (2006b) on the basis of the microphysical approach that as long as such inclusions are small and specific assumptions of ergodicity and spatial uniformity hold, one can still apply the classical VRTE in which the participating extinction and phase matrices are obtained by averaging the respective single-particle matrices over all the particles constituting the entire cloud. However, this result may not necessarily apply to clouds with larger inhomogeneities.

Apart from Mishchenko (2006b), the problem of multiple scattering in stochastic media composed of widely separated yet correlated particles has been analyzed so far by using the motley concepts of the phenomenological RT theory, including the fictitious "photons" (e.g., Pomraning 1991; Cairns et al. 2000; Kostinski 2001; Petty 2002; Barker et al. 2003; Borovoi 2006; Davis 2006; Davis and Marshak 2010, and references therein). We have seen in Section 1.18 that the extinction and phase matrices appear in the standard VRTE as a consequence of well-defined assumptions and approximations and only as ensemble-averaged quantities. In the phenomenological stochastic RT theory, the extinction and phase matrices are taken for granted and are postulated to be the primary optical attributes of the individual particles. In this sense the phenomenological theory belongs to the category of "dishonest" methods. Clearly, a systematic application of the "honest" microphysical approach is necessary to determine whether and to what extent the concepts of extinction and phase matrices can be applied to correlated particles.

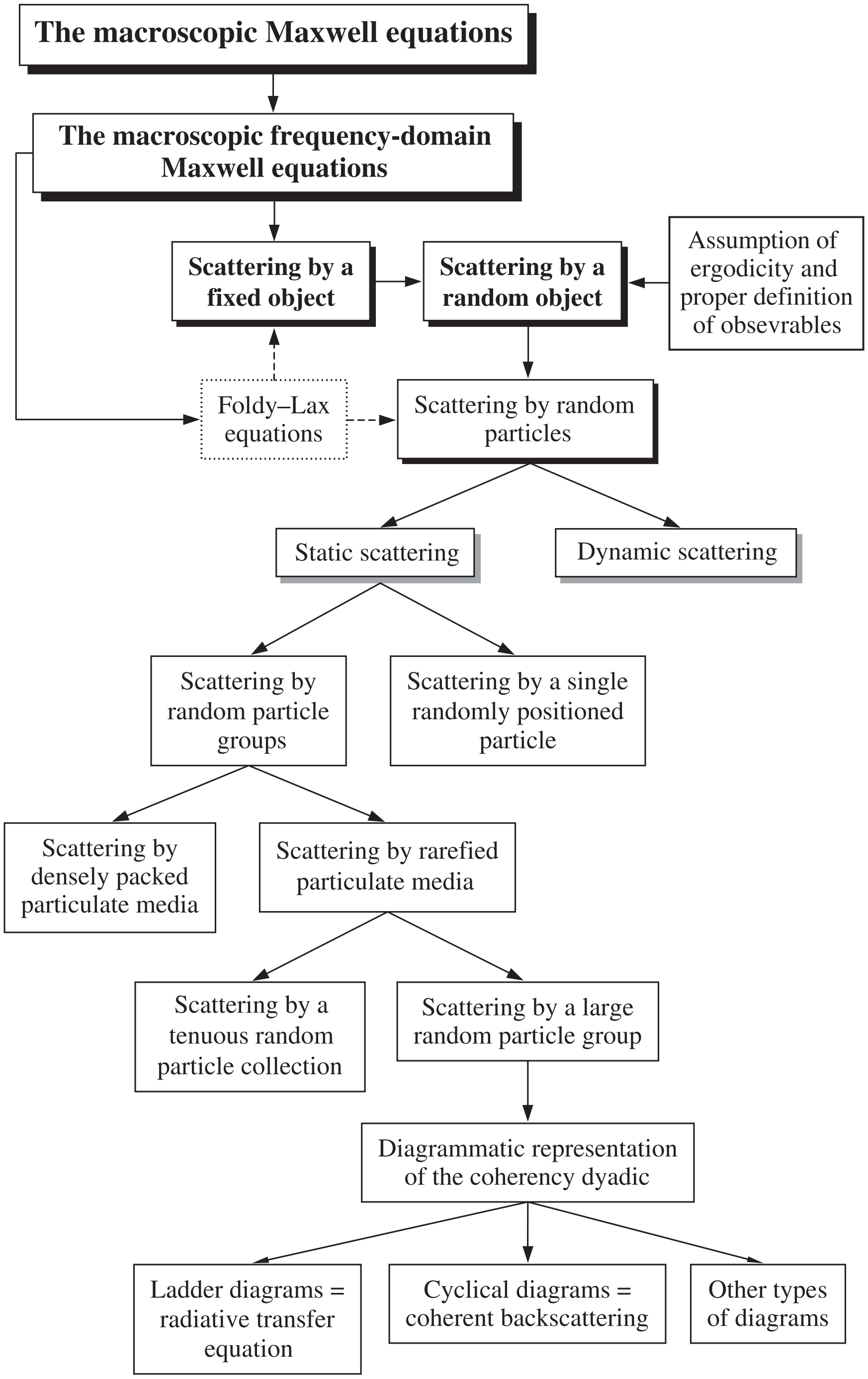

**Fig. 1.19.** Classification of electromagnetic scattering problems.

Many geophysical scattering media consist of densely packed and strongly correlated particles. Typical examples are snow (Wiscombe and Warren 1980; Dozier and Painter 2004; Kaasalainen et al. 2006), soil and regolith surfaces (Hapke 1993; Shkuratov et al. 2007), and vegetation. There is a rapidly growing number of publications in which numerical solutions of the RTE are used to model directional reflectance and transmittance characteristics of the various densely packed particulate media (see, e.g., Leroux et al. 1999; Mishchenko et al. 1999a; Petrova et al. 2001; Kokhanovsky 2004; Liang 2004; Okin and Painter 2004; Xie et al. 2006, and references therein). The reader can recall that the formal applicability of the RT theory rests on the assumption that scattering particles are located in each-other's far-field zones and are uncorrelated. The obvious violation of these assumptions in the case of densely packed particles can lead to significant deviations from the numerical predictions based on the RTE. Therefore, it is important to analyze both theoretically and experimentally to what extent the classical RT theory can be applied to densely packed particulate media. Some progress in this direction has been reported by Sergent et al. (1998), Hespel et al. (2003), Painter and Dozier (2004), Pitman et al. (2005), and Zhang and Voss (2005, 2008).

The rigorous *analytical* theory of electromagnetic energy transport in densely packed particulate media is still in progress (e.g., Иванов и др. 1988; Tsang and Kong 2001; Tishkovets 2007; Tsang et al. 2007; Тишковец 2007, 2009 and references therein). There are several "dishonest" phenomenological approaches to this problem which start with the notion of "independent scattering" and attempt to predict the modification of the phase and extinction matrices by effects of packing density (Hapke 1993), but the heuristic value of such approximations is limited and their range of validity is unknown if not quite questionable. The same is true of the phenomenological approaches based on the use of the geometrical optics approximation and Monte Carlo ray tracing, even if electric fields rather than Stokes column vectors are traced (e.g., Stankevich et al. 2007). Fortunately, the ever increasing power of scientific workstations and the availability of efficient numerical techniques have led to the emergence of an accurate quantitative approach to this complex problem based on direct computer solutions of the Maxwell equations (Mackowski 2006; Tseng et al. 2006; Mishchenko and Liu 2007, 2009; Mishchenko et al. 2007d, 2009b,c).

Another important problem is electromagnetic scattering by a medium composed of randomly positioned particles and adjacent to a random rough boundary such as the ocean surface. Although problems like this one are important in practice and have been treated using various phenomenological approaches, microphysical treatments based on consistent application of the Maxwell equations have been extremely scarce.

## 1.25. Weak localization in plane-parallel discrete random media

We have mentioned in Section 1.19 that it is rather straightforward to derive general analytical expressions for the cyclical coherency dyadic and the cyclical specific coherency dyadic. Unfortunately, the use of these expressions in practical computations, either analytical or numerical, still remains highly problematic. Nevertheless, there are two rigorous particular solutions of the problem which have already found extensive applications. The aim of this section is to briefly discuss these solutions and their consequences.

Consider the reflection of a plane electromagnetic wave $\mathbf{E}_0^{\text{inc}} \exp(\mathrm{i}k_1 \hat{\mathbf{n}}_0 \cdot \mathbf{r})$ (or a parallel quasi-monochromatic beam of light of infinite lateral extent propagating in the direction of the unit vector $\hat{\mathbf{n}}_0$) by a plane-parallel layer of sparse discrete random medium. A well-collimated polarization-sensitive detector of electromagnetic energy is located at a distant observation point and scans a range of upward propagation directions $\hat{\mathbf{n}}$ including the exact backscattering direction given by $\hat{\mathbf{n}} = -\hat{\mathbf{n}}_0$. The response of this detector can be described in terms of the $4\times4$ reflection matrix $\mathbf{R}$ according to $\widetilde{\mathbf{I}}(\hat{\mathbf{n}}) = |u_0|\mathbf{R}(\hat{\mathbf{n}}, \hat{\mathbf{n}}_0)\mathbf{I}_0/\pi$, where $\mathbf{I}_0$ is the Stokes column vector of the incident light and $u_0$ is the cosine of the angle between the unit vector $\hat{\mathbf{n}}_0$ and the upward normal to the upper boundary of the slab. In agreement with the previous discussion of WL in Sections 1.17.3 and 1.19, it is useful to consider the following three particular situations:

- The incoming propagation direction $\hat{\mathbf{n}}$ is relatively far from the exact backscattering direction $-\hat{\mathbf{n}}_0$. Then the cyclical specific coherency dyadic vanishes, and the detector response is fully determined by the slowly changing diffuse specific coherency dyadic $\overleftrightarrow{\Sigma}_{\mathrm{d}}(\hat{\mathbf{n}}) \approx \overleftrightarrow{\Sigma}_{\mathrm{d}}(-\hat{\mathbf{n}}_0)$. This means that the response of the detector can be fully quantified in terms of the diffuse reflection matrix $\mathbf{R}^{\text{diff}}(\hat{\mathbf{n}}, \hat{\mathbf{n}}_0)$ obtained by solving the VRTE (Section 1.18; see Fig. 1.20):

$$\mathbf{R}^{\text{diff}}(\hat{\mathbf{n}}, \hat{\mathbf{n}}_0) \equiv \mathbf{R}^{\text{RT}}(\hat{\mathbf{n}}, \hat{\mathbf{n}}_0) \approx \mathbf{R}^{\text{RT}}(-\hat{\mathbf{n}}_0, \hat{\mathbf{n}}_0), \tag{1.113}$$

$$\widetilde{\mathbf{I}}^{\text{diff}}(\hat{\mathbf{n}}) \equiv \widetilde{\mathbf{I}}_{\mathrm{d}}(\hat{\mathbf{n}}) \approx \widetilde{\mathbf{I}}_{\mathrm{d}}(-\hat{\mathbf{n}}_0). \tag{1.114}$$

- The detector registers reflected light propagating in the exact backscattering direction. Then the effects of CB can be expected to be maximal and must be taken into account by computing the respective full reflection matrix $\mathbf{R}(-\hat{\mathbf{n}}_0, \hat{\mathbf{n}}_0)$ and the specific Stokes column vector $\widetilde{\mathbf{I}}(-\hat{\mathbf{n}}_0)$ (see Fig. 1.20). We shall demonstrate in Section 1.25.1 that, somewhat unexpectedly, this can also be done in terms of the solution of the VRTE.
- As the detector axis deviates more and more from the exact backscattering direction the effects of CB can be expected to weaken and gradually disappear (see Fig. 1.20). The computation of the angular profile of the detector response in this transition region of incoming directions is a very complex problem that has been solved only in a few particular cases. We will discuss the consequences of one such solution in Section 1.25.2.

### *1.25.1. Exact backscattering direction*

As explained above, let us express the full reflection matrix as the sum of diffuse and cyclical components,

$$\mathbf{R}(\hat{\mathbf{n}}, \hat{\mathbf{n}}_0) = \mathbf{R}^{\text{diff}}(\hat{\mathbf{n}}, \hat{\mathbf{n}}_0) + \mathbf{R}^{\text{C}}(\hat{\mathbf{n}}, \hat{\mathbf{n}}_0) = \mathbf{R}^{\text{RT}}(\hat{\mathbf{n}}, \hat{\mathbf{n}}_0) + \mathbf{R}^{\text{C}}(\hat{\mathbf{n}}, \hat{\mathbf{n}}_0), \tag{1.115}$$

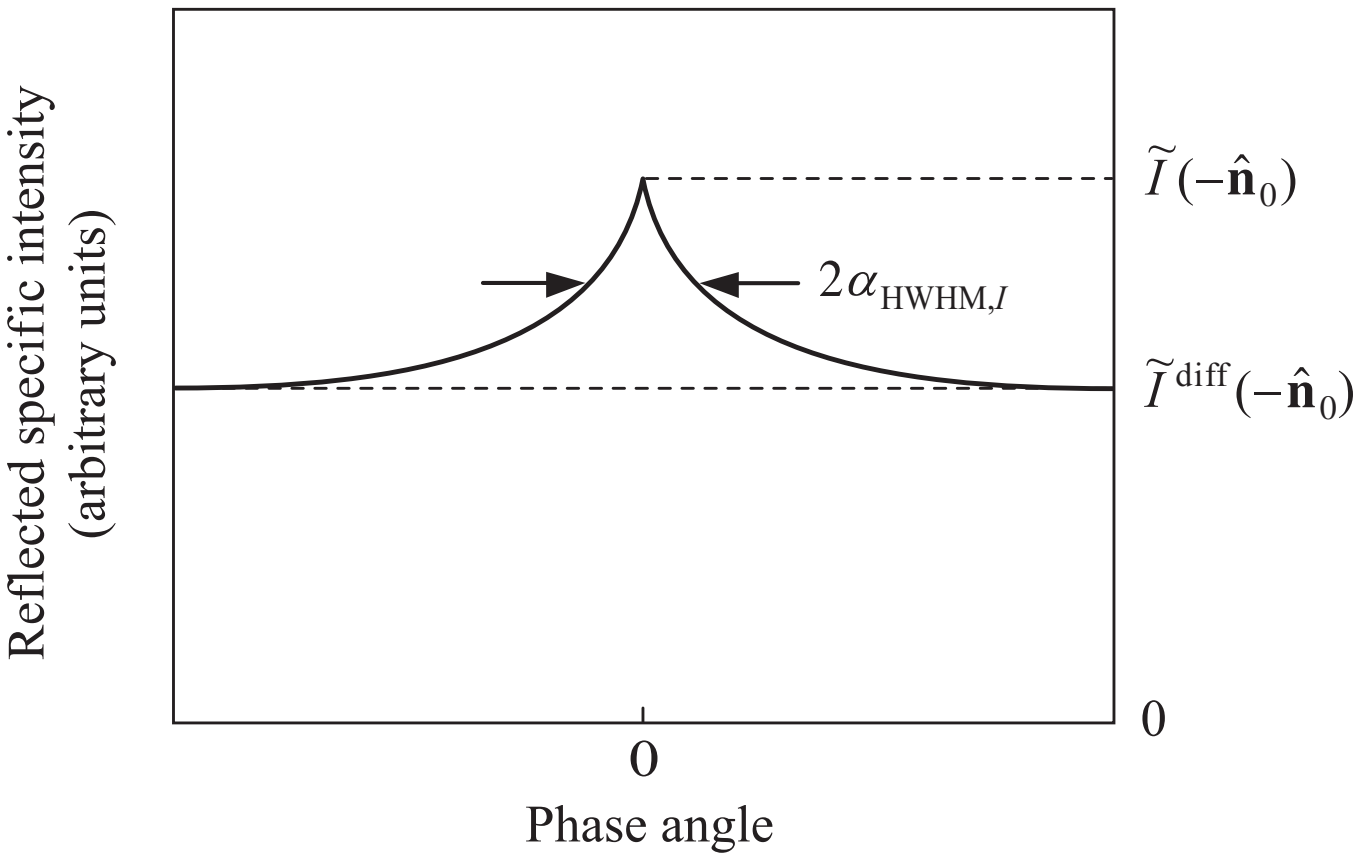

**Fig. 1.20.** Coherent enhancement of backscattered intensity. The phase angle is defined as the angle between the directions "scattering object → source of illumination" and "scattering object → detector of light".

and expand the diffuse component into the first-order-scattering (superscript 1) and multiple-scattering (superscript M) contributions:

$$\mathbf{R}(\hat{\mathbf{n}}, \hat{\mathbf{n}}_0) = \mathbf{R}^1(\hat{\mathbf{n}}, \hat{\mathbf{n}}_0) + \mathbf{R}^{\mathrm{M}}(\hat{\mathbf{n}}, \hat{\mathbf{n}}_0) + \mathbf{R}^{\mathrm{C}}(\hat{\mathbf{n}}, \hat{\mathbf{n}}_0). \tag{1.116}$$

Both $\mathbf{R}^1(\hat{\mathbf{n}}, \hat{\mathbf{n}}_0)$ and $\mathbf{R}^{\mathrm{M}}(\hat{\mathbf{n}}, \hat{\mathbf{n}}_0)$ can be found by solving the VRTE. Analogously, the specific Stokes vector is given by

$$\tilde{\mathbf{I}}(\hat{\mathbf{n}}) = \tilde{\mathbf{I}}^1(\hat{\mathbf{n}}) + \tilde{\mathbf{I}}^{\mathrm{M}}(\hat{\mathbf{n}}) + \tilde{\mathbf{I}}^{\mathrm{C}}(\hat{\mathbf{n}}), \tag{1.117}$$

where $\tilde{\mathbf{I}}^1(\hat{\mathbf{n}}) + \tilde{\mathbf{I}}^{\mathrm{M}}(\hat{\mathbf{n}}) = \tilde{\mathbf{I}}_{\mathrm{d}}(\hat{\mathbf{n}})$.

One of a few rigorous results of the theory of WL has been obtained by applying the Saxon's reciprocity relation for the scattering dyadic (Saxon 1955b) to the expressions for the diffuse and cyclical components of the specific coherency dyadic corresponding to the exact backscattering direction (Mishchenko 1991a). The final result is a remarkably simple relation between the corresponding matrices $\mathbf{R}^{\mathrm{M}}(-\hat{\mathbf{n}}_0, \hat{\mathbf{n}}_0)$ and $\mathbf{R}^{\mathrm{C}}(-\hat{\mathbf{n}}_0, \hat{\mathbf{n}}_0)$. In the particular case of a macroscopically isotropic and mirror-symmetric scattering medium this relation takes the following form (Mishchenko 1992a):

$$\mathbf{R}^{\mathrm{C}} = \begin{bmatrix} R_{11}^{\mathrm{C}} & R_{12}^{\mathrm{M}} & 0 & 0 \\ R_{12}^{\mathrm{M}} & R_{22}^{\mathrm{C}} & 0 & 0 \\ 0 & 0 & R_{33}^{\mathrm{C}} & R_{34}^{\mathrm{M}} \\ 0 & 0 & -R_{34}^{\mathrm{M}} & R_{44}^{\mathrm{C}} \end{bmatrix}, \tag{1.118}$$

where

$$R_{11}^{\mathrm{C}}(-\hat{\mathbf{n}}_0, \hat{\mathbf{n}}_0) = \frac{1}{2}[R_{11}^{\mathrm{M}}(-\hat{\mathbf{n}}_0, \hat{\mathbf{n}}_0) + R_{22}^{\mathrm{M}}(-\hat{\mathbf{n}}_0, \hat{\mathbf{n}}_0) - R_{33}^{\mathrm{M}}(-\hat{\mathbf{n}}_0, \hat{\mathbf{n}}_0) + R_{44}^{\mathrm{M}}(-\hat{\mathbf{n}}_0, \hat{\mathbf{n}}_0)], \tag{1.119}$$

$$R^{\mathrm{C}}_{22}(-\hat{\mathbf{n}}_0,\hat{\mathbf{n}}_0) = \frac{1}{2}[R^{\mathrm{M}}_{11}(-\hat{\mathbf{n}}_0,\hat{\mathbf{n}}_0) + R^{\mathrm{M}}_{22}(-\hat{\mathbf{n}}_0,\hat{\mathbf{n}}_0) + R^{\mathrm{M}}_{33}(-\hat{\mathbf{n}}_0,\hat{\mathbf{n}}_0) - R^{\mathrm{M}}_{44}(-\hat{\mathbf{n}}_0,\hat{\mathbf{n}}_0)], \tag{1.120}$$

$$R^{\mathrm{C}}_{33}(-\hat{\mathbf{n}}_0,\hat{\mathbf{n}}_0) = \frac{1}{2}[-R^{\mathrm{M}}_{11}(-\hat{\mathbf{n}}_0,\hat{\mathbf{n}}_0) + R^{\mathrm{M}}_{22}(-\hat{\mathbf{n}}_0,\hat{\mathbf{n}}_0) + R^{\mathrm{M}}_{33}(-\hat{\mathbf{n}}_0,\hat{\mathbf{n}}_0) + R^{\mathrm{M}}_{44}(-\hat{\mathbf{n}}_0,\hat{\mathbf{n}}_0)], \tag{1.121}$$

$$R^{\mathrm{C}}_{44}(-\hat{\mathbf{n}}_0,\hat{\mathbf{n}}_0) = \frac{1}{2}[R^{\mathrm{M}}_{11}(-\hat{\mathbf{n}}_0,\hat{\mathbf{n}}_0) - R^{\mathrm{M}}_{22}(-\hat{\mathbf{n}}_0,\hat{\mathbf{n}}_0) + R^{\mathrm{M}}_{33}(-\hat{\mathbf{n}}_0,\hat{\mathbf{n}}_0) + R^{\mathrm{M}}_{44}(-\hat{\mathbf{n}}_0,\hat{\mathbf{n}}_0)]. \tag{1.122}$$

The importance of this rigorous result is hard to overstate. Indeed, it demonstrates that although the RT theory is based on the neglect of all cyclical diagrams, all observable characteristics of WL at the *exact backscattering direction* can still be calculated by solving the VRTE. Furthermore, this can be done by characterizing the scattering medium in terms of actual physical parameters such as the optical thickness of the slab and the size, shape, and refractive index of the constituent particles.

Let us consider two applications of Eqs. (1.116) and (1.118). In the case of unpolarized incident light (e.g., sunlight) the specific intensity of light reflected in exactly the backscattering direction is given by $\widetilde{I}(-\hat{\mathbf{n}}_0) = |u_0| R_{11}(-\hat{\mathbf{n}}_0,\hat{\mathbf{n}}_0) I_0/\pi$ where $I_0$ is the incident intensity. On the other hand, the diffuse background is given by $\widetilde{I}^{\,\mathrm{diff}}(-\hat{\mathbf{n}}_0) = |u_0|[R^{1}_{11}(-\hat{\mathbf{n}}_0,\hat{\mathbf{n}}_0) + R^{\mathrm{M}}_{11}(-\hat{\mathbf{n}}_0,\hat{\mathbf{n}}_0)] I_0/\pi$ (see Fig. 1.20). Therefore, we can define the corresponding "unpolarized" CB enhancement factor as

$$\zeta_I = \frac{\widetilde{I}}{\widetilde{I}^{\,\mathrm{diff}}} = 1 + \frac{R^{\mathrm{M}}_{11} + R^{\mathrm{M}}_{22} - R^{\mathrm{M}}_{33} + R^{\mathrm{M}}_{44}}{2(R^{1}_{11} + R^{\mathrm{M}}_{11})}, \tag{1.123}$$

where the angular arguments $(-\hat{\mathbf{n}}_0,\hat{\mathbf{n}}_0)$ on the right-hand side are omitted for the sake of brevity. In the case of the incident light circularly polarized in the anticlockwise sense as viewed by an observer looking in the direction of propagation, the corresponding circular polarization ratio is given by

$$\mu_{\mathrm{C}} = \frac{\widetilde{I}_{\mathrm{sh}}}{\widetilde{I}_{\mathrm{oh}}} = \frac{R^{1}_{11} + R^{1}_{44} + 2R^{\mathrm{M}}_{11} + 2R^{\mathrm{M}}_{44}}{R^{1}_{11} - R^{1}_{44} + R^{\mathrm{M}}_{11} + R^{\mathrm{M}}_{22} - R^{\mathrm{M}}_{33} - R^{\mathrm{M}}_{44}}, \tag{1.124}$$

where "sh" and "oh" denote the "same-helicity" and "opposite-helicity" components of the reflected radiation. The enhancement factor $\zeta_I$ is a quantity which can potentially be determined from telescopic observations of Solar System objects from the ground or from spacecraft, while the ratio $\mu_{\mathrm{C}}$ is often measured during polarimetric radar or lidar observations of terrestrial and celestial targets.

Plates 1.12 and 1.13 show the results of numerically exact computer calculations of $\zeta_I$ and $\mu_{\mathrm{C}}$ for a semi-infinite particulate layer composed of polydisperse spherical particles as functions of the particle effective size parameter and $|u_0|$ for several values of the real, $m_{\mathrm{R}}$, and imaginary, $m_{\mathrm{I}}$, parts of the relative refractive index (Mishchenko 1996). The matrices $\mathbf{R}^{1}(\hat{\mathbf{n}},\hat{\mathbf{n}}_0)$ and $\mathbf{R}^{\mathrm{M}}(\hat{\mathbf{n}},\hat{\mathbf{n}}_0)$ were found by solving numerically the so-called Ambarzumian nonlinear integral equation as described in de Rooij (1985) and Mishchenko (1996). These and similar numerical results have been used to interpret the results of controlled laboratory measurements reported by van Albada et al. (1988) and Wolf et al. (1988) (see Mishchenko 1991a), the results of photometric observations of Saturn's rings (Mishchenko and Dlugach 1992; Mishchenko 1993b) and E-type asteroids (Mishchenko and Dlugach 1993), and the results of radar observations of planets, planetary satellites, and Saturn's rings (Mishchenko 1992c, 1996; Mishchenko and Dlugach 2008a). The practical importance of these results will be further discussed in Chapter 3.

It is important to emphasize that unlike the scalar approximation of multiple scattering, the rigorous formula (1.123) involves all the diagonal elements of the matrix $\mathbf{R}^{\mathrm{M}}(\hat{\mathbf{n}}_0, \hat{\mathbf{n}}_0)$ rather than only its (1,1) element. It has been found that numerical predictions of the unpolarized enhancement factor $\zeta_I$ based on the scalar approximation can differ from the corresponding exact results by as much as 25% (Mishchenko and Dlugach 2008b).

### *1.25.2. Angular dependence of weak localization: Rayleigh scattering*

In this section we use an exact vector solution of the WL problem obtained by Ozrin (1992) and Amic et al. (1997) in order to illustrate the angular distribution of the reflected intensity and polarization in the vicinity of the exact backscattering direction (Mishchenko et al. 2000b). Let us assume that a homogeneous semi-infinite slab is composed of nonabsorbing particles with sizes much smaller than the wavelength. The slab is illuminated by an unpolarized quasi-monochromatic beam of light incident perpendicularly to the slab boundary. Figure 1.21 depicts the corresponding unpolarized backscattering enhancement factor $\zeta_I(q)$ as a function of the dimensionless so-called angular parameter $q$ defined as $q = k_1 l\alpha$, where $\alpha$ is the phase angle (defined as the angle between the directions "scattering object → source of illumination" and "scattering object → detector") and $l$ is the mean free path of light in the scattering medium. The latter is given by

$$l = \frac{1}{n_0 \langle C_{\text{ext},1} \rangle_\xi}, \tag{1.125}$$

where, as before, $n_0$ is the particle number density and $\langle C_{\text{ext},1} \rangle_\xi$ is the extinction cross section per particle averaged over particle states. In other words, $l$ is the geometrical distance corresponding to a unit optical path-length. The dotted curve demonstrates the renowned coherent intensity peak centered at exactly the opposition. The amplitude of the peak is $\zeta_I(0) \approx 1.537$ and its half-width at half-maximum is $q_{\text{HWHM},I} = k_1 l \alpha_{\text{HWHM},I} \approx 0.597$ (cf. Fig. 1.20). Thus the relationship between the half-width at half-maximum of the backscattering intensity peak and the mean free path for the particular case of conservative Rayleigh scattering and unpolarized normal illumination is given by

$$\alpha_{\text{HWHM},I} \approx \frac{0.597}{k_1 l}. \tag{1.126}$$

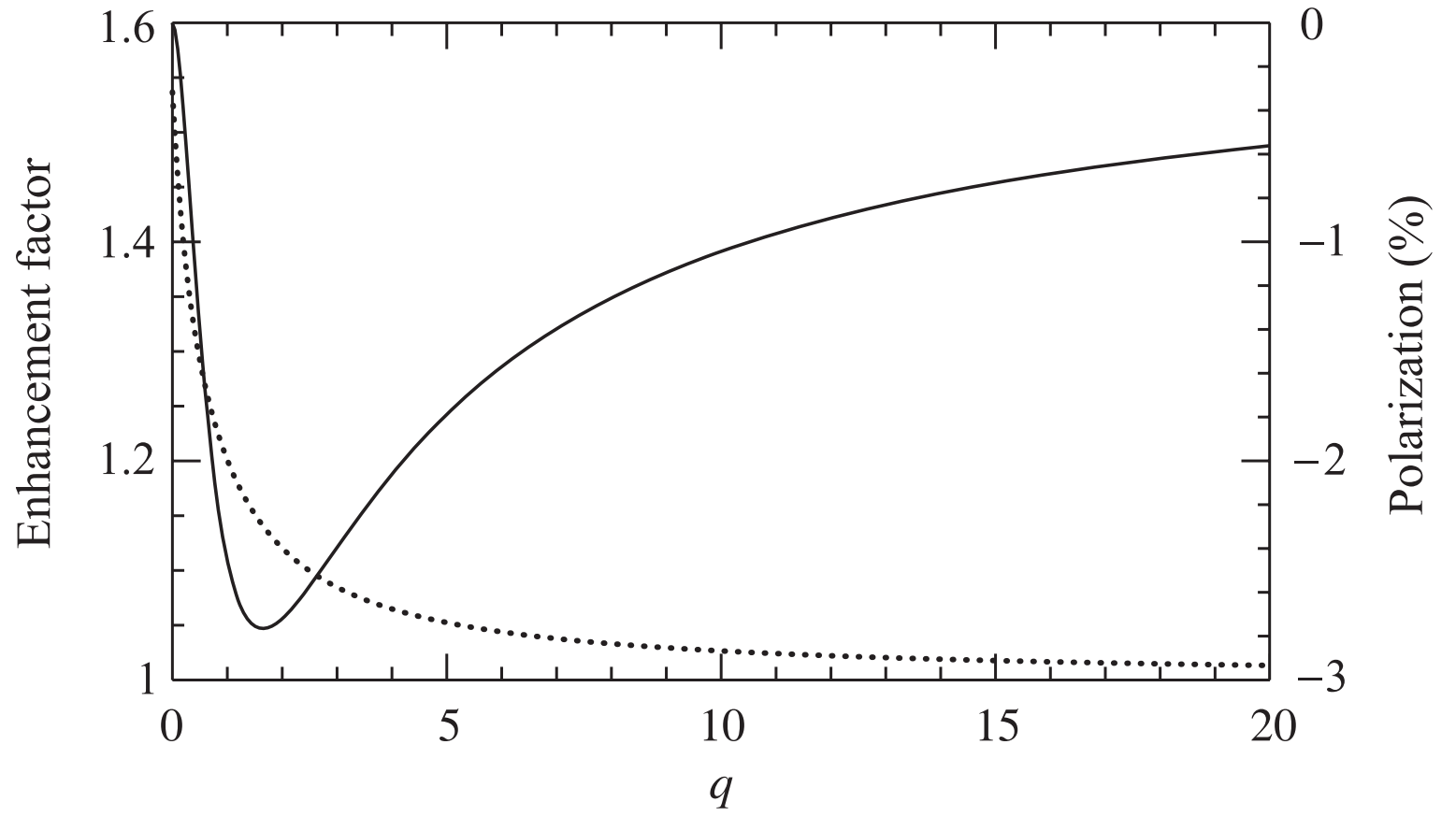

**Fig. 1.21.** Angular profiles of the unpolarized enhancement factor (dotted curve) and the degree of linear polarization for unpolarized incident light (solid curve).

The degree of linear polarization of the reflected light for unpolarized incident light is equal to minus the ratio of the second element of the total reflected specific intensity column vector to the total reflected specific intensity:

$$P(q) = -\frac{\widetilde{Q}(q)}{\widetilde{I}(q)} = -\frac{R_{21}^{1}(q) + R_{21}^{\text{M}}(q) + R_{21}^{\text{C}}(q)}{R_{11}^{1}(q) + R_{11}^{\text{M}}(q) + R_{11}^{\text{C}}(q)}. \tag{1.127}$$

Both $R_{21}^{1}(0)$, $R_{21}^{\text{M}}(0)$, and $R_{21}^{\text{C}}(0)$ vanish, the latter two quantities as a consequence of azimuthal symmetry in the case of normal illumination and unpolarized incident light. Furthermore, both $R_{11}^{1}(q)$, $R_{11}^{\text{M}}(q)$, $R_{21}^{1}(q)$, and $R_{21}^{\text{M}}(q)$ change with reflection direction much more slowly than $R_{11}^{\text{C}}(q)$ and $R_{21}^{\text{C}}(q)$ within the range of reflection directions affected by CB. Consequently,

$$P(q) \approx -\frac{R_{21}^{\text{C}}(q)}{R_{11}^{1}(0) + R_{11}^{\text{M}}(0) + R_{11}^{\text{C}}(q)}. \tag{1.128}$$

This quantity is shown in Fig. 1.21 by the solid curve. It is seen indeed that polarization is zero at the exact backscattering direction. However, with increasing $q$, polarization becomes negative, rapidly grows in absolute value, and reaches its minimal value $P_{\min} \approx -2.765\%$ at a reflection direction very close to opposition ($q_P \approx 1.68$). This depression is highly asymmetric so that the half-minimal value –1.383% is first reached at $q_{P,1} \approx 0.498$, which is even smaller than the value $q_{\text{HWHM},I} \approx 0.597$ corresponding to the half-width at half-maximum of the backscattering intensity peak, and then at a much larger $q_{P,2} \approx 7.10$. This unusual behavior of polarization at near-backscattering angles was called in Mishchenko (1993b) the polarization opposition effect (POE). This theoretical prediction is very important to astrophysics of atmosphereless Solar System bodies (ASSBs) and will be further discussed in Chapter 3. The specific optical mechanism causing the POE is explained qualitatively in Muinonen (1993), Shkuratov et al. (1994), and Muinonen et al. (2002).

### *1.25.3. Angular dependence of weak localization: wavelength-sized scatterers*

In the case of nonabsorbing Rayleigh scatterers considered above, it is assumed that the particle size is much smaller than the wavelength. As a consequence, the various manifestations of the WL effect depend on the particle size parameter, refractive index, and shape implicitly, via the formula (1.125) for the mean free path of light. The calculation of the angular distribution of the reflected intensity and polarization for a medium composed of arbitrary particles must be based on a more general approach not imposing restrictions on particle physical characteristics. The method of summation of cyclical diagrams described in Section 1.19 can be used to obtain formulas describing WL which are valid under the same conditions as the VRTE.

Below we consider the scattering by a plane-parallel layer consisting of randomly positioned and randomly oriented particles. As in the preceding subsection, the incident plane wave is assumed to propagate normally to the upper boundary of the medium. A detailed derivation of the equations describing the coherent and incoherent (diffuse) parts of the backscattered radiation was presented by Tishkovets (2002), Tishkovets and Mishchenko (2004), and Тишковец (2009). This derivation is based on the rigorous theory of electromagnetic scattering by particle groups. The formulas derived allow a detailed study of the dependence of the brightness (or photometric) opposition effect (BOE) and the POE on the microphysical properties of the scatterers (particle size parameter, real and imaginary parts of the relative refractive index, shape, number density, etc.).

The reflection matrix is decomposed according to Eq. (1.115). Then the nonzero components of the cyclical matrix $\mathbf{R}^{\mathrm{C}}$ in Eqs. (1.115) and (1.116) can be written as follows (e.g., Tishkovets and Mishchenko 2009):

$$R_{11}^{\mathrm{C}} = U\sum_{pn} S_{pnpn}^{\mathrm{C}}, \qquad R_{12}^{\mathrm{C}} = R_{21}^{\mathrm{C}} = -U\sum_{pn} S_{pn-pn}^{\mathrm{C}}, \tag{1.129}$$

$$R_{22}^{\mathrm{C}} = U\sum_{pn} S_{pn-p-n}^{\mathrm{C}}, \qquad R_{33}^{\mathrm{C}} = U\sum_{pn} S_{pn-p-n}^{\mathrm{C}}\,\mathrm{i}^{p-n}, \tag{1.130}$$

$$R_{44}^{\mathrm{C}} = U\sum_{pn} S_{pnpn}^{\mathrm{C}}\,\mathrm{i}^{p-n}, \qquad R_{34}^{\mathrm{C}} = -R_{43}^{\mathrm{C}} = -\mathrm{i}U\sum_{pn} S_{pnp-n}^{\mathrm{C}}\,\mathrm{i}^{p-n}. \tag{1.131}$$

Here, $U = (\pi n_0)^2/k_1^6$; $n, p = \pm 1$; and the matrix $\mathbf{S}^{\mathrm{C}}$ has the form (Tishkovets 2002; Tishkovets and Mishchenko 2004)

$$S_{pn\mu\nu}^{\mathrm{C}} = \sum_{qq_1LM} (-1)^L\,\zeta_{LM}^{*(q_1\mu)(qp)} \int_0^{k_1Z_0} \mathrm{d}z\,\exp(-\tilde{\varepsilon}z)\,\beta_{LM}^{(z)(qn)(q_1\nu)}, \tag{1.132}$$

where, as before, the asterisk denotes complex conjugation; $n, p, \mu, \nu, q, q_1 = \pm 1$; $Z_0$ is the geometrical thickness of the layer;

$$\tilde{\varepsilon} = \mathrm{Im}(m_{\mathrm{eff}})\left(1 - \frac{1}{\cos\Theta}\right) + \mathrm{i}\,(1+\cos\Theta)\left(\frac{\mathrm{Re}(m_{\mathrm{eff}})-1}{\cos\Theta} + 1\right); \tag{1.133}$$

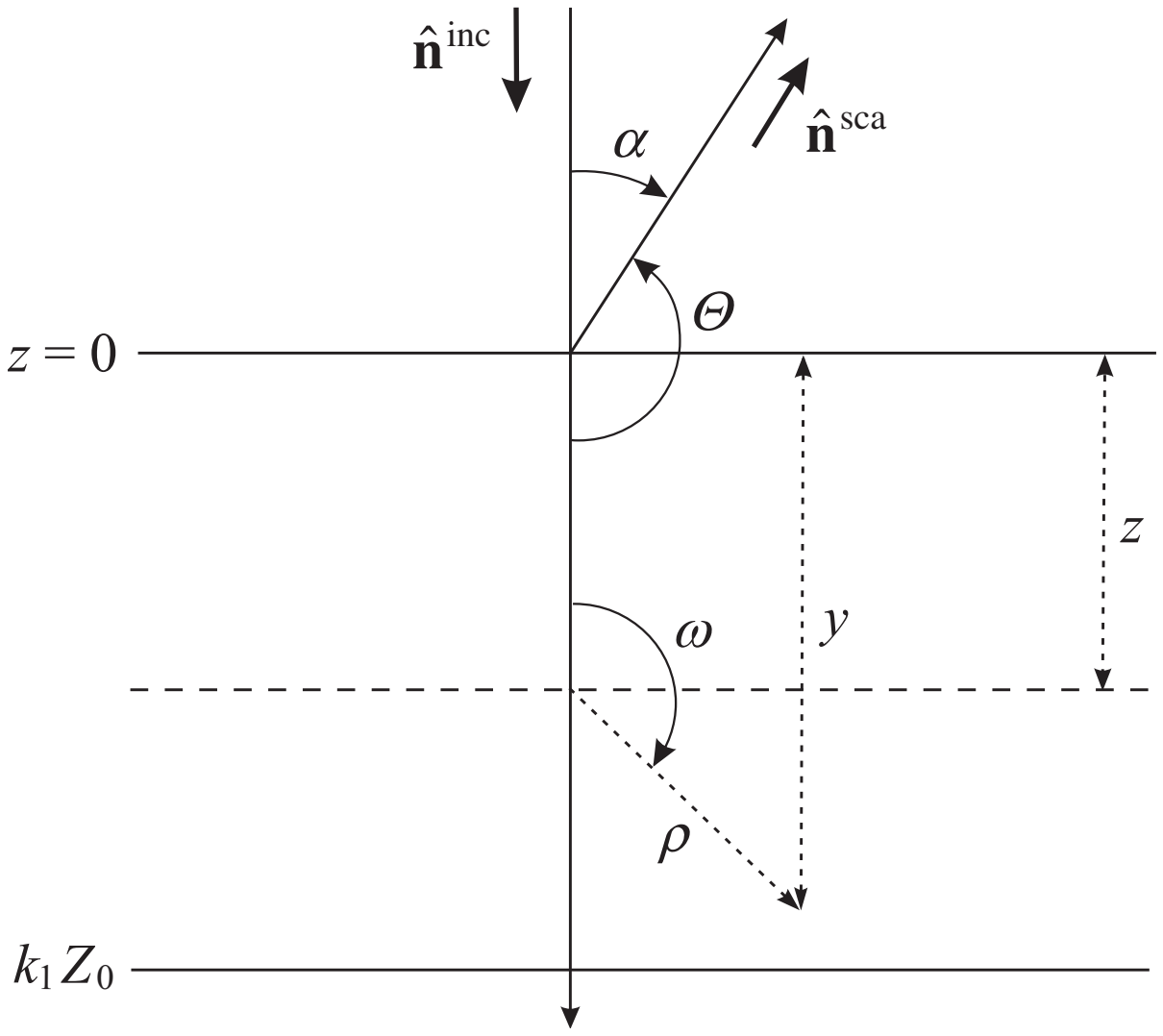

**Fig. 1.22.** Geometry of scattering by a particulate slab. The incident plane wave propagates normally to the upper boundary of the medium ($z = 0$).

and $m_{\rm eff}$ is the complex effective refractive index of the medium (see Fig. 1.22). The coefficients $\beta_{LM}^{(z)(pn)(\mu\nu)}$ are determined from the following system of equations:

$$\beta_{LM}^{(z)(pn)(\mu\nu)} = \exp(-\widetilde{\varepsilon}^* z)\, B_{LM}^{(z)(pn)(\mu\nu)} + \frac{2\pi n_0}{k_1^3} \sum_{qq_1 l_1 m_1} \chi_{l_1}^{(pq)(\mu q_1)}\, \mathrm{i}^{M-m_1}$$

$$\times \int \mathrm{d}\omega \mathrm{d}\rho\, \beta_{l_1 m_1}^{(y)(qn)(q_1\nu)} \exp(-\tau\rho)\, d_{MN_0}^{L}(\omega) d_{m_1 N_0}^{l_1}(\omega) J_{m_1 - M}(\rho \sin\Theta \sin\omega) \sin\omega, \tag{1.134}$$

where $\tau = 2\,\mathrm{Im}(m_{\rm eff})$, $N_0 = \mu - p$, $y = z - \rho\cos\omega$, $d_{MN}^{L}(\omega)$ are Wigner $d$ functions (Varshalovich et al. 1988), and $J_m(x)$ are Bessel functions. The angle $\omega$ (see Fig. 1.22) is measured from the exact backscattering direction given by $-\hat{\mathbf{n}}^{\rm inc}$. Furthermore,

$$B_{LM}^{(z)(pn)(\mu\nu)} = \sum_{l_1 m_1} \zeta_{l_1 m_1}^{(pn)(\mu\nu)}\, \mathrm{i}^{M-m_1}$$

$$\times \int \mathrm{d}\omega\, \mathrm{d}\rho \exp(-\tau_1\rho) d_{MN_0}^{L}(\omega) d_{m_1 N_0}^{l_1}(\omega) J_{m_1 - M}(\rho \sin\Theta \sin\omega) \sin\omega, \tag{1.135}$$

where $\tau_1 = 2\,\mathrm{Im}(m_{\rm eff}) - \widetilde{\varepsilon}^* \cos\omega$. The coefficients $\chi_L^{(pn)(\mu\nu)}$ and $\zeta_{LM}^{(pn)(\mu\nu)}$ are expressed in the amplitude scattering matrices of the particles forming the medium. In particular, for spherical particles they are given by (Tishkovets 2002; Tishkovets and Mishchenko 2004)

$$\chi_{L_1}^{(pn)(\mu\nu)} = \sum_{Ll} \frac{(2L+1)(2l+1)}{4} \langle a_L^{(pn)} a_l^{*(\mu\nu)} \rangle_\xi \, C_{L-n\,l\nu}^{L_1M_0} \, C_{L-p\,l\mu}^{L_1N_0}, \tag{1.136}$$

$$\zeta_{L_1M_1}^{(pn)(\mu\nu)} = \sum_{Llm_1} \frac{(2L+1)(2l+1)}{4} \langle a_L^{(pn)} a_l^{*(\mu\nu)} \rangle_\xi \, (-1)^{l+m_1} \, C_{L-n\,l-m_1}^{L_1M_1} \, C_{L-p\,l\mu}^{L_1N_0} \, d_{m_1\nu}^{l}(\Theta). \tag{1.137}$$

Here, $a_L^{(pn)(\mu\nu)} = a_L + pnb_L$, $a_L$ and $b_L$ are the standard Lorenz–Mie coefficients (see, e.g., Mishchenko et al. 2002a), the angular brackets denote averaging over particle microphysical properties (states), $M_0 = \nu - n$, and the $C$s are Clebsch–Gordan coefficients (Varshalovich et al. 1988). The integrals in Eqs. (1.134) and (1.135) are defined as

$$\int = \int_0^{\pi/2} \mathrm{d}\omega \sin\omega \int_0^{z/\cos\omega} \mathrm{d}\rho + \int_{\pi/2}^{\pi} \mathrm{d}\omega \sin\omega \int_0^{(z-k_1Z_0)/\cos\omega} \mathrm{d}\rho, \tag{1.138}$$

where we use the notation explained in Fig. 1.22.

The formulas given above describe CB for a homogeneous and macroscopically isotropic particulate slab composed of arbitrary scatterers. The meaning of these equations is as follows. The coefficients given by Eq. (1.135) characterize the interference of double-scattered waves. They determine the solution of the system of equations (1.134) which yields the coefficients $\beta_{LM}^{(z)(pn)(\mu\nu)}$ as functions of $z$. If one keeps in Eq. (1.134) only the first term then the matrix of Eq. (1.132) describes CB in the double-scattering approximation (Tishkovets et al. 2002a, Tishkovets and Mishchenko 2004). The results of this approximation are useful for semi-quantitative or qualitative analyses of the effect of particle microphysical properties on various characteristics of WL. The particular case of a semi-infinite particulate slab was studied in Tishkovets et al. (2002a), Tishkovets and Mishchenko (2004), and Litvinov et al. (2007).

The discussion of WL below pertains to the case of particulate slabs consisting of spherical particles or small clusters of spherical particles. Figure 1.23 depicts the degree of linear polarization $P$ and the normalized specific intensity $\tilde{I}(\Theta)/\tilde{I}(\pi)$ for radiation exiting a semi-infinite layer composed of identical spherical particles under the assumption that the incident beam is unpolarized. The scattering angle $\Theta$ is given by $180° - \alpha$, where, as before, $\alpha$ is the phase angle (Fig. 1.22). The results are shown for different values of the filling factor $\tilde{\xi} = 4\pi r^3 n_0/3$, where $r$ is the particle radius. The effective refractive index of the medium is given by $m_{\mathrm{eff}} = 1 + \mathrm{i}n_0 C_{\mathrm{ext}}/2k_1$, where $C_{\mathrm{ext}}$ is the extinction cross section per particle. The degree of linear polarization of light singly scattered by particles forming the medium, $P_0$, is shown in the bottom diagrams of Fig. 1.23.

One can see that the POE is more sensitive to particle properties than the BOE (i.e., a sharp peak of intensity centered at the exact backscattering direction). This sensitivity of polarization can be explained as follows. It has been demonstrated that in the case of positive single-scattering polarization for unpolarized incident light, the interference of double-scattered waves results in negative polarization at backscattering angles. Such a behavior of polarization of double-scattered light has been predicted by Шкуратов (1991) and Muinonen (1990). However, uniformly positive single-scattering polarization is a specific property of Rayleigh particles with sizes much smaller than the wavelength. The angular dependence of the single-scattering polarization $P_0$ for wavelength-sized particles becomes significantly more complex (see Plate 1.2a). Therefore, the interference of waves scattered twice can result in positive as well as in negative backscattering polarization.

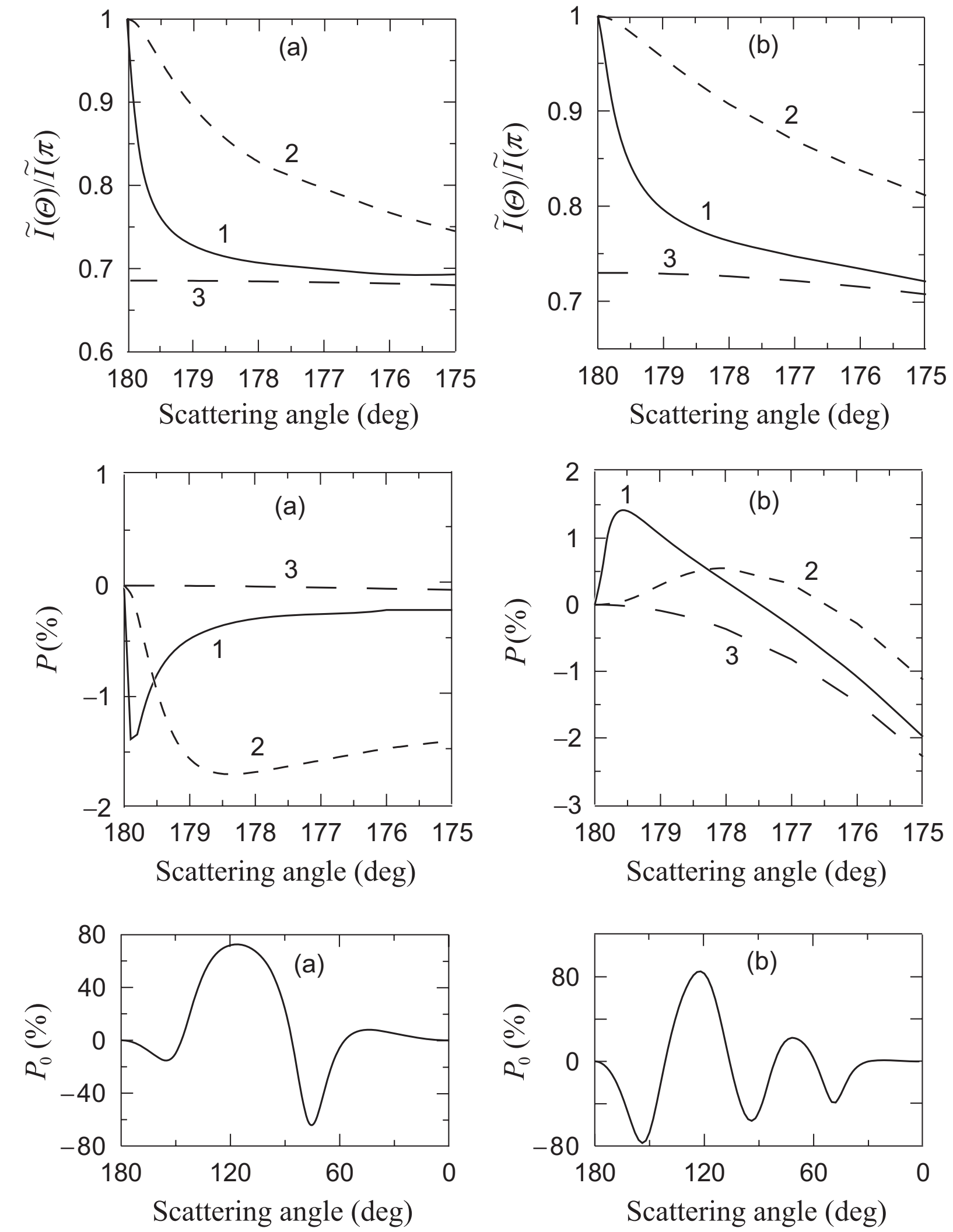

**Fig. 1.23.** Reflection of light by a semi-infinite slab composed of monodisperse spherical particles with $\{k_1r = 3, m = 1.35\}$ (left-hand column) and $\{k_1r = 4.5, m = 1.33\}$ (right-hand column). Curves 1 and 2 show the sum of the diffuse and cyclical components; curves 3 show the diffuse component. Curves 1 correspond to $\tilde{\xi} = 0.001$, curves 2 correspond to $\tilde{\xi} = 0.01$. $P_0$ denotes the single-scattering polarization.

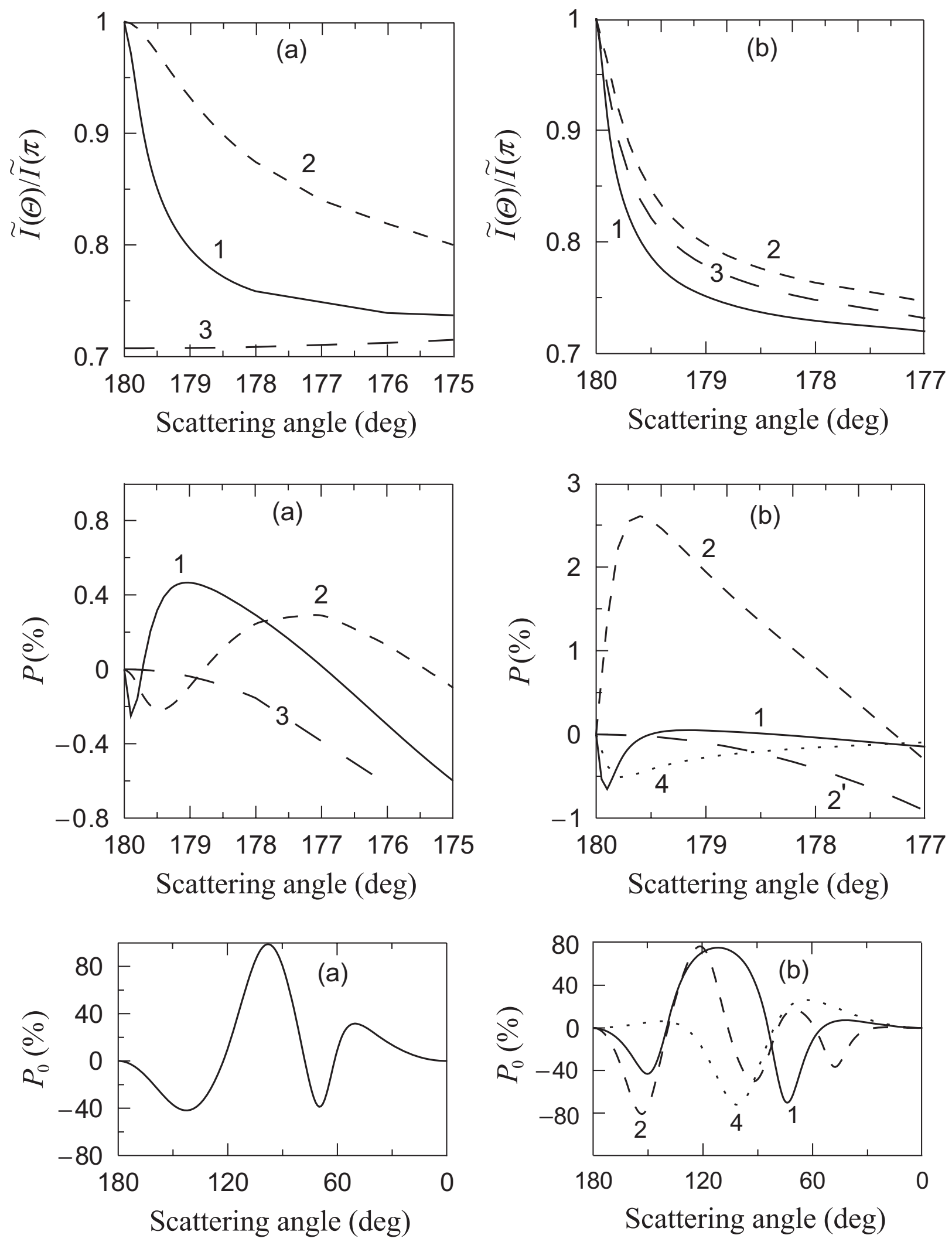

**Fig. 1.24.** Reflection of light by a semi-infinite slab composed of monodisperse spherical particles. The left-hand column corresponds to $\{k_1 r = 3,\ m = 1.5 + i0.5\}$. Curves 1 and 2 show the sum of the diffuse and cyclical components for $\tilde{\xi} = 0.001$ and $\tilde{\xi} = 0.01$, respectively, while curves 3 show the diffuse component. The right-hand column demonstrates the effects of WL on the particle properties for $\tilde{\xi} = 0.001$. Curves 1 correspond to $\{k_1 r = 3.075,\ m = 1.35\}$, curves 2 correspond to $\{k_1 r = 4.5,\ m = 1.35\}$, curves 3 correspond to $\{k_1 r = 4.5,\ m = 1.59\}$, and curves 4 correspond to $\{k_1 r = 2.5,\ m = 1.35\}$. Curve 2′ depicts the polarization of the incoherent component corresponding to the parameters of curves 2. The values of this incoherent polarization for the other particles are close to 0 (not shown). $P_0$ is the single-scattering polarization.

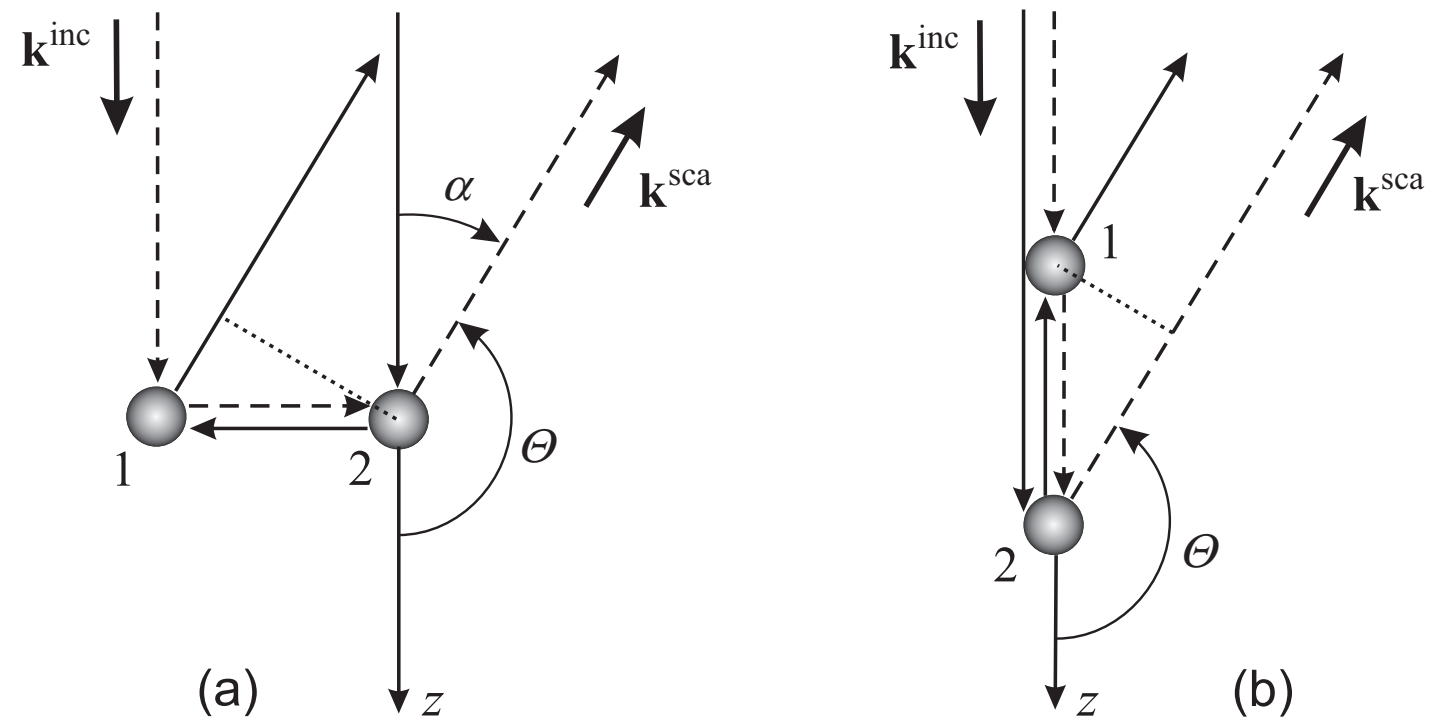

**Fig. 1.25.** On the dependence of the width of the CB intensity peak on the particle size. The wave vector $\mathbf{k}^{\text{inc}}$ specifies the propagation direction of the incident wave, the wave vector $\mathbf{k}^{\text{sca}}$ indicates the propagation direction of the scattered wave.

The left-hand column of Fig. 1.23 corresponds to spherical particles with the size parameter $k_1 r = 3$ and relative refractive index $m = 1.35$. The phase function for these particles is strongly asymmetric, most of radiation being scattered at angles $\Theta < 60°$. Therefore, the polarization of the coherent part of the reflected radiation is determined by the behavior of $P_0$ in the scattering-angle ranges $\Theta < 60°$ and $\Theta > 120°$, where the average value of $P_0$ is positive (see the bottom panel of the left-hand column of Fig. 1.23). Hence, the interference of second-order scattering paths yields negative polarization near the backscattering direction.

The right-hand column of Fig. 1.23 corresponds to particles with the size parameter $k_1 r = 4.5$ and relative refractive index $m = 1.33$. The phase function for these particles is strongly asymmetric as well, but the average value of $P_0$ in the scattering-angle ranges $\Theta < 60°$ and $\Theta > 120°$ is predominantly negative. In this case the interference of the second-order-scattering paths results in a positive peak of polarization near the backscattering direction.

As with the case of Rayleigh scatterers (see Eqs. (1.125) and (1.126)), the half-width at half-maximum of the interference intensity peak is a linear function of the particle packing density $\tilde{\xi}$ or $2\,\text{Im}(m_{\text{eff}})$, as discussed in a number of studies (e.g., Barabanenkov et al. 1991 and references therein). The degree of linear polarization of the cyclical component displays a similar behavior, which is especially obvious when the polarization of the diffuse light tends to zero at opposition. In this case the angle between the direction $\Theta = \pi$ and that of the polarization extreme is a linear function of $\tilde{\xi}$ (see the left-hand columns of Fig. 1.23 and 1.24). Otherwise, the angular dependence of the element $R_{12}^{\text{diff}}$ is superimposed on the angular dependence of $R_{12}^{\text{C}}$, and the linearity may not be observed.

The width of the interference intensity peak depends not only on the number density of wavelength-sized particles, but also on their physical properties. For example, the sizes and refractive indices of the particles corresponding to curves 1–3 in Fig. 1.24 have been selected in such a way that the values of $2\,\text{Im}(m_{\text{eff}})$ are ap-proximately the same, yet the widths of the backscattering intensity peaks are different. This effect can be explained as follows (Tishkovets et al. 2002b). Let us calculate the difference between the phases of two waves double-scattered by the pairs of particles shown in Figs. 1.25a and 1.25b. In both cases one wave propagates along the trajectory "source → particle 1 → particle 2 → detector" (dashed line), while the second wave propagates in the reverse direction: "source → particle 2 → particle 1 → detector" (solid line). It is assumed that the distance between the particle centers is the same in both cases. It is easily seen that the phase difference $\Delta$ in Fig. 1.25a is proportional to $\sin\alpha \sin\varphi$, where $\varphi$ is the azimuth angle of the line connecting the particle centers. Then $\alpha \to 0$ yields $\Delta \approx \alpha$. In Fig. 1.25b, $\Delta$ is proportional to $1 - \cos\alpha$, and so $\alpha \to 0$ yields $\Delta \approx \alpha^2/2$. Therefore, the interference of the waves in Fig. 1.25b leads to a wider interference peak than that in Fig. 1.25a. This simple geometrical interpretation allows one to derive the following qualitative conclusions valid under the assumption of a fixed $\mathrm{Im}(m_{\mathrm{eff}})$:

- An increase in the particle size leads to an increase in the half-width of the CB intensity peak because of the strengthened single-particle scattering in the forward direction.
- Increasing the imaginary part of the relative refractive index of wavelength-sized particles also leads to an increase in the half-width of the CB peak because of the reduced single-particle side scattering.
- An increase in the real part of the particle refractive index leads to a reduced half-width of the CB peak because of the strengthened single-particle side scattering.
- The half-width of the interference peak becomes minimal for Rayleigh particles scattering radiation almost isotropically.

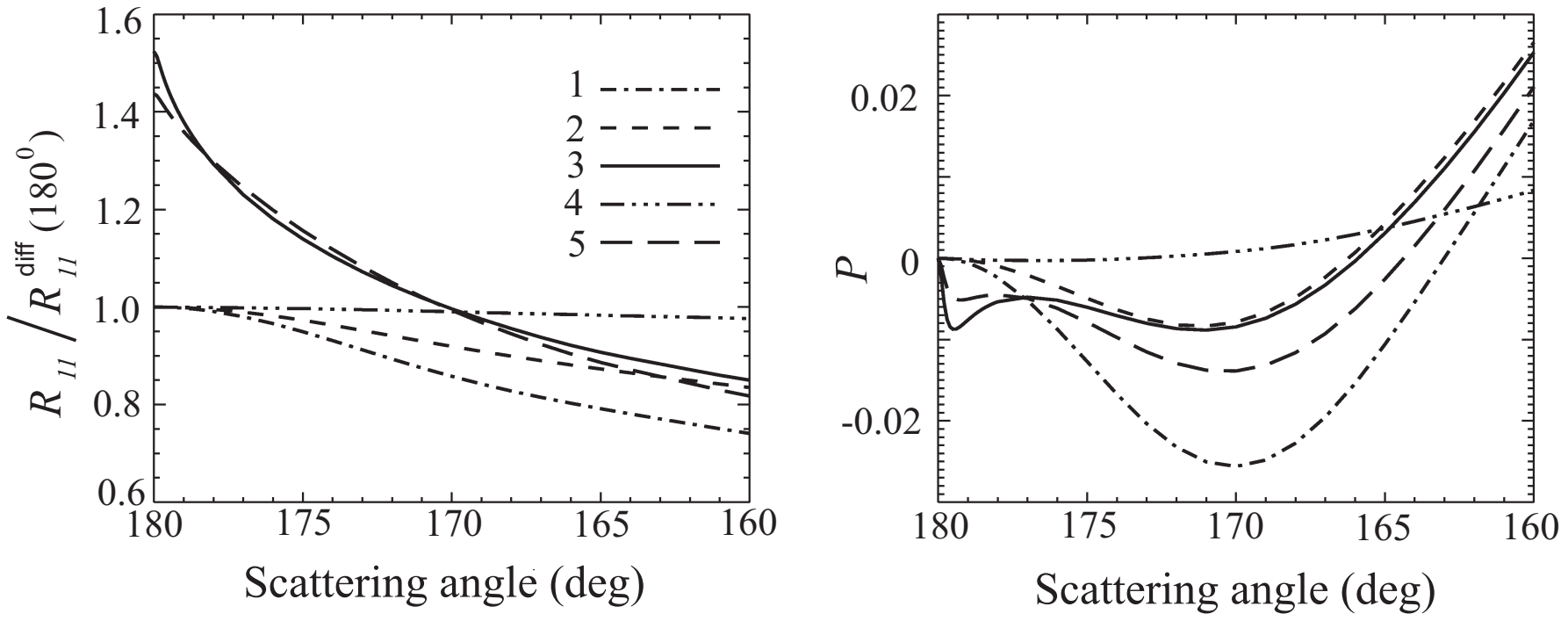

**Fig. 1.26.** The intensity and degree of linear polarization of radiation scattered by a semi-infinite medium composed of clusters of spherical particles.

Figure 1.26 shows the results of calculations of the normalized intensity, $R_{11}(\Theta)/R_{11}^{\mathrm{diff}}(\pi)$, and the degree of linear polarization $P$ of the reflected radiation for a semi-infinite layer composed of randomly oriented fractal-like clusters consisting of 334 identical spherical particles (see Plate 1.7h). The filling factor of the medium is $\tilde{\xi} = 4\pi R^3 n_0/3 = 0.026$, where $R$ is the radius of the smallest circumscribing sphere of the cluster. The size parameter of the constituent monomers is $k_1 r = 1.2$, while their relative refractive index is $m = 1.5 + \mathrm{i}0.001$. The clusters were created using the numerical procedure described by Mackowski (1995). The fractal dimension of the cluster is 2.5 and its prefactor is 8.

Curves 1 in Fig. 1.26 correspond to single scattering by an individual cluster. Curves 2 describe the angular dependence of the incoherent component $\mathbf{R}^{\mathrm{diff}}$ in the third-order-scattering approximation wherein the single scattering by an individual cluster as well as the second and third orders of scattering between different clusters are taken into account. Curves 3 show the cyclical component $\mathbf{R}^{\mathrm{C}}$ in the third-order-scattering approximation. To further illustrate the effects of different orders of scattering, curves 4 show the diffuse matrix $\mathbf{R}^{\mathrm{diff}}$ obtained by solving the VRTE (and thereby including all orders of scattering), while curves 5 show the matrix $\mathbf{R}^{\mathrm{diff}} + \mathbf{R}^{\mathrm{C}}$ obtained by keeping the first two orders of scattering.

The BOE and POE calculated using the third-order-scattering approximation manifest themselves mainly in the range of scattering angles 175°−180° (compare curves 2 and 3). The interference of scattered waves leads to a sharp depression in polarization because the single-scattering polarization for the individual clusters is predominantly positive in the entire range of scattering angles $\Theta \in [0°, 180°]$. Since CB is caused by multiple scattering between different clusters, the BOE and POE become more pronounced with including more orders of scattering (compare curves 3 and 5).

The results shown in Fig. 1.26 can be important for the interpretation of observations of certain ASSBs demonstrating a bimodal dependence of the degree of linear polarization on the phase angle (see Chapter 3). Indeed, the polarization computed for a slab of randomly oriented clusters in Fig. 1.26 has two minima. One of them (a wide branch of negative polarization) is caused by single scattering on individual clusters, while the second one (a narrow branch of negative polarization at very small phase angles) is caused by the interference of multiply scattered waves. As seen from Fig. 1.26, upon including more orders of scattering the wide branch of negative polarization degrades, whereas the narrow branch amplifies. This behavior of polarization suggests the existence of situations in which both branches of polarization could exist. For example, this could be possible for media composed of particles exhibiting a deep branch of negative single-scattering polarization. It is thus reasonable to hypothesize that the bimodal phase-angle dependences of negative polarization observed for certain Solar System objects could arise from the interference of multiply scattered waves coupled with a pronounced enough branch of negative polarization exhibited by the individual regolith grains. Of course, this explanation does not necessarily exclude other optical effects potentially causing bimodal branches of negative polarization.

## 1.26. Near-field effects

In the preceding sections we have mostly discussed the scattering of light by sparse particulate media, in which case a partial wavelet scattered by a particle becomes spherical upon its arrival at any other particle (Section 1.9). Furthermore, the radius of curvature of this wavelet becomes so large that the wavelet can be considered as a locally homogeneous plane wave. As a consequence, the VRTE (1.102) and the formulas (1.129)–(1.132) describing the cyclical component of the reflection matrix involve such single-scattering quantities as the phase, extinction, and amplitude scattering matrices and the optical cross sections. In the case of densely packed media, the far-field assumption can be violated, and certain near field and wave inhomogeneity effects can be observed. It is, therefore, important to analyze the conditions under which these effects can become pronounced as well as to study how they can influence the interpretation of remote-sensing observations of Solar System objects.

To clarify the role of the near-field effects, it is necessary to consider the structure of the scattered wavelet in the close vicinity of a scatterer and the peculiarities of the scattering of this wavelet by a neighboring particle. The total electric field at a point close to the scatterer can be represented as a vector sum (see Eq. (1.5))

$$\mathbf{E} = \mathbf{E}^{\mathrm{inc}} + \mathbf{E}^{\mathrm{sca}}, \tag{1.139}$$

where $\mathbf{E}^{\mathrm{inc}}$ is the electric vector of the incident wave and $\mathbf{E}^{\mathrm{sca}}$ is the electric vector of the scattered field. In the far-field zone of the particle the scattered field becomes an outgoing spherical wave. However, in the immediate vicinity of the particle the field described by Eq. (1.139) is strongly inhomogeneous (Tishkovets 1998; Тишковец и Литвинов 1999).

Let us assume that the incident wave propagates along the $z$-axis and is polarized linearly in the $xz$-plane (see Fig. 1.27). The spatial distribution of the near field can be calculated using the well-known formulas of the Lorenz–Mie theory (e.g., Mishchenko et al. 2002a) assuming that the particle is spherical. Figure 1.27 shows the section of the three-dimensional surfaces $\max(\mathrm{Re}|\mathbf{E}|)$ by the $xz$-plane as well as the local directions of the electric field vector for a particle with $k_1 r = 4$ and $m = 1.32 + i0.05$ (Tishkovets 1998; Tishkovets et al. 1999, 2004). For a homogeneous field the surfaces $\max(\mathrm{Re}|\mathbf{E}|)$ represent the surfaces of constant phase. For the specified polarization of the incident wave and for an observation point in the *xz*-plane, the electric field vector (1.139) lies in this plane and has nonzero components $E_\theta$ and $E_r$, while $E_\varphi = 0$. For an observation point in the *yz*-plane, the electric field vector is parallel to the vector $\mathbf{E}^{\mathrm{inc}}$ and has a nonzero component $E_\varphi$, while $E_\theta = E_r = 0$.

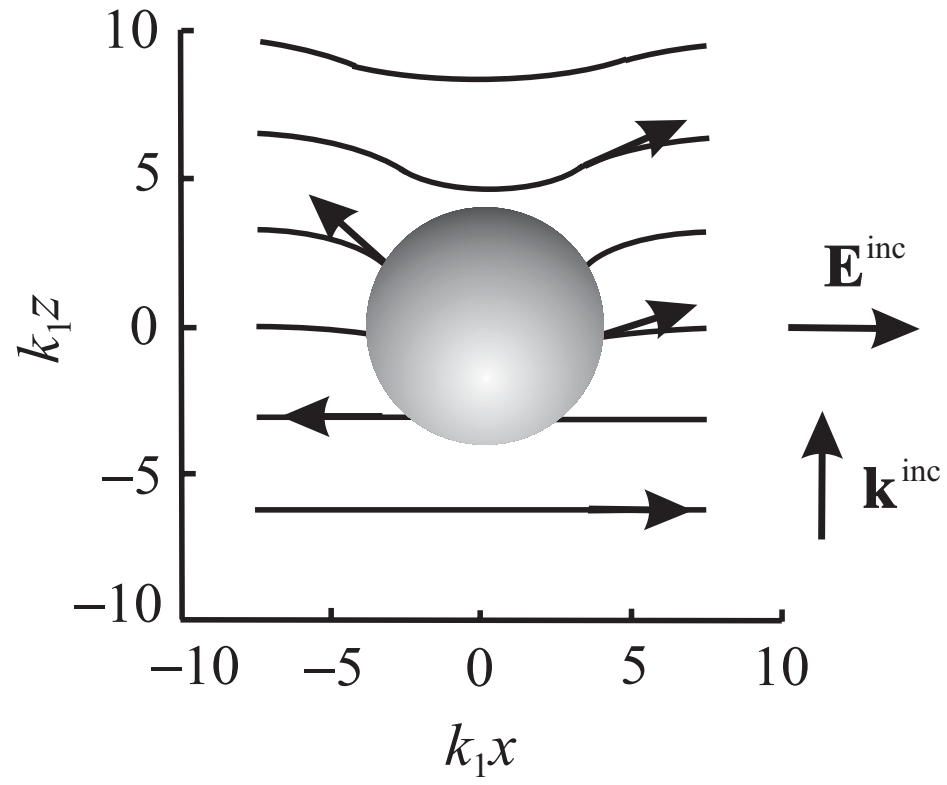

**Fig. 1.27.** Structure of the electric field in the close vicinity of a spherical particle. The incident wave propagates along the $z$-axis (as indicated by the wave vector $\mathbf{k}^{\mathrm{inc}}$) and is polarized along the $x$-axis. The particle size parameter is $k_1 r = 4$ and the relative refractive index is $m = 1.32 + \mathrm{i}0.05$.

Let us summarize the main features of the field (1.139) in the close vicinity of a particle:

1. The effect of interference of the light singly scattered by the particle and the light scattered first by a neighboring particle and then by the first particle is distinctly nonzero. It should be emphasized that under the assumption of sphericity of the secondary wavelets propagating between the scatterers the effect of this interference would be zero since the propagation directions of the incident and scattered waves do not coincide.

2. The rotation of the electric field vector given by Eq. (1.139) relative to the field vector of the incident wave causes a non-zero component $E_z$. Since this component always lies in the scattering plane, the light scattered by the next particle experiences an increase in the intensity component polarized in the scattering plane. Ultimately this can lead to the negative branch of linear polarization observed for some atmosphereless celestial bodies and cometary dust.

3. The dimensions of the volume inside which the fields of the incident and scattered waves strongly interfere are of the order of the wavelength. This implies that the near-field interference effects are essential for wavelength-sized and smaller particles and are insignificant for particles with sizes much greater than the wavelength.

Under such conditions, neighboring particles experience the influence of the inhomogeneous field and, consequently, scatter light in a different way than predicted by the theory involving only plane waves. To examine this problem in more detail, let us consider Rayleigh test scatterers using the coordinate system shown in Fig. 1.28 (Petrova et al. 2007; Петрова и др. 2009). Let us first analyze the case when the incident field $\mathbf{E}^{\mathrm{inc}}$ is polarized in the *xz*-plane serving here as the scattering plane. If the test particles are far from each other and from any other particle, they are subjected to a homogeneous electromagnetic field (Fig. 1.28a). However, if the test particles surround a wavelength-sized particle then the dipole moment **p** induced in test particles 1 and 3 has a non-zero component in the *z*-direction (Fig. 1.28b). In another case, when the incident wave $\mathbf{E}^{\mathrm{inc}}$ is polarized perpendicularly to the scattering plane, the roles of test particles 1 and 3 as well as 2 and 4 are interchanged, the *z*-component of the dipole moment again becomes non-zero, while the *y*-component decreases relative to the case of a homogeneous field. Consequently, for an unpolarized incident beam the appearance of a *z*-component lying in the scattering plane increases the negative polarization of the scattered radiation and diminishes the polarization in the side direction. This situation can be illustrated by simple formulas if we assume that the moment **p** of dipole 3 in Fig. 1.28b is oriented exactly along the positive *z*-axis, the moment **p** of dipole 1 is oriented in the opposite direction, and the absolute values of the moments are the same for all the dipoles. The total intensity $I$ and linear polarization $P$ of the scattered light for an unpolarized incident beam in the case of a homogeneous field (Fig. 1.28a) are described by the formulas

$$I = 4p^2(1+\cos\alpha), \tag{1.140}$$

$$P = \frac{1-\cos\alpha}{1+\cos\alpha}, \tag{1.141}$$

whereas in the case of an inhomogeneous field (Fig. 1.28b) they are described by the formulas

$$I = 2p^2(2+\sin^2\alpha), \tag{1.142}$$

$$P = -\frac{\sin^2\alpha}{2+\sin^2\alpha}. \tag{1.143}$$

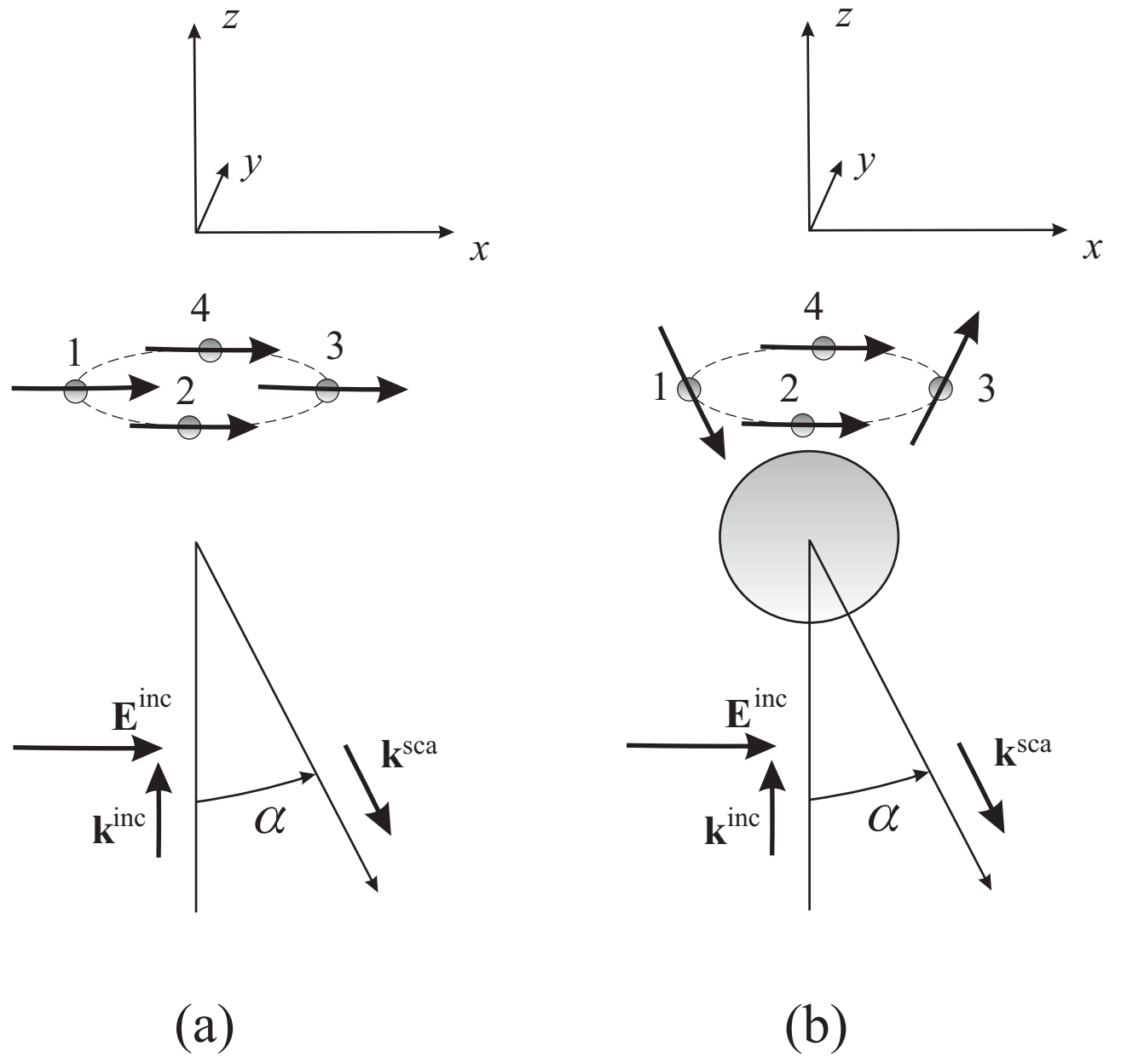

**Fig. 1.28.** Light scattering by a system of dipoles embedded in a homogeneous (a) and inhomogeneous (b) field. The incident wave propagates along the positive *z*-axis and is polarized in the *xz*-plane. In the homogeneous wave, all dipole moments point in the *x*-direction. In the inhomogeneous wave (see also Fig. 1.27), the distortion of the wave front causes a non-zero *z*-component of the moments of dipoles 1 and 3.

Figure 1.29 shows plots of intensity and polarization for the case of a homogeneous (left) and inhomogeneous (right) wave. It is seen that the rotation of the field vector in the particle vicinity reduces the scattered intensity in the back- and forward-scattering directions and induces negative polarization.

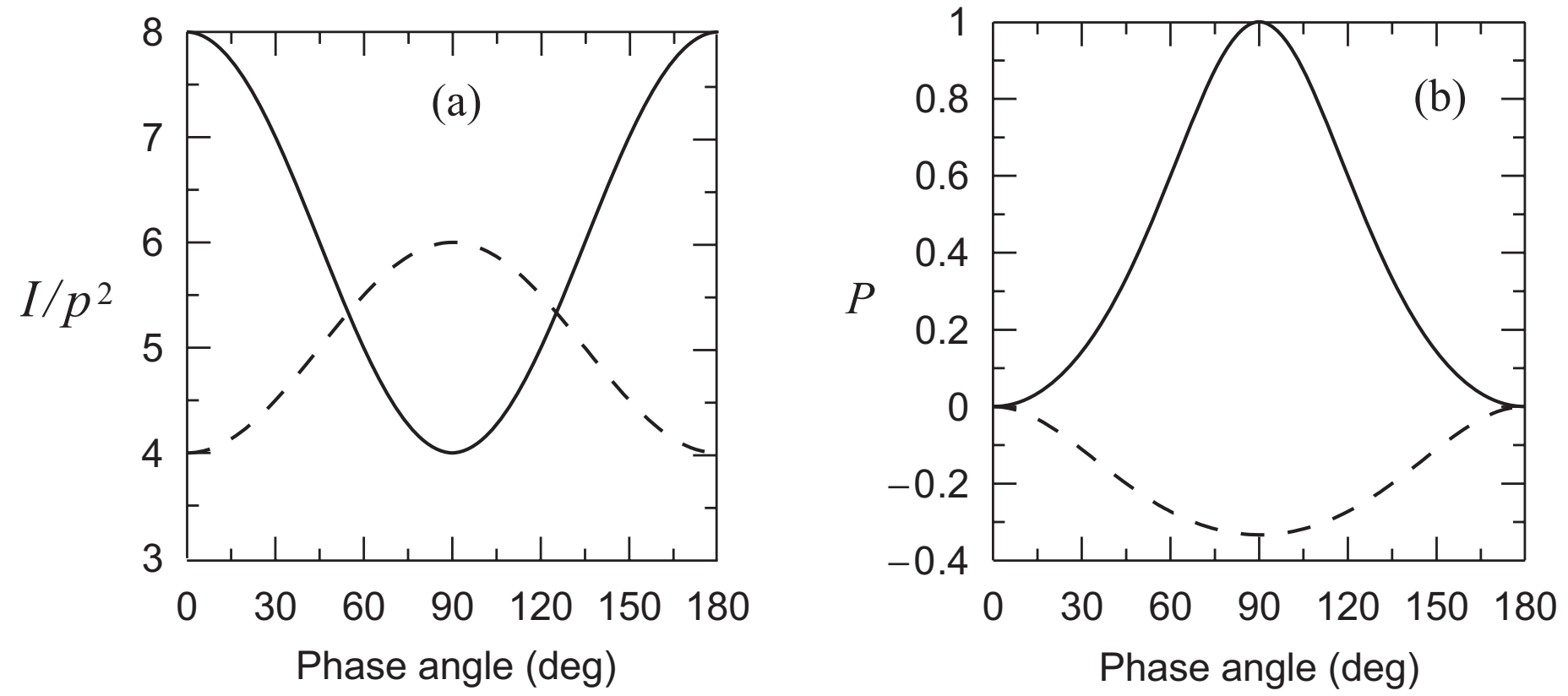

**Fig. 1.29.** Phase curves of intensity $I$ and polarization $P$ of light scattered by dipoles illuminated by unpolarized incident light.

Although the qualitative arguments given above are based on the assumption that the moments **p** of dipoles 3 and 1 in Fig. 1.28b are oriented along the positive $z$-axis and in the opposite direction, respectively, this line of thought can be easily generalized to the case of the orientation of the moments shown in Fig. 1.28b. The corresponding formulas become more complicated, but the scattering features discussed above remain the same. In particular, accounting for the shielding of dipole 1 by the large particle (Тишковец 2008; Tishkovets 2008) will yield the direction of the polarization minimum nearly perpendicular to the moment **p** of dipole 3. This can be seen indeed from the angular dependence of polarization for the light scattered by the cluster shown in Fig. 1.30 (Petrova et al. 2008). The parameters of the largest cluster component are the same as those of the scatterer in Fig. 1.27, i.e., $k_1r = 4$ and $m = 1.32 + \mathrm{i}0.05$. The eight smaller particles surrounding the large one have the following characteristics: $k_1r = 1.5$ and $m = 1.5 + \mathrm{i}0.1$. The coordinates of the small particles in the spherical system centered at the large particle (see the inset in Fig. 1.30) are $k_1R_i$ = 5.5, 6.5, 5.8, 6.7, 6.2, 5.9, 7.1, 6.8; $\theta_i$ = 75°, 70°, 65°, 60°, 55°, 65°, 75°, and 55°; and $\varphi_i$ = 0, 45°, 90°, 135°, 180°, 225°, 270°, and 315°. The scattering characteristics of the entire cluster are averaged over a range of equally probable configurations wherein the zenith angles $\theta_i$ are varied simultaneously within the ±5° range and the entire cluster is rotated around the $z$-axis.

Since the structure of this cluster is similar to that of the cluster in Fig. 1.28b, the corresponding results of calculations of the scattered polarization can be regarded as a quantitative corroboration of the near-field effects discussed above qualitatively. The dotted curve in Fig. 1.30 describes the result of single scattering combined with the interference of waves scattered once. The dashed curve represents the addition of the contribution of the doubly scattered light consisting of the incoherently scattered light and the interference of waves scattered twice. Thus this approximation takes into account the diffuse component and the WL effect in the double scattering approximation. One can see that this curve reveals a shallow negative polarization branch. The dotted-dashed curve represents the further incorporation of the interference of singly and doubly scattered waves possible only in dense media, i.e., the near-field effects. Now the absolute value of negative polarization exceeds 20%. As we have noted above, the interference of singly and doubly scattered waves arriving from different particles contributes significantly to the increasingly negative polarization. The solid line is the result of taking into account all scattering orders.

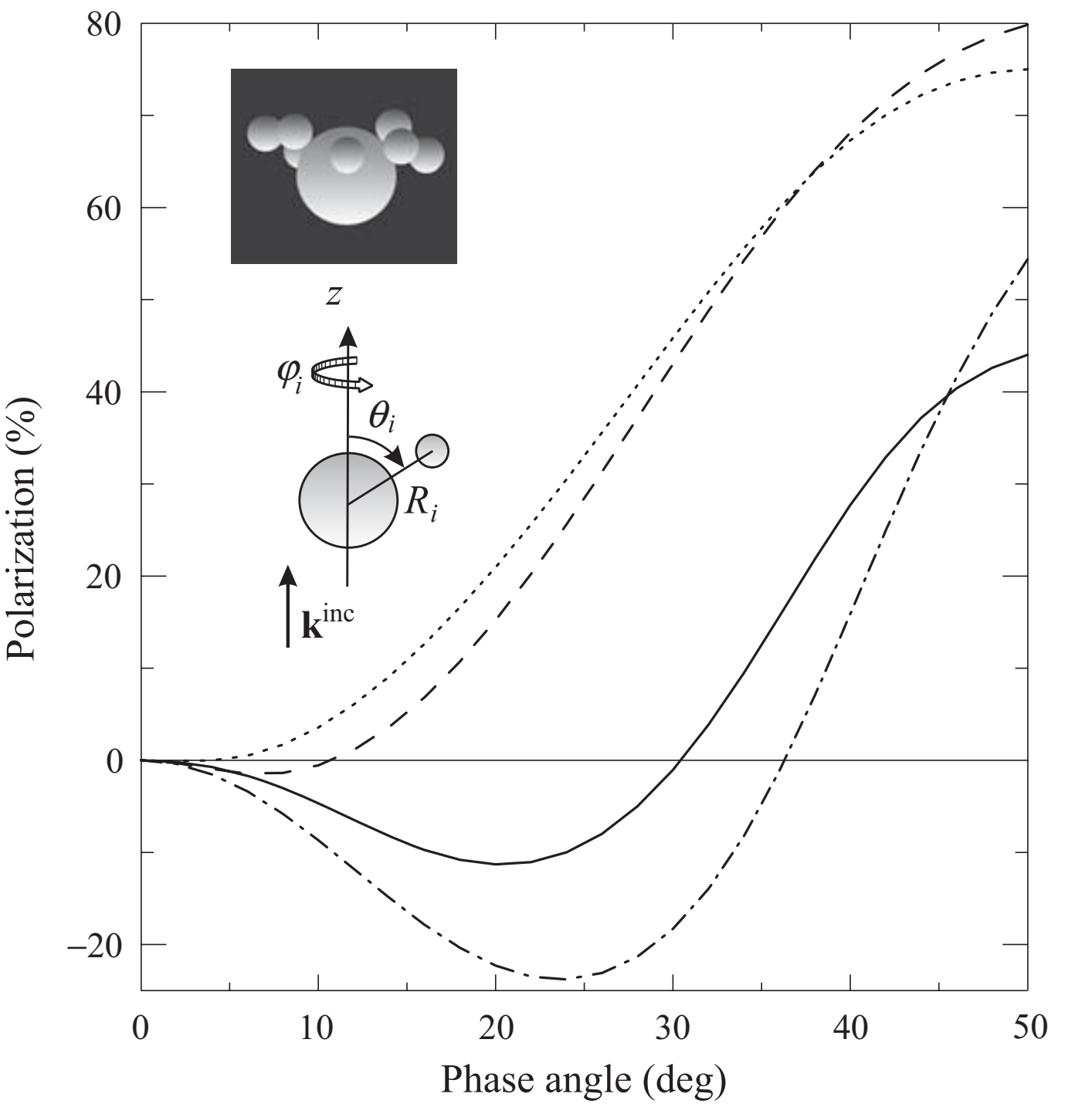

**Fig. 1.30.** Linear polarization of light scattered by the cluster shown in the inset. The incident light propagates in the vertical direction, as in Fig. 1.28b.

Figure 1.30 demonstrates indeed that theoretical computations of the scattered polarization for densely packed groups of particles must explicitly account for the near-field effects. Since the scale of wave inhomogeneities caused by scattering is comparable to the wavelength, these effects can be neglected for media consisting of very large scatterers. If the particles are small relative to the wavelength, the inhomogeneity scale is also small, and the near-field effects can be neglected as well. Therefore, the near-field effects are most pronounced for media composed of wavelength-sized particles or for strongly polydisperse media in which larger particles create wave inhomogeneities affecting the scattering of light by nearby smaller particles.

## 1.27. Scattering of quasi-monochromatic light

One of the primordial assumptions made in Section 1.1 is that any significant morphological changes in the scattering object occur over time intervals much longer than both the period of time-harmonic oscillations of the electromagnetic field and the typical temporal scale of variability of the amplitudes $\mathbf{E}_0^{\mathrm{inc}}(t)$ and $\mathbf{H}_0^{\mathrm{inc}}(t)$ of a quasi-monochromatic beam of light. It is then straightforward to show that most of the results of this chapter describing the transformation of the Stokes parameters of radiation upon scattering remain valid even when the incident radiation is a parallel quasi-monochromatic beam rather than a plain electromagnetic wave (Mishchenko et al. 2006b). For example, Eqs. (1.20) and (1.21) take the form

$$\langle \mathbf{Signal\,2} \rangle_t = \frac{\Delta S}{r^2} \mathbf{Z}(\hat{\mathbf{n}}^{\mathrm{sca}}, \hat{\mathbf{n}}^{\mathrm{inc}}) \langle \mathbf{I}^{\mathrm{inc}} \rangle_t, \quad \hat{\mathbf{n}}^{\mathrm{sca}} \neq \hat{\mathbf{n}}^{\mathrm{inc}}, \tag{1.144}$$

$$\langle \mathbf{Signal\,1} \rangle_t = \Delta S \langle \mathbf{I}^{\mathrm{inc}} \rangle_t + \frac{\Delta S}{r^2} \mathbf{Z}(\hat{\mathbf{n}}^{\mathrm{inc}}, \hat{\mathbf{n}}^{\mathrm{inc}}) \langle \mathbf{I}^{\mathrm{inc}} \rangle_t - \mathbf{K}(\hat{\mathbf{n}}^{\mathrm{inc}}) \langle \mathbf{I}^{\mathrm{inc}} \rangle_t, \tag{1.145}$$

where $\langle ... \rangle_t$ denotes an average over a time period much longer than $2\pi/\omega$. Similarly, the VRTE and the formulas describing observable manifestations of the CB effect also remain intact provided that the Stokes column vector of the incident plane electromagnetic wave is replaced by the time-averaged Stokes column vector of the incident quasi-monochromatic beam. This allows one to study diffuse multiple scattering and CB of partially polarized and even unpolarized incident light. Indeed, it follows from Eq. (1.18) that any plane electromagnetic wave is fully polarized, i.e.,

$$I^2 \equiv Q^2 + U^2 + V^2, \tag{1.146}$$

whereas the polarization state of a quasi-monochromatic beam of light is controlled by the relatively soft inequality

$$\langle I \rangle_t^2 \geq \langle Q \rangle_t^2 + \langle U \rangle_t^2 + \langle V \rangle_t^2 \tag{1.147}$$

(e.g., Mishchenko et al. 2002a), which includes the possibility for $\langle Q \rangle_t^2$, $\langle U \rangle_t^2$, and $\langle V \rangle_t^2$ to vanish simultaneously.

# 2

# Theoretical basis, methods, and hardware implementations of polarimetric remote sensing

The first polarimetric observations of the terrestrial atmosphere and the Moon were performed by François Arago 200 years ago (Arago 1854−57, Coulson 1988). Pioneering contributions to observational and laboratory polarimetry were made by another French astronomer, Bernard Lyot (Lyot 1929). However, really widespread applications of polarimetry in remote sensing and astrophysics began in the late 1950s−early 1960s. This resurgence of polarimetry can largely be attributed to the advent of such sensitive detectors of radiation as photomultipliers, which enable one to measure weak fluxes of radiant energy with very high accuracy. The more recent development of panoramic photoelectric detectors (charge coupled devices, or CCDs) has made possible simultaneous and accurate measurements of polarization characteristics of a large number of image pixels or spectral elements. As a result, the number of instruments intended for polarimetric observations of celestial objects either from the ground or from space has been growing rapidly.

Ukrainian scientists have been leaders in the development and implementation of polarimetric techniques in astrophysics and remote sensing. They and their international colleagues have made fundamental contributions to the practice of advanced polarimetric studies, including the development of sophisticated hardware, detection and registration methodologies, and polarization analysis techniques. The main objective of this chapter is to summarize these results in a systematic and self-contained form.

## 2.1. Theoretical foundations of polarimetry as an efficient means of remote characterization of terrestrial and celestial objects

Detailed knowledge of physical characteristics of particles either suspended in a planetary atmosphere, making up a regolith surface, or forming a cometary tail allows one to solve a wide range of science problems. It is, therefore, not surprising that one of the main customers of the rigorous theory of electromagnetic scattering by particles outlined in the first chapter is such an important branch of science as remote sensing of Solar System objects using ground-based, aircraft, and spacecraft instruments.

For example, the profound influence of aerosol and cloud particles suspended in the terrestrial atmosphere on global as well as regional weather, climate, and en-vironment is universally recognized (e.g., Kondratyev 1999; Hansen et al. 2005). The global nature of this influence and the global interdependence of the processes controlling the distribution of aerosols and clouds explain the need to continuously monitor detailed aerosol and cloud particle characteristics from orbital satellites (Mishchenko et al. 2007b; Kokhanovsky and de Leeuw 2009). Unfortunately, this problem is not addressed easily. Remote sensing of any other Solar System object often encounters similar obstacles (Videen and Kocifaj 2002; Videen et al. 2004).

The main cause of difficulties in solving inverse remote sensing problems are the extreme complexity, diversity, and variability of scenes typically captured by the field of view of a telescope. Quite often the radiation impinging on the detector has been scattered by a vertically and horizontally inhomogeneous atmosphere and/or a horizontally inhomogeneous surface. The atmosphere can consist of several layers composed of a mixture of gas and suspended particulates. The latter can be spherical liquid droplets or nonspherical and morphologically complex particles of dust, soot, sea salt, or ice (see, e.g., Plate 1.1). The diversity of solid surface types encountered in the Solar System defies a simple classification altogether, and even the scattering properties of the ocean surface can change profoundly with changing speed and/or direction of the local wind. What makes the problem especially severe is that all the model parameters that describe the exceedingly complex atmosphere–surface system captured by the field of view of a telescope must be determined simultaneously by means of a theoretical analysis of measured characteristics of the scattered electromagnetic radiation.

The above implies that a systematic and comprehensive approach to the solution of a specific geophysical or astrophysical problem through the use of remote sensing must include the following primary tasks (see the diagram on page 15):

1. A theoretical analysis intended to identify specific physical characteristics of aerosol and cloud particles and/or particles forming the surface that must be retrieved from remote observations as well as the requisite retrieval accuracy;

2. A theoretical analysis intended to identify specific optical measurements required to accomplish task 1;

3. An engineering study intended to identify the instrument design that would enable the solution of task 2, followed by a practical implementation of this design;

4. The development of theoretical and modeling tools required for the solution of the inverse remote-sensing problem; and finally

5. The application of these analysis tools to the acquired observational data.

Needless to say, these tasks are intimately and inherently interrelated.

Our contributions to the development of requisite theoretical analysis tools (task 4) have been described in Chapter 1. In this chapter we will focus on addressing tasks 1–3. Specific practical applications of polarimetry (task 5) will be discussed in Chapters 3 and 4.

There are two general classes of remote-sensing instruments (Stephens 1994). Passive instruments measure the reflected solar or emitted thermal radiation (Sharkov 2003). Active instruments rely on an artificial source of illumination, such as a laser or a transmitting antenna. The following discussion will be limited to pas-sive techniques and focus on the retrieval strategy and instrument design that allow one to maximize the information content of a passive remote-sensing observation.

The basic physical principles of polarimetric remote sensing are independent of the specific type of the particulate medium in question. Therefore, in this section we will give a practical example of addressing tasks 1 and 2 using as a template tropospheric aerosols in the terrestrial atmosphere, while keeping in mind that all major results and conclusions of our analysis are applicable to remote sensing of virtually any other Solar System object.

### *2.1.1. Retrieval requirements*

Tropospheric aerosols play a crucial role in climate and can cause a climate forcing directly by absorbing and reflecting sunlight, thereby cooling or heating the atmosphere, and indirectly by modifying cloud properties (e.g., Kondratyev 1999; Seinfeld and Pandis 2006). The indirect aerosol effect may include increased cloud brightness, as increasing the number of aerosols (serving as cloud condensation nuclei) leads to a larger number of smaller cloud droplets (the so-called Twomey effect), and increased cloud cover, as smaller droplets inhibit rainfall and increase cloud lifetime.

Unlike well mixed greenhouse gases, aerosols have a short lifetime in the atmosphere. After they are produced owing to various natural and anthropogenic events, they tend to mix with other agents, are transported within the troposphere both vertically and horizontally, and within about a week tend to disappear through sedimentation, rain out, etc. It is believed that the amount of anthropogenic aerosols has been increasing since the beginning of the industrial era.

Sulfate particles produced by volcanos or as a result of burning sulfur-bearing fossil fuels reflect the solar radiation out into space and are believed to cause cooling of the terrestrial atmosphere. Carbonaceous aerosols result from biomass burning and industrial combustion. They absorb the solar radiation and re-radiate it at infrared wavelengths; as such they are expected to contribute to global warming. Deposits of soot particles reduce the albedo of snow and ice surfaces and facilitate the process of melting. There are several other types of aerosols such as sea-salt particles from the ocean, mineral particles including desert dust, and various organic particulates. Whether they cool or warm the atmosphere depends on their microphysical properties as well as on their vertical location. Aerosol can also play a critical role in precipitation, but again some species of aerosols may increase precipitation, while others may inhibit it.

While it is recognized that tropospheric aerosols play a key role comparable to that of the greenhouse gases, the complexity and poorly understood variability of their composition and distribution in the atmosphere make it exceedingly difficult to quantify their effect on climate and weather: hotter or cooler, more rain or less, etc. Overall, the cumulative effect of the direct and indirect aerosol forcings may represent the largest uncertainty about future climate change caused by various anthropogenic activities (e.g., Hansen et al. 2005).

The complexity, heterogeneity, and strong variability of their global distribution make tropospheric aerosols a very difficult object of study. Because of unavoidable gaps in spatial and temporal coverage, the data collected with satellite, *in situ*, and ground based instruments will never be sufficient for a direct global assessment of the long-term aerosol effect on climate. This is especially true of the aerosol indirect effect since it is almost impossible to monitor the distribution and properties of aerosols inside clouds. Therefore, the actual role of measurements in climate studies has been and will be to provide as accurate, reliable, and comprehensive a constraint on aerosol parameterizations in chemical transport and global circulation models as possible. This makes continuous satellite observations with their quasi-global coverage and sustained, self-consistent, and uniform accuracy an indispensable component of any systematic approach to understanding and quantifying aerosol climate impacts (King et al. 1999; Kokhanovsky and de Leeuw 2009).

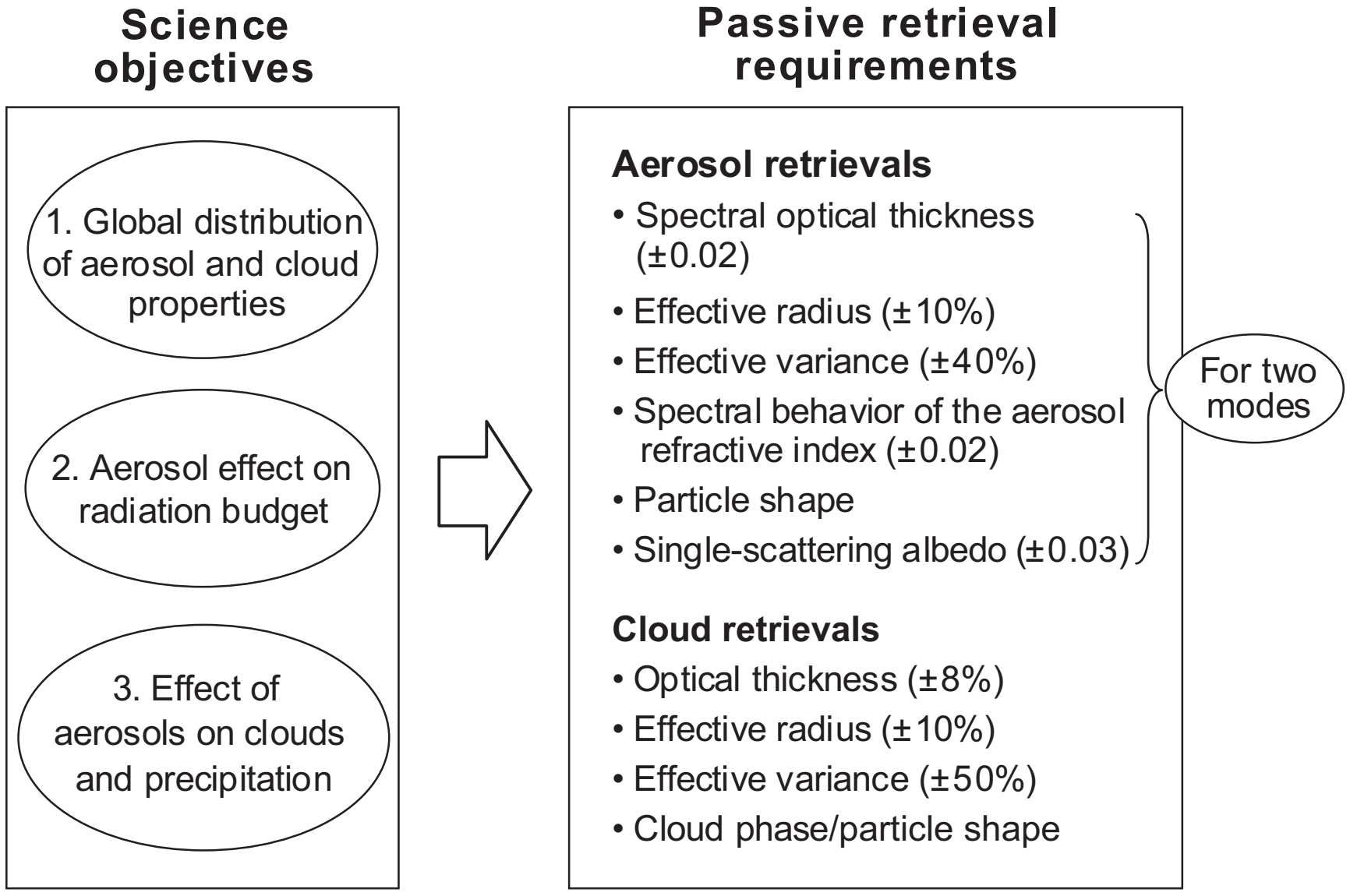

**Fig. 2.1.** Flowdown of geoscience objectives into specific retrieval requirements for a passive aerosol/cloud satellite instrument.

The left-hand panel of Fig. 2.1 summarizes the overall scientific objectives of a coordinated and systematic approach for dramatically improving our understanding of aerosol climate impacts and environmental interactions (Seinfeld et al. 2004). To achieve these objectives, one needs advanced models coupled with a comprehensive set of accurate constraints in the form of *in situ* measured and remotely retrieved aerosol and cloud distributions and properties. Because passive and active instruments have complimentary capabilities, a future comprehensive aerosol/cloud space mission should include instruments of both types. Accordingly, the right-hand panel of Fig. 2.1 lists the minimal set of aerosol and cloud parameters that must be contributed by a passive satellite instrument in order to facilitate the global quantification of the direct and indirect aerosol effects on climate (Mishchenko et al. 2004a).

The aerosol measurement requirements include the retrieval of the total column optical thickness (OT) and average column values of the effective radius and effective variance, the real part of the relative refractive index, and the single-scattering albedo. The effective radius has the dimension of length and provides a measure of the average particle size, whereas the dimensionless effective variance characterizes the width of the size distribution (Hansen and Travis 1974). Since the aerosol population is typically bimodal (e.g., Dubovik et al. 2002; Maring et al. 2003), all these parameters must be determined for each mode. The refractive index must be determined at multiple wavelengths in a wide spectral range, e.g., 400−2200 nm, since this is the only means of constraining aerosol chemical composition from space. An integral part of the retrieval procedure must be the detection of nonspherical aerosols such as dust-like and soot particles because, if ignored, nonsphericity can significantly affect the results of optical thickness, refractive index, and size retrievals (e.g., Lacis and Mishchenko 1995; Mishchenko et al. 1995a, 2003; Dubovik et al. 2006; Nousiainen 2009).

The respective minimum cloud measurement requirements include the retrieval of the column cloud OT and the average column cloud droplet size distribution as well as the determination of the cloud phase and detection of cloud particle nonsphericity.

The criteria for specifying the corresponding measurement accuracy requirements in the right-hand panel of Fig. 2.1 are dictated by the need to

- detect plausible changes in the direct and indirect aerosol radiative forcings estimated to be possible during the next 20 years; and
- determine quantitatively the contribution of this forcing to the planetary energy balance

(Hansen et al. 1995). For example, the estimated plausible 20-year change of the global mean aerosol OT is 0.04, whereas the global OT change required to yield a 0.25 W/m$^2$ flux change is 0.01. These numbers drive the very high specified accuracy for the aerosol OT measurement which, in turn, serves as a critical driver of the instrument design. The accuracies specified for the aerosol and cloud size distribution retrievals follow from the requirement to determine the aerosol and cloud number concentrations with an accuracy sufficient to detect and quantify the indirect aerosol effect (Schwartz 2004). Another factor is the strong sensitivity of the droplet-nucleation efficiency to aerosol particle radius (Dusek et al. 2006; Rosenfeld 2006). The accuracy requirement for the aerosol refractive index follows from the need to constrain the aerosol chemical composition and thereby facilitate the discrimination between natural and anthropogenic aerosol species.

### *2.1.2. Hierarchy of passive remote-sensing instruments*

Although several satellite instruments are currently used to study aerosols and their climatic effects on the global and regional scales, the retrieval of accurate particle characteristics from space remains a very difficult task. The main cause of the problem is the extreme complexity and variability of the atmosphere–surface system and the need to characterize this system by a large number of model parame-ters, all of which must be retrieved simultaneously. More often than not, the requisite number of unknown model parameters exceeds the number of *independent* (i.e., complimentary in terms of their information content) measurements provided by a satellite instrument for a given scene location, thereby making the inverse problem ill-posed. The retrieval procedure then yields a range of model solutions which are all equally acceptable in that they all reproduce the measurement data equally well within the measurement errors (Mishchenko and Travis 1997a,b; Mishchenko et al. 2007a, 2009a, 2010). The only way to ameliorate the ill-posed nature of the inverse problem is to increase the number of independent measurements per scene location until it significantly exceeds the number of unknown model parameters. Furthermore, the measurements must be inherently sensitive to specific particle properties in question. Then the retrieval procedure based on a minimization technique is likely to become stable and yield a unique solution.

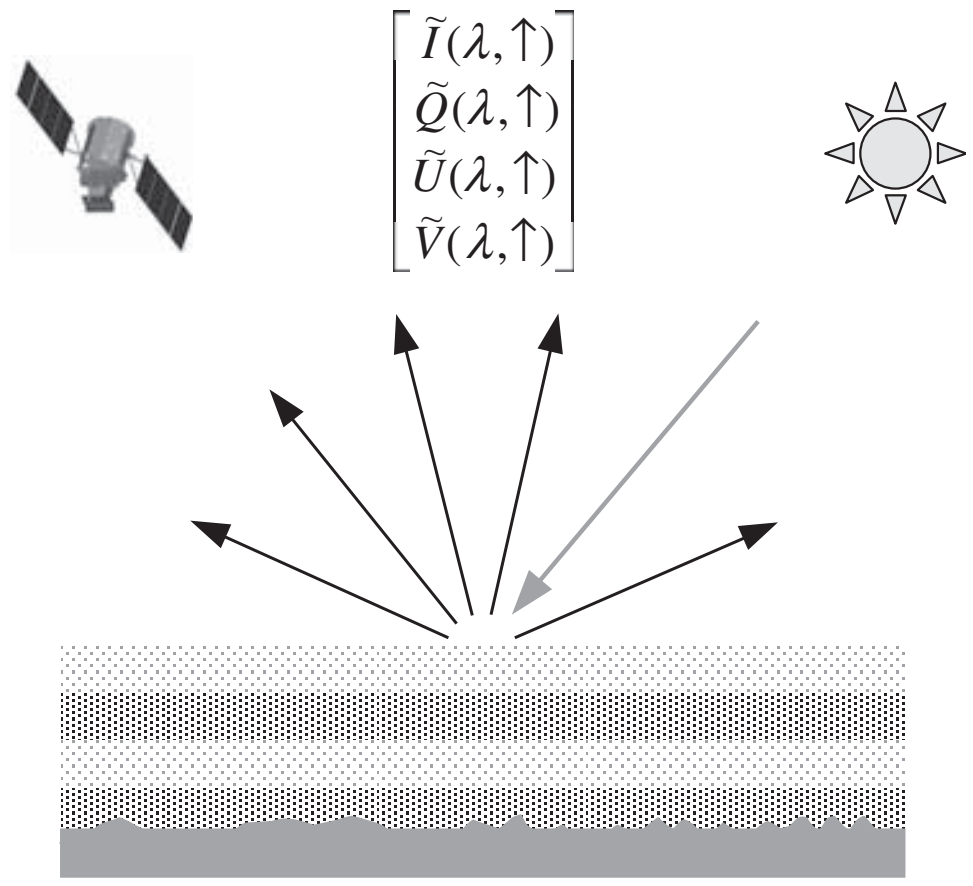

**Fig. 2.2.** Classification of passive satellite instruments measuring various characteristics of the reflected sunlight. The elements of the specific Stokes column vector $\tilde{I}$, $\tilde{Q}$, $\tilde{U}$, and $\tilde{V}$ of the reflected light vary with wavelength, $\lambda$, and scattering direction, $\uparrow$.

We have seen in Chapter 1 that polarization of scattered light, in particular as a function of scattering angle and wavelength (e.g., Plate 1.2a), is especially sensitive to particle physical properties such as size, shape, and refractive index. Therefore, the above discussion suggests that passive remote-sensing instruments intended for aerosol remote sensing can be classified according to the following criteria (see Fig. 2.2 taken from Mishchenko et al. 2004a):

- whether the instruments measures not only the specific intensity $\tilde{I}$, but also the other Stokes parameters describing the polarization state of the reflected radiation (i.e., $\tilde{Q}$, $\tilde{U}$, and $\tilde{V}$);
- the number of spectral channels and the total spectral range covered;
- the number and range of viewing directions from which a scene location is observed; and
- the measurement accuracy, especially for polarization.

At the bottom of this hierarchy is the NOAA's Advanced Very High Resolution Radiometer (AVHRR), which measures only the specific intensity in two closely spaced visible and near-IR spectral channels for only one viewing direction. The Moderate Resolution Imaging Spectroradiometer (MODIS) (King et al. 1992) and the Multiangle Imaging Spectroradiometer (MISR) (Diner et al. 1998) occupy intermediate positions. Indeed, both instruments measure only intensity, but have a significantly wider spectral range, especially MODIS. Furthermore, MISR has the capability to look at the same ground pixel from several viewing directions. The Polarization and Directionality of the Earth's Reflectance (POLDER) instrument (Deuzé et al. 2000) has the capability to measure the Stokes parameters $\tilde{Q}$ and $\tilde{U}$ as well as $\tilde{I}$, which makes it more advanced than the above-mentioned instruments. However, its spectral coverage is less wide that that of MODIS and the measurement accuracy for polarization is less than ideal.

Obviously, at the top of this hierarchy would be an instrument providing high-precision measurements of all four Stokes parameters in multiple spectral channels covering a spectral range from near-UV to short-wave IR wavelengths and at multiple viewing directions covering a wide angular range.

### *2.1.3. Sensitivity assessment of polarimetry as a remote-sensing tool*

To illustrate the hierarchy of aerosol retrieval strategies outlined in the previous section, we will now analyze theoretically the ability of passive satellite instruments to provide an accurate retrieval of the requisite aerosol characteristics, in particular, the aerosol column number density. For simplicity, we will consider only retrievals over the ocean surface since the ocean reflectance is low and can be rather accurately characterized. Our sensitivity analysis is based on numerically accurate calculations of polarized radiative transfer in a realistic atmosphere–ocean model (Mishchenko and Travis 1997b) and theoretical simulations of several types of aerosol retrievals utilizing single-channel radiance and/or polarization measurements of reflected sunlight.

We follow the approach developed in Mishchenko et al. (1997a) and use pre-computed radiance, $\tilde{I}$, and normalized second and third Stokes parameters, $q = \tilde{Q}/\tilde{I}$ and $u = \tilde{U}/\tilde{I}$ (%), for a large set of "candidate" aerosol models with effective radii $r_{\text{eff}}$ varying from 0.005 to 0.8 μm in 0.005-μm increments, refractive indices $m$ varying from 1.35 to 1.65 in steps of 0.005, and optical thicknesses $\tau$ ranging from 0 to 0.4 in steps of 0.005. The aerosol is assumed to be nonabsorbing, single-component, and monomodal with radii obeying a gamma size distribution with a fixed effective variance $v_{\text{eff}} = 0.2$. The analysis is restricted to a single near-IR wavelength $\lambda = 865$ nm. The ocean surface roughness corresponds to the global average of the long-term annual mean wind speed (7 $\text{ms}^{-1}$).

We consider two strategies of single-channel satellite measurements, viz., what can be called the AVHRR strategy (reflectance and/or polarization measurements of a scene are performed at only one viewing angle) and the Aerosol Polarimetry Sensor (APS) strategy (employing multiple-viewing-angle radiance and/or polarization measurements of a scene; see Mishchenko et al. 2007a). The illumination and view-ing directions are specified by the absolute value of the cosine of the solar zenith angle $|\cos\theta_0|$ (fixed at 0.8), the cosine of the satellite zenith angle $\cos\theta$, and the relative satellite-sun azimuth angle $\varphi$. For the APS type of measurements, $\cos\theta$ varies from 0.2 to 1 in steps of 0.2 in the satellite orbit plane specified by $\varphi = 60°$ and $-120°$, thereby yielding nine viewing directions covering the range of scattering angles from 82° to 148°. For the AVHRR type of measurements, the viewing direction is given by $\cos\theta = 0.6$ and $\varphi = -120°$, thereby implying a scattering angle 103.9° in the same orbit plane.

Computer-simulated data includes the quantities $\widetilde{I}$, $q$, and $u$ for a “standard” aerosol model with $\tau_0 = 0.2$, $r_{\text{eff}} = 0.3$ μm, and $m = 1.45$. We then assume that the aerosol model is unknown and attempt to reconstruct the “unknown” optical thickness, effective radius, and refractive index by comparing the reflectivity and/or polarization computed for the standard model with those for each of the candidate models from the large precomputed set.

We use three acceptance criteria which are intended to model retrievals using radiance measurements only (criterion A), polarization measurements only (criterion B), and radiance and polarization measurements combined (criterion C). These criteria select those candidate models for which the computed radiance and/or polarization values do not deviate from those for the standard model by more than the assumed measurement errors. For the APS type of measurements, the candidate–standard model deviations are averaged over the nine viewing directions. All candidate models that pass the acceptance criteria are equally good solutions so that none can be preferred as the unique retrieval.

The fact that the measurement errors in radiance and polarization never vanish results in multiple acceptable solutions. This is demonstrated in panels (a)–(f) of Plate 2.1 computed for the AVHRR and APS types of retrievals assuming radiance and polarization accuracies of 4% and 0.2%, respectively. The intersection of the white dashed lines in panels (c) and (d) indicates the standard model (i.e., the correct solution). The green color shows all candidate ($\tau$, $r_{\text{eff}}$, $m$)-combinations (for $m = 1.35$, 1.45, and 1.6) that passed the radiance-only acceptance criterion (A), the magenta color shows the result of applying the polarization-only criterion (B), and the intersections of the green and magenta areas show the result with criterion (C). The performance of the AVHRR type radiance-only algorithm is especially poor, the errors in the retrieved aerosol parameters $\tau$, $r_{\text{eff}}$, and $m$ being unacceptably large. This is not an unexpected result since it is hard to anticipate that an algorithm based on a single measurement can retrieve three unknown parameters with a high accuracy. That is why the actual NOAA AVHRR algorithm (Ignatov and Nalli 2002) is based on assuming rather than retrieving the aerosol model and retrieves only $\tau$. However, panels (a), (c), and (e) clearly show that assuming wrong $r_{\text{eff}}$ and $m$ can result in very large errors in the retrieved optical thickness.

The use of nine measurements in the APS-type radiance-only algorithm (green areas in panels (b), (d), and (f)) improves the retrieval significantly but still does not constrain all three aerosol parameters with the requisite accuracy. The latter is fully achieved only with the multiangle polarization algorithm (magenta area in panel (d)). The absence of magenta areas in panels (b) and (f) demonstrates the sensitivity of the multiangle polarization algorithm to refractive index. The com-bined use of multiangle radiance and polarization (the intersection of green and magenta areas in panel (d)) further improves the retrieval accuracy, but not much.

As we have mentioned above, errors in the retrieved optical thickness and assumed/retrieved aerosol model inevitably lead to errors in the retrieved aerosol column number density and, thus, to inaccurate assessments of the indirect aerosol effect. The blue-and-red background in each panel of Plate 2.1 shows the possible range of the ratio of the retrieved to the actual aerosol column number densities for the different types of aerosol retrievals. Note that the color bar is strongly nonlinear and that the white color marks the regions where the ratio *N*(retrieved)/*N*(actual) does not deviate from unity by more than ±10%.

Panels (a), (c), and (e) show that the region of possible *N*(retrieved)/*N*(actual) values for the AVHRR type intensity-only algorithm spans many orders of magnitude, thereby indicating that this type of retrieval is unsuitable for a reliable determination of the cloud condensation nucleus column number density. The APS type intensity-only algorithm provides a much better retrieval. However, the errors in the retrieved column number density are still much larger than in the other aerosol characteristics and can exceed a factor of 5. Only the multiangle polarization APS algorithm determines the aerosol model with such an accuracy that the retrieved aerosol column density is constrained to ±10%, panel (d).

An obvious limitation of this sensitivity analysis is that we have considered only single-channel retrieval algorithms. Using multispectral data from instruments like MODIS and MISR may be expected to improve the accuracy of radiance-only aerosol retrievals. However, our comparison of different remote-sensing strategies under exactly the same conditions clearly illustrates the tremendous improvement brought about by using high-precision polarimetric data in addition to radiance data. This result is fully corroborated by more detailed sensitivity studies (e.g., Hasekamp and Landgraf 2007) as well as by analyses of actual observational data (Chowdhary et al. 2001, 2002, 2005, 2006; Cairns et al. 2009; Waquet et al. 2009a,b,c). Again, this conclusion is quite general and applies to remote sensing of other Solar System objects (Мороженко 1985, 2004).

## 2.2. Basic concepts and principles of polarization measurements

Before discussing the main principles of polarimetric measurements, it is appropriate to introduce several basic concepts and definitions establishing the link between the formal mathematical account of polarization in Chapter 1 and Section 2.1 and working formulas used to describe the actual measurements with different types of polarimeters.

A plane electromagnetic wave given by Eq. (1.1) is the simplest type of electromagnetic radiation and is well represented by a perfectly monochromatic and perfectly parallel laser beam. Let us consider an arbitrary point within such a wave. It is straightforward to demonstrate that during each time interval $2\pi/\omega$, the endpoint of the real electric field vector $\mathrm{Re}(\mathbf{E})$ describes an ellipse in the plane normal to the propagation direction (hereinafter called the figure plane). The sum of the squares of the semi-axes of this ellipse multiplied by a constant factor yields the total intensity of the wave $I$. The ratio of the semi-axes, the orientation of the ellipse, and the sense in which the electric vector rotates (clockwise or counter-clockwise, when looking in the direction of propagation) can be derived from the other three Stokes parameters of the wave $Q$, $U$, and $V$ (Mishchenko et al. 2002a). Importantly, any plane electromagnetic wave is fully polarized (see Eq. (1.146)).

Quasi-monochromatic beams of light described by Eq. (1.2) are encountered much more often than perfectly monochromatic beams. As explained in Section 1.27, the Stokes parameters of a quasi-monochromatic beam are obtained by averaging the right-hand side of Eq. (1.18) over a time interval much longer than $2\pi/\omega$ (the angular brackets $\langle ... \rangle_t$ will thereafter be omitted) and satisfy the inequality (1.147). In general, the end-point of the vector $\mathrm{Re}(\mathbf{E})$ of a quasi-monochormatic beam does not describe a well-defined polarization ellipse. Still one can think of a "preferential" ellipse with a "preferential orientation", "preferential elongation", and "preferential handedness". As mentioned earlier, a quasi-monochromatic beam can be partially polarized and even unpolarized. The latter means that the temporal behavior of $\mathrm{Re}(\mathbf{E})$ is completely "erratic", so that there is no "preferential ellipse".

The parameters of the "preferential ellipse" of a parallel quasi-monochromatic beam of light can be viewed as quantitative descriptors of the asymmetry in the directional distribution and/or rotation direction distribution of the vector $\mathrm{Re}(\mathbf{E})$ in the figure plane. A common type of polarization of light coming from Solar System bodies is partial linear polarization, which implies that the Stokes parameter $V$ is negligibly small. This type of polarization is often described by the total intensity $I$, degree of linear polarization $P_{\mathrm{lp}}$, and angle $\vartheta$ specifying the orientation of the plane of preferential oscillations of the real electric field vector (hereinafter called the polarization plane) with respect to a prescribed direction. For example, it is customary in observational astrophysics to measure the angle $\vartheta$ from the direction "object → North Pole" on the celestial sphere in the counter-clockwise sense when looking in the direction of propagation (i.e., through the East). In theoretical data analyses, $\vartheta$ is usually specified with respect to the scattering plane (this will be consistently done throughout Chapters 3 and 4). The intensity of the partially linearly polarized beam can be decomposed into a fully linearly polarized intensity $I_{\mathrm{lp}}$ and an unpolarized intensity $I_{\mathrm{u}}$. The degree of linear polarization is then given by

$$P_{\mathrm{lp}} = \frac{I_{\mathrm{lp}}}{I} = \frac{I_{\mathrm{lp}}}{I_{\mathrm{lp}} + I_{\mathrm{u}}}, \qquad 0 \le P_{\mathrm{lp}} \le 1, \tag{2.1}$$

where

$$I_{\mathrm{lp}} = \sqrt{Q^2 + U^2}. \tag{2.2}$$

In order to characterize the degree of asymmetry in the directional distribution of the electric vector in the figure plane, we can also define the so-called signed degree of linear polarization as

$$P = \frac{I_{\perp} - I_{\parallel}}{I} = \frac{I_{\perp} - I_{\parallel}}{I_{\perp} + I_{\parallel}}, \tag{2.3}$$

where $I_\perp$ and $I_\parallel$ are the intensity components with real electric vector oscillations perpendicular and parallel to the scattering plane, respectively. As before, the latter is defined as the plane containing the incidence and scattering directions. This definition allows negative values of $P$ if $I_\perp < I_\parallel$, which means that oscillations of the real electric field vector in the scattering plane dominate those in the perpendicular plane. A positive value of $P$ implies that oscillations of the real electric field vector in the plane perpendicular to the scattering plane dominate those in the scattering plane. We can alternatively write

$$P = -Q/I \tag{2.4}$$

provided that the meridional plane of the propagation direction coincides with the scattering plane. This definition of the degree of linear polarization is consistent with that used in Sections 1.25 and 1.26.

If $V \neq 0$ then, by analogy with Eqs. (2.3) and (2.4), the (signed) degree of circular polarization can be written as

$$P_{\mathrm{cp}} = \frac{I_{\mathrm{R}} - I_{\mathrm{L}}}{I_{\mathrm{R}} + I_{\mathrm{L}}} = -\frac{V}{I}, \tag{2.5}$$

where the right- and left-handed intensity components

$$I_{\mathrm{R}} = \frac{1}{2}(I - V) \quad \text{and} \quad I_{\mathrm{L}} = \frac{1}{2}(I + V) \tag{2.6}$$

are polarized in the clockwise sense and in the counter-clockwise sense, respectively, when looking in the direction of propagation.

In general, the incoming light is partially elliptically polarized and is described by all four Stokes parameters. Alternatively one can use the full intensity $I$, the degree of linear polarization $P_{\mathrm{lp}}$, the orientation (or position) angle $\vartheta$, and the degree of circular polarization $P_{\mathrm{cp}}$. It is straightforward to derive the following relations:

$$\begin{aligned} Q &= IP_{\mathrm{lp}} \cos 2\vartheta, \\ U &= IP_{\mathrm{lp}} \sin 2\vartheta, \\ V &= -IP_{\mathrm{cp}}, \end{aligned} \tag{2.7}$$

$$\vartheta = \frac{1}{2} \operatorname{arctg}(U/V). \tag{2.8}$$

Besides the usual Stokes parameters $Q$, $U$, and $V$ having the dimension of intensity, it is customary to use normalized dimensionless parameters (cf. Section 2.1.3):

$$q = Q/I = P_{\mathrm{lp}} \cos 2\vartheta, \quad u = U/I = P_{\mathrm{lp}} \sin 2\vartheta, \quad v = V/I = -P_{\mathrm{cp}}. \tag{2.9}$$

If a partially polarized beam passes through a rotating polarizer (e.g., a Polaroid) then the intensity of the transmitted radiation $I'$ will depend on the orientation angle of the polarizer $\phi$ measured with respect to the fixed reference frame in the counter-clockwise sense when one is looking in the direction of propagation:

$$I'(\phi) = \frac{1}{2}(I + Q\cos 2\phi + U \sin 2\phi) = \frac{1}{2} I [1 + P_{\mathrm{lp}} \cos 2(\phi - \vartheta)], \tag{2.10}$$

where $I$, $Q$, and $U$ are the initial Stokes parameters. Alternatively one can use a half-wave plate $\lambda/2$. Equation (2.10) implies that the determination of the degree of linear polarization requires the measurement of the ratio of transmitted intensities corresponding to two mutually orthogonal orientations $\phi$ of the polarizer. In order to determine the intensity, the degree of linear polarization, and the position angle of the polarization plane $\vartheta$ of the beam, one needs to measure the outgoing intensity for at least three orientation angles of the polarizer with 60° increments. In practice, it is customary to take the measurements for four orientation angles $\phi = 0°$, 45°, 90°, and 135° and then determine the polarization parameters from Eqs. (2.1), (2.2), and (2.8) coupled with

$$\begin{aligned} I &= I_0 + I_{90} = I_{45} + I_{135}, \\ Q &= I_0 - I_{90} = I_{\mathrm{lp}} \cos 2\vartheta, \\ U &= I_{45} - I_{135} = I_{\mathrm{lp}} \sin 2\vartheta, \end{aligned} \tag{2.11}$$

where $I_0$, $I_{45}$, $I_{90}$, and $I_{135}$ are the corresponding intensity values obtained from Eq. (2.10).

According to Serkowski (1974), the modulation formula describing the intensity of the transmitted beam after the passage through a rotating polarizer (a quarter-wave plate $\lambda/4$) and a fixed Polaroid (analyzer) is as follows:

$$I'(\psi_0) = \frac{1}{2}(I + \frac{1}{2}Q + \frac{1}{2}Q\cos 4\psi_0 + \frac{1}{2}U\sin 4\psi_0 - V\sin 2\psi_0), \tag{2.12}$$

where $\psi_0$ is the angle between the optical axes of the analyzer and the polarizer measured with respect to the fixed reference frame in the counter-clockwise sense when one is looking in the direction of propagation. The determination of only the Stokes parameter $V$ requires measurements of the transmitted intensity for four orientation angles of the polarizer 0°, 45°, 90°, and 135°. The simultaneous measurement of all four Stokes parameters necessitates measurements for 8 orientation angles with 22.5° increments.

More often than not, the brightness of astrophysical objects is low, and polarization measurements are accompanied by noise caused by nonzero polarization of the background skylight, poor seeing, and the detector. This necessitates (near) simultaneous detection of the intensity components with orthogonal polarization directions. The selection of different harmonics (Stokes parameters $Q$, $U$, and/or $V$) according to Eq. (2.10) or (2.12) is achieved using the method of synchronous detection. A very high rotation frequency of the polarizer ensures a quasi-simultaneous measurement of the polarization parameters. This technique is widely used in aperture polarimetry.

## 2.3. Development of observational polarimetry and polarimetric instrumentation in Ukraine

In the mid 1950s, the director of the Crimean Astrophysical Observatory (CrAO) G. A. Shain indicated that the non-random orientations of filaments of diffuse nebulae could be caused by a global magnetic field of our Galaxy. By that time, it had already been known that direct starlight can be polarized by nonspherical interstellar grains preferentially oriented by the galactic magnetic field. A comparison of the filament orientations and the orientation distribution of polarization planes for different stars allowed Shain to prove the existence of the galactic magnetic field. Jointly with his colleagues, G. A. Shain demonstrated, for the first time, that the radiation emitted by the Crab Nebula is polarized, polarization being different for different parts of the nebula. This provided an independent demonstration of the synchrotron nature of the emitted radiation and the complex morphology of the magnetic field of the nebula (Шайн и др. 1955). These studies marked the beginning of active applications of polarimetry in astrophysics and ensured the leadership of Ukrainian astrophysicists in the subsequent advancement of polarimetric methodologies.

### *2.3.1. Aperture polarimeters*

Regular large-scale polarimetric observations in CrAO began in the early 1960s after the development of the first integrating polarimeter shown in Plate 2.2a (Шаховской и Димов 1962). Polarimetric parameters were measured using the differential technique and employing a low rotation frequency of the polarizer (one rotation per second) synchronized with switches of 20 RC-integrators. After a large number of rotations the charges accumulated by the capacitors were proportional to the average intensities of the polarization components corresponding to the different orientations of the polarizer. The accumulation of the charges was accompanied by a continuous registration of the total intensity of the incoming light. The constant component of the photocurrent was proportional to $I$, while the amplitude of the oscillating component was proportional to $IP_{\mathrm{lp}}$, in agreement with Eq. (2.10). The solution of the resulting 20 equations yielded $P_{\mathrm{lp}}$ and $\vartheta$.

The very first surveys of stars conducted in CrAO resulted in several discoveries such as the detection of intrinsic polarization variability of eclipsing stellar systems, Be stars, pulsating stars, and stars of the RV Tau type (Шаховской 1963, 1964a,b). These studies also led to the development of methodologies for the measurement and analysis of variable polarization useful in studies of small Solar System bodies. However, it became clear that the differential method of polarization measurements and the low rotation frequency of the polarizer limited the utility of the instrument. It was, in fact, demonstrated (Ксанфомалити и Шаховской 1968) that observations with large telescopes require high modulation frequencies (at least 20 rotations per second), which helps to reduce significantly the effect of atmospheric scintillations and improve the measurement accuracy as well as to perform polarization observations of objects with rapidly varying brightness (e.g., flare stars).

The goal of improving the measurement accuracy, measuring all four Stokes parameters, and studying polarization of fainter and rapidly varying astrophysical objects necessitated the development of advanced polarimetric methodologies and next-generation polarimeters. The characteristics of CrAO polarimeters developed over the past five decades are briefly summarized in Table 2.1. The CrAO telescopes currently used in polarimetric observations are shown in Plate 2.3.

**Table 2.1.** Polarimeters developed and modernized in CrAO

| Polarimeter type (modernization) | Telescope and diameter (m) | Period of use | Astrophysical objects (main types) |
|---|---|---|---|
| Integrating polarimeter | 0.4, 0.7 | 1960–68 | Different types of variable stars |
| Single-channel polarimeter for measurements of linear polarization | ZTSh, 2.6<br>ZTE, 1.25 | 1969–80 | Flare stars, polars, eruptive stars, novae and supernovae, galaxies with active nuclei, comets |
| Five-channel polarimeter of the Helsinki University Observatory (software registration of Stokes parameters) | AZT-11, 1.25 | 1981–present | Symbiotic stars, binary systems, young stars with circumstellar dust disks, asteroids, comets, planetary satellites |
| Polarimeter with a dual acousto-optical modulator for measurements of circular polarization in the wings of hydrogen lines | ZTSh, 2.6 | 1990 | Ap stars, polar AM Hercules |
| Single-channel polarimeter for measurements of linear and circular polarization (improved registration of the Stokes parameters) | ZTSh, 2.6 | 1997–present | Polars, asteroids, comets, planetary satellites |

***Photopolarimeters of the 2.6-m CrAO telescope.*** In the 1970s, polarimetric observations were continued with the 2.6-m CrAO telescope named after Shain (hereinafter ZTSh) as well as with the 1.25-m Engelgart telescope (hereinafter ZTE) of the South Station of the Shternberg State Astronomical Institute using a new single-channel polarimeter with a rapidly rotating polarizer (33 rotations per second). The signal is registered using the pulse-counting technique developed by N. M. Shakhovskoy and Y. S. Efimov. The original design of the instrument and its modification were described in Шаховской и Ефимов (1972) and Шаховской и Ефимов (1976), respectively. The measurement is based upon using four pulse counters registering photoelectrons coming from the photomultiplier. Special electronic keys switch pulse counters on and off synchronously with the rotations of the polarizer. During each revolution of the polarizer each pulse counter is switched on twice (for a period of time equal to 1/4 of the period of revolution $T$), the switch-on moment for one pair of counters being shifted by $T/2$ relative to that for the other pair of counters. This technique ensures a quasi-simultaneous measurement of the polarization parameters $q$ and $u$. The latter are expressed in the numbers of pulses accumulated by the four counters $N_1$, $N_2$, $N_3$, and $N_4$ as follows:

$$q = \frac{\pi}{2}\frac{(N_1 - N_2)}{(N_1 + N_2)}, \qquad u = \frac{\pi}{2}\frac{(N_3 - N_4)}{(N_3 + N_4)}. \tag{2.13}$$

Important advantages of this instrument are its high efficiency in observations of faint objects as well as the virtual independence of the measured polarization parameters of changes in the incoming intensity caused by potential intrinsic variability of the object and/or by atmospheric scintillations. As a result, the list of astrophysical objects studied had been significantly expanded and included, for the first time, selected Solar System bodies. Among the most important results were independent polarization observations of comet West by N. M. Shakhovskoy and Y. S. Efimov (Нарижная и др. 1977) which confirmed the existence of negative polarization for this comet. These observations were stimulated by N. N. Kiselev who was the first to discover a negative polarization branch in the phase dependence of the degree of linear polarization of light scattered by comets (Киселев и Чернова 1976).

The above measurement approach is still in use, albeit with several enhancements such as the use of modern computers and pulse registration devices. The registration and storage of information have become much more reliable, while the efficiency of polarimetric observations has become substantially higher owing to the increased rate of transferring data between the instrument and the computer.

An important stage in the progress of astrophysical polarimetry was the development of instruments capable of measuring circular polarization. The first instrument of this type built in CrAO was a dual-beam polarimeter with an acousto-optical modulator (Вайсс и др. 1990). The incoming beam passes in succession a modulator inducing a phase difference between two intensity components with orthogonal polarization states and a Wollaston prism creating two outgoing beams. Each beam is registered by two photomultipliers and pulse counters which are switched synchronously with the periodic phases of modulation. The instrument allowed one to measure polarization of incoming light in the wings of the hydrogen spectral band $H_\beta$, which was used to normalize the values of the magnetic fields of several Ap stars and the polar AM Hercules. This method of polarization measurements was patented (Батюк и др. 1981).

More recently, circular polarization was measured with the modified single-channel polarimeter of the 2.6-m telescope. The number of pulse counters remained the same, whereas the rotating Polaroid was replaced with a rotating achromatic phase plate $\lambda/4$ followed by a fixed Polaroid (Шаховской и др. 1998, 2001). This improvement became possible in 1991–92 owing to the emergence of achromatic phase plates designed by V. A. Kucherov (Кучеров 1986a,b).

The most recent modernization of this polarimeter allows one to measure all four Stokes parameters simultaneously according to the modulation formula (2.12). The intensity components used to compute the Stokes parameters *Q, U*, and *V* are selected by eight virtual counters (one physical counter) which are switched on sequentially and accumulate the signal for 1/16 of the rotation period of the quarter-wave phase plate.

The resulting improvement of the measurement approach, signal registration hardware, and software have made possible the first accurate measurements of the linear and circular polarization across the comas of comets Hale–Bopp (Rosenbush et al. 1997b; Розенбуш и др. 1999), S4 (LINEAR) (Rosenbush et al. 2007b), and Q4 (NEAT) (Rosenbush et al. 2007a). These results were used to derive, for the first time, the phase dependence of the circular polarization of light scattered by comets (Розенбуш 2006). It has been found that for all the comets studied, the circular polarization is preferentially left-handed (Rosenbush et al. 2008a). These studies as well as measurements of polarization of starlight during occultations of stars by comets (Rosenbush et al. 1994, 1997b; Розенбуш и др. 1999) have led to the discovery of oriented particles in cometary atmospheres.

***Five-channel photopolarimeter of the CrAO 1.25-m telescope.*** The installation of the telescope AZT-11 and the availability of the five-channel photopolarimeter of the Helsinki University Observatory (Piirola 1988; see Plate 2.2b) have dramatically expanded the range of polarimetric observations in CrAO. Polarization measurements can be performed simultaneously in five spectral channels U, B, V, R, and I with effective wavelengths 370, 440, 530, 690, and 830 nm, respectively.

The optical scheme of this polarimeter is shown in Fig. 2.3. The calcite prism serves as the analyzer and splits the incoming beam into two components with orthogonal linear polarizations. The corresponding images are formed inside two diaphragms in the focal plane. The signals are modulated using a rotating half-wave phase plate in measurements of linear polarization and a rotating quarter-wave phase plate in measurements of circular polarization. Inside the diaphragms, the background skylight with orthogonal polarization directions is superposed on the images. This is convenient in observations of point-like objects, because then there is no need to account for the background polarization. However, this renders the instrument not suitable for observations of extended objects. Therefore, comets are observed using an original technique based on the insertion, before the calcite prism, of an additional Polaroid with a specific orientation intended to cancel one of the polarization components. The software package used in the registration of data from the five-channel photopolarimeter was further improved by Бердюгин и Шаховской (1993). This enabled simultaneous measurements of polarization parameters as well as the brightness of the object in a wide spectral interval ranging from the ultraviolet to the near infrared.

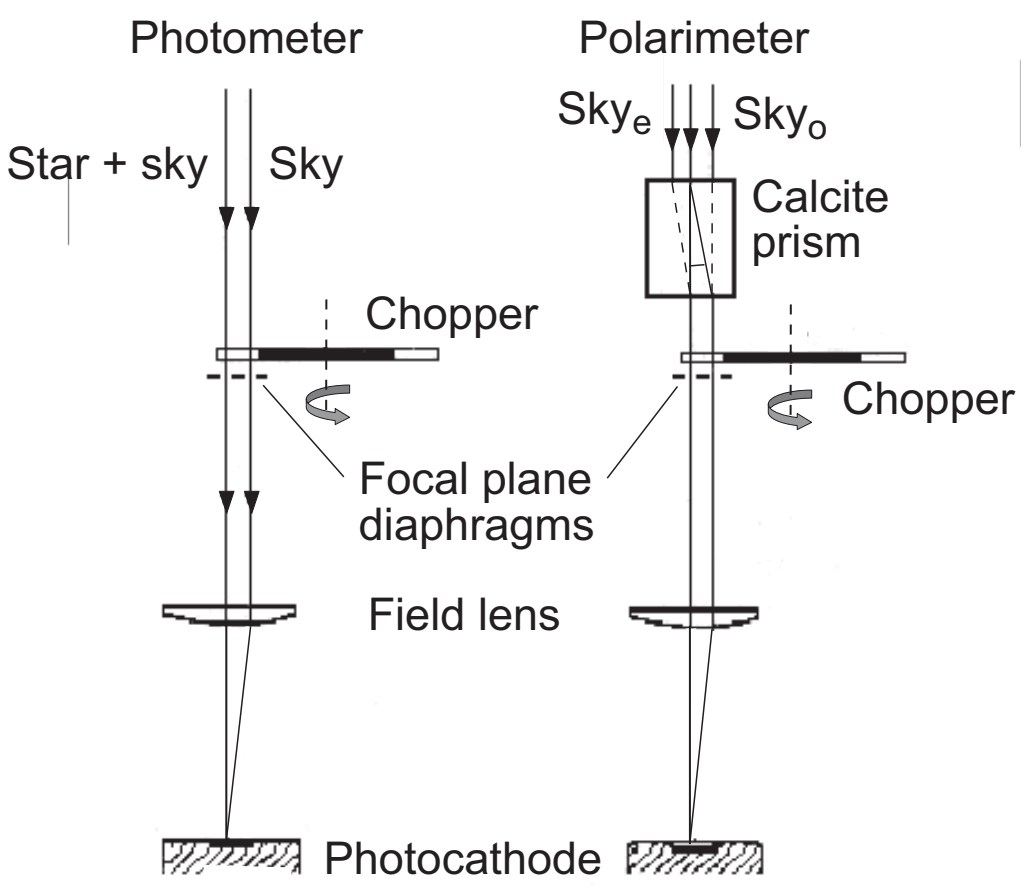

**Fig. 2.3.** Principle of operation of the five-channel photopolarimeter.

The observational program was expanded to include asteroids and comets. Unique results obtained for stars surrounded by dust envelopes and disks are also quite relevant to the studies of small Solar System bodies. Polarimetric observations of the star R CrB during deep brightness minima performed by Y. S. Efimov have revealed a relationship between the brightness and polarization variations (Ефимов 1980, 1990). This phenomenon exhibited by R CrB was first discovered for young Ae/Be Herbig stars with circumstellar dust disks (so-called UXORs) during a cooperative program of patrol observations in CrAO (V. P. Grinin and his colleagues) as well as in Bolivia and the Sanglok Observatory in Tajikistan (N. N. Kiselev and his colleagues) (Вощинников и др. 1988). Both types of stars reveal quite similar changes of polarization with brightness. However, despite the significant similarity of observed optical phenomena, there are profound physical differences between these two types of stars. For stars of the R CrB type, the decrease in brightness is caused by the ejection of a cloud of gas and dust along the line of sight. The expansion of the cloud is accompanied by the growth of the dust grains, both causing changes in the photometric and polarimetric characteristics of the stars. According to the model by Grinin (1988), the decrease in brightness of UXORs is caused by the passage across the line of sight of dust clouds of variable density (comet-like objects) which rotate along Keplerian orbits in the plane of the circumstellar dust disk. This model implied that during a brightness minimum there should be a significant increase in polarization caused by the scattering of light by dust grains forming the circumstellar envelope. This prediction has been confirmed by observations of a number of UXORs (Grinin et al. 1991).

Telescope AZT-11 has been upgraded by implementing a special software package which allows one to perform offset guiding after moving objects such as asteroids, comets, and planetary satellites. Among the most important results thus obtained have been the following: (i) the discovery of the theoretically predicted POE (Mishchenko 1993b) for the high-albedo asteroid 64 Angelina (Rosenbush et al. 2005b) and the bright hemisphere of the Saturnian satellite Iapetus (Розенбуш 2006; Розенбуш и др. 2007); (ii) the determination of the spectral dependence of minimal negative polarization for different types of asteroids and the discovery of a number of new effects and unusual properties exhibited by certain asteroids (Бельская и др. 1991; Лупишко 1998a; Belskaya et al. 2005); and (iii) the discovery of an anomalous spectral dependence of polarization for the unique comet Holmes as well as for the B and C fragments of the nucleus of comet Schwassmann–Wachmann 3 (Розенбуш 2006; Kiselev et al. 2008; Rosenbush et al. 2008b, 2009b). A detailed discussion of polarimetric observations of Solar System objects will be presented in Chapters 3 and 4.

***Collaborative programs.*** The success of the CrAO program of polarimetric observations stimulated the spread of this powerful remote-sensing technique in other observatories of the former Soviet Union. In particular, CrAO provided technical and operational support in the development of polarimeters for the 0.7-m telescope of the Gissar Astronomical Observatory in Tajikistan, the 1-m telescope of the Sanglok Observatory, and the 0.7-m telescope of the Institute of Astronomy of the Kharkiv V. N. Karazin National University (KhNU). The technique based on the measurement of linear polarization using a rapidly rotating Polaroid (33 rotations per second) and the synchronous signal registration with four pulse counters was implemented with further improvements (Шаховской и Ефимов 1976). Specifically, photometric observations of bright objects are usually performed using the polarimetric mode of the photopolarimeter. However, for faint objects, it was necessary to reduce intensity losses inside the Polaroid, which can be as significant as $\sim 1.5^{m}$ in the spectral band U. Therefore, the optomechanical unit of the new photopolarimeters was modernized, thereby making possible a virtually instantaneous transition from polarimetric observations to purely photometric ones by removing the Polaroid from the optical tract, and vice versa. The new mechanical element ensures a reliable fixation of the rotating analyzer without any noticeable change in the instrumental polarization or in the position of the polarization plane of the Polaroid (Киселев 2003). The first two telescopes were used to initiate a program of systematic observations of comets and asteroids, which was eventually joined by the CrAO and KhNU astronomers. Comprehensive polarimetric studies of asteroids, comets, and planetary satellites in Ukraine have been continued ever since and have involved scientists from CrAO, the KhNU, and the Main Astronomical Observatory of the Ukrainian National Academy of Sciences (MAO, Kyiv).

### *2.3.2. Panoramic polarimetry*

A significant disadvantage of aperture polarimeters has always been their low angular resolution. As a consequence, the measured polarization parameters are often the result of averaging over the entire solid angle subtended by a large object such as a planet, a planetary satellite, or a comet. Panoramic polarimetry was initially based on the photographic technique, but the resulting accuracy was relatively low. The advent of CCDs has led to a substantial progress in polarimetric observations of extended objects owing to two critical advantages: a high angular resolution and a high photometric sensitivity over a broad spectral range.

The simplest measurement approach is to register a sequence of images obtained with Polaroids which are first oriented at the 0° and 90° angles, and then at the 45° and 135° angles (see Eq. (2.11)). Alternatively, one can register simultaneously two images of the object obtained with orthogonally polarized beams. The latter approach usually involves a Wollaston prism. The Stokes parameters $Q$ and $U$ are obtained via two expositions corresponding to the 0° and 45° orientations of the instrument or to the 0° and 22.5° orientations of a half-wave plate inserted before the Wollaston prism. A significant drawback of these techniques is that the measurements for two (or four) orientation angles are not simultaneous.

An advanced four-beam CCD-polarimeter was designed and built at the Max-Planck Institute for Solar System Research in Katlenburg-Lindau (Germany). An active role in the instrument design, development of observation methodologies, and actual observations has been played by N. N. Kiselev (Geyer et al. 1996). The measurement principle is based on a simultaneous registration of four images of the object using beams with polarization directions 0°, 45°, 90°, and 135° obtained with a special composite Wollaston prism. The latter is inserted in the parallel beam of the collimator and consists of two pairs of Wollaston prisms (Fig. 2.4). The prisms forming the first pair were cut and glued in such a way that they split a half of the parallel light beam into two orthogonally polarized components (0° and 90°) in the horizontal plane. The other pair of prisms splits the remaining half of the original beam in the vertical plane such that the resulting orthogonally polarized components have polarization directions 45° and 135°. The use of a special mask in the focal plane allows one to form four images of the background sky simultaneously with the four images of the object.

An important advantage of this instrument is that it does not involve moving parts. The normalized Stokes parameters $q$ and $u$ are obtained simultaneously and from the same image, which shortens the exposition time by a factor of two. On the other hand, only half of the incoming energy flux is used for the determination of each parameter, which makes the efficiency of this instrument essentially the same as that of the other panoramic polarimeters.

As an example, Plate 2.4 shows images of comet Ashbrook–Jackson and the background sky obtained with the advanced four-beam panoramic polarimeter. This instrument has been used to perform extensive polarimetric observations of asteroids, planetary satellites, and comets with the 2-m telescope of the Terskol observatory. These results will be discussed in Chapters 3 and 4.

Panoramic detectors are ideally suitable not only for obtaining images in polarized light, but also for obtaining spectra. Scientists from MAO and CrAO have proposed an innovative design of a low-spectral-resolution ultraviolet spectropolarimeter (UVSPEPOL) without a slit intended for use in space missions (Kucherov et al. 1997; Yefimov 2004). The instrument design involves original optical elements (see Plate 2.5). Unlike the conventional classical scheme, the rotating superachromatic quarter-wave phase plate is followed by a composite Wollaston prism with two off-centered elements and deformed surfaces. This prism functions as the dispersive element as well as the polarization analyzer. This allows one to obtain two spectra with orthogonal polarization directions on one CCD image.

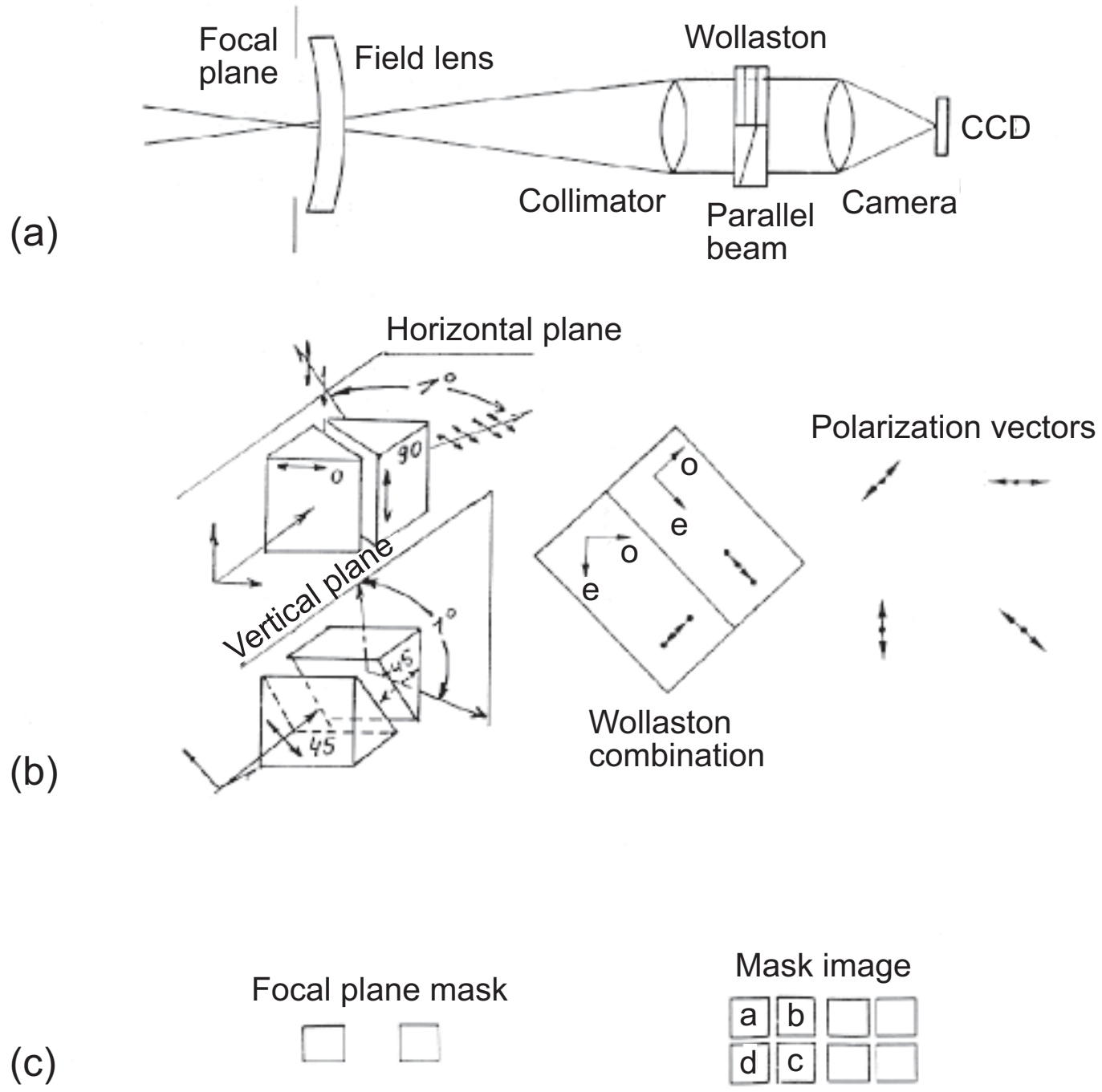

**Fig. 2.4.** (a) Optical scheme of the four-beam CCD-polarimeter. (b) Special Wollaston prism intended for a simultaneous registration of four images and the background sky with polarization directions 0°, 45°, 90°, and 135°. (c) Focal mask and its images.

## 2.4. Passive photopolarimeter for remote sensing of terrestrial aerosols and clouds from orbital satellites

The profoundly important role played by tropospheric aerosols in the formation of the terrestrial climate has been mentioned in Section 2.1.1. The knowledge of the global distribution of aerosols and their properties is also required for ecological monitoring purposes. Unlike the AVHRR mentioned in Section 2.1.2 and all other orbital instruments currently used to study aerosols (Mishchenko et al. 2007b), the photopolarimeter APS (see Plates 2.6 and 2.7) was specifically designed for accurate and detailed retrievals of aerosol and cloud properties. Owing to its unique technical characteristics, the APS is at the very top of the hierarchy of passive remote-sensing instruments formulated in Section 2.1.2. It is intended to be part of the NASA orbital mission called Glory and should be launched in 2010 (Mishchenko et al. 2007a).

The unique APS design is intended to maximize the microphysical retrieval capability of the instrument. It allows one to take full advantage of the extreme sensitivity of high-accuracy polarization data to aerosol and cloud particle microphysics coupled with the advanced modeling capabilities, as discussed in Chapter 1 and Section 2.1, and thereby ensures the retrieval of all the quantities listed in the right-hand panel of Fig. 2.1.

The flowdown of APS measurement capabilities into the requisite aerosol and cloud retrieval capabilities is summarized in Table 2.2. The key measurement requirements for the retrieval of aerosol and cloud properties from photopolarimetric data are *high* (i.e., *fine*) *accuracy*, a *broad spectral range*, and observations from *multiple angles*, including a method for reliable and stable *calibration* of the meas-urements. The APS built for the Glory mission by Raytheon (Peralta et al. 2007) meets these measurement requirements.

**Table 2.2.** Flowdown of APS measurement characteristics into specific retrieval capabilities

| Measurement characteristic | Retrieval capability |
|---|---|
| Precise and accurate polarimetry (~0.1%) | Particle size distribution, refractive index, shape |
| Wide scattering angle range for both intensity and polarization | Particle size distribution, refractive index, shape |
| Multiple ($\gtrsim$30) viewing angles for both intensity and polarization | (i) Particle size distribution, refractive index, shape<br>(ii) Ocean surface roughness |
| Multiple ($\gtrsim$60) viewing angles for polarization | Cloud particle size via rainbow angle |
| Multiple ($\gtrsim$30) viewing angles and accurate polarimetry | Aerosol retrievals in cloud-contaminated pixels |
| Wide spectral range (400–2200 nm) for both intensity and polarization | (i) Separation of submicron and supermicron particles<br>(ii) Spectral refractive index $\Rightarrow$ chemical composition |
| 1370 nm channel for both intensity and polarization | Detection and characterization of thin cirrus clouds and stratospheric aerosols |
| 2200 nm polarization channel | Characterization of the land surface contribution at visible wavelengths |
| 910 nm channel | Column water vapor amount |

The measurement approach required to ensure *high accuracy* in polarimetric observations employs a Wollaston prism to make *simultaneous* measurements of orthogonal intensity components from the same scene (Travis 1992), as illustrated in Plate 2.8. The field stop constrains the APS instantaneous field of view (IFOV) to $8 \pm 0.4$ mrad which, at the nominal Glory altitude (705 km), yields a geometric IFOV of 5.6 km at nadir. The spatial field is defined by the relay telescope and is collimated prior to the polarization separation provided by the Wollaston prism. This method guarantees that the measured orthogonal polarization states come from the same scene at the same time and allows the required polarimetric accuracy of 0.2% be attained. To measure the Stokes parameters that define the state of linear polarization ($I$, $Q$, and $U$), the APS employs a pair of telescopes with one telescope measuring $I$ and $Q$ and the other telescope measuring $I$ and $U$. This provides a redundant measurement set that increases the reliability of the APS data.

The *broad spectral range* of the APS is provided by dichroic beam splitters and interference filters that define nine spectral channels centered at the wavelengths $\lambda$ = 410, 443, 555, 670, 865, 910, 1370, 1610 and 2200 nm, as shown in Plate 2.9. Blue enhanced silicon detectors are used in the visible and near-infrared channels, while HgCdTe detectors, passively cooled to 160 K, are used in the short-wave infrared channels and offer the very high signal-to-noise ratio required to yield a polarimetric accuracy better than 0.3% for typical clear sky scenes over the dark ocean surface.

All spectral channels but the 910- and 1370-nm ones are free of strong gaseous absorption bands. The 1370-nm exception is centered at a major water vapor absorption band and is specifically intended for characterization of thin cirrus clouds and stratospheric aerosols. The locations of the other APS spectral channels are consistent with an optimized aerosol retrieval strategy because they take advantage of several natural circumstances such as the darkness of the ocean at longer wavelengths in the visible and near-infrared, the lower land albedo at shorter visible wavelengths, and the potential for using the 2200-nm band to characterize the land surface contribution at visible wavelengths. The 910-nm band provides a self-contained capability to determine column water vapor amount.

The critical ability to view a scene from *multiple angles* is provided by scanning the APS IFOV along the spacecraft ground track (Plate 2.7) with a rotation rate of 40.7 revolutions per minute with angular samples acquired every $8 \pm 0.4$ mrad, thereby yielding ~250 scattering angles per scene. The polarization-compensated scanner assembly includes a pair of matched mirrors operating in an orthogonal configuration and has been demonstrated to yield instrumental polarization less than 0.05%. The APS viewing angle range at the earth is from $+60°$ to $-80°$ with respect to nadir.

The scanner assembly also allows a set of *calibrators* to be viewed on the side of the scan rotation opposite to the Earth. The APS on-board references provide comprehensive tracking of polarimetric calibration throughout each orbit, while radiometric stability is tracked monthly to ensure that the aerosol and cloud retrieval products are stable over the period of the mission.

## 2.5. Methodology of polarimetric observations

According to the definitions introduced in Section 2.2, the majority of polarization measurements are reduced to the measurement of the difference between two intensity components with different polarization states and are a particular case of photometric observations. However, polarimetry has several distinctive features, the main one being that the difference between two intensities can be very small compared to the intensities themselves. Furthermore, one needs to correct very carefully for instrumental and atmospheric effects distorting the measurement results. There are also significant differences in the measurement process and data analysis.

### *2.5.1. Analysis of random and systematic errors*

***Random errors of polarimetric measurements.*** As with any radiometric measurement, the accuracy of polarimetric measurements is inherently limited by random fluctuations of the number of photoelectrons (pulses) owing to the quantum nature of the photoelectric effect. Almost always this quantum fluctuations are accompanied by random fluctuations of the signal caused by atmospheric (scintillations and image jitter) and instrumental (detector noise) effects. Furthermore, some polarization measurement techniques have their specific sources of random errors. For a prudent planning of observations and data analysis, one needs to estimate the ultimate accuracy feasible in specific conditions. Ефимов (1970) and Шаховской (1971) presented a detailed analysis of common sources of random errors and their effects on the resulting polarimetric accuracy.

*Atmospheric scintillations.* Relative changes in the brightness of a point-like object caused by scintillations depend on the diameter $D$ of the telescope used for observations. According to Ефимов (1970), the corresponding error in the degree of linear polarization is given by

$$\sigma_P \approx 1.1 D^{-1} (\Delta t)^{-1/2}, \tag{2.14}$$

where $D$ is measured in centimeters, while the cumulative observation time $\Delta t$ is expressed in seconds. For example, $D = 100$ cm and $\Delta t = 100$ s yield $\sigma_P = 0.11\%$, which agrees very well with published data on the minimal polarization error achieved with single-channel polarimeters in observations of bright stars. The contribution due to scintillations to the cumulative signal fluctuations becomes significant for telescopes with diameters 50–100 cm and objects brighter than $10^{\mathrm{m}}$. For objects brighter than $6^{\mathrm{m}}-7^{\mathrm{m}}$, scintillations are the main source of random errors.

*Shot noise.* According to the Poisson law, the square-root error in the registration of $N$ photoelectrons equals $N^{-1/2}$, which implies that the error in the degree of linear polarization caused by the shot noise of the detector is $\sigma_P \approx N^{-1/2}$. Let us assume that the energy losses in the atmosphere, telescope, and instrument are 50% and that the quantum yield of the photomultiplier is 0.1. Then the average polarization error in the spectral band B is given by

$$\sigma_P \approx 1.45\times10^{-2} \times 10^{0.2m} \times D^{-1} (\Delta t)^{-1/2}, \tag{2.15}$$

where $m$ is the magnitude of the object. For example, $m = 12^{\mathrm{m}}$, $D = 100$ cm, and $\Delta t = 100$ s yield $\sigma_P = 0.4\%$. For telescopes with diameters 50–100 cm, the shot noise dominates the total polarization error for objects with $m = 8^{\mathrm{m}} - 14^{\mathrm{m}}$. Polarization errors for brighter objects are dominated by scintillations, while those for fainter objects are dominated by the background skylight and background current.

*Shot noises due to background skylight and dark current.* The contribution of these sources of noise to the cumulative random error dominates if the ratio of the useful signal $N_*$ to the background skylight signal $N_{\mathrm{b}}$ is very small. Then $\sigma_P \approx N_{\mathrm{b}}^{-1/2} N_*^{-1}$. Assuming the above parameters of the atmosphere, the telescope, and the instrument as well as a 10″ diaphragm and a 50 pulse/s dark current, and also assuming that $\Delta t$ is the same for the separate polarization measurements of the object and the background skylight, we have:

$$\sigma_P \approx 10^{-3} \times D^{-1} (\Delta t)^{-1/2} [9.7 \times 10^{-5} + 1.9 D^{-2}]^{1/2} 10^{0.4m}. \tag{2.16}$$

The first and second terms in the square brackets correspond to the background sky and the dark current, respectively. For example, $m = 16^{\mathrm{m}}$, $D = 100$ cm, and $\Delta t = 100$ s yield $\sigma_P = 4\%$.

*Total random error*. Assuming again the above parameters of the atmosphere, the telescope, and the instrument, we have for the cumulative random error of the measured degree of polarization:

$$\sigma_P \approx D^{-1} (\Delta t)^{-1/2} [1.1 + 2.1 \times 10^{-4} \times 10^{0.4m} + 9.7 \times 10^{-11} \times 10^{0.8m} + 1.9 \times 10^{-6} \times D^{-2} \times 10^{0.8m}]^{1/2}. \tag{2.17}$$

The four terms inside the square brackets correspond to atmospheric scintillations, the detector shot noise, the background skylight, and the dark current, respectively.

Let us now assume that the measurement is based on counting the photoelectron pulses and involves a continuously rotating polarizer. Then the random polarization error (without the contribution of atmospheric scintillations) can be estimated from the following formula (Шаховской и Ефимов 1976):

$$\sigma_P = \frac{\pi}{2} \sqrt{\frac{1}{\sum N_*} \left( 1 + \frac{1 + \tilde{t}}{R} \right)}, \tag{2.18}$$

where $\sum N_*$ is the number of photoelectron pulses accumulated for the object, $\tilde{t} = \sum \Delta t_* / \sum \Delta t_{\mathrm{b}}$ is the ratio of the accumulation times for the object and the background sky, and $R = \bar{n}_* / \bar{n}_{\mathrm{b}}$ is the ratio of the average pulse rates for the object and the background sky. This formula or its straightforward modifications are used by the majority of observers all over the world. It gives the mean-square error of the degree of polarization defined by the pulse statistics for the object and the background sky. Equation (2.18) yields the following commonly used formula for the optimal ratio of the accumulation times for the object and the background sky:

$$\tilde{t}_{\mathrm{opt}} = \left( \frac{\sum \Delta N_*}{\sum \Delta N_{\mathrm{b}}} \right)_{\mathrm{opt}} = \sum (1 + R) = \sqrt{\frac{\bar{n}_* + \bar{n}_{\mathrm{b}}}{\bar{n}_{\mathrm{b}}}}. \tag{2.19}$$

***Systematic errors of polarimetric measurements.*** High-quality polarimetric observations of astronomical objects must be based on a careful analysis of and correction for all systematic errors. The main sources of systematic errors are the following: polarization of the background skylight; instrumental polarization; instrumental depolarization; relative errors in the radiometric calibration of different channels (for differential measurement techniques); imperfections of the polarization analyzer; errors in the zero point of position angles; and some others.

1. Polarization of the background skylight must be accounted for in all types of polarization measurements. Because of the additivity of the Stokes parameters *I*, *Q*, and *U*, one could subtract the polarization parameters measured separately for the background sky from the cumulative polarization parameters measured for the object and the background sky together. However, the relatively small energy flux coming from the background sky causes low measurement accuracy and a significant scatter of the measured polarization values. It is, therefore, advantages to use techniques based on a direct measurement of the difference between the respective polarization parameters.

2. Instrumental polarization is a false polarization registered in polarimetric observations of unpolarized sources of light. This error has many components such as polarization by the optical elements, polarization sensitivity of the detector, false modulation caused by the rotation of the polarizer, etc. Most of these components and their specific effects were discussed by Шаховской (1971). In practice, it is advantageous to correct for the cumulative effect of these sources. It is important to make sure that the instrumental polarization is stable and small.

3. Instrumental depolarization manifests itself as a decrease in the degree of linear polarization of the incoming light after it has passed through the optical tract of the telescope and polarimeter. The main cause of depolarization is the conversion of linear polarization into elliptical polarization owing to an artificial phase shift between the linear polarization components induced by various optical elements. The magnitude of depolarization depends on the orientation of the polarization plane of the incoming light. Unfortunately, it is difficult to decouple instrumental depolarization from other systematic errors.

4. A systematic error in the degree of polarization can be caused by radiometric calibration errors (for differential types of measurements) or by a less then 100% polarizing efficiency of the polarizer (for absolute types of measurements). In order to identify and correct for these errors, it is necessary to compare the observed polarization parameters for select standard stars with their catalog values (see Heiles (2000) and references therein).

5. Systematic polarization errors can also be caused by errors in the zero point of position angles specifying the orientation of the polarization plane. The relation between polarization parameters computed with respect to reference systems with different zero points of orientation angles is described by the well known rotation transformation law:

$$q = q'\cos 2\delta - u'\sin 2\delta, \qquad u = q'\sin 2\delta + u'\cos 2\delta, \tag{2.20}$$

where $\delta$ is the angle between the respective zero points.

### 2.5.2. Reduction of polarimetric data

In general, the relation between the actual polarization parameters and those distorted by the cumulative effect of the various systematic errors can be written as follows:

$$q = A_1 q' + B_1 u' + C_1, \qquad u = A_2 q' + B_2 u' + C_2. \tag{2.21}$$

By forming a sufficiently large number of such equations using observations of standard stars with different polarization parameters and then solving the resulting system of equations using the least squares technique, one can find the most probable values of the coefficients $A_1$, $B_1$, $C_1$, $A_2$, $B_2$, and $C_2$. If the instrumental depolarization is equal to zero then $A_1 = B_2$ and $A_2 = -B_1$, whereas assuming that the error in the zero point of position angles is zero yields $A_1 + B_2 = k$ and $A_2 = B_1$. In this case one has to determine only four unknown coefficients. A detailed description of this reduction technique was given by Шаховской (1971). He also listed formulas relating the coefficients of Eqs. (2.21) with the individual sources of systematic errors.

In practice, it is recommended to use the following data deduction procedure. During the first iteration, it is assumed that the requisite correction for the zero point of position angles is zero: $\Delta\vartheta = 0°$. By observing stars with zero (or small, ~0.02%) degree of polarization, one determines approximate average parameters of the instrumental polarization $\overline{q}'_{\rm ins}$ and $\overline{u}'_{\rm ins}$ as well as their respective mean-square errors. Then by observing standard stars with large polarization one finds their approximate polarization parameters according to

$$q'_i = q_i^{\rm obs} - \overline{q}'_{\rm ins}, \qquad u'_i = u_i^{\rm obs} - \overline{u}'_{\rm ins}, \tag{2.22}$$

where $q_i^{\rm obs}$ and $u_i^{\rm obs}$ are the observed polarization parameters of standard star $i$. The individual corrections for the zero point of orientation angles are determined from

$$\Delta\vartheta_i = \vartheta_i^{\rm cat} \pm \vartheta_i^{\rm obs}, \tag{2.23}$$

where $\vartheta_i^{\rm cat}$ is taken from the polarimetric catalog of standard stars and $\vartheta_i^{\rm obs} = \mathrm{arctg}\,(u_i^{\rm obs}/q_i^{\rm obs})/2$, in accordance with Eq. (2.8). The sign in Eq. (2.23) is chosen such that the scatter of the $\Delta\vartheta_i$ values for the standard stars is minimized. The computation of the average value of the approximate correction $\overline{\Delta\vartheta'}$ is followed by a second iteration yielding the final values of the instrumental polarization parameters $\overline{q}_{\rm ins}$ and $\overline{u}_{\rm ins}$ and the correction for the zero point of orientation angles $\Delta\vartheta$. The reduction of operational polarization measurements is then performed according to Eqs. (2.22) and (2.23).

### 2.5.3. Statistical analysis of polarimetric observations

It was demonstrated by Шаховской (1994) that the correct analysis of polarimetric observations should be based on the normalized Stokes parameters $q$ and $u$ (and sometimes on the original Stokes parameters $I$, $Q$, and $U$) rather than on the conventional parameters $P_{\rm lp}$ and $\vartheta$ since the latter do not obey the normal law.

Therefore, the statistical analysis of the parameters $P_{lp}$ and $\vartheta$ requires special precautions.

It has been shown above that the directly measured polarimetric quantities are instrumental readings proportional to the numbers of registered photoelectrons corresponding to different orientations of the polarization analyzer. For example, according to Eq. (2.13), $q = \varkappa(N_1 - N_2)/(N_1 + N_2)$, where $\varkappa$ characterizes the analyzer efficiency and the type of modulation. Similar expressions can be written for the parameters $u$ and $v$. According to the Poissonian law, the parameters $q$, $u$, and $v$ would be normally distributed, and the average value of each parameter, calculated according to $\langle q \rangle = \sum q_i / n$ (and analogously for $u$ and $v$), would be equal to the corresponding most probable value. The dispersion of each parameter would be determined by the dispersion of the numbers $N_1$ and $N_2$ and, in the absence of the background skylight, would be given by $\sigma_{q,u,v} = \varkappa N^{-1/2} = (N_1 + N_2)^{-1/2}$. However, this is strictly valid only for large values of $N$ and small values of $q$, i.e., when $(N_1 - N_2) \ll (N_1 + N_2)$. In general, the value of $q$ is computed as the ratio of two normally distributed quantities and thus is not normally distributed. This is important since the normal law overestimates the probability of large deviations from the average value exceeding $2\sigma$ and $3\sigma$. The discrepancies between the actual and normal distributions become especially significant in the presence of a non-Poissonian component of the noise (e.g., that caused by atmospheric scintillations). Therefore, it is preferable to average not the values of the parameters $q$, $u$, and $v$ but rather the read-ings $N_1$ and $N_2$ themselves. Then the best estimate of the parameter $q$ is given by the following ratio:

$$\langle q \rangle = \frac{\sum N_1 - \sum N_2}{\sum N_1 + \sum N_2}. \tag{2.24}$$

Equation (2.24) is equivalent to the computation of $q$ as a weighted average of the values $q_i$, the corresponding weights being given by the sum of the accumulated pulses ($w_i = N_i = N_{1i} + N_{2i}$):

$$\langle q \rangle = \sum q_i w_i / \sum w_i. \tag{2.25}$$

The distribution law for $P_{lp}$ depends not only on the dispersions $\sigma_q$ and $\sigma_u$, but also on the actual value of the degree of linear polarization $P_{lp}^{a}$. Therefore, if one needs the average values of the degree of linear polarization and the orientation (position) angle $\vartheta$ then they must be determined from

$$\langle P_{lp} \rangle = |\langle q \rangle^2 + \langle u \rangle^2|^{1/2}, \qquad \langle \vartheta \rangle = \frac{1}{2}\mathrm{arctg}|\langle u \rangle / \langle q \rangle|, \tag{2.26}$$

where the average values of the normalized Stokes parameters are found from Eq. (2.24). The value of $\langle P_{lp} \rangle$ calculated according to Eq. (2.26) still differs from the actual value of the degree of linear polarization $P_{lp}^{a}$. Therefore, the use of the parameter $P_{lp}$ in a statistical analysis of polarization measurements is justified only if the ratio $P_{lp}$ to the corresponding standard deviation $\sigma_P$ is sufficiently large.

The distribution law for the polarization orientation angle $\vartheta$ is even more complex and is determined by many parameters depending on both $P_{lp}^{a}$ and $\vartheta$. In prac-tice, the corresponding error $\sigma_\vartheta$ can be estimated in the first approximation by assuming that $\sigma_q = \sigma_u = \sigma_P$ and using the following formulas:

$$\sigma_\vartheta = \pi/12^{1/2}\,\text{rad} \approx 52^\circ \quad \text{for} \quad P_{\text{lp}} \ll \sigma_P, \tag{2.27a}$$

$$\sigma_\vartheta = \sigma_P/2\langle P_{\text{lp}}\rangle \approx 28.65\,\sigma_P/\langle P_{\text{lp}}\rangle \quad \text{for} \quad P_{\text{lp}} \gg \sigma_P. \tag{2.27b}$$

The members of the CrAO scientific school of polarimetric studies have always paid special attention to the careful determination of errors in the measured polarization parameters. As a rule, the errors are estimated in two ways. First of all, the errors $\sigma_q = \sigma_u = \sigma_P$ are evaluated from the statistics of the accumulated pulses using Eq. (2.18). Another estimate of the polarization error is provided by the mean-weighted values of the mean-square errors of the parameters $\overline{q}$ and $\overline{u}$. These errors are defined in terms of deviations of the corresponding parameters for separate measurements from their average values computed for the entire series of observations:

$$\sigma_{\overline{q}} = \sqrt{\frac{\sum (q_i - \overline{q})^2 N_{*i}}{(n-1)\sum N_*}}, \qquad \sigma_{\overline{u}} = \sqrt{\frac{\sum (u_i - \overline{u})^2 N_{*i}}{(n-1)\sum N_*}}. \tag{2.28}$$

Here, $q_i$ and $u_i$ are the normalized Stokes parameters calculated for the $i$th measurement of a series; each weighting factor is given by the corresponding number of pulses $N_{*i}$; $\overline{q}$ and $\overline{u}$ are average values of these parameters calculated for the entire series of $n$ observations; and $\sum N_*$ is the total number of pulses accumulated during the entire series of measurements. The final step is to calculate the average values of the degree of linear polarization and position (orientation) angle of the polarization plane,

$$\overline{P}_{\text{lp}} = \sqrt{\overline{q}^2 + \overline{u}^2}, \qquad \overline{\vartheta} = \frac{1}{2}\,\text{arctg}\,\frac{\overline{u}}{\overline{q}}, \tag{2.29}$$

as well as their respective average errors,

$$\sigma_{\overline{P}} = \sqrt{0.5(\sigma_{\overline{q}}^2 + \sigma_{\overline{u}}^2)}, \qquad \sigma_{\overline{\vartheta}} = 28.65\,\frac{\sigma_{\overline{P}}}{\overline{P}_{\text{lp}}}. \tag{2.30}$$

The greater of the two average errors serves as the final estimate of the error of polarization measurements.

### *2.5.4. Methods of polarimetric observations of small Solar System bodies*

Usually polarimetric observations are performed according to the sequence "background sky – object, …, object – background sky". The observation of the object is preceded and followed by the calibration of the instrument in order to make sure that the registration times for all counting channels are exactly the same. Observations of bright objects during a moonless night usually require only one observation of the background sky. Observations of faint objects at large zenith angles and/or during moonlit nights necessitate frequent observations of the back-ground sky. The latter are then interpolated for the specific moments of object observations. The requisite accumulation time in the observation of an object depends on the required polarimetric accuracy. The optimal accumulation time for the background sky with respect to that for the object is specified by Eq. (2.19).

***Specific aspects of polarimetric observations of Galilean satellites.*** The Galilean satellites of Jupiter are relatively bright ($\sim 5.0^{m} - 5.7^{m}$), which allows one to perform polarimetric observations at a 0.7-m telescope with a ~0.02% accuracy and a ~20 min accumulation time. However, the measured degree of linear polarization depends strongly on the type of procedure for the measurement of the background skylight polarization. Indeed, the proximity of Jupiter makes the background skylight very inhomogeneous and strongly polarized. Therefore, the search for subtle effects at small phase angles requires the use of adequate observation methodologies. In particular, the background sky must be observed before and after the observation of the object on both sides of the object and at the same distance in the direction perpendicular to the direction "planet – satellite" because the intensity and polarization variations of the background skylight are minimal in the radial direction relative to the center of the planet.

***Practical aspects of measurements of linear and circular polarization for comets.*** Polarimetric observations of comets are performed using the following technique. The focal diaphragm of the polarimeter is positioned with respect to the photometric center of the comet in such a way that owing to its proper movement the cometary nucleus passes exactly through the center of the diaphragm. Repeating this procedure several times while keeping the instrument in the automatic measurement regime allows one to obtain a continuous temporal profile of intensity for each registration channel. The resulting signal-to-noise ratio can be improved by repeating the entire measurement cycle. By superposing the intensity maxima of the different profiles and factoring in the proper velocity of the comet, one can perform a time-to-distance conversion. As a result, one can determine the distribution of the intensity in each registration channel along the line through the nucleus in the direction of the proper velocity vector as well as the corresponding distributions of the total intensity and the parameters of linear and circular polarization.

***Observations of stellar occultations by comets.*** Polarimetric and photometric observations of stellar occultations must include the measurement of the total radiation from the star and the coma as well as the measurement of the radiation from the same part of the coma in the absence of the star. The latter measurement was not performed during the stellar occultation by comet Levy (Rosenbush et al. 1994). Therefore, the brightness of the coma along the trajectory of the star was estimated based on photometric observations of the comet with different diaphragms centered at the nucleus.

The combined radiation coming from the comet and the star is usually weakly polarized because the strong and virtually unpolarized radiation flux from the star is superposed on the polarized yet weak radiation flux from the comet. Therefore, careful planning of polarimetric observations of stellar occultations by a comet should include accounting for the ratio of the brightness of the star to that of the comet.

## 2.6. Multicomponent superachromatic retarders: theory and implementation

One of the most important functional units of an astronomical polarimeter is an optical retarder which can include phase plates, prisms, electrooptical modulators, etc. Among the different types of retarders, phase plates are distinguished by their compactness, simplicity of design, and the absence of beam shifts. In order to provide broad spectral coverage, a retarder must have an achromatic design. Achromatic combinations based on different spectral dependences of the refractive index for the individual components are constrained by the actual properties of natural crystalline materials available. As a consequence, such systems usually involve three or fewer materials and have limited achromatic properties. It is, therefore, impossible to overstate the importance of the theory of multicomponent retarders developed by V. A. Kucherov. This theory can be used to design systems with any degree of achromatism by including a sufficiently large number of combinations of optical elements (Кучеров 1985, 1986a,b).

### *2.6.1. Extension of the Pancharatnam system to an arbitrary number of components*

In the general case, a system built of $N$ plates is described by a product of the matrix describing a phase plate with parameters equivalent to those of a simple retarder and a rotation matrix $\mathbf{R}_{\Omega}$, where the angle $\Omega$ depends on the parameters of the individual components. This means that an arbitrary multicomponent retarder has, in general, different orientations $\Phi_1$ and $\Phi_2$ of the equivalent optical axes at the entrance and exit points, respectively. The transmission matrix of a multicomponent plate is given by

$$\mathbf{M}(\Phi_2, \Delta, \Phi_1) = \mathbf{R}_{\Phi_2}\mathbf{L}_{\Delta}\mathbf{R}_{\Phi_1}. \tag{2.31}$$

Here, $2\Delta$ is the cumulative phase shift,

$$\mathbf{L}_{\delta} = \begin{bmatrix} e^{i\delta} & 0 \\ 0 & e^{-i\delta} \end{bmatrix}$$

is the transmission matrix of a simple phase plate with a phase shift $2\delta$ with respect to its optical axis, and $\mathbf{R}_{\phi}$ describes the transformation of the electric field components upon a rotation of the reference frame around the propagation direction by an angle $\phi$ measured counter-clockwise if looking in the propagation direction:

$$\mathbf{R}_{\phi} = \begin{bmatrix} \cos\phi & -\sin\phi \\ \sin\phi & \cos\phi \end{bmatrix}.$$

The transition from the reference frame of the incoming beam to that of the outgoing beam is accomplished by changing the sign of the corresponding rotation angle of the optical axis of the retarder. For example, for a simple phase plate one has:

$$\phi_2 = -\phi_1. \tag{2.32}$$

Let us now consider the problem of achromatization. Since the equivalent parameters of a multicomponent system are functions of the parameters of the individual components, it may be possible to select the latter in such a way that the former has a minimal spectral dependence in a requisite wavelength interval. We will consider only systems in which all components are made of the same material. Let us assume that as a result of external factors (e.g., a change in the incoming wavelength $\lambda$ or in the ambient temperature) the phase shifts of the individual components have changed by the same relative amount. This deviation is conveniently characterized by a so-called achromatization parameter $x$:

$$x = \frac{\delta_1}{\tilde{\delta}_1} = \ldots = \frac{\delta_N}{\tilde{\delta}_N}, \tag{2.33}$$

where the tildes denote the nominal parameters of the components at the central wavelength $\tilde{\lambda}$ of the requisite spectral interval.

In the first approximation, the phase shift of a separate plate is inversely proportional to the wavelength: $x \approx \tilde{\lambda}/\lambda$. In order to render a multicomponent system with the requisite equivalent parameters achromatic, one needs to ensure that a number of $x$-derivatives of the equivalent parameters vanish at $x = 1$. In the matrix form, we have:

$$\begin{aligned} \mathbf{M}(\Phi_2, \Delta, \Phi_1)\Big|_{x=1} &= \mathbf{R}_{\tilde{\Phi}_2} \mathbf{L}_{\tilde{\Delta}} \mathbf{R}_{\tilde{\Phi}_1}, \\ \mathbf{M}^{(j)}(\Phi_2, \Delta, \Phi_1)\Big|_{x=1} &= \mathbf{0}, \end{aligned} \tag{2.34}$$

where $\mathbf{0}$ is a $2\times2$ zero matrix, and $j = 1, 2, \ldots, j_{\max}$ is the order of the derivative. A system built of $N$ plates has $2N$ variable parameters, which must satisfy $3(1 + j_{\max})$ conditions. Therefore, for a given $N$ we have $j_{\max} \le \mathrm{E}\{2N/3 - 1\}$, where $\mathrm{E}\{a\}$ is the integer part of $a$. This means that the minimal number of components required to zero out the first three derivatives of the equivalent parameters of the system is three.

Let us subdivide an arbitrary system of phase plates into three nominal parts, each having an arbitrary number of components, and denote them $a$, $b$, and $c$. Then, according to Eq. (2.31),

$$\mathbf{M}(\Phi_2, \Delta, \Phi_1) = \mathbf{R}_{2a}\mathbf{L}_a\mathbf{R}_{1a}\mathbf{R}_{2b}\mathbf{L}_b\mathbf{R}_{1b}\mathbf{R}_{2c}\mathbf{L}_c\mathbf{R}_{1c}. \tag{2.35}$$

Now we can evaluate the first derivative of this matrix by taking into account that the differentiation of the matrices $\mathbf{R}$ and $\mathbf{L}$ is equivalent to the action of the following matrix operators:

$$\begin{aligned} \mathbf{R}'_\phi &= \mathbf{H}\phi'\mathbf{R}_\phi, \\ \mathbf{L}'_\delta &= \mathrm{i}\mathbf{G}\delta'\mathbf{L}_\delta, \end{aligned} \tag{2.36}$$

where

$$\mathbf{G} = \begin{bmatrix} 1 & 0 \\ 0 & -1 \end{bmatrix}, \quad \mathbf{H} = \begin{bmatrix} 0 & -1 \\ 1 & 0 \end{bmatrix}.$$

Differentiating the matrix $\mathbf{M}(\Phi_2, \Delta, \Phi_1)$ with respect to $x$, equating the result to zero, left- and right-multiplying the matrix $\mathbf{M}'$ by nondegenerate inverse matrices according to

$$\mathbf{R}_{2b}^{-1}\mathbf{R}_{1a}^{-1}\mathbf{L}_a^{-1}\mathbf{R}_{2a}^{-1}\mathbf{M}'\mathbf{R}_{1c}^{-1}\mathbf{L}_c^{-1}\mathbf{R}_{2c}^{-1}\mathbf{R}_{1b}^{-1}\mathbf{L}_b^{-1} = \mathbf{0}, \tag{2.37}$$

and using commutation properties of the participating matrices yields (Кучеров 1985):

$$\begin{aligned}\phi'_{2a}\mathbf{R}_{2b}^{-1}\mathbf{R}_{1a}^{-1}\mathbf{L}_a^{-2}\mathbf{R}_{1a}\mathbf{R}_{2b}\mathbf{H} + \mathrm{i}\delta'_a\mathbf{R}_{1a}^{-2}\mathbf{R}_{2b}^{-2}\mathbf{G} + \phi'_{1a}\mathbf{H} + \phi'_{2b}\mathbf{H} + \mathrm{i}\delta'_b\mathbf{G} + \phi'_{1b}\mathbf{L}_b^2\mathbf{H}\\ + \phi'_{2c}\mathbf{L}_b^2\mathbf{H} + \mathrm{i}\delta'_c\mathbf{L}_b\mathbf{R}_{1b}^2\mathbf{R}_{2c}^2\mathbf{L}_b^{-1}\mathbf{G} + \phi'_{1c}\mathbf{L}_b\mathbf{R}_{1b}\mathbf{R}_{2c}\mathbf{L}_c^2\mathbf{R}_{2c}^{-1}\mathbf{R}_{1b}^{-1}\mathbf{L}_b\mathbf{H} = \mathbf{0}\end{aligned} \tag{2.38}$$

Let us now assume that the end plates of the system are simple. This means that the condition (2.32) is satisfied, and the following relations are valid:

$$\phi'_{2a} = \phi'_{1a} = \phi'_{2c} = \phi'_{1c}, \tag{2.39}$$

$$\delta'_a = \tilde{\delta}_a, \tag{2.40}$$

$$\delta'_c = \tilde{\delta}_c. \tag{2.41}$$

Equations (2.40) and (2.41) follow from Eq. (2.33). In what follows, the tildes will be omitted for the sake of brevity. Let us introduce the notation $2(\phi_{1a}+\phi_{2b}) = \alpha_{ab}$ and $2(\phi_{1b}+\phi_{2c}) = \alpha_{bc}$. Then Eq. (2.38) takes the following form:

$$\mathrm{i}\delta_a\mathbf{R}_{\alpha_{ab}}\mathbf{G} + \phi'_{2b}\mathbf{H} + \mathrm{i}\delta'_b\mathbf{G} + \phi'_{1b}\mathbf{L}_b^2\mathbf{H} + \mathrm{i}\delta_c\mathbf{L}_b\mathbf{R}_{\alpha_{bc}}\mathbf{L}_b^{-1}\mathbf{G} = \mathbf{0}. \tag{2.42}$$

By separating the real and imaginary parts of Eq. (2.42) and making several straightforward transformations, we obtain the following three independent equations:

$$\phi'_{2b} + \phi'_{1b}c_{2\delta_b} + \delta_c s_{2\delta_b}s_{\alpha_{bc}} = 0, \tag{2.43a}$$

$$\delta_a c_{\alpha_{ab}} + \delta'_b + \delta_c c_{\alpha_{bc}} = 0, \tag{2.43b}$$

$$\delta_a s_{\alpha_{ab}} + \phi'_{1b}s_{2\delta_b} - \delta_c c_{2\delta_b}s_{\alpha_{bc}} = 0, \tag{2.43c}$$

where $c_\alpha = \cos\alpha$ and $s_\alpha = \sin\alpha$. These imply that the simplest necessary achromatization conditions are the following:

$$2\delta_b = \pi, \tag{2.44}$$

$$\phi_{2b} - \phi_{1b} = \text{constant}. \tag{2.45}$$

Let us now consider the matrix equation (2.34) and assume that the internal element is a half-wave plate. Substituting Eqs. (2.35) and (2.44) and denoting

$$\phi_{2a} - \phi_{1a} = \beta_a, \tag{2.46}$$

$$\phi_{2b} - \phi_{1b} = \beta_b, \tag{2.47}$$

$$\phi_{2c} - \phi_{1c} = \beta_c \tag{2.48}$$

yields the following matrix relations:

$$s_{\delta_a} c_{\delta_c} \mathbf{R}_{\beta_a - \beta_b} + c_{\delta_a} s_{\delta_c} \mathbf{R}_{\beta_b - \beta_c} = -c_{\Delta} \mathbf{R}_{\Phi_2 + \Phi_1}, \tag{2.49a}$$

$$c_{\delta_a} c_{\delta_c} \mathbf{R}_{\beta_b} - s_{\delta_a} s_{\delta_c} \mathbf{R}_{\beta_a - \beta_b + \beta_c} = s_{\Delta} \mathbf{R}_{\Phi_2 - \Phi_1}. \tag{2.49b}$$

Four independent scalar equations of this system along with Eq. (2.43) form a system of equations for seven parameters defining the achromatization problem: two parameters for each end plate and three parameters for the internal composite element $\phi_{1a}$, $\delta_a$, $\phi_{1c}$, $\delta_c$, $\phi_{1b}$, $\delta_b$, and $\phi_{2b}$.

Let us assume for simplicity that the entire system of plates as well as the internal composite plate are equivalent to simple plates, i.e., satisfy the condition (2.32). Then isolating the $(2,1)$ matrix element from the matrix relation (2.49) and using Eqs. (2.32) and (2.46)–(2.48) yields the following scalar equation:

$$s_{\delta_a} c_{\delta_c} s_{\alpha_{ab}} + c_{\delta_a} s_{\delta_c} s_{\alpha_{bc}} = 0. \tag{2.50}$$

This implies that $\delta_a = \delta_c$, while Eqs. (2.43c) and (2.44) yield one more symmetry condition: $\alpha_{ab} = \alpha_{bc}$.

Hence, to satisfy both conditions (i.e., the achromatization requirement $\mathbf{M}' = \mathbf{0}$ and the symmetry $\Phi_2 = -\Phi_1$) it is sufficient under normal conditions that the internal plate be equivalent to a simple plate with a half-wave phase shift and a constant (up to the first derivative) orientation of the axis, while the end plates have identical phase shifts and orientations.

The requirement of a symmetric accommodation of the end components with respect to the central plate constitutes a significant simplification of the achromatization problem. Furthermore, the internal half-wave plate can be built of an odd number of half-wave components. For the specific case of symmetric composite phase plates, V. A. Kucherov derived the following generalized expression for a phase plate consisting of an arbitrary number of symmetrically positioned elements:

$$\mathbf{M}(N) = A_k \mathbf{J} + \mathrm{i}\mathbf{G}\mathbf{R}_{2\phi_k}(B_k \mathbf{J} + C_k \mathbf{H}), \tag{2.51}$$

where

$$\mathbf{J} = \begin{bmatrix} 1 & 0 \\ 0 & 1 \end{bmatrix}.$$

The equivalent phase shift and the orientation of the optical axis are given by

$$\Delta_k = \arccos A_k, \qquad \Phi_k = \phi_k + \frac{1}{2}\,\mathrm{arcctg}\,\frac{B_k}{C_k}. \tag{2.52}$$

The coefficients $A_k$, $B_k$, and $C_k$ can be found from the recurrent relations

$$\begin{bmatrix} A_k \\ B_k \end{bmatrix} = \mathbf{R}_{\tau_k} \begin{bmatrix} A_{k-1} \\ D_k \end{bmatrix}, \tag{2.53}$$

$$\begin{bmatrix} D_k \\ C_k \end{bmatrix} = \mathbf{R}_{\alpha_k} \begin{bmatrix} B_{k-1} \\ C_{k-1} \end{bmatrix}, \tag{2.54}$$

where

$$\tau_k = 2\delta_k, \qquad \alpha_k = 2(\phi_{k-1} - \phi_k), \tag{2.55}$$

supplemented by the initial conditions (Кучеров 1985)

$$\begin{aligned} A_0 &= c_{\delta_0}, \\ B_0 &= s_{\delta_0}, \\ C_0 &= 0. \end{aligned} \tag{2.56}$$

### *2.6.2. Practical implementation of the theory of multicomponent superachromatic retarders*

Based on the above theory, V. A. Kucherov performed a structural analysis of three-, five-, and seven-element composite systems for different values of the phase shift and determined their parameters. The five-element systems are considered to be the most promising from the perspective of their practical implementation and have been analyzed specifically in terms of optimizing their optical properties. In particular, the analysis has yielded optimal orientation angles for the individual components $\tau_i$ serving to broaden significantly the spectral range of achromatization. Furthermore, it has been shown that prior achromatization of the individual components improves the achromatization of the entire system. The errors caused by imperfections of the phase plates have also been analyzed. It has been found that in order to achieve the requisite accuracy of astronomical observations, the tolerance for the thickness of the individual components must be 100−200 nm and that for the orientation of their optical axes must be 1′−2′. As a result, it has become possible to build unique superachromatic retarders with optical parameters stable to

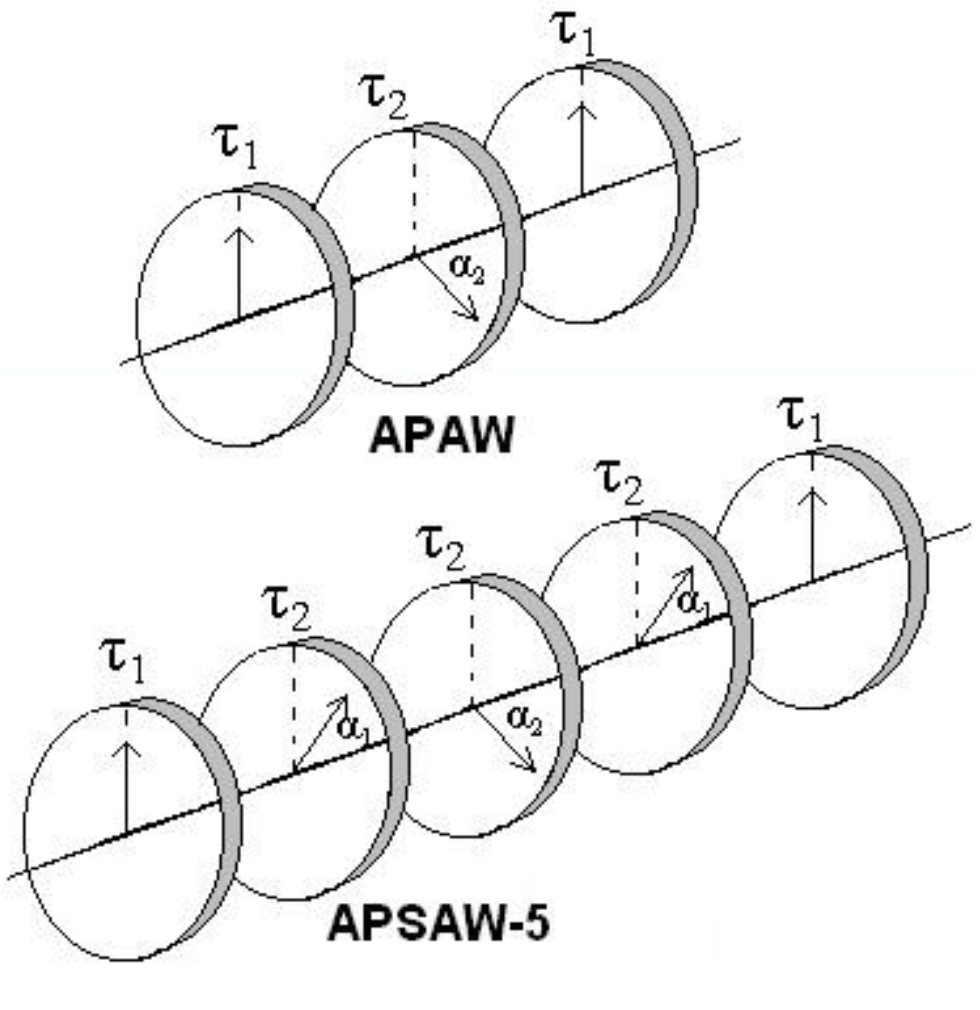

**Fig. 2.5.** A three-component achromatic phase plate (top panel) and a five-element superachromatic phase plate (bottom panel).

**Table 2.3.** Parameters of phase plates built by the company "Astroprylad"

| | |
|---|---|
| Anisotropic material | Set of anisotropic polymeric plates |
| Material of the exterior windows | Well-annealed optical glass |
| Phase shift | $\lambda/4$, $\lambda/2$, $\lambda/2.83$ (90°, 180°, 127°) |
| Accuracy of the phase shift | $\pm\lambda/100$ |
| Distortions of the wave front | $0.5\lambda$ per 1 cm at $\lambda = 632.8$ nm |
| Operational temperature range | From –20° to +50° |
| Beam shifts | < 5 arc sec |
| Working diameter | 15, 20, 25, 30, 40, 50, and 60 mm |
| Light diameter | $\geq 90\%$ of the outer diameter |

±2% or ±3% with respect to their nominal values over the entire spectral range 330−1400 nm. Figure 2.5 illustrates the positioning of the constituent optical elements of a three-component achromatic phase plate and a five-element superachromatic phase plate built according to the Kucherov theory.

Another important result concerns the design of a special plate with a phase shift of $\lambda/2.83$ ($\approx 127°$). Equation (2.12) implies that the efficiencies of measuring the parameters $q$, $u$, and $v$ are different, which affects the measurement accuracy in certain types of observation. A plate with the 127° phase shift allows one to perform measurements of linear and circular polarization with the same efficiency (Кучеров 1986b). Such a plate has been specifically designed, built, and used in the MAO polarimeter called Planetary Patrol (Бугаенко и Гуральчук 1985).

The theoretical studies by V. A. Kucherov have enabled an efficient practical implementation of achromatic and superachromatic $\lambda/4$ and $\lambda/2$ retarders as well as special retarders with the $\lambda/2.83$ phase shift. The unique properties of these phase plates built by the company "Astroprylad" (Samoylov et al. 2004) are summarized in Table 2.3. These products are highly valued and have been used in polarimetric observations worldwide, in particular in CrAO, the Institute of Astronomy of the KhNU, MAO, and the Astronomical Observatory of the Odessa University as well as in astronomical observatories of Bulgaria, Brazil, Chile, Finland, Germany, Italy, Japan, Korea, The Netherlands, Russia, Sweden, and Switzerland.

# 3

# Photopolarimetric observations of atmosphereless Solar System bodies and planets and their interpretation

In the Introduction and Section 2.1 we have formulated the five primary steps involved in a systematic and comprehensive approach to the solution of a specific geophysical or astrophysical problem by means of remote sensing. Steps 1 through 4 have been addressed in the first two chapters. In this and the following chapter we discuss the final step, i.e., the acquisition of extensive remote-sensing data with ground-based, aircraft, and spacecraft photopolarimeters and their interpretation using the previously developed theoretical analysis tools.

## 3.1. Planets and satellites

### *3.1.1. Terrestrial aerosols*

***Global Aerosol Climatology Project (GACP).*** We have already mentioned in Section 2.1.2 that AVHRR is not an instrument specifically designed for accurate retrievals of aerosol characteristics from a satellite orbit. Indeed, this radiometer measures only intensity. Furthermore, it does that in only two closely spaced visible and near-infrared spectral channels (centered at $\lambda \approx 650$ and 850 nm) and for only one scattering geometry per scene. The latter depends on the season, orbit, and scene geographical coordinates. This type of measurement necessitates the simplest retrieval algorithm yielding estimates of only the column optical thickness of aerosols $\tau$ and their average size (Mishchenko et al. 1999b). All other model parameters are assumed to be "known" and have to be fixed at values providing the best overall accuracy of the aerosol optical thickness (AOT).

However, despite their obvious limitations, AVHRR instruments on board of NOAA weather satellites remain a unique source of climatological information about aerosol properties owing to the extensive duration of their combined data record and the global coverage of the Earth (Li et al. 2009). Having this in mind, we have developed an aerosol retrieval algorithm based on the analysis of channel-1 and -2 AVHRR data over the oceans (Mishchenko et al. 1999b, 2003; Geogdzhayev et al. 2002, 2004, 2005) and applied it to the so-called International Satellite Cloud Climatology Project (ISCCP) DX radiance dataset composed of calibrated and sampled AVHRR radiances (Rossow and Schiffer 1999). Data from the following sun-synchronous polar-orbiting platforms have been included: NOAA-7 (August 1981–January 1985), NOAA-9 (February 1985–October 1988), NOAA-11 (November 1988–September 1994), NOAA-14 (February 1995–June 2001), and NOAA-16 (October 2001–present). The resulting public-domain aerosol database developed in the framework of the NASA/World Climate Research Program's GACP is available on-line at http://gacp.giss.nasa.gov.

The GACP retrieval algorithm yields the AOT and the Ångström exponent $A$ for each cloud-free ISCCP pixel by minimizing the difference between two AVHRR radiances measured in the 650- and 850-nm channels at the specific illumination and observation angles determined by the satellite orbit, on the one hand, and the radiances computed theoretically for a realistic atmosphere–ocean model, on the other hand. The Ångström exponent characterizes aerosol particle size in terms of the spectral dependence of aerosol extinction and is defined as

$$A = - \left. \frac{\mathrm{d}[\ln \langle C_{\mathrm{ext}}(\lambda) \rangle_{\xi}]}{\mathrm{d}(\ln \lambda)} \right|_{\lambda = \lambda_1}, \tag{3.1}$$

where $\lambda_1 = 650$ nm is the nominal wavelength of the AVHRR channel 1 and $\langle C_{\mathrm{ext}} \rangle_{\xi}$ is the ensemble-averaged extinction cross section per particle. The GACP aerosol product consists of the AOT reported at $\lambda = 550$ nm and the constrained Ångström exponent (Geogdzhayev et al. 2002). It is limited to areas over large water bodies such as oceans, seas, and lakes, in which case the surface reflectance is often low and can be parameterized with sufficient accuracy.

We have already indicated that with only two spectral radiance values per pixel available, the GACP retrieval algorithm cannot be expected to yield any additional parameters besides the AOT and $A$ and thus must rely on a number of assumptions which fix all other atmosphere and surface model parameters globally and permanently. In particular, GACP retrievals are based on the assumption that aerosol particles are homogeneous spheres with a wavelength-independent refractive index 1.5 + i0.003 and obey a monomodal modified power law size distribution given by

$$n(r) = \begin{cases} C, & r \le r_1, \\ C\left(\dfrac{r}{r_1}\right)^{-\tilde{\alpha}}, & r_1 < r \le r_2, \\ 0, & r > r_2 \end{cases} \tag{3.2}$$

with $r_1 = 0.1\,\mu\mathrm{m}$, $r_2 = 10\,\mu\mathrm{m}$, and $\tilde{\alpha} \in [2.5, 5]$. The normalization constant $C$ is chosen such that

$$\int_0^\infty \mathrm{d}r\, n(r) = 1, \tag{3.3}$$

while the above range of power-exponent values translates into a representative range of $A$. The radiance contributions of the upwelling radiation from within the ocean body and of the white caps are modeled with a small constant Lambertian component, while the surface slope distribution is determined using the Cox–Munk relation and corresponds to a globally uniform wind speed value of 7 m/s.

The single-scattering properties of the aerosol polydispersion are computed with the Lorenz–Mie code described by Mishchenko et al. (2002a). The multiple-scattering computations are performed using the scalar version of the adding–doubling technique (Hansen and Travis 1974). The corresponding computer code incorporates the reflectance of the rough ocean surface via the so-called modified Kirchhoff approximation (Mishchenko and Travis 1997b) and the effect of water vapor, oxygen, and $CO_2$ absorption via the *k*-distribution technique (Lacis and Oinas 1991). For a complete description of the assumptions used in the algorithm and the cloud screening procedure we refer to Mishchenko et al. (1999b, 2003) and Geogdzhayev et al. (2002, 2004, 2005).

The GACP retrieval approach outlined above is consistent with the goal of minimizing long-term statistical errors in the global aerosol parameters derived by the algorithm. As such, it cannot preclude regional biases in areas dominated by aerosol types significantly different from the global model assumptions, for example, nonspherical dust and soot aerosols (Mishchenko et al. 2003; Liu and Mishchenko 2005; Dubovik et al. 2006). However, the primary GACP objective is to study long-term changes in the retrieved aerosol characteristics, and so we expect that the main conclusions drawn will hold despite potential systematic biases in the retrieved parameters.

Plate 3.1 shows the global and hemispherical monthly averages of the AOT and Ångström exponent over the oceans for the period August 1981 – June 2005 derived with the latest version of the GACP retrieval algorithm. Note that two periods of unavailable or unreliable AVHRR data between July 1994 and February 1995 and between April (January for the Ångström exponent) and August 2001 are excluded. For reference, the blue curve depicts the Stratospheric Aerosol and Gas Experiment (SAGE) record of the globally averaged stratospheric AOT. A specific analysis of the effects of stratospheric aerosols on the GACP aerosol climatology and a detailed comparison of the GACP and SAGE global aerosol records can be found in Geogdzhayev et al. (2004).

The two major AOT maxima in Plate 3.1 are caused by the stratospheric aerosols generated by the El Chichon (March 1982) and Mt Pinatubo (June 1991) eruptions, whereas the quasi-periodic oscillations in the GACP curves are the result of inter-annual aerosol variability. The overall behavior of the GACP AOT during the quiescent period from January 1986 to June 1991 hardly reveals any statistically significant tendency. Specifically, a linear model of the form $y = a_1x + b_1$ fitted to the clear pre-Pinatubo period of data suggests that the global column AOT value just before the eruption was close to 0.145. After the eruption, the GACP curves reveal complex tropospheric and stratospheric AOT temporal variations as well as a clear long-term decreasing trend in the tropospheric AOT first identified by Mishchenko et al. (2007c) (see also Zhao et al. 2008).

To analyse the changes during this latter period, Mishchenko and Geogdzhayev (2007) calculated a linear fit of the form $y = a_2x + b_2$ using data from January 1996 to June 2005 such that its extrapolation back to June 1991 yielded the previously found clear-period AOT value. This two-step procedure allowed them to exclude from the tendency analysis the period affected by the Pinatubo eruption. As a result, they found that the residual decrease in the global tropospheric AOT during the 14-year period from June 1991 to June 2005 was close to 0.033 (~ 0.0024 per year). The corresponding chi-square values indicated that the fitted parameters were reliable at a very high confidence level of over 99%.

Although there have been significant drifts in the equator crossing times for the individual AVHRR instruments (Brest et al. 1997; Ignatov et al. 2004), Plate 3.1 shows no obvious artifacts potentially attributable to the drifts. The only exception is the end of the NOAA-14 record when two factors caused the retrievals to become unreliable. First, the use of the pre-launch calibration of channel-2 radiances coupled with the likely strong degradation of the channel-2 sensitivity resulted in the spurious increase of the Ångström exponent. Second, the strong drift of the NOAA-14 orbit resulted in a partial loss of data and in a loss of coverage for much of the Southern Hemisphere, thereby causing a bias in the global and Southern-Hemisphere averages.

Overall, however, the GACP AOT record appears to be self-consistent, with no drastic intra-satellite variations, and is obviously consistent with the SAGE record. This seems to testify to the robustness of the ISCCP channel-1 radiance calibration and the GACP retrieval algorithm. This conclusion is reinforced by the close correspondence of the calculated and observed top-of-the-atmosphere solar fluxes (Zhang et al. 2004). Furthermore, GACP AOT retrievals have been successfully validated against precise sun-photometer data taken from 1983 through 2004 by employing a special procedure which, by design, tested the entire retrieval process as well as the radiance calibration (Liu et al. 2004; Smirnov et al. 2006).

According to Plate 3.1, the Ångström exponent appears to have decreased by ~ 0.3 over the duration of the GACP record. However, the retrieval accuracy for $A$ is expected to be worse than that for $\tau$ because the central wavelengths of AVHRR channels 1 and 2 are rather close (Mishchenko et al. 1999b). In addition, the Ångström exponent retrieval often "saturates" by yielding a value equal to either the upper or the lower boundary of the $A$ range afforded by the nominal $\alpha \in [2.5, 5]$ range of power exponent values in Eq. (3.2) (Geogdzhayev et al. 2002). Such retrievals are not included in the computation of monthly averages, which reduces the number of useful data points and the accuracy of the final result. Furthermore, the accuracy with which $A$ is retrieved depends on the relative calibration of channel-1 and -2 radiances. Since we use the ISCCP calibration of channel-1 radiances and the NOAA calibration of channel-2 radiances, additional studies may be necessary before the decreasing Ångström exponent trend is accepted as a definitive result.

The visualization of potential long-term regional aerosol trends is not as straightforward as that of the global averages because of the much larger amount of data to be displayed. Perhaps the simplest approach is to plot three-year averages computed for two quiescent, volcano-free periods: the one preceding the Mt Pinatubo eruption and the one covering the most recent period of the GACP record (Mishchenko and Geogdzhayev 2007). The results computed for the cumulative annual $1^\circ \times 1^\circ$ averages are depicted in Plates 3.2 and 3.3.

The bottom panel in Plate 3.2 reveals pronounced regional AOT changes that appear to have occurred between the late 1980s−early 1990s and the early 2000s.

Although there is a significant seasonal variability in these changes, the most obvious regional AOT trends can be summarized as follows (Mishchenko and Geogdzhayev 2007; Zhao et al. 2008):

- a significant decrease over much of Europe and especially over the Black Sea (cf. Geogdzhayev et al. 2005);
- a significant decrease over the part of the Atlantic Ocean most affected by dust aerosols originating in Africa;
- a noticeable increase along the part of the western coast of Africa most affected by biomass burning events; however, this trend appears to have strong seasonality and is even replaced by a decreasing tendency during the autumn;
- a significant increase along the southern and south-east coasts of Asia, especially for the summer months; and
- a significant increase over the 45°S−60°S latitudinal belt

These trends seem to be superposed on a fairly uniform global decrease of AOT causing the global and hemispheric downward trends in Plate 3.1.

Plate 3.3c also reveals noticeable regional changes in the Ångström exponent:

- a significant decrease over the areas of the Atlantic Ocean affected by the African dust and biomass burning aerosols;
- a substantial decrease over the Northern and Central Indian Ocean;
- a significant decrease over the equatorial part of the Pacific Ocean; and
- a noticeable increase over the 15°S−45°S latitudinal belt.

The strong AOT decrease over the large Atlantic Ocean area between the equator and 30°N appears to indicate a reduced productivity of the African sources of dust aerosols. This finding is quite consistent with the results of Herrmann et al. (2005) and Olsson et al. (2005) indicating that the amount of rain and vegetation in the Sahel has increased significantly since the late 1980s. An alternative (or additional) explanation may be that the dust particles have become larger and, as a result, spend less time in the atmosphere. The latter explanation would be consistent with the concurrent decrease of the Ångström exponent in this area, Plate 3.3.

The overall AOT increase along the western coast of Africa from the equator to about 20°S may be explained by intensified biomass burning. Importantly, the significant inter-annual variability of the AOT in this region perfectly correlates with expected changes in the number and extent of fires during the local dry and rainy seasons.

The noticeable AOT increase along the southern and south-east coasts of Asia can be the expected result of rapidly growing regional economies coupled with the widespread use of technologies that are not environmentally clean (Stern 2006; Streets et al. 2006; Ohara et al. 2007).

The elevated amounts of aerosols throughout the 40°S−60°S latitudinal belt seen in Plates 3.2a and 3.2b correlate almost perfectly with the Special Sensor Microwave/Imager (SSM/I) surface wind speed data. It is well known that increasing wind speed can lead to increased sea-salt aerosol production (Lewis and Schwartz 2004) as well as cause increased ocean surface reflectivity by both making the ocean surface more rough and generating more white caps. Since the GACP retrieval algorithm is based on a constant wind speed value and a constant diffuse component of the ocean surface reflectance, all of these factors would have the same effect on the GACP-derived AOT, thereby potentially making a part of it artificial. Partial cloud contamination of "clear-sky" pixels used for aerosol retrievals is also a constant concern in this cloud-dominated region. Obviously, this remote area of the globe represents a challenging problem in terms of aerosol retrievals; the solution of this problem may require much more capable satellite instruments such as the APS (Mishchenko et al. 2007a) or even a combination of passive and active sensors.

The significant long-term AOT decrease over much of Europe is quite consistent with the supposed reversal from increasing to decreasing anthropogenic sulfur and black carbon emissions owing to the enactment of clean air legislation in many countries (Stern 2006; Streets et al. 2006; Vestreng et al. 2007). The enactment of similar legislation in the USA may have led to a similar decrease of AOT over much of the US territory not necessarily revealed by the GACP data. The downfall of the local economies throughout the territory of the former Soviet Union had also resulted in a dramatic and well documented decrease of aerosol emissions and AOT (Terez and Terez 2002; Geogdzhayev et al. 2005; Makhotkina et al. 2005; Gorbarenko et al. 2006). These three factors, coupled with long-range aerosol transport, may have had a significant global impact, potentially causing much of the long-term AOT trend revealed by Plate 3.1. An alternative (or additional) explanation of the global downward AOT trend is that it may be due to rapid global warming, approximately 0.28°C per decade (Hansen et al. 2006), and resulting moistening of the atmosphere (Hansen et al. 2007).

We believe that the totality of these results demonstrates the potential of satellite remote sensing to identify long-term aerosol trends. Nevertheless, more work still needs to be done in terms of both verification and potential improvement of the AVHRR radiance calibration and comparisons of GACP aerosol retrievals with potentially more accurate retrievals afforded by the newer satellite instruments (Mishchenko et al. 2007b).

***Research Scanning Polarimeter retrievals.*** Although the APS has not been flown in space yet, there exists an aircraft prototype of this instrument, called the Research Scanning Polarimeter (RSP; see Plate 3.4), which can be expected to provide a close model of the future APS performance (Cairns et al. 1999, 2003). Numerous RSP retrievals have been documented and validated by Chowdhary et al. (2001, 2002, 2005, 2006), Cairns et al. (2009), Waquet et al. (2009a,b), and Litvinov et al. (2010). Examples of the fidelity of the AOT, size distribution, and absorption estimated from the APS type of remote-sensing measurement during seven different RSP flights are shown in Plate 3.5. Panel 3.5a demonstrates that the spectral AOT values retrieved from polarimetric measurements agree well with those measured by ground-based sunphotometers over an AOT range from 0.05 to more than 1. The absence of spectrally-dependent biases in these retrievals also demonstrates the reliability of the size distribution estimate for both small and large modes of a bimodal aerosol distribution. Comparisons have also been made between *in situ* and retrieved size distributions and have also been found to agree extremely well (difference in aerosol effective radius of less than 0.04 μm).

The aerosol single-scattering albedo can also be estimated from polarimetric measurements because of the differing sensitivities of polarized and unpolarized reflectances to aerosol absorption. In Plate 3.5b (Chowdhary et al. 2005), the single-scattering albedo derived from polarimetry is compared with *in situ* (Magi et al. 2005) and ground-based sky radiance (Dubovik et al. 2002) estimates. The discrepancy between these estimates may be related to the loss of particles in the sampling system for *in situ* measurements, humidification of the *in situ* extinction (but not the absorption) coefficients, and uncertainties in the single-scattering albedo retrieval from Aerosol Robotic Network (AERONET) data that may be caused by horizontal variability in the aerosol burden. Nonetheless the polarimetric estimate of the single-scattering albedo is consistent with the other measurements given their inherent uncertainties. Overall, Plate 3.5b illustrates the complexity of retrieving the single-scattering albedo from both *in situ* and remote-sensing measurements and suggests that the validation of APS retrievals of this quantity will be a challenging task.

### 3.1.2. Imaging polarimetry of Mars with the Hubble Space Telescope

The polarization of sunlight scattered by martian aerosols and particulate surfaces contains information about particle sizes, compositions, shapes, and orientation. The degree of polarization and the orientation of the polarization plane are functions of the phase angle $\alpha$ and wavelength $\lambda$. For the majority of particulate surfaces and aerosol clouds at small phase angles ranging from 0° to $\alpha_{\mathrm{inv}}$ (the so-called inversion angle), the polarization plane is close to the scattering plane, thereby defining the so-called negative polarization branch (NPB; see Fig. 3.1). The polarization minimum $P_{\mathrm{min}}$ is reached at the phase angle $\alpha_{\mathrm{min}}$. For phase angles greater than $\alpha_{\mathrm{inv}}$, the polarization plane is usually normal to the scattering plane, thereby defining the so-called positive polarization branch (PPB). The disk-integrated and low-resolution disk-resolved polarimetric phase curves for Mars have been obtained mostly with ground-based telescopes and are thus limited to phase angles from 0° to ~45°. Space-borne polarimetric measurements of Mars were carried out during the Mars-5 mission (Santer et al. 1986). They were limited to large phase angles ~90° and provided high-resolution regional data.

The surface scatterers are soil particles of different sizes and aggregates of small grains. The atmospheric scatterers are air density fluctuations (Section 1.22), very small particles (permanent sub-micron dust haze) typical of the "clean" atmosphere, mists and clouds consisting of ice crystals, and relatively large suspended particles resulting from dust storms. The typical parameters of the martian NPBs are $|P_{\mathrm{min}}| \approx 1\%$, $\alpha_{\mathrm{min}} \approx 12°$, and $\alpha_{\mathrm{inv}} \approx 25°$ (see Fig. 3.1).

***Polarimetric data collected during the 2003 opposition.*** During the 2003 martian opposition, an international team of American and Ukrainian astrophysicists carried out joint observations with the Hubble Space Telescope (HST, NASA) (program HST-GO-9738 "Spectroscopy and Polarimetry of Mars at Closest Approach") (Shkuratov et al. 2005). These were the first polarimetric observations of Mars on the HST. The observations in August and September 2003 took advantage of the closest Earth–Mars encounter as Mars passed within 0.372 AU of the Earth. The angular diameter of the apparent martian disk at opposition was 25.1″. Five series of images of Mars were taken on August 24, just before the closest approach, and on September 5, 7, 12, and 15. The phase angles at the observation moments were 6.4°, 8.2°, 9.7°, 13.6°, and 15.9°, respectively. The observations were timed to allow imaging of the same hemisphere of Mars at all five phase angles (disk center at 19°S, 20°W–35°W), including Valles Marineris and the contrasting albedo details of Terra Meridiani and its surroundings (see Plate 3.6). This perihelion observation period corresponded to the martian summer in the Southern Hemisphere (the areocentric longitude of the Sun $L_S$ = 247°–261°). The atmosphere was relatively free of both dust and water ice clouds.

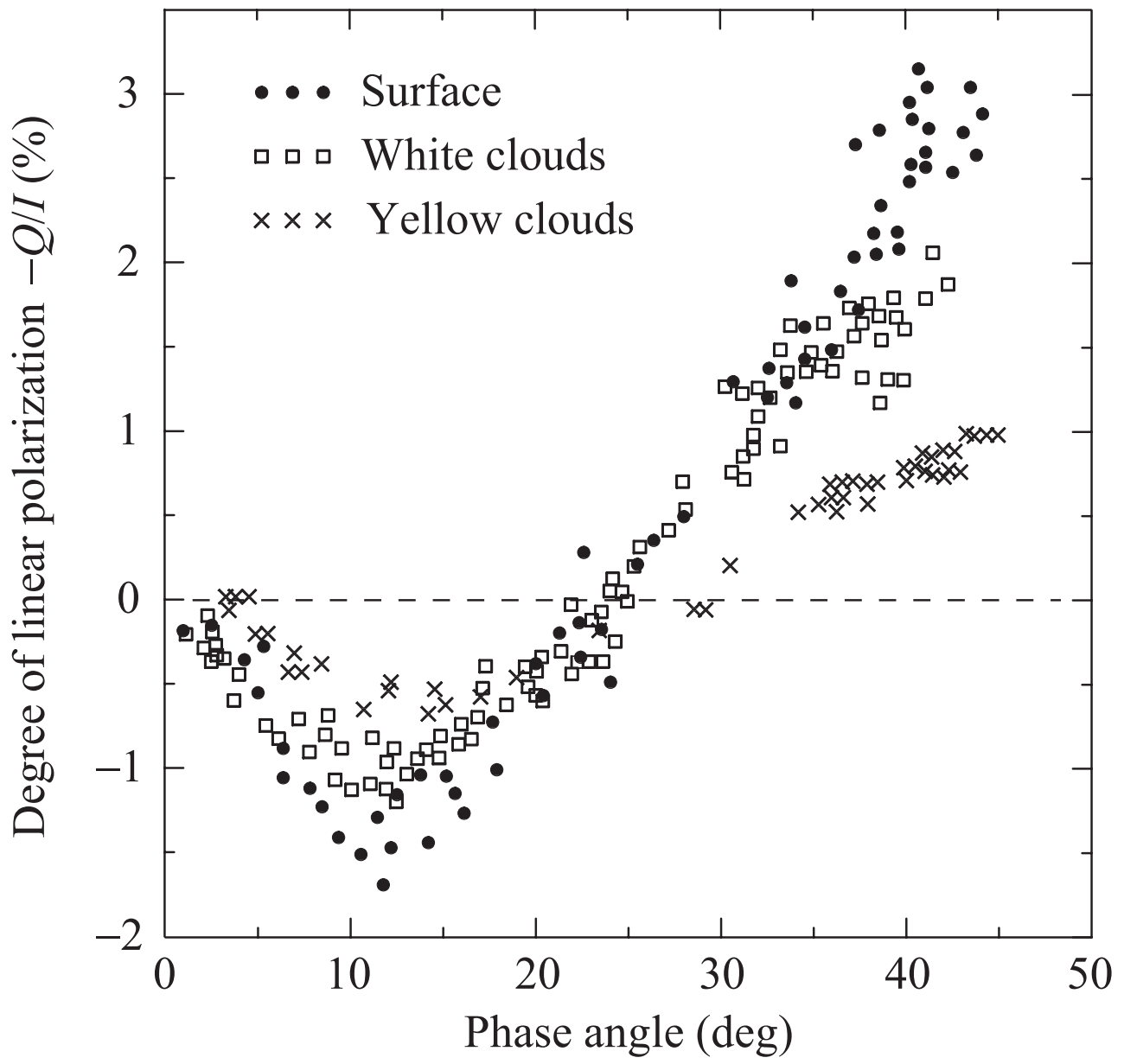

**Fig. 3.1.** Polarimetric phase curves of the martian surface (at $\lambda$ = 610 nm) as well as white (at $\lambda$ = 610 nm) and yellow (at $\lambda$ = 590 nm) clouds (after Ebisawa and Dollfus 1993).

The images were taken with the High-Resolution Camera (HRC) of the Advanced Camera for Surveys (ACS) (Pavlovsky et al. 2002). The diameter of Mars in the HRC field of view decreased from ~1010 to ~950 pixels from the first to the last observation series. The scale of the image was about 7 km/pixel in the center of the disk. These observations have the highest spatial resolution ever achieved in observations from the Earth. Each observation series (taken on each day) consisted of sets of images acquired with different broad-band spectral filters (F250W, F330W, and F435W; the number in each filter name approximately characterizes the effective wavelength in nm). Each filter set contained 3 images taken with 3 broad-spectral-band polarization filters (POL0UV, POL60UV, and POL120UV). Also, several series of images taken without polarization filters were obtained using visible and IR filters (F502W, F658W, and F892W; see Plate 3.6).

The standard pipeline calibration procedure (Pavlovsky et al. 2002), including corrections for dark current, flat field, and geometric distortion, was applied routinely to each image by the HST data retrieval facility. Afterwards, cosmic ray tracks were identified and removed from the images. Finally, the three polarimetric component images within each set were co-registered to calculate the degree of linear polarization. The calibration of the observations had been performed previously as part of the HST programs 9586, 9661, 10055 to account for instrumental polarization of the ACS HRC and the actual orientation of the polarizers. This involved observations of two standard unpolarized stars and one standard polarized star with all filter combinations used for the polarimetric observations of Mars. Our estimates show that the derived calibration with the filters F435W and F330W is quantitatively reliable, with a characteristic absolute accuracy of the degree of linear polarization better than 0.5% in the center of the field of view. Small-amplitude (<0.2%) local variations of polarization can, however, be readily detected. The accuracy with the F250W filter is probably worse. The polarization of the polarized standard star at this wavelength is unknown and was estimated by extrapolating longer-wavelength observations. The extrapolation is reliable for the orientation of the polarization plane, but the bias in the degree of linear polarization can be significant.

***Results and general interpretation.*** The calibrated HST images of Mars were used to produce maps of the intensity and the Stokes parameter ratios $-Q/I$ and $U/I$. The maps obtained with the spectral band F330W at phase angles 6.4°, 8.2°, 9.7°, 13.6°, 15.9° are depicted in Plate 3.7. Note that these are the first near-UV polarization maps of Mars. The scale for the ratios $-Q/I$ and $U/I$ in each map is the same. Almost everywhere the ratio $U/I$ is close to zero, and its spatial variations are noticeably weaker than those of the ratio $-Q/I$. The total brightness of the disk decreases during the time series owing to the increase of the phase angle (the opposition effect). The Mars–Sun distance did not change significantly during the observation time span, and this change did not influence the light flux from Mars. The bright seasonal polar cap is clearly visible in the southern (bottom) part of each image. The cap shrinks with time during the observation time span.

A bright cloud belt is clearly seen in the northern (top) portion of the disk along the limb. Outside this belt a number of surface albedo features and the largest topographic features (Valles Marineris, Argyre Basin) can be identified. One can also see faint semitransparent clouds, especially in the western (left) part of the disk and towards south-east, close to the polar cap. These clouds change dramatically between observation dates. We found that the contrast of all surface features outside the northern cloud belt, including the southern polar cap, is much lower in the UV than in the visible part of the spectrum. This well-known effect is due to the fact that in the visible, surface scattering is the major contributor to the reflected light, whereas gas and aerosol scattering dominates in the UV. Semitransparent cloud features are denser and more abundant in the UV also, since scattering by fine aerosols is more effective at shorter wavelengths. Furthermore, the albedo contrast of mare and highland surface materials is much lower in the UV than in the visible.

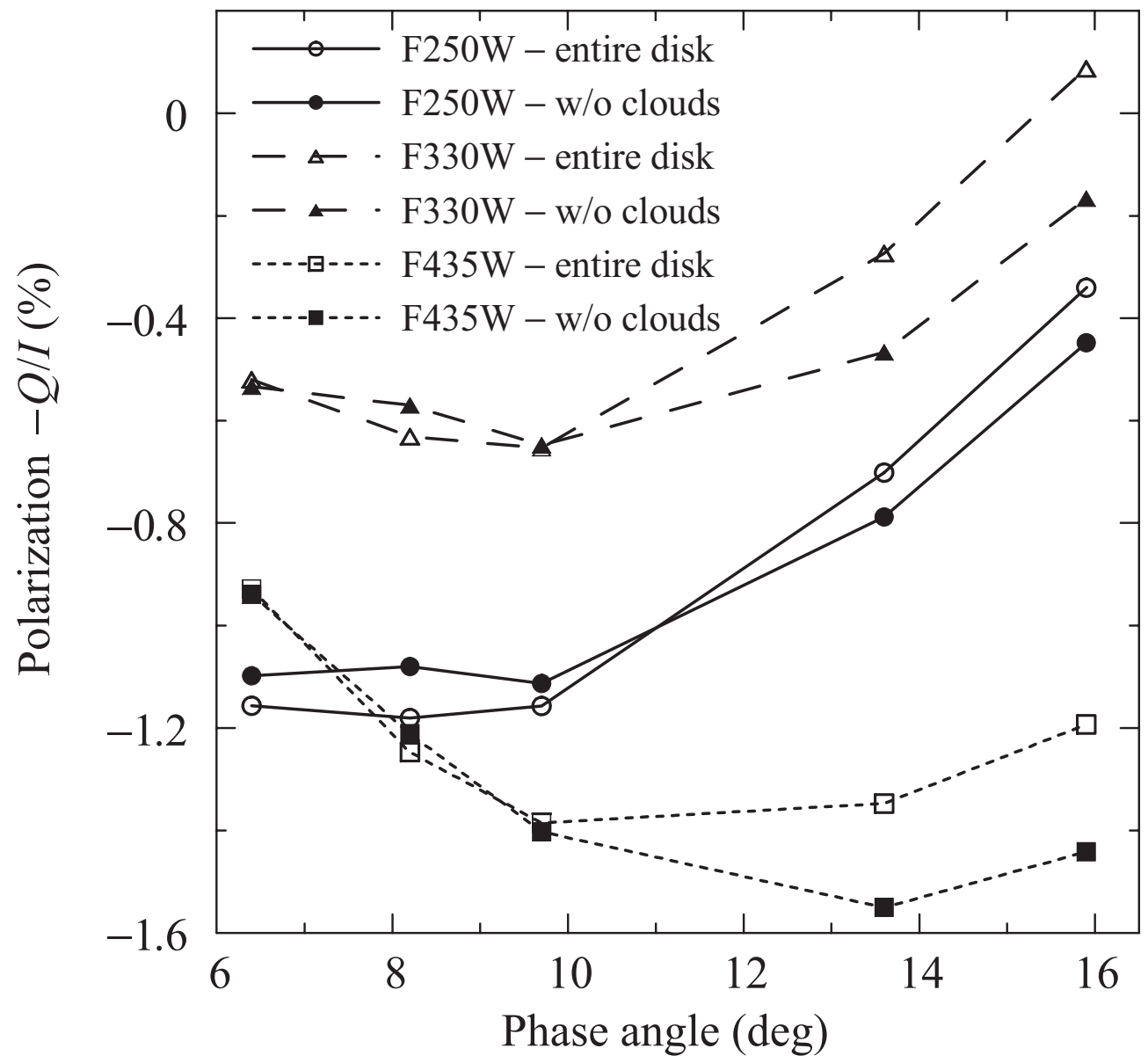

**Fig. 3.2.** Polarization phase curves for Mars obtained by averaging over the entire disk (open symbols) and by averaging over all pixels outside the visible clouds and the southern seasonal polar cap (filled symbols).

To illustrate the general behavior of polarization, we averaged the ratio $-Q/I$ over the entire disk for each date and filter. The results are plotted in Fig. 3.2 versus phase angle $\alpha$ using open symbols. In blue light (filter F435W), the polarization is on the order of 1%. These data show good quantitative agreement with the ground-based measurements by Ebisawa and Dollfus (1993) (see Fig. 3.1) in both the degree of linear polarization and the position of the polarization minimum $\alpha_{\text{min}}$. The minimum in $-Q/I$ is shallower in the F330W band than in the visible. The corresponding angles $\alpha_{\text{min}}$ and $\alpha_{\text{inv}}$ are also smaller in the UV than in the visible. This result is expected and can be explained as follows.

In the first approximation, the light reflected by Mars is the sum of the surface and atmosphere contributions (i.e., we neglect the interactive scattering between the surface and the atmosphere). The light scattered by the surface has a pronounced NPB with rather large $\alpha_{\text{min}}$ and $\alpha_{\text{inv}}$. The scattering in a clear atmosphere is dominated by the gaseous contribution and obeys the Rayleigh law: the polarization is positive and approaches zero quadratically as the phase angle tends to zero. Therefore, the surface-to-atmosphere scattering ratio is smaller in the UV than in the visible, thereby causing the above tendency.

Although aerosols make the situation somewhat more complicated, they do not negate the above explanation. Indeed, dense clouds have weaker negative polarization than the surface (e.g., see Fig. 3.1), and their relative contribution is greater in the UV. The tendency "the shorter the wavelength, the shallower the NPB and the smaller the angle of minimal polarization" thus holds.

This optical mechanism predicts weaker polarization at shorter wavelengths. However, we observe stronger negative polarization in the short-wavelength filter F250W than in the F330W filter, whereas $\alpha_{\min}$ is approximately the same. As mentioned previously, the absolute polarimetric calibration for the F250W filter is unreliable, and this unexpected effect may be an artifact of calibration errors.

To eliminate the polarization uncertainty related to the aerosol contribution, we superposed a spatial mask that included the southern seasonal polar cap, the northern cloud belt, and the wider areas of apparent clouds, as seen in the image taken with the F435W filter. Prior to the calculation of $-Q/I$, we averaged $Q$ and $I$ separately for the unmasked areas and derived the phase-angle dependence of the polarization produced jointly by the surface and the clear atmosphere. The corresponding phase curves are shown in Fig. 3.2 using filled symbols. It is seen that these cloud-free areas exhibit stronger negative polarization with a polarization minimum at larger phase angles.

Non-zero values of the ratio $U/I$ indicate a deflection of the polarization plane from the photometric equator. We have found that this deflection is much stronger in the UV than in the visible. Near the limb the polarization plane tends to deflect from the photometric equator toward the local radius, as can be expected. For powdery surfaces the deflection is virtually absent. For an optically thick Rayleigh atmosphere, the RT theory predicts the radial orientation of the polarization plane and stronger polarization toward the limb (Chandrasekhar 1950). The deflection appears to be stronger in the UV where the atmospheric contribution is greater. This effect is more pronounced along the northern limb where the stronger-scattering clouds serve to enhance it.

The polarization of the southern seasonal polar cap is weaker than that of its surroundings and is close to zero in the visible. This is a manifestation of the surface contribution: the degree of polarization of light reflected by particulate surfaces at small phase angles usually anticorrelates with their albedo (e.g., Bowell and Zellner 1974; Shkuratov et al. 2004a). In the UV, the polarization contrast between the polar cap and the other parts of the disk is much weaker than in the visible, and at the shortest wavelength the cap is almost invisible in polarization. Again, this can be explained by the increased contribution of the atmospheric scattering in the UV.

***Transient polarization features.*** The most interesting phenomena revealed by these observations are transient high-polarization features. The maps in Plate 3.7 show several such features with polarization magnitudes exceeding 2%. The polarization contrast of these features is the highest in the near-UV filter F330W. They are especially prominent in the western (left) part of the disk. Although being absent on August 24, these features became clearly visible and had distinctively different shapes and locations on September 5 and 7 (as indicated by the arrows in Plate 3.7) before eventually fading away. One such feature can be seen in detail in the full-resolution color map shown in Plate 3.8. The shades of red indicate areas with high polarization and clearly exhibit cloud fields.

The spatial distribution of polarization differs from filter to filter, but the locations of major transient polarization maxima remain the same. There are clouds associated with the areas of strong polarization in the UV images. These clouds are semitransparent so that surface albedo patterns and the Valles Marineris topography remain recognizable. The densest clouds are still seen in the blue filter F430W, although many clouds "disappear". It is interesting that in the same part of the martian disk there are similar clouds that do not exhibit the strong polarization effect, apparently because these clouds consist of different particles. Similar polarization features are noticeable at the north-western (left) limb (see Plates 3.7 and 3.8).

It is instructive to compare the above results with data obtained with the Mars Global Surveyor Thermal Emission Spectrometer (TES), in particular with the atmospheric dust opacity (at 9.7 μm) and ice opacity (at 12.1 μm) retrievals (Pearl et al. 2001). During the observation period, close to the perihelion, the atmosphere was relatively dusty compared to the aphelion season. According to the TES data, the dust was evenly distributed throughout the lower atmosphere. The TES data did not show any pronounced dust-lifting events collocated with our observations. The average TES-derived ice-cloud opacity in the western part of the visible martian disk at low southern latitudes was distinctly greater than in the eastern part at the same latitudes. This result confirms the somewhat higher ice opacity in that same part of the disk where the polarizing clouds were identified in the HST images.

The strong polarization of light scattered by these optically thin clouds indicates that they consisted of strongly polarizing particles. Irregular crystals of ~1 μm size could cause such strong polarization (Zubko et al. 2006). This size corresponds to Type 1 particles according to Wolff and Clancy (2003) who have shown that particles of this type constitute high-altitude hazes through various locations and seasons. Perhaps, highly polarizing clouds can be formed in the very beginning of the nucleation of $H_2O$ ice crystals on submicron dust. Measurements of synoptic cloud dynamics and comparisons of the derived values of the wind speed with the European Mars Climate Database have suggested that the polarizing clouds are located at 30−40 km above the martian surface (Kaydash et al. 2006).

### *3.1.3. Polarimetry of Mercury*

Mercury is a very special planet. Indeed, owing to its close proximity to the Sun, the absence of a substantial atmosphere, and the slow rotation around its axis, Mercury experiences the largest surface temperature drops (from +452° to −183°C) among all Solar System planets and satellites. When passing through perihelia, Mercury is facing the Sun with one or the other hemisphere centered at the $0^{th}$ or the $180^{th}$ meridian, often called the "hot" meridians. The "cold" $90^{th}$ and $270^{th}$ meridians face the Sun when Mercury is at the respective aphelion points of its orbit.

Among the Earth-type planets, Mercury remains the least studied. At the same time it is a focus of substantial interest, primarily because, along with asteroids, it is one of the most pristine relicts of the period during which the giant planets were formed. Characteristic features of the composition and internal structure of Mercury as well as its magnetic field, surface topography, etc. are undoubtedly related to the specific processes that affected the formation and evolution of the planet during the period of post-accretion bombardment by large bodies about 3.9 billion years ago. Hence, studies of Mercury can be expected to yield key information about the formation and evolution of the other planets as well as the Solar System at large. This explains the decision to make Mercury the target of a dedicated space mission (Langevin 1997; Curtis et al. 1998).

The results of ground-based and space-borne CCD photometry (Mallama et al. 2002) reveal brightness variations of Mercury with an amplitude of $\sim 0.05^m$ depending on the central longitude of the illuminated disk. The brightness minima correspond to the longitudes around ~0° and ~180°, whereas the brightness maxima occur for the longitudes around ~90° and ~270°. Thus the areas of the surface of Mercury centered at the "hot" meridians are darker than those centered at the "cold" meridians. These differences could be caused by the countless day–night temperature drops which may have led to the accumulation of the results of soil baking (fusion), thereby affecting the surface structure and its optical (in particular, polariza-tion) properties. Furthermore, these differences could be caused by different rates of maturation of the Mercurian soil owing to differences in the illumination by the sunlight, solar wind, and solar cosmic rays, the intensity of which is 2.3 times greater at perihelia than at aphelia. By using an empirical relation between $P_{max}$ and the albedo (Section 3.2.8), one readily finds that $P_{max}$ must vary with the longitude of the central meridian with an amplitude of ~0.5%, which is large enough for a reliable detection. Therefore, one could expect the existence of a double wave in the dependence of polarization on the longitude of the central meridian. This hypothesis motivated our polarimetric observations of Mercury.

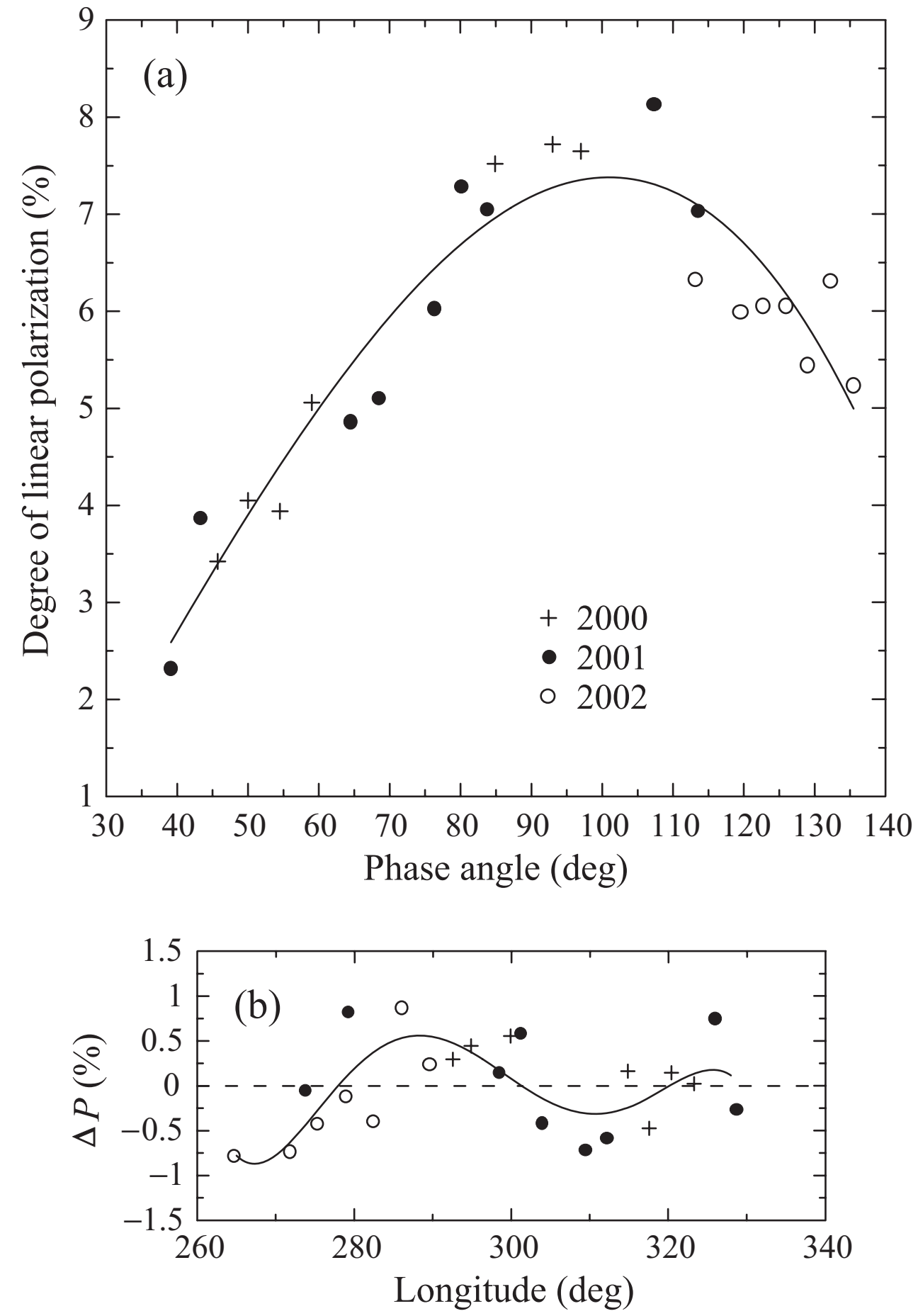

**Fig. 3.3.** Dependence of the degree of linear polarization on the phase angle (a) and on the longitude of the central meridian (b; the symbols are the same) of Mercury according to the 2000–02 observations with the GC and V filters.

The observations were performed during three periods of visibility in 2000–02 using the 0.7-m reflector of the Chuguev Observation Station of the KhNU (Lupishko and Kiselev 2004). In 2000, the observations were performed with narrow-band filters BC, GC, and RC. Because of the extremely limited observation time, the accuracy of polarization measurements in 2001 and 2002 was improved by observing with only the GC and the V filter, respectively.

Figure 3.3a depicts the phase-angle dependence of the reflected polarization in the green part of the spectrum. The solid curve approximates the measurement data. One can see a significant scatter of data points with respect to the approximation curve reaching ±1%, which exceeds the measurement errors by a factor of several. This can be interpreted as an indication of a possible longitudinal effect in the variations of polarization over the Mercurian surface. Figure 3.3b shows the deviation of the measured polarization from the approximation curve as a function of the longitude of the central meridian corresponding to the illuminated part of the observed Mercurian surface. Even though the observed range of longitudes was close to 65°, one can indeed clearly identify polarization variations with an amplitude of ~1.5%. The actual existence of these variations is corroborated by the match of the results obtained during different observation periods.

### *3.1.4. Optical properties of the lunar surface*

***Photometric phase curve of the lunar surface and optical mechanisms of its formation.*** The brightness of the lunar surface depends on the illumination–observation geometry. This dependence is described by a photometric function of three angles: the photometric latitude, the photometric longitude, and the phase angle (for definitions, see, e.g., Kreslavsky et al. 2000). The dependence on the phase angle provides the best microphysical diagnostics. This component of the photometric function is called the phase function (which should not be confused with the single-particle phase function defined by Eq. (1.84)). Figure 3.4 illustrates the average phase function of the lunar surface $f(\alpha)$ at $\lambda = 430$ nm compiled from various data sets (Shkuratov et al. 2003). This plot clearly reveals the opposition effect in the form of a sharp nonlinear rise of brightness as the phase angle approaches zero. Earth-based studies of the lunar opposition effect at very small phase angles are impossible because the Moon cannot be observed from the Earth at phase angles smaller than 1° out of eclipse. Unique observations of the lunar opposition effect at extremely small phase angles were obtained with the UVVis camera onboard the Clementine spacecraft (Shkuratov et al. 2004b).

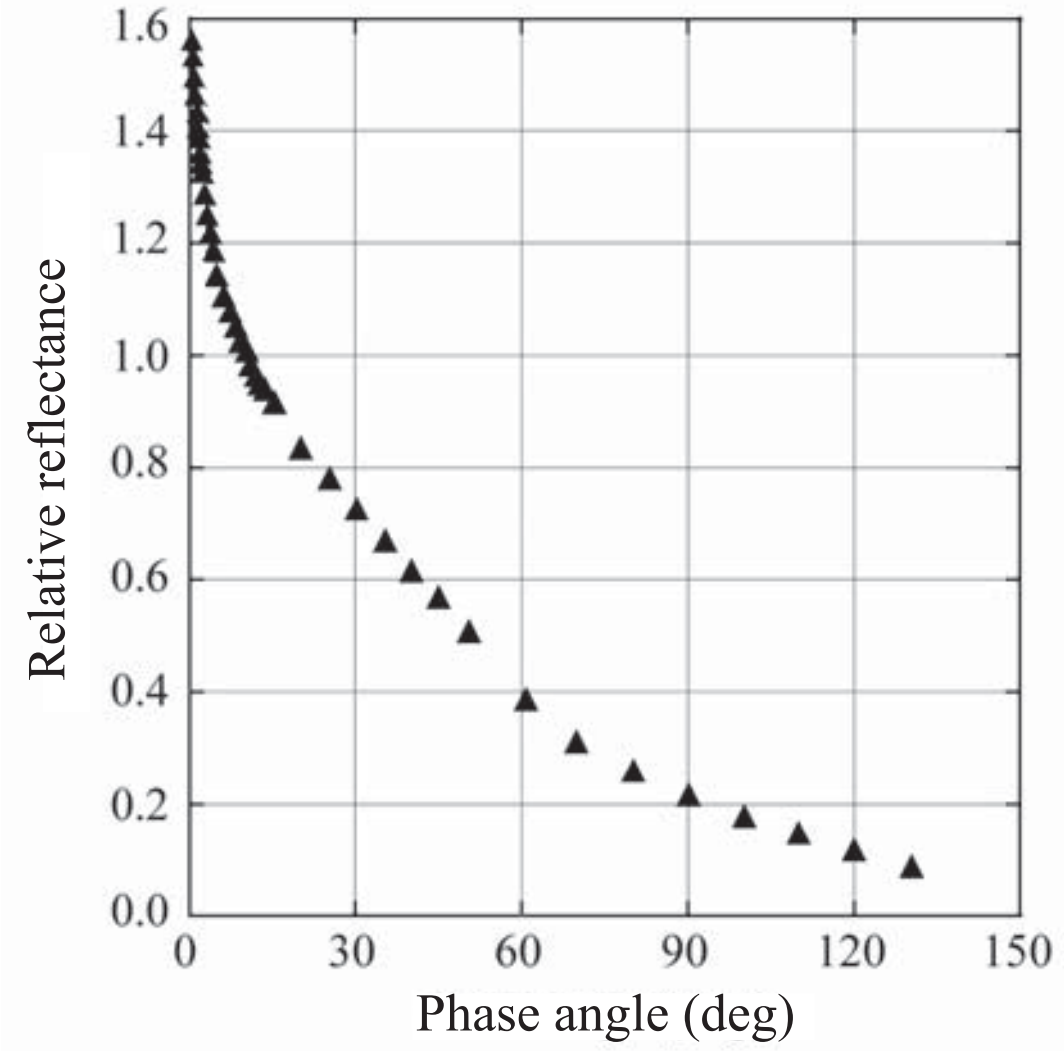

**Fig. 3.4.** Average photometric phase curve of the lunar surface at $\lambda$ = 430 nm.

In the entire range of phase angles in Fig. 3.4, the average phase function of the lunar surface $f(\alpha)$ increases as the phase angle $\alpha$ decreases. The amplitude of the opposition spike in the range of phase angles 0.2°–10° is about 60% at the wavelength $\lambda$ = 430 nm. For individual lunar regions, the deviations of the respective phase functions from the average one depend on the local structural properties of the surface and its albedo. Different phase-angle ranges of the phase function are affected by different scales of the regolith morphology. At very small phase angles, $\alpha \lesssim 4°$, the WL effect (Sections 1.17.3, 1.19, and 1.25) may contribute to the enhanced backscattering by bright areas. At larger phase angles, $\alpha \lesssim 40°–50°$, the dominant contribution likely comes from the shadowing effect caused by irregularities of the order of ~100 μm. In this phase angle range, the phase function depends on the surface albedo owing to partial illumination of the shadows by light passing through partly transparent regolith particles (Hapke 1993). When $\alpha$ increases, the meso- and macrorelief also starts to have a noticeable effect on the phase function. Thus, from the spatial distribution of the phase-function parameters (e.g., the slope in Fig. 3.4), one can try to infer specific features of the lunar regolith structure.

For moderate- and low-albedo surfaces, the shape of the single-particle phase function can directly influence the phase function of the surface through the first-order scattering (Kreslavsky and Shkuratov 2003). Our laboratory measurements and calculations show that particles and roughness at the scale of the illuminating wavelength, after averaging over orientations and shapes, cause an enhancement of backscattering at phase angles smaller than 30°–40° (Shkuratov et al. 2007). For samples of terrestrial volcanic ashes, the backscattering effect is caused by the first-order scattering. The first-order-scattering contribution should also be essential for dark mare sites on the Moon. We have observed the CB effect for bright particulate surfaces. Surges in backscatter with amplitudes ~50 % have been detected in laboratory measurements at extremely small phase angles. We note that for the Moon, this effect may appear only for surface areas with high albedos; these are cites with non-mature highland soils (young craters and ray systems).

Thus space-borne photometry enables us to clarify which primary scattering mechanisms affect the photometric properties of the lunar surface and how these mechanisms are related to the structural properties of the lunar regolith.

***Photometric opposition effect.*** Among the images of the lunar surface taken by the UVVis and NIR cameras during the Clementine mission, there are ones capturing the zero phase-angle point or, in other words, the spacecraft shadow point. This provides a unique opportunity to study the lunar opposition spike in the 400–1500 nm spectral range for several lunar sites. Plate 3.9 shows images with the opposition point, which is clearly seen as a diffuse bright spot superimposed on the local albedo variations. As a consequence of the spacecraft movement along the lunar orbit, the zero-phase point moved in concert with the sub-spacecraft point.

To infer the phase function from the small-phase-angle Clementine images, several methods have been used (Кайдаш и др. 2003):

*Simple averaging.* To estimate the phase function at very small phase angles, one can perform a simple averaging over different images containing the opposition spot, thereby smoothing out the regional albedo variations and separating out the brightness pattern associated with the opposition effect.

*Using phase-angle ratio ~0°/30°.* There are lunar sites near the equator that were imaged by Clementine twice at $\alpha \approx 0°$ and $\alpha \approx 30°$. A phase-angle ratio image is obtained by dividing the small- and large-phase-angle images of the same region. This provides an efficient correction for major albedo variations, thereby isolating the opposition brightness spot. After averaging this ratio for a number of concentric zones of constant phase angle in the opposition image, one can estimate the phase function for a given site.

*Differential method.* The sub-spacecraft point and hence the zero-phase-angle point shifts only slightly over the surface between two consecutive Clementine images taken with different spectral filters (see, e.g., Plates 3.9a,b). If two consecutive spectral filters have close central wavelengths then by computing the ratio of the corresponding intensities one can suppress the spatial albedo variations, whereas the opposition spot is not quenched completely (Plate 3.9c). The phase-angle ratio in this case is proportional to the logarithmic derivative of the phase function $\mathrm{d}[\ln f(\alpha)]/\mathrm{d}\alpha$. If the phase angle corresponding to each point of both images is known then one can calculate the logarithmic derivative for each point of the frame. After averaging the derivative over all points with the same phase angle, the integration of the derivative yields the photometric phase curve.

*Using parameterized models of the phase function.* The differential method cannot be applied to a pair of images obtained with the same spectral filter since the shift of the opposition point between the images is too large. In this case one can compute the ratio of the respective intensities for every point of the common area of the two images. To model the photometric phase dependence of the Moon in a nar-row interval of phase angles, one can use a simple parameterized mathematical function such as $f(\alpha) \approx \exp(-k\alpha)$. Then the phase function is found by minimizing the differences between the modeled and real intensity ratios over the common area of the two images.

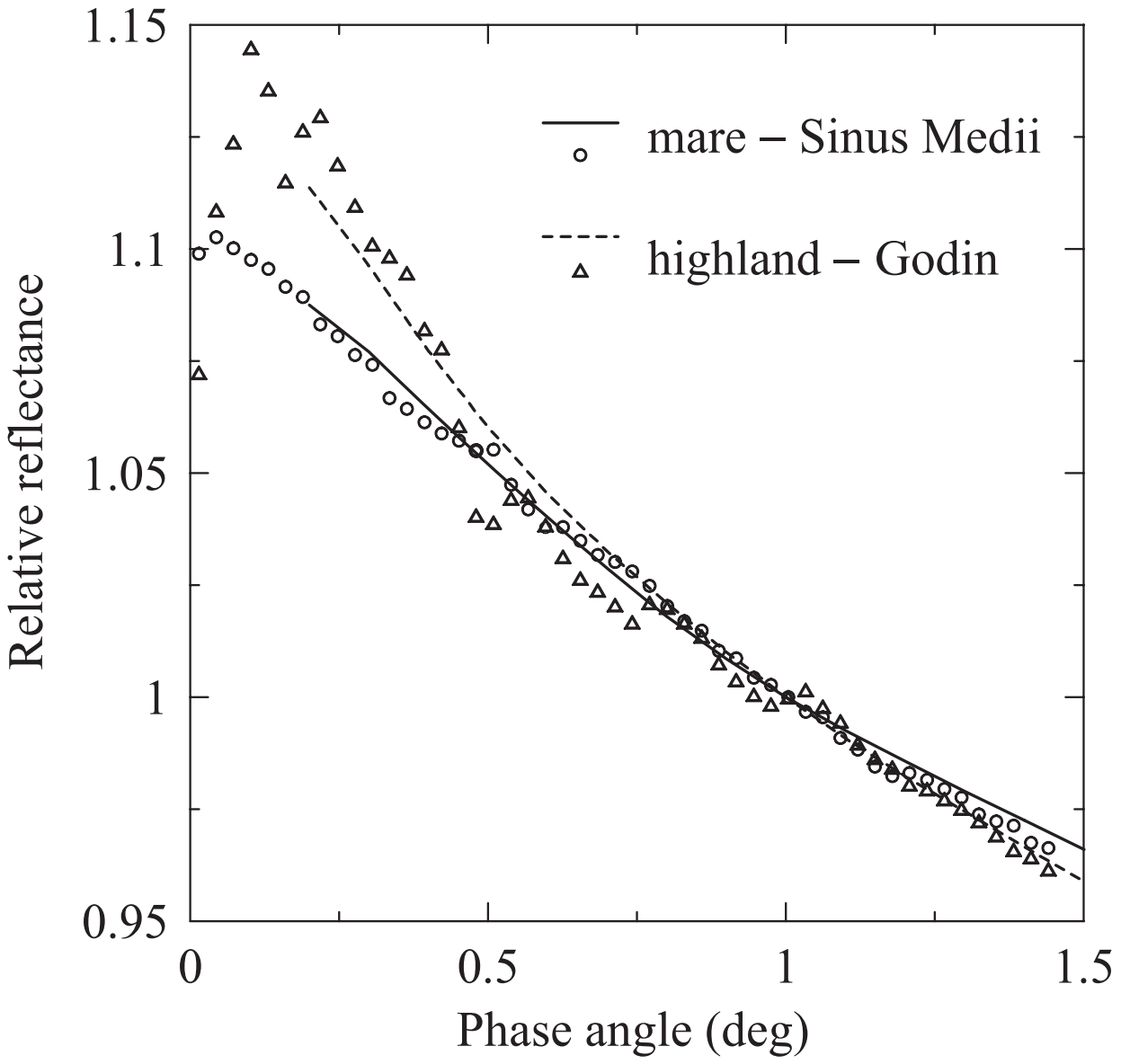

**Fig. 3.5.** The lunar opposition effect for a highland site near crater Godin (triangles and dotted curve) and a mare site Sinus Medii (circles and solid curve) inferred using two methods: the phase-angle ratio imaging (symbols) and the differential method (curves).

***Results.*** We have found that the opposition brightening of the lunar sites studied is almost linear in the phase-angle range 0.2°–1.5° (see Fig. 3.5). There is no wavelength dependence of the opposition spike for this phase-angle range in the visible and near-IR parts of the spectrum (400–2000 nm). Different parts of the lunar surface are characterized by different values of the height-to-width ratio computed for the opposition brightness spike. The phase slope for highland regions is slightly steeper then that for mare sites (see Fig. 3.5).

There is no obvious systematic dependence of the height-to-width ratio on the surface albedo. This suggests that the surface albedo for the regions studied is too low to cause a significant contribution from the CB effect, and the opposition peak is dominated by the shadow hiding effect. The latter is weakly dependent on the wavelength of the incident radiation and the surface albedo.

***Mapping the parameters of the photometric function.*** An important branch of modern lunar science is the mapping of parameters describing the photometric (or polarimetric) phase-angle dependence (see, e.g., Shkuratov et al. 2008; Kaydash et al. 2009; Опанасенко и др. 2009 and references therein). The quantities usually se-lected for mapping are the parameters of theoretical formulas which model experimental data or the parameters of suitable empirical approximation formulas. Regional distributions of such parameters carry information on certain structural features of the lunar surface. The simplest approximation function adequately describing the phase dependence of brightness in the phase angle range 5°−50° is the exponential law $\exp(-\tilde{\tau}\alpha)$. If a specific region of the Moon is imaged several times at different phase angles then the combined analysis of the resulting images will allow one to map the parameter $\tilde{\tau}$, which depends on the surface roughness and albedo.

We have suggested that variations of the phase-function slope can be interpreted in terms of changes in the small-scale regolith structure (0.1−10 mm). To validate this interpretation, we have mapped the parameter $\tilde{\tau}$ over the landing site of the lunar module Apollo-15 (1971), the reason being that there are reliable estimates of the ambient geological environment, data on the ambient crater morphology, and estimates of the impact of the Apollo landing module on the ambient lunar regolith. This region was imaged by the UVVis camera onboard the Clementine spacecraft. We have performed a photometric analysis of this site using a total of 52 images taken at phase angles from 26° to 55°. The spatial resolution of the images was about 100−200 m per pixel, depending on the spacecraft distance to the imaging region and perspective distortions. Plate 3.10a shows one of the images from this series taken at $\lambda$ = 750 nm with the camera axis pointing at nadir.

***Photometric anomalies.*** Plate 3.10b shows the spatial distribution of the parameter $\tilde{\tau}$ for the region shown in Plate 3.10a. The former reveals a number of features unrelated to the variations in the phase dependence of brightness, i.e., to the variations in $\tilde{\tau}$. These primarily include the large-scale relief features: the Apennine Mountains, Hadley Rille, and certain spatially resolved morphological details (craters, hills). Despite these drawbacks, the image of $\tilde{\tau}$ contains much useful information. In particular, one can identify several large diffuse features unrelated to the resolvable morphological features of the lunar surface. These primarily include the diffuse halos around young craters (see the upper and left arrows in Plate 3.10b). The fact that these craters are young follows from their morphology visible in the high-resolution images obtained by Lunar Orbiter V at low elevations of the Sun. Young craters often stand out in a usual brightness image, because the soil excavated during the impact is usually brighter than the surrounding material (see Plate 3.10a). The halos with anomalously low values of $\tilde{\tau}$ have appreciably larger sizes than the albedo halos. This suggests that the anomaly in the photometric properties cannot be completely explained by an enhanced albedo but rather is attributable to the surface structure.

In the range of phase angles considered, the slope of the photometric phase dependence is determined primarily by the shadow hiding and albedo effects. If two regions of a particulate surface have the same albedo then the region with a flatter phase function (dark shades in Plate 3.10b) has a smoother surface and/or a higher packing density at 0.1−10 mm scales. Several factors can be responsible for the unusual smoothness or the higher density of the top regolith layer, local seismic oscillations emerging at the location of an impact event being among them. In some cases, dense subsurface structures can cause large-amplitude surface oscillations (Sato and Fehler 1995), which shake the ambient regolith and, as a result, can lead to smoother microtopography of the top layer.

***Landing site of the lunar module Apollo-15.*** The small dark spot indicated by the lower right arrow in Plate 3.10b deserves special attention. It is not associated with any crater but closely coincides with the Apollo-15 landing site. This spot cannot be an artifact, because an examination of all 52 images used for our analysis has revealed their high quality in this region. This photometric feature is probably indicative of an actual structural detail of the lunar soil associated with the impact of gaseous jets from the lunar module engine. The radius of disruption of the regolith is 50–150 m. This range is consistent with the dust sputtering radius estimated previously from studies of the degree of dustiness of the surface details of the Surveyur-3 lander located 180 m away from the Apollo-12 lander (Benson et al. 1970). Helfenstein and Shepard (1999) performed a quantitative photogrammetric analysis of the stereomicrophotographs of the lunar surface taken by astronauts during the Apollo-11, -12, and -14 missions. The regolith at the Apollo-14 landing site was found to be much rougher on scales from 0.1 mm to several centimeters. One of the possible explanations of this phenomenon is the blowing of the soil by gases from the jet engines of the lander during the landing. Indeed, the microphotographs at the Apollo-14 landing site were taken at a distance of 180 m from the lander, while the microphotographs at the Apollo-11 and -12 landing sites were taken at a much smaller distance. It is very likely that Helfenstein and Shepard (1999) identified the difference in the regolith microstructure that causes the photometric anomaly indicated in Plate 3.10b.

This result demonstrates that photometric anomalies can provide valuable information about various processes affecting the lunar surface. For example, a statistical study of craters with anomalous halos may yield data on typical processing time scales for the top regolith layer subject to micrometeoritic bombardment. It is also hoped that geologically recent seismic events can be identified based on the resulting structural changes of the lunar surface in the form of specific photometric anomalies.

### *3.1.5. Phase, longitudinal, and spectral dependences of polarization for the Galilean satellites of Jupiter and Iapetus*

Polarimetric studies of such distant objects as satellites of Jupiter and Saturn are quite difficult because of the narrow phase-angle range accessible in the ground-based observations. Nevertheless, previous polarization measurements of the Galilean satellites (Dollfus 1975; Veverka 1977) had shown that these objects differ in terms of their surface microtexture and albedo. Furthermore, the Galilean satellites and Iapetus exhibit certain differences between the polarization curves measured at eastern and western elongations. This asymmetry arises from the fact that these satellites rotate around their own axes and orbit their respective planets synchronously, and so areas with different reflectivities come into view. Different polarization curves observed for the two hemispheres and a significant scatter of data points for either hemisphere indicate variations in satellite polarization with the orbital longi-tude which may be caused by intrinsic differences in surface characteristics between the hemispheres.

In the 1960s and 1970s, large volumes of measurements of the degree of linear polarization for the Galilean satellites at different phase angles were obtained by Veverka (1971), Dollfus (1975), and Gradie and Zellner (1973) in the visible spectral range. These measurements had allowed the authors to determine the main polarization parameters of the satellites and to discover that Callisto is polarimetrically quite different from the other Galilean satellites and that its leading hemisphere differs considerably from the trailing one. To explain the difference between the two hemispheres, Dollfus (1975) suggested that the leading hemisphere of Callisto is covered with a thick, dusty, lunar-like regolith, whereas the trailing hemisphere contains larger grains or large areas of clean, dust-free rocky surface. It had also been found that the polarization of Ganymede near zero phase does not tend to zero, as might be expected, but rather has a distinctly nonzero value. A comprehensive review of polarimetric observations performed in the 1960s and 1970s can be found in Veverka (1977).

The first series of multicolor polarimetric observations of the Galilean satellites in six broadband filters from 390 to 685 nm was performed by Ботвинова и Кучеров (1980). These authors concluded that the polarization measured for all four satellites was essentially wavelength independent. They also confirmed that there is a strong dependence of polarization for Callisto on its orbital position.

Detailed multicolor polarimetric observations of the Galilean satellites in four spectral bands (BVRI) sampling the spectral range from 420 to 700 nm and at phase angles ranging from ~12° to 0.2° were performed in 1981−88 by Чигладзе (1989). Contrary to the results of Ботвинова и Кучеров (1980), he found that the minimal polarization value $P_{\min}$ was rather strongly dependent on the wavelength of observation. He also concluded that for all four Galilean satellites, the polarization at phase angles $\alpha < 1°$ changed sign from negative to positive and reached positive values of about 0.2%−0.3%. However, it has recently been demonstrated (Kiselev et al. 2009) that the polarization at small phase angles remained negative but had the same absolute value as that measured by Чигладзе (1989).

The above-mentioned earlier papers exhaust the list of publications in which polarimetric investigations of the Galilean satellites had been carried out. The available polarimetric data, even for the bright satellites of Jupiter, were rather limited and even mutually contradictory. Furthermore, despite the fact that most of the NPB for the satellites was measured in the 1980s, there were no reliable polarization data at phase angles smaller than 2°. Therefore, we initiated a program of low-phase-angle observations of different objects, including the Galilean satellites of Jupiter and the Saturnian satellite Iapetus, in order to investigate in detail the behavior of polarization near opposition. In this section, we present and analyze the most recent polarimetric observations of the Galilean satellites and Iapetus as well as revisit the results of previous observations.

***Summary of observations.*** Systematic polarimetric observations of the Galilean satellites of Jupiter in the U, B, V, and R filters and in the entire accessible range of phase angels from ~11.8° down to almost zero were initiated by these au-thors in 1988 in Tarija, Bolivia and continued in 1989−91 on the Maydanak mountain in Uzbekistan. In both cases nearly identical spectropolarimeters installed in the Cassegrain focus of similar 0.6-m reflecting telescopes were used. The errors in the degree of polarization due to uncertainties in the instrumental polarization and reductions to the standard system varied from ±0.04% to ±0.06% in the B, V, and R filters and to ±0.1% in the U filter. The average error in the measurements of the position angle of the polarization plane was <10°. More recently, polarimetry of these satellites near opposition was carried out with a CCD polarimeter installed on the 0.7-m reflector of the Chuguev Observation Station of the KhNU; however, those results had a lower accuracy of ~0.05%−0.1%.

In 2000 and 2007, the main purpose of our observation programs was to study in detail the behavior of polarization for the Galilean satellites near opposition and to determine the shape of the secondary polarization minimum predicted by the theory of WL (Mishchenko 1993b; see Section 1.25.2) and discovered by Rosenbush et al. (1997a). The weakness of polarization at very small phase angles ($|P| \sim$ 0.1%−0.3%) necessitates measurement accuracies better than ~0.03%, which means that the instrumental polarization must be known to better than 0.02%. In fact, the errors in the parameters of instrumental polarization determined from our observations of unpolarized standard stars did not exceed 0.01%. The calibration of the orientation angle of the polarization plane was performed using observations of standard stars with well-known large polarization. The zero-point of position angles was determined with an accuracy of ±0.7°. All polarimetric standard stars were taken from the database compiled by Heiles (2000). The cumulative error in our measurements of the degree of linear polarization included the mean square error averaged over each night and the error of the instrumental system. In most cases the cumulative error did not exceed 0.03%−0.04% and was carefully accounted for. The zero-point of position angles was stable to within ~1°. We have never noticed any significant deviation of the orientation angle of the polarization plane from 0° with respect to the scattering plane (the plane through the Earth, the Sun, and the object), which corresponds to consistently negative polarization for all four satellites.

In 1998, polarimetric observations of the unique Saturnian satellite Iapetus were initiated with the same general objective. This satellite is by far the most suitable object for the study of polarimetric and photometric phase-angle effects and their dependence on surface reflectivity since its leading hemisphere has an albedo of $p_V \approx 0.05$, while its trailing hemisphere is much brighter, with $p_V \approx 0.6$.

It is obvious that in order to arrive at a reliable conclusion regarding the existence of the POE for the Galilean satellites, it is necessary to determine detailed phase-angle polarization curves. A summary of our observational program is given in Table 3.1. Further details and specific results can be found in Rosenbush et al. (1995), Rosenbush et al. (1997a), Розенбуш и Аврамчук (1999), Rosenbush et al. (2000), Rosenbush (2002), and Rosenbush and Kiselev (2005).

***Phase-angle dependence of polarization for Io, Europa, and Ganymede.*** Figure 3.6 shows the phase-angle dependence of the degree of linear polarization in the U, B, V, R, and I filters for the leading (filled symbols) and trailing (open symbols) hemispheres of Io, Europa, and Ganymede in the entire phase-angle range accessible from the Earth (i.e., from almost 0.2° up to 12° relative to the exact backscattering direction). The branch of negative polarization as well as the branch of positive polarization are essentially flat over wide intervals of phase angle. The significant scatter of data points may be explained by significant systematic differences in the phase-angle dependences of polarization for the leading and trailing hemispheres as well as by differences in the measurement accuracy. It is clear, however, that at phase angels $\alpha \lesssim 2°$ the measured polarization deviates from the expected typical polarization curve. For the first time, we discovered this effect during the 1988−89 observations in Bolivia (Rosenbush et al. 1995; Rosenbush et al. 1997a). After the theoretical prediction by Mishchenko (1993b), it became obvious that the detected secondary minimum of negative polarization located at very small phase angles is the POE caused by CB.

**Table 3.1.** Polarimetric observations of the Galilean satellites and Iapetus

| Satellite | Observation periods | Filter(s) | $\alpha$ (deg) | Number of nights |
|---|---|---|---|---|
| Io | 02.10.1988−14.03.1989 | UBVR | 0.58−10.64 | 13 |
| | 02.11.1989−01.10.1990 | UBVR | 9.01−9.67 | 5 |
| | 09−25.09.1998 | V | 1.41−2.28 | 15 |
| | 19.11−07.12.2000 | BVR | 0.26−2.19 | 7 |
| | 06−08.06.2007 | V | 0.22−0.62 | 3 |
| Europa | 04.09.1988−28.03.1989 | UBVR | 0.53−11.55 | 25 |
| | 02.11.1989−16.02.1991 | UBVR | 3.91−9.67 | 10 |
| | 09−25.09.1998 | V | 1.41−2.28 | 15 |
| | 19.11−07.12.2000 | BVR | 0.20−2.22 | 8 |
| | 05−08.06.2007 | V | 0.14−0.63 | 4 |
| Ganymede | 06.08.1988−23.02.1989 | UBVR | 0.54−11.55 | 24 |
| | 02.11.1989−19.02.1991 | UBVR | 3.74−9.67 | 13 |
| | 09.09−25.09.1998 | V | 1.41−2.28 | 15 |
| | 20.11−07.12.2000 | UBVR | 0.19−2.18 | 6 |
| | 06−08.06.2007 | V | 0.22−0.61 | 3 |
| Callisto | 08.08.1988−24.04.1989 | UBVR | 0.54−11.52 | 35 |
| | 02.11.1989−19.02.1991 | UBVR | 3.74−9.67 | 16 |
| | 09.09−25.09.1998 | V | 1.41−2.28 | 15 |
| | 20.11−07.12.2000 | BVR | 0.28−2.18 | 4 |
| | 05−08.06.2007 | V | 0.13−0.61 | 3 |
| Iapetus | 21.11−25.11.1998 | BR | 3.5 | 2 |
| | 11.10.1998−18.01.1999 | UBVRI | 1.38−6.10 | 20 |

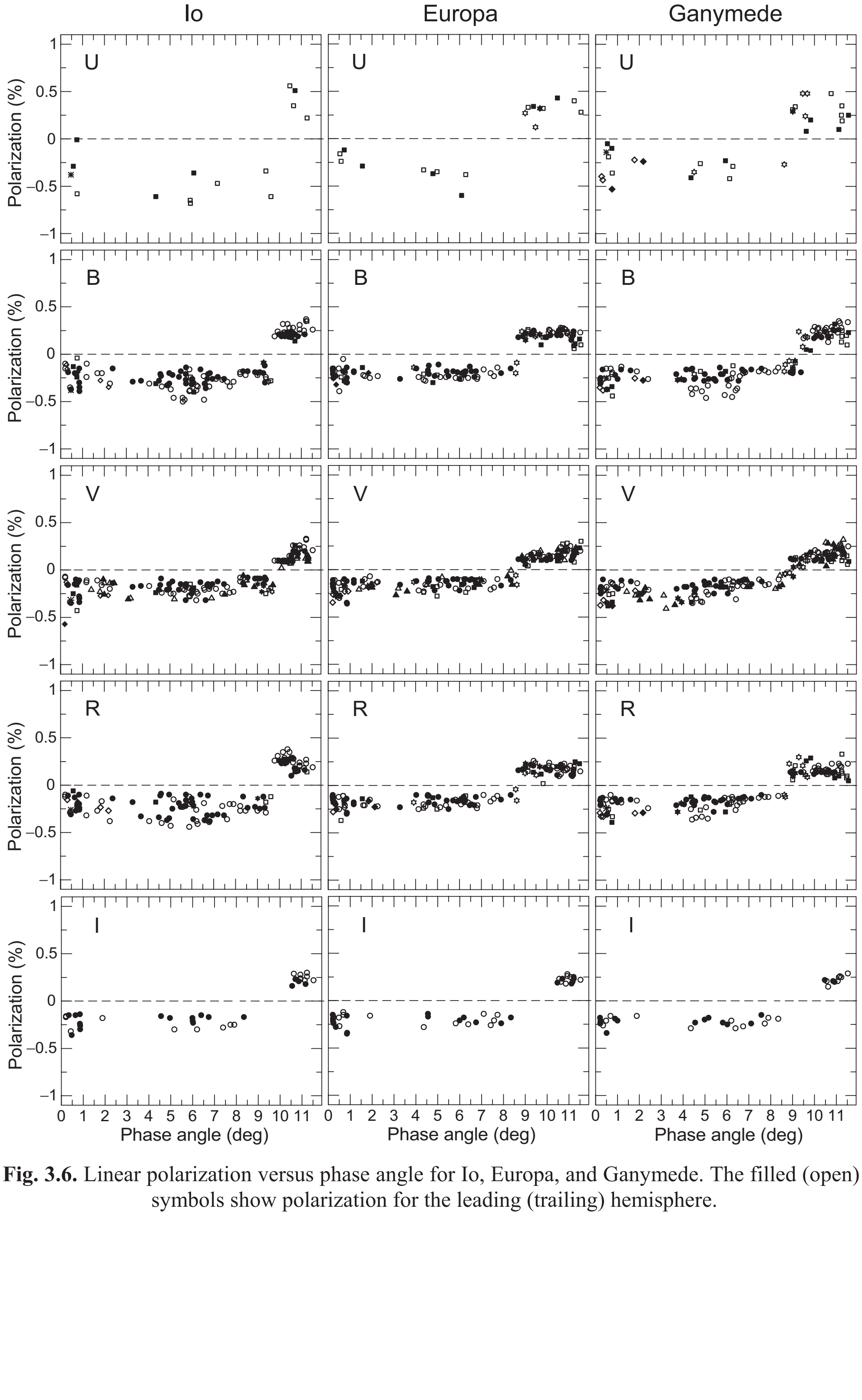

**Fig. 3.6.** Linear polarization versus phase angle for Io, Europa, and Ganymede. The filled (open) symbols show polarization for the leading (trailing) hemisphere.

***Phase-angle and longitudinal dependences of polarization for Callisto.*** Polarimetric measurements by Veverka (1971) and Dollfus (1975) showed for the first time that Callisto is polarimetrically quite distinct from the other Galilean satellites in that its leading hemisphere (in Callisto's phase-locked orbital motion around Jupiter) differs from the trailing one. The polarization curve for the leading face of Callisto (i.e., at its eastern elongation, when the orbital longitude is 90°) is characterized by $P_{min} \approx -0.9\%$, whereas at the western elongation (i.e., at the orbital longitude L = 270° corresponding to the trailing hemisphere), the polarization minimum is about −0.6%. The longitudinal dependence of polarization (LDP) was determined only for two phase-angle intervals: 9.7°−11.3° (Gradie and Zellner 1973; Dollfus 1975) and 10.7°−11.8° (Ботвинова и Кучеров 1980).

To distinguish between the polarization variations caused by changing orbital position of a satellite and those caused by changing its phase angle is a difficult problem. For a long time, this problem had not been fully resolved, primarily because observations for a wide range of longitudes at fixed phase angles had been lacking. To investigate the phase-angle dependence of polarization (PDP), measurements pertaining to different longitudes of the central meridian had been used: 0° ≤ L ≤ 180° for the leading hemisphere and 180° ≤ L ≤ 360° for the trailing one. Such aggregation leads to a significant scatter of polarization-versus-phase-angle data points significantly exceeding measurement errors. This scatter could be attributed to local inhomogeneities of the satellite surface coupled with global differences between polarimetric properties of the leading and trailing hemispheres.

A detailed study of the linear polarization measured for Callisto in the U, B, V, and R filters was carried out by Rosenbush (2002) using the results of the author's own observations as well as the totality of previously accumulated data. In particular, variations of polarization with phase angle, longitude, and wavelength were analyzed. This analysis revealed systematic discrepancies between different sets of observations, which had to be corrected for, thereby yielding a unified polarization dataset. Figures 3.7a and 3.7b summarize all available V-filter measurements. It is evident that the observed polarization depends on the phase angle as well as on the longitude. Furthermore, the amplitude of longitudinal variations of polarization also depends on the phase angle. This suggests that the observed degree of polarization can be approximated as a superposition of two functions:

$$P(\alpha, \mathrm{L}) = P_1(\alpha) + P_2(\alpha, \mathrm{L}). \tag{3.4}$$

The function $P_2(\alpha, \mathrm{L})$ can be expressed as

$$P_2(\alpha, \mathrm{L}) = A(\alpha)P(\mathrm{L}), \tag{3.5}$$

where $P(\mathrm{L})$ describes a quasi-sinusoidal dependence of polarization on the orbital longitude at a given phase angle and $A(\alpha)$ is the phase-angle-dependent amplitude vanishing at $\alpha = 0°$. $P_1(\alpha)$ is the longitude-independent polarization component common to both hemispheres.

The separation of the PDP and LDP was achieved using the method of successive iterations. If the differences between the longitudinal dependences for the two hemispheres are corrected for, a common longitudinal dependence at $\alpha = 6°$ can be

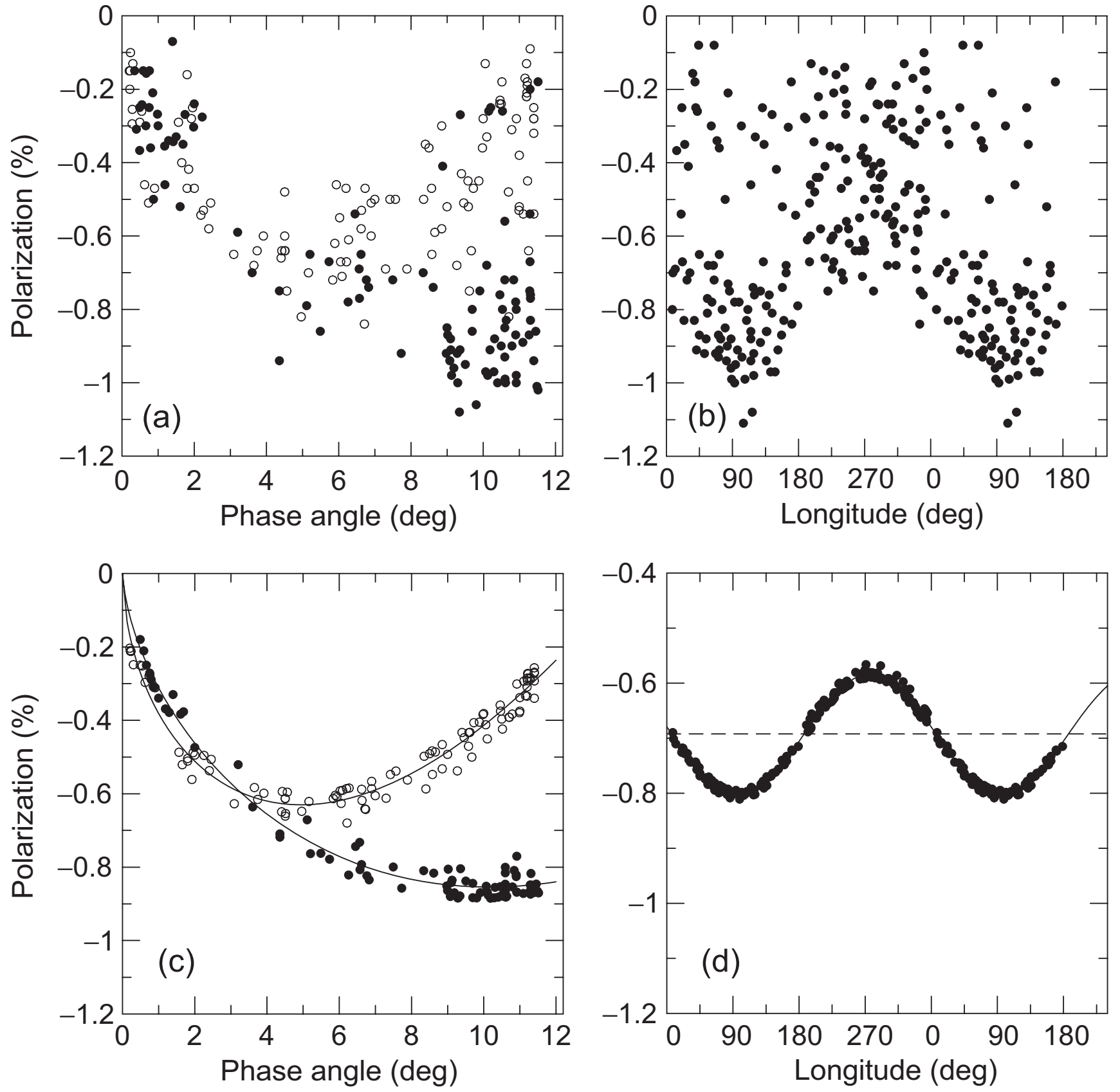

**Fig. 3.7.** Degree of linear polarization for Callisto versus phase angle (a) and longitude of the central meridian (b). In panel (a), open (filled) circles show measurements for the trailing (leading) hemisphere. In panel (b), the data points at L = 180° replicate those at L = 0°. (c) PDP for the leading (filled circles) and trailing (open circles) hemispheres of Callisto after the correction for the orbital latitudinal variations. (d) LDP at the phase angle $\alpha = 6°$ after the correction for the PDP. All data were taken in the V filter.

derived. The latter is approximated by the following function:

$$P(\alpha = 6°, \mathrm{L}) = P^1 \sin(\mathrm{L} - \mathrm{L}_0) + P^2, \tag{3.6}$$

where $P^1$ is the amplitude of the unified longitudinal dependence at $\alpha = 6°$, $P^2$ is the corresponding longitudinal shift, and $L_0$ is the longitude of equal polarization for the two hemispheres. The resulting phase-angle and longitudinal dependences of polarization in the V filter are shown in Figs. 3.7c and 3.7d along with the corresponding best-fit curves.

Comparison of Eqs. (3.4)–(3.6) shows that $P^1(\alpha) \equiv A(\alpha)$ and $P^2(\alpha) \equiv P_1(\alpha)$. Hence, the polarimetric observations of Callisto for any phase angle and any longi-tude can be described by the following general expression:

$$P(\alpha, L) = A(\alpha)\sin(\mathrm{L} - \mathrm{L}_0) + P_1(\alpha). \tag{3.7}$$

Aside from the factor $\sin(\mathrm{L} - \mathrm{L}_0)$ representing the LDP, Eq. (3.7) specifies the PDP $P(\alpha)$. The parameters of this relationship have been derived for each of the U, B, V, and R filters.

The PDP curves derived for Callisto reveal no features near opposition (i.e., at $\alpha \lesssim 2°$) exceeding in magnitude the measurement errors. For both hemispheres, the observed polarization at $\alpha \approx 1°$ ranges from about −0.3% to −0.4%. At smaller phase angles, the degree of polarization tends to zero, thereby exhibiting no POE.

The orbital variations of polarization are indicative of a heterogeneous distribution of materials with different physical properties over the surface of Callisto. As one can see in Fig. 3.7a, there is a cluster of data points with very low values of polarization for the leading hemisphere, from about −0.2 to −0.3%. A close inspection of this cluster reveals that these data were obtained at phase angles between 8.88° and 11.52° and longitudes between 14° and 42°. A comparison of the topographic maps obtained from Voyager with our data shows that the Valhalla ring system, which has a diameter of ~4000 km, with a high-albedo ~600-km-diameter palimpsest in its center, may have caused this particularly low polarization. The large-scale irregularities of the surface can have a profound influence on the total flux of the sunlight scattered by the particular area at a given phase angle. At small phase angles, the total brightness of the satellite is governed largely by the surface albedo, while at larger phase angles it is also affected by shadows cast by large-scale surface irregularities. Since the degree of linear polarization is the ratio of the linearly polarized flux and the total flux, the low polarization value detected for the leading hemisphere at phase angles ~10° is likely related to the impact basin Valhalla.

Both the solar-phase and the rotational-phase curves of brightness for Callisto exhibit the hemispherical dichotomy: at phase angles $\alpha \approx 2°–12°$ the trailing hemisphere is brighter than the leading one (Buratti 1991; Аврамчук и Шавловский 1998). However, the leading hemisphere has a more pronounced BOE ($\Delta m \approx 0.23^m–0.30^m$) than the trailing one ($\Delta m \approx 0.15^m–0.18^m$) (Аврамчук и Шавловский 1988). The leading and trailing hemispheres of Callisto are especially dissimilar in terms of their polarization properties such as the parameters of the NPB $P_{\min}$, $\alpha_{\min}$, and $\alpha_{\mathrm{inv}}$ and their spectral dependence. There is a correlation between the values of $|P_{\min}|$ and $\Delta m$ as well as between their spectral trends. The similarity of the geometric albedos and their spectral trends for both hemispheres may be indicative of the generally uniform surface composition of Callisto. Therefore, the differences in both the NPB and the amplitude of the BOE are, most likely, caused by the same factors, viz., differences in the surface microstructure of the leading and the trailing hemisphere. The well-developed NPB for the leading side indicates that the surface is not only microscopically rough but is also covered with dark grains. High-resolution images obtained from the Galileo spacecraft have confirmed that the surface of Callisto is strongly pitted and much of it is covered with dark, ice-depleted material. These traits have been interpreted as the result of erosion of the surface caused by sublimation of volatile compounds (Klemaszewski et al. 2001).

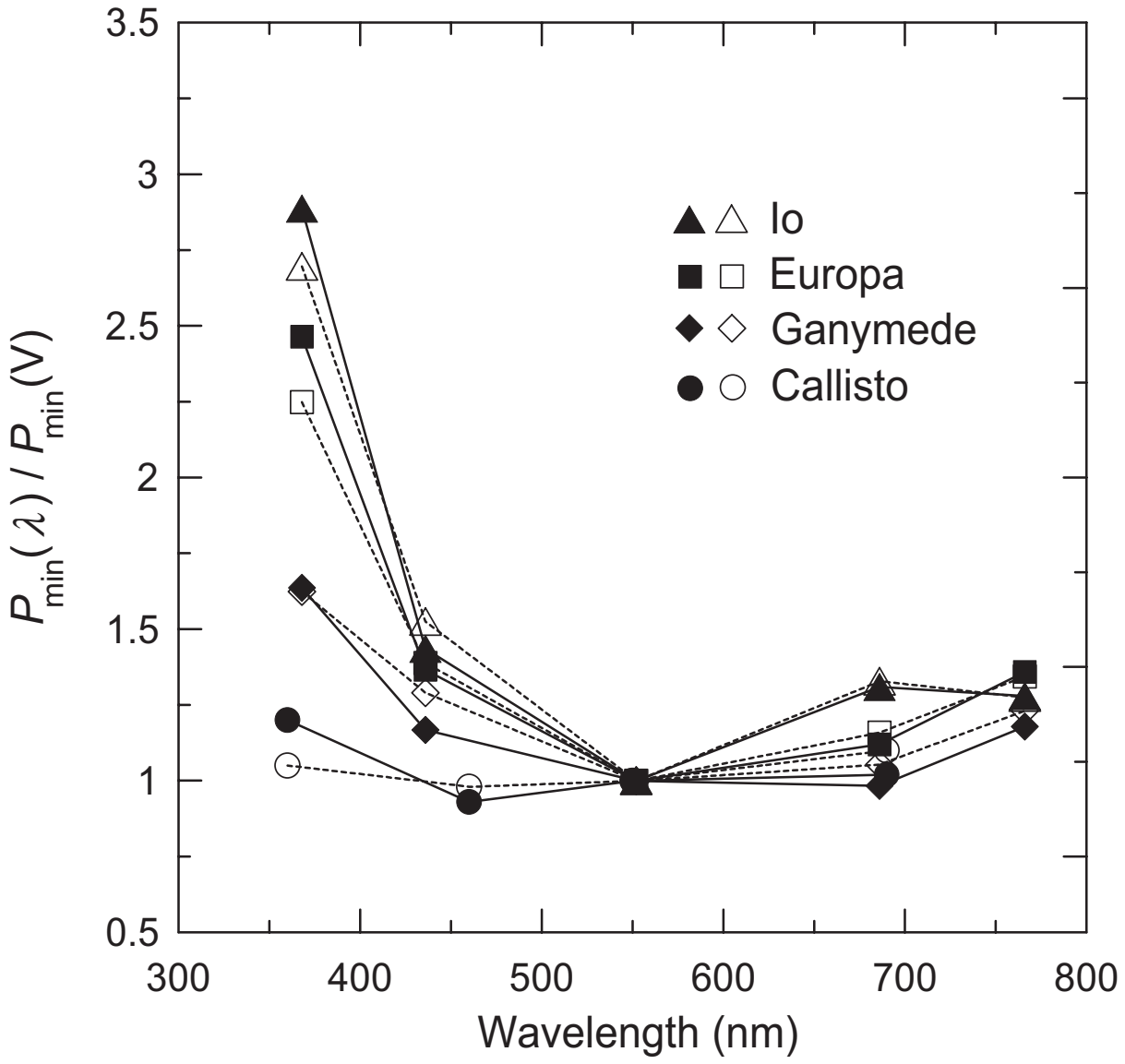

**Fig. 3.8.** Spectral dependence of normalized negative polarization for the Galilean satellites. Filled (open) symbols refer to the leading (trailing) hemispheres.

***Spectral dependence of polarization for the Galilean satellites.*** According to data from the Galileo and Pioneer 11 spacecraft (Martin et al. 2000), the PPB slightly weakens with wavelength in the red spectral range. Since the albedo of the Galilean satellites is smaller in the blue than in the red, this implies the applicability of the Umov law to the satellite surfaces. Figure 3.8 depicts the normalized spectral dependences $P_{min}(\lambda)/P_{min}(V)$ for the four Galilean satellites and allows one to compare these objects in terms of the spectral dependence of the depth of their NPBs. The latter dependence is twofold for all the satellites:

- the absolute value of $P_{min}$ decreases with wavelength in the UV, albeit at a different rate, the latter being the largest for Io and the smallest for Callisto;
- the absolute value of $P_{min}$ slightly increases with wavelength in the red.

The leading and trailing hemispheres differ not only in terms of the $P_{min}$ values but also in terms of their spectral dependences. Specifically, the leading hemispheres of the satellites show a somewhat larger rate of spectral change in the UV than the trailing hemispheres. The spectral dependence of polarization is consistent with that of brightness, which suggests that the Umov law applies to the NPBs as well as to the PPBs of the Galilean satellites.

***Behavior of polarization at the inversion point.*** As we have already discussed in Section 3.1.2, essentially for all ASSBs the polarization plane at phase angles $\alpha < \alpha_{inv}$ coincides with the scattering plane, thereby implying negative polarization, whereas at phase angles $\alpha > \alpha_{inv}$ the polarization plane is perpendicular to the scattering plane, thereby implying positive polarization. In order to study how polarization switches its sign, we have analyzed all available observations of the satellites Io, Europa, and Ganymede at phase angles within $\pm 1°$ of the inversion point (Fig. 3.9). This analysis has led to the following conclusions:

- In a narrow yet finite range of phase angles centered at $\alpha_{inv}$, the absolute value of polarization does not vanish completely but rather remains significant;
- The switch of the polarization sign occurs within a very narrow interval of phase angles: $9.67° \le \alpha_{inv} \le 9.78°$ for Io, $8.62° \le \alpha_{inv} \le 8.69°$ for Europa, and $8.84° \le \alpha_{inv} \le 8.90°$ for Ganymede;
- At phase angles $\alpha_{inv} - 1° \le \alpha \le \alpha_{inv} + 1°$, the absolute value of $P$ is essentially independent of $\alpha$, but exhibits a spectral dependence similar to that of $P_{min}$;
- Near the inversion point, when $|\alpha - \alpha_{inv}| \le 1°$, the position angle $\vartheta$ of the polarization plane relative to the scattering plane for Io deviates from both 90° and 0° by amounts exceeding the measurement errors. In fact, the polarization plane rotates smoothly within a very narrow range of phase angles.

The radiation coming from a Jovian satellite and captured by the telescope is a superposition of two components: (i) the sunlight scattered by the satellite, and (ii) the sunlight scattered first by Jupiter and then by the satellite. Theoretical estimates show that at the inversion point ($\alpha \approx 10°$) the absolute value of the degree of linear polarization of the second component is $\sim 0.002\% - 0.004\%$, which is significantly smaller than the measurement errors. Thus, the significant residual polarization observed for Io in the vicinity of the inversion point is caused by an optical and/or structural asymmetry of the scattering surface.

***Phase-angle and longitudinal dependences of polarization for Iapetus.*** Figuratively speaking, one hemisphere of Iapetus is coal-black, whereas the other one is brightly white. This satellite has an irregular shape (Plate 3.11). Both hemispheres are strongly pitted by craters, the crater density being quite comparable to those of Mercury, Callisto, and the lunar terrae. The equatorial regions are dominated by dark materials with albedos $\sim 4\%$, but at latitudes $> 40°$ one can observe a transition to much brighter ice-covered areas. Near the equator the dark areas can extend to the bright trailing side (Buratti and Mosher 1995). The brightest ice-covered areas have albedos exceeding 60%; however, they are not uniform and have thin darker stripes. Analyses of high-resolution images of Iapetus suggest the presence of fine-grained and porous surface materials.

There are several hypotheses intended to explain the striking dissimilarity of the two hemispheres of Iapetus. Spectrophotometric properties of the dark surface of Iapetus resemble those of the satellite Phoebe and are indicative of organic inclusions in carbonaceous chondrites. Hence one hypothesis implies the contamination of the leading side of Iapetus by dust ejected from Phoebe. According to another hypothesis, emissions of water vapor from sub-surface layers had caused the condensation of frost over the entire surface of Iapetus, but the interaction of the leading hemisphere with the Saturnian magnetosphere has led to the regional sublimation of ice. One way or the other, the two thoroughly different hemispheres of Iapetus are of exceptional interest from the standpoint of polarimetric studies.

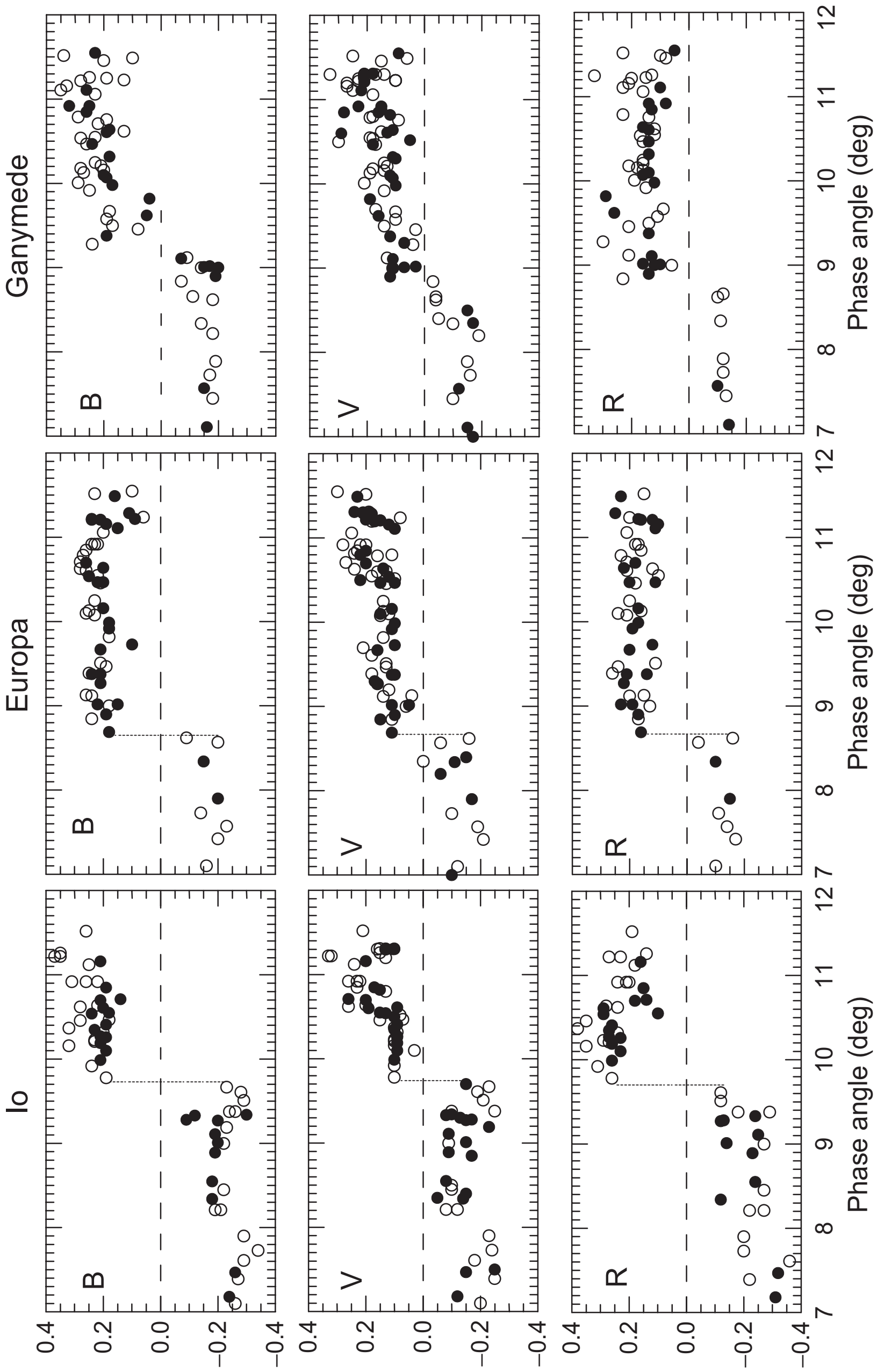

**Fig. 3.9.** Changes of the degree of linear polarization for the Galilean satellites Io, Europa, and Ganymede in the vicinity of the inversion point measured in the B, V, and R filters. Filled (open) symbols show the results for the leading (trailing) hemispheres.

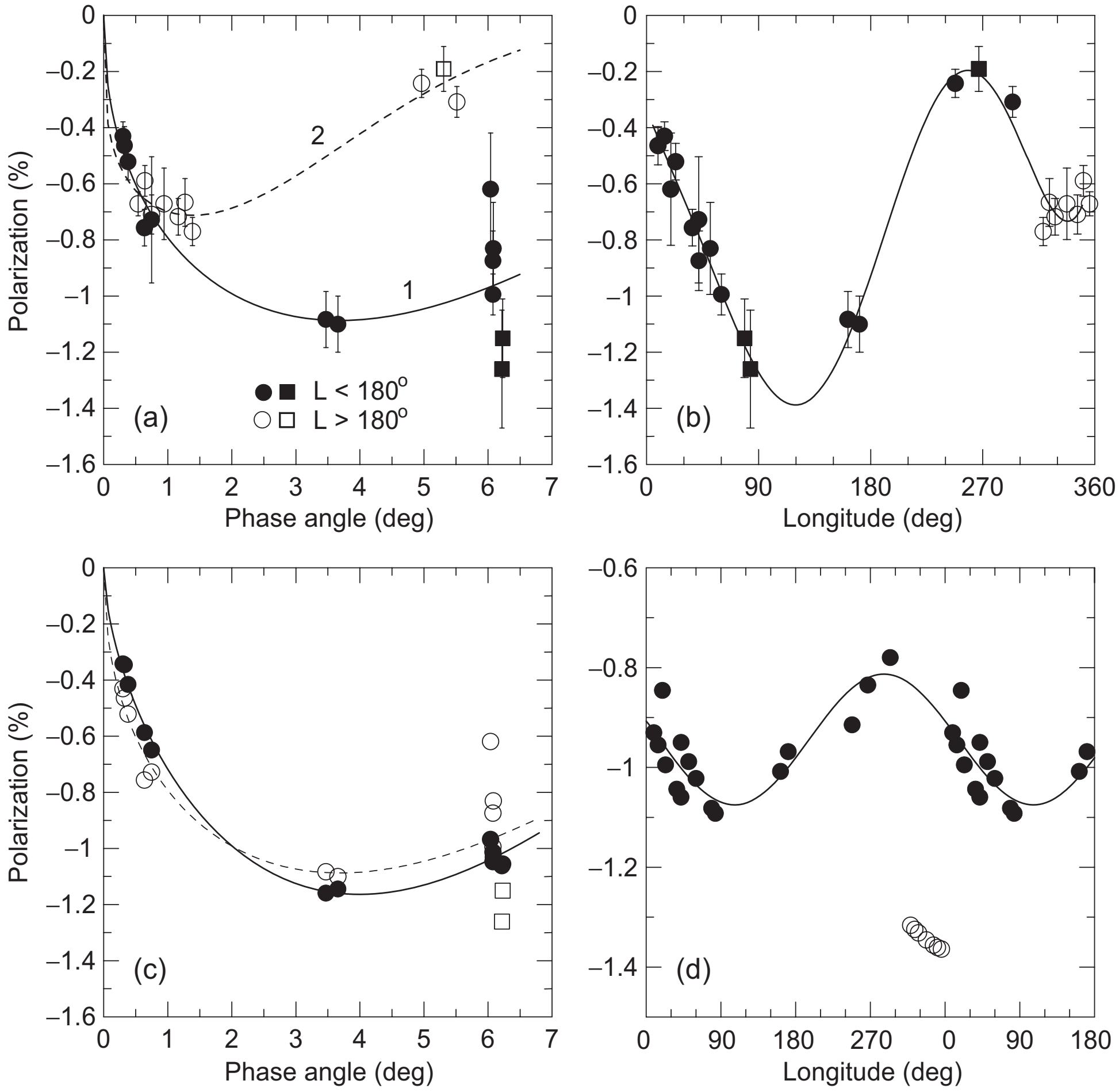

**Fig. 3.10.** (a) PDP for the leading (solid symbols) and trailing (open symbols) hemispheres of Iapetus observed in the R filter. The squares show data from Zellner (1972). (b) As in (a), but for the LDP. (c) PDP for the leading hemisphere. The filled circles demonstrate the result of correction for the LDP. (d) LDP at the phase angle $\alpha = 2°$.

The observed PDP and LDP in the R filter for the leading dark side (0° ≤ L ≤ 180°) and the trailing bright side (180° ≤ L ≤ 360°) are shown in Figs. 3.10a and 3.10b. One can readily see significant PDP differences between the two hemispheres. The leading-hemisphere polarization changes abruptly from about −0.4 to about −0.8% within the phase-angle range from ~0.3° to ~1°. At $\alpha \approx 6°$ polarization is close to that at $\alpha \approx 1°$. Therefore, the PDP for the leading hemisphere of Iapetus in the phase-angle range 0.3°−6.1° can be represented by an asymmetric parabola with a minimum at $\alpha \approx 3°–4°$ without any hint of a secondary minimum.

The separation of the PDP and LDP has been done only using R-filter observations, for which the measurement accuracy is the best, using the same technique as that applied to Callisto (Rosenbush 2002). In order to compare measurements taken under the same illumination conditions, the discrepancies between the model phase curves and the measured polarization resulting from the aggregation of longitudinal dependences observed at different phase angles have been reduced to the reference phase angle 2° (Figs. 3.10c and 3.10d). The correction for the LDP yields more consistent PDP points for the leading hemisphere compared to the original data (cf. the open and filled circles in Fig. 3.10c). The polarization phase curve has a minimum of $P_{min} = -1.16\% \pm 0.1\%$ at $\alpha_{min} = 4.0° \pm 0.2°$ and an inversion angle of $\alpha_{inv} \approx 13°$. It is worth noticing that asteroids with albedos as low as that of the leading hemisphere of Iapetus have minimal polarization at $\alpha_{min} \approx 10°$.

The LDP correction for the trailing hemisphere of Iapetus yields a phase dependence which is distinctly different from that for the leading hemisphere (compare the dashed and solid curves in Fig. 3.10a). This result is corroborated by Fig. 3.10d, in which the open circles correspond to the secondary minimum of polarization at small phase angles and deviate significantly from the regular LDP. We have also found that the absolute value of polarization reaches its maximum at L ≈ 106° for the leading hemisphere and at L ≈ 187.6° for the trailing hemisphere, thereby indicating that the dark and bright sides of Iapetus do not coincide precisely with its leading and trailing hemispheres, respectively.

### *3.1.6. Opposition effects for planetary satellites*

***Galilean satellites.*** High-accuracy observations of 2000 (Rosenbush and Kiselev 2005) fully corroborated our previous conclusion (Rosenbush et al. 1997a) that Io, Europa, and Ganymede exhibit a pronounced and narrow POE at very small phase angles (0.4°−0.7°) superposed on a regular NPB. We have also established that there is no POE for Callisto (Rosenbush 2002). Figure 3.11 summarizes the measurements of polarization for Io, Europa, and Ganymede in different filters during the 1988, 2000, and 2007 oppositions and reveals significant differences between the POEs for the three Galilean satellites. The POE for Europa has the shape of a sharp and asymmetric minimum with an amplitude of ~0.35% centered at $\alpha_{min,POE} \approx 0.2°$. The POE for Ganymede is wider ($\alpha_{min,POE} \approx 0.6°$) and, apparently, slightly deeper ($P_{min,POE} \approx -0.4\%$). The POE for Io appears to be more symmetric and shallow ($P_{min,POE} \approx -0.2\%$) and is centered at $\alpha_{min,POE} \approx 0.5°$. In addition, the POEs for Europa and Ganymede deepen with decreasing wavelength.

***Iapetus.*** The results of our polarization measurements in the B, V, R, and I filters for the trailing hemisphere of Iapetus along with the V-filter data by Zellner (1972) are shown in Fig. 3.12. The trailing-hemisphere polarization at phase angles 5°−6° is close to −0.3% (as opposed to the respective leading-hemisphere polarization values close to −1%), which is a typical $P_{min}$ value for the regular NPB of high-albedo asteroids such as 64 Angelina and 44 Nysa or ice-covered satellites such as Europa and Ganymede. However, at $\alpha \lesssim 1.5°$ one can clearly see a local minimum with a maximal depth of $P \approx -0.75\%$ at $\alpha \approx 1.1°$. Despite the absence of data at phase angles between 2° and 5°, one can conclude with confidence that the polarization measured at phase angles 5°−6° corresponds to the regular NPB whereas that at small phase angles exhibits a secondary local minimum centered at $\alpha \approx 1°$. Because of the scatter of data points at $\alpha \approx 5°–6°$, one cannot exclude a strongly asymmetric polarization phase curve for the trailing hemisphere of Iapetus similar to those observed in the laboratory measurements for particulate MgO (Lyot 1929) and $Al_2O_3$ (Geake and Geake 1990) surfaces.

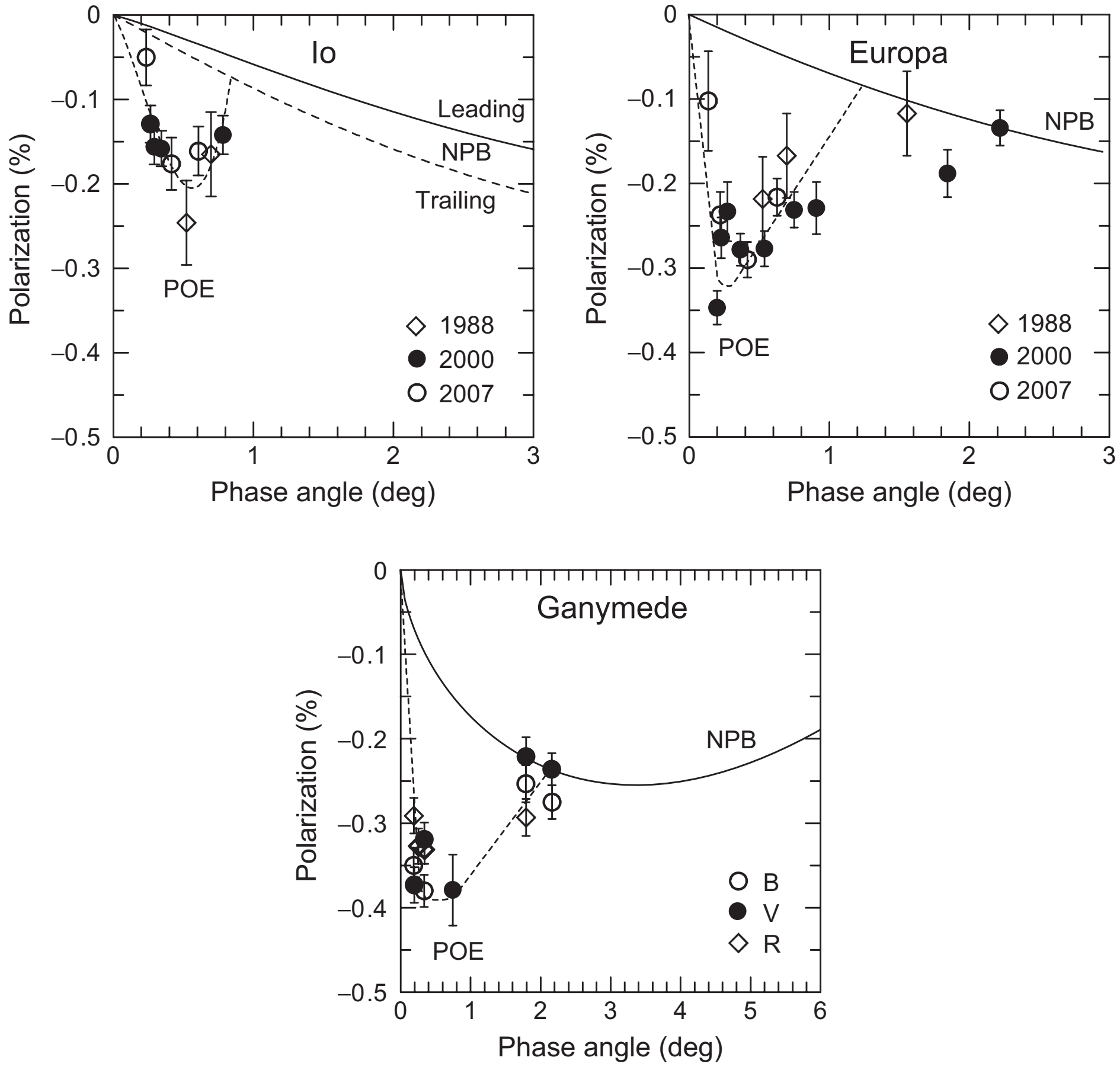

**Fig. 3.11.** POEs for the Galilean satellites Io, Europa, and Ganymede. The respective NPBs are shown schematically for the whole disks of Europa and Ganymede and for the leading and trailing hemispheres of Io.

### *3.1.7. Relation between observed optical effects and physical processes controlling the formation of the surface layer*

Exogenic processes such as the bombardement by high-energy particles and micrometeorites strongly affect the chemical composition of the surfaces of the icy Galilean satellites (e.g., by producing materials strongly absorbing UV light, in particular $SO_2$, $O_3$, $H_2O_2$) and their structure, thereby affecting the scattering properties of the surfaces and, ultimately, their photometric and polarimetric characteristics (Розенбуш 2006). A comparative analysis of the properties of the satellite surfaces and the processes controlling the formation of the optically active layer on the one hand and the observed spectral and phase dependences of the brightness and polarization on the other hand leads to the following conclusions:

- The discovery of the pronounced and narrow BOE and POE for the high-albedo satellites suggests that a significant fraction of the satellite surfaces is covered with fine-grained material causing strong CB;
- The detection of stronger polarization for the trailing hemispheres of Io, Europa, and Ganymede compared to their leading hemispheres is consistent with the observed albedo dichotomy caused by different formation processes ultimately defining the scattering properties of the opposite hemispheres. The reverse trait of Callisto corroborates this conclusion;
- Europa has the shallowest NPB (−0.15% in the V filter), whereas Callisto has the deepest NPB (−0.85% for the leading and −0.63% for the trailing hemisphere). These differences imply that the main process controlling the scattering properties of Europa is the bombardment by charges particles, whereas the defining processes for Callisto are the bombardment by micrometeorites and the fusion of particles;
- All four Galilean satellites are characterized by a nonlinear spectral dependence of polarization summarized by Fig. 3.8, the leading hemispheres exhibiting a stronger spectral gradient of polarization in the UV than the trailing hemispheres. The spectral dependence of polarization is consistent with that of the reflectivity of the Galilean satellites;

- The decrease of the absolute value of polarization with wavelength can be caused by a surface material strongly absorbing in the UV, potentially in combination with a surface-texture effect;
- The observed non-zero polarization of light scattered by Io, Europa, and Ganymede at the inversion point is indicative of an optical asymmetry caused by different scattering properties of the leading and trailing hemispheres.

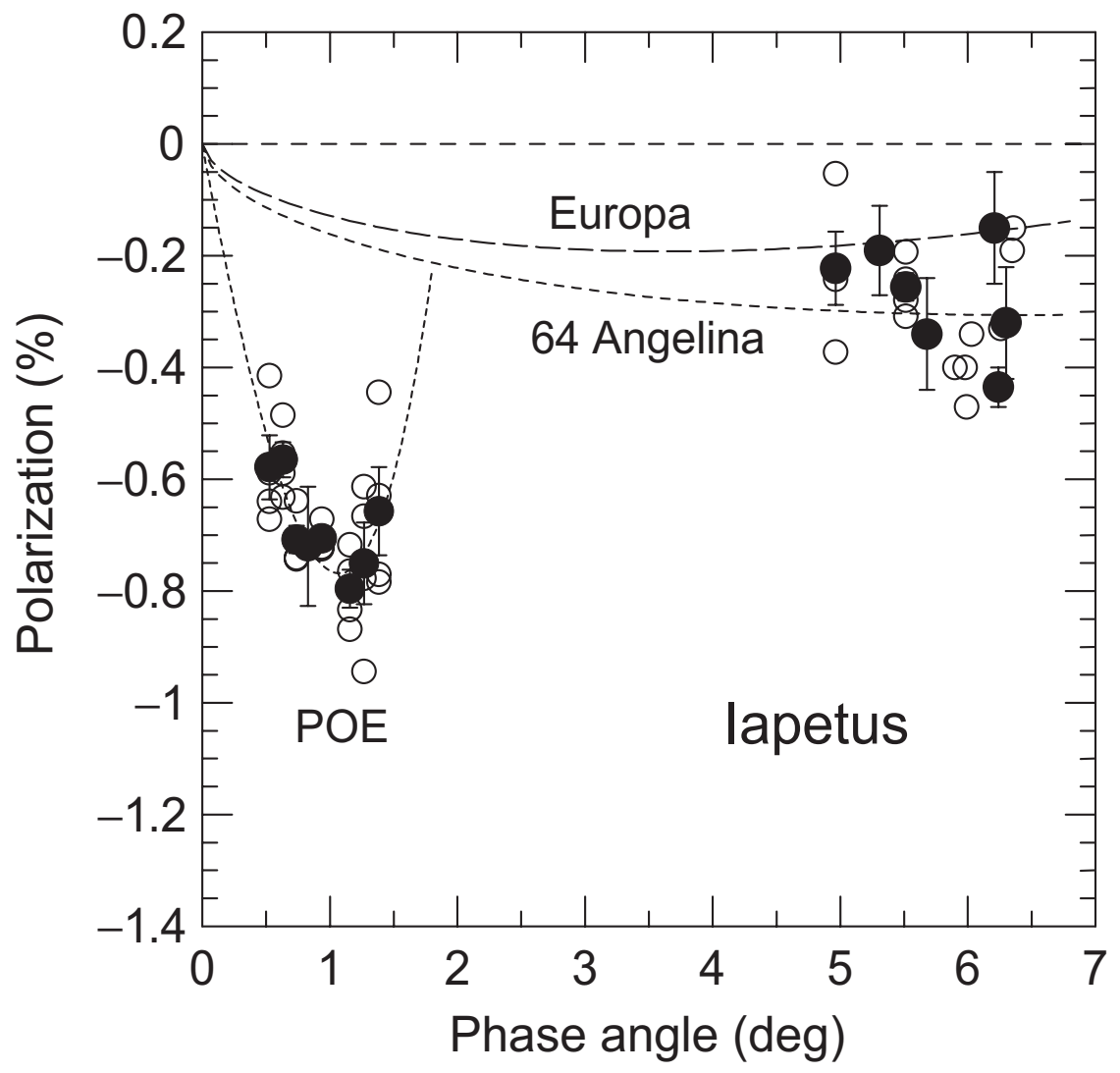

**Fig. 3.12.** PDP for the trailing hemisphere of Iapetus. Open circles show the results of individual measurements in the B, V, R, and I filters, while filled circles show filter-averaged data. The local minimum at $\alpha \approx 1°$ is the POE. The data at larger phase angles are close to the average PDP for Europa and 64 Angelina.

## 3.2. Asteroids

### *3.2.1. Observations*

Regular and systematic polarimetric observations of asteroids were initiated by these authors in 1983 and have been carried out in cooperation with different observatories and using different reflecting telescopes, including the following ones:

- 0.6 m, Soviet–Bolivian Observatory, Tarija (Bolivia);
- 0.7 m, Gissar Astronomical Observatory (Tajikistan);
- 0.7 m, Institute of Astronomy of the Kharkiv V. N. Karazin National University (Ukraine);
- 1.0 m, Sanglok Observatory (Tajikistan);
- 1.25 m, Crimean Astrophysical Observatory, Nauchny (Ukraine);
- 1.25 m, Abastumani Astrophysical Observatory (Georgia);
- 1.82 m, Asiago Astrophysical Observatory (Italy);
- 2.0 m, Ukrainian–Russian Observatory on Peak Terskol (Russia);
- 2.15 m, Complejo Astronomico El Leoncito (Argentina);
- 2.6 m, Crimean Astrophysical Observatory, Nauchny (Ukraine);
- 8.2 m, European Southern Observatory (ESO, Chile).

Identical UBVRI (370–830 nm) photometers–polarimeters designed by V. Piirola at the Helsinki University Observatory (Finland) have been used with the 1.25-m (Crimea) and 2.15-m (Argentina) telescopes. The 1.8-m (Italy), 2.0-m (Peak Terskol) and 8.2-m (Chile) telescopes were equipped with CCD-polarimeters based on Wollaston prisms, while the other telescopes were equipped with single-channel polarimeters based on a rapidly rotating analyzer.

During the past 25 years, we have accumulated polarimetric observations of ~200 asteroids of different sizes and types, of which 10 are near-Earth asteroids (NEAs). These numbers represent ~70% of all asteroids and ~70% of all NEAs studied by means of polarimetry. The main directions of research have been the measurement of polarization phase curves and the study of optical and geometrical characteristics of asteroids, asteroid type classification, analyses of the spectral behavior of polarization, observations of NEAs over a wide range of illumination–observation geometries, the identification and analysis of asteroids with anomalous characteristics, etc.

### *3.2.2. Phase-angle dependence of polarization*

In addition to the NPB at small phase angles, asteroids also exhibit a PPB at larger phase angles. Therefore, the phase dependence of polarization is typically described by the parameters $P_{min}$, $\alpha_{min}$, $\alpha_{inv}$ (see Section 3.1.2), polarimetric slope $h$ at the inversion phase angle, $P_{max}$, and $\alpha_{max}$. Polarization phase curves determined for a wide range of phase angles provide the most comprehensive information on the polarizing properties of asteroid surfaces. They are used for the classification of asteroids into types, the study of optical, structural, and mineralogical characteristics of asteroid surfaces, and for the determination of asteroid albedos and diameters. However, to obtain the polarization phase curve for an asteroid requires an extended period of observations and, in general, much observation time. In the case of ground-based observations of main-belt asteroids located between the orbits of Mars and Jupiter, the phase angles only reach values $\alpha \leq 30°$, which does not allow one to determine the parameters $P_{max}$ and $\alpha_{max}$. On the other hand, NEAs can provide an opportunity to obtain data at phase angles in the range of maximal positive polarization ($\alpha \sim 100°$) which yield additional valuable information (see Section 3.2.4).

As an example, Fig. 3.13 shows the polarization phase curve for asteroid 4 Vesta compiled from observations during several oppositions. Its parameters in the V filter (Розенбуш 2006) are the following: $P_{min} = -0.66\% \pm 0.05\%$, $\alpha_{min} = 7.0° \pm 0.2°$, $\alpha_{inv} = 21.8° \pm 0.2°$, $h = 0.066 \pm 0.002$. The $P_{min}$ and $h$ usually correlate with the surface albedo (see Section 3.2.8), which allows one to estimate the geometric albedo of Vesta at $p_V = 0.245$. This value is in a good agreement with an earlier estimate $p_V = 0.255$ (Lupishko and Mohamed 1996) but smaller than in Cox (1999).

We have also compiled detailed phase dependences of polarization for more than 20 asteroids of different types. Among them are seven NEAs, for which the phase dependences have been determined in a wide range of phase angles (see Section 3.2.4). These results, combined with data from the Asteroid Polarimetric Database (APD; see Section 3.2.11), allow us to derive and analyze the average phase curves of basic asteroid types (Бельская 2007). The respective parameters obtained by using a linear-exponential approximation function are summarized in Table 3.2.

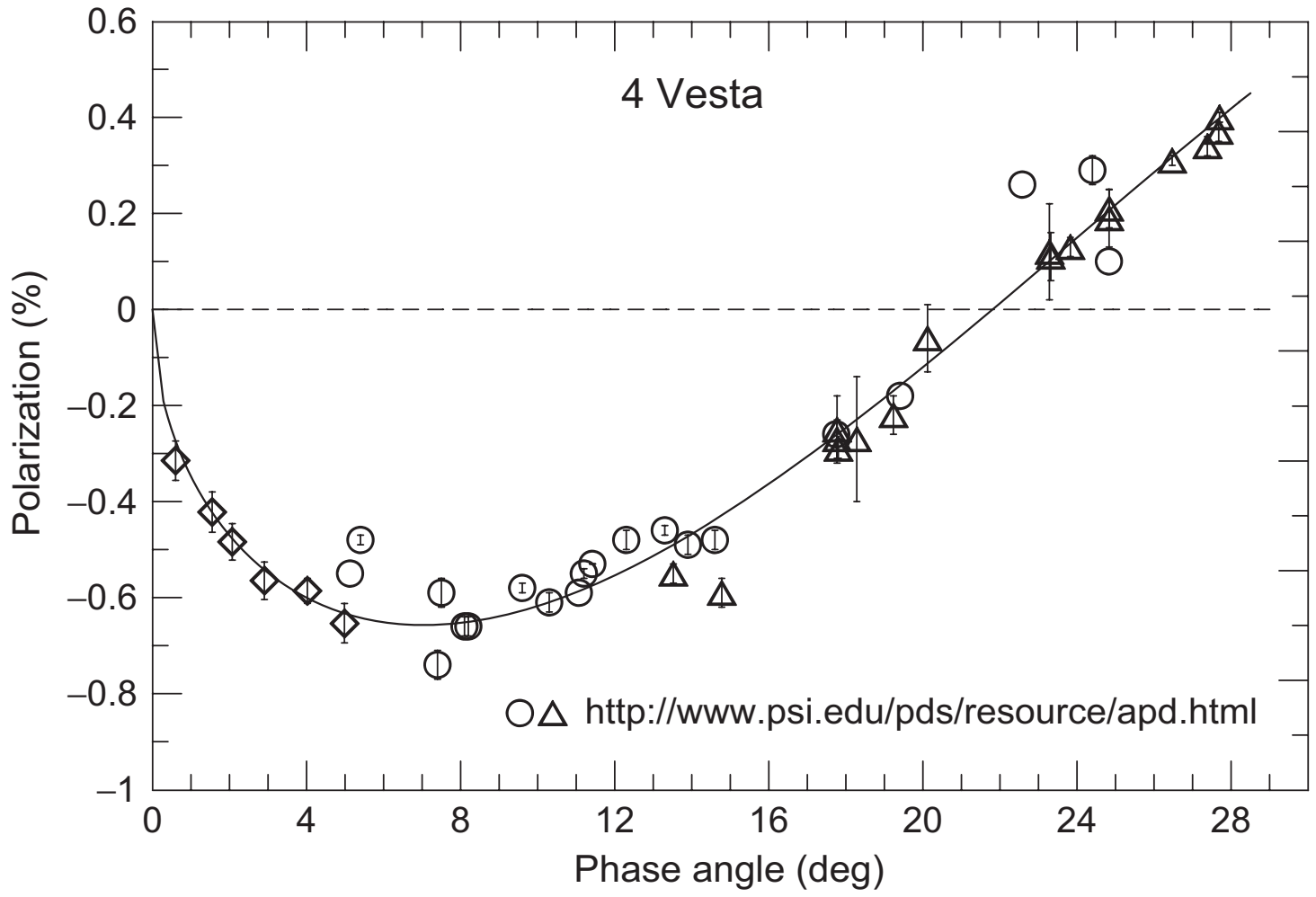

**Fig. 3.13.** Polarization phase curve of asteroid 4 Vesta in the V filter. The diamonds are data from Розенбуш (2006).

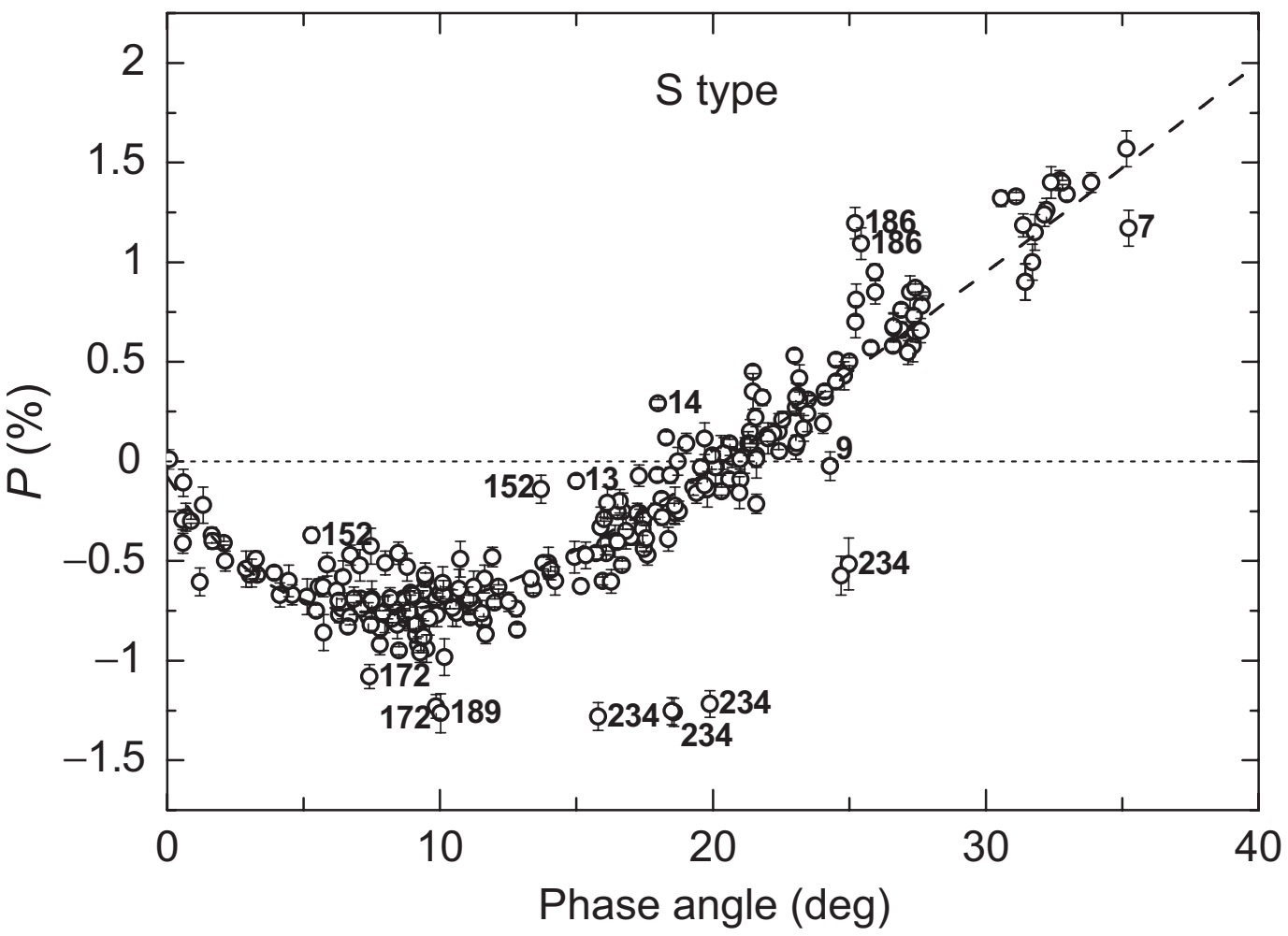

**Fig. 3.14.** Average phase-angle dependence of polarization for S-type asteroids. The data are compiled from Cellino et al. (2005, 2006) and the APD.

**Table 3.2.** Albedos and parameters of average polarization phase dependences of the main asteroid types in the V filter

| Type | Albedo | $\|P_{\min}\|$ (%) | $\alpha_{\min}$ (deg) | $\alpha_{\text{inv}}$ (deg) | $h$ (%/deg) |
|---|---|---|---|---|---|
| F | 0.05 | 1.15 ± 0.10 | 7.0 ± 1.7 | 15.5 ± 1.5 | 0. 327 ± 0.037 |
| C | 0.07 | 1.55 ± 0.55 | 8.7 ± 2.1 | 19.7 ± 1.5 | 0. 369 ± 0.039 |
| M | 0.15 | 1.08 ± 0.25 | 8.4 ± 1.3 | 22.0 ± 2.0 | 0. 170 ± 0.010 |
| S | 0.20 | 0.77 ± 0.20 | 8.0 ± 1.2 | 20.6 ± 2.0 | 0. 107 ± 0.005 |
| A | 0.42 | 0.40 ± 0.10 | $<7$ | 18.1 ± 1.5 | 0. 044 ± 0.008 |
| E | 0.51 | 0.31 ± 0.05 | 4.7 ± 1.3 | 18.0 ± 1.5 | 0. 042 ± 0.013 |

The average phase dependences are useful for the analysis of individual phase curves and identification of objects with unusual properties. Figure 3.14 exemplifies an average polarization phase curve derived for asteroids of the S type. This figure is also used to identify S-type asteroids with polarization properties considerably deviating from the average phase curve. The greatest deviation is exhibited by asteroid 234 Barbara. In fact, this asteroid has polarimetric properties distinctly different not only from the S-type asteroids, but also from all other asteroids for which polarimetric data are available (see Section 3.2.5).

### *3.2.3. Spectral dependence of polarization*

We have already mentioned that data on the spectral behavior of polarization characteristics of asteroid surfaces had been largely absent prior to the beginning of our observational program. However, such data are important both for the study of the composition and morphology of the surfaces of different asteroid types and for the development of the theory of negative polarization of light reflected by surfaces with complex structure, such as regolith surfaces. Since 1983, we have collected polarimetric observations of ~100 asteroids of different types. Initially our observations were carried at the 1.25-m CrAO reflector using the five-color UBVRI-polarimeter designed for simultaneous measurements in different spectral bands (Table 2.1) and at the 1-m telescope of the Sanglok Observatory with UBVR filters. More recently, five-color polarimetric observations were performed on the 2.15-m telescope of the Complejo Astronomico El Leonsito equipped with a similar photo-polarimeter (Cellino et al. 1999, 2005). The overwhelming majority of observations pertain to main-belt asteroids and, consequently, are limited to phase angles corresponding to the NPB.

Analyses of these data (Бельская и др. 1987b; Лупишко 1998a; Киселев 2003; Belskaya et al. 2009b) have shown that in the spectral interval 370–830 nm, the spectral variability of $|P_{\min}|$ can range from ~0.1% (2 Pallas, 20 Massalia, 349 Dembowska) up to 0.4%−0.5% (51 Nemausa, 324 Bamberga, 704 Interamnia), the magnitude of variability not obviously correlating with the value of $P_{\min}$ or with the measurement accuracy. For low-albedo asteroids (1 Ceres, 2 Pallas, 324 Bamberga, 704 Interamnia and others), the absolute value of $P_{\min}$ decreases with wavelength, whereas for moderate-albedo asteroids of the S, M, and V types (4 Vesta, 16 Psyche, 21 Lutetia, 27 Euterpe, 55 Pandora and others), $P_{\min}$ increases with $\lambda$. Similar conclusions have been reached based on analyses of the average polarization phase dependences for these asteroid types in the blue (B) and green (V) filters. Thus, the wavelength dependence of $P_{\min}$ is primarily defined by the asteroid type, although the variability within a specific asteroid type can be significant and usually exceeds the measurement uncertainties.

The general pattern of the spectral behavior of $P_{\min}$ for asteroids appears to be similar to that of meteorites of appropriate types. Бельская и др. (1989) measured $P_{\min}(\lambda)$ in the laboratory for samples of 15 different meteorites including all basic types of chondrites and achondrites as well as for 4 samples of terrestrial silicates (olivine, bronzite, hypersthene, and hedenbergite). The average particle size in the samples was $\lesssim$50 μm, which corresponds to many existing estimates of the mean size of particles covering the surfaces of asteroids and other ASSBs. The albedos of the samples, depending on their composition, were in the range 0.07−0.52.

The measurement results have shown that for the samples of ordinary and enstatite chondrites, achondrites, and terrestrial silicates, the NPB monotonously deepens with wavelength, which is similar to the polarization behavior of moderate-albedo asteroids of the S, M, and V types. The carbon chondrites show a different behavior of $P_{\min}$: their $|P_{\min}(\lambda)|$ dependences show a maximum at 500−600 nm and then a decrease with wavelength. Low-albedo asteroids exhibit a similar behavior of $|P_{\min}(\lambda)|$.

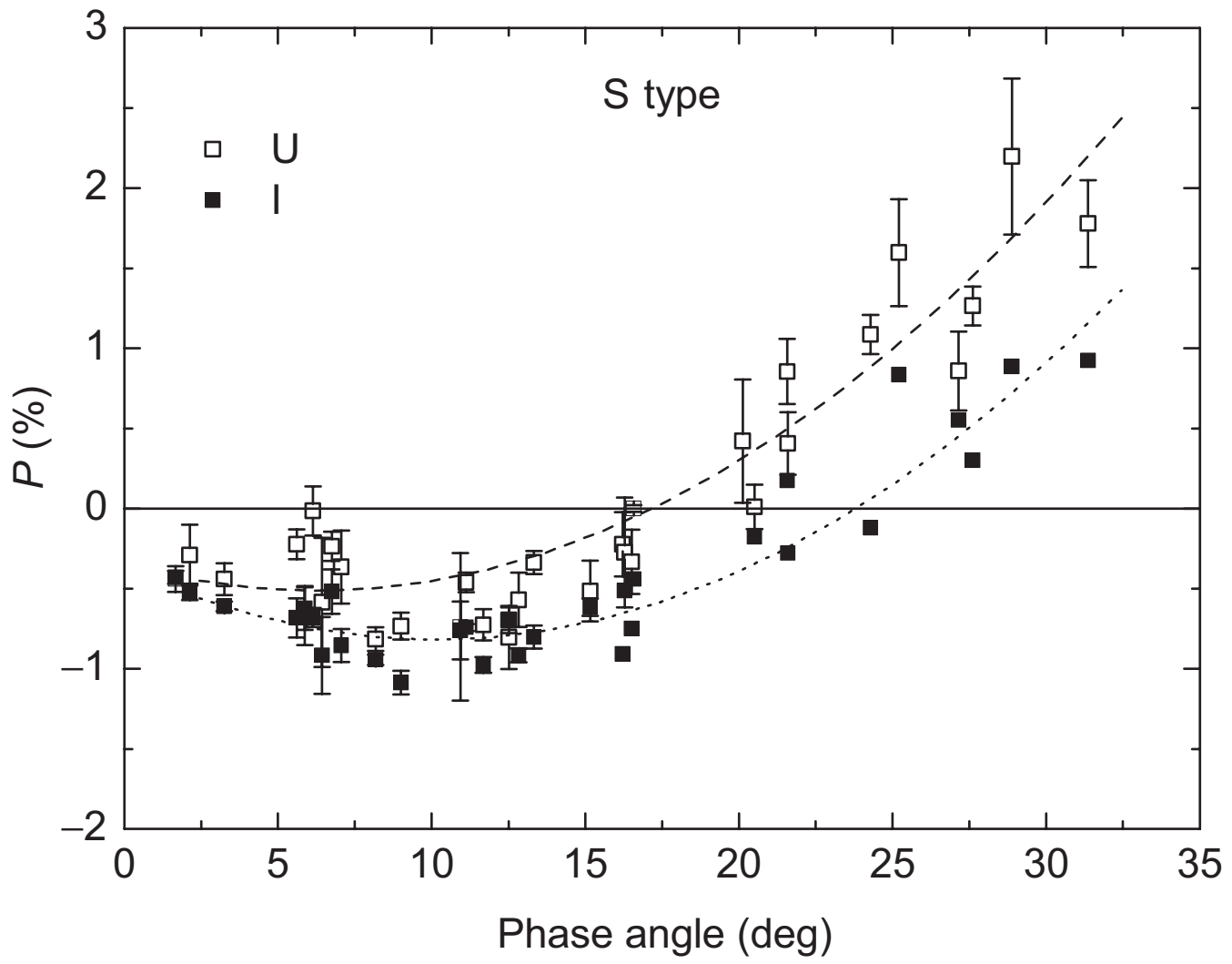

**Fig. 3.15.** Composite polarization phase dependence of S-type asteroids in the U and I filters.

Thus, one can conclude that asteroids of the basic types and their meteoritic analogues have similar spectral dependences of $P_{\min}$. This result is important both for establishing a genetic relationship between asteroids and meteorites and for a more reliable classification of asteroids into types.

Analyses of these data have also led to the conclusion that the spectral behavior of the surface albedo, $p(\lambda)$, does not affect noticeably the spectral behavior of $P_{\min}$. It is worth mentioning that according to the model proposed by Шкуратов (1987), the dependence $P_{\min}(\lambda)$ is largely defined by differences in the spectral behavior of the single-scattering albedos of different particle types forming a multi-component particulate surface. This theoretical prediction is corroborated by laboratory photopolarimetric measurements of a 1:1 mixture of the high-albedo colorless MgO powder and the intensively red $Fe_2O_3$ substance. The measurements have shown that the NPB reaches its maximal depth (more than 2%) in the blue part of the spectrum, where the contrast in the component single-scattering albedos is also maximal.

The spectral dependence of the PPB for asteroids is less well known since this information can come only from observations of NEAs. Unfortunately, because of their small brightness and fast visible movement, NEAs are much more difficult to observe with requisite accuracy. We have performed measurements of the PPB for NEAs 1036 Ganymed and 1685 Toro in the B, V, and R bands, 4179 Toutatis and 33342 1998 WT24 in the U, B, V, R, and I filters, and 2100 Ra-Shalom in the B and V filters. It turns out that while the absolute value of $P_{\min}$ for S-type asteroids increases with wavelength, the positive polarization becomes substantially weaker. This inversion of the spectral dependence of polarization occurs at phase angles close to $\alpha_{\text{inv}}$, as can be seen from Fig. 3.15. On the other hand, the spectral de-pendence of polarization for C-asteroids 2100 Ra-Shalom (Kiselev et al. 1999) and 1580 Betulia (Tedesco et al. 1978) reveals no inversion: both the NPB and the PPB become slightly weaker with wavelength.

### *3.2.4. Polarimetry of near-Earth asteroids*

Recently, the interest in the NEAs has strongly increased owing to the recognition of the potential asteroid hazard problem. Furthermore, the NEAs are considered to be potential sources of metals and other raw materials in the close vicinity of the Earth. In many cases the polarimetric method represents essentially the only means of obtaining meaningful estimates of the albedo and size of an asteroid. Table 3.3 summarizes the results of our analyses and lists the parameters of polarization phase curves of NEAs derived from observations in the V filter as well as estimates of the asteroid albedos $p_V$ and diameters $D$.

***1036 Ganymed, 1627 Ivar, and 1685 Toro.*** The first two asteroids belong to the Amor group and were observed by us polarimetrically for the first time, while Toro belongs to the Apollo group. We observed these NEAs using the 1.0-m reflector of the Sanglok Observatory. 1036 Ganymed was also observed with the 0.6-m reflector of the Soviet–Bolivian Observatory (Киселев и др. 1994). These observations took much time: 34 nights for Ganymed (B, V, and R filters), 12 nights for Ivar (V filter), and 21 nights for Toro (B, V, and R filters). As a result, we have derived the corresponding polarization phase dependences in specific phase-angle ranges $\Delta\alpha$; the latter ultimately define which numerical parameters of the polarization phase curves can be determined (see Table. 3.3).

In 1988, the phase angle for 1685 Toro reached 110°, which corresponds to the location of the positive polarization maximum; this range of phase angles for asteroids had never been captured before. Our observations (Киселев и др. 1990) have thus allowed us to estimate, for the first time, the parameters of maximal positive polarization: $P_{max} = 8.5\% \pm 0.7\%$ at $\alpha_{max} = 110° \pm 10°$ (Table 3.3). The comparison of the polarization phase curves for Toro and Ivar shows that at $\alpha = 70°$, $P$ for Toro is equal to 5.3% while that for Ivar is equal to 6.0%. Taking into account the difference in the polarimetric slopes $h$ for these asteroids, one can estimate that $P_{max}$ for Ivar (assuming equal values of $\alpha_{max}$) is 10%−10.5%. Thus, on the basis of our observations one can conclude that for S-type asteroids in the V filter, $P_{max}$ can reach values 8.5%−10.5% (for Mercury and the Moon, the maximal positive polarization is equal to 8.2% and 8.6%, respectively).

By applying the Geake and Dollfus (1986) calibration, one can use the measured parameters $h$ and $P_{max}$ to estimate the average size of Toro's regolith particles at $d \approx 30$ μm. Analogous estimates for the fine fraction of the lunar regolith yield $d \approx 10$ μm (see Киселев и др. 1990 for more details).

***4179 Toutatis and 1620 Geographos.*** The UBVRI polarimetry of these NEAs was carried out using the 1.25-m CrAO reflector equipped with the five-color polarimeter. 4179 Toutatis was observed during six nights in the range of phase angles 15.8°−51.4° (Lupishko et al. 1995). A simple linear approximation of the resulting polarization phase curves yielded the corresponding values of the parameters $h$ and $\alpha_{inv}$ in the U, B, V, R, and I filters and allowed these authors, for the first time, to analyze their spectral changes.

**Table 3.3.** Parameters of polarization phase dependences of NEAs in the V filter and polarization-based estimates of albedos and diameters

| Asteroid | $\Delta\alpha$ (deg) | $P_{min}$ (%) | $h$ (%/deg) | $P_{max}$ (%) | $\alpha_{max}$ (deg) | Albedo | $D$ (km) | Reference |
|---|---|---|---|---|---|---|---|---|
| 1036 Ganymed | 9.7−37.7 | –0.84 | 0.112 | – | – | 0.16 | 36.3 | Киселев и др. (1994) |
| 1627 Ivar | 34.4−64.1 | – | 0.131 | (10.3) | – | 0.14 | 9.5 | Киселев и др. (1994) |
| 1685 Toro | 47.0−106.3 | – | 0.099 | 8.5 ± 0.7 | 110 ± 10 | 0.18 | 3.0 | Киселев и др. (1990) |
| 4179 Toutatis | 15.8−51.4 | – | 0.111 | – | – | 0.16 | 3.1 | Lupishko et al. (1995) |
| 23817 $2000PN_9$ | 90.7−115.0 | – | – | 7.7 ± 0.5 | 103 ± 12 | 0.24 | 1.6 | Belskaya et al. (2009a) |
| 33342 $1998WT_{24}$ | 11.9−83.3 | –0.25 | 0.039 | 1.7 | 72 | 0.43 | 0.38 | Kiselev et al. (2002b) |
| 144898 $2004VD_{17}$ | 26.3−76.8 | – | 0.037 | 2.3 | 80 | 0.45 | 0.32 | De Luise et al. (2007) |

The analysis of the Toutatis data has revealed a totally unexpected result. Specifically, when the second normalized Stokes parameter $q = P_{lp} \cos 2\vartheta$ is relatively large ($q$ = 1%–4% during three nights with $\alpha > 30°$), the third normalized Stokes parameter $u$ is equal to zero to within the measurement errors. However, in the vicinity of the inversion point, where the absolute value of $q$ is close to zero, the parameter $u = P_{lp} \sin 2\vartheta$ deviates from zero significantly. This is clearly seen from the measured dependence of the position angle on the phase angle (Fig. 3.16) expressed in terms of the angle $\vartheta_r$ between the polarization plane and the normal to the scattering plane. The results obtained during three different nights with $\alpha$ = 15.8°–18.6° show that $\vartheta_r$ differs from 0° or 90° and is close to 45°. This effect is present in all five bands and exceeds considerably the corresponding measurement errors, thereby indicating a significant deviation of the preferential vibration plane of the electric field vector from the directions parallel or perpendicular to the scattering plane. Another NEA 1620 Geographos (also of the S type) was observed in September of 1994 during four nights at the same telescope and with the same polarimeter at phase angles close to and bracketing the inversion point (Васильев и др. 1996). However, no noticeable deviation of the polarization plane from the directions parallel or perpendicular to the scattering plane was detected. Among the possible explanations of this unusual polarization trait of Toutatis could be the effect of the extremely complex and irregular shape of this asteroid (as revealed by radar observations) on its disk-integrated scattering properties. Note that a similar polarization effect has been observed for Io, Europa, and Ganymede (see Section 3.1.5) as well as for comets Halley and Hale–Bopp (see Section 4.7).

***23817 (2000 PN$_9$).*** This object belongs to the Apollo group and is classified as a potentially hazardous asteroid (PHA). It approached the Earth to within 0.02 AU in March of 2006 and became observable with moderately sized telescopes. Our polarimetric observations of this object were performed with the 1.82-m Asiago telescope during two nights at phase angles 90° and 115° (Belskaya et al. 2009a). Note that the latter is the largest phase angle value ever achieved in asteroid polarimetry. The measured maximal polarization value $P$ = 7.7% agrees with observations of two other S-type NEAs 1685 Toro and 4179 Toutatis (see Table 3.3). The angle of maximal positive polarization for the S-type asteroids was confirmed to occur in the range 90°–115°. The general agreement of the polarization data for the three asteroids points to the similarity of their surface textures.

***33342 (1998 WT$_{24}$).*** In December of 2001, the PHA asteroid 33342 1998 WT24 from the Aten group passed at a distance of 0.0125 AU of the Earth. Simultaneous photometric and polarimetric observations of this object were carried out using three reflecting telescopes (the 0.7-m telescope of the Institute of Astronomy of the KhNU, the 1.25-m CrAO telescope, and the 2.0-m telescope of the Terskol Observatory) during 9 nights in the range of phase angles $\alpha$ = 12°–83°. The resulting brightness and polarization phase curves allow a reliable identification of this object as a high-albedo E-type asteroid as well as the determination of its basic characteristics (rotation period, maximal and minimal sizes 420 and 330 m, and albedo $p_V = 0.43$). In combination with available polarimetric data for main-belt E-type asteroids 44 Nysa, 64 Angelina, and 2867 Steins, one can derive a complete polarization phase dependence of E-type asteroids in the V filter (Fig. 3.17). This composite polarization phase curve shows a relatively low value $P_{max} = 2.00\% \pm 0.05\%$ (compared to 8.5% for the S-type asteroid 1685 Toro and 25%–30% for dusty comets) and an unexpectedly small phase angle of maximal positive polarization $\alpha_{max} = 76° \pm 7°$ (compared to ~110° for S-type asteroids and 95° for dusty comets; see Kiselev et al. 2002b; Киселев 2003). The application of the Geake and Dollfus (1986) calibration to the measured values of $h$ and $P_{max}$ yielded an estimate of the average size of surface regolith particles of ~25 μm (Kiselev et al. 2002b).

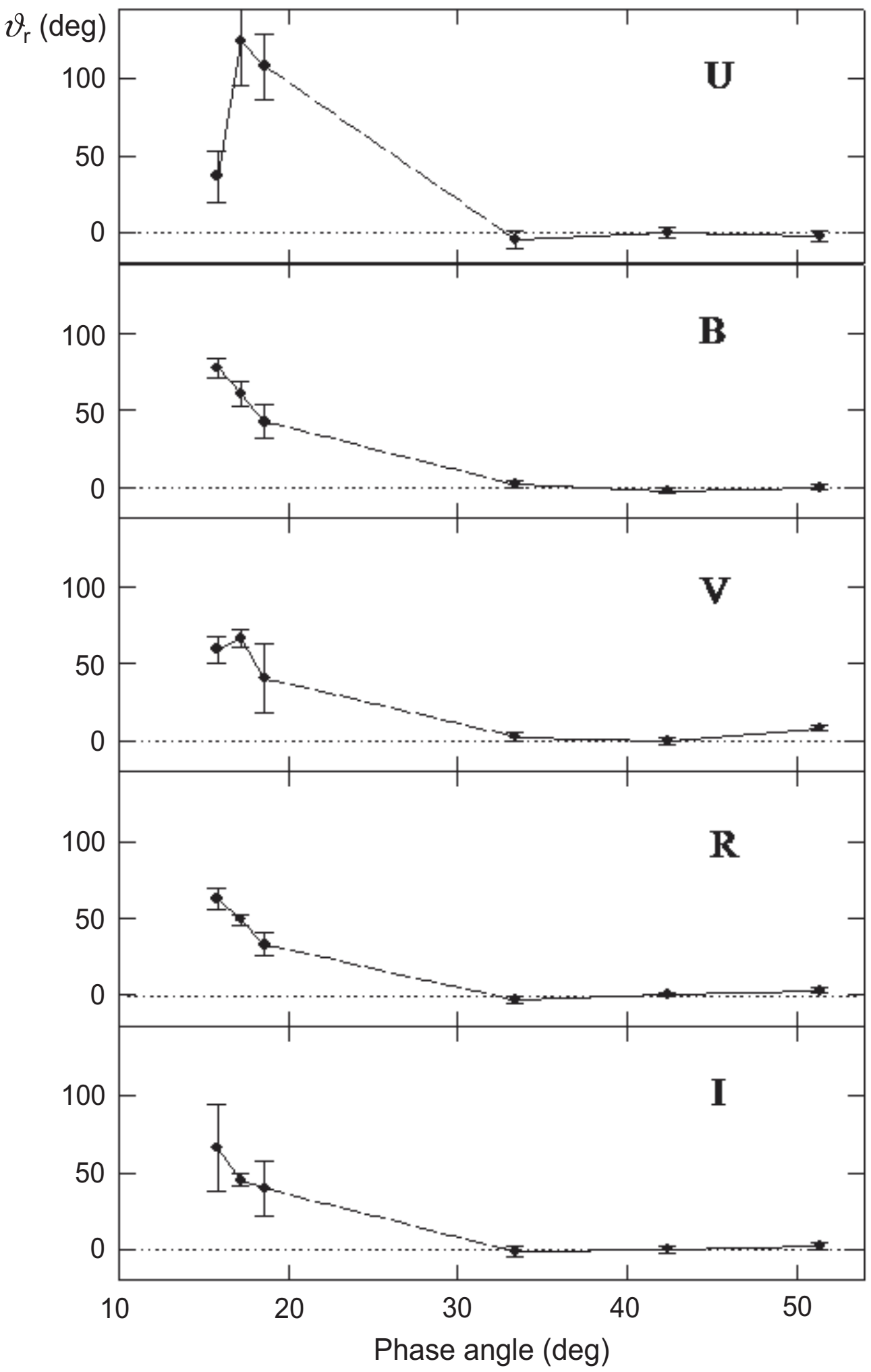

**Fig. 3.16.** Phase-angle dependence of the angle $\vartheta_r$ for asteroid 4179 Toutatis.

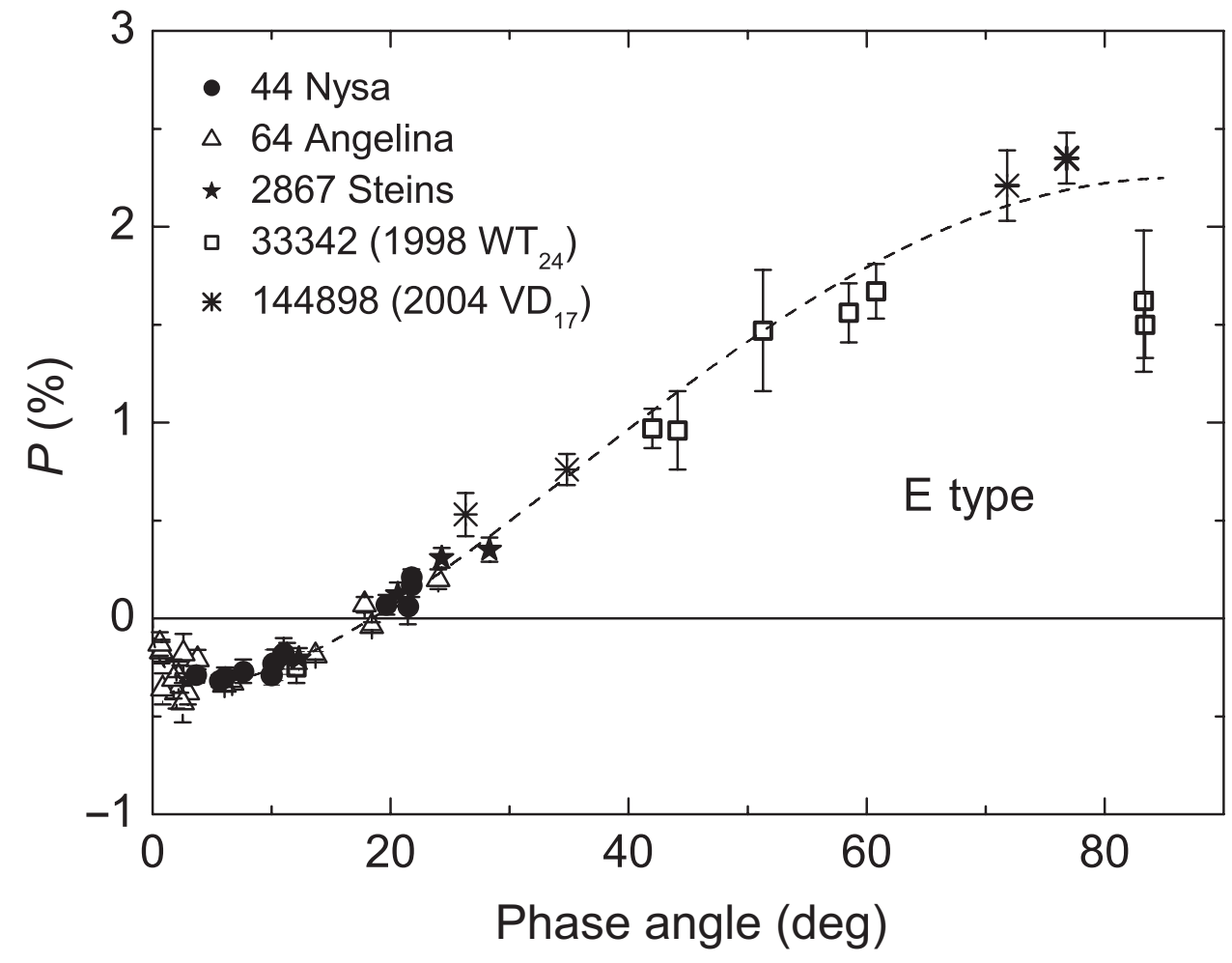

**Fig. 3.17.** Composite polarization phase dependence of E-type asteroids in the V band (see text for references).

***144898 (2004 VD$_{17}$).*** This member of the Apollo group was discovered in 2004 during its close approach to the Earth to within 0.09 AU and is classified as a PHA. The main objective of our observations was to estimate its albedo and size. Polarimetric observations were performed with the 8.2-m ESO telescope during four nights in April–May of 2006 in the so-called service mode (De Luise et al. 2007). The measured value of the polarimetric slope (see Table 3.3) and the relatively small polarization $P$ = 2.35% measured at a large phase angle of $\alpha$ = 76.8° are indicative of a high-albedo E-type asteroid. These results agree well with data available for another NEA 33342 1998 WT24 (see Fig. 3.17).

### *3.2.5. Identification and study of asteroids with anomalous polarimetric characteristics*

A long-term cooperative program of polarimetric observations of selected asteroids was carried out in 1996–2006 at three observational sites in Ukraine, Argentina, and Italy. One of its primary goals was to study the diversity of polarimetric properties of asteroids. The results thus obtained have helped us increase considerably the number of asteroids studied polarimetrically and derive the average polarimetric characteristics of the main compositional types of asteroids.

In general, asteroids from the same compositional type show very similar values of the respective polarimetric parameters (see Section 3.2.2). Only 10% of all asteroids studied so far have been found to exhibit significant deviations of their polarization phase curves from the average polarization phase curves characteristic of their respective compositional types. Whenever such deviations were discovered for a particular object, we continued polarimetric observations of not only this asteroid but also of asteroids of similar composition in order to understand possible causes of polarimetric anomalies. As a result, we have found that the inversion angles of asteroid polarization phase curves cover a wider range of phase angles than had been accepted previously. We have discovered asteroids with uniquely small (~14°) and uniquely large (~28°) inversion angles. Furthermore, we have found that the minimal polarization values $P_{\min}$ for these asteroids violate significantly the well-known correlation "$P_{\min}$ – albedo". Below we analyze these unique cases in more detail.

***Asteroids with small inversion angles.*** The first polarimetric observations of asteroid 419 Aurelia carried out in 1997 at phase angles close to $\alpha_{\min}$ showed a relatively shallow NPB with $|P_{\min}| \sim 1\%$. This value is considerably smaller than those typically measured for low-albedo asteroids (see Section 3.2.2). The next suitable opposition occurred in 2001. The small value of $|P_{\min}|$ was confirmed. In addition, we found that the likely value of the inversion angle was unusually small. We continued observations of Aurelia in 2003 and found that its polarization inversion angle was ~14°, which is the smallest value ever observed for an asteroid. The polarization phase dependence of Aurelia measured in the V and R spectral bands is shown in Fig. 3.18a.

All low-albedo asteroids studied previously by means of polarimetry had been found to have a standard phase curve with $|P_{\min}| \approx 1.5\%-1.7\%$ and $\alpha_{\text{inv}} \approx 20°$ (see Section 3.2.2). The only known exception had been asteroid 704 Interamnia with $\alpha_{\text{inv}} \approx 16°$ and an uncertain $P_{\min}$ (Лупишко и др. 1994). The small inversion angle of Interamnia was interpreted as an indication of a bare rock surface or of some peculiar surface composition which remained unknown at that time (Dollfus and Zellner 1979). More recently, this asteroid was classified as a member of the rather rare F type (Tholen 1989). We assumed that the small inversion angle could be related to some specific surface properties of F-type asteroids and continued polarimetric observations of such objects (Belskaya et al. 2005; Fornasier et al. 2006b; Бельская 2007). As a result, observations of seven asteroids classified by Tholen (1989) as F-type objects have revealed that five of them have anomalous polarimetric properties with shallow NPBs and small inversion angles. The other two asteroids, namely 762 Pulcova and 3200 Phaethon, exhibited usual polarimetric properties typical of dark asteroids. Further analyses have shown that the latter asteroids had likely been misclassified because of lack of good-quality data (Halliday 1988).

Thus, all known cases of small inversion angles pertain to members of the F type, which is one of the most puzzling taxonomic classes of asteroids. Unlike other very dark primitive asteroids, F-type objects are abundant in the inner asteroid belt. A large fraction of them may have been produced in energetic collisional events (the Polana family), although they may also be genetically related to extinguished cometary nuclei (Weissman et al. 2002). We have proposed a possible interpretation of the peculiar polarimetric properties of asteroid 419 Aurelia and other F-type asteroids. It has been shown that high degree of optical homogeneity of the asteroid regolith microstructure on scales of the order of the visible wavelength may be responsible for a relatively shallow NPB and a small inversion angle (Belskaya et al. 2005). One of the plausible origins of such a microstructure is the deposition of carbon on the surfaces of F-type asteroids which may have made them optically more homogeneous compared to other types of low-albedo asteroids. This explanation is corroborated by nearly flat spectra characteristic of the F-type asteroids.

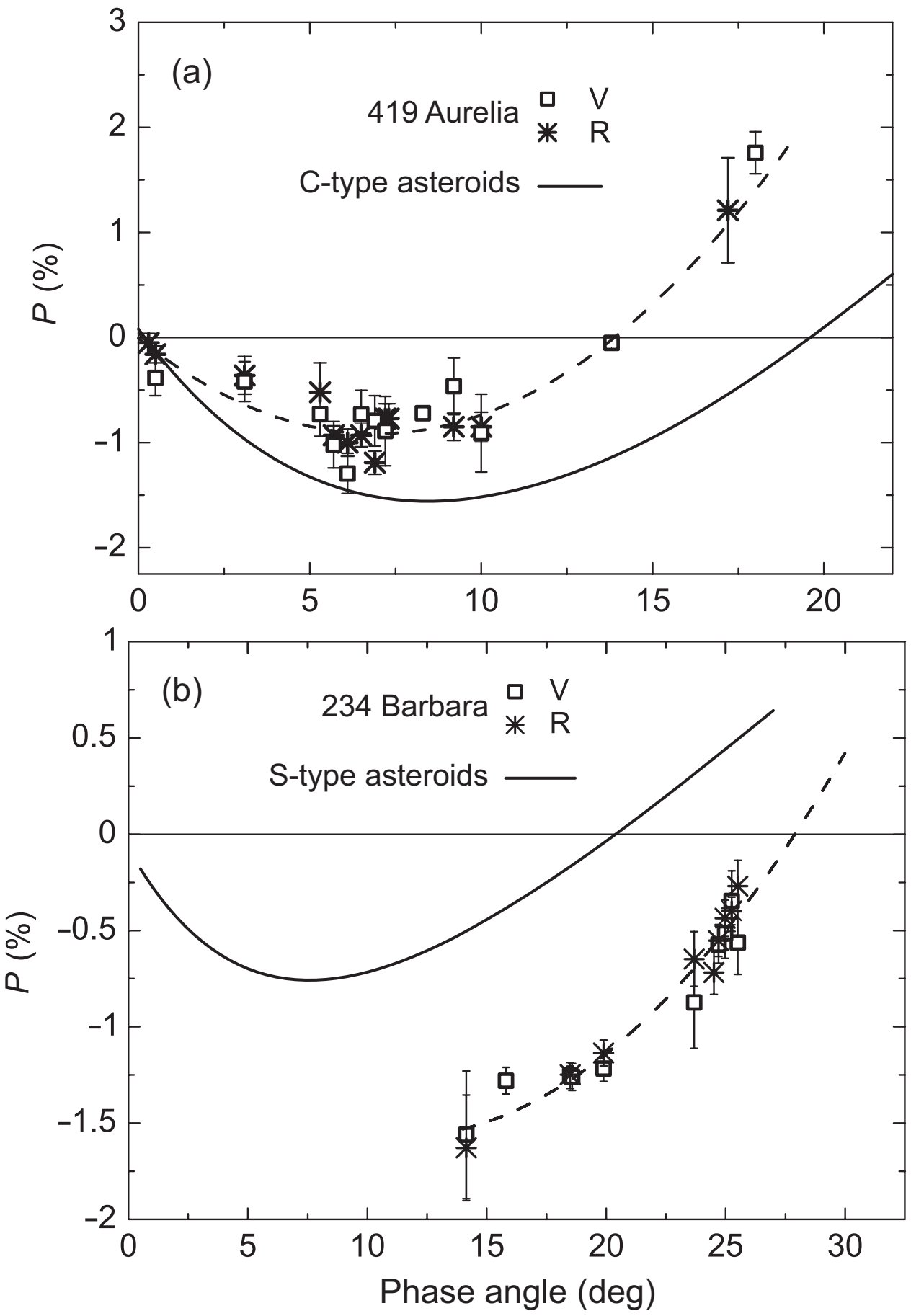

**Fig. 3.18.** Polarization phase dependences of asteroids 419 Aurelia (a) and 234 Barbara (b) according to Belskaya et al. (2005) and Cellino et al. (2006, 2007). For comparison, the average polarization phase curves of the C- and S-type asteroids are also shown.

***Asteroids with large inversion angles.*** The first polarimetric observations of the moderate-albedo asteroid 234 Barbara at phase angles $\alpha \geq 20°$ revealed significant negative polarization of order −1.2% (Cellino et al. 2006). This result was surprising since all other previously observed asteroids had exhibited negligible or positive polarization at such large phase angles close to the typical inversion angle (see Table 3.2). Further observations of 234 Barbara during different oppositions and with different telescopes have confirmed the presence of a wide NPB (Бельская 2007; Cellino et al. 2007). The polarization phase dependence measured for asteroid Barbara in the V and R filters is shown in Fig. 3.18b. It covers a phase an-gle range from 14.1° to 25.5°. Unfortunately, larger phase angles for this asteroid cannot be reached in ground-based observations. The approximation of the measurements by a second-degree polynomial function shown in Fig. 3.18b yields an inversion angle of $\alpha_{\rm inv} = 28° \pm 1°$, while using a linear fit to the data at $\alpha > 18°$ yields an estimate $\alpha_{\rm inv} = 29° \pm 1°$. Although there is some remaining uncertainty in the estimated values of the inversion angle for asteroid Barbara, it undoubtedly represents the largest value ever measured for an asteroid. Previously, large inversion angles were measured for the M-type asteroids 16 Psyche and 21 Lutetia: 23.1° and 24.5°, respectively. The inversion angles of all other asteroids had never exceeded 22°. Besides the large inversion angle, the polarization phase curve of asteroid 234 Barbara is characterized by a deep NPB which is atypical of moderate-albedo surfaces. More recently it has been shown that the spectral dependence of negative polarization for Barbara in the wavelength range 370−830 nm is also different from that typical of moderate-albedo asteroids (Belskaya et al. 2009b).

The anomalous polarimetric properties of asteroid 234 Barbara could be related to its specific surface composition. According to available spectral data, this asteroid belongs to the rather rare Ld subclass of taxonomic type L separated from the S-type asteroids in the asteroid classification by Bus and Binzel (2002). This factor has stimulated intensive polarimetric observations of asteroids of the *L* class and its subclasses. So far, five more asteroids have been found to have wide NPBs with large inversion angles similar to that measured for Barbara (Gil-Hutton et al. 2008; Masiero and Cellino 2009). Certain spectral features of these asteroids are indicative of high concentrations of calcium- and aluminum-rich inclusions associated with spinel (Sunshine et al. 2008). The aluminum-rich inclusions have been found in some types of carbonaceous chondrites and are considered to be the most primitive composition preserved from the early stages of the evolution of the Solar System.

### *3.2.6. Polarimetry of asteroids selected as targets of space missions*

Among approximately 200 asteroids studied by us using the polarimetric method, four are targets of ongoing space missions. Two of them, asteroids 1 Ceres and 4 Vesta, are the targets of the NASA space mission “Dawn” successfully launched on 27 September 2007. The other two, 21 Lutetia and 2867 Steins, are targeted by the Rosetta mission of the European Space Agency launched on 2 March 2004.

Ceres and Vesta are the largest and most massive bodies in the main belt, comprising about 35% of its total mass. These objects are thoroughly different in terms of their composition and evolutionary history and are unique in terms of their internal structure. Our observations of these asteroids during several oppositions have shown that both objects are also polarimetrically unique. Being the largest and most massive bodies of the main belt, they could resist near-catastrophic collisions during the entire history of formation and evolution of the asteroid belt and have remained largely intact. That is why a detailed study of these objects is very important from the point of view of cosmogony of the asteroid belt and may facilitate the understanding of the earliest stages of the Solar System formation.

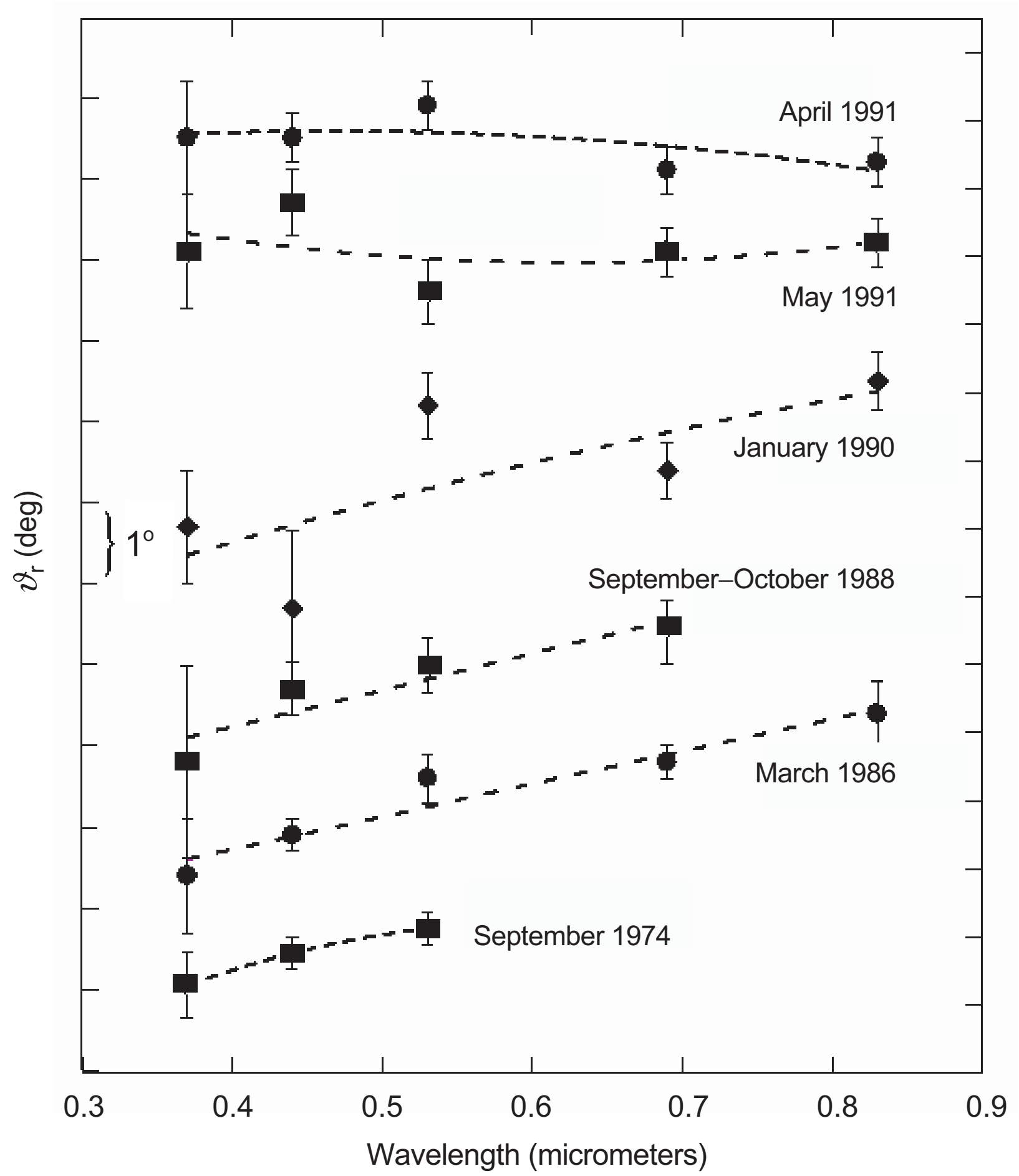

**Fig. 3.19.** Spectral dependence of the angle $\vartheta_r$ for Ceres during different oppositions.

***1 Ceres.*** Our UBVRI-polarimetry of Ceres in 1986 (Бельская и др. 1987b) revealed an absolutely unique feature: a monotonous increase of the angle $\vartheta_r$ with wavelength. The value of the increase over the wavelength interval 370–830 nm is 2°, which exceeds measurement errors by a factor of several. In fact, the same increase of the angle $\vartheta_r$ could be identified in the highly accurate UBV-observations of Ceres in 1974 by Zellner and Gradie (1976).

The observations carried out in 1988 by V. K. Rosenbush (Розенбуш 2006) and in 1990 by A. Cellino with colleagues (Cellino et al. 2005) have confirmed the spectral increase of the angle $\vartheta_r$ for Ceres (Fig. 3.19). However our more recent observations in 1991 with the same 1.25-m CrAO telescope and five-color polarimeter did not shown any increase in the angle $\vartheta_r$ (Лупишко 1998a) (note that the accuracy of these newer observations was somewhat worse than those in 1974 and 1986). Thus, highly accurate additional observations are necessary to study this effect. Lupishko et al. (1991) hypothesized that the rotation of the Ceres' polarization plane with wavelength can be related to a thin gaseous (e.g., water loss?) or gas-dust shell around Ceres. However, it seems more likely that the observed polarization effect is related to the existence of $H_2O$ hoarfrost on the Ceres surface near one of its poles. The position of this hoarfrost on the visible disk and temporal (e.g., seasonal) variations in its extent could explain the differences in the spectral dependence of the polarization position angle observed during the different oppositions.

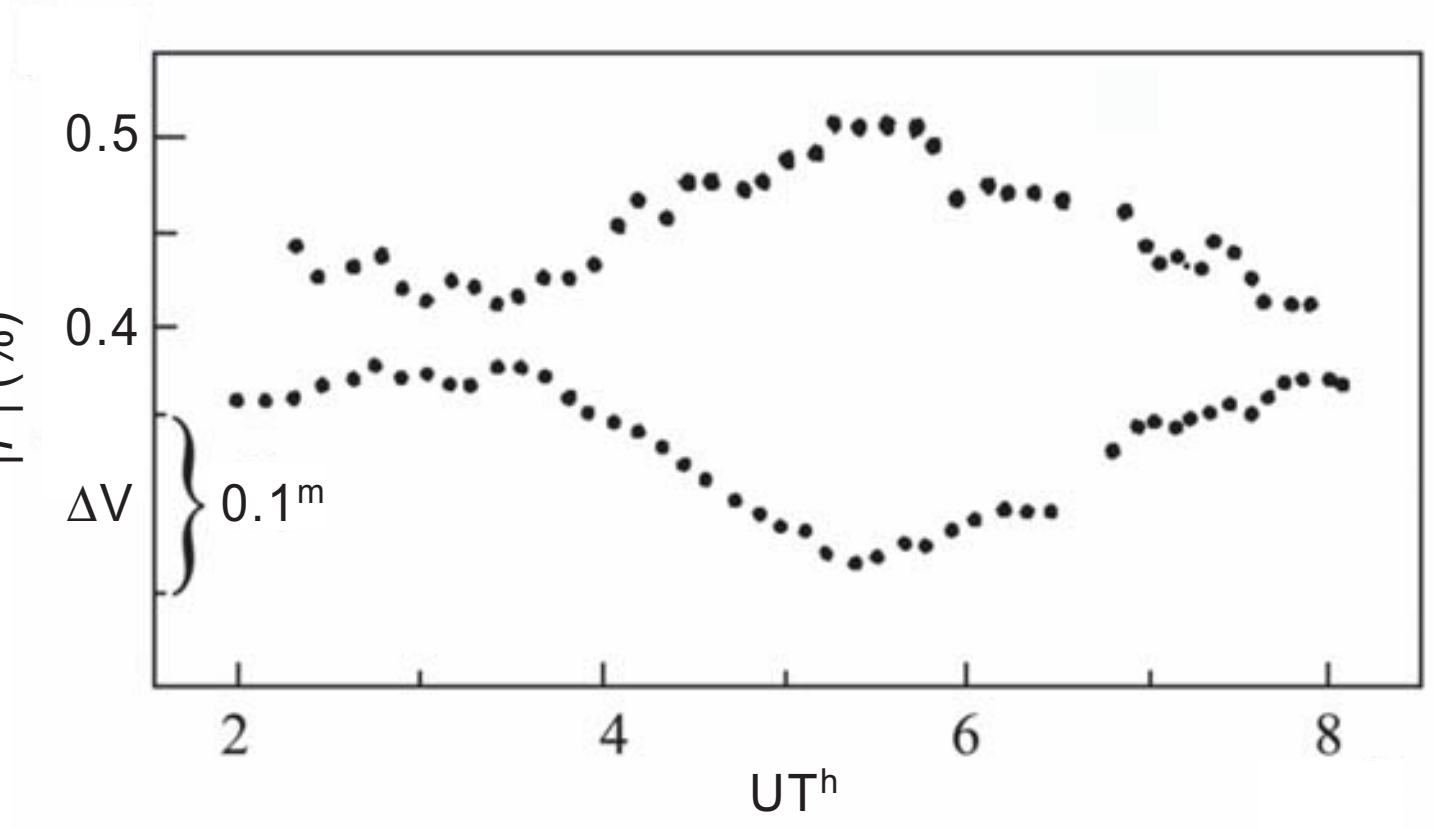

**Fig. 3.20.** Variations of polarization and brightness over the surface of asteroid Vesta in the V filter according to observations in 1986 at the Soviet–Bolivian Observatory.

***4 Vesta.*** A well-known property of Vesta is the significant variability of the degree of linear polarization over its surface caused by albedo variations. This property has been known since the 1970th and was confirmed by our observations in 1986 in the framework of the program Vesta-86 (Лупишко и др. 1988) as well as by our more recent observations. In the context of this program, N. N. Kiselev and A. V. Morozhenko used the 0.6-m reflector of the Soviet–Bolivian Observatory to perform simultaneous photometric and polarimetric observations of Vesta which, for the first time, revealed an inverse correlation of the degree of negative polarization and the asteroid brightness (Fig. 3.20). It is recognized now that Vesta's lightcurve is controlled primarily by the spatial distribution of the surface albedo. Therefore, the observed inverse polarization–brightness correlation for Vesta extends the validity of the well-known Umov law to the case of negative polarization.

As a result of our program of cooperative observations of Vesta in 1986 at several observatories, highly accurate measurements of the NPB ($\alpha$ = 5.4°–13.3°) were performed which, along with the previously available data on the PPB (Zellner and Gradie 1976), have allowed us to derive the detailed phase dependence of polarization with parameters $P_{\min} = -0.61\% \pm 0.03\%$ and $h = 0.065 \pm 0.004$ (Лупишко и др. 1988). These parameters were used to estimate the albedo of Vesta at $p_V = 0.254$. The UBVRI-observations of Vesta in 1990 and 1996 with the 1.25-m

CrAO reflector confirmed the above inverse correlation (Fig. 3.20) and revealed a new polarization effect for Vesta, viz., variations of the position angle of the polarization plane over the asteroid surface. This effect will be discussed in detail in Section 3.2.9.

In December 2005−January 2006, new observations of Vesta near opposition ($\alpha < 5°$; Розенбуш 2006) yielded a more detailed polarization phase dependence (see Section 3.2.2).

***21 Lutetia.*** The first polarimetric observations of this asteroid had revealed a rather large inversion angle, which was interpreted as an indication of a metal composition of its surface, similar to that of iron meteorites (Dollfus et al. 1979). We performed extensive polarimetric observations of Lutetia in 1985−2006 (Бельская и др. 1987a; Fornasier et al. 2006a; Belskaya et al. 2009b) at phase angles which had not been captured by the previous observations. This has allowed us to determine more accurately the main polarimetric parameters of this asteroid. We interpreted these data based on the results of laboratory measurements of meteorites (see Section 3.2.9). The best polarimetric analogues of the Lutetia's surface are carbonaceous chondrites of the CV3 и CO3 types (Belskaya and Lagerkvist 1996; Birlan et al. 2004), which are characterized by a low carbon content and thus a relatively larger albedo when compared to other carbonaceous chondrites. An additional evidence in favor of the primitive surface composition of Lutetia has been provided by spectral observations in the 1200 to 3500 nm region. They have revealed the presence of an absorption feature at 3000 nm diagnostic of hydrated minerals (Rivkin et al. 1995). It is inconsistent with the surface composition similar to iron meteorites and indicates a relatively primitive material like enstatite chondrites or salt-rich carbonaceous chondrites.

There is a rather large discrepancy in the existing values of the surface albedo for Lutetia. Its IRAS albedo has been found to be rather high, $p_V = 0.22$, which completely rules out a composition similar to carbonaceous chondrites. However, the polarimetric albedo has been found to be smaller, 0.10−0.11 (Lupishko and Mohamed 1996; Fornasier et al. 2006b), which is consistent with a primitive surface composition. More recent radiometric observations of Lutetia have yielded an albedo of $p_V = 0.13$ (Carvano et al. 2008), close to the polarimetric estimate. There is no doubt that new data on Lutetia's surface properties expected as an outcome of the space mission Rosetta in 2010 will serve as an important test of the polarimetric prediction.

***2867 Steins.*** Dedicated studies of the physical properties of this asteroid were initiated only in 2004, when it was selected as another target of the space mission Rosetta. The first spectral observations in the visible and near-IR spectral regions revealed similarities of Steins to E-type asteroids (Barucci et al. 2005). However, for a definitive taxonomic classification and a proper understanding of its surface properties and size determination, a knowledge of the surface albedo was needed. First polarimetric observations of Steins were carried out with the 8.2-m Very Large Telescope (VLT) of the ESO (Chile) in the service mode during six nights in June to August 2005 (Fornasier et al. 2006a). The data were obtained in the phase angle range from 10.3° to 28.3° in the V and R spectral bands. They permit the de-termination of the inversion angle and the slope of the polarization phase curve at the inversion angle: $\alpha_{\text{inv}}$ =17.3° ± 1.5° and $h$ = 0.037 ± 0.003 %/deg in the V band; $\alpha_{\text{inv}}$ = 18.4° ± 1.0° and $h$ = 0.032 ± 0.003 %/deg in the R band. These values are consistent with those typical of E-type asteroids.

The albedo of Steins was determined using the empirical relation "polarimetric slope – albedo" (Section 3.2.8). The geometric albedo in the V band was estimated to be $p_V$ = 0.45 ± 0.10 (Fornasier et al. 2006a). The uncertainty in the albedo was evaluated using both the uncertainty in the slope determination and that in the constants of the "slope – albedo" relation. Recent preliminary analyses of data from the Rosetta mission collected in 2008 have shown that the albedo of this asteroid should be close to 0.4, which confirms the polarimetric estimate.

### *3.2.7. Opposition effects of asteroids*

The theoretical prediction of the POE by Mishchenko (1993b) stimulated systematic observations of high-albedo asteroids intended to verify the presence of a narrow secondary minimum of polarization near the exact backscattering direction accompanying an equally narrow BOE peak. Until then polarimetric observations of asteroids at the smallest phase angles ($\lesssim$2°) had been very rare, thereby making the detailed behavior of polarization at opposition poorly understood. Note also that the magnitude of polarization at such angles is often weak (several tenths of a percent), thereby necessitating very high measurement accuracy and making most of the previously available data poorly suitable for the study of opposition effects.

Below we summarize the results of the most recent polarimetric observations of asteroids 64 Angelina, 44 Nysa, 214 Aschera, and 4 Vesta near opposition. These asteroids were specifically selected for the search of the POE since they are high-albedo objects for which a spike-like BOE had been found. Furthermore, there had been different opinions as to the shape of the polarization phase dependence at small phase angles, ranging from a sharply asymmetric NPB to a secondary minimum of negative polarization distinctly separate from the main minimum. Therefore, our observations were also intended to investigate the shape of the POE and determine its main parameters.

***Asteroid 64 Angelina.*** 64 Angelina belongs to a rare taxonomic class of E-type asteroids characterized by a high surface albedo (Tholen 1989). The geometric albedo and diameter of this asteroid according to the radiometric data are 0.43 and 60 km, respectively (Tedesco and Veeder 1992). The albedo obtained from polarimetric data is somewhat larger, ~0.48 (Lupishko and Mohamed 1996). According to Harris et al. (1989), the photometric phase curve for 64 Angelina as well as that for another high-albedo asteroid 44 Nysa exhibit a strong and unusually narrow surge of brightness, of about $0.25^{\text{m}}$, at small phase angles. Harris et al. called this phenomenon the opposition spike. The photometric observations were carried out only in the V filter.

Zellner and Gradie (1976) were the first to perform polarimetric observations of 64 Angelina, using the B and G(V) spectral filters. However, the range of phase angles covered by their observations was 3.7°–24°, i.e., relatively far from opposition.

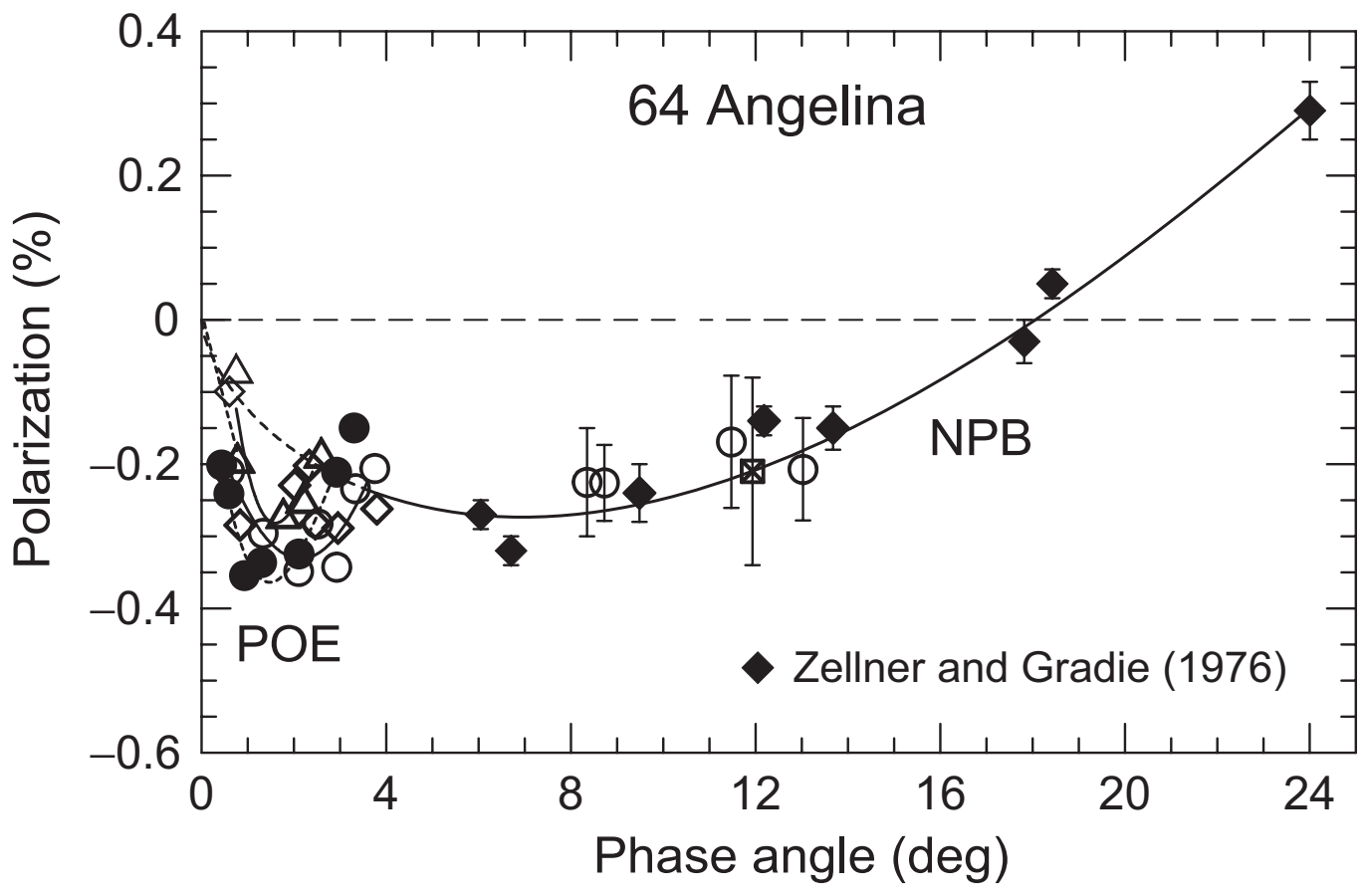

**Fig. 3.21.** Phase-angle dependence of polarization for asteroid 64 Angelina in the R filter obtained during three apparitions: Δ – 1995; ● and ○ – 1999 (observations before and after opposition, respectively), ◇ – 2000/01. Filled diamonds depict data by Zellner and Gradie (1976). The solid curve represents the best fit to the B- and V-filter data.

The first near-opposition polarimetric observations of 64 Angelina, at phase angles from 0.43° to 13.02°, were carried out with the 1.25-m CrAO reflecting telescope during the 1995, 1999, and 2000/01 apparitions. During the 1999 apparition, photometric CCD observations of 64 Angelina in the B and R filters were carried out with the 2-m telescope of the Peak Terskol Observatory and in the U, V, and R filters with the 0.7-m telescope-reflector of the KhNU.

These observations have shown that in all the spectral channels, there is a secondary minimum of negative polarization at a phase angle of $\alpha_{\min} \approx 1.8°$, which has a ~0.4% depth in the R channel and, along with the regular NPB, forms a bimodal phase curve of polarization. As an example, Fig. 3.21 shows the phase dependence of polarization for 64 Angelina observed in the R filter; these measurements have the highest accuracy (~0.03%–0.07%).

***Asteroid 44 Nysa.*** This is the largest member of the class of high-albedo E-type asteroids. Tedesco et al. (2002) estimated its albedo (0.55) and diameter (70.6 km) from IRAS data (see Section 3.2.8), whereas the albedo estimated by Lupishko and Mohamed (1996) from polarimetric data is somewhat smaller (0.41). The photometric phase curve for this asteroid measured by Harris et al. (1989) in the V band, as well as that for asteroid 64 Angelina, showed the opposition spike effect within the range of phase angles from 2° to 0°.

Zellner and Gradie (1976) were the first to carry out polarimetry of 44 Nysa in the B and G (V) spectral bands within the range of phase angles 3.7°–21.8°. During the 1987 apparition, Kiselev and Lupishko carried out polarimetric observations of 44 Nysa in the U, B, V, R, and I filters at phase angles 7.6°–11°. Thus, the existence of the NPB for 44 Nysa had been a well-known fact. However, no polarimetric observations had been performed at phase angles smaller than 3.7°.

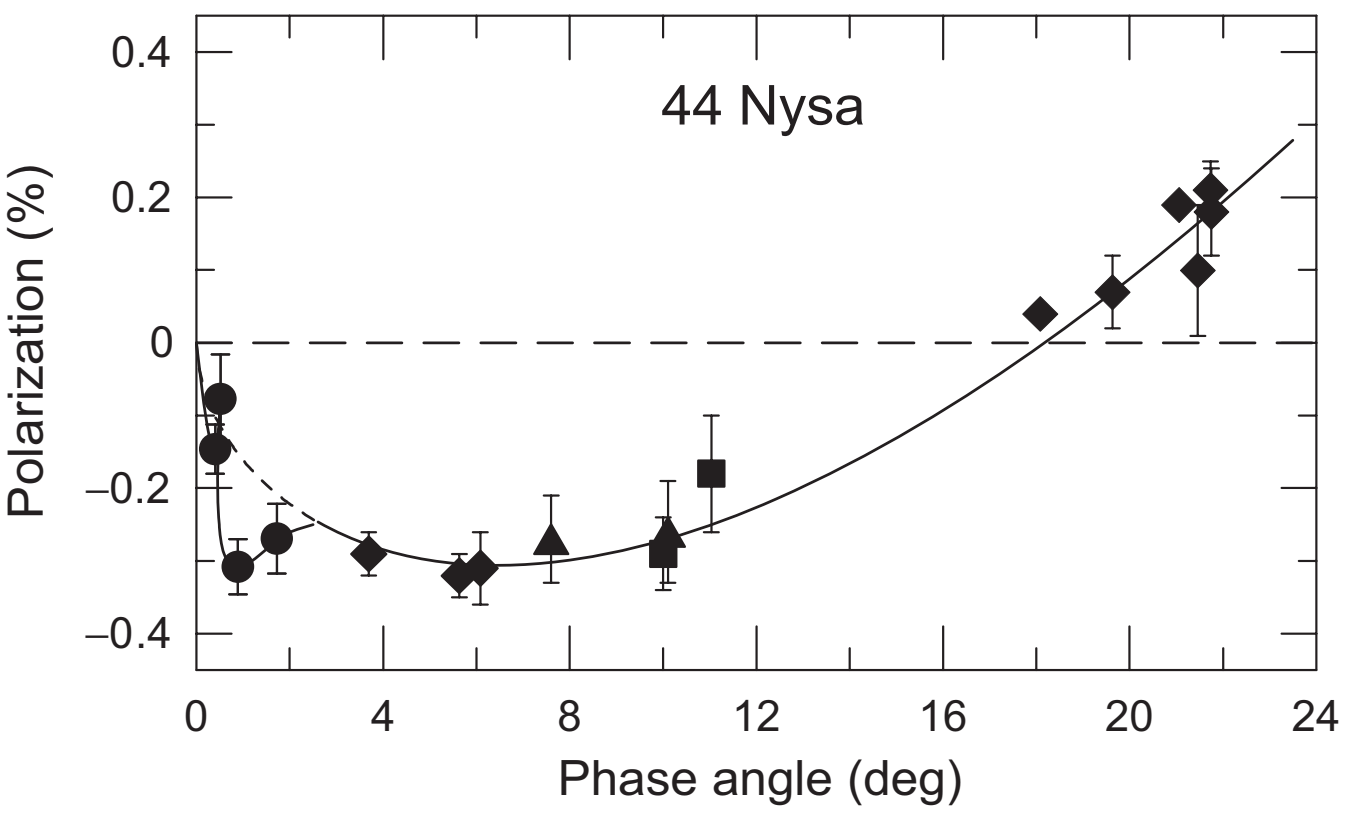

**Fig. 3.22.** Phase-angle dependence of polarization for asteroid 44 Nysa consists of the common NPB and a second minimum near the backscattering direction. The data obtained during the 2005 apparition are denoted by filled circles. The diamonds depict the data by Zellner and Gradie (1976). The triangles and squares show the data obtained by Kiselev and Lupishko in 1987 (http://www.psi.edu/pds/resource/apd.html). The curve shows the approximation of the NPB data with a trigonometric polynomial proposed by Penttilä et al. (2005).

V-filter polarimetric observations of 44 Nysa near opposition were carried out by these authors on 10–14 August 2005, when the minimal phase angle reached 0.4°. The one-channel photoelectric photometer-polarimeter mounted in the Cassegrain focus of the 2.6-m CrAO telescope was used. The most important result of these observations was the detection of a bimodal phase-angle dependence of polarization, which consists of the POE in the form of a narrow minimum of negative polarization with parameters $\alpha_{\mathrm{min,POE}} \sim 0.8°$ and $P_{\mathrm{min,POE}} = -0.31\% \pm 0.04\%$ superposed on a much broader NPB with $P_{\mathrm{min}} \approx -0.30\%$ at $\alpha_{\mathrm{min}} \sim 5.8°$ (Fig. 3.22).

Comparison of the photometric phase curves (after correction for the linear trend) and the PDPs for 44 Nysa and 64 Angelina (Fig. 3.23) reveals certain differences potentially attributable to differences in the morphology and/or composition of the reflecting surfaces. In particular, Nysa exhibits a significantly narrower POE than Angelina.

The depth of the POE for Angelina in the V filter varied significantly with apparition: $P_{\mathrm{min,POE}} \approx -0.3\%$ in 1995 and $-0.45\%$ in 1999 (with no account for low-accuracy pre-perihelion observations). Depending on apparition, the difference between the measured polarization and extrapolated NPB value at $\alpha = \alpha_{\mathrm{min,POE}}$ is ~0.12%–0.25% for Angelina and ~0.12% for Nysa. One can also notice differences between the respective NPBs: $P_{\mathrm{min}} = -0.30\% \pm 0.02\%$ and $\alpha_{\mathrm{min}} = 5.8° \pm 0.1°$ for Nysa and $P_{\mathrm{min}} = -0.26\% \pm 0.02\%$ and $\alpha_{\mathrm{min}} = 6.9° \pm 0.1°$ for Angelina. It is important, however, that for either asteroid the characteristic angular widths of the respective BOE and POE are essentially the same, which is consistent with a common WL origin of these opposition phenomena.

***Asteroid 214 Aschera.*** The E-type asteroid 214 Aschera has an estimated diameter of 23 km and a geometric albedo of 0.52 (Tedesco et al. 1992). The first po-larimetric observations of this rather faint asteroid were performed during the 1996–98 apparition (Belskaya et al. 2003) using the 1.25-m CrAO telescope equipped with the five-channel UBVRI photopolarimeter. The measurements were taken at phase angles 0.72°, 5.34°, and 11.49°. Significant measurement errors for Aschera comparable with its own polarization in combination with poor phase-angle sampling do not allow one to determine the NPB minimum and identify a secondary polarization minimum in individual spectral channels or to characterize the spectral variability of polarization. Therefore, in order to determine the phase-angle dependence of polarization for Aschera more accurately, we averaged data for this asteroid in the B, V, R, and I filters using the respective errors $\sigma_P$ as weights. As a result, we have found that Aschera exhibits a rather significant degree of polarization, $P = -0.38\% \pm 0.09\%$, at the phase angle $\alpha = 0.72°$.

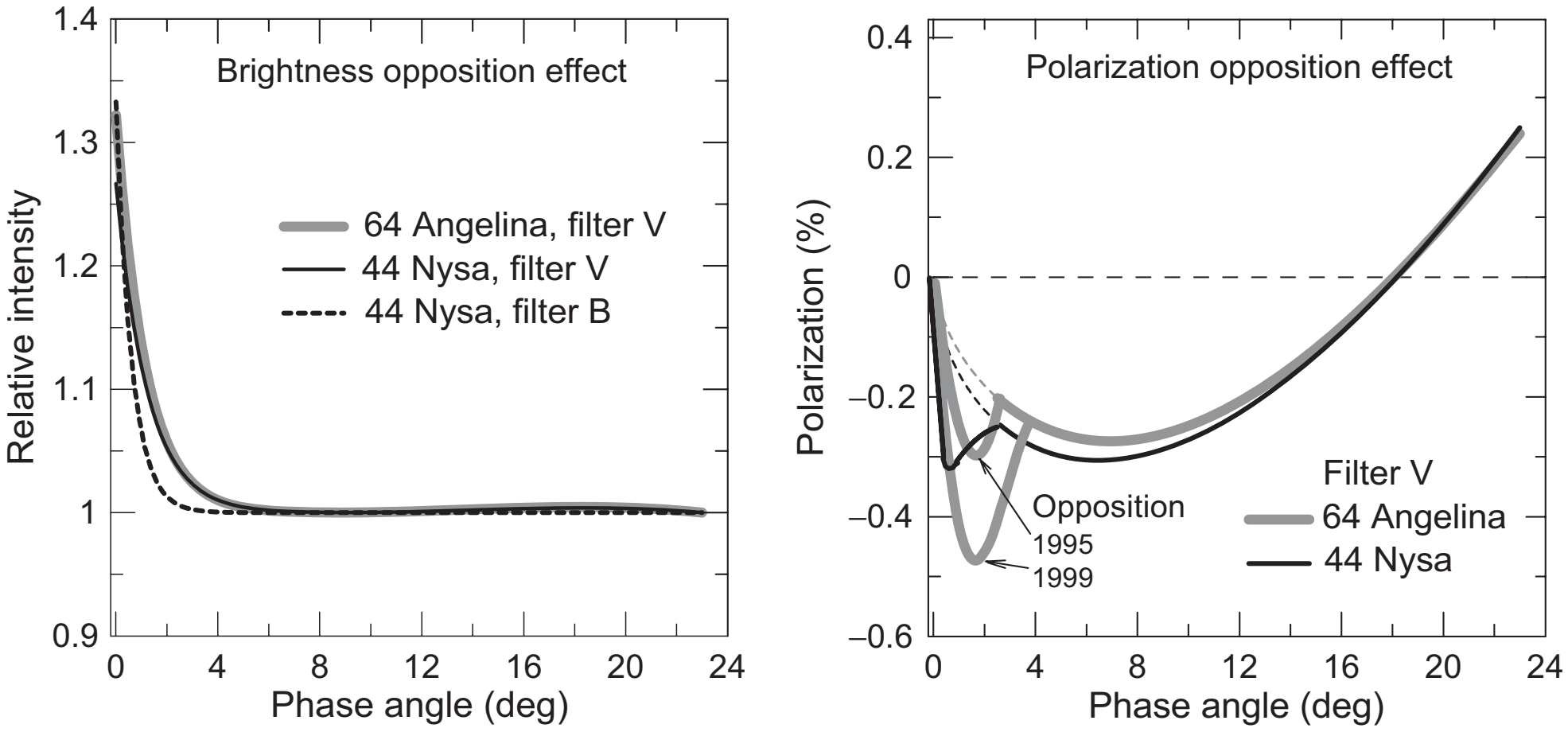

**Fig. 3.23.** Comparison of BOEs and POEs for high-albedo asteroids 64 Angelina and 44 Nysa.

Figure 3.24 summarizes all polarization measurements for E-type asteroids. The composite phase-angle dependence of polarization for 44 Nysa, 64 Angelina, and 214 Aschera rigorously proves the presence of the POE in the form of a sharp minimum of negative polarization centered at a phase angle of ~1° with an amplitude of ~0.35%. Of course, additional high-accuracy polarimetric observations of E-type asteroids are required to confirm this conclusions.

***Asteroid 4 Vesta.*** According to the high-precision system of ephemeredes HORIZONS (http://ssd.jpl.nasa.gov), the diameter of the V-type asteroid 4 Vesta is 530 km, while its albedo according to the IRAS data (Tedesko et al. 2002) is $p_V = 0.423$, i.e., is quite comparable to those of E-type asteroids. During the 2005–06 apparition, Vesta reached a minimal phase angle of 0.12°. Since all previous polarimetric observations of Vesta had been performed at phase angles $\alpha > 5.13°$ and all photometric observations at phase angles $\alpha > 1°$, the obvious objective of our observations was to measure Vesta's brightness and polarization at very small phase angles and analyze the potential opposition effects for this high-albedo V-type asteroid. Photopolarimetric observations of Vesta in the U, B, V, R, and I fil-ters were performed in December 2005–January 2006 using the 1.25-m CrAO telescope over the range of phase angles from 0.6° to 5.0°.

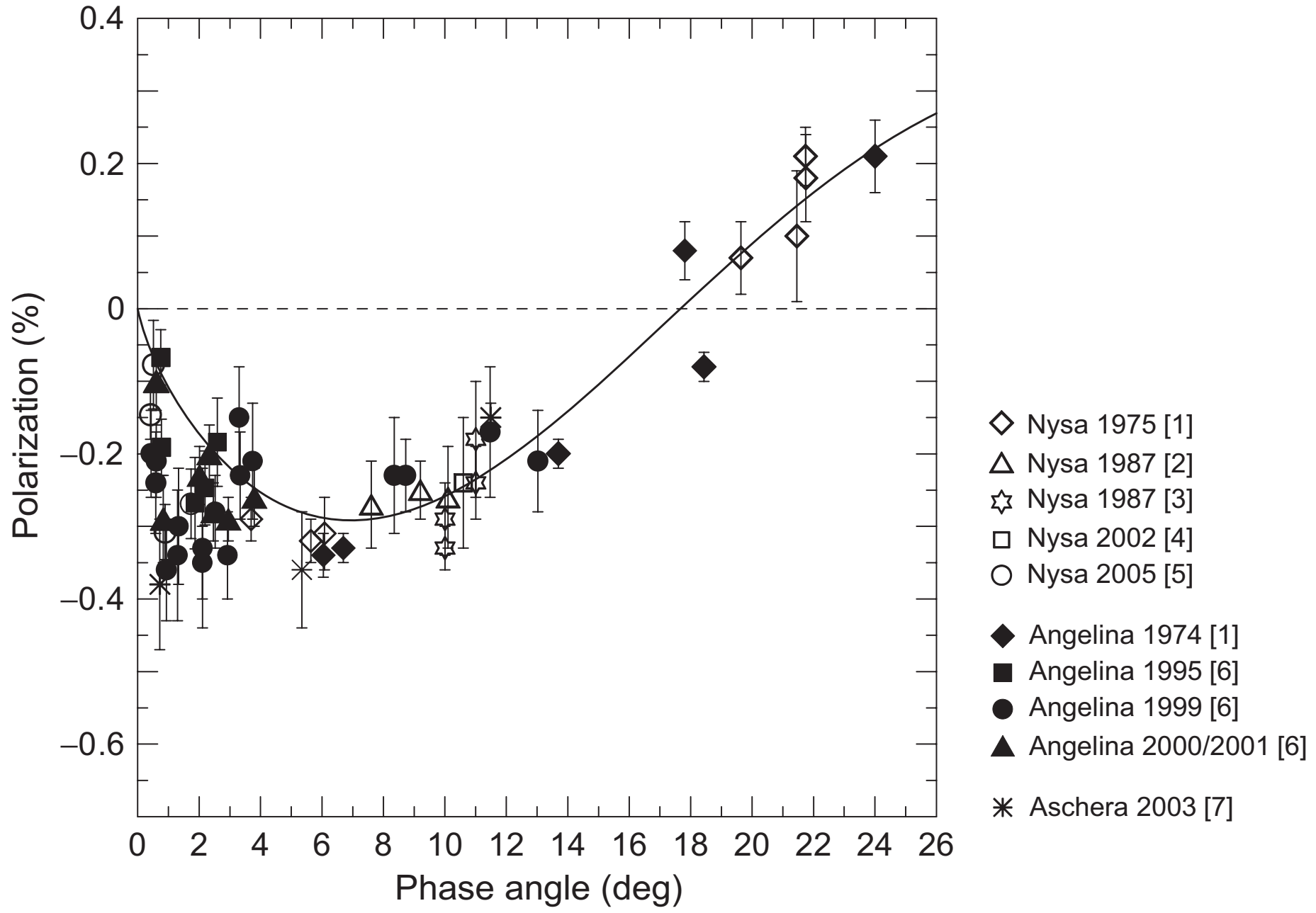

**Fig. 3.24.** Composite phase-angle dependence of polarization for E-type asteroids based on all available data: [1] Zellner and Gradie (1976); [2,3] Data by N. N. Kiselev and D. F. Lupishko, respectively, from http://www.psi.edu/pds/resource/apd.html; [4] A. Cellino, personal communication; [5] Rosenbush et al. (2009a); [6] Rosenbush et al. (2005a,b); [7] Belskaya et al. (2003).

One can see from Fig. 3.13 that within the measurement errors, the V-filter polarimetry of Vesta shows no secondary minimum at opposition. This result is significant because 64 Angelina is only slightly brighter (with an IRAS albedo of $p_V = 0.48$) and yet exhibits a pronounced POE. Among the reasons for this distinction one might think of differences in the mineralogical composition. However, the absence of a second polarization minimum may actually imply that Vesta is not a high-albedo asteroid. We have used the most recent polarimetric observations to obtain improved estimates of the parameters $h$ и $P_{min}$ and derived an average value of $p_V = 0.24$, which is smaller than the IRAS-based albedo by a factor of almost two. The low albedo of Vesta is indirectly confirmed by photometric data (Розенбуш 2006). Thus, the absence of the POE for Vesta may not, after all, be surprising.

### *3.2.8. Determination of albedos and sizes of asteroids on the basis of polarimetric and photometric observations*

The polarimetric method for the determination of asteroid albedos has already been mentioned several times in previous sections. It is based on empirical linear relations between the polarimetric slope $h$ or the depth of the NPB $P_{\min}$ on one hand and the asteroid geometric albedo $p_V$ on the other hand:

$$\log p_V = A \log h + B, \qquad \log p_V = C \log |P_{\min}| + D. \tag{3.8}$$

The values of the constants $A$ and $B$ (–0.93 and −1.78, respectively) were determined as early as in the mid-1970s using laboratory measurements of meteorite samples (Zellner et al. 1977). The constants $C$ and $D$ (− 1.31 and − 0.9, respectively) were determined from an approximation of the asteroid albedo dependence on $P_{\min}$ (Zellner and Gradie 1976). Unfortunately, those albedos were obtained from the "slope $h$ – albedo" relation, but now it is recognized that the parameters $h$ and $P_{\min}$ correlate linearly with each other.

In the 1990s other estimates of asteroid albedos became available (viz., so-called IRAS-based albedos, occultation albedos, and ground-based radiometric albedos), which allowed us to test the correspondence of the polarimetric albedo scale calibrated using meteorite samples to the real surfaces of asteroids. Furthermore, since 1984 we have carried out polarimetric observations of many asteroids and have determined reliably the parameters $P_{\min}$ and $h$ for ~40 additional asteroids. All these data have allowed us to attempt a new calibration of the relations (3.8) using a quite different method in order to: (i) test the accuracy of the old calibration, and (ii) try to determine the constants $A$, $B$, $C$, and $D$ more accurately.

The gist of the new calibration procedure is based on the assumption that the aggregation of the above-mentioned independent datasets of asteroid albedos provides a more reliable information than each of these datasets individually. Under this assumption, the three available sets of asteroid albedos (i.e., the IRAS-based albedos, occultation albedos, and ground-based radiometric ones) have been correlated with the parameters $P_{\min}$ and $h$. The corresponding relations were approximated by simple linear functions (primarily because the coefficients of the quadratic terms were found to be small and statistically insignificant) and yielded the coefficient values listed in Table. 3.4 ($n$ is the number of asteroids for which data are available). The resulting accuracy of the constants $A$, $B$, $C$, and $D$ is different for different data sets. In fact, the accuracy is the best when one uses the albedo values retrieved from observations of star occultations, despite the fact that such albedo estimates are available for only seven asteroids.

In order to average the results derived from the three albedo datasets, one can formally use weights proportional to $\sigma^{-2}$. Unfortunately, the corresponding $\sigma$ values for the three sets of constants $A$, $B$, $C$, and $D$ account only for random errors of a dataset but not for its systematic errors. However, systematic differences between the ground-based radiometric and occultation datasets are about twice as large as those between the IRAS and occultation datasets. Therefore, we have averaged the constants $A$, $B$, $C$, and $D$ using the following weights: 2.5 for the occultation dataset, 2.0 for the IRAS dataset, and 1.0 for the ground-based radiometric dataset. Although there is some arbitrariness in these weight assignments, they are believed to represent the real situation better than the formal weights proportional to $\sigma^{-2}$.

The new empirical relations thus obtained (Lupishko and Mohamed 1996) have already been widely used to determine the albedo of asteroids in the standard V band from polarimetric observations and are as follows:

**Table 3.4.** Values of linear-approximation constants

| Albedo data | $h$ – albedo | | | $P_{min}$ – albedo | | |
|---|---|---|---|---|---|---|
| | $n$ | $A$ | $B$ | $n$ | $C$ | $D$ |
| IRAS | 44 | −1.084 ± 0.093 | −1.742 ± 0.073 | 106 | −1.332 ± 0.133 | −0.891 ± 0.027 |
| Radiometry | 51 | −1.175 ± 0.086 | −1.988 ± 0.079 | 83 | −1.378 ± 0.143 | −1.019 ± 0.029 |
| Occultation | 7 | −0.826 ± 0.069 | −1.619 ± 0.053 | 7 | −1.074 ± 0.090 | −0.904 ± 0.016 |
| Mean-weighted | – | −0.983 ± 0.082 | −1.731 ± 0.066 | – | −1.223 ± 0.118 | −0.920 ± 0.023 |
| Used peviously | – | −0.93 | −1.78 | 52 | −1.31 | −0.9 ± 0.1 |

$$\log p_V = -0.98 \log h - 1.73, \qquad \log p_V = -1.22 \log |P_{min}| - 0.92. \qquad (3.9)$$

Using these relations and the APD (see Section 3.2.11), we have compiled a more complete and accurate set of polarimetric albedos for 127 asteroids (Lupishko and Mohamed 1996). Both relations (3.9) have been applied, and the results have been averaged with the weights 2.0 and 1.0, respectively, based on the better accuracy of the albedo determination from the polarimetric slope (Колоколова и Яновицкий 1988). Having the asteroid polarimetric albedo and its absolute magnitude derived from photometric observations, one can estimate the asteroid diameter. This has been done quite frequently in studies of selected asteroids (see, e.g., Киселев и др. 1994; Лупишко и др. 1994; Kiselev et al. 2002b).

Comparisons of the new polarimetric albedos (Lupishko and Mohamed 1996) with the old ones (Morrison and Zellner 1979) have shown that, in general, they do not differ by much. This means that the empirical relations "$h$ – albedo" and "$P_{min}$ – albedo" are sufficiently robust and apply equally well to the laboratory analogues and to the surfaces of real asteroids. However, as demonstrated by Лупишко (1998a), the new constants in the relations (3.9) provide much closer polarimetric albedo values estimated separately from $h$ and $P_{min}$.

It was in 1983 that the Infrared Astronomical Satellite (IRAS) carried out radiometric observations of about 2000 asteroids. By using those data as well as ground-based measurements of the respective magnitudes in the standard V band, a dataset of asteroid geometric albedos and diameters was created (i.e., the IRAS-based dataset mentioned above). The latest version of this database was published by Tedesco and Veeder (1992) and Tedesco et al. (2002). Comparisons of the IRAS-derived albedos with other albedo estimates for the same asteroids (i.e., polarimetric, ground-based radiometric, and occultation albedos) have revealed the following (Лупишко 1998b):

- systematic errors in the IRAS-based geometric albedos are proportional to the albedo values themselves; for example, for $p_{IRAS} = 0.3$, $\Delta p_V/p_V = 0.3$ and 0.5 for asteroids of the S and M types, respectively;
- the polarimetric albedo estimates are intermediate between those based on the IRAS and ground-based radiometric data and show the best agreement with the most accurate occultation albedos.

Based on this result, polarimetric albedos of 127 asteroids (Lupishko and Mohamed 1996) have been used to quantify systematic errors in the IRAS-based albedos and thereby derive more accurate values of the asteroid albedos and diameters compared to the original IRAS dataset (Лупишко 1998b).

### *3.2.9. Structure of asteroid surfaces according to polarimetric data*

Polarimetry as well as radiometry had provided convincing indications of the presence of regolith on asteroid surfaces long before the space missions. As we have demonstrated in Section 3.2.4 using asteroids 1685 Toro and 33342 WT24 as examples, the polarimetric method can often be used to estimate the average size of particles forming a planetary surface (Киселев и др. 1990). Furthermore, polarimetric data can also be used to get an improved understanding of the surface

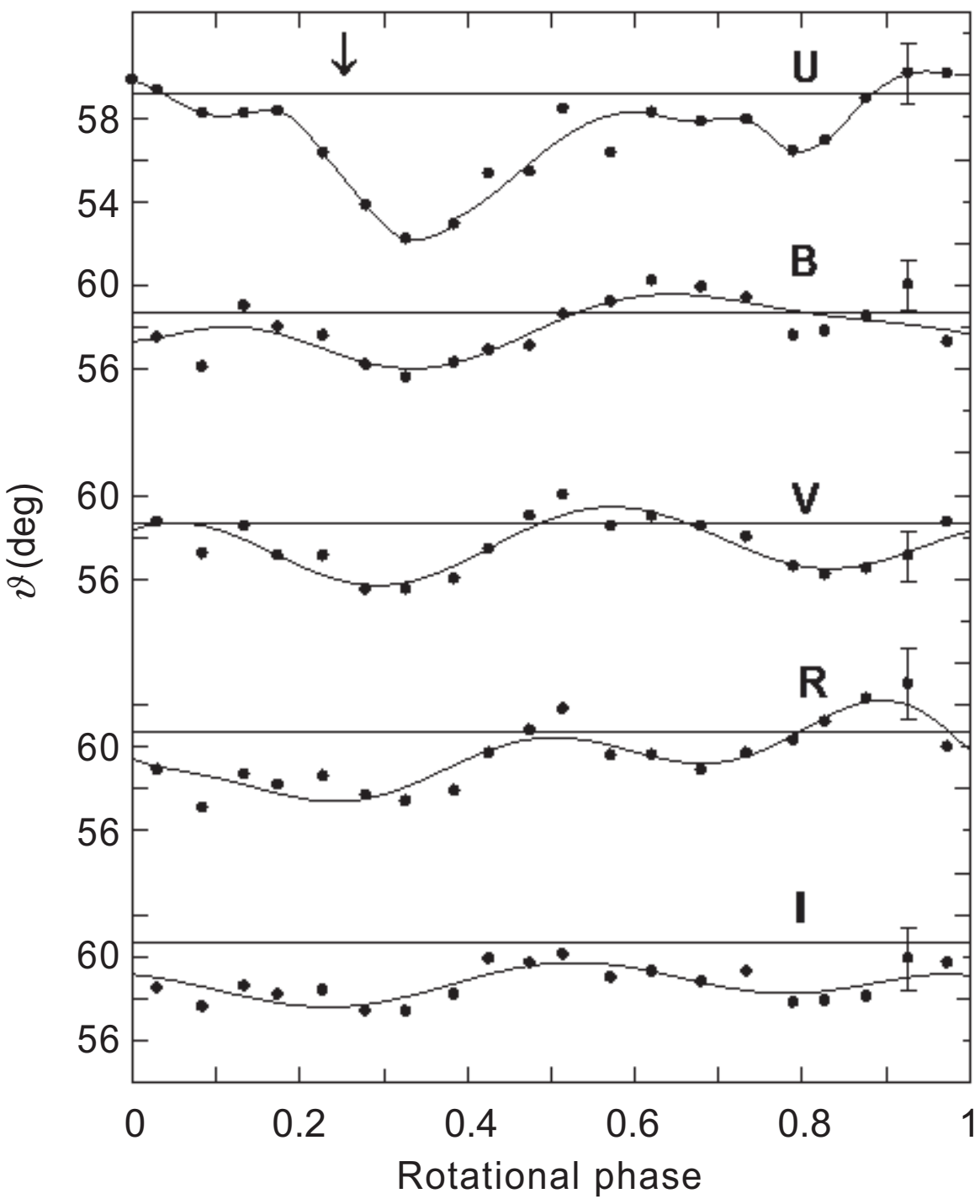

**Fig. 3.25.** Variations of the polarization position angle with Vesta's rotational phase according to observations on 6 December 1990. The arrow indicates the rotation phase of the center of Vesta's largest crater.

macrostructure. We shall demonstrate this by using the main-belt asteroid 4 Vesta as an example.

The UBVRI-polarimetry of Vesta for its entire rotation period (5.342 hrs) was carried out in December 1990 with the 1.25-m CrAO reflector (Лупишко и др. 1999). The corresponding variations of the parameters of the NPB were measured with good accuracy in all five spectral bands. In the V filter, the relative variations of polarization were found to be as large as $\Delta P/P = 0.24$. These variations were larger than those measured previously and, in fact, the largest possible for Vesta, since they were measured at its equatorial aspect ($\psi = 89°$). Furthermore, we discovered an absolutely unique effect, viz., variations of the polarization position angle $\vartheta$ with asteroid rotation phase. These variations are maximal in the U filter ($\Delta\vartheta = 8°$) and minimal in the I filter (2.5°), as is readily seen in Fig. 3.25. The magnitude of the spectral and rotation-phase variability of the polarization position angle exceeds the measurement errors (shown by an error bar for each filter) significantly, thereby providing strong evidence in favor of the actual existence of these variations. Further observations of Vesta in April 1996 at virtually the same equatorial aspect ($\psi = 87°$) confirmed this phenomenon.

More recent observations of Vesta were carried out in 2005–06 in the CrAO at the time when the equatorial aspect of Vesta was close to 128° (Розенбуш 2006). They confirmed the results of the previous studies and led to the conclusion that the variations of the degree of polarization and those of the polarization position angle correlate with each other and are distinctly antiphased. Furthermore, the polarization position angle regularly deviates from 90°, which means that there are some distinctly oriented surface structures causing rotations of the polarization plane as a consequence of peculiar scattering geometries. The minimum of the degree of polarization corresponds to the longitude of the central peak of a gigantic 460-km-diameter crater detected in images taken with the HST (Thomas et al. 1997).

How can the position-angle variations over the surface of Vesta be explained? Since the albedo of the crater is significantly lower than that of the surrounding surface, one can conclude that the variations of the degree of polarization with asteroid rotational phase are related to the albedo heterogeneity of the asteroid surface. Topographic maps of Vesta show that in its Southern Hemisphere, the drops in the relief height reach 24 km, perhaps as a consequence of the near-catastrophic collision that led to the formation of the crater with the size comparable to the diameter of the asteroid (525 km). This collision could result in large-scale oriented surface features such as grooves and slopes, thereby causing a significant global asymmetry of the relief and the resulting deflections of the polarization plane from the scattering plane.

### *3.2.10. Asteroid composition according to polarimetric data*

The most general characteristic of the surface of an asteroid is its composition type or, in other words, its taxonomic class. The basic types of asteroids are the following:

- C type (and its varieties P, D, F, B, and G) — low-albedo asteroids analogues to carbon chondrite meteorites;
- S and Q types — moderate-albedo silicate objects;
- M type — moderate-albedo “metal” asteroids; and
- E type — high-albedo asteroids with compositions similar to that of enstatite chondrite meteorites.

Following the pioneering work by Bowell et al. (1978), Zellner (1979) and others, the results of asteroid polarimetry have been widely used for the classification of asteroids into types.

In 1984–87 we carried out a program of polarimetric observations of so-called CMEU asteroids. Previously, these objects either had no unequivocal classification and could pertain to the C, M, or E type or were considered unclassifiable (U). The common trait of these asteroids is that their spectra have no appreciable absorption features in the visible part of the spectrum and are rather flat. On the other hand, these objects can have vastly different albedos. For the majority of CMEU asteroids their color indices had been known, and so the measurement of the albedo or $P_{min}$ for a CMEU object could be sufficient for a definitive classification into a specific class.

The observations were carried out on the 1-m reflector of the Sanglok Observatory with the standard V filter. The values of $P_{min}$ for 12 main-belt asteroids were measured and, for the first time, their polarimetric albedos were estimated. On the basis of these data, nine of these asteroids were classified as members of the M class, one as a member of the C class, and two as members of the CM class (Бельская и др. 1987a, 1991). As a result, the total number of M-type asteroids increased to 24. Furthermore, it was confirmed with confidence that, on average, M-type asteroids rotate faster than S- and C-type asteroids (Бельская и др. 1987a).

The most important results were obtained when both the polarimetric and the photometric technique were applied to study the composition of M-type asteroids. M-class ("metal") objects were originally singled out because of their reddish spectra and lack of noticeable absorption features in the spectral range 300−1100 nm. Such spectra are characteristic of iron meteorites or enstatite chondrites which contain abundant free iron. Asteroids of the M type deserve special attention primarily because they could be remaining cores of differentiated celestial bodies which had lost their silicate mantles as a result of collisions. If this is the case then they provide a unique opportunity to study the interiors of differentiated bodies, attempt the reconstruction of their internal structure, and simulate breakup processes. These problems are quite relevant to the cosmogony of the asteroid belt and the entire Solar System. Furthermore, as we have already mentioned, it is important to study the M-type asteroids as potential sources of raw materials in the near-Earth space.

An extensive program of polarimetric and photometric observations of the largest M-type asteroids (16 Psyche, 21 Lutetia, 22 Kalliope, 55 Pandora, 110 Lydia and others) was carried out in the Institute of Astronomy of the KhNU in 1978−91. In order to facilitate the interpretation of these observations, we also performed laboratory polarimetric and photometric measurements of certain meteoritic and terrestrial samples of different compositions but similar structure (a total of 22 samples representing: all main types of meteorites; terrestrial silicates pyroxenes, olivine, and enstatite; and technogenic metals Fe, Ni, Al, Pb). The objective of these measurements was to clarify the effect of composition on the photometric and polarimetric properties of reflecting surfaces and to identify possible meteorite analogues of the M-type asteroids.

Analyses of these data as well as supplementary spectral data have shown that the surfaces of large M-type asteroids are not purely metallic (contrary to what was claimed by Dollfus et al. (1979)) but rather contain a significant fraction (about 50%) of silicate materials (Lupishko and Belskaya 1989). It has also been shown that the most suitable analogues of the M-type asteroids can be stony–iron meteorites (mesosiderates) and enstatite chondrites of the E4 type. The latter result has been corroborated by spectroscopic data collected by our colleagues in the US and Russia (Rivkin et al. 1995, 1997; Busarev 1998).

### *3.2.11. Asteroid Polarimetric Database*

The rapid increase in the volume of planetary observations makes it imperative to create, maintain, and improve publicly accessible computer databases. At present, the most widely-known electronic depository is the NASA database "Plane-tary Data System" created originally at the Institute of Planetary Sciences in Tucson (AZ, USA). A subset of this database is composed of the results of observations of asteroids.

By the mid-1990s, polarimetric data for over 170 asteroids had been accumulated. Because of differences in data formatting by the various authors, sharing those results was rather difficult. However the total volume of observational data at that time was already large enough to warrant systematic analyses and comparisons, especially by combining data obtained using alternative observational and/or reduction methodologies. It thus had become necessary to reprocess all existing data and store them, along with new data, using a single, unified format. This important objective has been accomplished by creating the APD (Лупишко и Васильев 1997).

The initial sources of data were the following:

- polarimetric observations of asteroids carried out in the USA until the mid-1970s (Zellner and Gradie 1976);
- a large volume of observations (mostly UBVRI polarimetry) accumulated in the 1980−90s by these authors (Лупишко и Васильев 1997 and references therein);
- a substantial amount of polarimetric data accumulated by these authors but not yet published in peer-reviewed literature.

From the very outset, it has been expected that the APD will contain all the necessary information on the measured polarimetric parameters as well as the dates and conditions of observations, and will also allow one to link each observation to the corresponding publication (to the extent possible). On the other hand, the database has been supposed to be user-friendly and adaptable to other systems and databases. Therefore, ASCII was selected as the preferred data format, and data have been stored in the tabular form. All the available data (the results of simultaneous UBVRI-observations or observations in individual spectral filters) are organized into one file which contains the following:

- the number and name of the asteroid;
- the date and the median time of the observation (UT);
- the spectral band;
- the phase angle of the asteroid;
- the measured values of the degree of linear polarization and the position angle of the polarization plane with respect to the equatorial coordinate system along with their respective errors;
- the calculated values of the degree of linear polarization and the position angle of the polarization plane with respect to the scattering plane;
- the computed value of the position angle of the scattering plane;
- the place of observation (observatory and telescope);
- reference(s) to the corresponding publication(s) (or the name of the observer if the data have not been published).

Since 2008, the APD has also included the results of polarimetric observations of several trans-Neptunian objects (TNOs) and Centaurs. In addition to the numeri-cal data, the APD includes a file containing over 190 references on polarimetry of specific objects.

The database is updated annually. As of today, it contains all published and the majority of unpublished results of polarimetric observations of about 280 asteroids. The APD is freely accessible to users worldwide via the Internet links http://PDS.jpl.nasa.gov and http://www.psi.edu/pds/resource/apd.html and has been extensively used in planetary astrophysics.

### 3.3. Trans-Neptunian objects and Centaurs

#### *3.3.1. Description and results of polarimetric observations*

The study of physical properties of TNOs is still at its initial stage, similar to the study of the main-belt asteroids in the early 1970s when various techniques and methodologies were being developed. The applicability of those techniques to the study of TNOs is limited, primarily because of the faintness of these distant bodies which necessitates the use of large telescopes from the 8−10-m class. Another restriction is the geometry of ground-based observations, which is limited to small phase angles owing to the small angular diameter of the Earth's orbit as viewed from a TNO or a Centaur. The maximal possible phase angle that can be reached for Centaurs is 7°−8°, while that for the TNOs never exceeds 2°. At such small phase angles, the surfaces of TNOs had been expected to exhibit relatively small polarization values which would be hard to detect even using the largest telescopes. However, the first polarimetric observations of a TNO revealed a pronounced NPB. Observations of (28978) Ixion on the 8.2-m VLT (ESO) showed negative polarization values of about −1% at a phase angle as small as 1° (Boehnhardt et al. 2004). These observations have demonstrated both the capability of the instrument to provide good-quality observations of faint objects ($\sim 20^m$) and the capability of the polarimetric technique to yield valuable information about distant objects even if they are observable only at very small phase angles.

In 2004−08, an international program intended to study the surface properties of TNOs and Centaurs by means of polarimetry was carried out (Bagnulo et al. 2006, 2008; Belskaya et al. 2008a,c). The research team included scientists from France, the United Kingdom, Italy, Germany, Ukraine, Finland, and the USA. The main purpose of the program was to probe surface properties of TNOs belonging to different dynamical groups, which are believed to have different evolutionary histories or different origins. The following criteria were used to select objects for this study:

- the visible magnitude of an object should be smaller than $21^m$ in order to enable polarization measurements with an accuracy better than 0.1% in less than two hours of observation time at a 8-m telescope;
- sufficient complementary information on the object's physical properties (e.g., albedo, composition from spectral data, etc.) must be available in order to facilitate comprehensive modeling of surface properties;
- selected objects should represent all dynamical groups of TNOs;

- it must be possible to sample the widest possible phase-angle range reachable for a TNO or a Centaur from ground-based observations.

Using these criteria, nine objects had been selected: dwarf planets Eris and Haumea, classical objects (20000) Varuna and (50000) Quaoar, a resonant object (38628) Huya, a scattered disk object 26375 (1999 $DE_9$), and Centaurs (2060) Chiron, (5145) Pholus, and (10199) Chariklo. To perform polarimetric observations of these objects, the international team has repeatedly applied for observing time on the VLT, which is presently the only telescope suitable for this purpose. Observing time was granted under a program 073.C-0561 led by H. Boehnhardt as well as under a program LP 178.C-0036 led by A. Barucci and devoted to the characterization of surface properties of the TNOs using different types of observation.

All observations were carried out in the so-called service mode using the Focal Reducer/Low Dispersion Spectrograph (FORS1) instrument mounted in the Cassegrain focus of Unit Telescope 2 (Kueyen) of the VLT. The measurements of linear polarization in the imaging mode were performed by using a remotely controlled rotatable half-wave retarder positioned in front of the Wollaston prism. The rotation of the half-wave plate in steps of 22.5° and the corresponding measurements of intensities of the two rays allow one to derive the values of the Stokes parameters $Q$ and $U$ and hence the values of the degree of linear polarization and polarization position angle. The latter were determined with an accuracy of 0.03% and 0.2°, respectively. A detailed description of the data reduction procedure can be found in Bagnulo et al. (2006). The flexibility offered by the VLT service observing mode was used to distribute observations throughout the entire observing period in order to cover the maximal possible range of phase angles and obtain a good phase-angle sampling.

The results of these observations are summarized in Table 3.5, which includes the dynamical and spectral classes of each object (Fulchignoni et al. 2008), its albedo according to Stansberry et al. (2008), the range of phase angles covered by the observations, and the minimal value of the degree of polarization for this phase-angle range in the R Bessel band along with its uncertainty.

Below, we briefly describe the specific results obtained for each object. The sizes of the objects are given according to the results of radiometric observations with the Spitzer Space Telescope (Stansberry et al. 2008).

***(2060) Chiron.*** Chiron is the first discovered Centaur. It has a diameter of about 233 km and a rather flat spectrum characterized by weak absorption bands of water ice. Chiron exhibits sporadic cometary activity not correlated with the object's distance to the Sun. A specific search for a coma around Chiron gave negative results. It was concluded that the coma was beyond the detection limit if it existed at all (Bagnulo et al. 2006). This implies that the results of observations can be attributed to the Chiron's surface and are not contaminated by a coma. Observations in the standard B, V, and R filters have revealed no spectral dependence of polarization beyond measurement errors. The polarization phase curve of Chiron in the R filter is shown in Fig. 3.26. It exhibits a deep NPB with a −1.4% minimum at a phase angle of ~1.5°. The inversion is estimated to occur at a phase angle between 6° and 8°.

**Table 3.5.** Summary of polarimetric observations of TNOs and Centaurs

| Object | Dynamical class | Spectral class | Albedo | Phase angle range (deg) | $P_{\min}$ (%) |
|---|---|---|---|---|---|
| (2060) Chiron | Centaur | BB | 0.08 | 0.51 – 4.23 | –1.40 ± 0.03 |
| (5145) Pholus | Centaur | RR | 0.08 | 0.93 – 2.58 | –1.99 ± 0.12 |
| (10199) Chariklo | Centaur | BR | 0.06 | 2.66 – 4.37 | –0.92 ± 0.06 |
| (20000) Varuna | Classical | IR | 0.16 | 0.14 – 1.30 | –1.18 ± 0.13 |
| (26375) 1999 $DE_9$ | Scattered | IR | 0.07 | 0.11 – 1.41 | –1.39 ± 0.12 |
| (38628) Huya | Resonant | IR | 0.05 | 0.61 – 1.98 | –1.61 ± 0.07 |
| (50000) Quaoar | Classical | IR–RR | 0.20 | 0.25 – 1.23 | –0.65 ± 0.04 |
| (136108) Haumea | Classical | BB | 0.84 | 0.99 | –0.68 ± 0.06 |
| (136199) Eris | Detached | BB | 0.70 | 0.15 – 0.60 | –0.33 ± 0.06 |

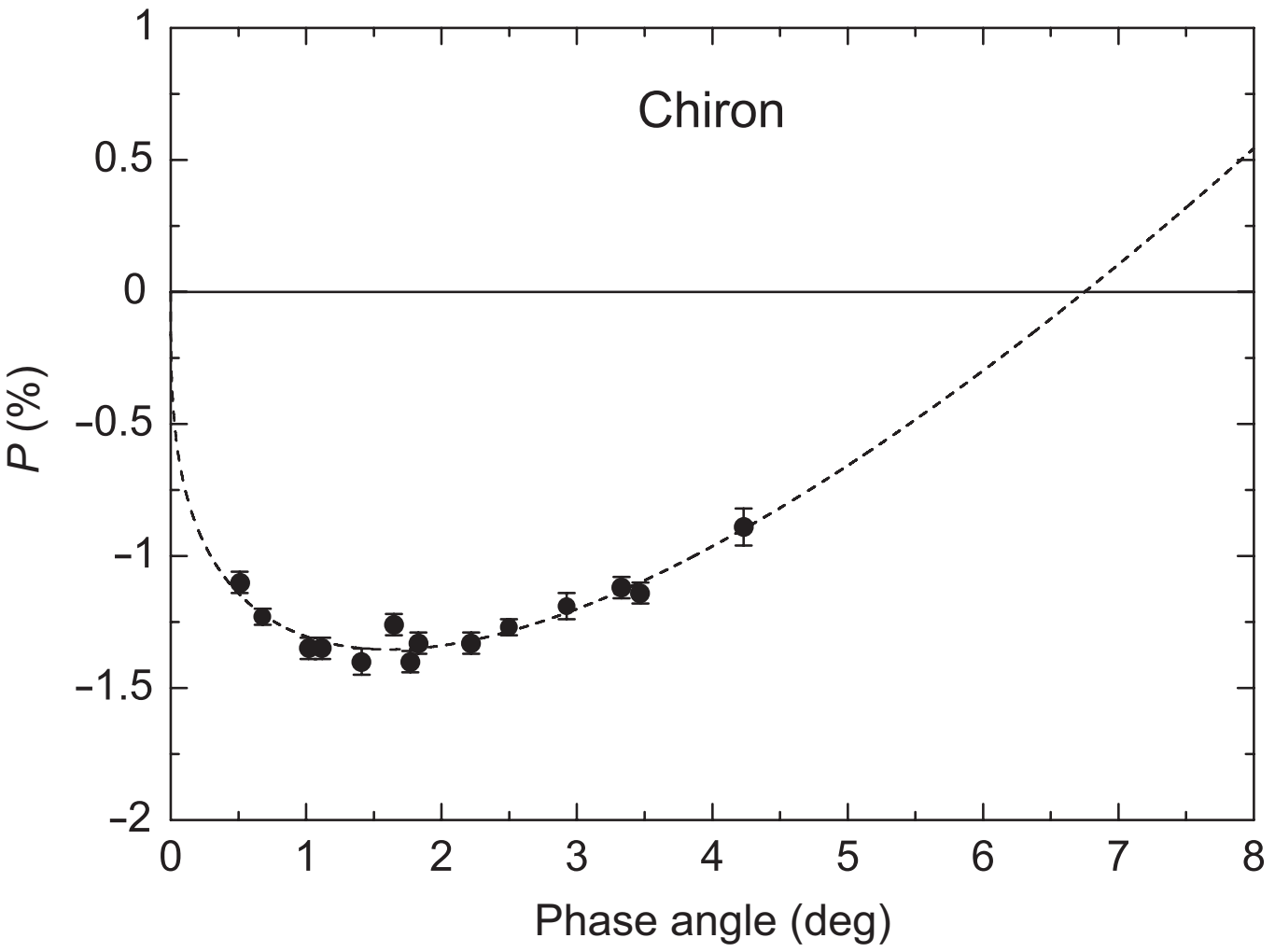

**Fig. 3.26.** Polarization phase dependence of (2060) Chiron in the R filter.

***(5145) Pholus.*** This object has a diameter of 140 km and represents the extreme case of a reflectance spectrum with the largest red slope among all Centaurs. A cometary activity has not been observed. The polarimetric measurements revealed a pronounced NPB with a −2% minimum at a phase angle as small as 2°.

***(10199) Chariklo.*** This is the largest Centaur discovered so far with a diameter of 260 km. Its spectral properties are intermediate between Chiron and Chariklo but closer to those of Chiron. A cometary activity has not been found for this object. The measured NPB is less pronounced than those for other Centaurs, although Chariklo has the darkest surface (see Table 3.5).

***(20000) Varuna.*** This object with a diameter of about 500 km belongs to the so-called classical population of TNOs, which includes objects having almost circular orbits with semimajor axes in the range of 39.4 to 47.8 AU. Photometric observations have revealed a pronounced opposition surge at phase angles smaller than 0.1° with an amplitude of $0.2^{m}$ relative to the extrapolation of the linear part of the magnitude phase dependence to the zero phase angle (Belskaya et al. 2006). Polarimetric observations have shown significant negative polarization rapidly changing with phase angle. The gradient of polarization changes in the observed range of phase angles is about 1% per degree.

***(26375) 1999 DE$_9$.*** This 460-km object has a highly eccentric orbit and thus belongs to the population of scattered-disk objects. Polarimetric observations have shown a NPB extending down to −1.4%. The phase-angle dependence of polarization is very similar to that observed for Varuna.

***(38628) Huya.*** This celestial body belongs to the dynamical class of resonant objects or Plutinos which have orbits in the 3:2 mean motion resonance with Neptune (the same as Pluto). Its diameter is about 530 km. Polarimetric observations have shown a deep NPB with $P = -1.6\%$ at a phase angle of 2°.

***(50000) Quaoar.*** This is a large classical object with a diameter of about 840 km and an almost circular orbit. Spectral measurements have indicated the presence of crystalline water ice and ammonia hydrate on the surface, which has been interpreted as an indication of a recent resurfacing of this object (Jewitt and Luu 2004). Polarimetric measurements have revealed a relatively shallow and flat NPB with $P$ close to −0.6%.

***(136108) Haumea.*** This is one of the largest TNOs with a diameter of 1200 km. Together with Pluto and Eris, it was identified as a member of a new category called dwarf planets. According to its orbital parameters Haumea belongs to the classical population. Polarimetric observations at a phase angle of 1° showed a negative polarization value close to that observed for Quaoar.

***(136199) Eris.*** This is the largest TNO with a diameter of 2600 km (i.e., greater than that of Pluto). Eris was discovered near its aphelion at 97 AU from the Sun and has an orbit with high eccentricity and inclination. According to its dynamical characteristics, it belongs to the so-called "detached objects" class with pericenters decoupled from Neptune. The spectral properties of Eris have been found to be similar to those of Pluto. For both objects, absorption features due to methane ice dominate the near-infrared spectrum (Brown et al. 2005). Polarimetric measurements have revealed rather weak negative polarization (from about −0.1 to about −0.3%) hardly changing with phase angle.

### *3.3.2. Surface structure and composition*

As of today, polarimetric measurements have been carried out for 12 TNOs including Pluto. Nine of these objects were observed in the framework of the international cooperative program described above. The resonant object (28978) Ixion was observed by Boehnhardt et al. (2004), while the scattered disk object (129981) 1999 $TD_{10}$ was measured by Rousselot et al. (2005). The observations of Pluto pertain to the Pluto–Charon system and show a notable similarity in the degree of polarization measured in different oppositions by different authors. All data and references can be found in the TNO polarimetric database at http://www.psi.edu/pds/resource/tnopol.html.

The resulting dataset allows one to draw the following conclusions regarding the polarimetric properties of surfaces of these distant bodies:

- negative polarization has been observed for each of these objects and ranges from about −0.3% to about −2%;
- two different polarimetric trends can be identified at small phase angles (Fig. 3.27): a slowly changing negative polarization for the largest objects (Pluto, Eris, Quaoar) and a rapidly changing negative polarization (a ~1 %/deg slope in the phase-angle range 0.1°–1°) for smaller TNOs (Ixion, Varuna, Huya, 1999 $DE_9$);
- the minimum of the NPB (reliably measured for the Centaur Chiron) occurs at a small phase angle of 1.5° with $P_{min} = -1.4\%$; the inversion angle is estimated to occur at phase angles between 6° and 8°;

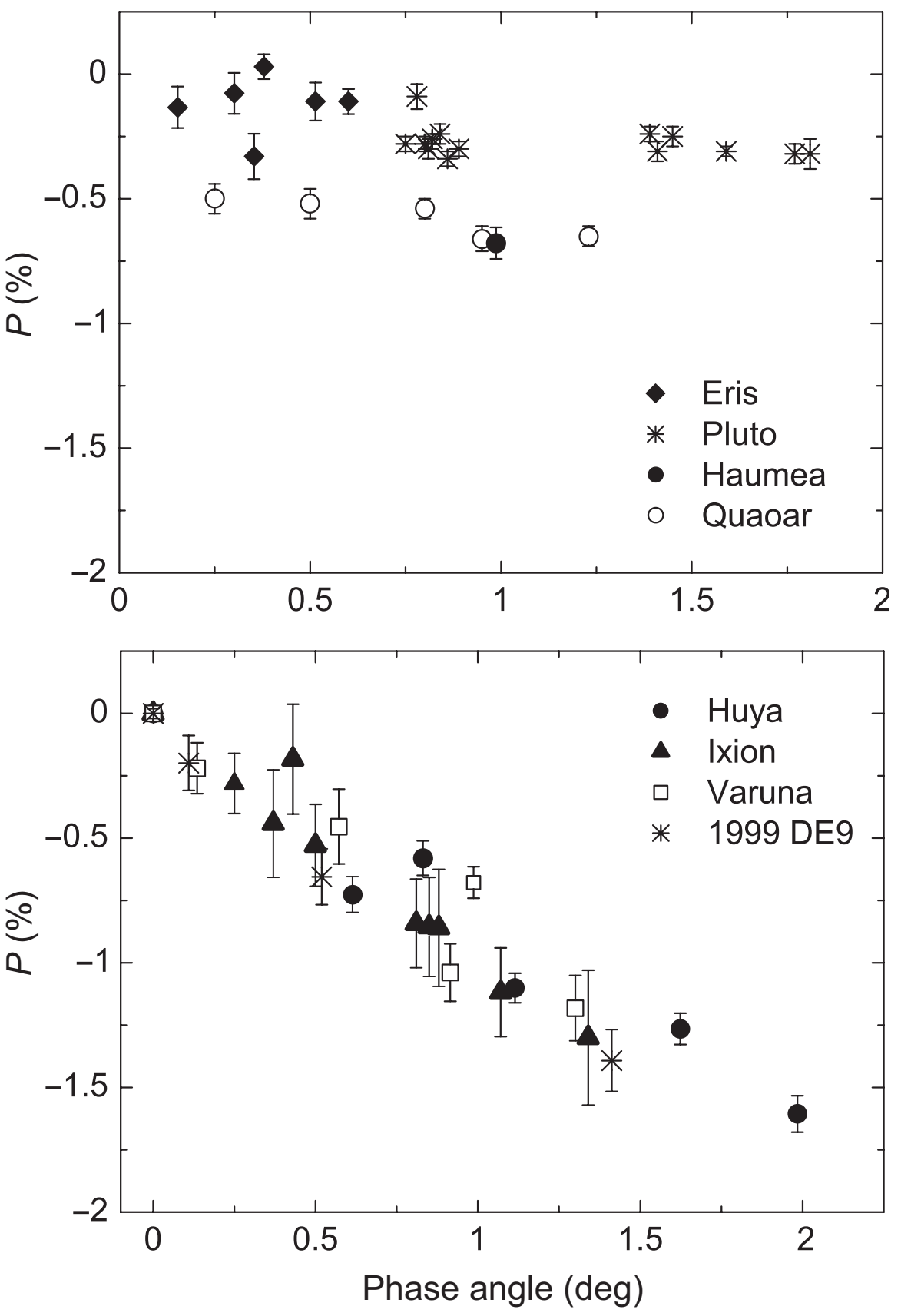

**Fig. 3.27.** Two types of polarization phase dependence for TNOs.

- Centaurs exhibit a great diversity in polarization properties;
- no wavelength dependence of the degree of polarization (B, V, and R filters) is observed for Chiron; a color effect (V and R bands) is suspected for 1999 $TD_{10}$, but its confirmation requires more observations.

A pronounced NPB deeper than −1% at phase angles close to 1° found for most of the observed objects is rather unique among Solar System bodies. At such small phase angles, the degree of polarization typically does not fall below −0.6% for a variety of minor bodies (asteroids, planetary satellites, cometary dust). Moreover, substantial variations in the phase dependence of polarization have been identified despite the small range of phase angles covered, whereas a noticeable diversity in polarization properties of asteroid surfaces becomes apparent only near the polarization minimum at phase angles 5°–12°. Thus, a possible explanation of the large range of polarization values measured for the TNOs and Centaurs is that their polarization minima occur at smaller phase angles compared to other Solar System bodies. This conclusion is confirmed by observations of the Centaur Chiron, for which the position of the polarization minimum is reliably identified at 1.5°. More-over, photometric observations of TNOs have shown that the BOE characterized by a non-linear increase in magnitude towards opposition also occurs at smaller phase angles: $< 1°$ as opposed to $4°–7°$ for main-belt asteroids (Belskaya et al. 2006). The narrower widths of both the BOE and the POE for the TNOs may be indicative of substantial differences between the surface microstructure of these distant objects and that of small bodies in the inner part of the Solar System. Indeed, according to the theory of WL (see Sections 1.19, 1.25, and 1.26), certain changes in the average size of regolith particles and/or a reduced particle packing density can result in significantly narrower opposition phenomena (e.g., Mishchenko 1992b).

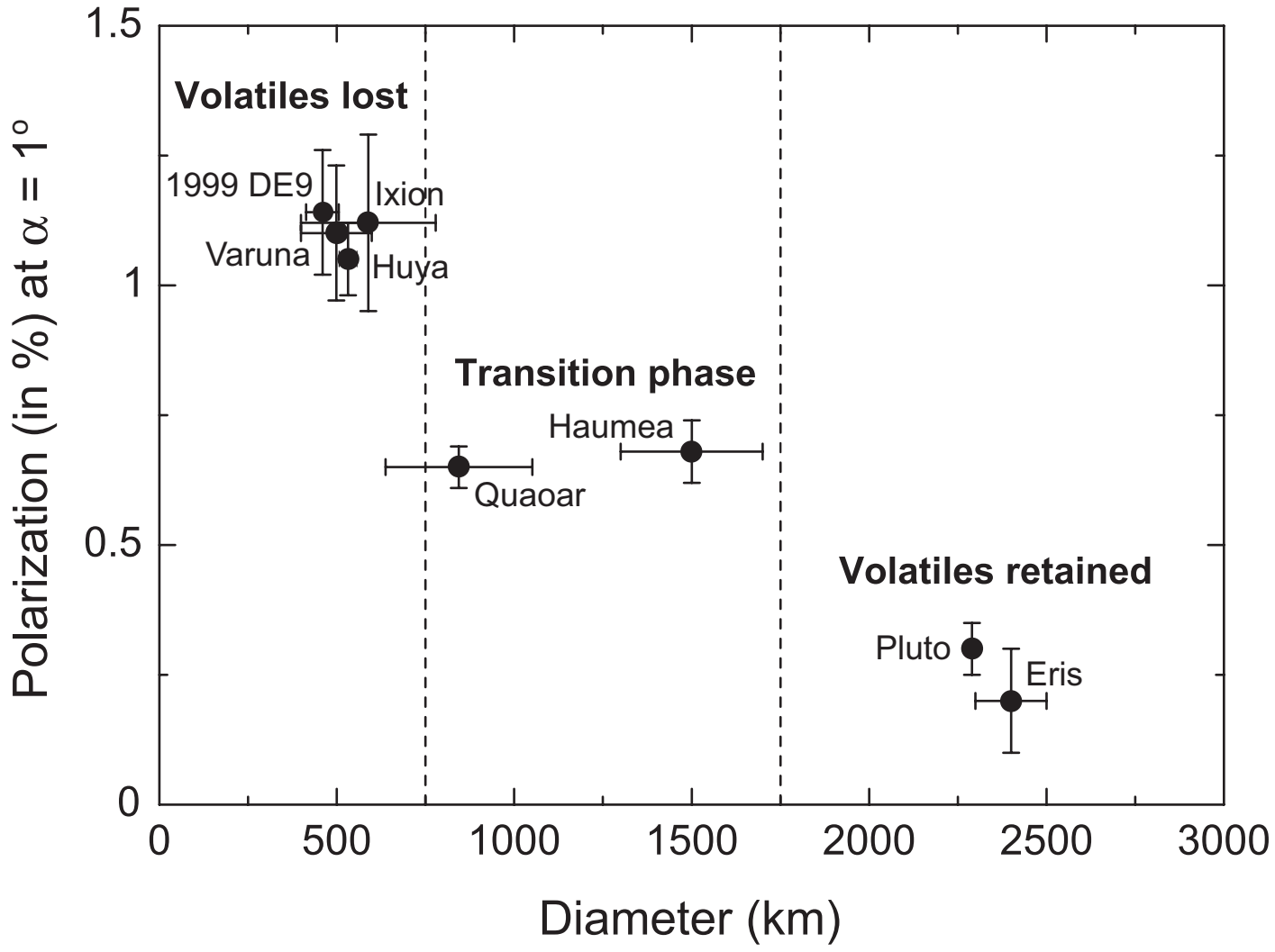

**Fig. 3.28.** Polarization at $\alpha \approx 1°$ versus diameter for several TNOs.

The discovery of two different types of behavior of the polarization phase dependence for TNOs (Belskaya et al. 2008b; Bagnulo et al. 2008) provides an evidence of considerable differences in surface properties of large and small TNOs. The two groups of objects with different polarimetric properties differ not only in size but also in surface albedo. The tendency for darker surfaces to exhibit stronger negative polarization is well-known for Solar System bodies. However the dependence of the polarization behavior on albedo cannot fully explain the observed differences. Two TNOs, Varuna and Quaoar, which both have similar albedos and orbital properties, show completely different polarization phase dependences (see Table 3.5).

Perhaps the most important difference between the surface characteristics of objects with different polarization properties is that the TNOs with small and nearly constant negative polarization are expected to be capable of retaining volatiles such as CO, $N_2$, and $CH_4$. The degree of polarization at $\alpha \approx 1°$ for TNOs with different diameters is plotted in Fig. 3.28. The limiting sizes of objects capable of retaining volatiles are shown by the vertical dashed lines according to Schaller and Brown (2007).

One can conclude that the surfaces of TNOs going through the same evolutionary phase exhibit very similar polarization properties. It is interesting to note that all TNOs with diameters smaller than 750 km possess similar polarization properties despite the fact that they have different albedos and belong to different dynamic groups.

## 3.4. Opposition phenomena exhibited by Solar System objects: a general perspective

The observed phase-angle dependences of brightness and polarization exhibit a great diversity (Rosenbush et al. 2002a; Аврамчук и др. 2007), as illustrated in Fig. 3.29. Both for high-albedo objects (e.g., icy satellites, Saturn's rings, asteroids 44 Nysa and 64 Angelina) and intermediate-albedo major satellites of Uranus (e.g., the darkest satellite Umbriel with albedo $p_V = 0.19$ and the brightest satellite Ariel with $p_V = 0.35$), a very narrow spike-like BOE has been observed. For some objects (e.g., Deimos, Phobos, Callisto, and comet Halley), the curvature of the photometric phase dependence changes rather slowly over a broad range of phase angles, which does not allow one to identify a linear part. For some dark objects (e.g., C- and P-type asteroids), the BOE is very weak, and one can see only the linear part of the phase curve. These traits are related to physical and chemical properties of the scattering regolith particles.

The shape of the polarization phase curves is also quite variable (see Fig. 3.29b). For certain high-albedo objects (including asteroids 44 Nysa and 64 Angelina, the Galilean satellites Io, Europa, and Ganymede, the trailing hemisphere of Iapetus, and Saturn's rings), for which a spike-like BOE has been detected, the PDP curves consist of the POE in the form of a separate peak of negative polarization at phase angles $\alpha < 2°$ superposed on the regular NPB. Very unusual asymmet-ric curves of polarization are exhibited by Chiron, Ixion, and the leading side of Iapetus. However, the range of phase angles covered does not allow one to determine unambiguously the entire shape of the respective PDPs. Nevertheless, we can conclude that the PDPs for objects in the inner and outer parts of the Solar System are very different.

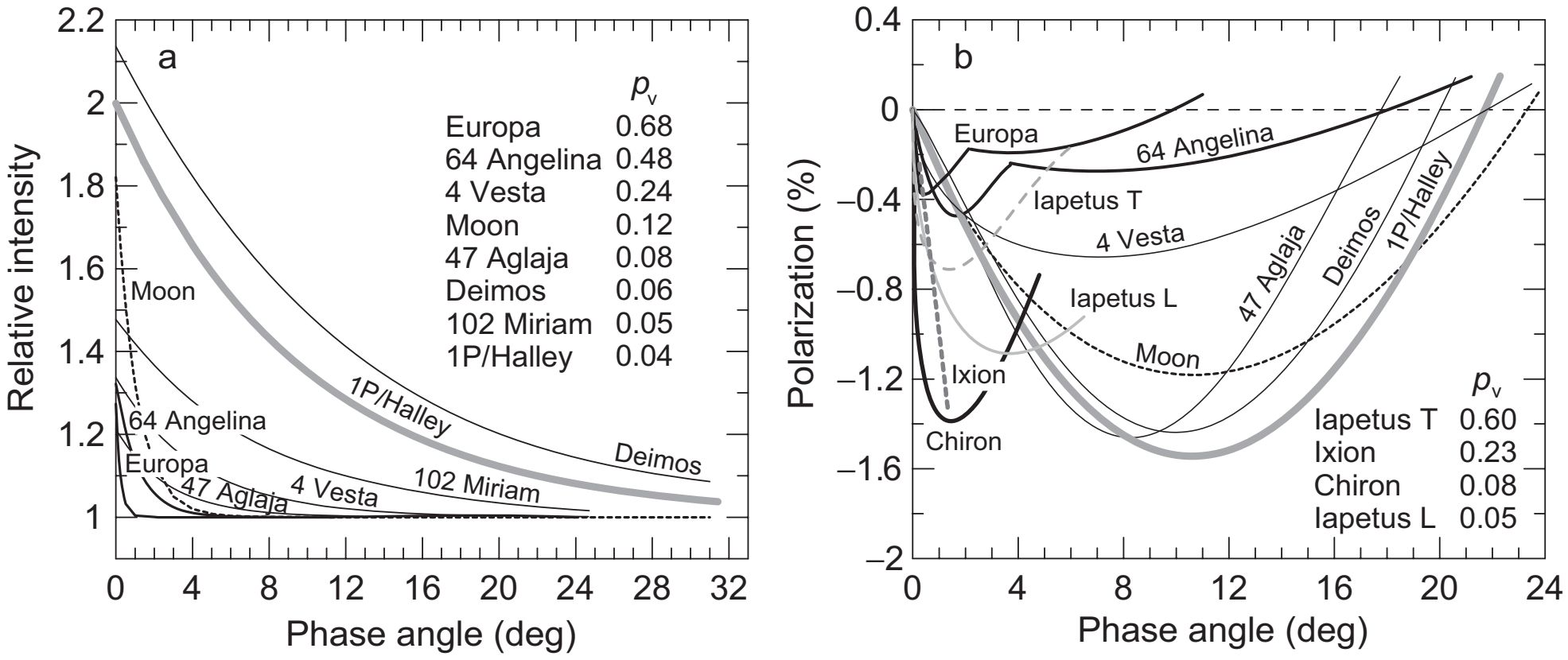

**Fig. 3.29.** Comparison of the photometric phase curves after the subtraction of the linear part (a) and polarization phase curves (b) for different objects. A second local minimum of polarization near opposition has been observed for the Jovian satellite Europa and asteroid 64 Angelina.

The photometric and polarimetric behavior of different ASSBs near opposition is highly variable. The amplitude of the BOE lies in the range 1.1–5.4; the half-width at half-maximum of the BOE varies from 0.08° to 6.1°; the phase angle at which the nonlinear BOE begins varies from 0.4° to ~30°; and the phase coefficient varies from 0.0005 to 0.06 mag/deg. The parameters of the NPB $|P_{\min}|$, $\alpha_{\min}$, and $\alpha_{\text{inv}}$ typically range between 0.2% and 2.1%, 2° and 12°, and ~7° and ~29°, respectively.

According to the scattering theory, CB and shadowing hiding are the main factors controlling the observed opposition phenomena. A significant role can also be played by the near-field effects which tend to decrease the intensity of the backscattered light and contribute to the NPB (see Section 1.26). The first-order scattering by individual particles can also contribute to the BOE and NPB. The relative contributions of these factors depend on the physical parameters (refractive index, packing density, particle morphology and size distribution, etc.) of the regolith layer and the scattering geometry. Therefore, a reliable detection of CB for an ASSB requires the observation of more than one manifestation of CB and a verification that the observations do not contradict theoretical predictions of the plausible ranges of measured parameters (Mishchenko 1993b; Mishchenko et al. 2006b; Tishkovets and Mishchenko 2010). According to Chapter 1, these theoretical predictions can be summarized as follows.

- By virtue of being the result of multiple scattering, CB is more likely to be observed for high-albedo ASSBs rather than for low-albedo objects such as the Moon. Indeed, the phase-angle distribution of the backscattered light for low-albedo objects is primarily controlled by the first-order scattering and near-field effects, including mutual particle–particle shielding and shadow hiding (Tishkovets 2008).
- Irrespective of particle size relative to the wavelength, CB causes the BOE as a narrow intensity peak centered at exactly the opposition. The observed angular width and amplitude of this peak must be in reasonable agreement with the results of theoretical computations of CB for the expected range of particle sizes, refractive indices, and packing densities (e.g., Wolf et al. 1988; Mishchenko 1992b,c; Mishchenko et al. 2006b, 2009b,c)
- Since the incident sunlight is essentially unpolarized, the BOE must be accompanied by the POE provided that the regolith grains have sizes comparable to or smaller than the wavelength (Mishchenko 1993b; Mishchenko et al. 2000b, 2009b,c). The angular width of the POE must be comparable to that of the BOE.

Even though these predictions are formulated intentionally in rather broad terms, their verification for specific ASSBs is extremely difficult. Indeed, the accumulation of a detailed data set with fine angular resolution and phase-angle cov-erage extending down to a small fraction of a degree typically takes several observation cycles separated by long periods (often lasting for years if not decades) during which the corresponding Sun–object–observer configurations are unsuitable. Furthermore, the photometric and polarimetric accuracy and precision of the instruments used must be very high.

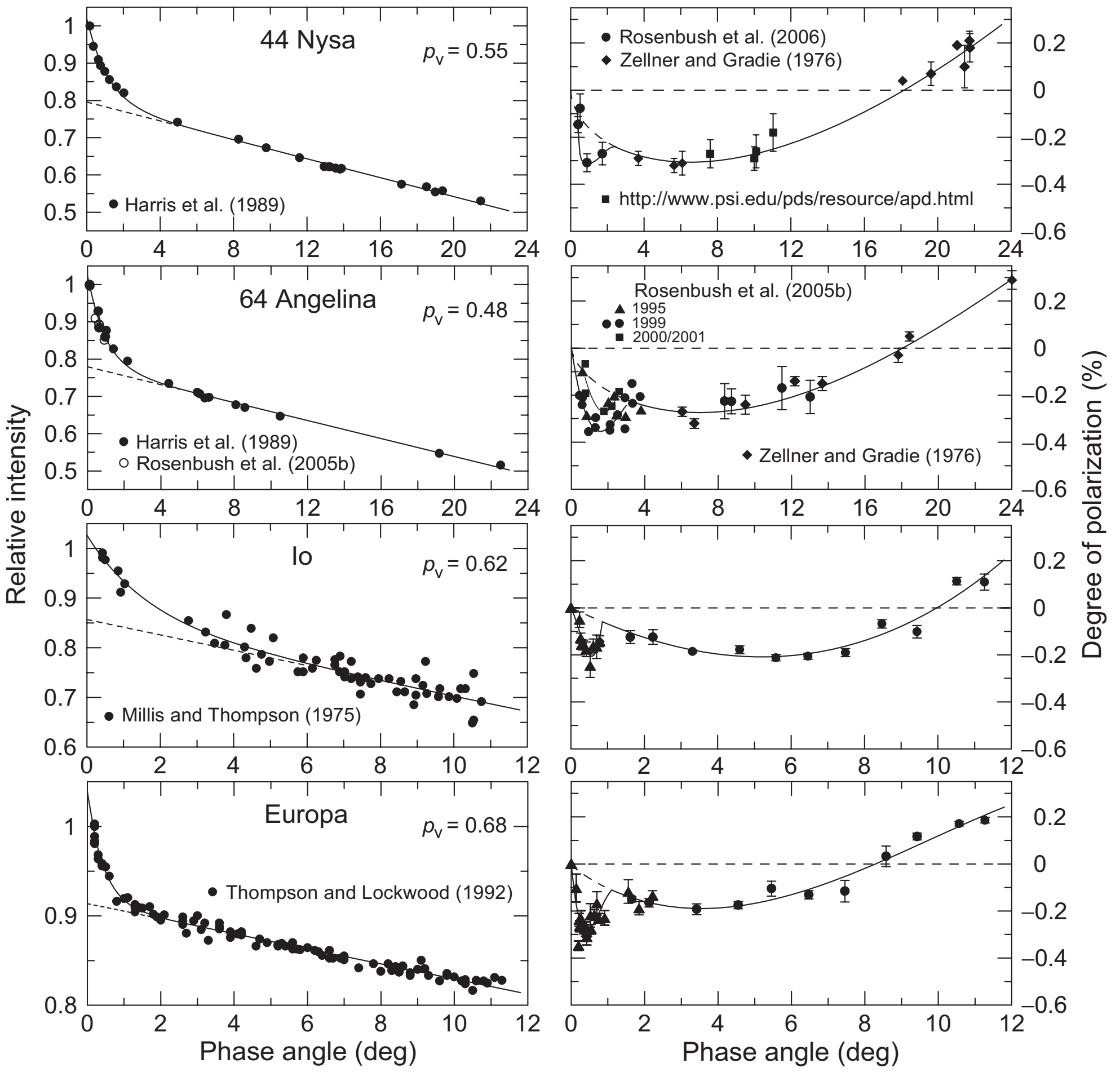

**Fig. 3.30** *(continued on p. 201)*

Despite the above-mentioned challenges, the accumulated body of high-quality long-term astronomical observations does allow one to identify a class of high-albedo ASSBs with unique opposition properties (Mishchenko et al. 2006a). This class includes the E-type asteroids 44 Nysa and 64 Angelina, the planetary satellites Io, Europa, Ganymede, and Iapetus, and the A and B rings of Saturn. The results of photometric and polarimetric observations of these objects in the visible spectral range are summarized in Fig. 3.30 and reveal the following main features.

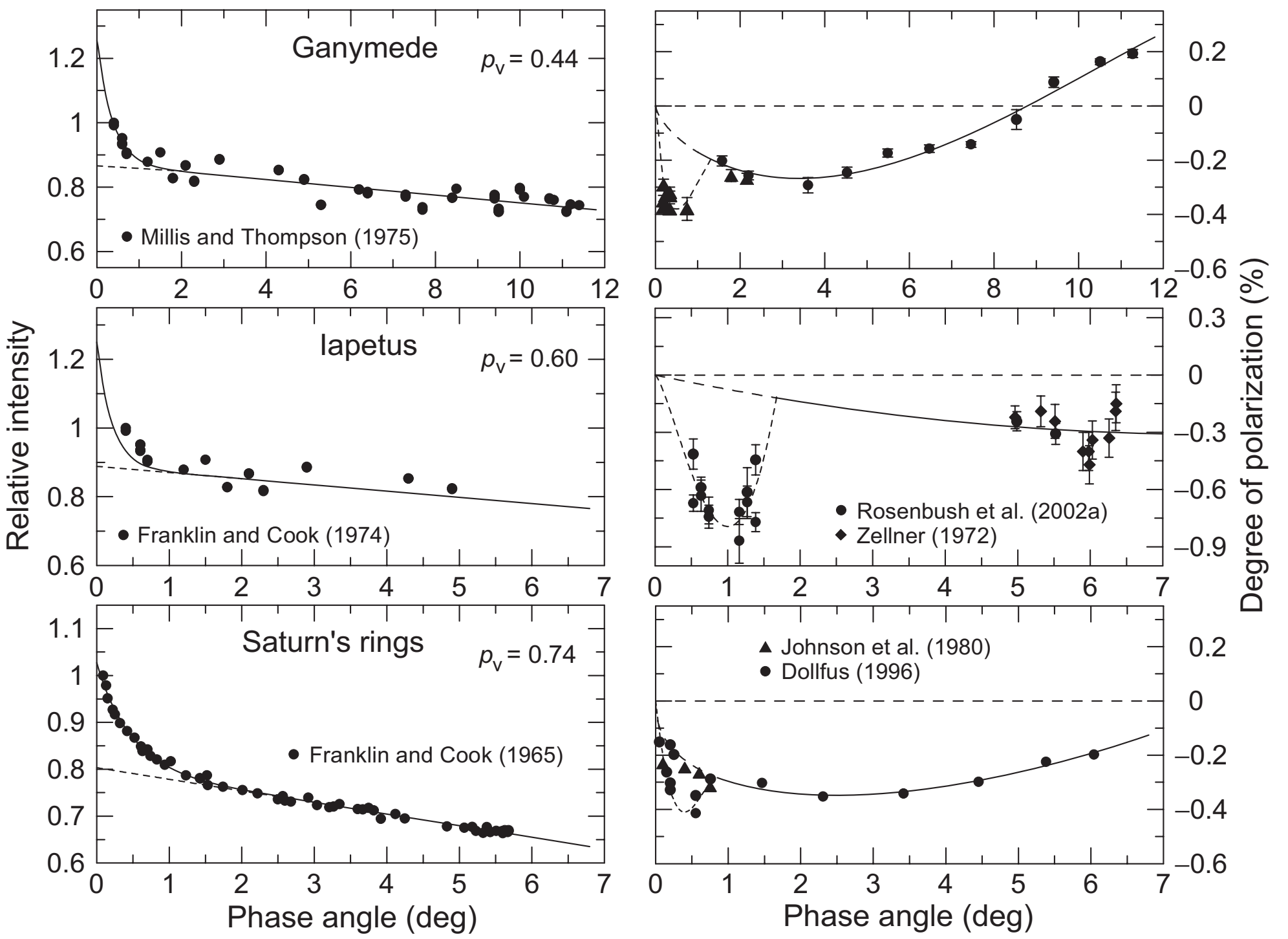

**Fig. 3.30.** Relative intensity and degree of linear polarization versus phase angle for high-albedo ASSBs. The intensity is normalized to unity at the smallest phase angle available. The polarization data for Io, Europa, Ganymede, and Saturn's rings at phase angles greater than 1° were obtained by averaging data from Dollfus (1975), Rosenbush et al. (1997a), and Rosenbush and Kiselev (2005) over 1° intervals with equal weights assigned to all observations. The intensity data were approximated by an exponential–linear function (Kaasalainen et al. 2003), whereas the polarimetric data were approximated by a trigonometric polynomial (Penttilä et al. 2005).

- Each photometric phase curve has a linear background with a superposed nonlinear peak centered at opposition (see the left-hand panel of Fig. 3.30). The backscattering intensity peaks are extremely narrow. Their actual angular widths are still uncertain because (i) the data points remain sparse, especially for Io, Ganymede, and Iapetus; (ii) the smallest phase angle in the actual astronomical observations is never equal to zero; and (iii) the Sun is a source of light with a nonzero angular width when viewed from an ASSB. However, it is obvious that the peaks are much narrower than those caused by shadow hiding in dark regoliths of low-albedo ASSBs such as the Moon (Section 3.1.4).
- Each polarization phase curve in the right-hand panel of Fig. 3.30 exhibits a narrow local minimum centered at a phase angle approximately equal to the angular width of the corresponding intensity peak. Each minimum is superposed on a much broader, nearly parabolic NPB which appears to be a ubiquitous trait of the majority of ASSBs. However, its physical origin remains uncertain (see, e.g., Videen et al. 2004 and references therein). Although it has been speculated that the NPB could be caused by CB from particulate media, the results of Mishchenko (1993b), Mishchenko et al. (2000b), and Tishkovets et al. (2002a,b) indicate that this conjecture is unlikely to be correct.

The angular widths and amplitudes of the backscattering intensity peaks in Fig. 3.30 are consistent with the results of theoretical computations of CB (Mishchenko 1992b, 1993b; Mishchenko and Dlugach 1993) for particle sizes of the order of the wavelength, packing densities ranging from several percent to approximately 40%, and particle compositions ranging from water ice (Europa, Saturn's rings) to silicates (44 Nysa and 64 Angelina). The amplitudes of the peaks are also consistent with the theory of CB (Mishchenko 1992c; Mishchenko et al. 2006b, 2009b,c) and the assumption that a significant fraction of the surface of these ASSBs is covered with a fine-grained material causing CB.

The shapes of the narrow backscattering polarization minima are quite similar to that of the POE caused by CB from a half-space of nonabsorbing Rayleigh scatterers (see Fig. 1.21). The magnitudes of the polarization minima are smaller than that in Fig. 1.21 and are different for the different ASSBs. This is not surprising since the actual regolith grains are not perfect Rayleigh scatterers and their sizes and refractive indices cannot be expected to be the same for all the ASSBs. Furthermore, the fraction of the visible surface causing CB can also vary with object, thereby changing the resulting polarization.

The theory of CB outlined in Chapter 1 allows one to estimate certain parameters of regolith particles, for example, those covering the surfaces of macroscopic bodies forming Saturn's rings. One can hypothesize that the observed bimodal negative-polarization PDPs (Fig. 3.30) can be caused by the interference of multiply-scattered waves provided that the individual particles have a pronounced NPB in their single-scattering polarization phase curves (Mishchenko et al. 2002b). Let us assume that the macroscopic ring bodies are covered by a particulate layer composed of clusters of small ice monomers with a refractive index $m \sim 1.31$ and estimate the packing density of these clusters (i.e., the product of the number density of the clusters and the average volume of the smallest circumscribing sphere of a cluster) using the following formula (Mishchenko et al. 2002b):

$$\alpha_{\mathrm{m}} \approx 4.39\,\mathrm{Im}(m_{\mathrm{eff}}) \approx \frac{2}{k_1 l}, \tag{3.10}$$

where $\alpha_{\mathrm{m}}$ is the phase angle of the POE minimum (see Fig. 3.30), $l$ is the mean free path given by Eq. (1.125), and $m_{\mathrm{eff}}$ is the complex effective refractive index of the medium (see Section 1.25.3). This formula is strictly valid only for Rayleigh particles, which scatter light almost isotropically. To account for the scattering anisotropy of phase functions typical of wavelength-sized and larger particles, $l$ must be replaced by the transport mean free path given by

$$l_{\mathrm{tr}} = \frac{1}{n_0 C_{\mathrm{ext}}(1 - \langle \cos\Theta \rangle)} \tag{3.11}$$

(Section 1.19). Here, $\langle \cos\Theta \rangle$ is the asymmetry parameter given by Eq. (1.87) and $n_0$ is the number density of the clusters. Equations (3.10) and (3.11) yield the following formula for the cluster packing density $\tilde{\xi}$:

$$\tilde{\xi} = \frac{2\pi x_{\text{eff}}^3 \alpha_{\text{m}}}{3k_1^2 C_{\text{ext}}(1 - \langle \cos\Theta \rangle)}, \tag{3.12}$$

where $x_{\text{eff}}$ is the effective size parameter of the clusters. For Saturn's rings, $\alpha_{\text{m}} \approx 0.5°$ (see Fig. 3.30). Computations for effective size parameters $5 \le x_{\text{eff}} \le 20$ yield the following range of packing densities: $0.02 \lesssim \tilde{\xi} \lesssim 0.3$. Clusters of spherical particles composed of ~100 monomers with size parameters ~1 and refractive indices ~1.31 also readily reproduce the wide NPB observed for Saturn's rings (Dollfus 1996).

A more detailed quantitative analysis of the observational data in terms of specific physical parameters of the regolith layer is hardly possible at this time given the limited nature of the observational dataset, the constrained theoretical ability to compute all radiometric and polarimetric characteristics of CB for realistic polydisperse particle models (Tishkovets et al. 2002a,b; Muinonen 2004), and the extreme morphological complexity and heterogeneity of the surfaces of the ASSBs. It is fundamentally important, however, that the observations exhibit both the BOE and the POE and are in a reasonable quantitative agreement with the existing theory. Furthermore, no other optical mechanism is currently known to produce simultaneously both opposition features with their unique morphological characteristics.

# 4

# Polarimetric observations of comets and their interpretation

Unlike other small Solar System bodies, comets are non-stationary objects rapidly evolving as they approach the Sun. Cometary nuclei are composed of ices and refractory materials, and most of them have not undergone a warm-up process. As such, they are relict objects with compositions most resembling that of the primordial protoplanetary cloud. The evolution of a cometary atmosphere and tail and the controlling physical processes are specific for each comet and unique for most long-period comets. In other words, each comet is somewhat of an experiment which allows one to study many unique physical processes (including light scattering) uncharacteristic of other small bodies. This makes it imperative to study each comet approaching the Sun by using all astrophysical techniques available in order to identify its individual traits as well as more typical properties. Such studies can be expected to yield eventually the requisite knowledge of the origin, physical nature, and evolution of the entire cometary population.

In general, sunlight scattered by a cometary atmosphere becomes elliptically polarized. The observed phase-angle and wavelength dependence of polarization as well as its spatial and temporal variations depend on the specific processes of light scattering, the microphysical parameters of the constituent dust particles, the distribution of dust and gas in the coma, and potential angular alignment of nonspherical dust grains. This makes polarimetry an essential tool for the determination of optical and microphysical properties (size, refractive index, morphology) of cometary dust particles. Another important functions of polarimetric observations are to facilitate taxonomic classification of comets and to help relate the properties of cometary dust to the origin and evolution of a comet.

## 4.1. Historical background

The history of polarimetric investigations of comets started with the invention of the visual polariscope by François Arago, who used this instrument to discover traces of polarization in the light coming from the Great Comet 1819 II (Arago 1854−57). His observations of comet Halley in 1835 confirmed the non-zero polarization of cometary light. Visual and, more recently, photographic observations of many bright comets in the late 19th−early 20th century had helped establish certain common properties of cometary polarization. It had been found that the polari-zation plane is usually perpendicular to the scattering plane. Variations of polarization across distinct parts of a comet (coma, tail) as well as among different comets had been detected. All these polarization features had been attributed to the diffuse reflection and scattering of sunlight.

The modern era of polarimetric studies of comets began with the work by Yngve Öhman (Öhman 1939, 1941). He was the first to observe separately the polarization in the continuum and in emission bands. The polarization of molecular emissions was explained as being due to resonance fluorescence. For this mechanism, Öhman found that the degree of linear polarization changes with phase angle according to the following law:

$$P(\alpha) = \frac{P_{90}\sin^2\alpha}{1 + P_{90}\cos^2\alpha}, \tag{4.1}$$

where $P_{90}$ is the maximal polarization at the phase angle $\alpha = 90°$. This relation has been confirmed by numerous observations of molecular bands for many comets (see, e.g., Blackwell and Willstrop 1957; Bappu et al. 1967). For the diatomic molecules CN and $C_2$, the results of observations (Le Borgne et al. 1986; Le Borgne and Crovisier 1987; Киселев 2003) agree well with the theoretical result $P_{90}$ = 0.077. However, no analytical formula had been known for the scattering of sunlight by cometary dust particles. Öhman and, more recently, others (Blackwell and Willstrop 1957; Bappu and Sinvhal 1960) modeled the phase-angle dependence of polarization in the continuum using the approximation (4.1), which works well in the range $\alpha = 40°–80°$. Somewhat surprisingly, there had been no observations of comets at phase angles smaller than 40° until 1975, apparently because of the belief that cometary polarization at large phase angles is more informative owing to its relatively large values. Neither the value and phase angle of maximal polarization nor the behavior of polarization at small phase angles had been established until the mid 1970s. In 1986, circular polarization of the light scattered by comets was detected, but with very low measurement accuracy (Metz and Haefner 1987; Dollfus and Suchail 1987; Мороженко и др. 1987). A more detailed historical account of cometary polarimetry can be found in Kiselev and Rosenbush (2004).

### 4.2. Observational data

Since the late 1960s, we have carried out regular polarimetric and photometric observations of comets. By now, aperture and CCD imaging measurements of linear and circular polarization have been performed for 52 comets (3 of them have been observed during two apparitions) with several different telescopes and using various polarimetric instrumentation. The observation periods, spectral filters used, ranges of helio- and geocentric distances and phase angles, and total numbers of observation nights $N$ are listed in Table 4.1. These polarimetric data (~ 1000 observations) constitute a major part of the cumulative international database (~2650 measurements of linear and circular polarization for 64 comets) currently used for analyses of phase-angle variations of cometary polarization (Section 4.11).

**Table 4.1.** Summary of polarimetric observations of comets

| Comet | Period of observations | Filter(s) | $r$ (AU) | $\Delta$ (AU) | $\alpha$ (deg) | $N$ |
|---|---|---|---|---|---|---|
| | | *Linear polarization (aperture polarimetry)* | | | | |
| C/1968 N1 (Honda) | 03.09.68 | B | 1.24 | 0.66 | 54.5 | 2 |
| C/1969 T1 (Tago–Sato–Kosaka) | 20.01 – 12.02.70 | B, V | 0.85–1.24 | 0.38–0.80 | 77.1–52.8 | 34 |
| C/1969 Y1 (Bennett) | 08.04 – 02.05.70 | V | 0.69–1.08 | 0.83–1.40 | 79.7–45.3 | 15 |
| C/1970 N1 (Abe) | 04.09 – 04.10.70 | V | 1.34–1.14 | 0.81–1.26 | 53.3–48.8 | 22 |
| C/1973 E1 (Kohoutek) | 01.12.73 – 14.02.74 | B, V | 0.78–1.32 | 0.92–1.37 | 43.7–63.8 | 11 |
| C/1974 C1 (Bradfield) | 23.05 – 26.05.74 | B, V | 1.46–1.50 | 1.36–1.39 | 41.8–40.9 | 3 |
| 45P/Honda–Mrkos–Pajdusakova | 04 – 12.12.74 | V | 0.74–0.66 | 1.05–0.94 | 64.0–73.5 | 2 |
| C/1975 N1 (Kobayashi–Berger–Milon) | 17.07 – 13.08.75 | Narrow bands,[1] V | 0.70–1.20 | 0.28–0.67 | 43–95 | 67 |
| C/1975 V1 (West) | 06.03 – 04.07.76 | Narrow bands,[1] V | 0.41–2.64 | 0.85–1.76 | 13.5–98 | 74 |
| 29P/Schwassmann–Wachmann 1 | 16.12.76 – 10.02.77 | V | 5.86–5.87 | 5.06–5.86 | 5.9–9.8 | 6 |
| C/1977 R1 (Kohler) | 10.09 – 15.10.77 | Narrow bands,[1] V | 1.42–1.09 | 1.49–1.14 | 40.4–53.0 | 12 |
| 101P/Chernykh | 20.09.77 – 02.01.78 | V | 2.74–2.53 | 1.74–2.39 | 3.0–22.8 | 6 |
| 49P/Arend–Rigaux | 01.01.78 | V | 1.48 | 0.86 | 61.3 | 1 |
| 47P/Ashbrook–Jackson | 12.07 – 27.11.78 | V | 2.32–2.48 | 1.98–1.76 | 0.4–25.2 | 19 |
| C/1979 S1 (Meier) | 01.08 – 25.09.79 | V | 1.78–1.47 | 1.47–1.50 | 34.7–39.6 | 5 |
| C/1982 M1 (Austin) | 25.08 – 26.09.82 | V | 0.65–0.95 | 0.63–1.44 | 45.3–103.5 | 12 |
| 67P/Churyumov–Gerasimenko | 25.11 – 26.11.82 | IHW | 1.32 | 0.38 | 27.3–27.7 | 2 |
| 22P/Kopff | 12.06 – 08.08.83 | IHW | 1.58–1.68 | 0.73–0.92 | 18.0–37.2 | 9 |
| 161P/Hartley–IRAS | 28.03 – 02.04.84 | IHW | 1.68–1.72 | 1.65 | 34.6–34.9 | 3 |
| 27P/Crommelin | 01.02 – 23.03.84 | IHW | 0.74–0.95 | 0.93–1.10 | 56.1–74.9 | 7 |
| 21P/Giacobini–Zinner | 21.07 – 12.09.85 | IHW, R | 1.03–1.68 | 1.03–1.21 | 57.2–73.3 | 6 |
| 1P/Halley | 13.09.85 – 15.05.86 | IHW, R | 0.84–2.56 | 0.62–5.83 | 1.6–61.1 | 52 |
| C/1987 P1 (Bradfield) | 19.12.87 – 09.02.88 | IHW | 1.15–1.81 | 0.85–1.05 | 32.8–56.5 | 4 |
| C/1988 A1 (Liller) | 18.05 – 23.05.88 | IHW | 1.21–1.28 | 1.23–1.24 | 47.2–49.0 | 4 |
| 23P/Brorsen–Metcalf | 09.08 – 14.08.89 | IHW, R | 0.81–0.89 | 0.63–0.66 | 82.5–87.0 | 5 |
| C/1989 X1 (Austin) | 25.05 – 16.06.90 | IHW | 1.16–1.55 | 0.24–0.66 | 14.8–46.5 | 2 |
| C/1989 Q1 (Okazaki–Levy–Rudenko) | 13.11.89 | BVR | 0.65 | 0.77 | 89.8 | 1 |
| C/1990 K1 (Levy) | 22.07.90 – 09.02.91 | IHW, UBVR | 1.17–2.10 | 0.43–1.22 | 13.4–58.4 | 23 |
| 4P/Faye | 30.11 – 09.12.91 | IHW | 1.60–1.61 | 0.71 | 23.4–26.5 | 4 |
| C/1991 T2 (Shoemaker–Levy) | 31.05 – 01.08.92 | IHW | 1.28–0.85 | 1.83–1.14 | 35.6–60.1 | 11 |
| C/1992 F1 (Tanaka–Machholz) | 03.05.92 | R | 1.27 | 1.72 | 35.6 | 1 |
| 109P/Swift–Tuttle | 25.10 – 17.12.92 | V | 1.24–1.00 | 1.24–1.57 | 50.3–31.8 | 9 |
| 47P/Ashbrook–Jackson | 17.09 – 10.12.93 | $R_x$ | 2.36–2.54 | 1.43–1.93 | 11.7–20.1 | 4 |
| 76P/West–Kohoutek–Ikemura | 09 – 10.12.93 | $R_x$ | 1.58 | 0.61 | 8.4 | 1 |
| 2P/Encke | 13.12.93 – 14.01.94 | $R_x$ | 0.71–1.25 | 0.80–0.95 | 51.1–80.5 | 2 |
| 31P/Schwassmann–Wachmann 2 | 11.12.93 | $R_x$ | 2.10 | 1.32 | 20.8 | 1 |
| C/1995 O1 (Hale–Bopp) | 18.06.96 – 24.04.97 | HB, 4430/44, 6420/26 | 4.05–0.91 | 3.08–1.32 | 4.8–49.1 | 30 |
| C/1996 B2 (Hyakutake) | 25.03 – 07.04.96 | IHW | 1.05–0.76 | 0.10–0.48 | 57.7–111.4 | 7 |
| D/1996 Q1 (Tabur) | 09 – 10.01.96 | 4430/44, 6420/26, 6620/59 | 0.96 | 0.42 | 83.3 | 2 |
| 21P/Giacobini–Zinner | 19.11.98 – 25.01.99 | R, 4430/44, 6420/26 | 1.03–1.37 | 0.85–1.22 | 44.3–73.8 | 5 |
| C/1998 U5 (LINEAR) | 24.11.98 | 4430/44, 6420/26 | 1.30 | 0.54 | 43.8 | 1 |
| C/1999 J3 (LINEAR) | 19.09.99 | 4430/44, 6420/26, 6620/59 | 0.98 | 0.99 | 61.6 | 1 |
| D/1999 S4 (LINEAR) | 24.06.00 – 29.07.00 | HB, WRC | 0.97–0.76 | 1.21–0.37 | 54.1–122.1 | 21 |
| C/2001 A2 (LINEAR) | 30.06 – 30.07.01 | HB | 1.05–1.45 | 0.24–0.51 | 26.5–74.8 | 17 |
| C/2000 WM1 (LINEAR) | 10.11 – 08.12.01 | Narrow bands | 1.55–1.10 | 0.61–0.32 | 13.1–61.1 | 70 |
| 2P/Encke | 17 – 24.11.03 | UBVRI | 0.88–1.00 | 0.26–0.28 | 80.8–105 | 5 |

**Table 4.1** *(continued)*

| Comet | Period of observations | Filter(s) | $r$ (AU) | $\Delta$ (AU) | $\alpha$ (deg) | $N$ |
|---|---|---|---|---|---|---|
| C/2002 T7 (LINEAR) | 21 – 25.11.03 | UBVRI, WRC, gc | 2.67–2.62 | 1.71–1.66 | 6.46–6.77 | 5 |
| C/2001 Q4 (NEAT) | 21 – 23.05.04 | R | 0.97 | 0.58–0.63 | 77.4–75.2 | 3 |
| 9P/Tempel 1 | 02 – 03.07.05 | R | 1.49 | 0.82 | 40.8 | 2 |
| 73P/Schwassmann–Wachmann 3 (nuclei B, C, G) | 22.04 – 06.05.06 | UBVRI | 1.17–1.04 | 0.21–0.09 | 37–66 | 18 |
| 17P/Holmes | 08 – 22.11.07 | UBVRI | 2.50–2.55 | 1.62–1.65 | 13.2–11.3 | 8 |
| *Circular polarization* | | | | | | |
| 1P/Halley | 09 – 28.04.86 | IHW | 1.37–1.65 | 0.45–0.80 | 21.1–34.8 | 10 |
| C/1995 O1 (Hale–Bopp) | 11.03.97 | 4845/10 | 2.78–0.91 | 3.02–1.32 | 46.7 | 1 |
| D/1999 S4 (LINEAR) | 28.06 – 22.07.00 | R | 0.77–0.94 | 0.36–1.10 | 61–122 | 8 |
| C/2001 Q4 (NEAT) | 21 – 23.05.04 | R | 0.97 | 0.58–0.63 | 77.4–75.2 | 3 |
| 8P/Tuttle | 10.01.08 | R | 1.06 | 0.30 | 68 | 1 |
| *Polarimetry during star occultation by comets* | | | | | | |
| C/1990 K1 (Levy) | 18 – 19.08.90 | R | 1.48 | 0.50 | 18.3–17.5 | 2 |
| 109P/Swift–Tuttle | 26.10 – 25.11.92 | V | 1.24–1.00 | 1.23–1.30 | 47.4–48.7 | 2 |
| C/1995 O1 (Hale–Bopp) | 09.03.97 | 4845/10 | 1.00 | 1.39 | 46 | 1 |

[1]Nonstandard narrow bands

## 4.3. Phase-angle dependence of linear polarization

The light coming from a cometary atmosphere is a superposition of radiation scattered by dust particles and that emitted by the gaseous constituent. The scattering by dust and the resonance fluorescence from the most abundant cometary molecules CN, $C_2$, $C_3$, and $NH_2$ generally produce fundamentally different phase-angle dependences of linear polarization. Since molecular, ionic, and atomic emissions contribute to all parts of the cometary spectrum, it is often difficult to identify the "uncontaminated" continuum and its intrinsic polarization. Solving this problem implies the need to perform spectropolarimetric observations with a high spectral resolution or, at least, simultaneous polarimetric and spectrophotometric observations. In addition, because of inherent scattering-geometry and brightness limitations in observations of comets, it is virtually impossible to determine the phase-angle and spectral dependence of polarization for a single comet in the entire phase-angle range $0° \lesssim \alpha \lesssim 130°$ accessible in ground-based observations. In practice, one usually derives a composite phase-angle dependence of polarization using observations of many comets.

In the final analysis, extensive polarimetric observations based on the use of filters separating continuum and molecular emissions have enabled planetary astrophysicists to derive a composite phase-angle dependence of continuum polarization for the entire range of phase angles $0°-130°$ and identify a number of specific characteristics of cometary polarization.

### *4.3.1. Negative polarization branch*

Our observations of comet West revealed a negative branch of linear polarization at phase angles $\alpha \lesssim 22°$ (Киселев и Чернова 1976, 1978; Нарижная и др. 1977). Subsequently, it has been demonstrated that this dependence is an inherent property of all comets and that the phase-angle dependence of polarization for cometary grains is similar to that for asteroids and interplanetary dust (Киселев 1981, 2003; Kiselev et al. 2002c; Kiselev and Rosenbush 2004). The corroboration of this empirical fact has been of utmost significance for the understanding of properties of cometary dust as well as of the nature of light scattering by dust particles covering or forming various types of Solar System bodies. The similarity of the polarization phase curves observed for different objects likely testifies to the similarity of the dust-particle morphologies. Figure 4.1 depicts the NPB observed for several comets. The data are approximated by the trigonometric expression

$$P(\alpha) = b\,(\sin\alpha)^{c_1}\,[\cos(\alpha/2)]^{c_2}\,\sin(\alpha-\alpha_0), \tag{4.2}$$

where $b$, $c_1$, $c_2$, and $\alpha_0$ are free parameters (Penttilä et al. 2005).

The linear polarization of light scattered by cometary dust is negative for backscattering geometries, $\alpha \lesssim 22°$, and positive at phase angles $\alpha \gtrsim 22°$. According to theory (Placzek 1934; Mrozowski 1936; Le Borgne and Crovisier 1987), the polarization of light emitted by diatomic molecules must be positive over the entire range of phase angles, the phase dependence of polarization $P(\alpha)$ being given by Eq. (4.1). Earlier observations of comets had confirmed this dependence for phase angles down to 40°. Our observations of comet Kopff (Chernova et al. 1993) showed that the polarization of light emitted by CN molecules is well described by Eq. (4.1) at phase angles down to 18°.

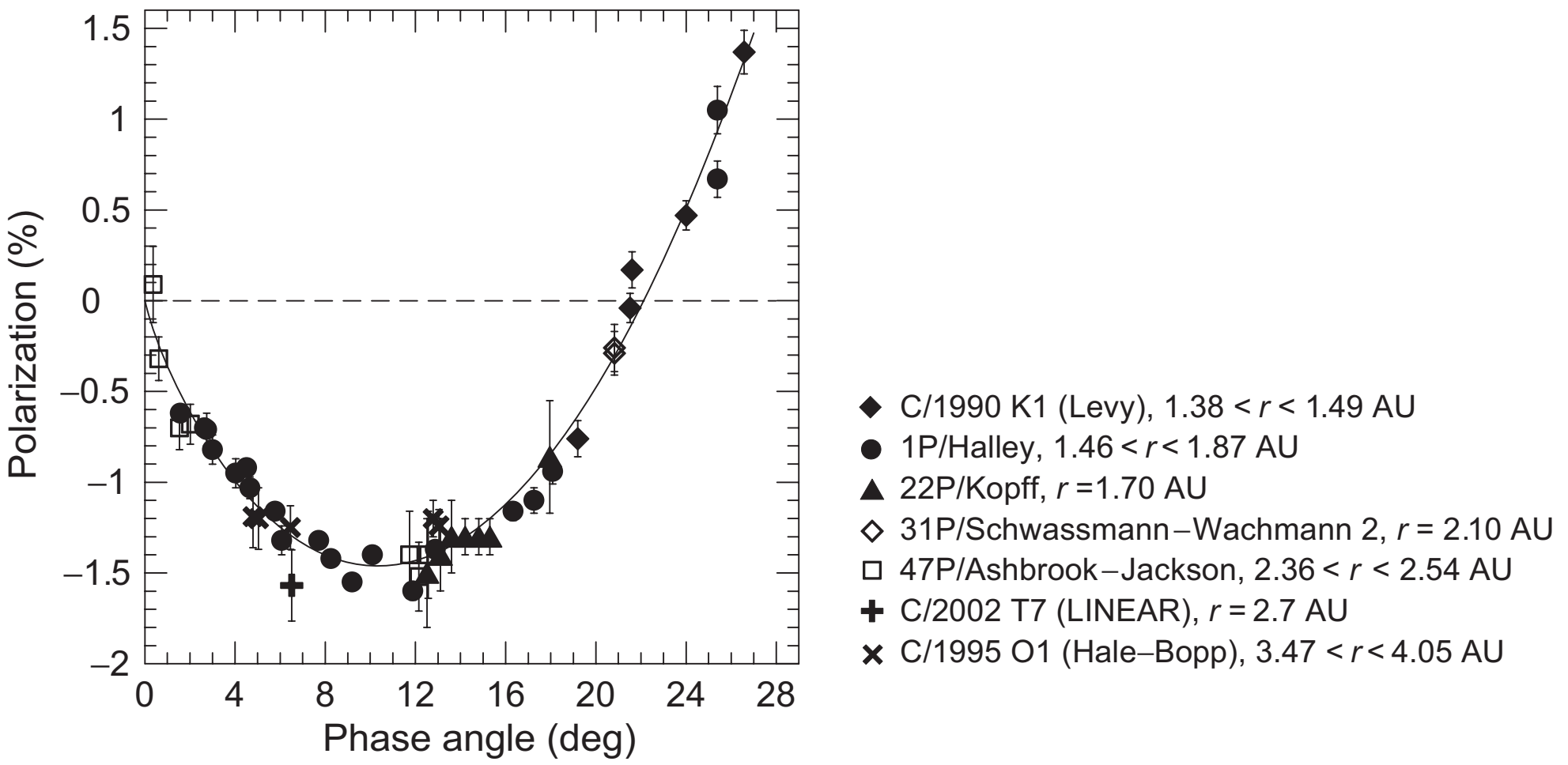

**Fig. 4.1.** NPB for comets at different heliocentric distances according to our observations.

### *4.3.2. Positive polarization branch*

We have already mentioned that Eq. (4.1) had often been used to approximate the results of polarimetric observations of comets in the continuum part of the spectrum, thereby implying that maximal polarization occurs at 90°. Our observations of comet Hyakutake and S4 (LINEAR) (Kiselev and Velichko 1998; Киселев и др. 2001a) as well as observations of comet Ikeya–Zhang (Velichko and Velichko 2002) at large phase angles allowed us to conclude that independently of the wavelength, the maximum of the cometary PPB occurs at $\alpha_{max} \approx 95°$. This value is substantially different from $\alpha_{max} \approx 110°$ for S-type asteroids and $\alpha_{max} \approx 70°$ for E-type asteroids (see Section 3.2.4).

In Fig. 4.2 we show the results of our polarimetric observations of comets with the narrow-band filters BC and RC adopted for the International Halley Watch program or with similar filters $\lambda$5300/50 Å, $\lambda$4430/44 Å, and $\lambda$6420/26 Å. The typical measurement errors are within the range 0.1%−1% depending on the brightness of a comet. One-sigma deviations of an individual phase curve from the average synthetic phase curve owing to measurement errors, diaphragm dependence, and varying dust properties can reach ~3% (Киселев 2003).

One can see that in the range of phase angles between ~80° and ~100°, the degree of linear polarization of the continuum is about 3 times greater that that of the CN, $C_2$, and $C_3$ cometary emissions. This result appears to suggest that all comets could be classified into two groups: (i) dust-rich comets exhibiting a strong continuum polarization, and (ii) gas-rich comets exhibiting a strong contribution from molecular emissions and a relatively weak contribution from continuum polariza-tion. This classification was first introduced by N. N. Kiselev (Киселев 1981; Добровольский и др. 1985; Dobrovolsky et al. 1986). Below we compare polarization properties of dust grains typical of these two groups of comets.

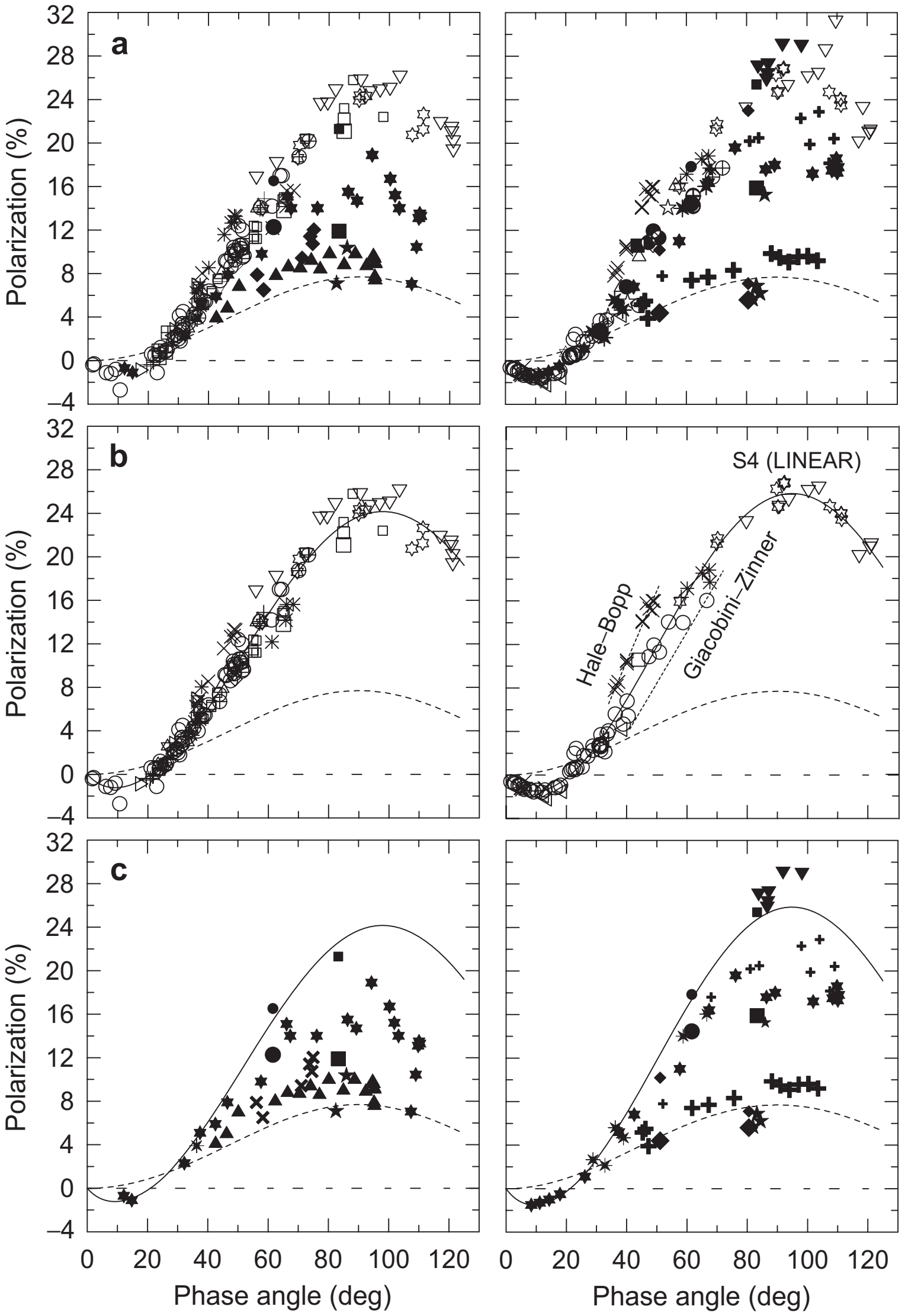

**Fig. 4.2.** Phase-angle dependence of polarization for comets in the blue (left-hand column) and red (right-hand column) parts of the spectral continuum: (a) all comets; (b) dust-rich comets; (c) gas-rich comets. Dust-rich comets: □ West; ◁ Churyumov–Gerasimenko; ▷ Kopff; ✵ Hartley–IRAS; ⊕ Giacobini–Zinner; ○ Halley; △ Bradfield (1987); ◇ Liller; + Levy; ✡ Faye; ✳ Shoemaker–Levy; ⊞ Ashbrook–Jackson; × Hale–Bopp; ✩ Hyakutake; ⊠ U5 (LINEAR); ▽ S4 (LINEAR). Gas-rich comets: ▲ Kobayashi–Berger–Milon; ✖ Crommelin; ★ Brorsen–Metcalf; ✚ Austin (1982); ▼ IRAS–Araki–Alcock; ✸ Austin (1989); ◆ Encke; ■ Tabur; ● J3 (LINEAR); ✷ A2 (LINEAR). The solid curve is the best fit for dust-rich comets; the short-dashed curve depicts theoretical results for diatomic molecules according to Eq. (4.1).

***Dust-rich comets***. The overwhelming majority of dust-rich comets display similar PPBs (see Fig. 4.2b) with $P_{max} \approx 25\%–28\%$. Therefore, one can derive an average polarization phase curve for this group by approximating the composite PDP with the trigonometric expression (4.2). The parameters of this average phase dependence are summarized in Table 4.2.

Figure 4.2b shows that comets Giacobini–Zinner, Hale–Bopp, and S4 (LINEAR) are notable exceptions. The degree of polarization measured for comet Hale–Bopp at phase angles 34°–49° is ~4% higher than that for any other comet observed so far (Kiselev and Velichko 1997; Hadamcik et al. 1997; Jockers et al. 1997). In contrast, comet Giacobini–Zinner exhibits a very low polarization at the same phase angles (Kiselev et al. 2000). A drastic increase of polarization was observed during a complete disruption of the nucleus of comet S4 (LINEAR) (Киселев и др. 2001a). The latter two comets belong to the group of "depleted" comets characterized by a low abundance of carbon and an anomalous spectral dependence of polarization. The existence of the three "outliers" in the group of dust-rich comets may be indicative of real differences in the physical properties of dust grains in various comets.

***Gas-rich comets.*** These objects reveal a significantly stronger scatter of data points than the dust-rich comets, so that $P_{max}$ varies between 5% and 20% (Fig. 4.2c). This factor precludes the construction of a single composite phase dependence of polarization for this group. It is important to note that significant polarization differences between the two groups of comets are observed at phase angles exceeding ~35°, i.e., where significant differences between the continuum polarization and the polarization of molecular emissions occur (Fig. 4.2c). The degree of linear polarization reaches a maximum at a phase angle ≈ 90°.

We will demonstrate below that both the strong scatter of data points at larger phase angles and the weaker contribution from the continuum polarization observed for gas-rich comets are caused by the residual contamination of the continuum filters by molecular emissions. However, it is still possible that this group of comets includes members with unusual dust properties.

### *4.3.3. Causes of observed polarimetric differences*

In our early publications (Киселев 1981; Добровольский и др. 1985; Dobrovolsky et al. 1986) we had emphasized that the separation of comets into two distinct groups is a consequence of two different optical mechanisms causing polarization as well as of the varying gas-to-dust ratio. These factors are especially relevant to gas-rich comets, for which the observed continuum contribution is usually rather weak. Arpigny (1995) and Brown et al. (1996) analyzed high-resolution CCD spectra of a number of comets, including Austin (1989) and Swift–Tuttle, and demonstrated the presence of numerous unidentified emissions. It is, therefore, possible that the weak continuum polarization typical of gas-rich comets is none other than a superposition of emission lines caused by minor chemical constituents. This finding provided a motivation for us to develop a technique to correct for the depolarizing effect of molecular emissions based on the observed polarization in two adjacent spectral intervals and on the dust-to-gas ratio estimated from photometric observations. This technique has been used to process the polarimetric data obtained for gas-rich comets Tabur, J3 (LINEAR), and Encke (Киселев и др. 2001b; Киселев 2003; Kiselev et al. 2004; Jockers et al. 2005). It has been demonstrated that the dust-to-gas ratio decreases significantly with distance from the nucleus. For example, the circumnuclear (<600 km) area of gas-rich comets Tabur and Encke is dominated by dust, whereas at nucleocentric distances >1000 km the dust-to-gas ratio is small, and the resulting polarization is dominated by the weakly polarized molecular emissions. This explains the observed substantial decrease of polarization with distance from the nucleus (i.e., with the increase of the fraction of the coma captured by the diaphragm in the focal plane of the telescope) (Kiselev and Rosenbush 2004).

**Table 4.2.** Average parameters of phase-angle dependences of polarization for dust-rich comets in three spectral bands located in the ultraviolet, blue, and red continuum

| Spectral band | $\alpha_{\min} \pm 1\sigma$ (deg) | $P_{\min} \pm 1\sigma$ (%) | $\alpha_{\text{inv}} \pm 1\sigma$ (deg) | $h \pm 1\sigma$ (%/deg) | $P_{\max} \pm 1\sigma$ (%) | $\alpha_{\max} \pm 1\sigma$ (deg) |
|---|---|---|---|---|---|---|
| UC | 12.2 ± 2.2 | −1.2 ± 0.3 | 25 ± 3 | | 20 | 92 |
| BC | 8.7 ± 1.0 | −1.42 ± 0.10 | 20.9 ± 0.7 | 0.26 ± 0.03 | 24 ± 1 | 96 ± 3 |
| RC | 9.6 ± 1.2 | −1.50 ± 0.10 | 21.8 ± 0.5 | 0.31 ± 0.03 | 26 ± 1 | 95 ± 3 |

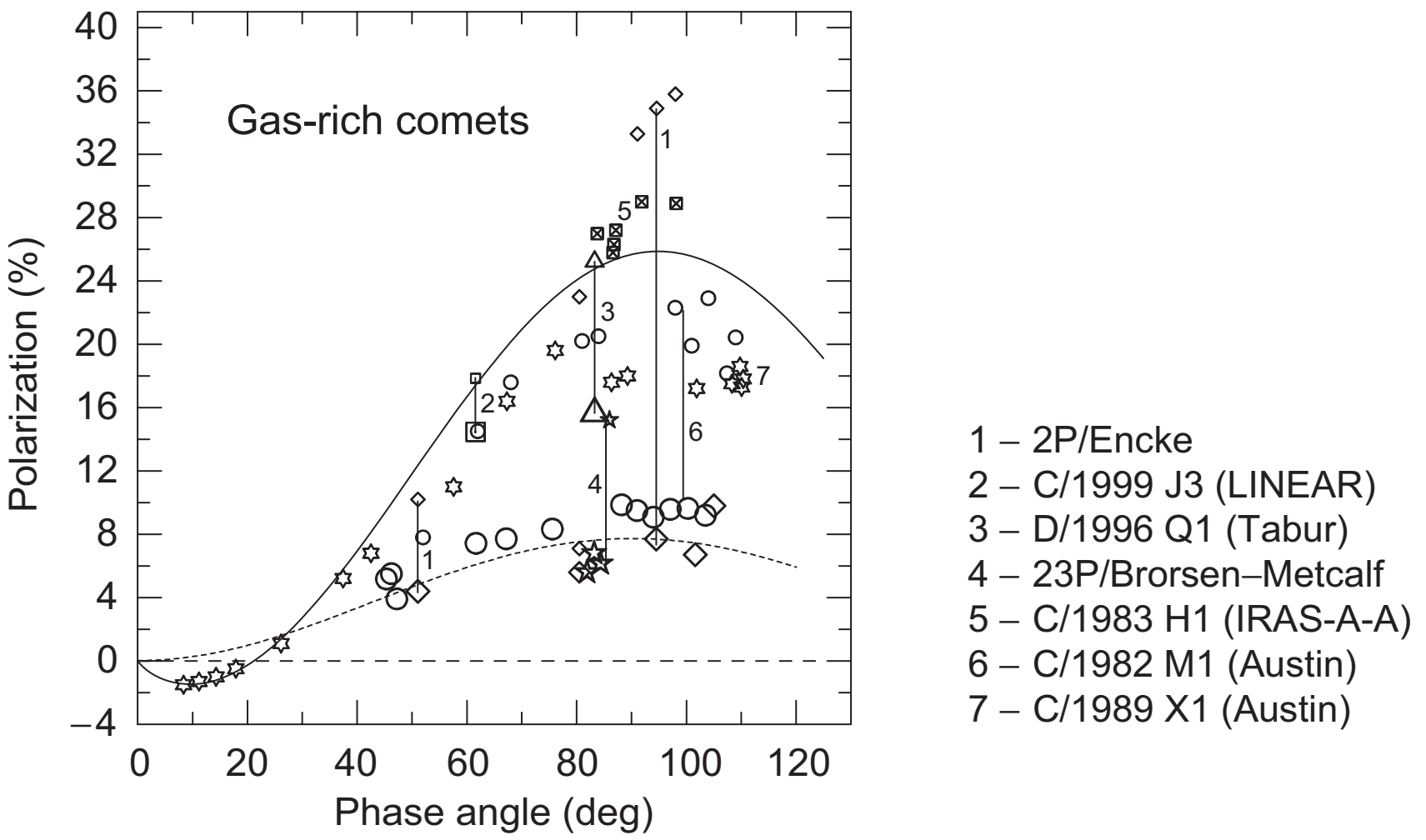

**Fig. 4.3.** Linear polarization versus phase angle for gas-rich comets observed with different diaphragms in the red spectral range is compared with theoretical results for diatomic molecules (short-dashed curve). The solid curve is the best-fit PDP for dust-rich comets. The polarization averaged over a large diaphragm for gas-rich comets (large symbols) is substantially smaller than that for dust-rich comets. The polarization for gas-rich comets measured in the vicinity of the coma and/or properly corrected for emission contaminations is significantly larger (small symbols). Vertical lines show the range of polarization change as the distance from the nucleus (i.e., the size of the diaphragm) is varied.

In the immediate vicinity of the nucleus, where the scattering by dust grains is dominant, the degree of polarization for gas-rich comets Tabur (~25%), Encke (~35%), and IRAS–Araki–Alcock (~30%) is comparable to (if not greater than) that for dust-rich comets, as Fig. 4.3 illustrates. Since comets are observed at different geocentric distances, the measured polarization often corresponds to varying fractions of the coma captured by the focal-plane diaphragm. It is this factor, in combination with significant variations of the dust-to-gas ratio with the nucleocentric distance, that explains low observed values of polarization and their substantial variations for gas-rich comets (see also Fig. 4.2c). In comparison with gas-rich comets, the dust production rate of dust-rich comets is greater by one or two orders of magnitude, thereby resulting in mighty dust atmospheres and weak variations of the dust-to-gas ratio with nucleocentric distance. The degree of polarization for these comets is high and is also nearly independent of the nucleocentric distance (see, e.g., Fig. 3 in Kiselev and Rosenbush 2004), the PDPs for different comets being very similar (see Fig. 4.2b).

It should be noted that the polarization caused by scattering on dust grains and that caused by resonance molecular emissions are nearly identical at phase angles between ~25° and ~30°. Therefore, the PDPs of different comets at these phase angles are very similar irrespective of the dust-to-gas ratio and nucleocentric distance.

Thus, the *apparent* separation of comets into two groups — dust-rich and gas-rich — based on polarimetric data at large phase angles is an artifact caused by the low spatial and spectral resolution of the instrumentation traditionally used in cometary observations. This conclusion, however, raises two new questions: (i) why are there two groups of comets with different distributions of dust and gas in their atmospheres, and (ii) how to reconcile the polarimetric and thermal properties of the cometary dust?

### 4.4. Similarity and diversity of comets: classification issues

The results of *in situ* studies of dust grains in comet Halley (Jessberger 1999), laboratory studies of interplanetary dust particles (Bradley 1994; Bradley et al. 1999), and recent findings in the IR spectroscopy of comets (Hanner and Bradley 2004) suggest that cometary dust may include particles from the protosolar nebula as well as pre-solar (interstellar) grains. Thus, the morphology and mineralogy of cometary dust particles can provide important information about the chemical and physical processes in the early Solar System. The radial gradients of temperature and chemical composition and the degree of mixing of materials from the warm interior and cold outer regions of the nebula at the time of comet formation may have caused differences in dust properties and chemical compositions of comets formed in different strata of the nebula. The main issue in the identification of common and individual traits of comets is the separation of the primordial properties of comets from the subsequent effects of their evolution. This explains the fundamental importance of the problem of cometary taxonomy.

Chernova et al. (1993) were the first to show that the observed separation of comets into two groups according to the polarization properties of cometary dust correlates with the separation of comets according to thermal properties of dust. Dust-rich comets (such as Halley, Bradfield (1987), Levy, Mueller, Hale–Bopp, and Hyakutake) exhibit strong excesses of color temperature relative to the equilibrium black-body temperature as well as silicate IR emissions. These traits could be interpreted in terms of the abundance of overheated micrometer-sized grains in cometary atmospheres, which would be consistent with the strong polarization ob-served in the visible and near-IR parts of the spectrum. For comets exhibiting a weaker continuum contribution (such as Kobayashi–Berger–Milon, Crommelin, Brorsen–Metcalf, IRAS–Araki–Alcock, and Encke) and a low degree of polarization measured over large parts of the coma, the color temperature excesses are weak or completely absent, which had been usually explained by the presence of large, $\gtrsim 5$ μm-radius, particles in their atmospheres.

While ignoring the effect of gaseous emission contaminations, some authors (e.g. Levasseur-Regourd et al. 1996) have concluded that the existence of two polarimetric classes of comets is most likely an indication of significant differences in the bulk properties (albedo, size distribution, porosity) of dust in different comets. More recently, Levasseur-Regourd and Hadamcik (2003) suggested that there are three classes of comets: (i) those with a low polarization maximum, from ~10% to ~15%; (ii) those with a higher maximum, from ~25% to ~30%; and (iii) comet Hale–Bopp. Thus, until quite recently there had been a widespread opinion that comets tend to form groups according to the value of maximal polarization (Levasseur-Regourd et al. 1996; Hanner 2003; Levasseur-Regourd and Hadamcik 2003). Indeed, the correlation between large maximal-polarization values, infrared color temperature, and silicate emissions observed for dust-rich comets and the absence of this correlation for gas-rich comets might be an indicator of real differences between the properties of dust typical of these two groups of comets.

However, we have already mentioned that there are gas-rich comets (such as Tabur, Encke, and IRAS–Araki–Alcock) with large dust particles (as evidenced by the absence of superheated grains and the lack of silicate features in the thermal IR) for which the observed degree of polarization is similar or even greater than that for dust-rich comets presumably characterized by much smaller (sub-micrometer) grain sizes. This raises the question of how to reconcile the polarization and thermal properties of dust in these gas-rich comets. To address this problem, Jockers et al. (2005) suggested that the strong polarization typical of both dust-rich comets and circumnuclear parts of gas-rich comets could be caused by the scattering of sunlight by large dust particles in the form of aggregates composed of sub-micrometer sized monomers (Петрова и др. 2004). The results of Kolokolova et al. (2007) show that increasing the size of aggregate particles from ~1 μm до ~$10^3$ μm is likely to increase drastically the relative contribution of fluffy aggregates to the scattered light and the emitted thermal radiation. This implies that particle porosity might be the decisive factor causing the separation of comets into two distinct groups. This conclusion would also be consistent with the different spatial distribution of dust in the dust-rich and gas-rich cometary atmospheres. Indeed, large compact aggregates presumably typical of gas-rich comets cannot be transported by radiation pressure far from the nucleus, whereas large fluffy aggregates presumably characteristic of dust-rich comets can be easily accelerated by radiation pressure and form extensive dusty atmospheres. This conclusion is consistent with the results of submillimeter observations (Jewitt and Matthews 1999) which revealed the presence of millimeter-sized particles in cometary atmospheres.

There is a significant correlation between the larger semi-axes of cometary orbits and the strengths of silicate emissions. The correlation coefficients for short-period and long-period comets are 0.751 and 0.662, respectively. This suggests that the observed differences between the two groups of comets are related to the surface properties of cometary nuclei which, in turn, are affected by the intensity and duration of solar insolation. Indeed, intense insolation over longer periods of time tends to make cometary particles more compact and facilitates the creation of crust on the surfaces of cometary nuclei.

In summary, the known polarization and thermal properties of cometary dust as well as the dynamical characteristics of comets suggest the following binary classification. Type I comets exhibit strong polarization of scattered light coming from the circumnuclear area, weak or no silicate emissions, and a concentration of compact aggregate dust particles in the vicinity of the nucleus. This class is populated by short-period comets which have experienced a relatively long exposure to intense solar insolation. Type II comets exhibit strong silicate emissions and strong polarization throughout their extended dusty atmospheres composed of fluffy aggregate particles. These are mostly "newcomers" which have spent relatively little time in the close vicinity of the Sun.

## 4.5. Spectral dependence of polarization

In the post-Halley era, when the use of narrowband cometary filters had become widespread, it was established (Kiselev et al. 2005) that the majority of the dust-rich comets exhibit an increase in the degree of polarization with wavelength at large phase angles, viz., 30°–80°, which means that the spectral gradient of polarization is positive (see, e.g., the data for comet Halley in Fig. 4.4). At longer wavelengths ($\lambda \geq 1$ μm), the gradient reverses its sign. An almost flat spectral dependence of polarization is observed in the vicinity of the inversion point (see data for $\alpha = 30°$ in Fig. 4.4).

In contrast to the PPB, the NPB exhibits an opposite spectral tendency (see the data for $\alpha = 10°$ and 17° in Fig. 4.4). The degree of polarization slightly decreases with wavelength in the visible, and then increases in the near-IR, the minimum occurring at 1.3–1.7 μm. Although the latter tendency is relatively weak, it makes the curves for $\alpha = 10°$ and 17° look differently from that for $\alpha = 30°$. A similar modest decrease of polarization with wavelength in the visible domain (0.484–0.684 μm) was observed by Ganesh et al. (1998) for comet Hale–Bopp at phase angles around 17°–19°.

However, this regular trait has exceptions. Figure 4.4 illustrates anomalous spectral dependences of polarization for comets Giacobini–Zinner and Schwassmann–Wachmann 3, with the spectral trend of polarization opposite to that usually observed for other comets. One can see that for the dust-rich comet Giacobini–Zinner, $\Delta P/\Delta\lambda \approx -1.46$ %/1000 Å (Kiselev et al. 2000). So far the atypical behavior of polarization has been found for six comets: Giacobini–Zinner, Austin (1989), S4 (LINEAR), Tempel 1, B and C nuclei of Schwassmann–Wachmann 3, and Holmes. These results do not appear to be observational artifacts and may justify the separation of such comets into a peculiar group with inherently unusual dust properties (Розенбуш 2006; Kiselev et al. 2008; Rosenbush et al. 2008b,c; Kolokolova et al. 2008).

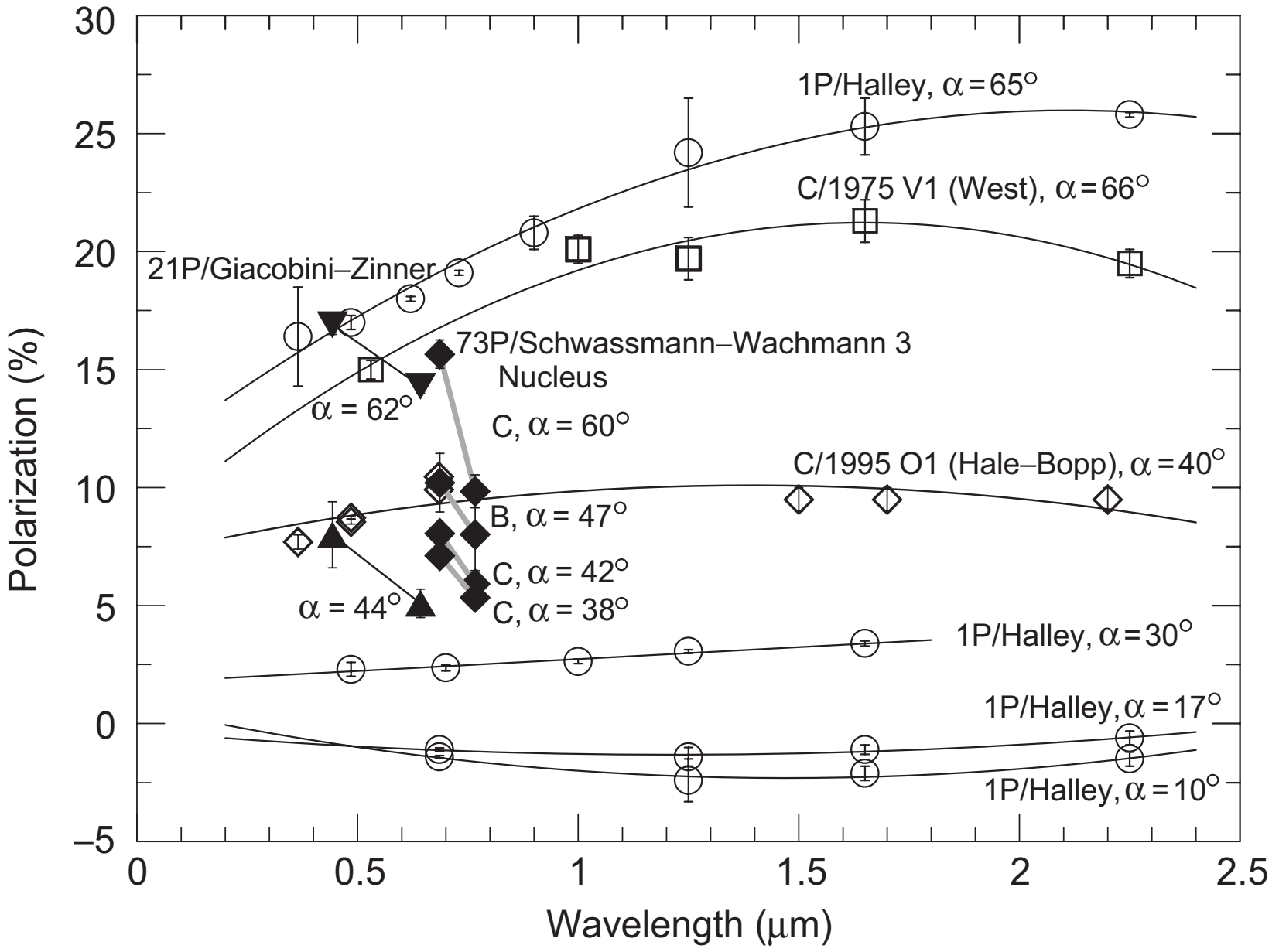

**Fig. 4.4.** Typical (West, Hale–Bopp, and Halley) and anomalous (Giacobini–Zinner, B and C subnuclei of Schwassmann–Wachmann 3) spectral dependences of polarization for comets at different phase angles. The curves are polynomial fits.

Spectral dependence of polarization is an important indicator of the composition and size of scattering particles (Kolokolova and Jockers 1997; Petrova et al. 2000, 2004) since it is primarily controlled by the spectral dependence of the imaginary part of the refractive index of the materials constituting cometary dust. In particular, as Kolokolova and Jockers showed, in the visible spectral range astronomical silicates cause a growth of polarization with wavelength, whereas organic materials cause a negative gradient of polarization. The size distribution of particles is another important factor. Indeed, Petrova et al. (2000) showed that the negative spectral gradient of polarization in comets similar to Giacobini–Zinner could be explained either by a stronger negative spectral gradient of Im($m$) than that for other comets or by a specific particle size distribution.

One may wonder if the specific properties of cometary dust result from the particular place of origin of a comet or from its specific evolutionary path. To address this question, we have summarized the results of observations of the anomalous spectral dependence of polarization for comets Giacobini–Zinner, S4 (LINEAR), Schwassmann–Wachmann 3, Tempel 1, Austin (1989), and Holmes and compared relevant physical and dynamical characteristics of the comets. We have found that the spectral gradients of polarization $\Delta P/\Delta\lambda$ for these comets are in the range from $-0.7$ to $-2.8$ %/1000Å. This result suggests indeed that these comets may form a separate group, although the specific nature of their individual peculiarity (e.g., the type of constituent organic materials) may be different for different comets.

Some, but not all, of the listed comets are characterized by a low abundance of carbon in the gas phase. This is consistent with the conclusion by A'Hearn et al. (1995) that not all Jupiter-family comets are carbon depleted. In our case, the exception is comet Tempel 1. On the other hand, some long-period comets originating in the Oort Cloud, are classified as "carbon depleted", comet S4 (LINEAR) being the only dynamically new member of this group. Farnham et al. (2001) suggested that comet S4 (LINEAR) had originated in the Kuiper Belt but then was gravitationally scattered, probably by Neptune, into the Oort Cloud. According to Mumma et al. (2001), the composition of this comet differs substantially from that of other long-period comets formed in the giant-planet region, e.g., comets Hyakutake and Hale–Bopp. Comets Austin and S4 (LINEAR) are CO depleted. The deficit of CO ices in these comets may imply that they had originated in the warmer region of the protosolar nebula, i.e., in the Jupiter–Saturn region. Note that the low abundance of volatile carbon-chain molecules does not necessarily mean that other carbon-bearing molecules are absent (Weaver et al. 1999) since some carbon may be part of organic grains (Hanner and Bradley 2004).

It thus appears that the majority of comets exhibiting a negative spectral gradient of polarization in the visible belong to the Jupiter family. The Jupiter-family comets are characterized by both compositional peculiarities (A'Hearn et al. 1995) and the dominance of larger dust particles (Lisse 2002; Sitko et al. 2004). However, the existing observational data are insufficient in order to determine which of these factors is primarily responsible for the negative polarization gradient. Thus, the definitive clarification of the "place-of-origin vs. evolution" dilemma necessitates further systematic spectropolarimetric observations of Jupiter-family comets.

### 4.6. Polarimetry of comets during stellar occultations

Comet Levy presented a unique opportunity to observe the occultation of a star by a cometary coma. We could measure, for the first time, the parameters of linear polarization of the star during its visible movement across the coma (Rosenbush et al. 1994). Figure 4.5a explains the geometric configuration and shows the light-curve of the star as a function of time. One can see that a substantial attenuation of the starlight occurred only at three observation points (i.e., points 2−4) corresponding to minimal observed angular distances between the star and the cometary nucleus. At these moments the degree of polarization of the total light from the star and the comet was $P_{\rm sum} = -0.5\%$, $-0.7\%$, and $-0.4\%$, respectively, within $\pm 0.1\%$ errors. The corresponding angles between the polarization plane and the scattering plane were $\vartheta_{\rm sum} = 6.5°$, $14.5°$, and $179.5°$. The intrinsic polarization parameters of this BD +11°4576 star are $P_{\rm s} = 0.080\% \pm 0.063\%$ and $\vartheta_{\rm s} = 62.6° \pm 19.1°$. According to our data, the optical thickness of the cometary coma along the line of sight at a 3500-km distance from the nucleus was $0.40 \pm 0.06$.

The total light from the star and the comet collected by the telescope consists of the exponentially attenuated starlight and the sunlight scattered by the cometary atmosphere towards the telescope. Let us consider the factors that may affect the resulting degree of polarization:

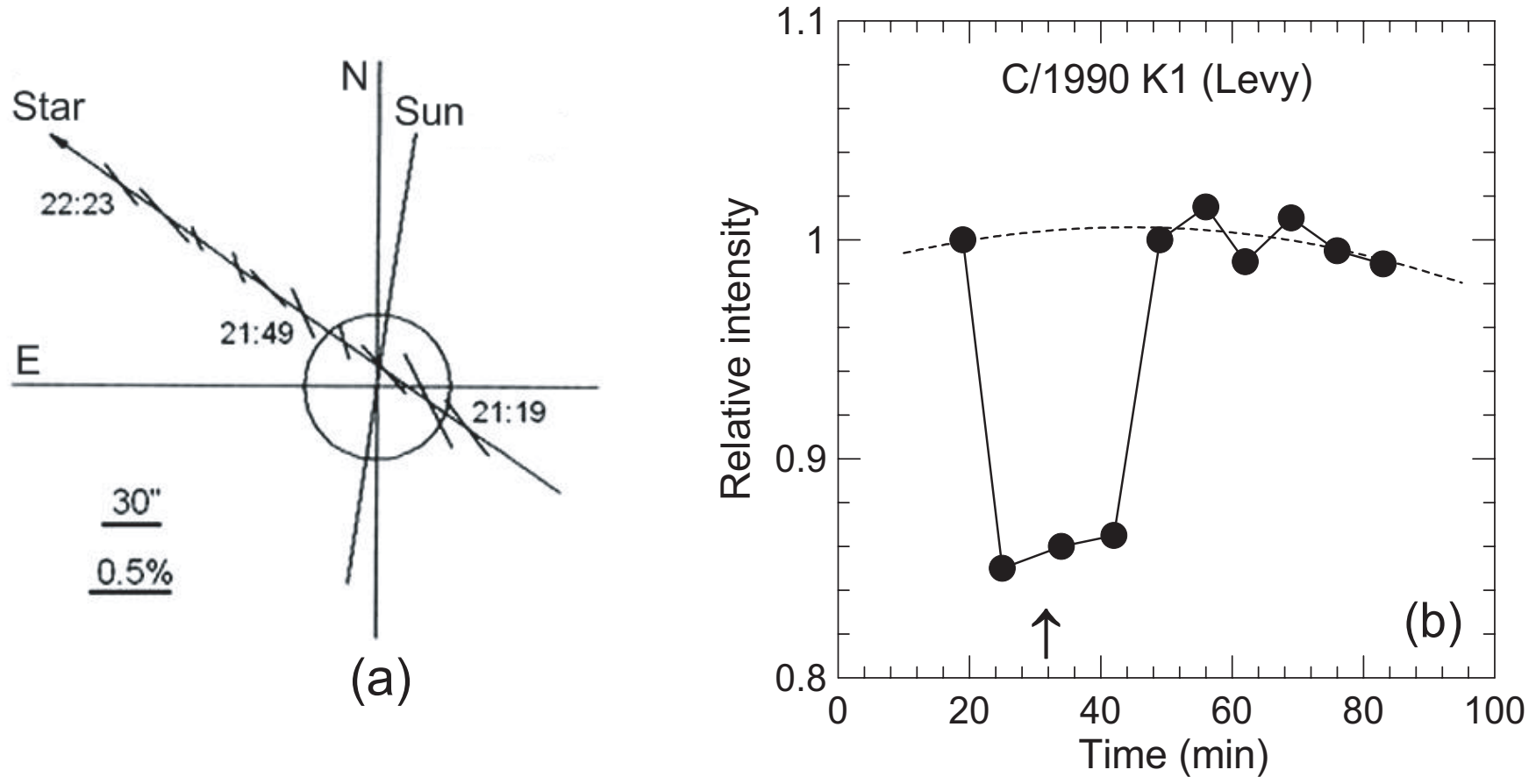

**Fig. 4.5.** (a) Visible movement of the star during the occultation of BD +11°4576 on 18 August 1990. The cometary nucleus is placed at the origin of the coordinate system. The circle represents the region of the coma in which substantial extinction of starlight was detected. The bars represent the degree of polarization and the orientation of the polarization plane. The scales are provided for the angular distance of the star from the nucleus (in arc seconds) and *P* (in %). (b) The measured intensity as a function of time.

- The cometary dust is a mirror-symmetric ensemble of randomly oriented particles and exhibits no dichroism (Sections 1.5 and 1.15). In this case the polarization of the total "star + comet" light will depend on the exponential attenuation of the starlight by the coma only if the intrinsic polarization of the star is substantial. This obviously follows from Eqs. (1.103) and (1.106) upon neglecting the second term on the right-hand side of the latter.
- Cometary dust particles are nonspherical and preferentially oriented (Section 1.14). In this case the polarization of the total light will depend on both the intrinsic polarization of the star and the extinction matrix of the dust according to Eq. (1.104).

Based on the known intrinsic brightness and polarization of the star and the occultation measurement results, we have concluded that the likely cause of the considerable modification of the polarization of the starlight is the presence of nonspherical particles partially aligned along the radial direction with respect to the nucleus.

This conclusion was corroborated by the observation of the occultation of star HD 211005 by the extensive dust coma of comet Hale–Bopp on 9 March 1997. Five polarimetric and three photometric observations of the star + coma combination and the corresponding areas of the coma without the star were performed with the 2.6-m and 1.25-m CrAO telescopes (Rosenbush et al. 1997b; Розенбуш и др. 1999). The smallest observed distance between the star and the photometric nucleus of the comet was about $10^5$ km. A theoretical analysis of the polarization data measured for the star + coma combination allowed us to estimate the upper limit of the optical thickness of the dust coma along the line of sight at $0.3 \pm 0.1$. This value was confirmed by the simultaneous photometric observations. The degree of polari-zation calculated for the coma alone proved to be slightly lower (by 0.2% on average) than that measured during the occultation. This implies that the initially unpolarized star radiation became partially polarized upon its passage through the coma containing partially oriented nonspherical grains.

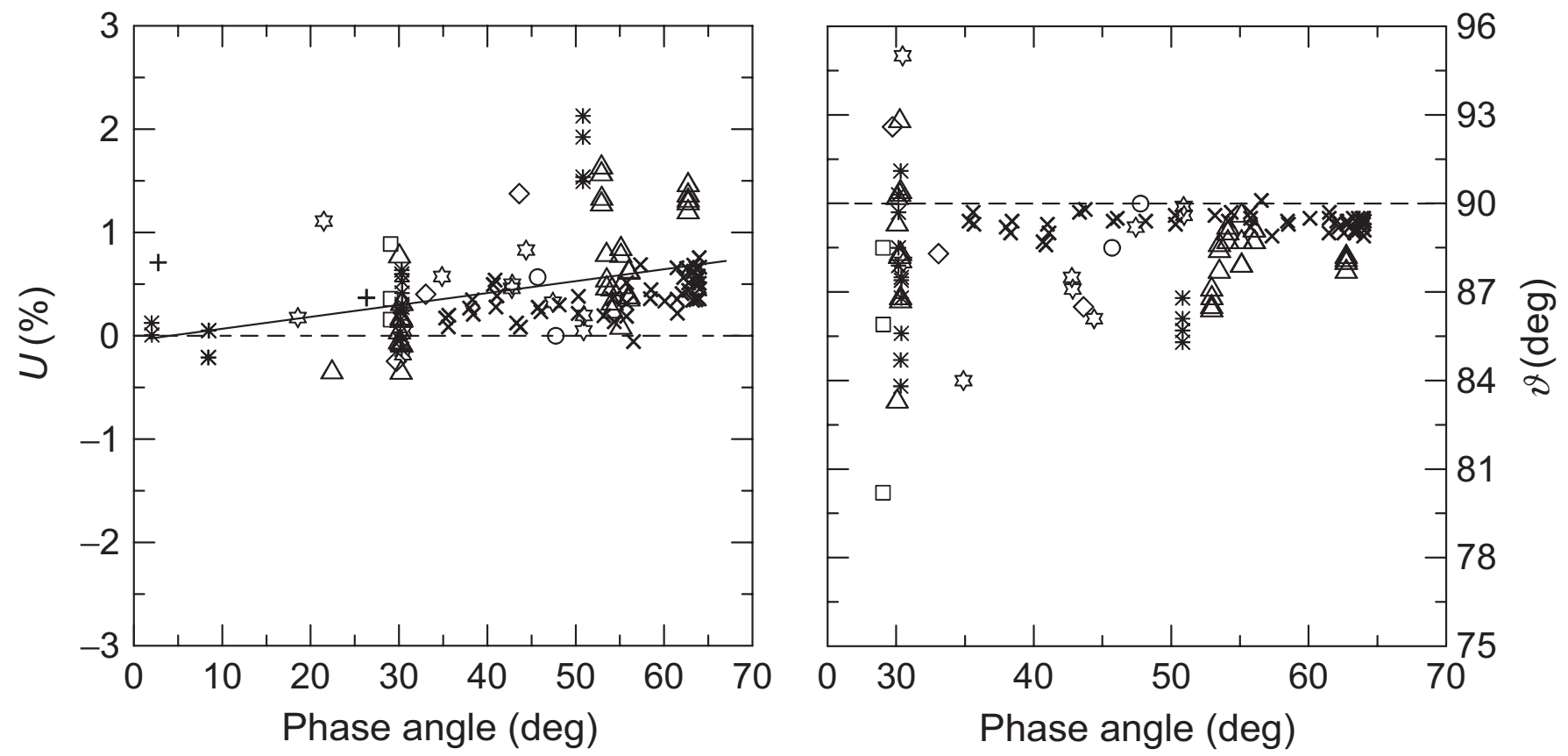

**Fig. 4.6.** Systematic deviations of the Stokes parameter $U$ from zero (left-hand panel) and deflections of the polarization plane from the normal to the scattering plane (right-hand panel) for comet Halley. Typical measurement errors for $\vartheta$, with the exception of phase angles $\alpha \approx 30°$, are between 0.1° and 1°.

## 4.7. Systematic deviations of polarization parameters from "nominal" values

The polarization plane is usually either perpendicular or parallel to the scattering plane, depending on the phase angle of a comet. However, there are indications based on aperture and surface polarimetry (see, e.g., Clarke 1974; Dollfus and Suchail 1987) of potentially significant deflections of the polarization plane from these orthogonal directions. This implies that the third Stokes parameter $U = IP_{\text{lp}} \sin 2\vartheta$ defined with respect to the scattering plane is not always equal to zero.

Unusual systematic deviations of the Stokes parameter $U$ from zero and the position angle of the polarization plane $\vartheta$ from 0° and 90° with respect to the scattering plane have been found for comet Halley. Our as well as all other available measurements for the entire observed range of phase angles in the red continuum filter RC with $P > 3\sigma_P$ (A'Hearn and Vanysek 1992) are summarized in Fig. 4.6. One can see that the average deviations are 0.48% ± 0.04 % for $U$ and −1.4° ± 0.2° for $\vartheta$. Furthermore, $U$ exhibits a nontrivial phase-angle dependence. Dollfus and Suchail (1987) did a similar study and found 0.2% ± 0.1% average deviations for $U$ and –1.9° ± 1.4° average deviations for $\vartheta$, although they could not exclude a systematic error in the position angle of the instrumental polarization plane as a possible cause of the deviations.

Systematic deviations of $\vartheta$ from 90° have also been found in the results of our polarimetric observations of different parts of the coma of comet Hale–Bopp with different filters (Kiselev and Velichko 1997; Розенбуш 2006). The value of $\vartheta$ measured for the shells was found to be greater than that measured between the shells by 1°–1.5°. During the same observation period, we detected, with high accuracy, significant circular polarization which also was stronger for the shells than for areas between the shells (see Section 4.8). Both results, coupled with the occurrence of strong dust jets facilitating particle alignment, are strong indicators of the presence of partially oriented nonspherical particles.

## 4.8. Circular polarization

Until the mid 1980s, measurements of circular polarization for comets had been highly irregular and had errors comparable to the measured polarization values. This explains the lack of efforts in the study of optical mechanisms causing circular polarization and mechanisms of grain alignment in cometary atmospheres.

So far circular polarization has been detected reliably only for seven comets: Halley, Hale–Bopp, S4 (LINEAR), Q4 (NEAT), Schwassmann–Wachmann 3, Tuttle, and Tempel 1 (Мороженко и др. 1987; Dollfus and Suchail 1987; Rosenbush et al. 1997b; Rosenbush et al. 2007a,b; Rosenbush et al. 2008a,b; Tozzi et al. 2007). The first five of them have been observed by these authors with the 2.6-m CrAO telescope. Our objectives have been not only to obtain accurate circular-polarization data across cometary comas but also to stimulate relevant theoretical research (Rosenbush et al. 2007a).

### *4.8.1. Measurement results*

The first high-accuracy measurements of circular polarization across a coma (see Fig. 4.7a), including the nucleus, shells, regions between shells, and tail, were done for comet Hale–Bopp by Rosenbush et al. (1997b). Only light with left-handed circular polarization (i.e., polarized in the counter-clockwise sense when looking in the direction of light propagation; see Section 2.2) with a minimal $P_{cp}$ value of $-0.26 \pm 0.02\%$ was detected for all observed areas of the coma on 11 March 1997. The polarization was close to zero ($|P_{cp}|$ smaller than 0.08%) in the vicinity of the nucleus. Spatial variations of up to 0.1% in the degree of circular polarization were detected. The region of stronger polarization coincided with shells. The weakest polarization was measured between the shells. Manset and Bastien (2000) performed similar measurements. The absolute value of $P_{cp}$ was found to be at the same level, ~0.3%, although they detected both left- and right-handed circular polarization during the period 2–16 April 1997. It should be noted that in general, the spatial distribution of circular polarization over the coma correlated with the distributions of the linear polarization and surface brightness.

The first measurements of the spatial distribution of circular polarization for a splitting comet were reported by Rosenbush et al. (2007a). During the observation period from 28 June to 22 July 2000, comet S4 (LINEAR) was in a very active state and its nucleus underwent multiple fragmentations culminating in a complete disruption around 20 July. The measured circular polarization was found to be rather strong and reached, on average, ~0.8% (in the absolute sense). Spatial variations along the sections through the coma and nucleus in the solar and tail directions up to distances of 6000 km were studied. As an example, the results obtained on 21 July 2000 are depicted in Fig. 4.7b. Although the absolute value of the degree of circular polarization across the coma sometimes reached 1%, that in the vicinity of the cometary nucleus was almost always close to zero. Left-handed as well as right-handed circular polarization was observed across the coma; the former was systematically found in the sunward part. Immediately after the complete disintegration of the nucleus, polarization in both directions of the section was mainly left-handed and substantially deviated from zero: its average absolute value was 0.41% ± 0.07%, thereby significantly exceeding $3\sigma$ (Fig. 4.7b).

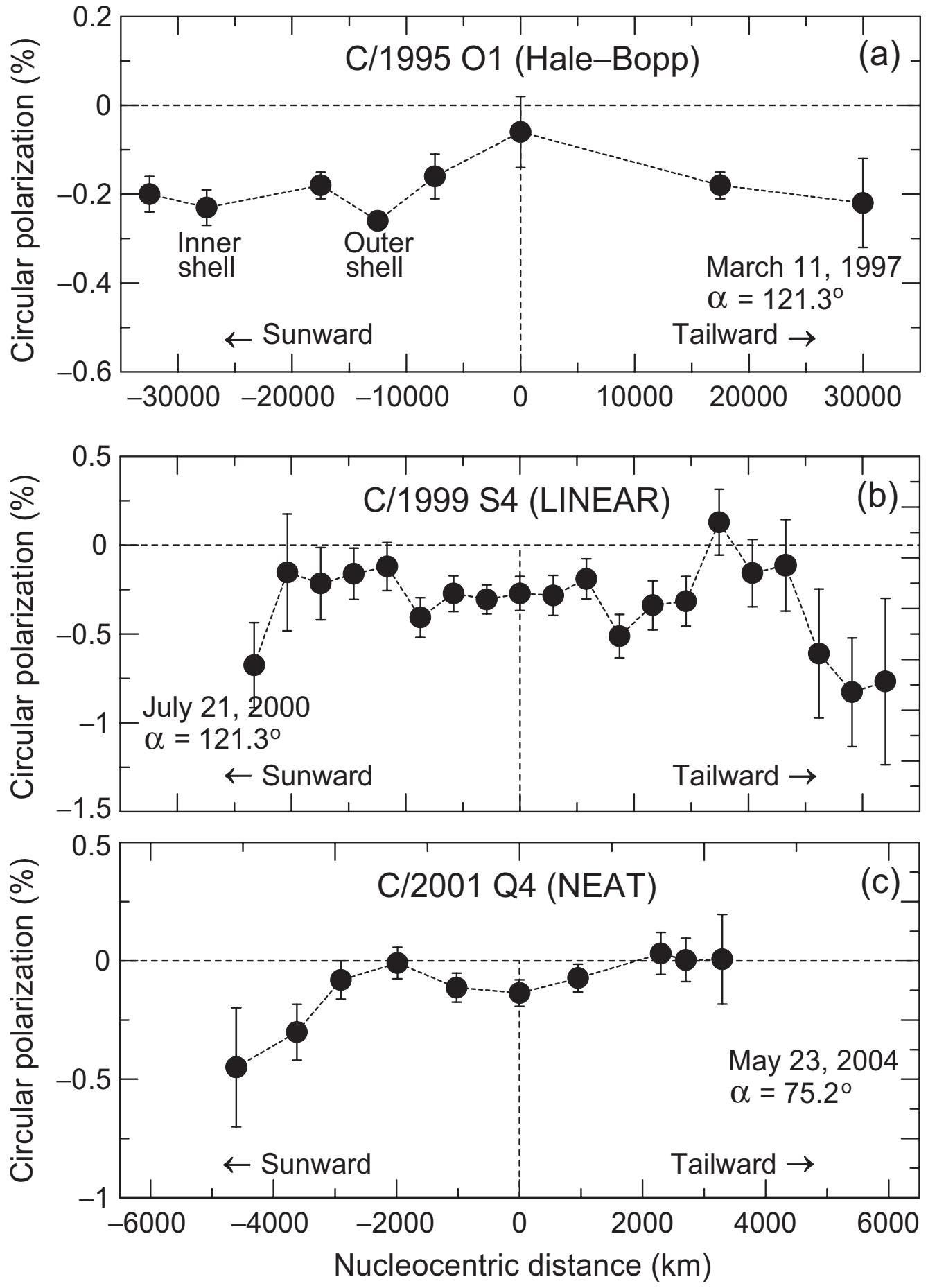

**Fig. 4.7.** Variations of the degree of circular polarization along the line through the coma and nucleus of comets Hale–Bopp (a), S4 (LINEAR) (b), and Q4 (NEAT) (c). The symbols in panel (b) show the running average fit to the data. Error bars indicate the standard error of the mean value of polarization at each point of the section.

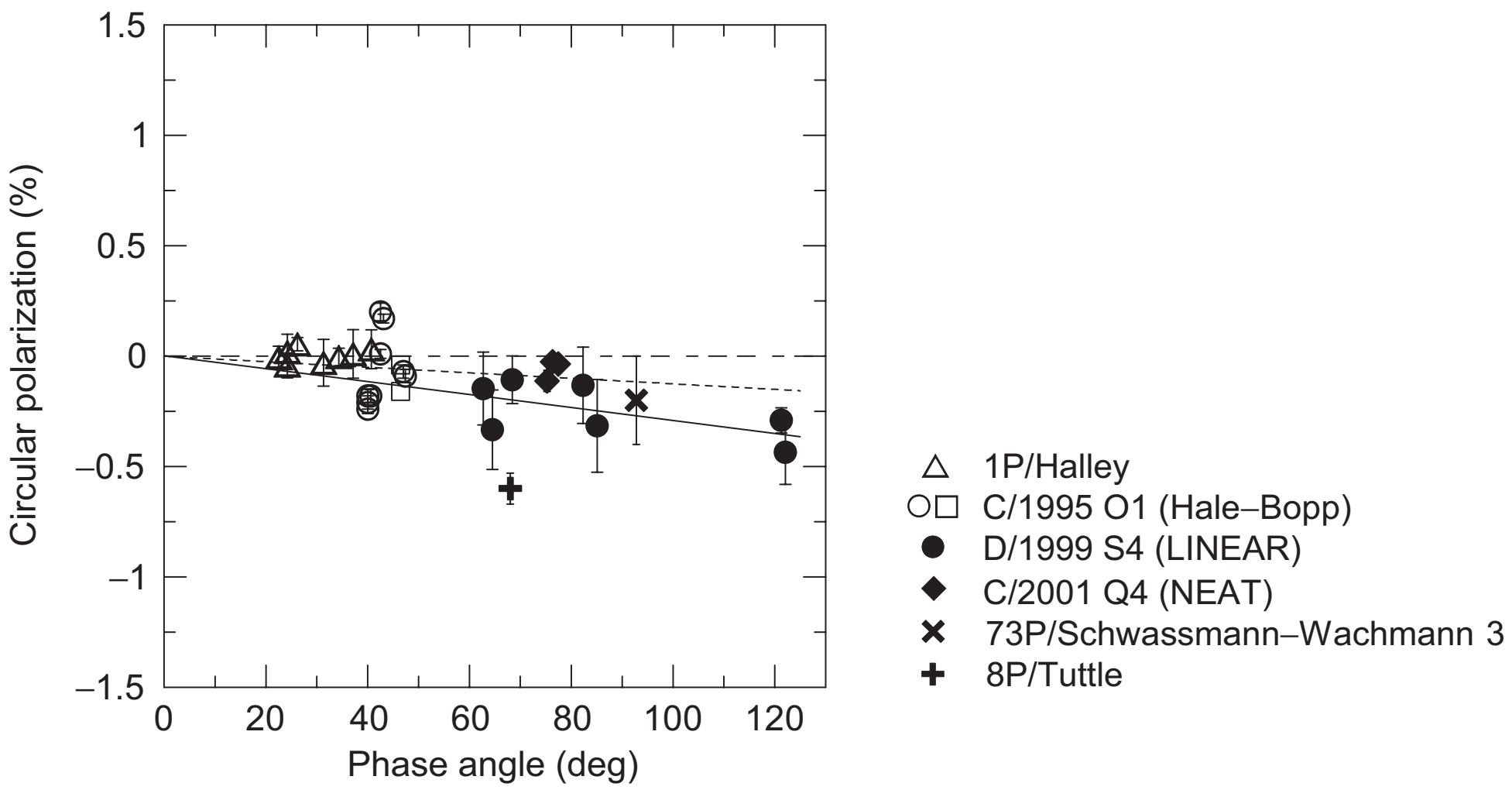

**Fig. 4.8.** Composite phase-angle dependence of circular polarization for comets Halley (Dollfus and Suchail 1987), Hale–Bopp (Rosenbush et al. 1997b; Manset and Bastien 2000), S4 (LINEAR) (Rosenbush et al. 2007a), Q4 (NEAT) (Rosenbush et al. 2007b), Schwassmann–Wachmann 3 (Tozzi et al. 2007), and Tuttle (Rosenbush et al. 2008b). The solid line is a linear fit to the observed data, while the short-dashed line depicts the results of calculations for optically active (chiral) spherical particles (Rosenbush et al. 2007a).

During these observations, the phase angle of the comet varied from 61° up to 122°, which allowed us to study variations of circular polarization with phase angle. A significant correlation between the degree of circular polarization, visual magnitude, water production rate, and linear polarization during the final fragmentation of comet S4 (LINEAR) in July 2000 was identified.

Simultaneous measurements of circular (Fig. 4.7c) and linear polarization along sections through the coma and nucleus and along the dust jet of comet Q4 (NEAT) were performed on 21–23 May 2004 at phase angles 77°–75° (Rosenbush et al. 2007b). There is a significant correlation between the circular polarization and the changes of parameters of linear polarization, which indicates that there is a substantial component of polarization that is not related to the scattering plane and can be explained by the inhomogeneity or anisotropy of the dusty medium in which particles are partially aligned.

Simultaneous measurements of circular and linear polarization for comet Tuttle were also performed along sections trough the coma on 10 January 2008 (Rosenbush et al. 2008b). The observed mean degree of circular polarization, –0.60% ± 0.07%, was relatively strong.

The results of all existing measurements are summarized in Fig. 4.8. They demonstrate significant circular polarization of light scattered by comets and reveal a systematic decreasing trend at phase angles from ~0° to ~120°. Furthermore, the observed circular polarization is predominantly left-handed.

### *4.8.2. Optical effects causing circular polarization*

Two general optical effects (or their combination) which may cause the circular polarization of light observed for comets have been discussed by Bandermann and Kemp (1973), Долгинов и Митрофанов (1975), Dolginov and Mytrophanov (1976), Beskrovnaja et al. (1987), and Dolginov et al. (1995). They are as follows:

- single scattering by dust grains which do not form a mirror-symmetric ensemble of randomly oriented particles (see Sections 1.14 and 1.15); and
- multiple scattering of light in any particulate medium.

Indeed, in the former case, the $(4,1)$ element of the phase matrix $\mathbf{Z}$ (see Eq. (1.20)) and that of the scattering matrix (1.27) do not vanish in general. In the latter case, the $(4,1)$ element of the diffuse reflection matrix describing the cumulative contribution of multiple scattering to the outgoing radiation is non-zero. It does not vanish even for the simplest model of a homogeneous plane-parallel slab (Hansen 1971), except in some very special cases of illumination and observation (Hovenier and de Haan 1985).

***Multiple scattering.*** In this case, neither special morphology nor special composition of particles is required in order to cause circular polarization. However, a substantial optical thickness of the dusty medium is necessary for multiple scattering to have a noticeable effect (Hansen and Travis 1974; Mishchenko et al. 2006b).

The presence of an optically thick atmosphere is doubtful for most comets (see, e.g., Jewitt 1989; Rosenbush et al. 1997b). Among the comets for which circular polarization has been observed, in particular Halley, Hale–Bopp, and S4 (LINEAR), the largest $\tau$ apparently existed for the coma of comet Hale–Bopp. Indeed, observations of a stellar occultation were indicative of optical thicknesses $\tau > 1$ at distances $\lesssim$100 km from the nucleus at ~3 AU (Weaver and Lamy 1997; Fernández et al. 1999). Analogous observations for the same comet by Rosenbush et al. (1997b) at 1.39 AU showed $\tau \approx 0.3$ at the distance ~$10^5$ km from the nucleus. *In situ* observations of jets in comet Halley (Keller et al. 1987) were indicative of the dust optical thickness ~0.28 at a distance of 6500 km, while Dollfus and Suchail (1987) visually estimated $\tau$ to be ~0.92 at a distance of ~500 km from the nucleus. Although a non-zero −0.3% circular polarization has been measured for the circumnuclear area of comet Hale–Bopp (Rosenbush et al. 1997b; Manset and Bastien 2000) as well as for comet Halley, in both cases the data on circular polarization and optical thickness were obtained for different areas of the comae, which makes it difficult to assess the likely effect of multiple scattering. There are no direct measurements of the dust optical thickness for comet S4 (LINEAR). However, since the radius of its nucleus was small, about 0.44 km (Farnham et al. 2001), it is difficult to expect a large optical thickness for its coma.

Most importantly, the results of representative, numerically exact computations of radiative transfer reported by Kawata (1978) suggest that one needs extreme values of the optical thickness to generate circular polarization values merely approaching 0.1% (in the absolute sense). Thus, it is highly unlikely that multiple scattering is a significant contributor to the substantial circular polarization measured for comets.

***Scattering by aligned nonspherical particles.*** The most frequently discussed optical mechanism potentially causing the observed circular polarization is light scattering by aligned nonspherical particles (see, e.g., Dolginov et al. 1995; Lazarian 2003; and references therein). The existence of oriented particles in cometary atmospheres was unambiguously revealed by our observations of star occultations by the comae of comets Levy and Hale–Bopp (Rosenbush et al. 1994, 1997b). An additional implicit indicator is the systematic deflection of the polarization plane from the scattering plane observed for comets Halley and Hale–Bopp (Розенбуш 2005, 2006). Obviously, a strong dynamical constraint on a candidate alignment mechanism is that it be extremely efficient. Indeed, a substantial degree of circular polarization was observed at distances of only 500−1000 km from the nucleus of comet S4 (LINEAR). Dust particles with velocities of hundreds of meters per second (Farnham et al. 2001) need just $\sim 3 \times 10^3$ s to reach these areas.

The most common explanation of grain alignment refers to the effect of paramagnetic relaxation first discussed by Davis and Greenstein (1951). However, the corresponding relaxation rates appear to be excessively low to make this effect particularly relevant to cometary atmospheres.

Two other factors that can contribute to the alignment of dust particles in cometary atmospheres are the relative gas–grain motion causing mechanical alignment and radiation torques. Outflowing gas may play an important role in the vicinity of a cometary nucleus, i.e., at distances not exceeding ~20 nucleus radii. In the outer parts of the coma and in the tail, the alignment caused by radiation torques and interactions with the solar wind is likely to be more important. The solar radiation tends to orient the longer axes of particles perpendicularly to the direction of the radiation flux, whereas collisions with gas atoms and molecules orient the longer axes parallel to the flow (Lazarian 2003). In other words, the mechanical orientation and that by the solar radiation are, in a sense, competing effects. Therefore, the residual orientation direction depends on the relative strengths of the two orientation effects and on the particle composition, size, and morphology. A detailed review of various alignment mechanisms can be found in Lazarian (2003) and Rosenbush et al. (2007a). Below we will discuss only those likely relevant to comet S4 (LINEAR).

The mechanical alignment mechanism can orient grains on time scales shorter than the gaseous damping time. This suggests that one may see mechanically aligned grains at distances shorter than $\sim 10^3$ km from the nucleus. In the case of comet S4 (LINEAR), the area of the coma where the interaction of the dust particles with the radially outflowing gas was most effective is approximately 50 km.

The radiation-torque alignment mechanism in application to comets was first discussed by Dolginov and Mytrophanov (1976). According to Eq. (2.6), the unpolarized incident sunlight can be represented as an incoherent superposition of fully righ- and left-hand polarized components. The difference in the particle extinction cross sections for the light components with different signs of circular polarization results in different amounts of left- and right-polarized light “extracted” from the incident beam. As a result of the recoil arising from the differential extinction, a nonspherical dust grain can become spun up. Dolginov and Mytrophanov suggested that special shapes of particles might be necessary for this mechanism to be effec-tive. However, numerical calculations by Draine and Weingartner (1996) showed that it might be sufficient for the grains to be irregular. The application of the radiation-torque alignment mechanism to comets requires certain additional considerations (Dolginov and Mytrophanov 1976; Lazarian 2003) such as the account of internal grain wobbling induced by thermal fluctuations within the grain material (Lazarian 1994).

To evaluate the efficiency of radiation torques in the case of comet S4 (LINEAR), we have performed numerical calculations for an irregular dust particle made of an astronomical silicate and considered the particle to be a ballistic aggregate consisting of 128 monomers each having a 100-nm radius (Rosenbush et al. 2007a). Our results show that in the specific comet environment, the radiation-torque mechanism can easily "outperform" the gas damping mechanism owing to the low density of gas. As a consequence, particles can become nearly completely aligned at distances $\sim 10^3$ km from the nucleus. This value is consistent with the distance at which substantial circular polarization was observed.

The efficiency of the radiation-torque mechanism increases rapidly with grain size (Cho and Lazarian 2005). However, it takes longer for larger grains to become aligned since their alignment time depends on the gas damping time, which is longer for larger particles. This may compensate for the enhanced efficiency of radiation torques, and so one might expect larger grains to be aligned farther away from the nucleus.

Lazarian (2003) concluded that the radiation-torque mechanism should be dominant in a typical comet environment. However, the joint action of two or more mechanisms might result in interesting and not yet explored transient alignment processes that could influence the circular-polarization patterns observed. Furthermore, if the alignment is produced only by the radiation torque, then the distribution of particle orientations is axially symmetric around the direction of the solar radiation flux. Since the solar light is unpolarized, the singly scattered light would have zero circular polarization. Therefore, an interplay of different alignment factors (such as radiation torques, gas outflows, and ambient magnetic and electrostatic fields; e.g., Neubauer 1987; Liu 1999; Hoang and Lazarian, personal communication) is necessary to explain the observed circular polarization in terms of preferentially oriented nonspherical particles.

There exists a variety of calculations of light scattering by aligned nonspherical particles (see, e.g., Beskrovnaja et al. 1987; Li and Greenberg 1997; Gledhill and McCall 2000). One can see that by varying the size, composition, and shape of the particles, it is possible to obtain a broad range of values of circular polarization, including those measured for comets Hale–Bopp, S4 (LINEAR), and Q4 (NEAT). However, further computations for realistic models of cometary dust grains are obviously required.

***Scattering by optically active particles.*** Circular polarization can also be caused by single scattering on preferentially or randomly oriented nonspherical and even spherical particles made of an optically active material with different refractive indexes for right- and left-handed circularly polarized light (see Section 1.14). Indeed, as a consequence of optical activity, the $(4, 1)$ elements of the extinction, phase, and scattering matrices become non-zero, thereby resulting in circular polarization of the singly-scattered light as well as in circular dichroism. Optical activity can be especially pronounced for organic molecules (Wolstencroft and Raven 2002).

Optical activity is caused by chirality (mirror asymmetry) of molecules (Barron 2004) and is a well known property of complex organic molecules existing in two forms: L (left-handed) and D (right-handed). They are identical in all respects except that they are "twisted" in opposite directions, looking like mirror images of each other. Non-living systems usually contain equal numbers of the L- and D-enantiomers of such molecules. However, the characteristic property of terrestrial bio-organic molecules is that their amino acids are left-handed, whereas the sugars are right-handed. This property is called "homochirality".

For a long time it had been believed that homochirality has terrestrial origin. However, more recently a significant excess of left-handed amino acids was found in carbonaceous materials of the Murchison and Murrey meteorites (Cronin and Pizzarello 1997; Pizzarello and Cooper 2001; Pizzarello 2004; and references therein), thereby suggesting its origin in the presolar nebula. This new paradigm is supported by the observation of high values of circular polarization in star-forming regions (Hough et al. 2001) and the subsequent discussion of its possible relation to fundamental problems of astrobiology and homochirality of biomolecules (Hough et al. 2001; Bailey 2000; Bonner and Bean 2000; Meierhenrich and Thiemann 2004; Nuevo et al. 2007; Sparks et al. 2009). It has been shown that the illumination of cosmic organics by circularly polarized light in protoplanetary nebulae may provide a viable mechanism for chiral selection of organic molecules. In that case enantiomeric excess in organics should be found not only in meteorites but also in other primitive bodies, including comets. The search for such an excess in cometary organics *in situ* is one of the goals of the Rosetta mission (Thiemann and Meierhenrich 2001).

Chiral organics may also be detectable remotely by studying circular polarization of light scattered by cometary dust. The existing observations of comets summarized in Fig. 4.8 reveal substantial values of circular polarization, a systematic phase-angle trend, and the dominance of left-handed polarization. Rosenbush et al. (2007b) assumed that the cometary particles contain a ~10% excess of chiral organics, and, as a result, are optically active. Their simulations of light scattering were based on the exact theoretical solution for optically active spheres (Bohren and Huffman 1983), the optical constants typical of the amino acids found in the Murchison meteorite (Cronin and Pizzarello 1986), and the power-law size distribution measured *in situ* for comet Halley (McDonnell et al. 1987). Rosenbush et al. were able to reproduce the observed phase-angle trend of circular polarization, albeit the modeled circular polarization did not reach absolute values greater than 0.15% at phase angles ~120°. One should, of course, expect that more realistic models of cometary dust particles in the form of aggregates of submicrometer-sized optically active grains would yield significantly stronger circular polarization.

Thus, the modeling results coupled with the spectroscopic detection of organically enriched chemical composition of comets (e.g., comet S4 (LINEAR); Mumma et al. 2001) allow us to put forth a reasonable hypothesis that the observed circular polarization implies the presence of prebiotic homochiral organics in comets (Rosenbush et al. 2007b, 2008a). This hypothesis is supported by the predominance of left-handed polarization for all seven comets observed in circularly polarized light.

The complete understanding of cometary circular polarization and its variability requires further research. The detection of the same sign of circular polarization for large parts of the coma and a common phase-angle dependence testify to the existence of a predominant circular polarization. On the other hand, the observed changes of the sign across the coma obviously indicate the presence of a locally variable circular polarization component.

### 4.9. Effects of non-steady processes on polarization

Brightness outbursts, sudden increases of dust outflows from the nucleus, and partial fragmentation or complete disintegration of a comet should be accompanied by the ejection of fresh material and, therefore, provide an opportunity to study the internal composition of the nucleus and the physical processes occurring in the atmosphere of a splitting comet. In particular, non-steady processes in comets can be expected to cause changes in the observed polarization parameters.

***Linear polarization.*** Temporal variations of the degree of linear polarization and the orientation of the polarization plane were registered during periods of outburst activity in comets Schwassmann–Wachmann 1 (Киселев и Чернова 1979) and Levy (Rosenbush et al. 1994). Dollfus et al. (1988) observed transient variations of linear polarization for comet Halley. Our own observations of comet Halley revealed temporary increases of linear polarization at phase angles 50°–65°. After the period of activity had ended, linear polarization returned to its normal value. The temporary deviations of polarization from normal values have been interpreted in terms of an increased contribution of small dust particles to the light scattered by a cometary atmosphere.

Kiselev et al. (2002a) performed a comparative analysis of polarimetric properties of dust in split and normal (non-split) comets. Comets West, Brooks 2, Liller, Tabur, S4 (LINEAR), and A2 (LINEAR) can be classified as split comets. Comets West, S4 (LINEAR) and A2 (LINEAR) were observed directly during the process of disintegration. The analysis showed that partial fragmentation of comets had not resulted in a systematic change of its polarization phase dependence. There was no obvious evidence of significant polarization differences between tidally split comets (e.g., Brooks 2), dissipating comets (e.g., Tabur), non-tidally split comets (e.g., West), and normal comets.

Comet S4 (LINEAR) provided a rare example of the total disintegration of a cometary nucleus. The fragmentation of this comet led to a dramatic increase of linear polarization, by ~4% (see Fig. 4.2), an anomalous spectral dependence of polarization, and a very high, >20 %/1000Å, color gradient (Киселев и др. 2001a; Bonev et al. 2002). The polarization plane was perpendicular to the scattering plane both before and after the disintegration. Note that until the very final disintegration stage, the phase-angle dependence of polarization showed no differences from the composite phase curve for dust-rich comets (see Fig. 4.2). Apparently the main cause of changes in the optical properties of dust at the last stage of disintegration was the ejection of small dust particles from the internal parts of the nucleus which resulted in the abundance of scatterers with atypical composition, morphology, and size distribution (Bonev et al. 2002).

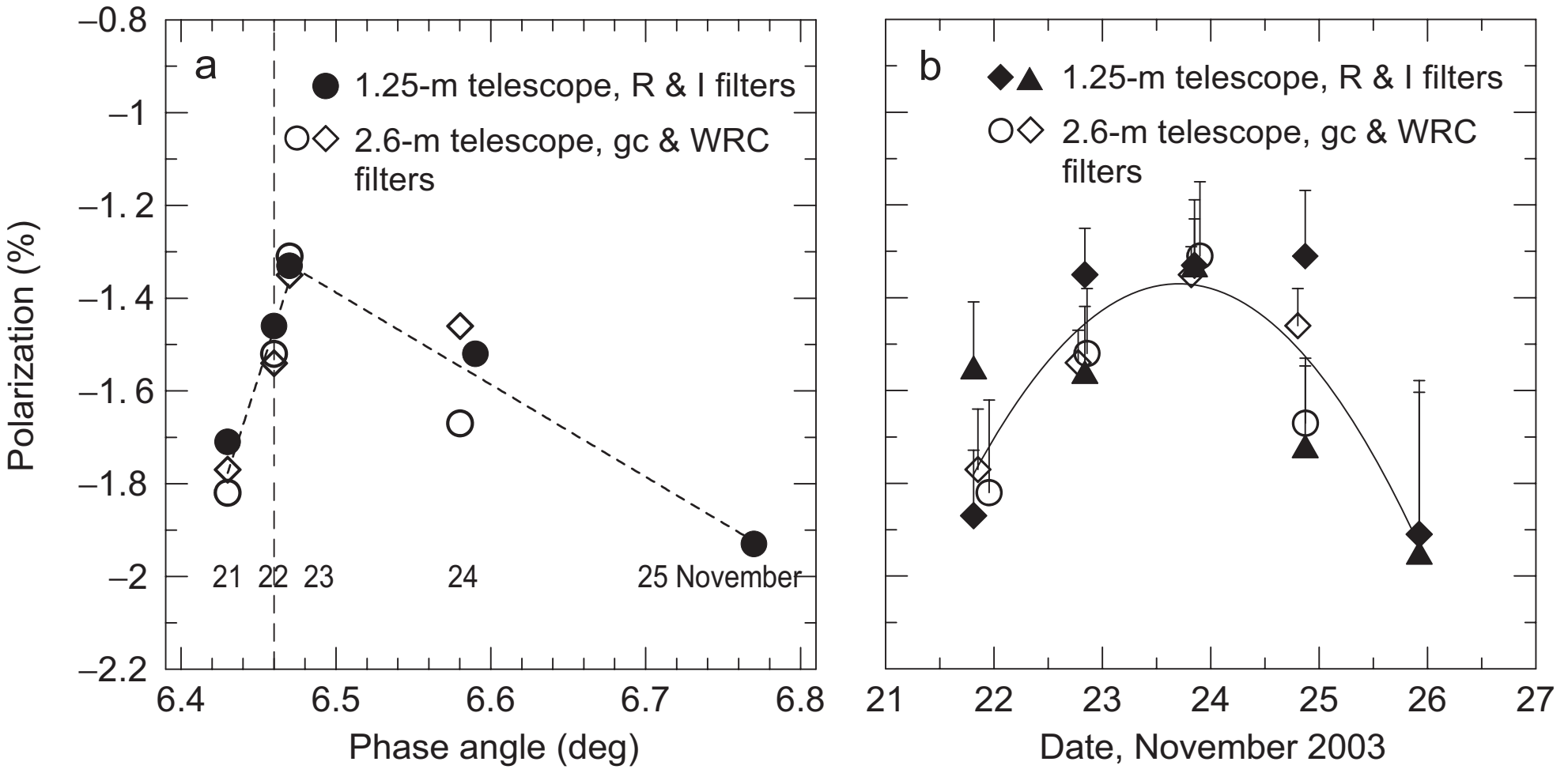

**Fig. 4.9.** Changes in the degree of polarization for comet T7 (LINEAR) within a small range of phase angles (a) and in time (b) observed at the 1.25-m and 2.6-m CrAO telescopes with R, I, GC, and RWC filters.

The split comet A2 (LINEAR) exhibited short-term variations of polarization on the time scale of a day (Rosenbush et al. 2002b). Inexplicable 90° rotations of the polarization plane and decreases in the degree of polarization (by up to 6%) were registered. This comet was observed in the active phase and displayed a series of sporadic outbursts due to the disruption of its nucleus. One cannot rule out short-time changes of dust properties owing to the separation of small short-lived splinters that could not be directly observed.

During polarimetric observations on 21–25 November 2003 (Розенбуш и др. 2006), the dynamically new comet T7 (LINEAR) showed significant variations in brightness (up to ~1.5$^m$) and in the degree of polarization (up to ~0.5%) in a very narrow phase angle range, approximately from 6.4° to 6.8° (Fig. 4.9). The comet was very active, although it was at a rather large distance (2.7 AU) from the Sun. Again, the sharp variations of polarization were probably caused by unusual physical properties of the dust newly ejected from inside the nucleus. In spite of the transient variability of the degree of polarization, the parameters of the phase-angle dependence of polarization were close to the average values typical of dusty comets: $P_{\min} = -1.63\%$, $\alpha_{\min} = 10.6°$, $\alpha_{\rm inv} = 22.7°$, and $h = 0.24$ %/deg.

***Circular polarization.*** Comets S4 (LINEAR) and Q4 (NEAT) exhibited short-term changes of circular polarization (Rosenbush et al. 2007a,b). During our observations, comet S4 (LINEAR) was an actively outgassing and highly fluctuating object which, while approaching the Sun, displayed sporadic outbursts caused by a partial and then a complete disintegration. As a result of the disintegration, the internal parts of the nucleus were exposed, the dust and gas production rates increased, and fresh dust particles were ejected into the cometary atmosphere. Moreover, the dust particles lifted from the nucleus were continually losing their volatiles and fragmenting, thereby changing their properties (sizes, composition, and morphology) with distance from the nucleus (Bonev et al. 2002). As a consequence, there were significant variations in the visible brightness and color. The outbursts mostly influenced the tailward side (Bonev et al. 2002), whereas the coma in the sunward direction was essentially unaffected. It is remarkable that temporal variations of circular polarization coincided in time with the outbursts caused by the fragmentation of the nucleus.

A comparison of the measured circular polarization with the magnitude of the comet in the visible (which is sensitive to the dust and gas production rates), the water production rate, and the degree of linear polarization is presented in Fig. 4.10. The circular polarization essentially follows the trend of the linear polarization. One can see that during the observation period, there were several large outbursts. The strongest one occurred around July 6–8, 16–18, and 22–23, with a number of weaker ones in between. These outbursts correlate with the intensified water production (Mäkinen et al. 2001): between July 18 and 23 the visible brightness of the comet increased by $\sim 1^m$, while the $H_2O$ production rate increased by a factor of $\sim 10$ (Bockelee-Morvan et al. 2001). It is during this period that the ultimate breakup of the comet occurred. Thus, there is a correlation between the time of fragmentation, the increase in brightness, and the gas and dust production rates.

The disintegration of the nucleus caused changes in the phase-angle dependence of polarization. Linear polarization sharply increased after the complete disintegration (Киселев и др. 2001a; Hadamcik and Levasseur-Regourd 2003). It is hard to expect a strong correlation between the variability of cometary brightness and the degree of polarization because the visible magnitude of a comet strongly depends on $C_2$ emissions rather than on only the strength of the continuum (Hadamcik and Levasseur-Regourd 2009). However, there was a consistency between the variations of linear polarization and the unfolding cometary activity. The measurements of circular polarization also support this view: one can see from Fig. 4.10 that the degree of circular polarization strongly correlated with that of linear polarization.

## 4.10. Comets with unique polarimetric properties

Based on photometric and spectroscopic investigations of comets carried out until the 1980s, there had been a common belief that most comets are rather similar in terms of their chemical composition, although their relative gas and dust production rates can vary widely. More recently, however, A'Hearn et al. (1995) introduced two principal taxonomical classes of comets by dividing them into (i) those with the "normal" composition, and (ii) those with a substantial deficit of carbon-chain species ($C_2$ and $C_3$). A'Hearn et al. also suggested that the taxonomic classes should be directly related to the origin of comets. According to A'Hearn et al., the majority of comets with origin in the Kuiper Belt are carbon depleted, have rela-tively small active areas of their nuclei, and are characterized by a significant asymmetry of their characteristics (such as brightness, production rates, etc.) with respect to the perihelion. Smaller differences can be found between comets with origin in the Oort Cloud, i.e., new or long-period comets. A'Hearn et al. had con-cluded that differences in the chemical composition of comets are largely caused by their origin rather than by the subsequent evolution.

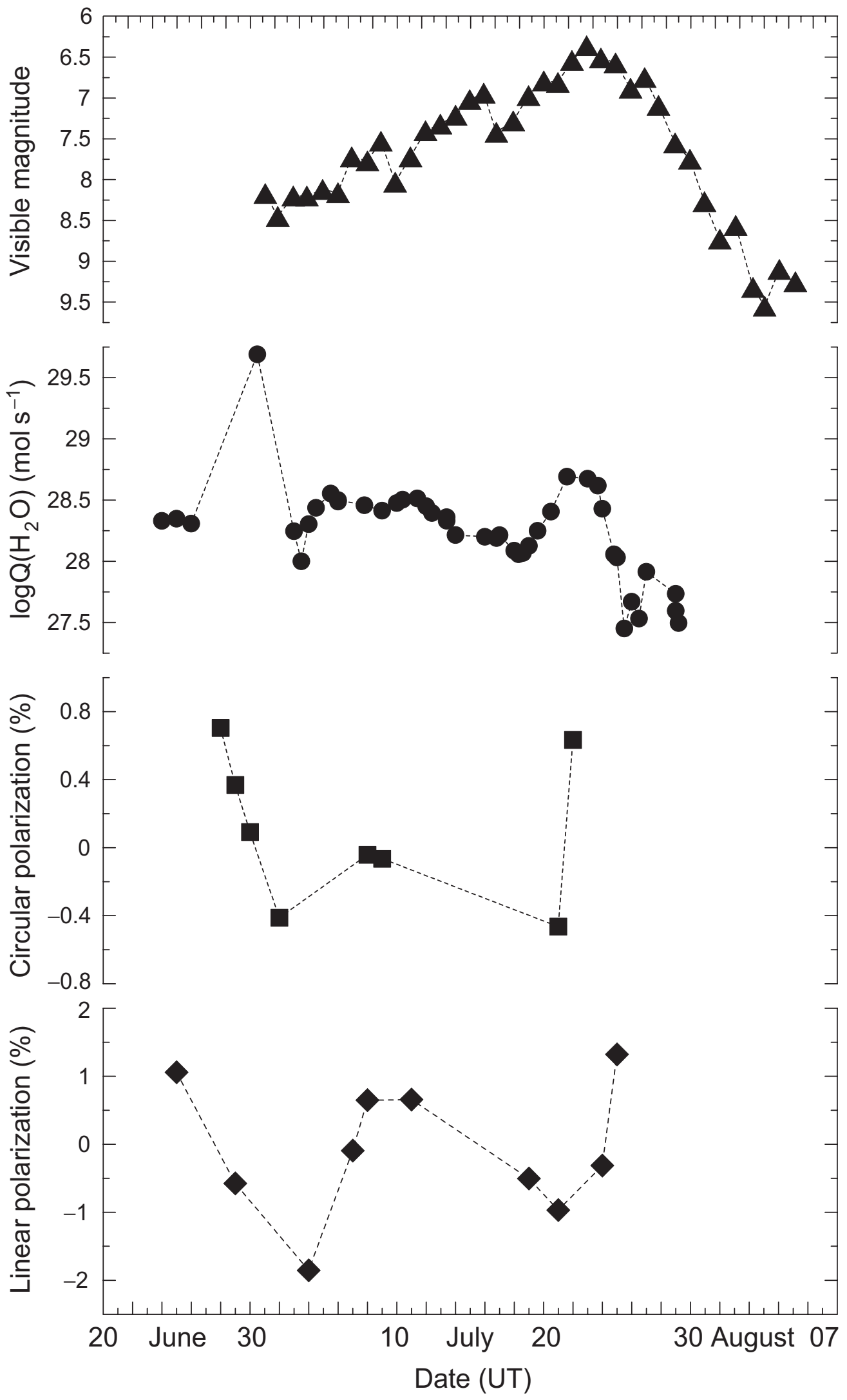

**Fig. 4.10.** Comparison of the degree of circular polarization with the visual magnitude (ICQ 2001), water production rate (Mäkinen et al. 2001; Bockelee-Morvan et al. 2001; Farnham et al. 2001), and the degree of linear polarization during the final fragmentation of comet S4 (LINEAR) in July 2000. Linear polarization is corrected for the regular phase-angle dependence.

Polarimetry can provide valuable information about the composition and size distribution of dust particles and, therefore, can complement spectrophotometric data in many respects. In particular, it has been demonstrated above that properties of cometary dust can serve as a basis for comet classification. Indeed, analyses of the observed angular and spectral characteristics of polarization for comets revealed specific regularities (see the reviews in Kolokolova et al. 2004, 2007; Levasseur-Regourd and Hadamcik 2003; Kiselev and Rosenbush 2004), including systematic polarization differences between dust-rich and gas-rich comets (Chernova et al. 1993; Levasseur-Regourd et al. 1996). Polarimetry can also be used to identify comets with unique dust properties. Among such objects are Encke, Giacobini–Zinner, Hale–Bopp, S4 (LINEAR), Schwassmann–Wachmann 3, Holmes, and Liller and Tabur (the latter two comets are debris of a single parent body).

In fact, we have found an anomalous (negative) spectral dependence of polarization for Giacobini–Zinner, S4 (LINEAR), subnuclei B and C of the split comet Schwassmann–Wachmann 3, and Holmes. All these comets have an atypical organic composition, a deficit of carbon species in the gas phase, and a dust population dominated by relatively large particles. Therefore, they may form a group with a particular origin and/or evolutionary path.

The specific interest in the gas-rich comet Tabur and the dust-rich comet Liller stems from the fact that they used to be parts of a single comet. Comet Tabur was discovered on 19 August 1996 and passed through its perihelion at 0.840 AU on 3 November 1996. Within two months of the discovery, the brightness of the comet rose in a typical way to $5.5^m$ but then began to decline catastrophically and fell by $4^m$ between 20 and 30 October. The comet was lost in early December. Since no flares and nucleus breakups were detected, it is reasonable to assume that the sudden disappearance of comet Tabur was caused by a complete deactivation of its nucleus. Kawakita et al. (1997) found that Liller was much more dusty than Tabur, which appears to suggest that the nucleus of the parent comet was highly heterogeneous. Analyses of polarimetric observations of the "sister" comets Liller (Chernova et al. 1993) and Tabur (Киселев и др. 2001b) have allowed us to conclude that, to within the measurement errors, there is no difference between the polarization properties of dust in the near-nucleus region of the gas-rich Tabur and that of the extremely dust-rich Liller (as well as other dusty comets).

Below we discuss in more detail the results of our observations of four comets demonstrating a wide variety of polarimetric properties. Analyses of similarities of and differences between the "typical" comets and comets with peculiar inherent properties of their dust should yield valuable insights into the origin and evolution of comets.

***Comet 2P/Encke.*** This is short-period comet with a period of 3.28 years and the smallest known perihelion distance of 0.33 AU, which suggests that Encke is probably the most evolved comet currently known. It does not display a dust tail extending in the anti-solar direction, but rather has a so-called "fan" in the form of a broad feature oriented at an angle with respect to the straight line through the comet and the Sun (Plate 4.1). The comet is rich in gas (according to Newburn and Spinrad (1989), its dust-to-gas mass ratio is < 0.1) and has weak silicate features (Gehrz et al. 1989) and large dust particles (Reach et al. 2000). Normally such traits would put this comet in the low-polarization class. However, the results of narrow-band imaging polarimetry over the range of phase angles 91°−105° obtained during the 2003 apparition in the visible (Jockers et al. 2005) showed that polarization at phase angles ~90° exceeded 30% after the correction for molecular emissions captured by the filter. Such values are even higher than those typical of dust-rich comets. On the other hand, the degree of linear polarization derived from broad-band aperture measurements was only ~8%, which demonstrates again that standard UBVR filters capture enough molecular emissions to reduce significantly the polarization measured for dust-poor comets. These results were confirmed by Jewitt (2004) who concluded that the new polarization data do not fit the standard view of cometary polarization according to which the least dusty comets show the weakest polarization (Levasseur-Regourd et al. 1996).

***Comet C/1995 O1 (Hale–Bopp).*** According to our data (Kiselev and Velichko 1997) confirmed by other authors, the gas and dust generation efficiencies in comet Hale–Bopp were one and two orders of magnitude greater, respectively, than those in comet Halley, which testifies to an extremely strong activity of the former. Hale–Bopp is the most dust-rich comet currently known. According to IR data (Williams et al. 1997; Mason et al. 2001), the atmosphere of this comet was dominated by dust particles with radii less than 0.4 μm, which appears to be consistent with the strong observed linear polarization and the complex morphology of the coma. The inner coma was optically thick (Fernández et al. 1999), which agrees with the results of our measurements (Rosenbush et al. 1997b; Розенбуш и др. 1999) indicative of optical thicknesses ~0.3 at nucleocentric distances $\sim(1-2)\times10^5$ km.

According to data of aperture and imaging polarimetry (Jockers et al. 1997; Kiselev and Velichko 1997; Rosenbush et al. 1997b), in March–April 1997 the degree of positive linear polarization for Hale–Bopp was ~4% greater than that for any other comet at similar phase angles. Furthermore, the polarization parameters varied over the coma: the degree of polarization observed for the shells (jets) was 1%−2% greater and the position angle of the polarization plane was 1°−1.5° smaller than those for areas between the shells. The polarization color (the ratio of the degree of polarization in the red filter to that in the blue filter) of the shells also systematically exceeded that of the surrounding areas. The spectral gradient of polarization for comet Hale–Bopp at phase angles ~45° was larger ($\Delta P/\Delta\lambda$ = 1.13 %/1000 Å) than the mean value typical of dust-rich comets (0.80 %/1000 Å).

The increase in the degree and the wavelength gradient of polarization in shells could be explained by the excess of particles with radii smaller than 0.1 μm (Jockers et al. 1997). However, such particles are highly susceptible to radiation force and would be swept into the tail before they could reach the observed locations in the shells. In fact, according to Lamy and Koutchmy (1979), the ratio of the radiation force to the gravity force for such particles is ~0.5, which corresponds to the radiation-pressure acceleration ~0.35 cm s$^{-2}$ at a heliocentric distance of 0.91 AU. As a consequence, this acceleration would create an additional tailward velocity of

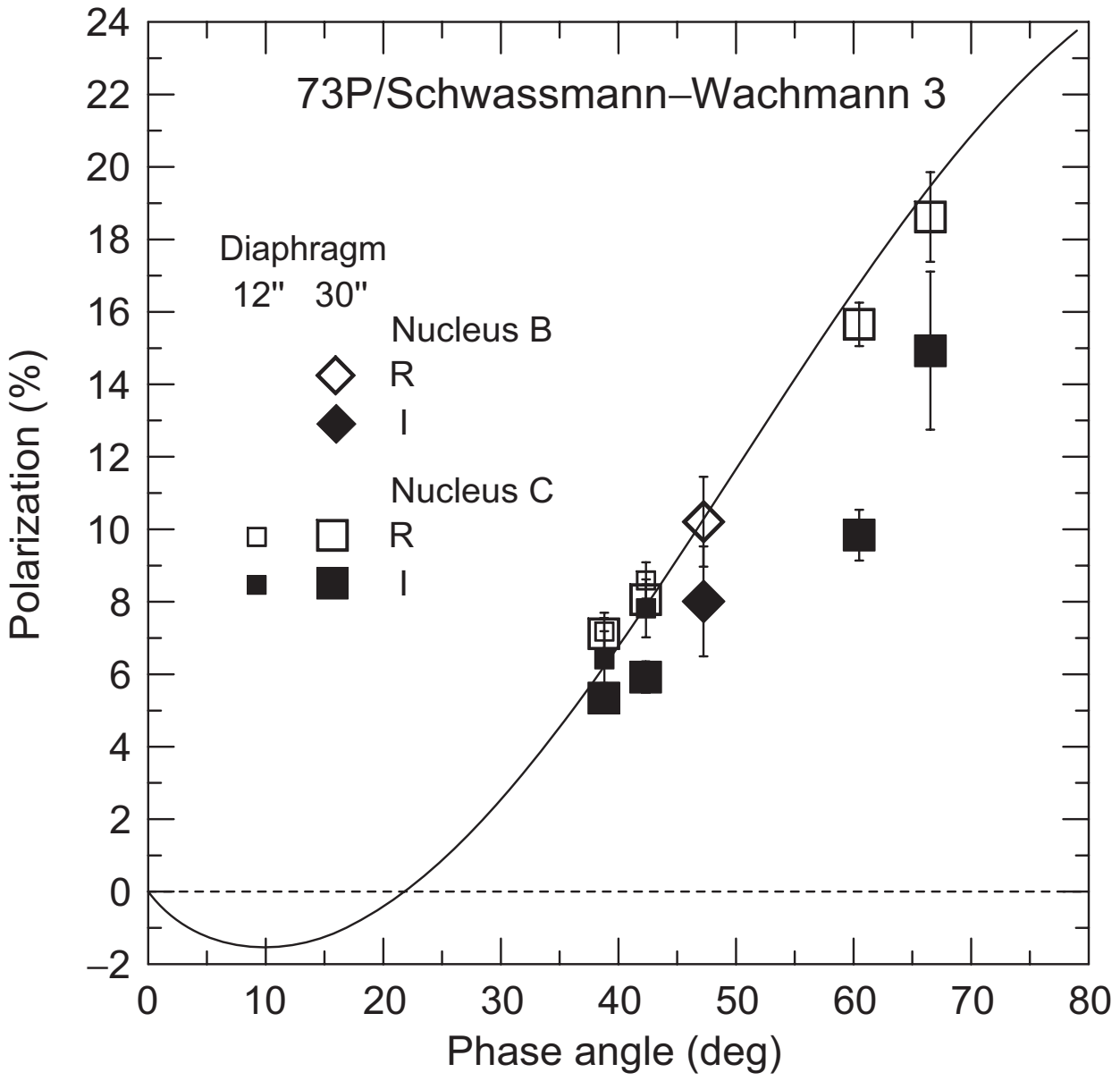

**Fig. 4.11.** Linear polarization measured for subnuclei B and C of comet Schwassmann–Wachmann 3 in filters I and R with different diaphragms. The curve depicts the average phase-angle dependence typical of dusty comets.

~150 m/s during just one rotation of the nucleus. As there were up to four nearly equidistant shells visible during this time, such a deceleration appears to be implausible. Therefore, it has been suggested that the observed polarization and its spectral gradient are caused by porous aggregates which are composed of monomers with radii smaller than 0.1 μm and are less susceptible to radiation pressure.

***Comet 73P/Schwassmann–Wachmann 3.*** This comet is interesting in many respects. In particular, it belongs to the group of "depleted" comets characterized by a low abundance of carbon (Fink and Hicks 1996), similar to comet Giacobini–Zinner, while its nucleus broke up into several subnuclei in 1995 (Lisse et al. 1998). During the 2006 apparition, this comet was an active and rapidly evolving complex consisting of several dozen fragments of subnuclei B, C and G (Plate 4.2). Therefore, the main objective of our polarimetric observations at large phase angles (Kiselev et al. 2008) was to facilitate a comparative study of different debris and to study the wavelength dependence of polarization.

Combined polarimetric (both aperture and CCD imaging) and spectral observations of the C, B, and G fragments were carried out in April–May 2006 as the phase angle increased from 38° to 67°. The results are summarized in Fig. 4.11. Polarization was averaged over ~4000–2000 km (the 30″ diaphragm) and ~1600–1500 km (the 12″ diaphragm) areas. One can see that for a range of phase angles, the degree of polarization measured with the I filter (solid symbols) is significantly smaller than that measured with the R filter (open symbols), i.e., both nuclei B and C exhibit the atypical spectral dependence of polarization. Furthermore, the observed degree of polarization increases systematically with size of the measured area of the coma. Polarization differences between the nuclei are within the measurement accuracy. In general, the results for the B and C fragments are consistent with the typical phase-angle dependence of polarization for dust-rich comets.

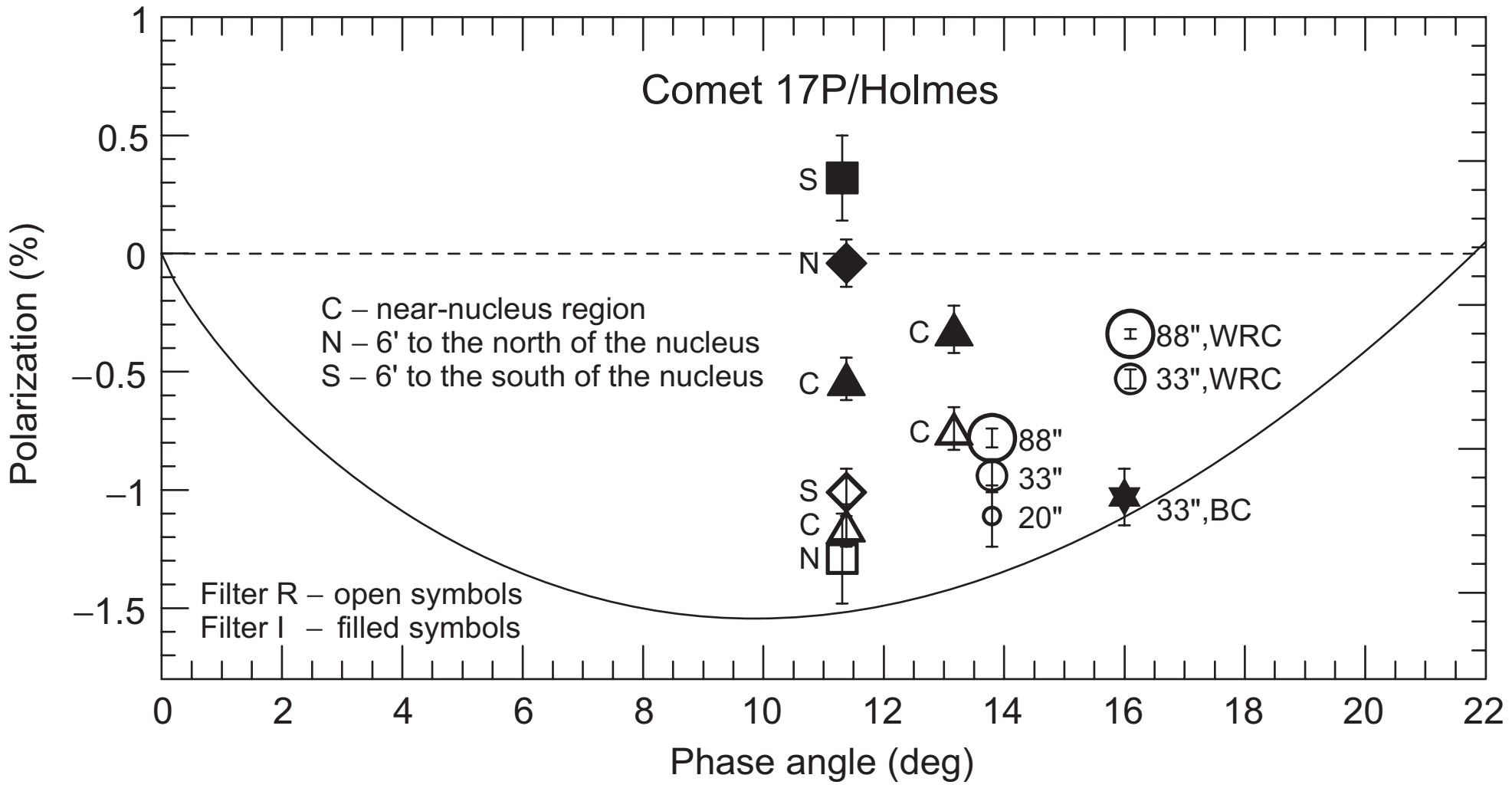

**Fig. 4.12.** Linear polarization measured for comet Holmes in different filters and with different diaphragms. The typical NPB for comets (solid curve) is the same as in Fig. 4.1.

***Comet 17P/Holmes.*** Despite a rather large distance from the Sun (2.8 AU), on 24 October 2007 this unique short-period comet suddenly increased its brightness almost million-fold, from $17^m$ to $2.8^m$, in just a few hours while doubling the brightness every 30 min. This outburst was truly unprecedented since it is quite unusual for a comet to become brighter as it is moving away from the Sun. To the best of our knowledge, there has been no other comet that would brighten dramatically at such a position in its orbit during such a short period of time and produce such an odd coma (see Plate 4.3). The results of our polarimetric observations during the outburst (Rosenbush et al. 2009b) are summarized in Fig. 4.12 and appear to be quite unique. Indeed, one can see that the degree of linear polarization depends on the location of the captured area in the coma; the absolute value of polarization is substantially smaller than that for typical comets; and the absolute value of polarization decreases with wavelength, i.e., the comet demonstrates the atypical spectral dependence of polarization.

As discussed in Section 4.3.1, the results of aperture polarimetry for all observed comets reveal remarkably similar NPBs with almost identical inversion angles as well as positions and values of minimal polarization (see Fig. 4.1). However, it is evident from Fig. 4.12 that irrespective of the measured area in the coma, the NPB for Holmes is much shallower than the average cometary NPB, with a $P_{min}$ of approximately −1.1% in the R filter and −0.5% in the I filter. According to the red-filter data (open symbols in Fig. 4.12), the expected inversion angle is also unusually small (~18°). Furthermore, even though the available data may be insufficient to draw a definitive conclusion, they appear to demonstrate that none of the previously observed comets has shown such a pronounced negative spectral gradient at small phase angles (Fig. 4.13) reaching about −0.77 %/1000Å near $\alpha_{\min}$.

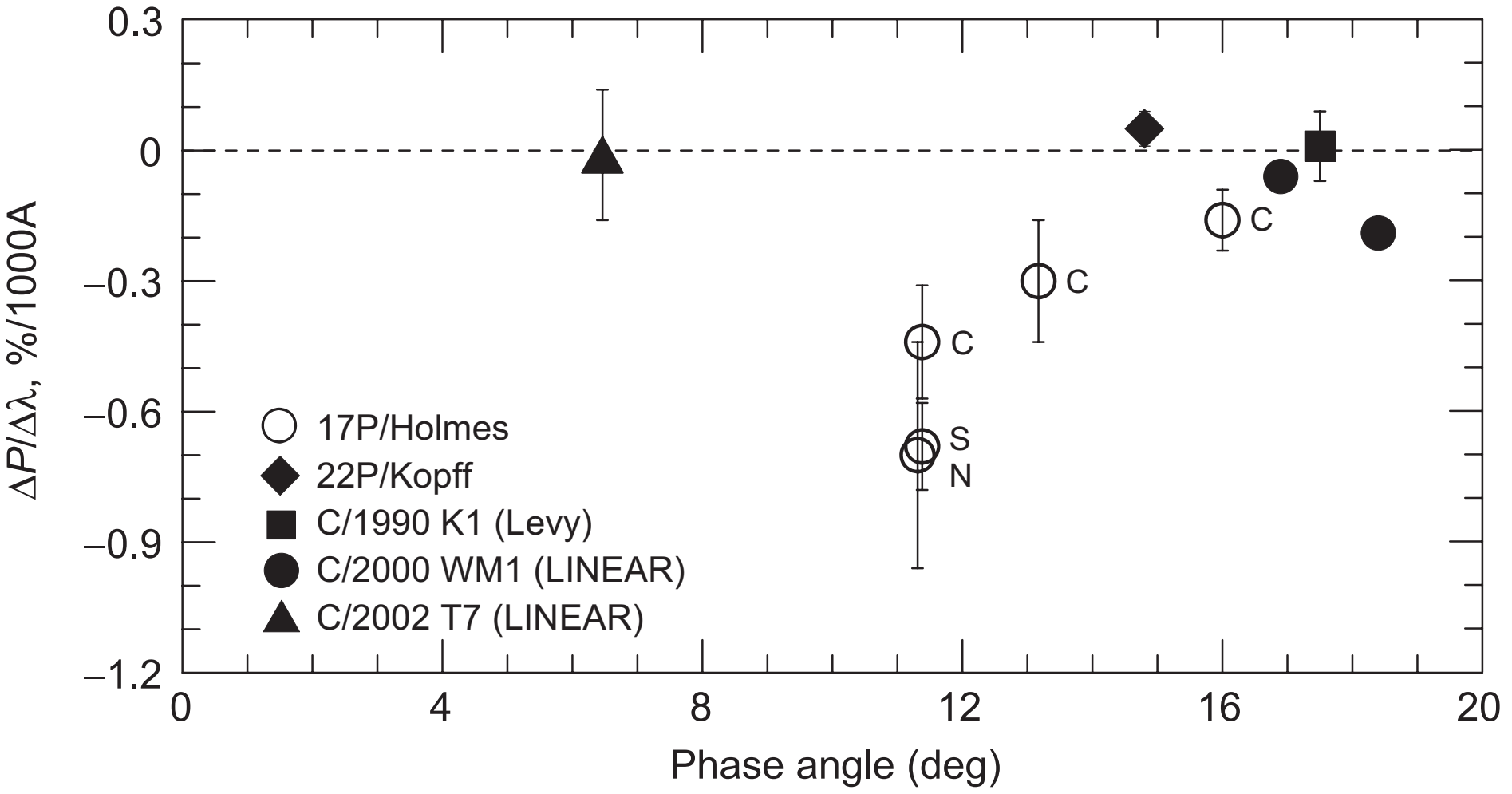

**Fig. 4.13.** Spectral gradient of polarization for comets Holmes, Kopff, Levy, and T7 (LINEAR).

There may be different reasons for the unusual polarimetric properties of comet Holmes (Rosenbush et al. 2009b). The depth of the NPB appears to strongly depend on the aggregate-particle size (the number of monomers) and the composition of the monomers (Petrova et al. 2004). In the case of comet Holmes, the dust ejected during the outburst could be dominated by very small particles, while their composition could be the primary factor that defined the spectral trend of polarization at small phase angles. An unusual mixture of compact silicate particles with aggregate particles composed of a mix of silicate and carbonaceous materials may provide another explanation of the peculiarities observed for comet Holmes. To conclude which explanation is more relevant, further modeling as well as more information about the dust from other types of observation, especially photometric and thermal, are necessary.

## 4.11. Database of Comet Polarimetry

The Database of Comet Polarimetry (DBCP) was initiated with the goal to archive all existing results of linear- and circular-polarization observations of comets and thereby facilitate comparative and individual analyses of polarization properties of comets, theoretical modeling of phase-angle and spectral dependences of polarization, taxonomy of comets, laboratory simulations, and selections of targets for future space missions. The DBCP is included in the NASA Planetary Data System (http://pdssbn.astro.umd.edu/holdings/ear-c-compil-5-db-comet-polarimetry-v1.0/dataset.html) and has already found extensive uses in planetary astrophysics.

The current version of the catalogue (Kiselev et al. 2006) includes the results of 2653 polarimetric observations of 64 comets performed between 1940 and 2005. The data for 48 comets have been contributed by these authors. The data cover the spectral range between 300 and 2200 nm, the phase-angle range between 0.4° and 122°, the range of heliocentric distances between 0.6 and 4.8 AU, and the range of geocentric distances between 0.03 and 4.9 AU.

The DBCP consists of two ASCII files. The first one tabulates the results of observations, while the second one provides references to relevant publications. The main table consists of 28 columns providing the following information:

- name of the comet;
- date of the observation (year, month, day) (UT);
- effective wavelength of the filter used (Å);
- filter bandpass (Å);
- name of the photometric band;
- diameter of the diaphragm (arc sec);
- projected linear diameter of the diaphragm at the comet (km);
- diaphragm location relative to the photocenter of the comet (arc sec or km);
- displacement of the captured area of the coma relative to the photocenter owing to the proper motion of the comet during a single exposure (arc sec or km);
- degree of linear polarization (%);
- estimated error in the degree of linear polarization (%);
- position angle of the polarization plane (deg);
- estimated error in the position angle (deg);
- degree of circular polarization (%);
- estimated error in the degree of circular polarization (%);
- phase angle (deg);
- position angle of the scattering plane (deg);
- deflection of the polarization plane from the direction perpendicular to the scattering plane (deg);
- heliocentric distance of the comet (AU);
- geocentric distance of the comet (AU); and
- reference(s).

The negative (positive) sign of linear polarization implies that the position angle of the average polarization plane is close to 0° (90°) relative to the scattering plane. Circular polarization is right-handed (positive) when the electric vector rotates predominantly in the clockwise sense as viewed by an observer looking in the direction of light propagation and is left-handed (negative) otherwise (Section 2.2).

Detailed descriptions of the observations, equipment used, data reduction procedures, and measurement errors can be found in the publications listed in the sec-ond file. In short, the polarimetric data included in the DBCP have been obtained using different instrumentation ranging from photographic plates to modern aperture and imaging polarimeters. Some results have been obtained with no filters or with broad-band filters. The estimated measurement accuracies for the degree of linear or circular polarization and the position angle of the polarization plane vary significantly depending on the brightness of the comet in question and the telescope–polarimeter combination used as well as on several other factors. This implies that the data quality is not uniform throughout the database and even for a single comet. However, the data for many comets are unique and cannot be obtained again.

As detailed in the preceding sections, we have used this database to analyze certain traits of light scattering by cometary dust particles (Kiselev et al. 2005) and to reveal similarities and differences in polarimetric properties of comets. We have shown that the division of comets into two classes based on polarimetric differences at large phase angles can be an artifact caused by weakly polarized molecular emissions captured by continuum filters and a low spatial resolution of aperture polarimeters. Linear polarization for several comets has been found to be a non-monotonous function of wavelength within a wide spectral range between 0.36 and 2.2 μm and exhibits different spectral trends at large and small phase angles. The detected circular polarization and its phase-angle dependence as well as systematic deviations of the Stokes parameter $U$ from zero and deflections of the polarization plane from the normal to the scattering plane indicate unequivocally the existence of aligned nonspherical dust particles in cometary atmospheres.

# Acronyms and abbreviations

## Acronyms

| | |
|---|---|
| ACS | Advanced Camera for Surveys |
| AERONET | Aerosol Robotic Network |
| AOT | aerosol optical thickness |
| APD | Asteroid Polarimetric Database |
| APS | Aerosol Polarimetry Sensor |
| ASSB | atmosphereless Solar System body |
| AVHRR | Advanced Very High Resolution Radiometer |
| BOE | brightness opposition effect |
| CB | coherent backscattering |
| CCD | charge coupled device |
| CrAO | Crimean Astrophysical Observatory |
| DBCP | Database of Comet Polarimetry |
| EBCM | Extended boundary condition method |
| ESO | European Southern Observatory |
| FLEs | Foldy–Lax equations |
| GACP | Global Aerosol Climatology Project |
| HB | International Hale–Bopp program |
| HRC | High-Resolution Camera |
| HST | Hubble Space Telescope |
| IFOV | instantaneous field of view |
| IHW | International Halley Watch program |
| IR | infrared |
| IRAS | Infrared Astronomical Satellite |
| ISCCP | International Satellite Cloud Climatology Project |
| KhNU | Kharkiv V. N. Karazin National University |
| LDP | longitudinal dependence of polarization |
| MAO | Main Astronomical Observatory |
| MDR | morphology-dependent resonance |
| MISR | Multiangle Imaging Spectroradiometer |
| MODIS | Moderate Resolution Imaging Spectroradiometer |
| NASA | National Aeronautics and Space Administration |
| NEA | near-Earth asteroid |
| NOAA | National Oceanic and Atmospheric Administration |
| NPB | negative polarization branch |
| OT | optical thickness |
| PDP | phase-angle dependence of polarization |
| PHA | potentially hazardous asteroid |
| POE | polarization opposition effect |
| POLDER | Polarization and Directionality of the Earth's Reflectance instrument |
| PPB | positive polarization branch |
| RSP | Research Scanning Polarimeter |
| RT | radiative transfer |
| RTE | radiative transfer equation |
| SAGE | Stratospheric Aerosol and Gas Experiment |
| SSA | single-scattering approximation |
| TES | Thermal Emission Spectrometer |
| TMM | *T*-matrix method |
| TNO | trans-Neptunian object |
| UV | ultraviolet |
| VIE | volume integral equation |
| VLT | Very Large Telescope |
| VRTE | vector radiative transfer equation |
| VSWF | vector spherical wave function |
| WL | weak localization |

**Comets**

| | |
|---|---|
| Halley | 1P/Halley |
| Encke | 2P/Encke |
| Faye | 4P/Faye |
| Tuttle | 8P/Tuttle |
| Tempel 1 | 9P/Tempel 1 |
| Brooks 2 | 16P/Brooks 2 |
| Holmes | 17P/Holmes |
| Giacobini–Zinner | 21P/Giacobini–Zinner |
| Kopff | 22P/Kopff |
| Brorsen–Metcalf | 23P/Brorsen–Metcalf |
| Crommelin | 27P/Crommelin |
| Schwassmann–Wachmann 1 | 29P/Schwassmann–Wachmann 1 |
| Schwassmann–Wachmann 2 | 31P/Schwassmann–Wachmann 2 |
| Honda–Mrkos–Pajdusakova | 45P/Honda–Mrkos–Pajdusakova |
| Ashbrook–Jackson | 47P/Ashbrook–Jackson |
| Arend–Rigaux | 49P/Arend–Rigaux |
| Churyumov–Gerasimenko | 67P/Churyumov–Gerasimenko |
| Schwassmann–Wachmann 3 | 73P/Schwassmann–Wachmann 3 |
| West–Kohoutek–Ikemura | 76P/West–Kohoutek–Ikemura |
| Chernykh | 101P/Chernykh |
| Swift–Tuttle | 109P/Swift–Tuttle |
| Ikeya–Zhang | 153P/Ikeya–Zhang |
| Hartley–IRAS | 161P/Hartley–IRAS |
| Honda | C/1968 N1 (Honda) |
| Tago–Sato–Kosaka | C/1969 T1 (Tago–Sato–Kosaka) |
| Bennett | C/1969 Y1 (Bennett) |
| Abe | C/1970 N1 (Abe) |
| Kohoutek | C/1973 E1 (Kohoutek) |
| Bradfield (1974) | C/1974 C1 (Bradfield) |
| Kobayashi–Berger–Milon | C/1975 N1 (Kobayashi–Berger–Milon) |
| West | C/1975 V1 (West) |
| Kohler | C/1977 R1 (Kohler) |
| Meier | C/1979 S1 (Meier) |
| Austin (1982) | C/1982 M1 (Austin) |
| IRAS–Araki–Alcock | C/1983 H1 (IRAS–Araki–Alcock) |
| Bradfield (1987) | C/1987 P1 (Bradfield) |
| Liller | C/1988 A1 (Liller) |
| Okazaki–Levy–Rudenko | C/1989 Q1 (Okazaki–Levy–Rudenko) |
| Austin (1989) | C/1989 X1 (Austin) |
| Levy | C/1990 K1 (Levy) |
| Shoemaker–Levy | C/1991 T2 (Shoemaker–Levy) |
| Mueller | C/1991 X2 (Mueller) |
| Tanaka–Machholz | C/1992 F1 (Tanaka–Machholz) |
| Hale–Bopp | C/1995 O1 (Hale–Bopp) |
| Hyakutake | C/1996 B2 (Hyakutake) |
| Tabur | D/1996 Q1 (Tabur) |
| U5 (LINEAR) | C/1998 U5 (LINEAR) |
| J3 (LINEAR) | C/1999 J3 (LINEAR) |
| S4 (LINEAR) | D/1999 S4 (LINEAR) |
| WM1 (LINEAR) | C/2000 WM1 (LINEAR) |
| Q4 (NEAT) | C/2001 Q4 (NEAT) |
| A2 (LINEAR) | C/2001 A2 (LINEAR) |
| T7 (LINEAR) | C/2002 T7 (LINEAR) |

**Spectral filters**

| | |
|---|---|
| U, B, V, R, I | Broadband filters with passbands from 3100 to 9000 Å approximating the standard Johnson–Cousins UBVRI system (Johnson 1966; Cousins 1976). |
| UC, BC, RC | International Halley Watch narrowband filters extracting cometary continuum and having central wavelengths/widths $\lambda/\Delta\lambda$ = 3650/80, 4845/65, and 6840/690 Å, respectively (Osborn et al. 1990). |
| GC | Hale–Bopp narrowband filter extracting cometary green continuum and having a central wavelength/width $\lambda/\Delta\lambda$ = 5260/56 Å (Farnham et al. 2000). |
| $R_x$ | Nonstandard broadband filter with a central wavelength/width $\lambda/\Delta\lambda$ = 6940/790 Å. |
| WRC | Nonstandard broadband filter with a central wavelength/width $\lambda/\Delta\lambda$ = 7228/1140 Å. |

# About the authors

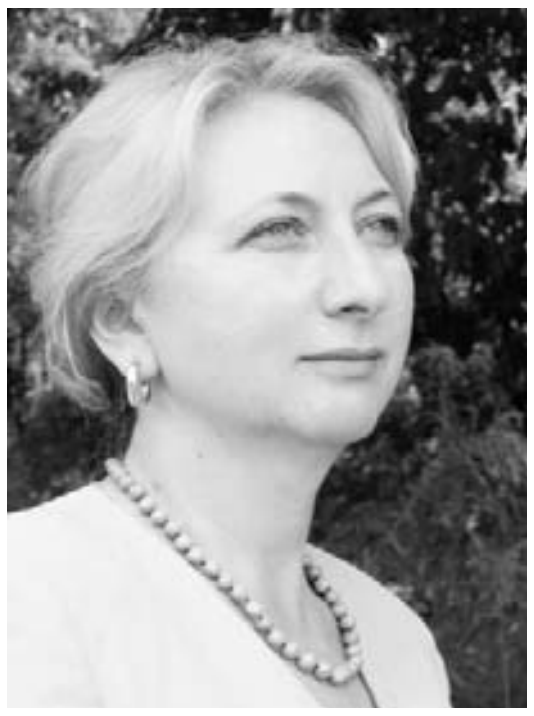

**Irina N. Belskaya** is a Leading Scientist at the Institute of Astronomy (IA) of the Kharkiv V. N. Karazin National University (KhNU). She graduated with honors from the KhNU in 1980 and has been a member of the IA staff thereafter. Her PhD thesis (1987) was on the study of physical properties of M-type asteroids by means of photometric and polarimetric observations. In 1992–93 she was a visiting scientist at the Uppsala Observatory. In 2002–04 she was a visiting scientist at the Paris Observatory and participated in an international program of studies of trans-Neptunian objects (TNOs). Dr. Belskaya gained a Habilitation Doctoral degree in 2007 for a dissertation on the optical properties of asteroids, Centaurs, and TNOs. She is a member of the International Astronomical Union (IAU), the European Astronomical Society, and COSPAR. As a member of the Task Group of the IAU Commission 15 "Physical studies of comets and minor planets", she has been actively involved in the revision and update of the photometric system of asteroids. Dr. Belskaya is a recipient of a number of awards and grants, including an award from the Ministry of Education and Science of Ukraine and a grant from the European Space Agency. To recognize her contributions to astronomy, asteroid 8786 was named "Belskaya". She is the author of more than 130 scientific publications, primarily dealing with physical properties of asteroids and TNOs.

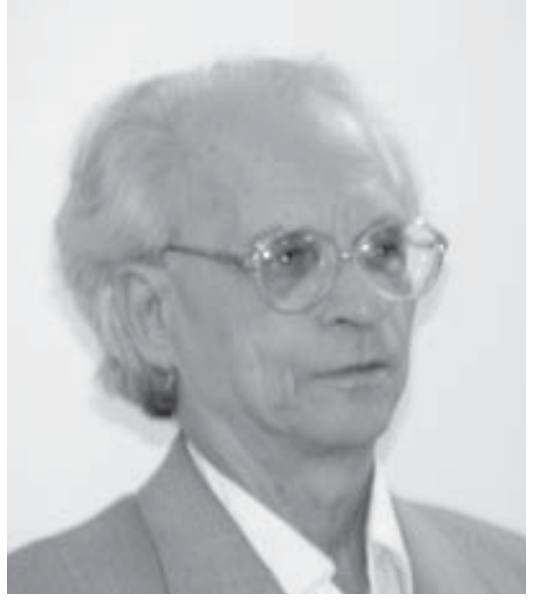

**Yuri S. Efimov** is a Leading Scientist at the Crimean Astrophysical Observatory (CrAO) in Nauchny, Crimea, Ukraine. After graduating from the Moscow State University in 1958, he gained his PhD degree in 1986. He has been with CrAO since 1958. Dr. Efimov is a member of the International Astronomical Union, the Euro–Asian Astronomical Association, and the Ukrainian Astronomical Association. His research interests have been the study of physical processes in various types of peculiar stars, circumstellar dust envelopes, interstellar polarimetry, and galaxies with active nuclei. Dr. Efimov is the (co)author of 160 peer-reviewed papers published in the leading Ukrainian, Russian, and International astronomical journals. His talks have been presented at major International conferences in Ukraine, Russia, Finland, Italy, Estonia, and Poland. To recognize Dr. Efimov's contributions to astronomy, asteroid 8781 was named "Yurka".

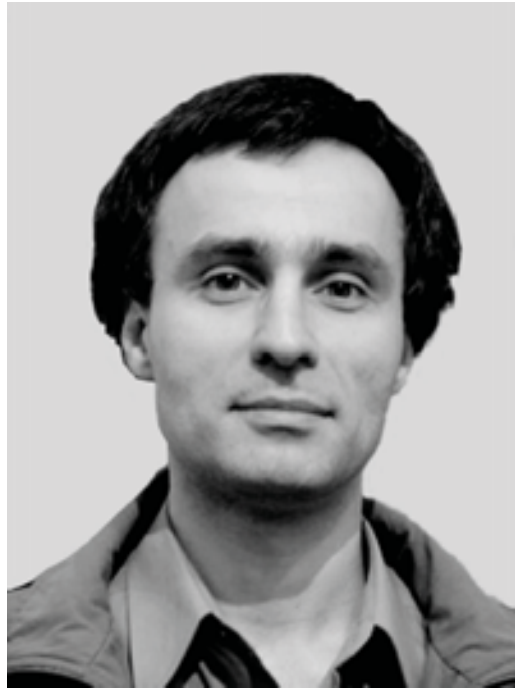

**Vadym G. Kaydash** is a Senior Research Scientist at the Institute of Astronomy of the Kharkiv V. N. Karazin National University. After graduating from the Kharkiv University in 1993, he gained his PhD degree in 1998 for a dissertation on the prognosis of the chemical composition and degree of maturity of the lunar surface from spectrophotometric data. Dr. Kaydash has been a co-author and investigator on several international research projects funded by INTAS and CRDF involving the analysis and interpretation of astronomical and spaceborne data on planetary surfaces. He has analyzed remote-sensing data obtained during lunar missions (Clementine, NASA–DoD; Lunar Prospector, NASA; SMART-1, European Space Agency) as well as lunar sample data produced by the Lunar Soil Characterization Consortium. He has also analyzed and interpreted the Hubble Space Telescope imaging polarimetry of Mars. Dr. Kaydash has 90 scientific publications on spectrophotopolarimery of planetary regoliths and atmospheres as well as on the evolution of planetary surfaces under the influence of different exogenous factors, such as impact processes and solar wind. He is the recipient of a Grant for young scientists from the Cabinet of Ministers of Ukraine and a Grant from the Ukrainian State Foundation for Fundamental Research. He was awarded the Prize of the Russian Academy of Sciences for the best science publication (1999). His research interests have included photopolarimetry of surfaces and atmospheres of Solar System objects and techniques to retrieve the information on the physical state, mineralogy, and chemical composition of planetary surfaces from remote-sensing data.

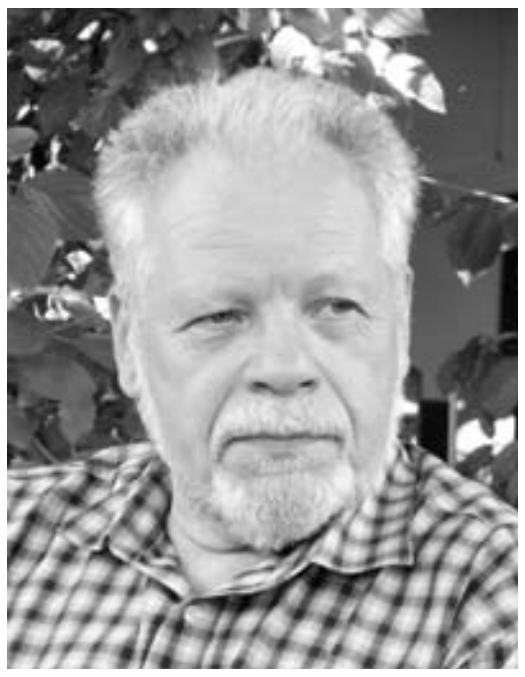

**Nikolai N. Kiselev** is a Principal Scientist at the Main Astronomical Observatory (MAO) of the Ukrainian National Academy of Sciences in Kyiv. After graduating from the St. Petersburg (formerly Leningrad) State University in 1966, he worked at the Institute of Astrophysics of the Academy of Sciences of Tadjikistan in Dushanbe. As Head of the Department of Variable Stars, he designed and built several polarimeters and led research in the field of photoelectric and imaging polarimetry and photometry of comets, asteroids, and stars. He was responsible for the building of the Sanglok Observatory (Tadjikistan) equipped with a 1-m reflecting telescope. He received his PhD degree in 1982 from the MAO for a thesis on polarimetry of comets and a Habilitation Doctoral degree in 2003 from the Kharkiv V. N. Karazin National University for a dissertation on "Light scattering by dust particles of comets, asteroids and circumstellar shells: observations and interpretation." He was the first to discover and investigate negative polarization and opposition optical effects for comets, the polarization opposition effect for E-type asteroids, and strong polarization in deep brightness minima for Herbig Ae/Be stars. Dr. Kiselev was a coordinator of the photometric and polarimetric network of the USSR component of the International Halley Watch Program. He is a recipient of the USSR "Plaque of Honor" (1986), the Medal of the Astronomical Council of the Academy of Sciences of the USSR (1984), and the Gold and two Bronze Medals of the Exhibition of Economic Achievements of the USSR (Moscow, 1987–1989). Dr. Kiselev is a member of the International Astronomical Union, the European Astronomical Society, the Astronomische Gesellschaft, and the Ukrainian Astronomical Association. To recognize his contributions to astronomy, asteroid 4208 was named “Kiselev”. Dr. Kiselev has authored/coauthored more than 200 peer-reviewed scientific publications.

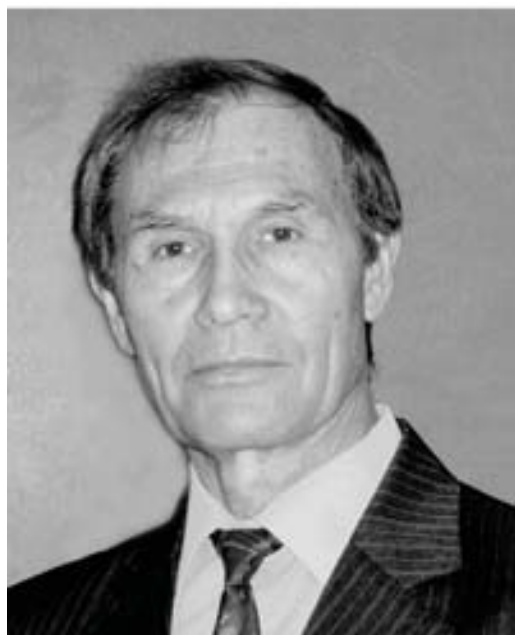

**Dmytro (Dmitrij) F. Lupishko** is a Leading Scientist at the Institute of Astronomy of the Kharkiv V. N. Karazin National University. After graduating from the Kharkiv University in 1969, he gained his PhD degree in 1975 for a dissertation on the photometry of Mars during the great opposition of 1971 and a Habilitation Doctoral degree in 1999 for a dissertation on the photometry and polarimetry of asteroids. In 2003 he was awarded the academic status of a Professor. He was the first astrophysicists of the former Soviet Union to begin systematic studies of the physical properties of asteroids. In 1984 he was invited to organize a standing workgroup “Asteroids” at the Astronomical Council of the USSR Academy of Sciences (Moscow) in order to coordinate asteroid research in the Soviet Union. He had served in this capacity until 1994, at which time the Astronomical Council was disbanded. In his home Institute, Prof. Lupishko created the Department of Asteroid and Cometary Physics, members of which have successfully defended seven PhD and three Habilitation Doctoral dissertations. Now the Kharkiv asteroid group is well known in the world and is among the most active in Europe. Prof. Lupishko has served as a member of Organizing Committees of many International conferences abroad and organized several International workshops and conferences in Kharkiv. He has authored/co-authored more than 200 scientific publications dealing with asteroids, Mars, the Moon, Mercury, and comets. He is a member of IAU and, for the second time, a member of the Organizing Committee of IAU Commission 15. He is also a member of the European Astronomical Society, a member of the Small Bodies Task Group of the Working Group for Planetary System Nomenclature, and a Vice-President of the Ukrainian Astronomical Association. Prof. Lupishko is a recipient of the N. P. Barabashov Award of the Ukrainian National Academy of Sciences, a Diploma of the Cabinet of Ministers of Ukraine, a Medal of the Ukrainian Ministry of Education and Science, and a Jubilee Medal awarded by the Mayor of Kharkiv. In 1987, asteroid 3210 was named “Lupishko”. Since January 2002 he has been an Honored Scientist of the Kharkiv National University. In May of 2009 Prof. Lupishko was awarded the title of an Honored Activist of Science and Technology of Ukraine.

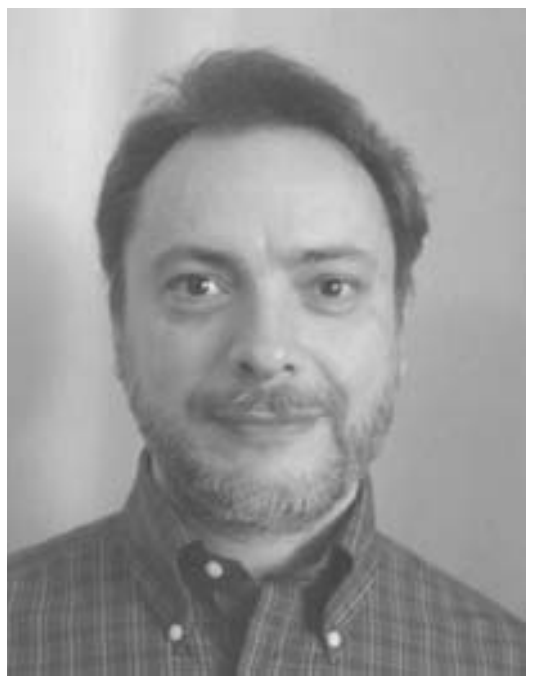

**Mykhailo (Michael) I. Mishchenko** is a Principal Scientist at the NASA Goddard Institute for Space Studies (GISS) in New York City. After graduating from the Moscow Institute of Physics and Technology in 1983, he gained his PhD and Habilitation Doctoral degrees from the Main Astronomical Observatory (MAO) of the National Academy of Sciences of Ukraine (NASU, Kyiv) in 1987 and 2008, respectively. He worked in MAO from 1987–92 and has been at GISS thereafter. Dr. Mishchenko has been Principal Investigator on several NASA, NOAA, and DoD research projects and has served as Topical Editor and Editorial Board member of such leading scientific journals as *Applied Optics*, *Kinematics and Physics of Celestial Bodies*, *Journal of the Atmospheric Sciences*, *Journal of Quantitative Spectroscopy and Radiative Transfer*, *Journal of Electromagnetic Waves and Applications*, and *Waves in Random and Complex Media*. Since 2006 he has been Editor-in-Chief of the *Journal of Quantitative Spectroscopy and Radiative Transfer*. Dr. Mishchenko was a Project Manager for the NASA/WCRP Global Aerosol Climatology Project (1998–2002) and has been the Project Scientist for the NASA Glory Satellite Mission since 2003. He has previously published four monographs (with Cambridge University Press, Academic Press, and Springer), 16 peer-reviewed book chapters, and 187 peer-reviewed journal papers. He has (co)organized 6 International conferences and served as Guest Editor for 7 special journal issues. Dr. Mishchenko is a recipient of the N. P. Barabashov Award of the NASU, the Henry G. Houghton Award of the American Meteorological Society, the Arthur S. Flemming Award of the George Washington University, the NASA Exceptional Scientific Achievement Medal, and 11 other NASA awards. In 2009, asteroid 22686 (1998 $QL_{53}$) was named "Mishchenko". Dr. Mishchenko is a Vice-President of the Ukrainian Astronomical Association, an elected member of the Council of the Ukrainian Academy of Arts and Sciences in New York, and an elected *Fellow* of the American Geophysical Union, the Optical Society of America, The Institute of Physics, and the American Meteorological Society. His research interests have included electromagnetic scattering, radiative transfer, and remote sensing.

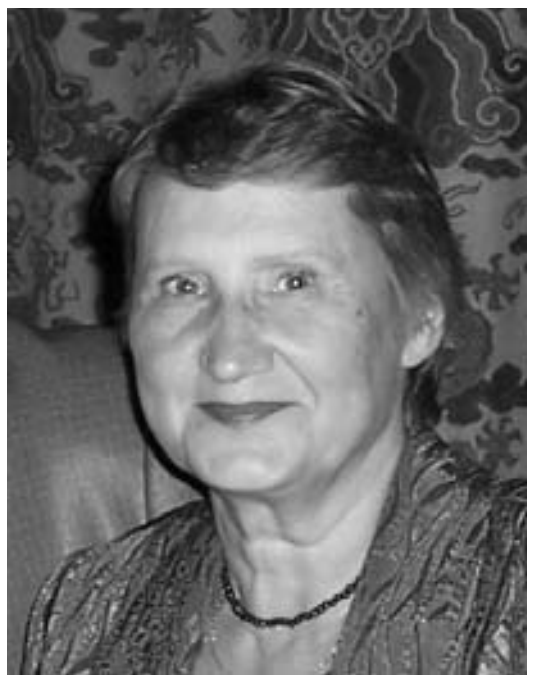

**Vera K. Rosenbush** is a Leading Scientist at the Main Astronomical Observatory (MAO) of the National Academy of Sciences of Ukraine in Kyiv. She graduated from the Kyiv Taras Shevchenko National University in 1971. Her tenure with MAO began in 1968, when she was a student, and has continued to this day. She defended a PhD thesis on spectral and photometric features of selected comets in 2004 and gained a Habilitation Doctoral degree for a dissertation on the properties of light scattered by small Solar System bodies in 2007. Her primary research interests have been polarimetry (both linear and circular) and photometry of selected celestial bodies, including comets, asteroids, and satellites, with emphases on unusual optical effects. She discovered and thoroughly investigated the po-larization opposition effect for high-albedo objects (such as the Jovian satellites Io, Europa, and Ganymede, the Saturnian satellite Iapetus, and E-type asteroids 64 Angelina and 44 Nysa) and thereby confirmed a remarkable prediction of the theory of coherent backscattering. She was the first to measure with high accuracy the circular polarization for comets, to determine its phase-angle dependence and spatial distribution across cometary atmospheres, and to establish the predominantly left-handed sense of polarization for all comets studied. Dr. Rosenbush has performed numerous unique polarimetric and photometric observations, including occultations of stars by comets, and has found a direct evidence of the alignment of dust particles in cometary atmospheres, an anomalous spectral dependence of linear polarization for comets, and transient variations of polarization parameters during non-steady-state processes in comets. She was also the first to separate phase-angle and longitude polarization curves for Callisto and Iapetus. She proposed a new classification of comets on the basis of polarimetric and thermal-emission properties of cometary dust. Dr. Rosenbush is a member of the Ukrainian Astronomical Association and the International Astronomical Union. Her contributions to astrophysics have been recognized by naming asteroid 18114 "Rosenbush". She is the author/coauthor of more than 160 scientific publications dealing primarily with physical properties of comets, asteroids, and planetary satellites.

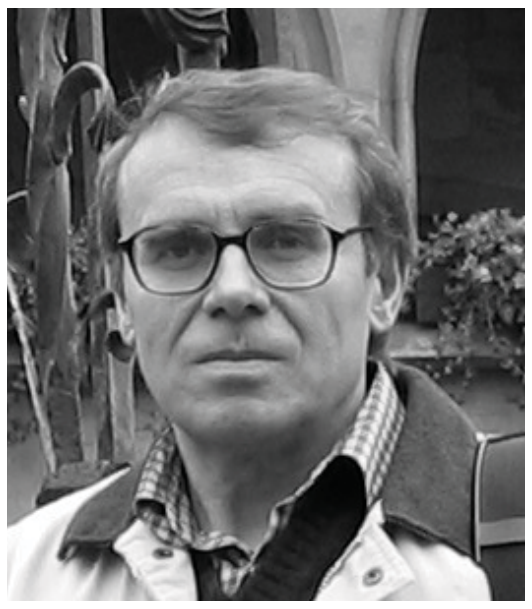

**Viktor P. Tishkovets** is a Senior Scientist at the Institute of Radio Astronomy (IRA) of the National Academy of Sciences of Ukraine (NASU) in Kharkiv. After graduating from the Kharkiv University in 1973, he gained his PhD degree in 1982 from the Main Astronomical Observatory of NASU and a Habilitation Doctoral degree in 2009 from the Kharkiv V. N. Karazin National University. He worked at the Institute of Astronomy of the Kharkiv University from 1975–2003 and has been at IRA thereafter. He is the author/coauthor of more than 70 scientific publications, including 3 peer-reviewed book chapters. Dr. Tishkovets has been a co-author and investigator on several research projects funded by INTAS and the Ukrainian Science & Technology Center. His research interests have included electromagnetic scattering and remote sensing. Dr. Tishkovets is a recipient of the Bronze Medal of the Exhibition of Economic Achievements of the USSR (Moscow, 1987).

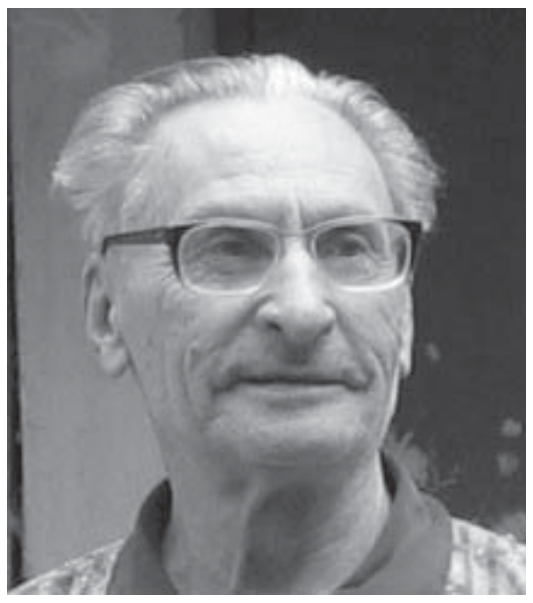

**Nikolay M. Shakhovskoy** is a Leading Scientist at the Crimean Astrophysical Observatory (CrAO) in Nauchny, Crimea. After graduating from the Moscow State University (MSU) in 1954, he worked at the Institute of Astrophysics of the Academy of Sciences of Tadjikistan in Dushanbe. In 1958–61 he was a post-graduate student at CrAO and, jointly with N. A. Dimov and A. F. Lagutin, designed and built an integrating stellar photoelectropolarimeter. He received a PhD degree from the MSU in 1965 for a thesis on the "Investigation of polarization of light of variable stars." He was the first to discover variable polarization for early-type eclipsing binary β Lyr stars. Dr. Shakhovskoy is a co-discoverer of many polarimetric effects for small Solar System bodies. His scientific career has been devoted to the development of theoretical foundations and methods of polarimetry and polarimetric instrumentation as well as to the study of various astronomica1 objects on the basis of extensive polarimetric and photometric observations. He authored a milestone chapter on "Methods of polarimetric observations" in the collective monograph on "Methods of investigation of variable stars" edited by Professor V. B. Nikonov (1972). He has published more than 240 peer-reviewed journal papers. For his pioneering work on the design of polarimeters, Dr. Shakhovskoy was awarded a USSR Author's Certificate and an Honorary Plaque "Inventor of the USSR". The international astronomical community has recognized his scientific achievements by giving his name to asteroid 4618 ("4618 Shakhovskoy").

# Index

НАЦІОНАЛЬНА АКАДЕМІЯ НАУК УКРАЇНИ
ГОЛОВНА АСТРОНОМІЧНА ОБСЕРВАТОРІЯ

НАЦІОНАЛЬНЕ УПРАВЛІННЯ АЕРОНАВТИКИ ТА ДОСЛІДЖЕННЯ КОСМІЧНОГО ПРОСТОРУ США
ҐОДДАРДІВСЬКИЙ ІНСТИТУТ КОСМІЧНИХ ДОСЛІДЖЕНЬ

МІЩЕНКО Михайло Іванович
РОЗЕНБУШ Віра Калениківна
КИСЕЛЬОВ Микола Миколайович
ЛУПІШКО Дмитро Федорович
ТИШКОВЕЦЬ Віктор Павлович
КАЙДАШ Вадим Григорович
БЕЛЬСЬКА Ірина Миколаївна
ЄФІМОВ Юрій Сергійович
ШАХОВСЬКИЙ Микола Михайлович

# ДИСТАНЦІЙНЕ ЗОНДУВАННЯ ОБ'ЄКТІВ СОНЯЧНОЇ СИСТЕМИ ПОЛЯРИМЕТРИЧНИМИ ЗАСОБАМИ

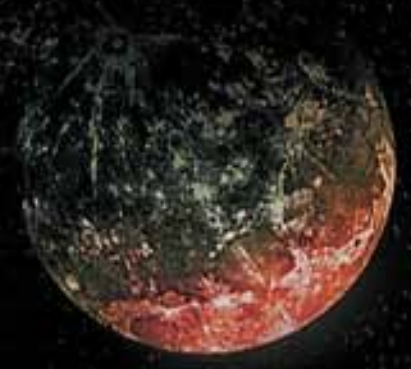

This book outlines the basic physical principles and practical methods of polarimetric remote sensing of Solar System objects and summarizes numerous advanced applications of polarimetry in geophysics and planetary astrophysics. The monograph is intended for science professionals, educators, and graduate students specializing in remote sensing, astrophysics, atmospheric physics, optics of disperse and disordered media, and optical particle characterization.

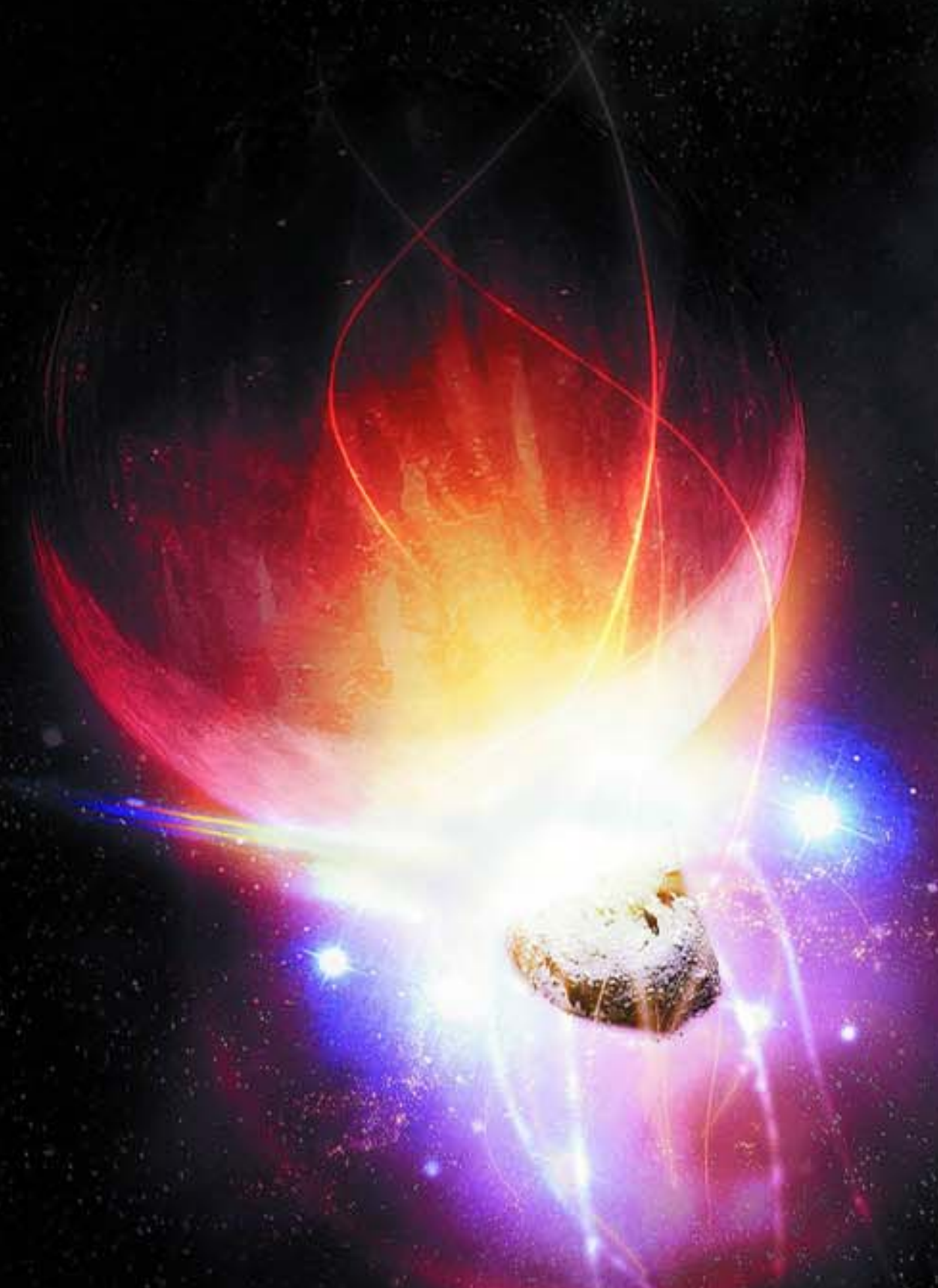